\documentclass[useAMS,usenatbib]{mn2e}

\pdfoutput=1 
\usepackage{times} 
\usepackage{graphicx}
\usepackage{rotating}
\usepackage{epsfig}
\usepackage{longtable}
\usepackage{lscape}

\newcommand{\HI}{H\,{\small I}}
\newcommand{\Msol}{$M_{\odot}$}
\newcommand{\be}{\begin{equation}}
\newcommand{\ee}{\end{equation}}
\newcommand{\kms}{\mbox{km s$^{-1}$}}
\newcommand{\sint}{S$_{\mathrm{int}}$}
\newcommand{\speak}{S$_{\mathrm{peak}}$}

\title[A blind \HI\ survey in the Canes Venatici region]{A blind \HI\ survey in the Canes Venatici region}
\author[K. Kova\v{c} et al.]
       {K.~Kova\v{c}$^{1,2}$\thanks{E-mail: kovac@phys.ethz.ch (KK)},
        T. A.~Oosterloo$^{3,1}$ and
       J. M.~van der Hulst$^{1}$ 
\vspace*{.25em} 
\\
$^1$Kapteyn Astronomical Institute, University of Groningen, Postbus 800, 9700 AV Groningen, The Netherlands\\
$^2$Institute of Astronomy, ETH Zurich, 8093 Zurich, Switzerland \\
$^3$Netherlands Foundation for Research in Astronomy, Postbus 2, 7990 AA Dwingeloo, The Netherlands} 

\pagerange{\pageref{firstpage}--\pageref{lastpage}}
\pubyear{XXXX}

\begin{document}

\maketitle

\label{firstpage}

\begin{abstract}
We have carried out a blind \HI\ survey using the Westerbork Synthesis
Radio Telescope to make an inventory of objects with small \HI\ masses
(between 10$^{6}$ and 10$^8$ \Msol)  and to constrain the low-mass end
of  the \HI\  mass function. The survey  has been  conducted  in a part of the
volume containing the  nearby Canes Venatici groups of  galaxies. The surveyed region covers an
area on  the sky of  about 86 square  degrees and a range  in velocity
from about -450  to about 1330 \kms. We find 70  sources in the survey
by applying  an automated searching algorithm. Two of the detections
have  not been  catalogued previously,  but  they can  be assigned  an
optical  counterpart,  based  on   visual  inspection  of  the  second
generation Digital Sky Survey images.  Only one of the \HI\ detections
is  without an  optical counterpart.  This object  is detected  in the
vicinity of NGC4822 and it  has been already detected in previous \HI\
studies.  Nineteen  of the  objects have been  detected for  the first
time in the  21-cm emission line in this  survey.  The distribution of the \HI\ properties of our detections confirms our ability to find low mass objects. 86\%  of the detections have profile  widths less than 130
\kms\ and can be considered dwarf  galaxy candidates.  The \HI\ fluxes
measured imply  that this survey goes  about 10 times  deeper than any
previous blind \HI\  survey. The \HI\ mass function and the optical 
properties of the detected sources will be discussed in future papers.
\end{abstract}


\section{Introduction}

In the currently favoured cosmological  models, based on the cold dark
matter (CDM)  paradigm, structure evolves from  the small, primordial,
Gaussian  fluctuations  by  gravitational  instability.   Dark  matter
haloes  grow in  a hierarchical  manner through  multiple  merging and
accretion  of smaller  systems  \citep[e.g.][]{White&Rees.1978}. In  the
framework depicted, galaxies form by cooling  of baryons captured inside of the
dark matter haloes.

Using the Local  Group as a cosmological probe,  a discrepancy between
the theory and the optical observations arises, the so called ``missing
satellites problem'': the number  of observed satellites known in the optical
is  an order of  magnitude smaller  than the  number of  small systems
predicted by  CDM models \citep{Kauffmann.etal.1993, Klypin.etal.1999,  
Moore.etal.1999}. Closely  related to this problem is a discrepancy
between the  slopes derived at the  faint end of  the observed optical
luminosity  functions \citep[e.g.][]{Blanton.etal.2003}  and the  \HI\ mass
functions \citep[e.g.][]{Zwaan.etal.2005} on one hand, and  the slope of the
halo mass  functions  calculated  from large  N-body  CDM  simulations
\citep[e.g.][]{Jenkins.etal.2001} or from an analytical framework such as
the Press-Schechter  formalism \citep{Press&Schechter.1974},  on the other
hand. The  faint-end slopes  from the observed distributions  are much
flatter when compared to the corresponding slopes of the theoretically
constructed functions.

The absence  of a more  numerous low-mass galaxy population  cannot be
straightforwardly   understood  and   points   out  a   lack  of   our
understanding  of those  systems. The  models proposed  to  solve this
discrepancy can be separated into  two categories. One type of models is
based on  the suppression  of the formation  of small haloes or on their
destruction, which can  be achieved by modifying the  properties of the
dark  matter.  These  type of  models include  allowing a  finite dark
matter   particles  self--interaction   cross--sections  
\citep{Spergel&Steinhardt.2000}, reducing the small--scale  power 
\citep[e.g.][]{Avila-Reese.etal.2001, Eke.etal.2001, Bode.etal.2001} 
or changing the shape of
the primordial power-spectrum \citep{Kamionkowski&Liddle.2000,
Zentner&Bullock.2002}. \citet{Kravtsov.etal.2004} suggest that low mass galaxies form in 
high mass halos that are tidally stripped to form low mass halos. The second set  of models proposes to suppress the star
formation  in  low-mass haloes.  Several  plausible baryonic  physical
processes may cause the depletion of the gas from the haloes and small
haloes  will remain  dark.  Such  processes may  be quenching  of star
formation \citep[e.g.][]{Gnedin&Kravtsov.2006}, photo-evaporation 
(see e.g. \citealt{Quinn.etal.1996} and  \citealt{Barkana&Loeb.1999}) 
and/or feedback from
supernovae  or galactic  winds (see  e.g. \citealt{Larson.1974}  and 
\citealt{Efstathiou.2000}).  Another possibility to suppress star formation 
in small haloes
is based on stability criteria.  During  the process of galaxy
formation the angular momentum of gas which settles into the disk will
be conserved and  a number of haloes with small  masses may never form
stars,  or form  them in  small numbers.   According to  one  of those
models,  all galaxies  with dark  matter halo  masses  below 10$^{10}$
\Msol\  will never  form  stars  \citep[e.g.][]{Verde.etal.2002}.   Still,
baryons  will remain  inside  these small  haloes.  Recently, 
\citet{Read.etal.2006}  argued that  there is a  sharp transition of  
the baryonic
content in the smallest haloes. Over the halo mass range  3 - 10 $\times$
10$^7$  \Msol\ at  z $\sim$  10 the  amount of  stellar mass  drops two
orders of magnitude  in these systems. Haloes below  the limiting mass
of $\sim$  2 $\times$ 10$^7$ \Msol\  will be almost  devoid of  gas and
  stars. Extrapolating these  results to  the present
redshift (using a combination of  arguments based on linear theory
and the  various literature results),  \citet{Read.etal.2006}  predict the
existence of many galaxies with  surface brightness about an order or
two orders of magnitude fainter than galaxies already detected.

Optical surveys are generally less sensitive to low luminosity and low
surface brightness (LSB) galaxies, which could be under-represented in
such surveys  \citep{Disney.1976, Disney&Phillipps.1987}. LSB galaxies
are found to be rich in neutral gas (Schombert et al. 1992, de Blok et
al. 1996). Given  that galaxies with a small  amount of stars compared
to  their \HI\  mass are  typically discovered  via \HI\  surveys, the
blind \HI\ surveys provide an  excellent probe to detect galaxies with
a  small  amount  of stars.   One  may  detect  even a  population  of
completely dark galaxies  using an \HI\ survey -  under the assumption
that dark galaxies contain \HI. If  one assumes that \HI\ makes up
a  few percent  of the  total mass  of a  galaxy, dark  galaxies would
contain \HI\ in the range  10$^7$ - 10$^8$ $M_{\odot}$ or less. Still,
one has to be  aware that a blind \HI\ survey will  miss that part of
the population    of     (dwarf)     galaxies     without    \HI\
 (e.g. \citealt{Geha.etal.2006}). Therefore, selection effects of  the \HI\ surveys
provide limits on the plausible galaxy formation scenarios which a blind \HI\  survey can  test.

To complete the story, recent searches for the missing satellites in the local Universe have been conducted by identifying the galaxies in the overdensities with respect to the Galactic foreground in the resolved stellar populations in the nearby optical and infrared surveys. In the photometric data of the Sloan
Digital Sky Survey  \citep[SDSS,][]{York.etal.2000, Abazajian.etal.2008}, 14 new companions of the Galaxy have been discovered, in which 9 objects are for sure identified as dwarf spheroidal galaxies \citep[e.g.][]{Willman.etal.2005, Zucker.etal.2006, Belokurov.etal.2008}. The galaxies discovered are among the lowest mass and lowest brightest galaxies being known. However, these galaxies have probably been very strongly affected by interactions with the Galaxy (tidal and
gaseous, e.g. \citealt{Grcevich&Putman.2009}) and likely remnants of larger galaxies. Their evolution is very complex and
highly uncertain. They do not tell the whole story, certainly not about small
galaxies in less dense environments.

\subsection{Blind \HI\ surveys}

In the last three decades several blind \HI\ surveys have been carried
out. The first blind survey in the 21-cm emission line was carried out
by \citet{Shostak.1977}  in driftscan mode, leading to 1, not clearly extragalactic, detection. \citet{Lo&Sargent.1979} surveyed three nearby groups of galaxies (including CVnI) with the Owens Valley Radio Observatory 40m telescope, without any discrete \HI\ detection. The higher sensitivity observations of the selected areas of the same groups with the Bonn 100 m telescope resulted in the detection of 6 \HI\ sources, from which 4 were previously uncatalogued dwarf galaxies \citep{Lo&Sargent.1979}. \citet{Krumm&Brosch.1984}
surveyed about  7\% of the Perseus-Pisces  void and about  19\% of the
Hercules        void, with no \HI\ detections.          After        that,        \citet[][also
\citealt{Henning.1992}]{Kerr&Henning.1987}  conducted   a  blind  \HI\
survey  by  observing a  series  of  pointings  on lines  of  constant
declination. The number of detected  objects was 37. Since
then, blind  \HI\ surveys have yielded  sufficient number of  detections to
describe the results in a  statistical manner. The main parameters of
the     major    blind    \HI\     surveys    are     summarised    in
Table~\ref{hisumtab}, adopted from http://egg.astro.cornell.edu/alfalfa/science.php.

\begin{table*}
\centering
\label{hisumtab}
\begin{tabular}{l l l l l l l l l}
\hline \hline
Survey & Area & Beam size & Velocity range & $Velocity resolution^{\mathrm{a}}$ & Detections & ${({\mathrm{min}} M_{\rm HI})}^{\mathrm{b}}$
      & Telescope & Ref \\
 & (deg$^2$) & (arcmin) & (\kms\ ) & (\kms\ ) &  (number)  & (10$^6$$\times$\Msol) & &  \\ \hline
AHISS & 65 & 3.3 & -700 -- 7400 & 16 & 65 & 1.9  & 305m Arecibo & 1  \\
Nan\c{c}ay CVn & 800 & 4 $\times$ 20 & -350 -- 2350 & 10 & 33 & 20 & Nan\c{c}ay & 2  \\
ADBS  & 430 & 3.3 & -650 -- 7980 & 34 & 265 & 9.9  & 305m Arecibo & 3  \\
HIJASS & 1115 & 12 & -1000 -- 4500 & 18 & 222 & 36 & 76m Jodrell Bank  & 4 \\
& & & 7500 -- 10000 & & & & & \\
WSRT WFS & 1800 & 49 & -1000 -- 6500 & 17 & 155 & 49 & WSRT & 5  \\
HIPASS & 21346 & 15.5 & 300 -- 12700 & 18 & 4315 & 36 & 64m Parkes & 6  \\ 
HIPASS Northern & 7997 & 15.5 & 300 -- 12700 & 18 & 1002 & x & 64m Parkes & 7  \\
extension & & &  & & & & & \\
ALFALFA & 7000 & 3.5 & -2000 -- 18000 & 11 & ($>$25000) & 4.4 & 305m Arecibo & 8 \\
& & &  & & & & (in progress) & \\
\hline
\end{tabular}

\begin{list}{}{}
\item[$^{\mathrm{a}}$] The given velocity resolution is after Hanning smoothing.
\item[$^{\mathrm{b}}$] Minimum detectable masses min M$_{\rm HI}$ are calculated at 10 Mpc, for 5$\sigma$ detections with velocity width 30 \kms.  
\end{list}

\caption{Parameters of major blind \HI\ surveys.
The references cited are as follows:
     $1$:\citet{Sorar.1994}; \citet{Zwaan.etal.1997},
     $2$:\citet{Kraan-Korteweg.etal.1999},
     $3$:\citet{Rosenberg&Schneider.2000},
     $4$:\citet{Lang.etal.2003}, 
     $5$:\citet{Braun.etal.2003}, 
     $6$:\citet{Meyer.etal.2004},
     $7$:\citet{Wong.etal.2006} - the authors do not give the $rms$ estimate, they claim that noise has increased by 31$\%$ in the Northern Extension of the HIPASS survey, particularly in the northernmost part (see their Figure 5),
     $8$:\citet{Giovanelli.etal.2005}.
     Most of the numbers in the table are based on the Arecibo Legacy Fast ALFA Survey (ALFALFA) webpage (http://egg.astro.cornell.edu/alfalfa/science.php). The surveys are ordered by the year of publishing results. ALFALFA is the last entered survey, as it is still ongoing.}
\end{table*}

The main conclusion  which can be drawn from  the (blind) \HI\ surveys
carried  out up  to  date is  that  there is  no essential  difference
between  the populations  of objects  detected in  \HI\  emission line
surveys  and the  population of  galaxies  detected at  optical or  at
infrared wavelengths, except that  \HI\ detected galaxies are more gas
rich   and   preferentially    of   the   late   morphological   types
\citep{Zwaan.etal.2005}.     A    new    population    of    isolated,
self-gravitating \HI\ clouds or  dark galaxies has not been revealed,
neither a  large population of  galaxies with low optical surface
brightness, which  would have gone  undetected in   optical surveys
\citep[e.g.][]{Zwaan.etal.2005}.  The distribution of  \HI\ selected
objects  follows  the  large-scale  structures  defined  by  optically
selected  galaxies  \citep{Koribalski.etal.2004, Zwaan.etal.2005proc},
but  these   objects  tend  to  populate  regions   of  lower  density
\citep{Ryan-Weber.2006, Basilakos.etal.2007}. However,  to be able 
to get  a more definitive answer to
the question  whether an additional number  of gas-rich low-luminosity
and LSB galaxies and/or a population of gas-rich dark galaxies, missed
in the  optical surveys  exists, it is  necessary for \HI\  surveys to
reach lower \HI\ mass limits.

Even though the minimum \HI\ masses which can be detected in the blind
\HI\   surveys   are  a   few   times   $\sim$   10$^6$  \Msol\   (see
Table~\ref{hisumtab}),  only a small number of galaxies  have been  detected with
such small  \HI\ masses, particularly   beyond the Local Group. Such detections include ESO 384-016 with the \HI\ mass $6 \times 10^6$ \Msol\ \citep{Beaulieu.etal.2006}, 4 galaxies in Sculptor with the \HI\ mass 2-9 $\times 10^5$  \citep{Bouchard.etal.2005}, and 4 galaxies in Centaurus with the \HI\ masses below $10^7$ \Msol\ \citep{Minchin.etal.2003}. All of these detections have 
optical counterparts. On
the other hand,  there is a population of  high-velocity clouds \citep[HVCs, e.g.][]{Wakker&vanWoerden.1991,  Braun&Burton.2000, deHeij.etal.2002} 
discovered  only in the 21-cm line (no optical counterparts
have yet been found).  These objects are distributed all over the sky,
either  as extended  complexes or  in  the form  of compact,  isolated
clouds (CHVCs).   The nature of the  (C)HVCs is a matter  of  debate,
despite nearly  four decades  of study.  The  main reason for  this is
the difficulty in estimating  distances to  the (C)HVCs. However, there are
a number of HVCs with well-constrained distance brackets via observations of
absorption lines towards stars in the Galactic halo [such as Complex A \citep{Wakker.etal.1996}, Complex C \citep{Wakker.etal.2007, Thom.etal.2008}, the Cohen Stream \citep{Wakker.etal.2008}, Complex GCP \citep[or Smith Cloud,][]{Wakker.etal.2008} and Complex WB \citep{Thom.etal.2006}]. The measurements place 
these complexes within about 10 kpc of the Sun, putting some constraints on the nature of the
HVC complexes. There is little doubt  that the Magellanic Stream, a $100^0 \times 10^0$ filament of gas extending within the Galactic halo, is produced
by interactions between the  Milky Way and companions, as a result
of either tidal  disruption or ram pressure stripping, or both 
\citep[e.g.][]{Putman.etal.1998, Putman.etal.2003, Putman&Gibson.1999}. 
Some of the extended HVCs can be explained as
the products  of Galactic fountains  \citep{Shapiro&Field.1976,
Bregman.1980, Bregman.1996}.  \citet{Blitz.etal.1999} proposed a dynamical  
model in which the
HVCs can be  explained as the gaseous counterparts  of the primordial
low-mass   haloes  predicted   by  $\Lambda$CDM   structure  formation
scenarios. This appeared  as a  very  attractive way  to resolve  the
discrepancy  on the number  of low-mass  systems discussed  above. The origin  of  CHVCs is  more  uncertain.  
One of  the
hypotheses that has received recent attention is that the CHVCs are of
primordial origin, residing  at typical distances of up  to 1 Mpc from
the  Milky Way  \citep{Oort.1966, Oort.1970,  Verschuur.1969,  
Kerr&Sullivan.1969}. The
recent observations \citep{Zwaan.2000, Pisano.etal.2004, Westmeier.etal.2005, Pisano.etal.2007}
 and simulations \citep{Sternberg.etal.2002, Kravtsov.etal.2004}
do not confirm the existence  of a circumgalactic population of 
CHVCs. The results by  \citet{Westmeier.etal.2005} suggest an upper
limit  of about  60 kpc  for  the distance  of CHVCs  from their  host
galaxies. This distance  would lead to a limiting  \HI\ mass for CHVCs
of 6 $\times$ 10$^4$ \Msol. Similarly, \citet{Pisano.etal.2007} infer a maximum distance of 90 kpc for the CHVCs, with average \HI\ mass smaller than or equal to $4 \times 10^5$ \Msol. So far, there is no observational evidence for a population of
\HI\ clouds more massive than $10^7$ \Msol\ that are not directly associated with a galaxy \citep{Sancisi.etal.2008}.

Recently,  there was  a lot  of debate  on the  nature of  a  few \HI\
detections without  an obvious optical counterpart -  whether they are
dark    galaxies    or   not    (e.g.     VIRGOHI 21   reported    by
\citealt{Davies.etal.2004}          and          confirmed          by
\citealt{Minchin.etal.2005} and  the   HVC  Complex  H  discussed  by
\citealt{Lockman.2003}         and         \citealt{Simon.etal.2006}).
\citet{Kent.etal.2007} report the discovery of the eight \HI\ features
lacking a  stellar counterpart (four of them  already known, including
the  VIRGOHI 21 object) detected  as  a  part  of the  ALFALFA  survey
\citep{Giovanelli.etal.2005}.  All of  these eight  \HI\  features are
withing the region of  Virgo cluster and if at the Virgo distance their \HI\ masses span a range between 1.9 $\times$ 10$^7$ and 1.1 $\times$ 10$^9$ \Msol \citep{Kent.etal.2007}. The
HVC Complex  H is at the distance $d$ of $27 \pm 9$ kpc from the Sun \citep{Lockman.2003} and with the \HI\ mass of $\sim 2.7 \times 10^4 d^2$ \Msol\ \citep{Wakker.etal.1998}. So  far, there is  no confirmation that  these detections are
gravitationally  bound objects,  residing within  a dark  matter halo.
Based on  the deeper ALFALFA data \citep{Haynes.etal.2007}  as well as
on modeling  of the WSRT data \citep{Duc&Bournaud.2008},  VIRGOHI 21 has
been reported to be a tidal feature of NGC 4254 with the \HI\ mass of $3 \times 10^7$ \Msol\ \citep{Minchin.etal.2007}.

The existing \HI\  surveys are incomplete in the  range of \HI\ masses ($10^7-10^8$ \Msol\ and below) which would correspond to the  majority of galaxies predicted to exist
with little or no stars. The few detected objects known in this mass range are all
   associated with nearby galaxies detected in the optical and do not
   represent the predicted class of small galaxies with gas but no 
    stars. To be able to address the question of whether such
   objects exist and in which numbers - a deeper blind \HI\ survey is needed, in which galaxies with \HI\ masses 
below 10$^8$ \Msol\ are a significant fraction of all detections.

We  carried out  a  new blind  \HI\  survey designed  to be  extremely
sensitive to  objects with \HI\ masses below  10$^8$ $M_{\odot}$.  The
inventory of  these objects allows us  to derive the  number density of
the low \HI--mass  objects and to constrain the  slope of the low-mass
end of the  \HI\ mass function about a decade  lower than any previous
study. We  leave the estimation of  the \HI\ masses  of the detections
and  the  \HI\   mass  function  for  a  follow-up   paper (see also 
 \citealt{Kovac.2007}). In  this work,  we  present  the  survey and  the
detections.    Paper   has    been   organised    as    follows.    In
Section~\ref{sec_survey} we  present the observational  setup and data
reduction. In  Section~\ref{sec_hidet} we describe the  method used to
search for  the signal  and the \HI\  parametrisation. We  present the
uncertainties of  the measured parameters and the  completeness of the
survey  in Section~\ref{sec_errors}  and we  discuss the  various \HI\
properties  of the observed  detections in  Section~\ref{sec_prop}. In
Section~\ref{sec_summary}  we give  a final  summary. At  the  end, in
Appendix~\ref{app_atlas}   we  provide   an  atlas   of   the  figures
emphasising the various properties  of the detections.  Throughout the
paper we express the coordinates of the objects in the J2000 system.

\section {Description of the survey}
\label {sec_survey}

Due to technical limitations  of the current cm-wave radio telescopes,
the  volumes probed  by \HI\  surveys  are much  smaller than  volumes
probed by optical and infrared surveys. Moreover, to date these sureys  have been  sensitive to
objects with small \HI\ masses (below $10^8$ \Msol) only  up to distances of a few (tens) Mpc (see
Table~\ref{hisumtab}). Therefore,  to be able  to make an  inventory of
objects with  small \HI\  masses in a  reasonable amount  of telescope
time,  only a  nearby  volume can be targeted for such search.

\subsection{The selected volume}

We have selected a part of the nearby volume containing galaxies residing in the Canes Venatici (CVn) groups (or clouds)
to  carry out  a  blind \HI\  survey. The  CVn groups  of
galaxies are concentrated in a  small area in the constellation of the
same name \citep[][constellation limits are 11$^h$30$^m < \alpha  <$ 13$^h$40$^m$ and 25$^0 < \delta <$
55$^0$]{Karachentsev.etal.2003a}, known to host a population of
small galaxies. Together with the Local Group (LG) and the loose group
in Sculptor, the  CVn galaxies extend along the line of  sight up to a
distance corresponding to $cz \approx  1200$ km s$^{-1}$ (or to about 17 Mpc assuming $H_0 = 70$ \kms\ Mpc$^{-1}$ and Hubble flow). The study of
the  velocity  flow  in  the  nearby  volume  of  the  CVn  groups  by
\citet{Karachentsev.etal.2003a} reveals  that galaxies in  this region
closely obey a  Hubble flow.  The  prospect of using  the Hubble
flow  to  estimate  distances  to  the objects  even  for  such  small
recession velocities makes the CVn  region an excellent target for the
\HI\ observations.

Two concentrations  can  be   distinguished in the CVn groups \citep{Tully&Fisher.1987}. The  redshift
distribution of  galaxies shows a peak at  $V_{LG} =  200-350$ \kms,
which corresponds to  the galaxies in the CVnI  cloud. Another peak is
seen in the range of $V_{LG}  = 500-650$ \kms\ and may correspond to a
more distant cloud CVnII  aligned along the Supergalactic equator. The
better studied CVnI  cloud is populated mostly by  late-type galaxies of
low luminosity, in  contrast to the groups in  the CVnI neighbourhood:
the M81,  Centaurus and Sculptor groups.  The  apparent overdensity of
the number of  galaxies seen in the CVn  direction exceeds $\delta$N/N
$\sim$ 7 \citep{Karachentsev.etal.2003a}.

We will refer to the volume covered by our survey as the CVn region from now on. The exact limits of the observed region are given in the following subsection.

\subsection{WSRT observations}
\label{obser_xs_ch1}

During  2001,  2002  and  2004  observations  comprising  a  total  of
 approximately  60 $\times$ 12 hr  have been performed for this survey  using
 the Westerbork Synthesis Radio Telescope (WSRT). The WSRT is an aperture
 synthesis interferometer with 14  antennas arranged in a linear array
 on a 2.7  km East-West (E-W) line.  Ten of  the telescopes are fixed,
 while  4 antennas are  movable on  2 rail  tracks.  The  antennas are
 equatorially mounted  25 m dishes.  In  a single 12 hr  time slot, 24
 fields were observed in mosaic mode.   These fields are on an E-W line
 and are separated by 15  arcmin in right ascension. On different days
 similar strips of constant declination were observed.  The separation
 in declination  between strips is 15  arcmin. Given that  the FWHM of
 the  WSRT primary beam  is 34  arcmin, we  obtained a  nearly uniform
 sensitivity over the whole observed  area with the 15 arcmin sampling
 used.  Each of the 24 fields in one strip of constant declination was
 observed for 100 sec before  moving to the next pointing, which gives
 18  different  $uv$ scans  per  field per  12  hr  period. Using
 interlaced sampling  on different days (the  pointings observed first
 during  two consecutive  nights  of observations  are  shifted by  15
 arcmin in right ascension) the $uv$  coverage improved to 36  $uv$ scans
 per observed  pointing.  The effective integration  time per pointing
 was 80.1  min (taking  the slew time  into account). The shortest spacing of the array used for the observation was 36 m. All structures larger than 20 arcmin will be filtered out completely. Structures smaller than 10 arcmin should, however, be recovered quite well.

The first 9 $\times$ 12  hr, performed in 2001, were observations with
 one 10 MHz wide band with 128 channels covering the velocity range of
 approximately -450 to  1450 \kms.  The rest of  the observations were
 carried out in two bands.  Then we used one band of 20 MHz width with
 512 channels covering  the velocity range from -750  to about 3250 km
 s$^{-1}$.  The second band  used was  20 MHz  wide with  512 channels
 covering roughly the interval from 3000 to 7000 \kms.  Only data with
 approximately $-450 < cz < 1330$ km s$^{-1}$ were used in the further
 data reduction and analysis process.

\begin{figure}
  \includegraphics[width=0.45\textwidth]{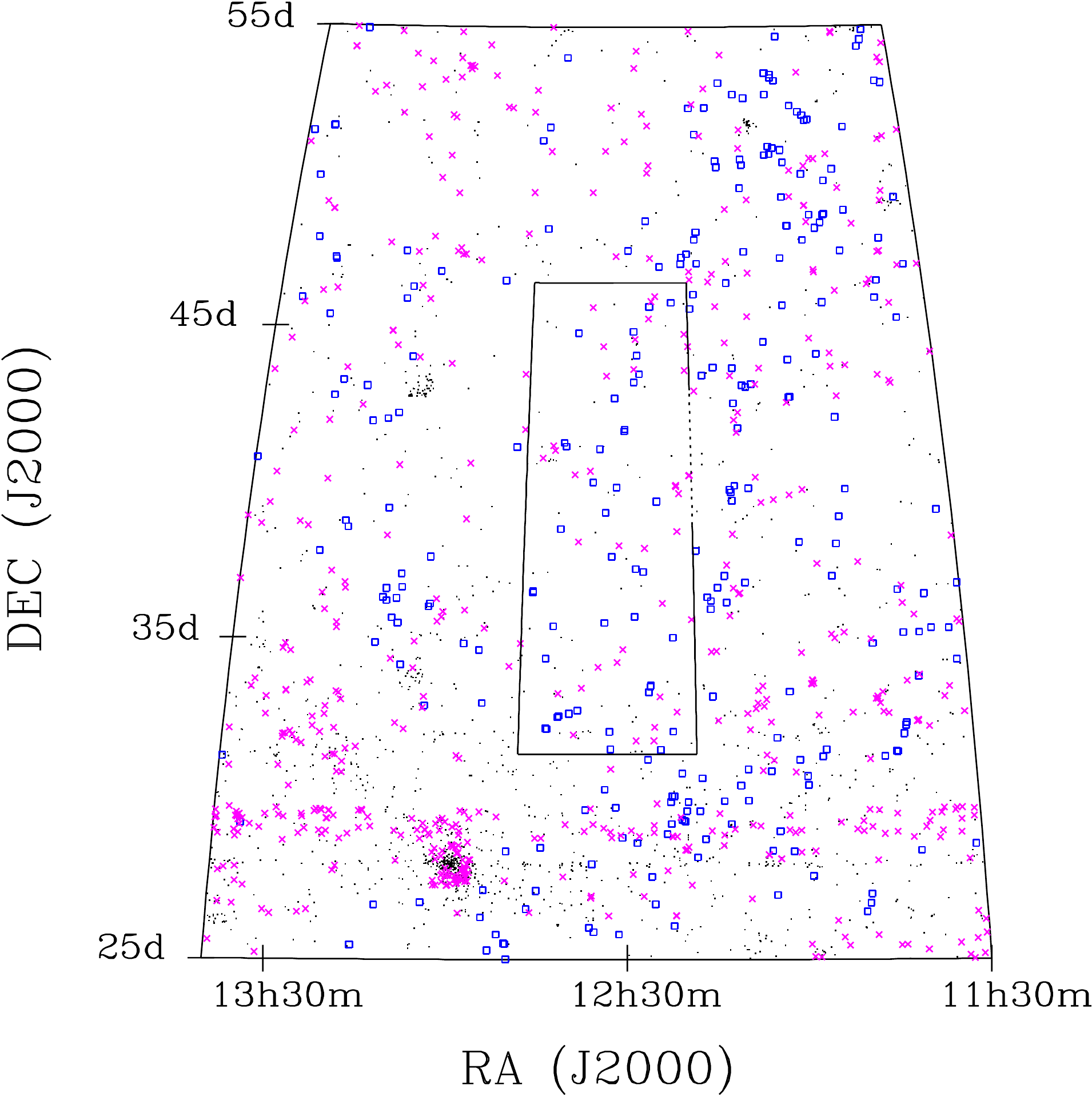}
  \caption{\small  Surface   distribution  of  galaxies   in  the  CVn
  region. Every point in the  plot corresponds to one detection in the
  CfA redshift survey catalogue \citep{Huchra.etal.1999}.  From these,
  galaxies with  the recession velocities $V_{LG} \le  2000$ \kms\ are
  presented  with squares. Detections  for which  redshift measurement
  has not been provided are  indicated by crosses. The area covered by
  the  WSRT CVn  survey  has been  marked  with the  combination of  a
  continuous and  a dotted line.   The dotted line corresponds  to the
  range in declinations for which  two pointings per strip of constant
  declination have not been observed. }

\label{cvn_env}
\end{figure}

All the pointings  for the survey were located within  the area on the
sky with limits 12$^h$19$^m$55.2$^s \le \alpha \le$ 12$^h$47$^m$2.4$^s$
for $\delta = 31^0  33^{'} 00^{''}$ and 12$^h$16$^m$42.9$^s \le \alpha
\le$ 12$^h$50$^m$3.0$^s$ for $\delta =  46^0 18^{'} 00^{''}$.  Due to a
human  error  in  the  observational  setup, two  of  the  first  four
pointings with the smallest right ascension have not been observed for
strips with declination between $38^0 48^{'} 00^{''}$ and $42^0 48^{'}
00^{''}$.   Therefore the  observed area  on the  sky is  approximately 86
deg$^2$   (instead   of    the   originally   planned   90   deg$^2$).
Figure~\ref{cvn_env} shows  the projected distribution  of galaxies in
the   CVn   region   \citep[region   limits  taken   from][are   given
above]{Karachentsev.etal.2003a}.  Detections are  taken  from the  CfA
redshift survey  catalogue \citep{Huchra.etal.1999}. 

\subsection {Data reduction}
\label{sub_dataredch2}

In  total  1372 (useful)  pointings  were  observed  for this  survey.
Scripts  were developed  to  automate the  processing  for this  large
number  of pointings.   The scripts  were based  on  MIRIAD programmes
\citep{Sault.etal.1995}   and  programmes   written  by   two   of  us
(T.A.O.  and K.K.). The  data reduction  process applied  is described
below.

The  $uv$ data  of each  pointing  were cross-calibrated and  Hanning
smoothed.  The few  first and last channels in  the observed band were
excluded  because  of  their  higher  noise. The  data  were  visually
inspected and obviously bad data were  flagged.  To be able to see the
\HI\  emission  in  the  observed   data,  the  continuum  had  to  be
subtracted.    The  continuum   $uv$  data   were  created   to  first
approximation by fitting a polynomial of second order to all available
channels for each pointing, excluding the obvious line emission. After
summing the continuum  emission observed over the whole  band into one
plane,  the data  were Fourier  transformed into  ($\alpha$, $\delta$)
continuum images  using standard MIRIAD programmes.  The final continuum
image   was  created  in   an  iterative   process  of   cleaning  and
self-calibration  of  the  continuum  data.  The line  $uv$  data  were
obtained by  copying the calibration coefficients  and subtracting the
modelled continuum emission from the observed data in the $uv$ domain.

The line $uv$ data were  processed into ($\alpha$, $\delta$, $V$) line
datacubes using the  MIRIAD programme INVERT. The mosaicing  mode of the
observations produced data  sampled very sparse in the  $uv$ plane. In
order to suppress  large, shallow wings of the synthesised  beam, a special
weighting was applied to  the $uv$ points.  This weighting corresponds
to natural weighting multiplied by radius in the $uv$ plane. The data
were smoothed spatially  by multiplying the $uv$ data  with a Gaussian
corresponding to a FWHM of 30  arcsec. From the first 9 $\times$ 12 hr
of observations,  216 line datacubes  each consisting of  115 channels
(i.e.  [$\alpha$,  $\delta$] images) were  obtained.  The rest  of the
data   were   processed   into    1156   line   datacubes   with   125
channels.  Additional   continuum  subtraction  was   applied  to  all
pointings by fitting the continuum with a polynomial of first order to
the   line  datacubes  excluding   line  emission,   and  subsequently
subtracting it from  them. The velocity spacing in  the line datacubes
produced is $\sim$ 16.5 km  s$^{-1}$ and the velocity resolution after
Hanning smoothing is $\sim$ 33 km  s$^{-1} $. The size of the image in
each of the channels is 512 $\times$ 512 pixels$^2$, with a pixel size
of  $8'' \times 8''$. The typical  spatial resolution  of  the datacubes
produced is $\sim 30'' \times 60''$.

\begin{figure*}
\centering
  \includegraphics[width=0.32\textwidth]{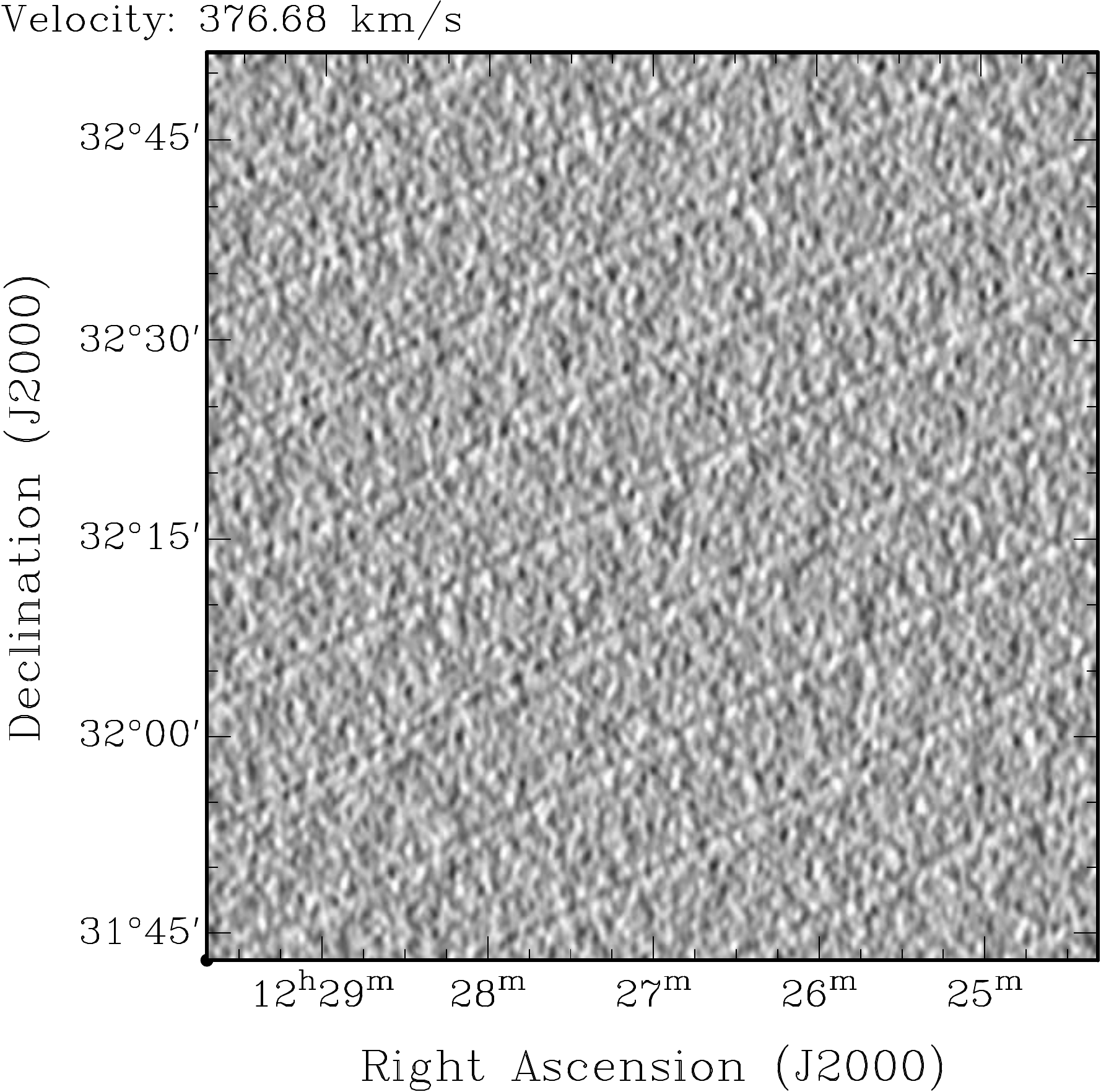}
  \includegraphics[width=0.32\textwidth]{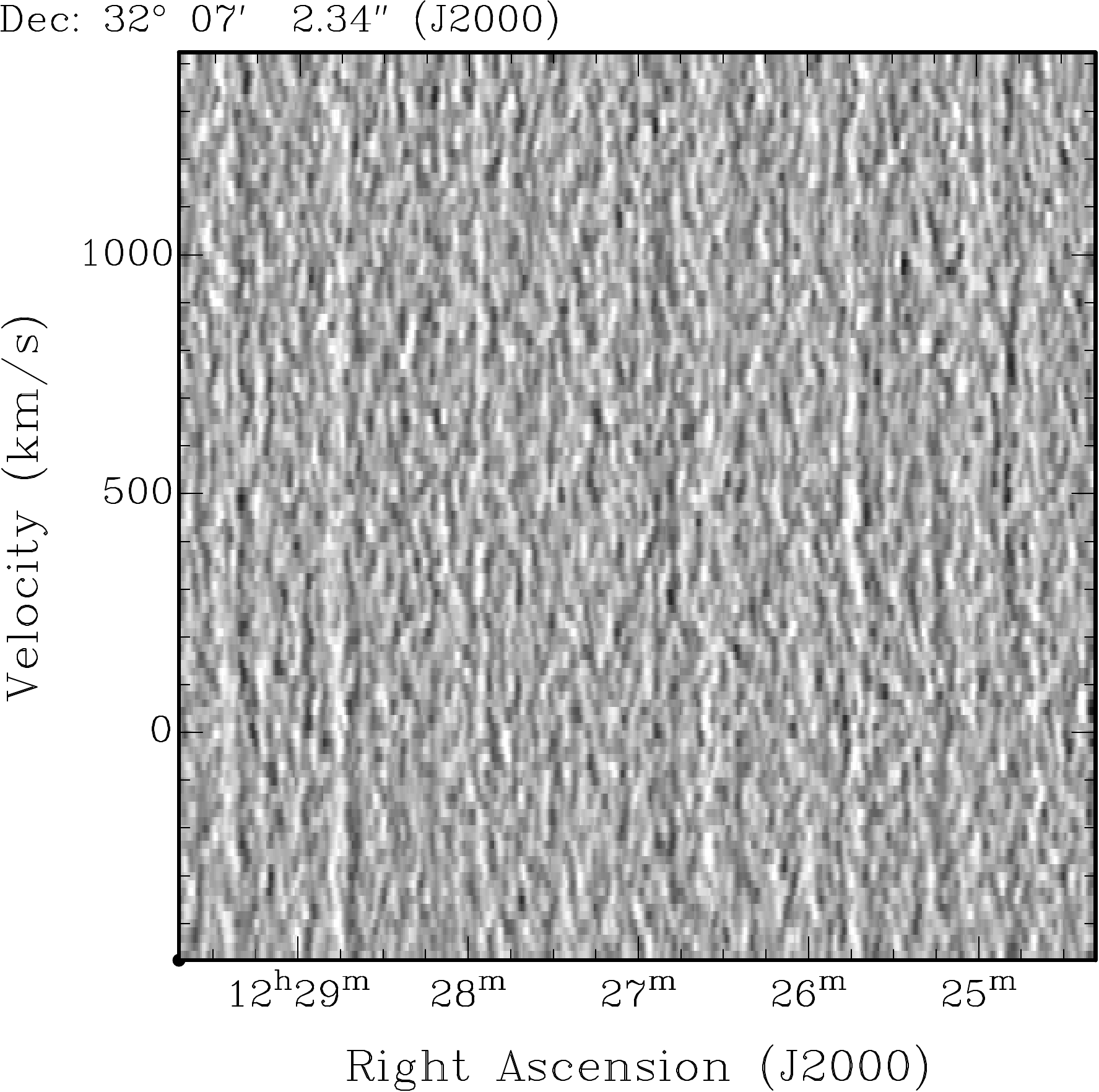}
  \includegraphics[width=0.32\textwidth]{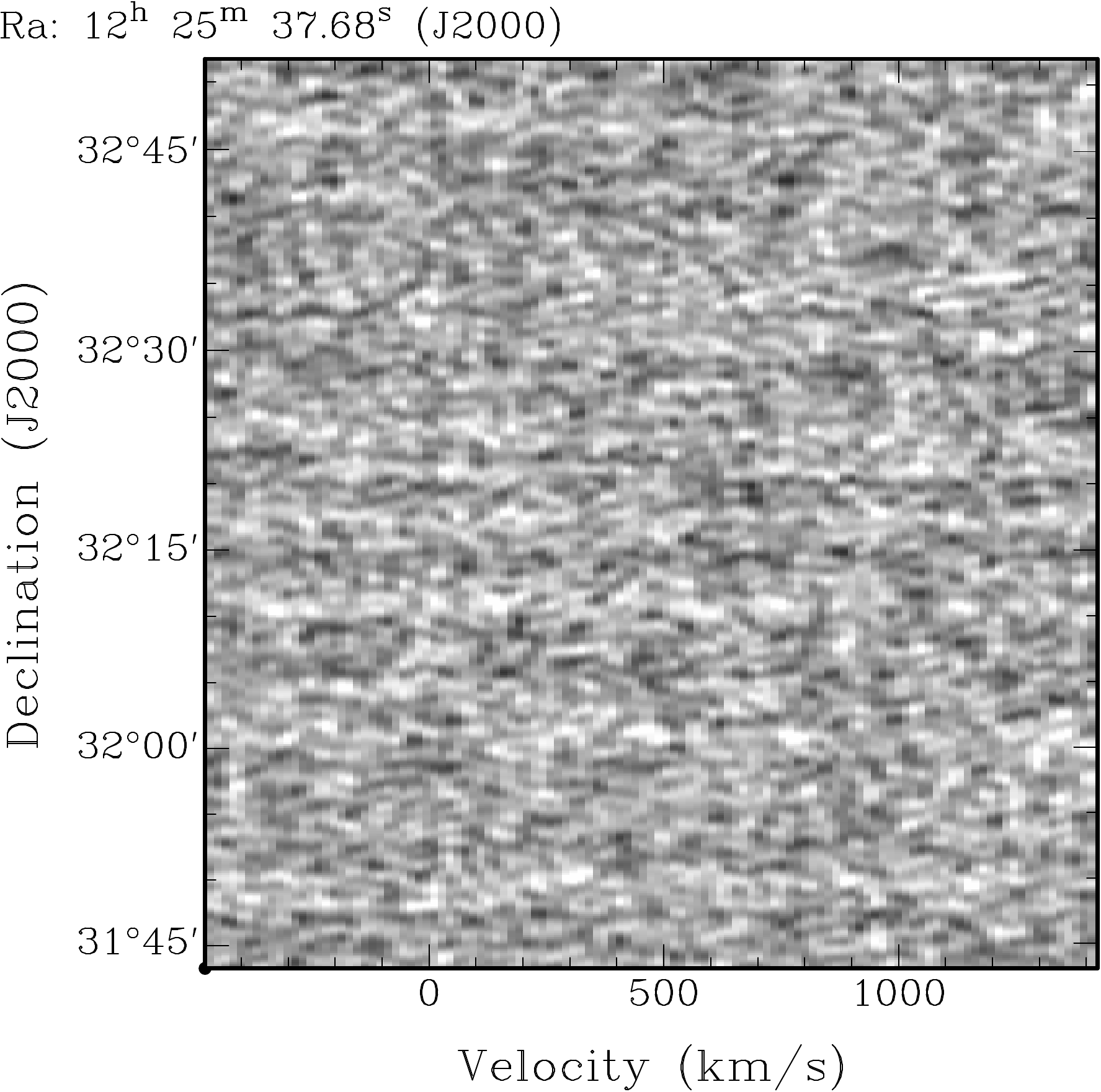}
\hskip 0.5cm
  \includegraphics[width=0.32\textwidth]{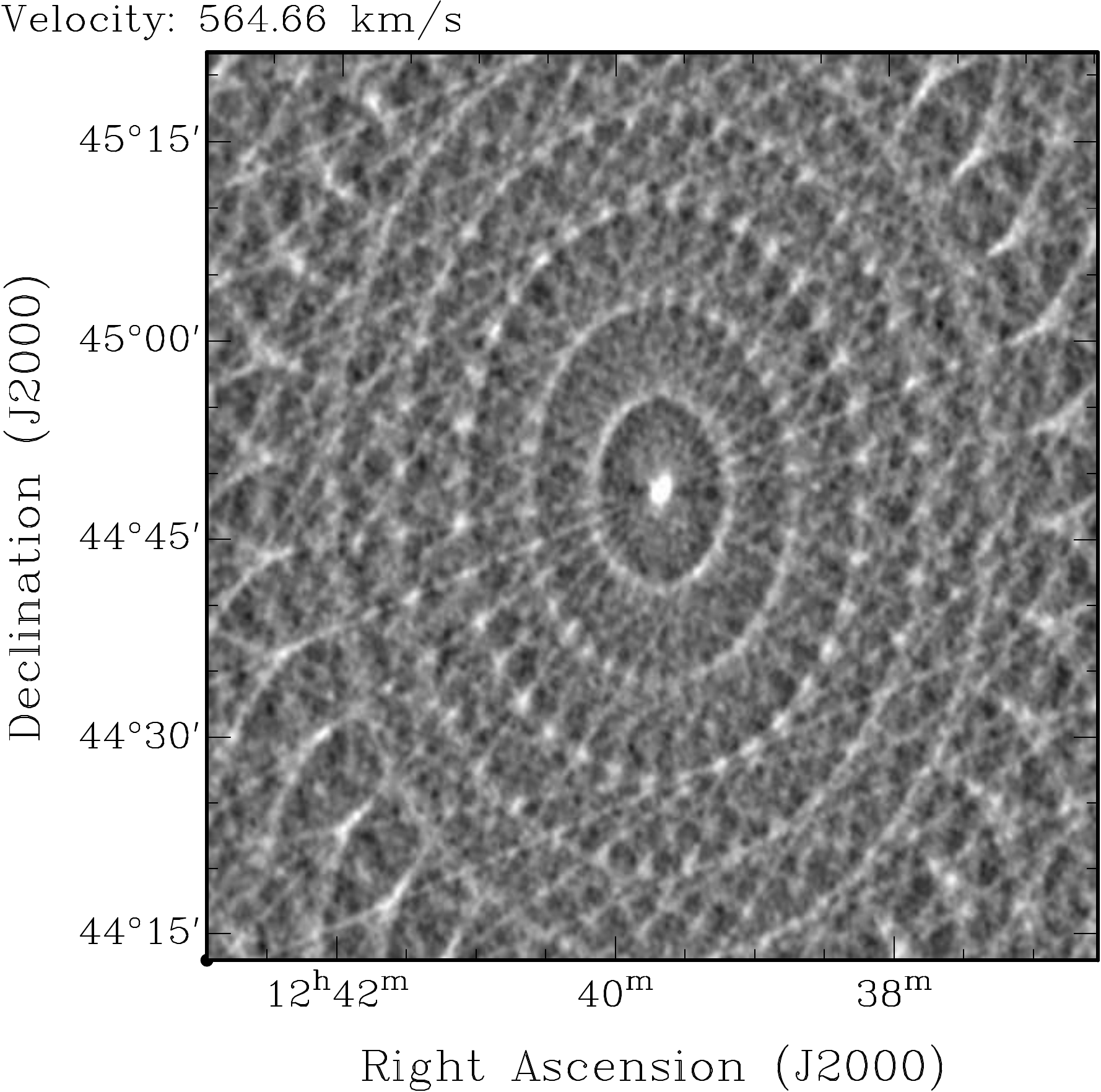}
  \includegraphics[width=0.32\textwidth]{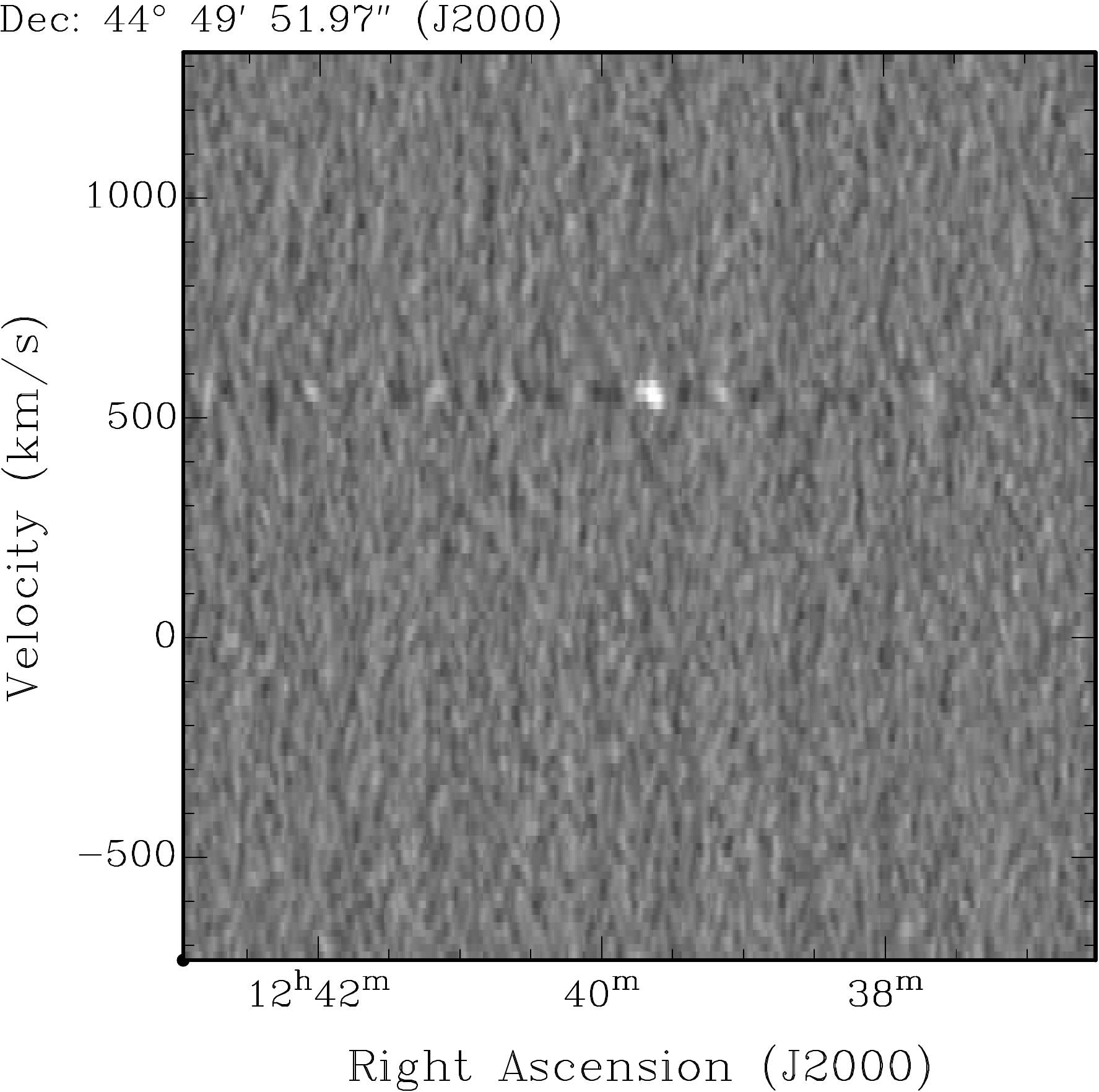}
  \includegraphics[width=0.32\textwidth]{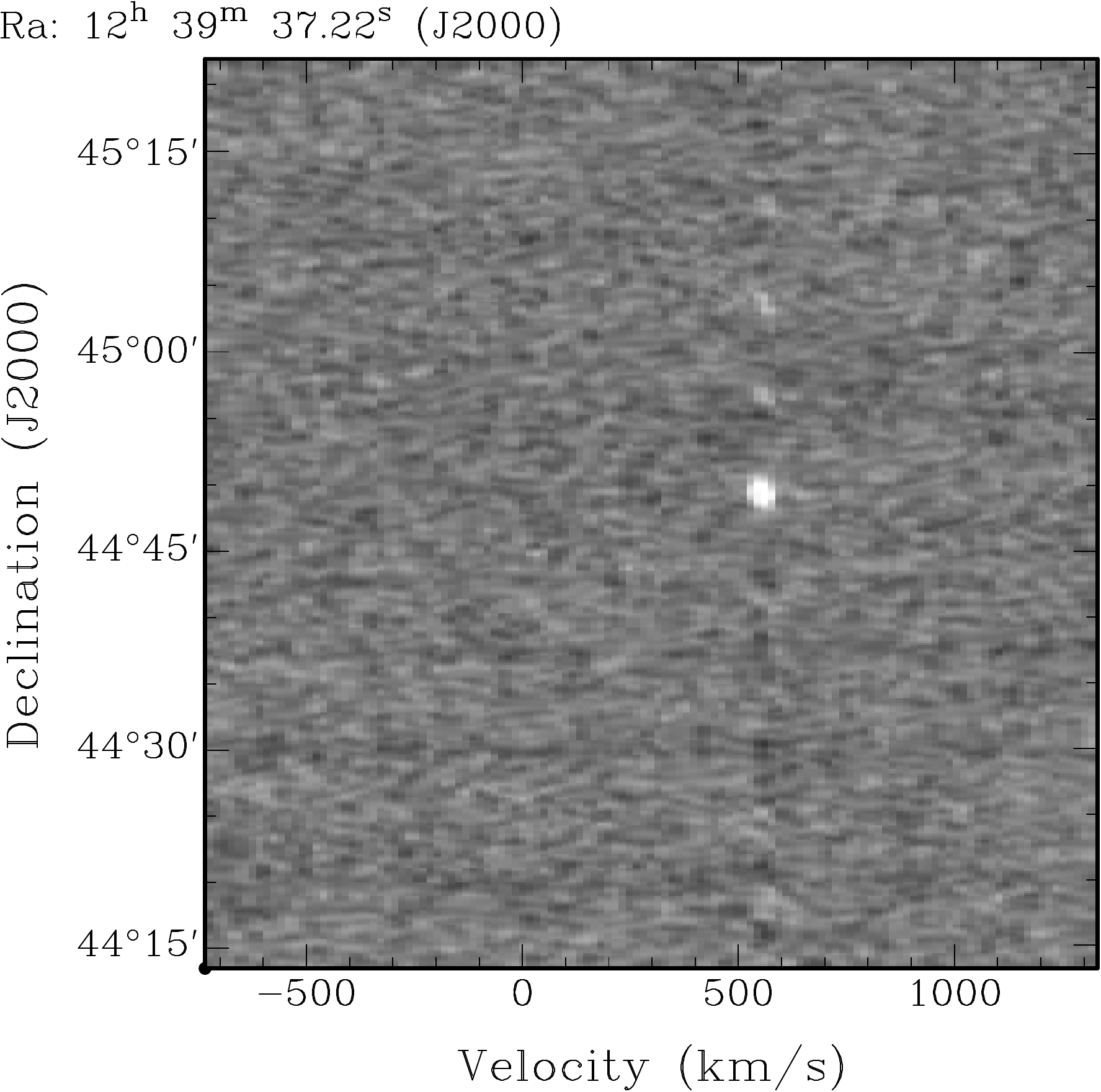}
\hskip 0.5cm
  \includegraphics[width=0.32\textwidth]{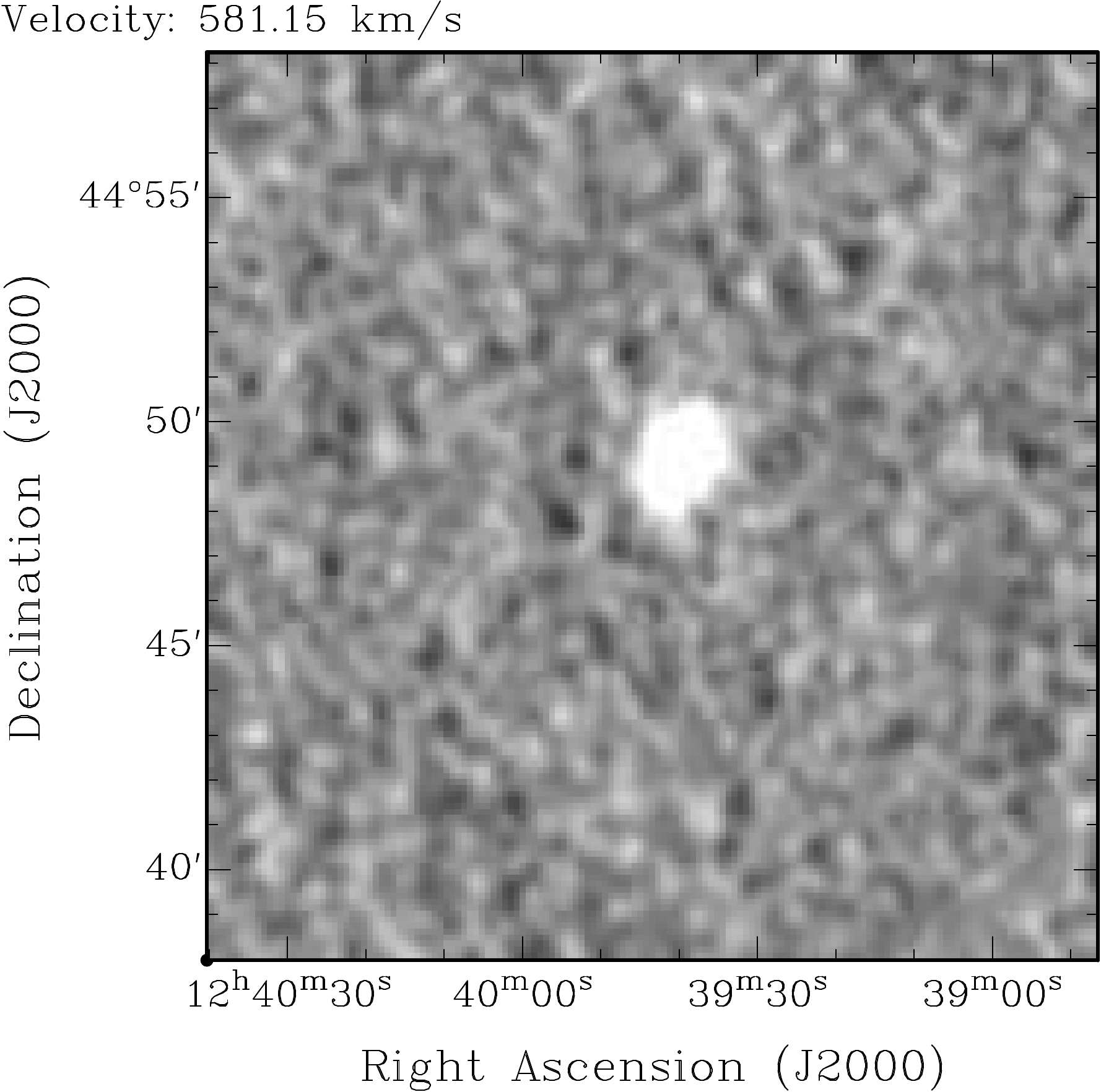}
  \includegraphics[width=0.32\textwidth]{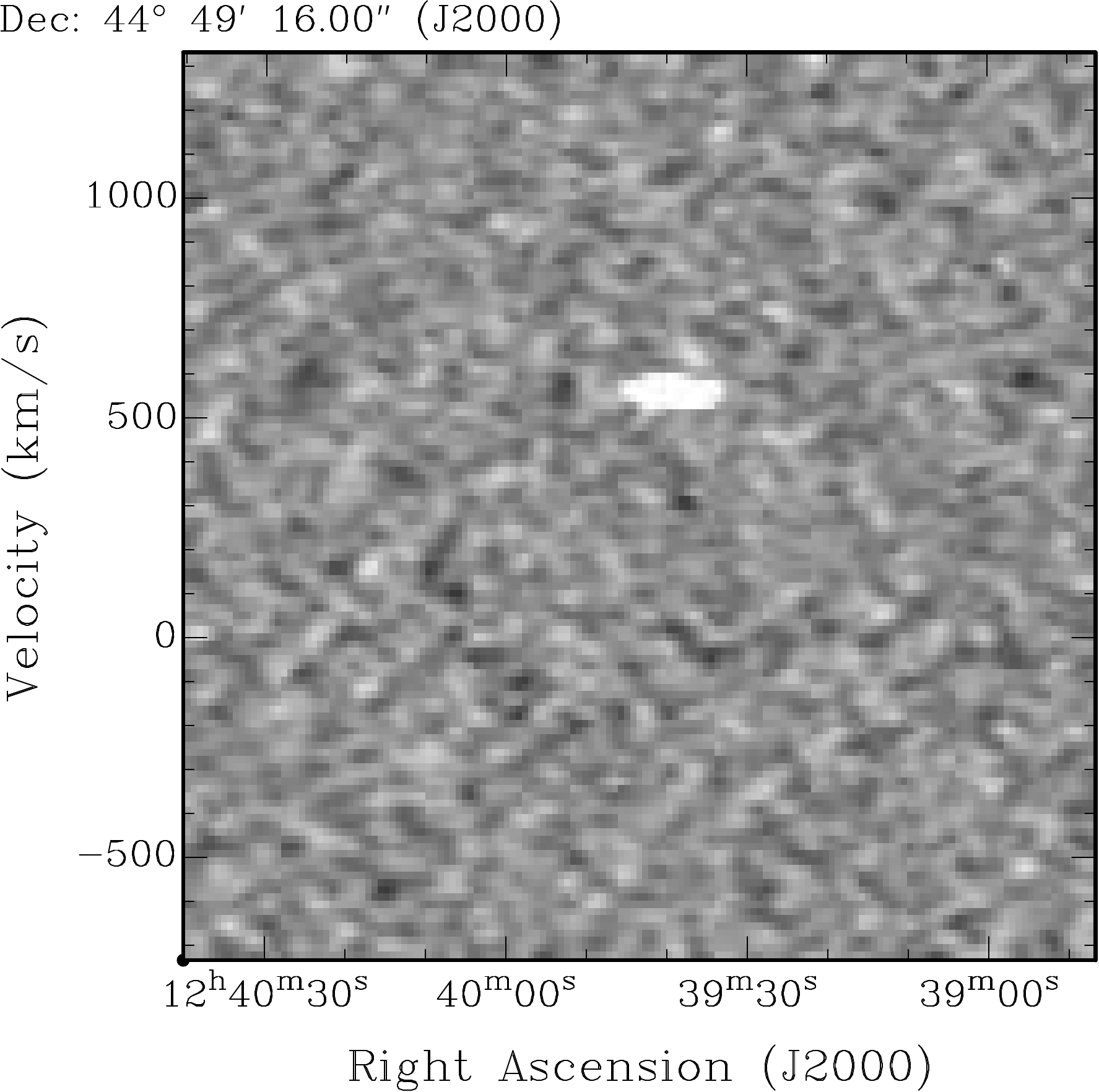}
  \includegraphics[width=0.32\textwidth]{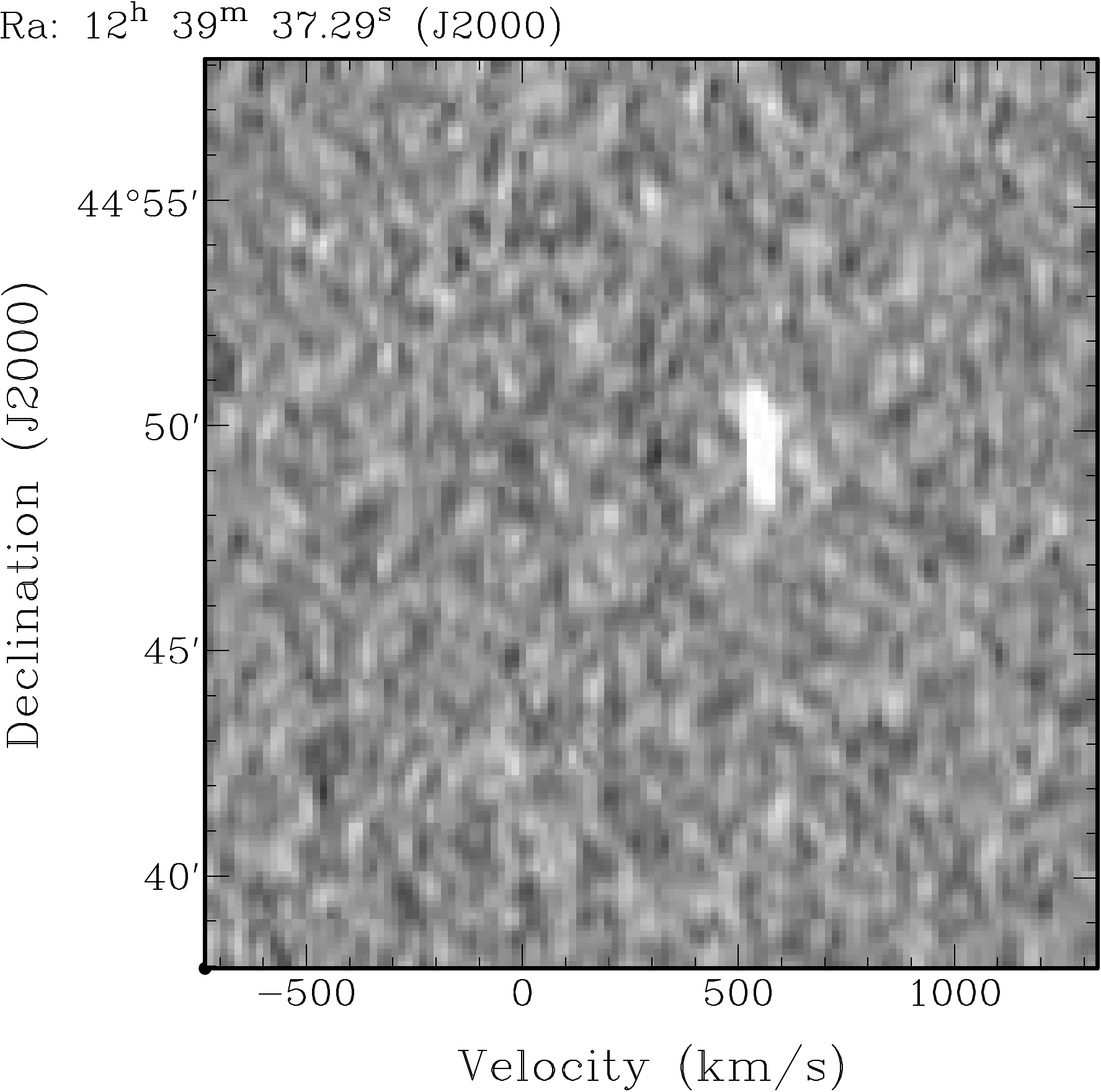}
\hskip 0.25cm
  \caption{\small Examples of the various features in the line datacubes. The
  panels in  the vertical direction  are made by extracting  one slice
  from   the  line  datacubes   in  $\alpha-\delta$,   $\alpha-V$  and
  $\delta-V$  planes, going  from  left to  right respectively.  Three
  uppermost  panels show  the appearance  of the  interference  in the
  datacubes.  The  middle panels  illustrate the grating  rings formed
  around an \HI\ source.  The  lowest panels show cuts through a final
  line  datacube produced  by combining  nine of  the  line datacubes,
  created from the data observed  for one pointing, and by cutting the
  central part of the combined datacube.}
  \label{cubes_examples}
\end{figure*}

A known problem  for detecting radio emission is  man-made and natural
interference.    No  good   automated  method   exists   for  removing
interference  from the  data,  especially not  from  data observed  in
mosaicing method. In  these kind of observations, marking  the data as
bad  outside an  interval of  data values  observed for  an individual
source (e.g. ``sigma-clipping'') will  not necessarily remove the data
affected by interference.  The scatter seen in the  data can be caused
both by  the interference and  by the observed source  itself, because
the $uv$ properties of  a source can  significantly change between
the two  consequent observations in  the mosaic mode of  that specific
pointing.    Therefore,  all   1372  line   datacubes   were  visually
inspected. Datacubes are composed of 36 different $XX$, $YY$ scans and
if  interference occurred  it  was easily  recognisable  in the  image
domain, where the interference appeared as a strong narrow stripe.  An
example  of  the  appearance  of  interference  in  the  datacubes  is
presented  in the three  upper panels  in Figure~\ref{cubes_examples}.
Using the MIRIAD task CGCURS, stripes induced by the interference were
marked  and  removed  by  flagging  the $uv$  scan  during  which  the
interference occurred.

As a  result of the  sparse sampling in  the $uv$ plane  the sidelobe
levels  and grating  rings  around the  strong  \HI\ sources  preclude
detecting  faint  \HI\  emission.  Channels with  \HI\  emission  were
CLEANed and RESTOREd  with a Gaussian beam with a  similar FWHM as the
synthesised beam corresponding to the  pointing.  In the  three middle
panels in  Figure~\ref{cubes_examples} we  show an example  of grating
rings produced around an \HI\ source.

To exploit the observing strategy with overlapping pointings, line
datacubes corresponding to the pointings with separation less or equal to
22 arcmin were  combined into one datacube. Smaller  datacubes of size
150  $\times$  150 $\times$  115  or  150  $\times$ 150  $\times$  125
pixel$^3$ in  $\alpha \times \delta \times  V$ directions respectively
were cut out of the central  part of each combined datacube, where the
size of  the third  dimension depends on  the observations. The $uv$ continuum subtraction worked well, leaving only minor residual effects in the image datacubes. These residual 
   continuum features were removed using a simple linear
   baseline fit to the spectra in the datacube excluding the channels
   with \HI\ emission. The small, combined line  datacubes have been used for all of
the  further  data analysis.   In  the  following  text they  will  be
referred to as  the full resolution datacubes. An  example of the full
resolution   datacube   is  presented   in   the three  lowest  panels   in
Figure~\ref{cubes_examples}.

For each of the combined datacubes the synthesised beam of the central
datacube used in the combining process  was chosen as the beam of that
datacube.  Theoretically,  all the  $uv$ data observed  in overlapping
pointings could  be used  jointly to build  a single large  data cube,
instead of applying the whole data reduction process on the individual
pointings and  then combining the  cleaned and interference  free line
datacubes from the individual pointings. In practice, the first method
would need  much more  computer time and  computer memory, and  at the
moment it is not affordable for such a large data set as ours.

\section{\HI\ detections}
\label{sec_hidet}

\subsection{Searching for detections}
\label{searching}

One of the most important  aspects of analysing the observations is to
define  what constitutes a  detection. In  surveys, regardless  of the
observed wavelength, the common way  is to consider a detection a real
object if the measured flux, or  part of it, of that particular object
exceeds     the     noise    by     a     certain    factor     \citep
[e.g.][]{Wall&Jenkins.2003}.

The  line datacubes, produced  as described  in the  previous section,
were  smoothed both in  the spatial  and velocity  domain in  order to
improve the detectability of extended objects with small signal to
noise ratios.  The datacubes were convolved with  Gaussians with FWHMs
such  that  the final  spatial  resolution  of  the produced  smoothed
datacubes was 1.5 and 2.0 times the original spatial resolution. In
the velocity domain, cubes  were Hanning smoothed by performing a weighted average of the fluxes over 5  and 7  neighbouring  channels.  Smoothing  in the  spatial  and
velocity domain was done separately.

\begin{figure}
  \centering

\includegraphics[width=0.47\textwidth]{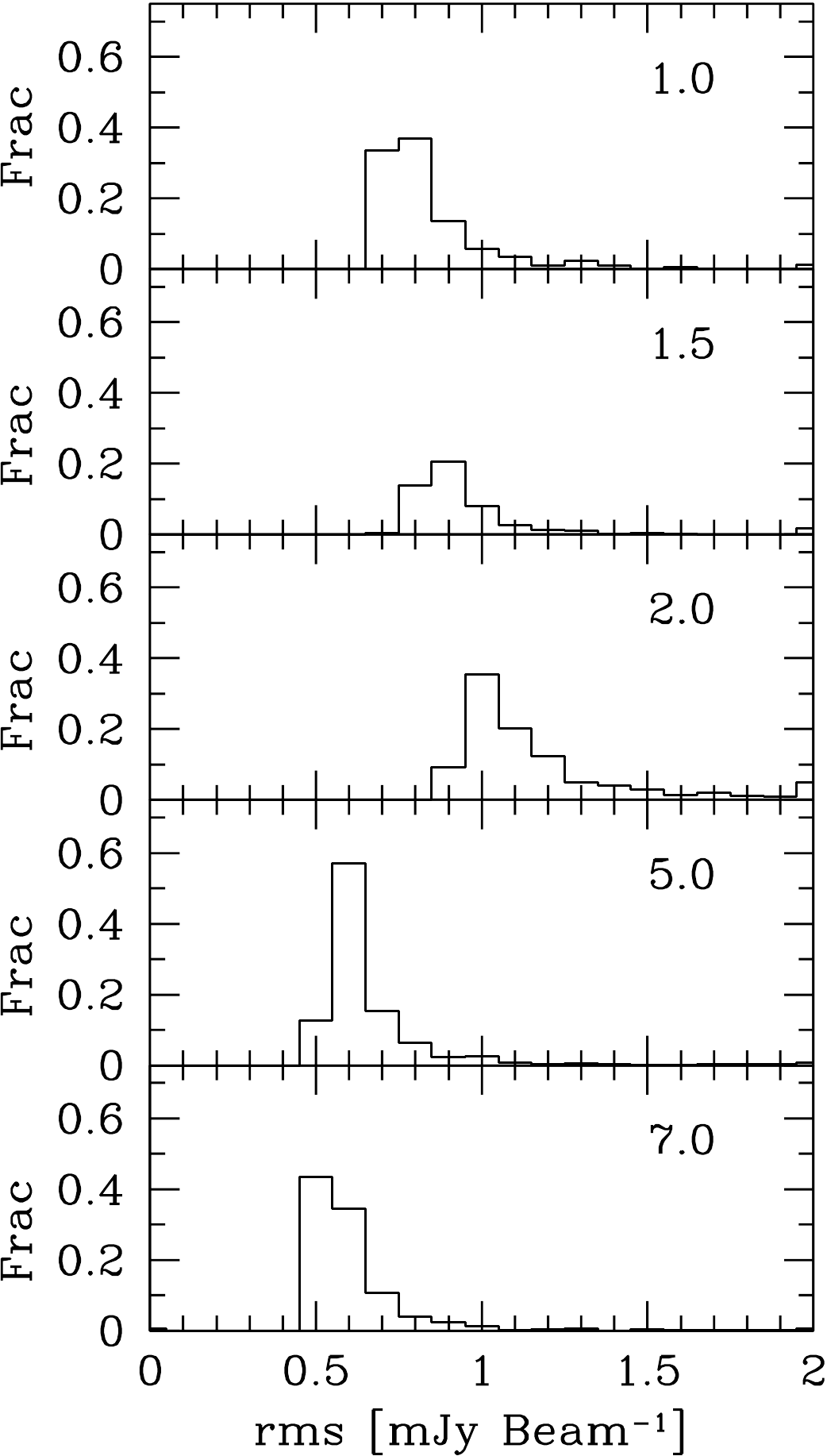}
\caption{\small Statistics of the line datacubes.  The panels show the
histogram  distributions of  rms  values measured  for 1372  datacubes
produced.  In the  first  row  results from  the  measurements in  the
datacubes with full resolution are presented. The second and the third
rows contain  the results  for the datacubes  smoothed in  the spatial
domain  to the  resolution 1.5  and 2  times the  original resolution,
respectively. The forth and the  fifth rows present the statistics for
the datacubes Hanning smoothed in the velocity domain averaging fluxes
over  5  and  7  neighbouring  channels, respectively. }
\label{statrms}
\end{figure}

To get a  good insight in the statistical properties  of the data, the
mean and  the rms values  of datacubes were  estimated for all  of the
1372 line  datacubes produced at the five  different resolutions (the full resolution, 2 smoothed in the spatial domain and 2 smoothed in the velocity domain).  The
mean and  the rms  values of the  individual datacubes  were estimated
from  the  pixels  with  absolute   flux  values  below  5  times  the
preliminary  rms  value  of  the  datacube. The  preliminary  rms  was
estimated  using  all  of  the  pixels in  the  datacube.  The  binned
distribution   of    the   final   rms   values    is   presented   in
Figure~\ref{statrms}, while the mean and the standard deviation of the
measured         rms    values     are     presented    in
Table~\ref{stat_all}.     For     reference,     we     include     in
Table~\ref{stat_all}  the typical spatial  and velocity  resolution of
the  specific types  of  datacubes,  as well  as  the limiting  column
density to detect an object with a profile width equal to the velocity
resolution (third column in Table~\ref{stat_all}) at the 5-sigma (5 
times the value in the forth
column  in Table~\ref{stat_all}) level.   The mean  noise value  in the
line datacubes with the full  resolution is 0.86 mJy Beam$^{-1}$.  For
an object  with a velocity  width of 30  \kms\ and an \HI\  mass of 10$^6$
\Msol\ this noise  limit would imply a maximum distance  of 5.7 Mpc at
which this object could be placed  and still be detected in the survey
at the 5-sigma level. In the same type of datacubes, the limiting column density to detect an object with a velocity  width of 30  \kms\ at the 5-sigma level is $7.9 \times 10^{19}$ atoms cm$^{-2}$. This is only a crude estimate of the detection 
limit of the survey. A more precise estimate of the detection limit, 
based on the Monte Carlo simulations, will be 
presented in Subsection~\ref{sub_compl}.

\begin{table*}
\centering
\begin{tabular}{l|c|c|c|c|c}
\hline \hline
 & \multicolumn{2}{|c|}{Resolution} & \multicolumn{2}{|c|}{rms} & N$_{\rm HI}$ \\
\hline
Datacube & spatial & velocity &     mean    &    rms  & [10$^{20}$ $\times$ \\
type  & [arcsec$^2$] & [\kms]  & [mJy Beam$^{-1}$]  & [mJy Beam$^{-1}$] & atoms cm$^{-2}$]\\
\hline
1.0 &  30 $\times$ 60  & 33    & 0.86 & 0.30  & 0.87 \\
G1.5 & 45 $\times$ 90  & 33   &  1.33 &  3.01 & 0.60 \\
G2.0 & 60 $\times$ 120 & 33   &  1.64 &  3.98 & 0.42 \\
H5 &   30 $\times$ 60  & 82.5  & 0.67 & 0.28  & 1.70 \\
H7 &   30 $\times$ 60  & 99   & 0.62 & 0.28   & 1.88 \\
\hline
\end{tabular}
\caption{\small Statistics of the 1372 line datacubes. The first row corresponds to the statistics of the full resolution datacubes. Statistics of the datacubes Gaussian smoothed in the spatial resolution by a factor 1.5 and 2 is given in the second and third row respectively. Statistics of the datacubes Hanning smoothed over 5 and 7 neighbouring velocity channels is presented in forth and fifth row respectively. See  text  for description of the columns.
\label{stat_all}}
\end{table*}

The distribution of the observed pointings in this survey was designed
 in such  a way  that the  noise in the  combined datacubes  is almost
 uniformly  distributed. Still,  the noise  in some  of  the datacubes
 shows  a  gradient  in  the  spatial  domain  along  the  declination
 axis. This  is due to  the fact that  the final datacube  is composed
 from   observations  typically   collected  during   three  different
 nights. The noise gradient reflects  the difference in quality of the
 data collected during each single 12 hr period of observations caused
 by,  for instance  the  loss of  one  of the  14  telescopes and  the
 difference in the data flagging.

To  overcome this  problem, the noise  in each pixel of the datacubes  was modelled  by averaging  the standard  deviations independently
estimated in  the spatial and  velocity projections in the following way.  First, the standard
deviation of the  flux values in the spectra at  the position of every
pixel in the spatial domain along the velocity was calculated. The standard deviation of
the  fluxes  was calculated  also  for each  of  the  channels in  the
datacube. For  these calculations, pixels in the  channels around zero
velocity,  where  the  Galactic  emission  can be  very  strong,  were
excluded. In  the second iteration, in  addition to the  pixels in the
region of Galactic  emission, pixels with flux values  larger or equal
than 5  times noise from the  first iteration were  excluded also from
the  calculations.   The ``characteristic  noise''  of  a  pixel in  a
datacube  was  defined  as  the  average value  of  the  two  standard
deviations calculated  in the second  iteration for the plane  and for
the  spectrum which  both contained  that  pixel. This  value will  be
referred to in  the text as the noise  ($\sigma$). The velocities with
Galactic emission  are in  the range from  approximately -50  \kms\ to
approximately   80  \kms\   and  from   approximately  -60   \kms\  to
approximately 85  \kms\ for the  datacubes produced from  the poitings
observed during  the first 9 $\times$  12 hr and  during the remaining
observations, respectively.

The next step was to inspect the line datacubes in order to detect the
presence  of \HI\  emission.   All line  datacubes  were searched  for
pixels with  an absolute flux value  above a given  limit expressed in
multiples of  the noise  in each pixel.  A procedure was  developed to
automate the process of searching for signal in the datacubes.

\begin{figure*}
 \centering
  \includegraphics[width=5.5cm]{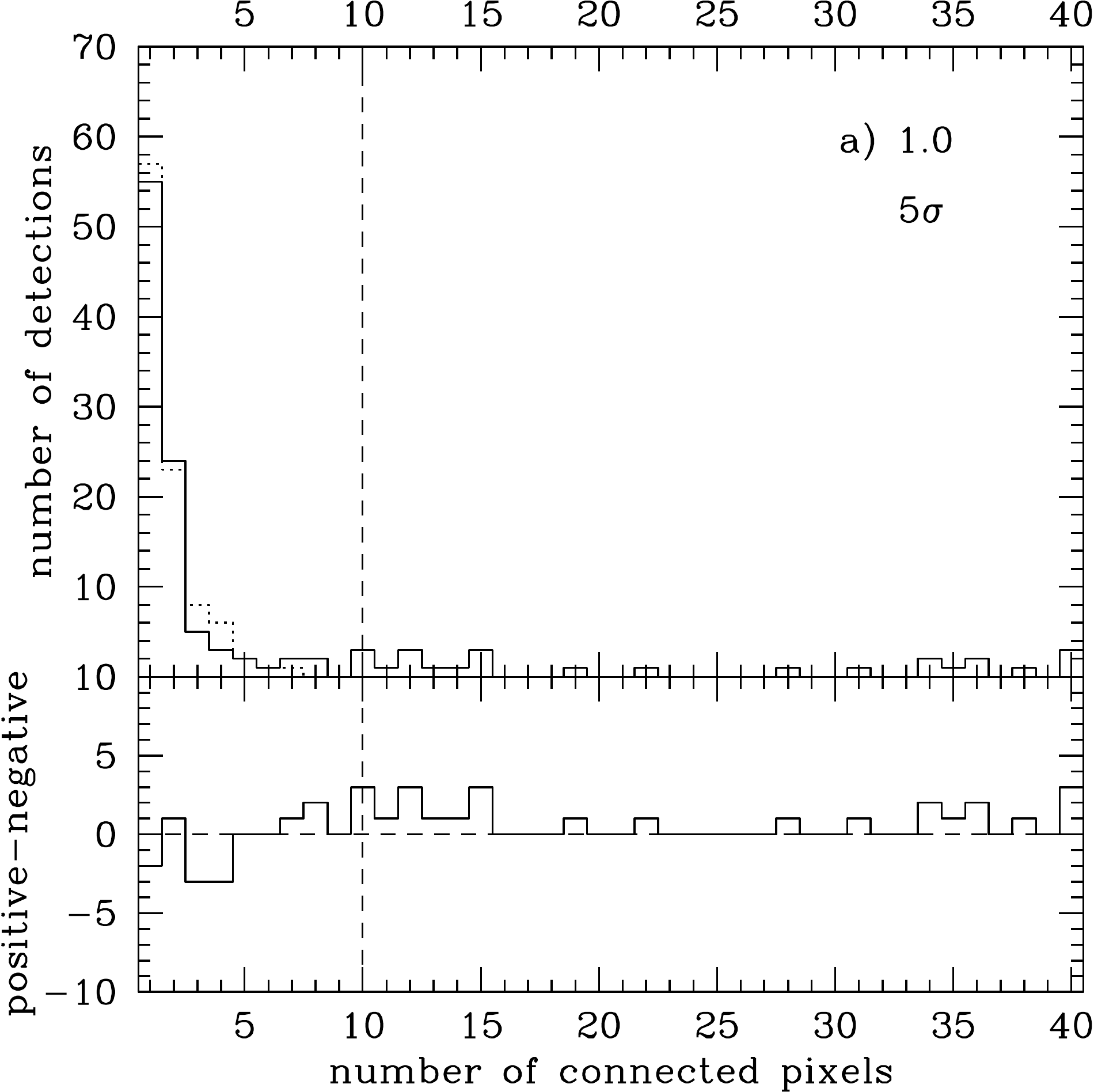}
 \includegraphics[width=5.5cm]{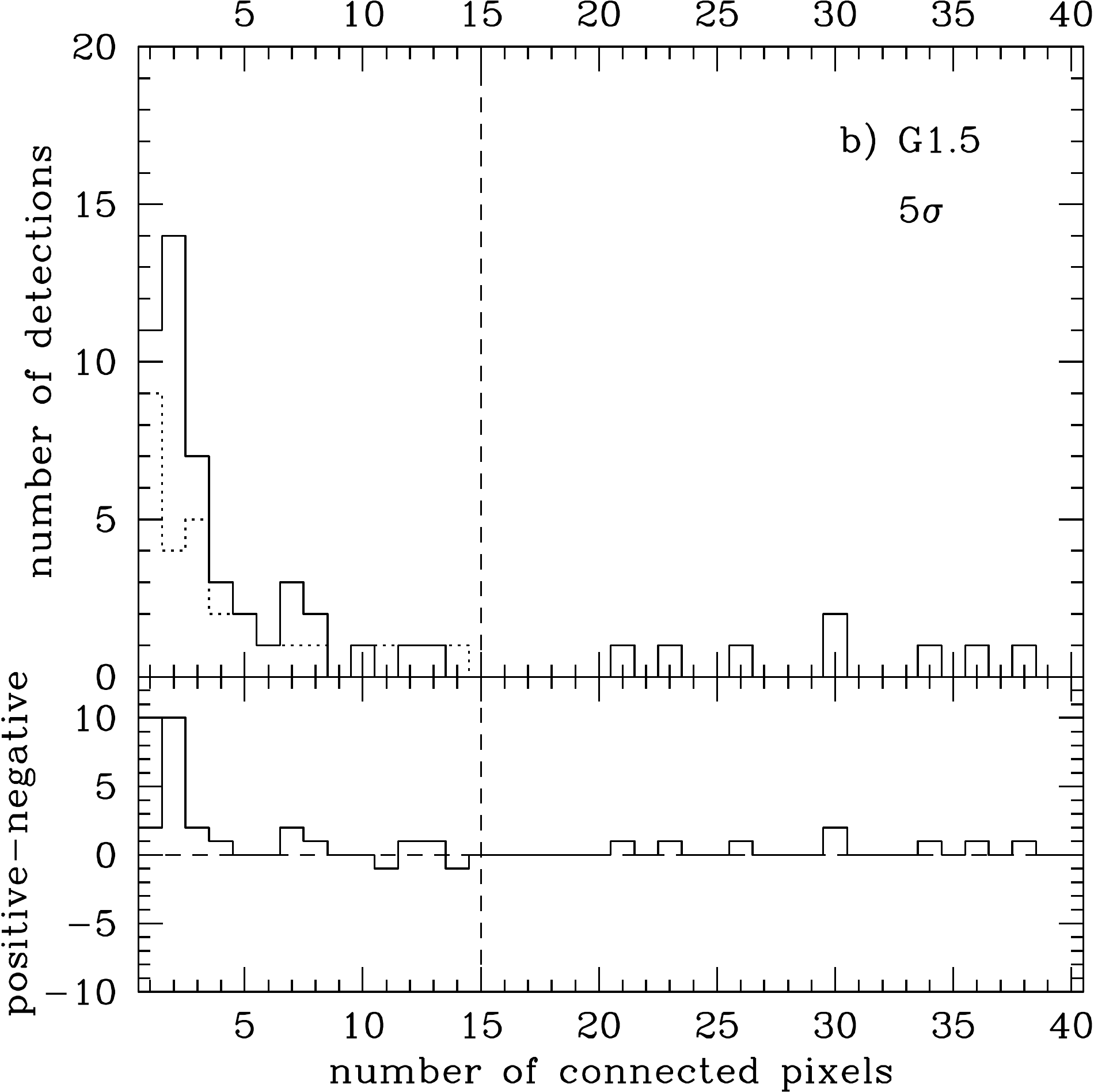}
 \includegraphics[width=5.5cm]{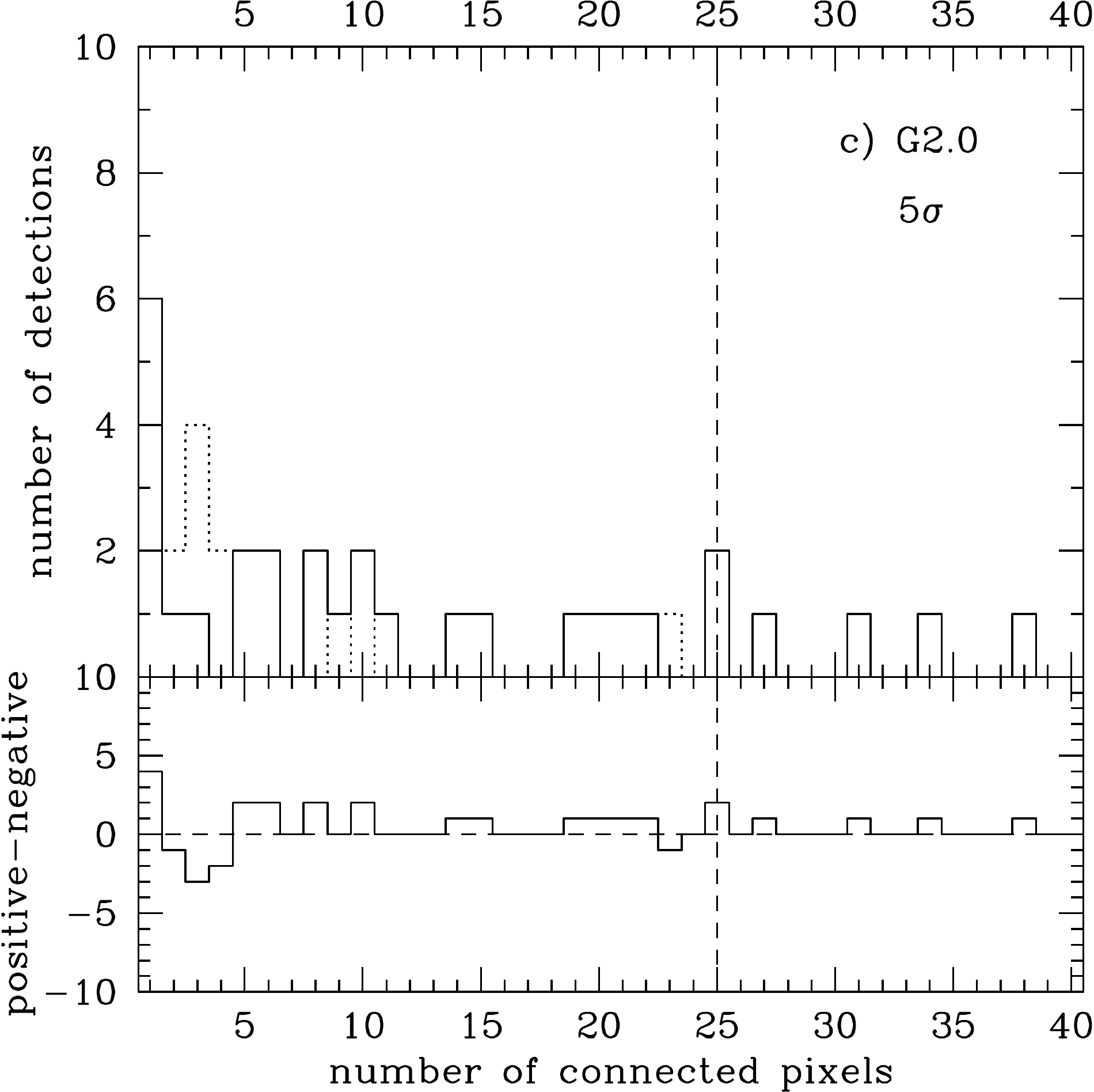}
 \includegraphics[width=5.5cm]{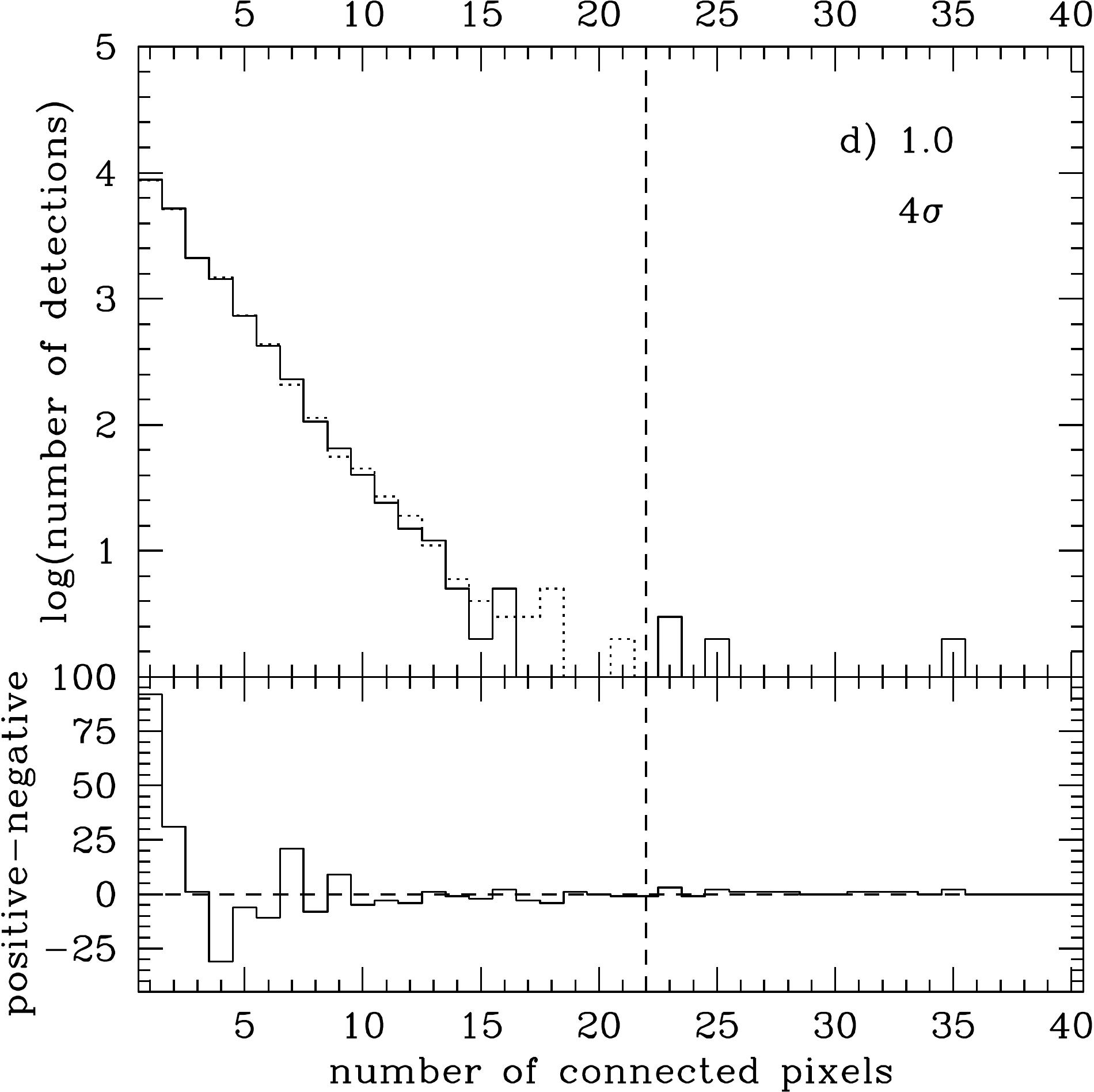}
 \includegraphics[width=5.5cm]{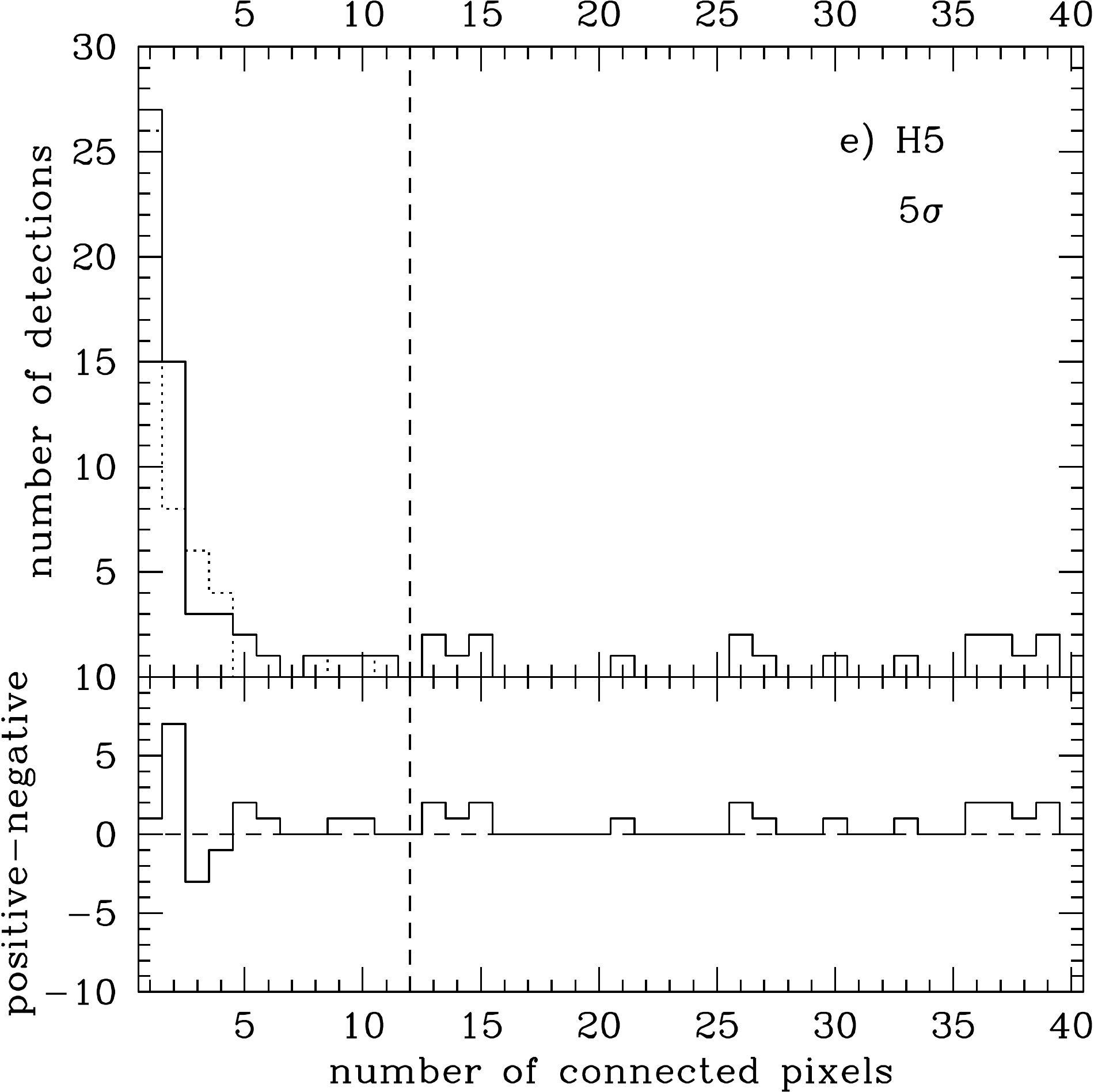}
 \includegraphics[width=5.5cm]{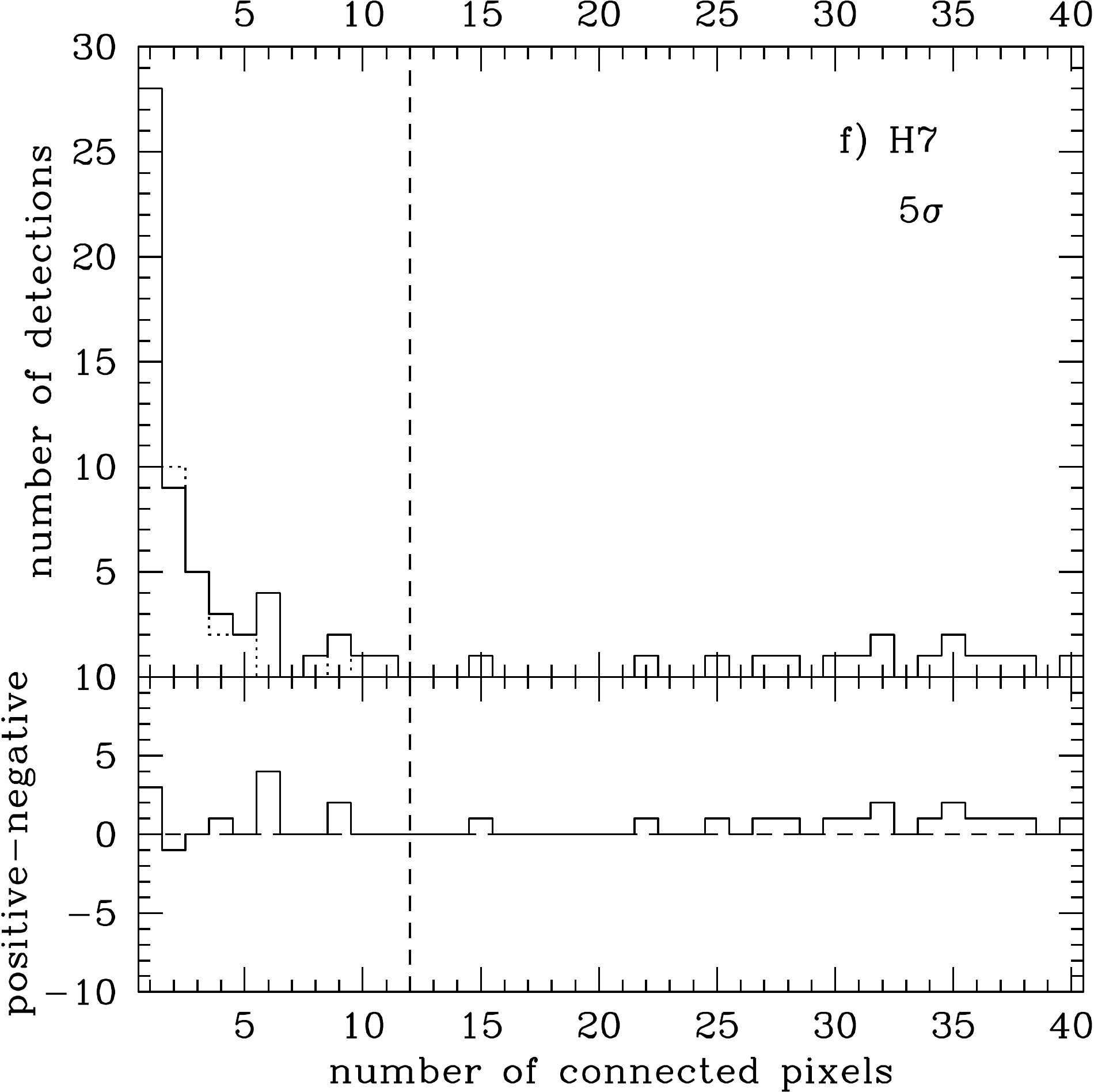}
\caption{\small Distributions of  counts of  connected pixels with  a certain
positive and  negative flux value  revealed in the  searching process.
The upper  parts of all panels show
the distribution  of counts  of the regions  of connected  pixels with
positive flux values above  the given threshold (full line histograms)
and negative flux values below $-$1 times the given threshold (dotted line
histograms).   The threshold is  $5\sigma$ for  the three  upper ($a$,
$b$, $c$)  and the last two  lower panels ($e$ and  $f$) and $4\sigma$
for panel  $d$.  The  vertical dashed line  marks the position  of the
minimum number of connected pixels  for a detection to be considered a
real object. The lower parts of  all panels present the difference between
the number counts of the regions of connected pixels with positive and
negative flux values  of a specific size over a  range of sizes, where
the size is expressed in pixels. The panels  $a$ and $d$ show  the results of the  searching process on
datacubes  with the  full  resolution.  Panels $b$  and  $c$ show  the
results of the searching  process done on datacubes spatially smoothed
with the Gaussians to the resolution 1.5 (panel $b$) and 2 (panel $c$)
times  the  original spatial  resolution.  Results  for the  datacubes
Hanning smoothed averaging over  5 neighbouring channels are presented
in panel $e$  and for the datacubes Hanning  smoothed averaging over 7
neighbouring channels are presented in panel $f$.}  

\label{histposneg}
\end{figure*}

First, the  procedure was used to  find all pixels  with absolute flux
values  above   5$\sigma$  in  the   datacubes  of  the   5  different
resolutions.  For a comparison of  the detections, the search was also
performed to  detect pixels with absolute flux  values above 4$\sigma$
for the line datacubes at the full resolution. The number of connected
pixels with flux values above  the given threshold (positive pixels in
the remaining text) or below  $-$1 times the given threshold (negative
pixels) was counted.  Pixels were  classified as connected if they had
at  least  one neighbouring  pixel  which  passed  the same  searching
criteria either  in the spatial or velocity  domain. Negative velocity
regions  were  searched for  galaxies,  but  the  velocity range  with
Galactic  emission  (the  same  velocity  intervals as  in  the  noise
calculations)  was excluded  from  this search.   The  results of  the
process  of  the  search  for  the regions  of  connected  pixels  are
presented in Figure~\ref{histposneg}.  The upper  parts of all panels show
the distribution  of counts  of the regions  of connected  pixels with
positive flux values above  the given threshold
and negative flux values below $-$1 times the given threshold. The lower parts of  all panels present the difference between
the number counts of the regions of connected pixels with positive and
negative flux values  of a specific size over a  range of sizes, where
the size is expressed in pixels.

The  thresholds  to  consider  a  detection  as  a  real  object  were
determined from the distributions  of the number of connected positive
and negative pixels for each  of the six explored cases.  The criteria
were  based   on  the  expectation  that  the   noise  is  distributed
symmetrically. Detections  were considered real objects  if the number
of connected  positive pixels  was larger than  the largest  number of
connected negative  pixels, which obviously corresponds  to noise. For
the datacubes with full resolution,  a detection then is a real object
if the  number of  connected pixels with  flux values larger  or equal
than 5$\sigma$ is larger or equal  than 10 pixels and  the number of
connected pixels with  flux values larger or equal  than 4$\sigma$ is
larger  or   equal  than  22   pixels.   The  typical  beam   size  is
approximately 32  pixels for the  datacubes with full  resolution. For
the datacubes smoothed  in the spatial resolution by  a factor 1.5 and
2, a detection would be considered real if it contains larger or equal
than 15 and larger or equal  than 25 connected pixels with flux values
larger  or equal  than  5$\sigma$  respectively.   For the  datacubes
Hanning  smoothed in  velocity over  5 and  7 channels  the  number of
connected pixels with flux values  larger or equal than 5$\sigma$ had
to  be  larger  or equal  than  12.   From  the distributions  of  all
connected pixels  with flux values  in a certain  interval, especially
from  the differences  between the  numbers of  positive  and negative
connected    pixels     with    the    same     number    of    pixels
(Figure~\ref{histposneg}),  it  is obvious  that  there  is no  hidden
distinct population of \HI\ sources with flux values at the sub--noise
levels. Such  a population  of missed objects  would have  been easily
recognisable  as a  systematic offset  of the  difference  between the
positive and negative pixels with  the same number of connected pixels
towards positive values.

Applying the determined criteria, a unique catalogue of \HI\ detections
 was  created by the union of the  6 catalogues  obtained by  applying the
 specific searching criteria on  the line datacubes of different type.
 In total,  our search  criteria reveal 70  \HI\ detections  which are
 considered real.   All 70 detections  were catalogued already  in the
 datacubes with  full resolution.  No additional  detection passed the
 ``real object''  criteria in the datacubes that  were smoothed either
 in the spatial or velocity domain. There are no detections with lower
 column densities  than the limiting column density  which has already
 been achieved in the line datacubes at the full resolution.

There  were  4 regions  detected  where  the  \HI\ emission  was  very
extended and the  objects detected in these regions  were of extremely
irregular shape.  For these cases the final decision what is an object
was made by eye, after consulting previous observations available from
the literature. These  \HI\ objects will be termed  extended from here
on.  The extended  objects are  the  objects with  the WSRT--CVn  id's
ranging from 63  and 68 including (with two  objects with the WSRT-CVn
id 67: 67A and 67B). More details on these and the rest of the objects
will be given in the following text.

\subsection{\HI\ parametrisation}
\label{sub_hiparam}

The  \HI\ parametrisation  of  the detected  objects  was carried  out
 combining programmes  developed for  this survey and  standard MIRIAD
 programmes.   The  cubes with  full  resolution  in  the spatial  and
 velocity domains  were used to  determine the parameters of  the \HI\
 detections.

The next task after the detecting  of the objects was to determine the
 total flux of  the object. Due to the uncertainty of the process, we have done this in two ways and taken the average of the two measurement as the total flux estimate. 

The first method was to select the pixels  which belong to an
 object (i.e. to  mask all pixels with  the signal). This was  done in two
 steps. The first step was developed in order to recover the total flux of an object, and the second step was developed to recover the shape of an object. Starting  from the  pixel  of the  detection with  the
 maximum  flux value,  the object  was enlarged  considering  that all
 connected pixels  with flux values  larger than or equal to 3.5$\sigma$
 belong to the  object, using our definition of  $\sigma$.  The
  3.5$\sigma$ limit  was obtained as the optimal limiting
 flux  value after  testing various  assumed limits to recover the total flux of an object in the INVERTED datacubes. For this test, we   used  the clean
 components  of  various objects  inserted  into  the line  datacubes,
 convolved with  a cleaned beam  of the datacube of  the consideration
 previous to the insertion.  The second step consisted of changing the
 shape of the masked pixels in each of the planes where the object was
 detected with significance above 3.5$\sigma$, to account for the fact that the detections in line datacubes are convolved with the beam of some finite size
 (which defines the  spatial resolution  element). First, to remove the detected pixels which are most probably only the noise, pixels with  more  than 3
 neighbouring pixels  which do  not belong to  the object  (their flux
 values  are smaller than  3.5$\sigma$)  were deleted.   After that,  remaining
 pixels in  the mask with at  least one neighbouring  pixel which does
 not belong  to the  object were marked  as the border  pixels.  For
 each  of the border  pixels an  area of  a beam  size centred  on the
 border  pixel was  inspected. All  pixels with  positive  flux values
 inside of  the beam  area studied were  added to the  detection.  The
 total  integrated flux  ($S_{\rm int, c}$)  of the  detections  was obtained  by
 summing the flux in pixels determined  to belong to the object in all
 channels and  dividing this value by the beam area.  The spatially
 integrated  peak flux  is  simply  the maximum  value  of the  fluxes
 integrated  in  the individual  channels  (maximum  in the  spectrum,
 $S_{\rm peak, c}$). From now on, we will use the term integrated flux instead of
 the total integrated  flux and the term integrated  peak flux instead
 of the spatially integrated peak flux.

The size  of the  detected  objects,  which  were not  classified  as
 extended  objects, was  estimated  using the  MIRIAD  task IMFIT. A
 two-dimensional  Gaussian was  fitted  to the  \HI\  map, created  by
 integrating  the flux  over the  velocity channels  contained  in the
 masked pixels. When possible, we estimated the size of an object from
 the  size  of an  ellipse  fitted to  a column density isophote of 
1.25  $\times$
 10$^{20}$  atoms cm$^{-2}$. This isophotal  level corresponds  to a
 value of 1 \Msol\ pc$^{-2}$. For the small objects (WSRT--CVn 7, 10, 11, 12,
 15, 19, 22, 25,  30, 31, 42, 43, 47 and 61) we  used the FWHMs of the
 fitted Gaussian along  the major and minor axis  as a {\it rough} indicator
 of the  angular size of an  object.  The FWHMs along  major and minor
 axis  and the positional  angles obtained  were deconvolved  with the
 beam.  The  exceptions are detections  with the WSRT--CVn id's  7 and
 47, which are  too small to be deconvolved.   For these two detections
 we  present  only the  values  of  the  FWHMs of  a  two-dimensional
 Gaussian  convolved with  a  beam  instead of  their  size. The  last
 parameters will be  used only as an indication  of the inclination of
 these two  objects.  For the  objects without a known  counterpart in
 the  literature,  we use  the  position of  the  peak  of the  fitted
 Gaussian as the position of that particular object (the cross-correlation with literature detections is described in Subsection~\ref{sub_crosscorr}. Most of our detections are very small and the estimated \HI\ sizes are very uncertain (see Table~\ref{prop2}; for the objects with an estimated size comparable to or smaller than the beam size we use an $``<"$ sign to indicate that these sizes are probably just upper limits). However, they are usefull for a comparison with the sizes estimated from the optical measurements.  We do not measure \HI\ sizes of the extended objects, as the noise distribution in the datacubes of these objects is much more inhomogeneous and we are not able to apply the masking method reliably for these objects.

\begin{figure}
  \centering
  \includegraphics[width=0.44\textwidth]{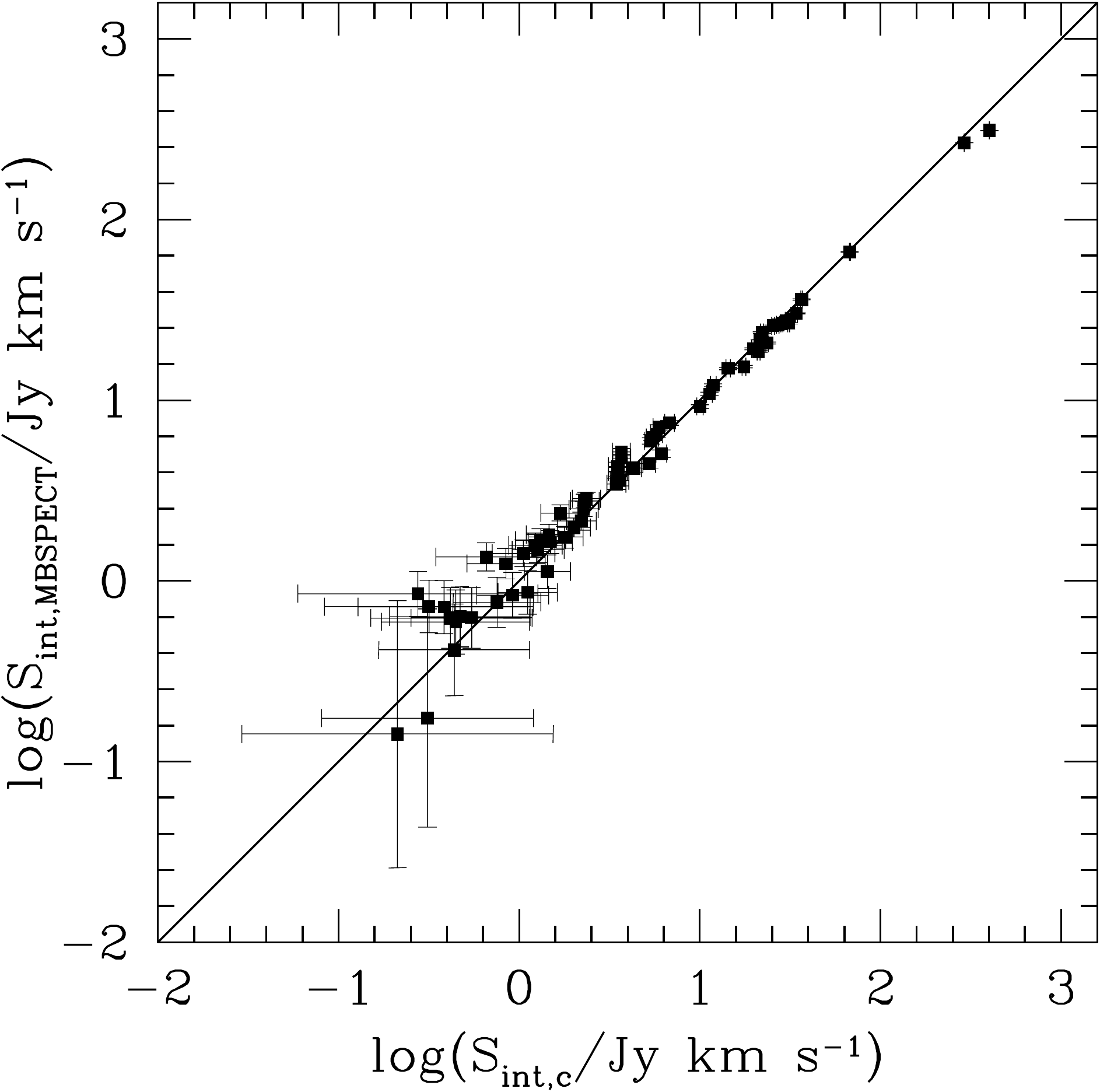}
  \includegraphics[width=0.44\textwidth]{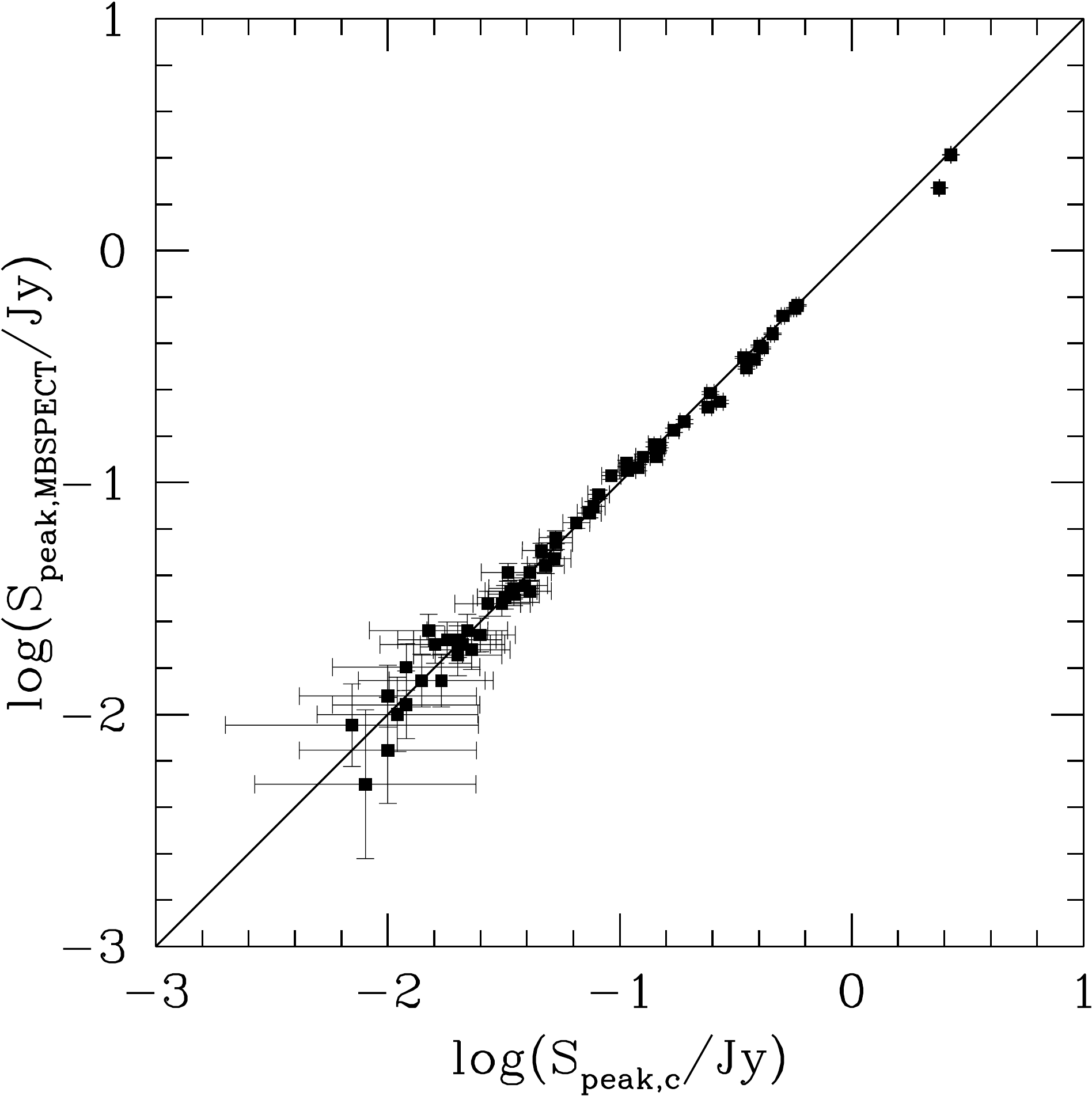}

\caption{\small Comparison  of  the   measured  integrated  flux  (top)  and
integrated   peak   flux    (bottom)   values   using   two   different
methods for objects for which we have both measurements. Horizontal  axis present the  measurements based on  the flux
values   in   the   pixels    defined   to   belong   to   an   object
(masking).  Vertical axis  present the  measurements using  the MIRIAD
task  MBSPECT, defining the  boxes around  detections. The estimated errors for the presented parameters are described in Subsection~\ref{sub_hiparunc}.}
\label{comp_s}
\end{figure}

As the second method to estimate the total flux of an object the  MIRIAD  programme  MBSPECT  was  used. The integrated
\HI\ spectrum of  each detection was made, summing  the flux in pixels
contained in a box placed around the object and weighting the sum with
the inverse  value of  the beam.   The size of  the box  was estimated
individually for each object based  on the extent of the \HI\ emission
and always slightly bigger than the object itself, both in the spatial
and in the velocity domain.  For the extended objects, which have id's
from 63 till  68, this was the only method used  to estimate these two
parameters. We mark the integrated fluxes obtained with MBSPECT $S_{\rm int, MBSPECT}$ and the corresponding peak values of the integrated
profiles $S_{\rm peak, MBSPECT}$. In Figure~\ref{comp_s}  the difference between integrated
fluxes and  peak fluxes measured using two  different methods [(1)masking by defining pixels which belong to an object  
and (2) summing in a  box using  MBSPECT] is  plotted. The  difference is
larger  for the  objects with  lower values  of integrated  fluxes and
integrated peak fluxes,  where the influence of the  noise in the flux
values is  relatively higher. Where  possible, the average  value from
the two methods  used to estimate integrated flux (\sint) and integrated peak
flux (\speak) will  be used as the  final estimate of these  two parameters for
the detected \HI\ objects.

The MIRIAD task MBSPECT was also used to parametrise the detections in
 velocity  space.   The  widths  of  the  profiles  of  the
 detections  were  measured  at  the 20\%  ($W_{20}^{obs}$)  and  50\%
 ($W_{50}^{obs}$) levels of  the peak flux in the  spectrum, using the
 methods  of  width maximisation  and  minimisation  in MBSPECT.   The
 maximisation  procedure measures  the line  widths starting  from the
 velocity limits given when specifying  the box around a detection and
 moves inward  till the  percentage of the  peak flux is  reached. The
 minimisation procedure  starts at velocity  at which the  profile has
 maximum and  searches outward. The  velocities at the centres  of the
 four  measured  profile  widths  were  also  estimated.   The  radial
 velocity of  a detection was estimated  as the average  of those four
 measured  velocities.  The  velocities in  the datacubes  and spectra
 were calculated  from the observed frequencies $\nu$  using the radio
 convention  $V_r  =  c  (1 -  \frac{\nu}{\nu_{0}})$.   The  estimated
 systemic  velocity  was recalculated  to  the  value  in the  optical
 convention $V_o = c (\frac{\nu_{0}}{\nu} - 1)$.  The velocities in the
 optical  convention  were recalculated  from  the  geocentric to  the
 barycentric  frame. In  addition, they  have been  corrected  for the
 motion of  the Sun around the  galactic centre and the  motion of the
 Galaxy in the Local Group using the expression \citep{Yahil.etal.1977}

\be  V_{LG} = V_o + 296 \sin l \cos b - 79 \cos l \cos b -36 \sin b  \ee

\noindent
which is  similar to  the IAU convention. In this formulae
$V_{LG}$  is the Local  Group velocity  and $l$  and $b$  are the galactic
coordinates of the detection.

The profile  widths measured from the data have been corrected for instrumental resolution. We
 used
method given by \citet{Verheijen&Sancisi.2001} to correct for broadening of  the global \HI\ profiles due to a
finite  instrumental velocity  resolution. Assuming a Gaussian line shape for the edge of the profile, the slopes of which are determined by the turbulent motion of the gas with a velocity dispersion of 10 \kms, this correction can be written in the form

\be    W_{20} = W_{20}^{obs} - 35.8 \left[ \sqrt{1 + \left(\frac{R}{23.5}\right)^2} - 1 \right]     
\label{eq_vs2001}
\ee
\be    W_{50} = W_{50}^{obs} - 23.5 \left[ \sqrt{1 + \left(\frac{R}{23.5}\right)^2} - 1 \right].  \ee   

\noindent
for  the   widths  at  20$\%$   and  50$\%$  of   the  peak  flux
respectively.  The observed  widths $W_{20}^{obs}$  and $W_{50}^{obs}$
were  calculated  by averaging  the  20$\%$  and  50$\%$ level  widths
measured in the maximisation and  minimisation procedure by  MBSPECT. The
instrumental velocity  resolution $R$ expressed in \kms\ was taken
to be 33 \kms.


\section{Parameter accuracy and completeness of the survey}
\label{sec_errors}

We use  an empirical approach to  assess the accuracy  of the measured
parameters of the  detections and the completeness of  the survey. Our
method is based  on the inserting a large  number of synthetic sources
throughout the selected survey data.  The major inputs to estimate
the accuracy of the  measured parameters are the recovered properties
of the synthetic sources.  The completeness of the survey is determined
from the rate at which the synthetic sources could be recovered.

As a basis  for the simulations, 10 line  datacubes of full resolution
were  selected from  the 1372  line  datacubes produced  in the  whole
survey. We refer to these datacubes as the basis datacubes. The datacubes selected were to our knowledge object free. Seven of the datacubes were selected from the central part of the area
covered by the  survey, while three of the  datacubes were selected to
be datacubes from  the edges of the area covered  by the survey;
two of them  are datacubes with one edge and one  is a corner datacube
with two edges.

The majority  of the objects  detected in the  WSRT CVn
survey are small \HI\ objects, and the uncertainties and the completeness of these detections were the main focus for designing these simulations.  
The synthetic objects were  created to resemble small \HI\ objects, and therefore our simulations are not optimal for all possible types of \HI\ objects. Five objects  of different sizes in the  spatial and velocity
domains with  different distributions of  flux were created  from the
CLEAN  components of  \HI\  objects detected  in  the survey.  Created
objects  are 2,3,4,5 and  6 channels  wide. The profiles of the all synthetic objects were also of triangular shape. In each of the basis datacubes
10   objects    were   inserted   and    distributed   quasi-randomly.
Quasi-randomness in this context means that objects were inserted only
in channels with positive velocities,  as are all the real detections,
and they were distributed in the  datacube such as not to overlap with
each  other. It is possible is that some overlapping sources could be partially accreted dark galaxies which show up only as asymmetries in single detections. Our small sources are too small with respect to the beam size to study their shapes in detail.  For the large sources, this remains an unexplored possibility. 

The  same synthetic objects were inserted at  the  same relative
positions  in all  of the  datacubes (same  $x$, $y$  and $z$  of  the three
dimensional  datacube), in order to emphasise the influence of the underlying noise in the datacube on the measured properties of inserted objects. Obviously, the noise distribution differs from datacube to datacube. Before  inserting  the   objects  into  the
datacube, their  flux was  rescaled and they  were convolved  with the
beam of the line datacube in which they were going to be inserted. Each  of the  synthetic objects was inserted  in two positions in
the datacube, with  two different flux values. In  total ten runs were
made, rescaling the maximum flux  values two times for the 5 different
model-objects  in  each  of  the  datacubes.   In  the  first  run  of
simulations, the  maximum value of different model-objects were  fixed at 1.0
and 3.0  mJy and in each of  the following 9 simulation  runs the peak
value was increased by 0.2 mJy.  These flux values were chosen in such
a way  to ensure that there  is a fraction of  synthetic sources which
will not be detected. In total 1000 different objects were inserted in
10 different datacubes.
 
The datacubes  with inserted synthetic objects were  then searched for
these synthetic sources in a manner identical to the searching process
used in  the WSRT  CVn    survey,     as    described    in
Subsection~\ref{searching}.  Detections which  satisfied  the criteria
for  real  objects  were  parametrised   the  same  way  as  the  real
detections, described  in Subsection~\ref{sub_hiparam}. The simulation
described above was used to estimate the uncertainties of the measured
\HI\ parameters  and the completeness of the WSRT CVn  survey
in  the following two subsections, respectively.

One of the shortcomings of our simulations is that all synthetic sources have profile widths of triangular shape. \citet{Zwaan.etal.2004} estimated parameter uncertainties and completeness of HIPASS (a single dish survey) using synthetic sources of Gaussian (e.g. triangular), double-horned and flat-topped profile shapes. Within the errors, the completeness of the survey is the same for all types of profiles. \citet{Zwaan.etal.2004} did not discuss the uncertainties in the measured parameters on the profile shape.

\subsection {Parameter uncertainties}
\label{sub_hiparunc}

\begin{figure*}
  \centering
  \includegraphics[width=5.5cm]{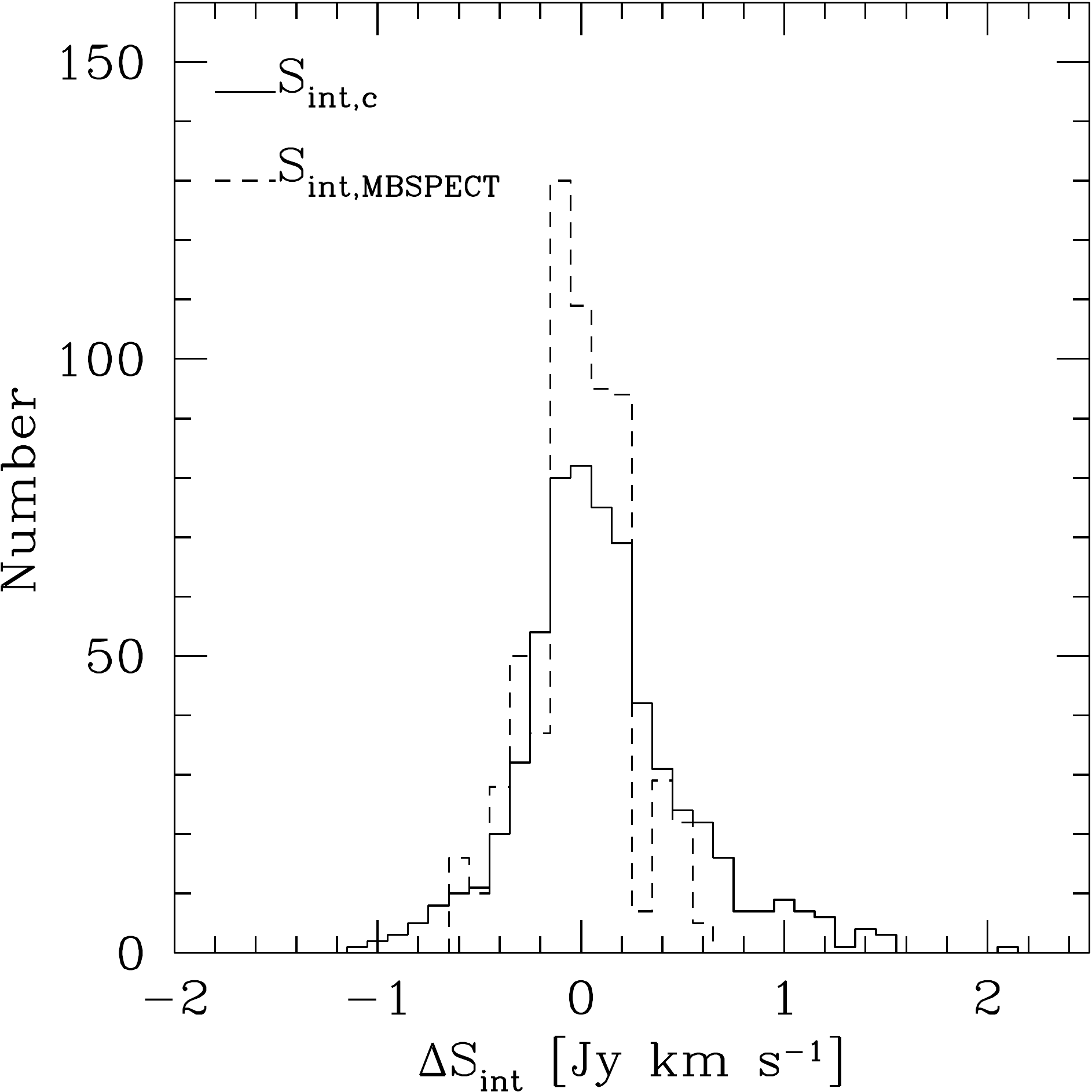}
  \includegraphics[width=5.5cm]{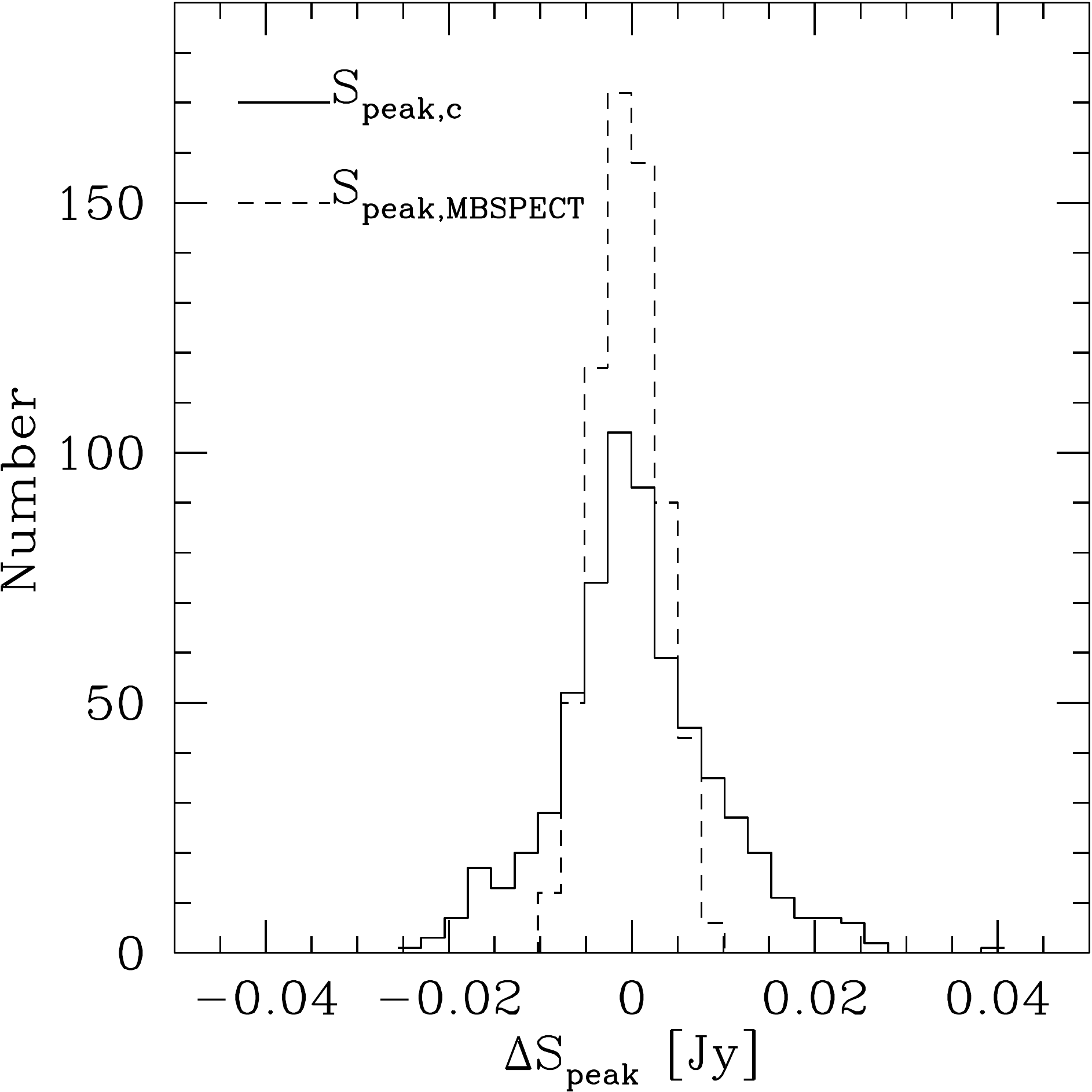}
  \includegraphics[width=5.5cm]{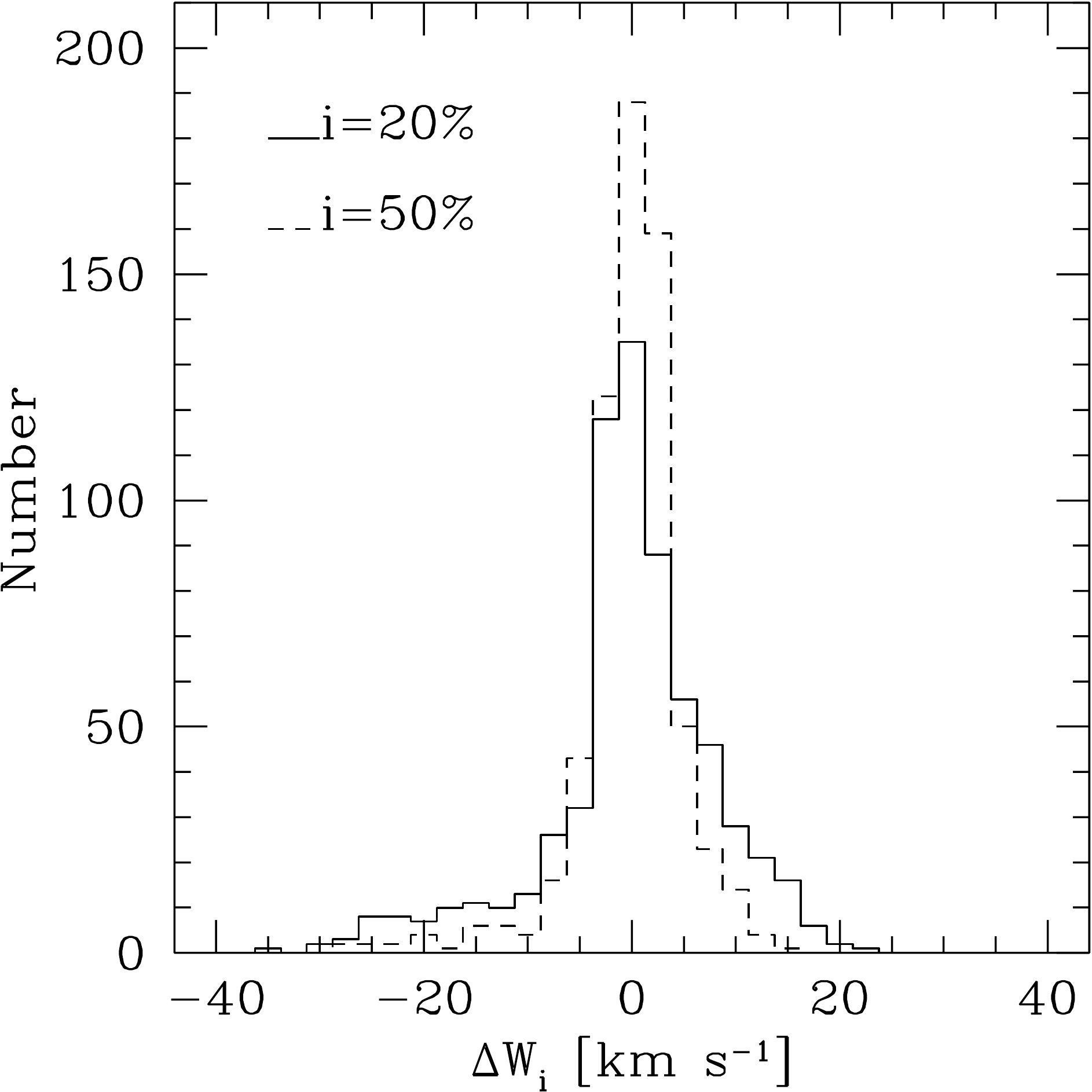}
  \caption{\small Comparison  between inserted  and measured  parameters. The
  first two panels show the  difference between the values of inserted
  parameters  and  parameters measured  by  forming  a  mask around  a
  detection (full line histogram) and parameters measured with MBSPECT
  (dashed  line  histogram).  The  first panel  shows  the  difference
  distribution for integrated flux  values. The second panel shows the
  distribution of  differences in  integrated peak flux  values. Third
  panel shows the difference  between the inserted profile widths and
  recovered profile widths at 20\% of the maximum in a spectrum (full
  line histogram)  and at  50\% of the  maximum in a  spectrum (dashed
  line histogram), measured with MBSPECT.  }
  \label{uncert}

\end{figure*}

The  uncertainties of  the \HI\  parameters  can be  estimated from  a
comparison of  the assigned and  measured properties of  the synthetic
objects  revealed  in the  simulated  datacubes.   In the  simulations
described  above, 794  of  the inserted  1000  synthetic sources  were
recovered using the searching  criteria defined for the datacubes with
the full  resolution: at  least 22 connected  pixels with  flux values
larger  or  equal  than  4$\sigma$.  Parametrisation  of  the  sources
detected in  the simulation was  carried out and the  distributions of
differences between  the real and the parameterised  properties of the
population  of  synthetic  objects  revealed  in  the  simulation  are
presented in Figure~\ref{uncert}.

From  the distributions  of differences,  the uncertainties  of \sint,
\speak\  and profile  widths measured  at 50$\%$  and 20$\%$  of  the line
maximum were calculated as the standard deviation in the corresponding
distributions. The uncertainties of  the measured integrated fluxes in
the WSRT  CVn survey are $\sigma =  0.421$ Jy \kms\ for  the case of
the flux summed  inside of the defined contour,  $\sigma = 0.242$ Jy
\kms\ for the flux inside of a box around a detection, while $\sigma =
0.240$ Jy \kms\ for the flux  of an object calculated as the average
value measured from the two  techniques used.  For the integrated peak
fluxes, uncertainties are  $\sigma = 8.8$ mJy, $\sigma  = 3.7$ mJy and
$\sigma = 4.7$ mJy for the three methods used, given in the same order
as  the \sint\  uncertainties  above.  Uncertainties  for the  profile
widths (as observed) are $\sigma = 5.1$ \kms\ and $\sigma = 8.4$ \kms\
for the profile widths measured at 50$\%$ and 20$\%$, respectively.

\begin{figure*}
  \centering
  \includegraphics[width=5.5cm]{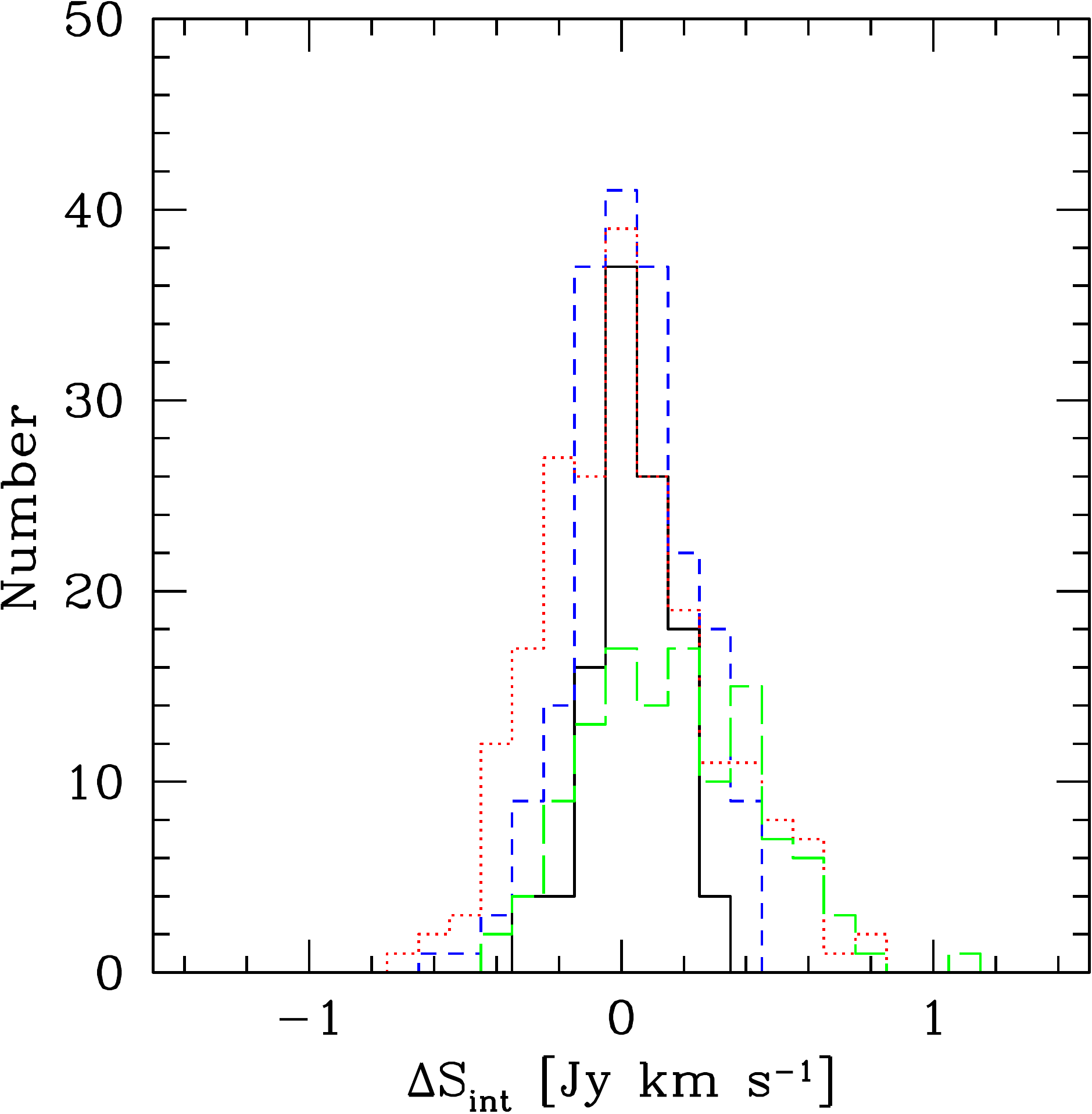}
  \includegraphics[width=5.5cm]{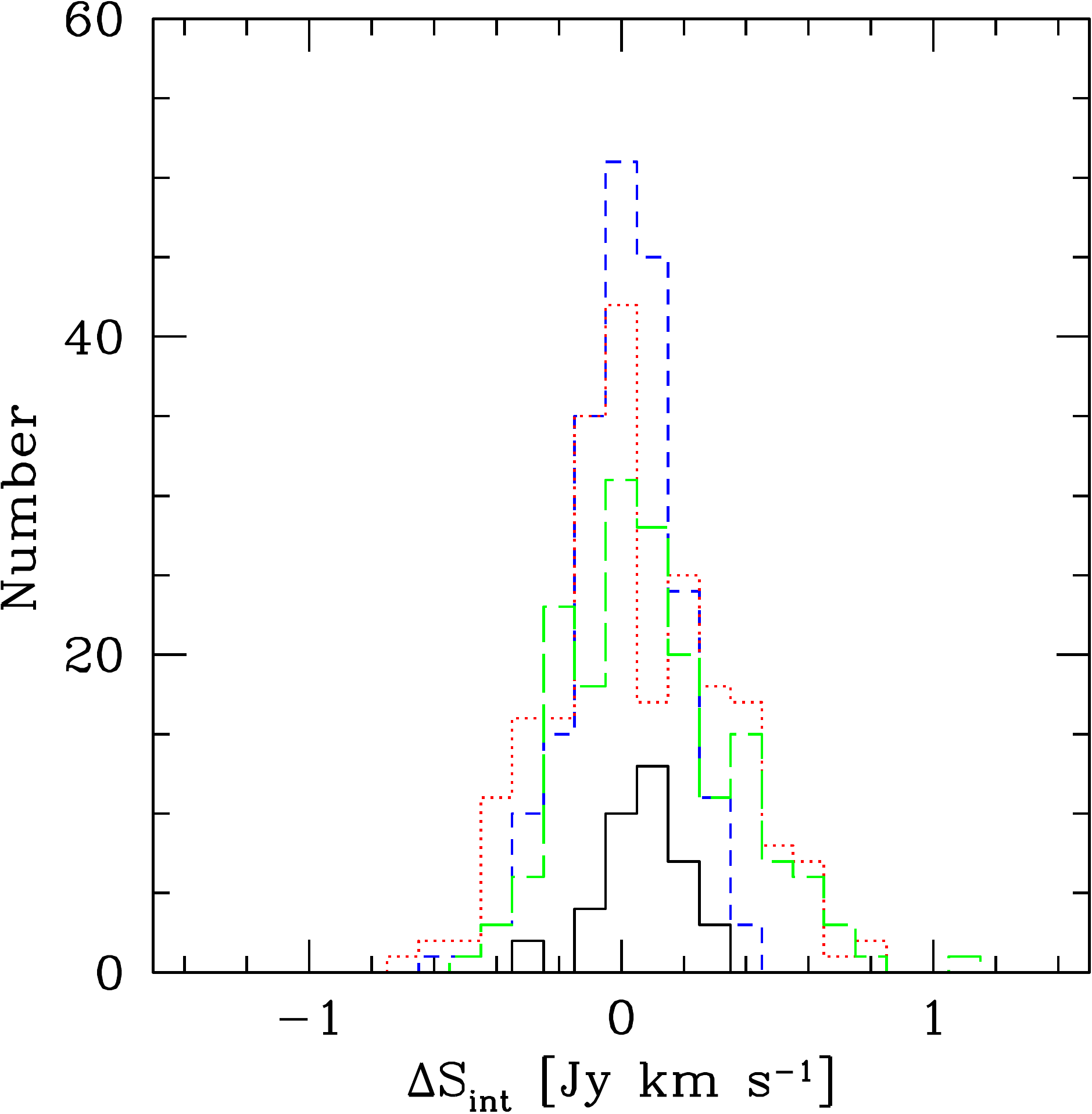}
  \includegraphics[width=5.5cm]{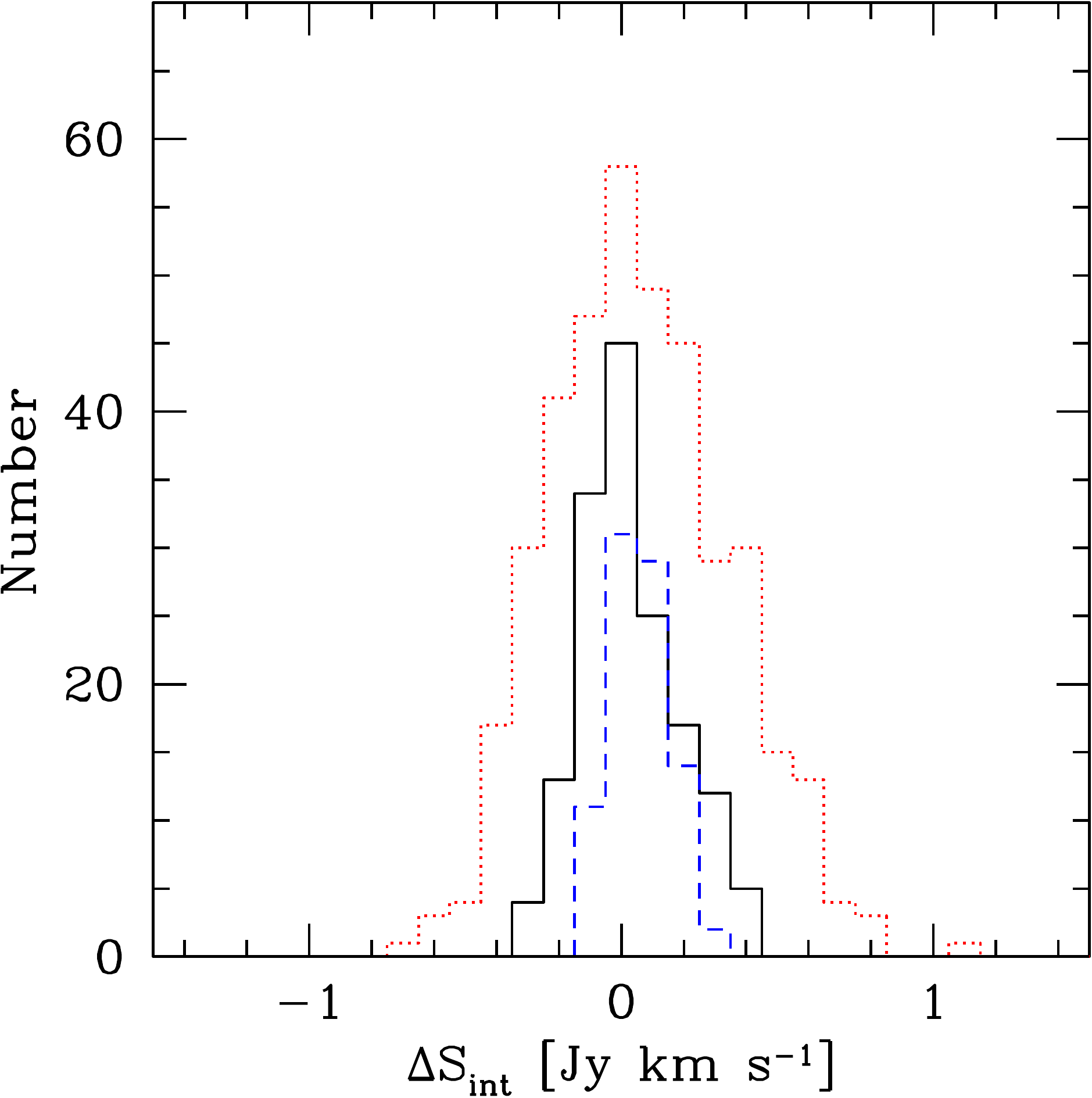}
\caption{\small  Distributions of  $\Delta$\sint,  differences between
the inserted  integrated fluxes and  detected integrated fluxes,  as a
function of  inserted \sint, \speak\ and profile  width.  The detected
\sint\ values  are calculated as  the average value of  the integrated
flux  calculated  by  defining  a  mask around  a  detection  and  the
integrated flux obtained by using  the MIRIAD task MBSPECT, defining a
box  around  a detection. In  the  left  panel the  continuous (black) line
corresponds  the the  distribution of  $\Delta$\sint\ for  \sint\ $\le
0.5$ Jy \kms,  the short dashed line (blue) is the  same type of distribution
for $0.5 <$ \sint\  $\le 1$ Jy \kms, dotted (red) line for  $1 <$ \sint\ $\le 2$
Jy \kms\ and long dashed (green) line for \sint\ $ > 2$ Jy \kms.  Middle panel
shows the distribution of $\Delta$\sint\  as a function of \speak. The
continuous (black) line  is for \speak\ $\le  0.01$ Jy, the  short dashed (blue) line
for $0.01 <$  \speak\ $\le 0.02$ Jy, dotted (red) line is  for $0.02 <$ \speak\
$\le 0.04$  Jy and long dashed (green) line is for  \speak\ $ > 0.04$  Jy. The
right panel shows the distribution  of $\Delta$\sint\ as a function of
$W_{20}$ for the  range of: $W_{20}^{obs} \le 45$  \kms\ -- continuous (black)
line,  $45 <  W_{20}^{obs}  \le 60$  \kms\  -- short  dashed (blue) line  and
$W_{20}^{obs} > 60$ \kms\ -- dotted (red) line. 
\label{dsintpar}}

\end{figure*}

The detectability of a 21-cm signal  depends not only on the flux, but
also on  how this flux is  distributed over the velocity  width of the
object  of  consideration.  There   is  probably  a  more  complicated
dependence of  the uncertainties of  the estimated \HI\  parameters on
the  intrinsic properties of  an object.   Given the  relatively small
number  of  detections in  the  WSRT CVn  survey,  we  neglect such  a
dependence in our  results.  We only demonstrate the  existence of the
additional  dependence of  measured  \sint\ values  on  the values  of
\sint,  \speak\  and profile  width.   The  results  are presented  in
Figure~\ref{dsintpar}.     Here,   the    difference   $\Delta$\sint\
corresponds to the difference between the value of the integrated flux
inserted and  the integrated flux measured, calculated  as the average
value  of the  integrated flux  obtained  by using  our programmes  --
defining a mask  around a detection, and the  integrated flux obtained
by  using  the  MIRIAD  task  MBSPECT  --  defining  a  box  around  a
detection. The uncertainty of \sint\ is 0.130 Jy \kms\ in the range of
true values  \sint\ $\le  0.5$ Jy \kms\  (continuous line in  the left
panel in Figure~\ref{dsintpar}), 0.188 Jy \kms\ for $0.5 <$ \sint\ $\le
1$ Jy \kms\ (short dashed line), 0.280 Jy \kms\ for $1 <$ \sint\ $\le 2$
Jy \kms\  (dotted line) and 0.278 Jy  \kms\ for \sint\ $>  2$ Jy \kms\
(long dashed line).   To test the dependence of  $\Delta$\sint\ on the
\speak\ value of  the inserted detection, we split  the \speak\ values
in  the four  arbitrary intervals:  \speak\ $\le  0.01$  Jy (continous
line),  $0.01 <$  \speak\ $\le  0.02$ Jy  (short dashed  line),  $0.02 <$
\speak\  $\le 0.04$  Jy (dotted  line) and  \speak\ $>  0.04$  Jy (long
dashed line). The  uncertainties are 0.134 Jy, 0.168  Jy, 0.278 Jy and
0.265 Jy  for the given  intervals, respectively. Finally,  we divided
$\Delta$\sint\  values  in  the   three  intervals  depending  on  the
$W_{20}^{obs}$  of the  object.   The uncertainty  in  \sint\ for  the
objects with:  $W_{20}^{obs} \le 45$ \kms\ (continuous  line) is 0.156
Jy \kms, $45 < W_{20}^{obs} \le 60$ \kms\ (short dashed line) is 0.091
Jy \kms\ and $W_{20}^{obs} >$ 60 \kms\ (dotted line) is 0.285 Jy \kms.

\subsection {Completeness}
\label{sub_compl}

Completeness of the  survey is the fraction of  galaxies detected in a
given volume  down to the  limiting sensitivity.  The  completeness of
the  blind  WSRT  CVn  survey  is  addressed  using  the  Monte  Carlo
simulations described.  It is defined  here as the ratio of the number
of synthetic  objects detected  in the simulations  and the  number of
synthetic objects inserted in the simulation.

\begin{figure}
  \centering
  \includegraphics[width=0.44\textwidth]{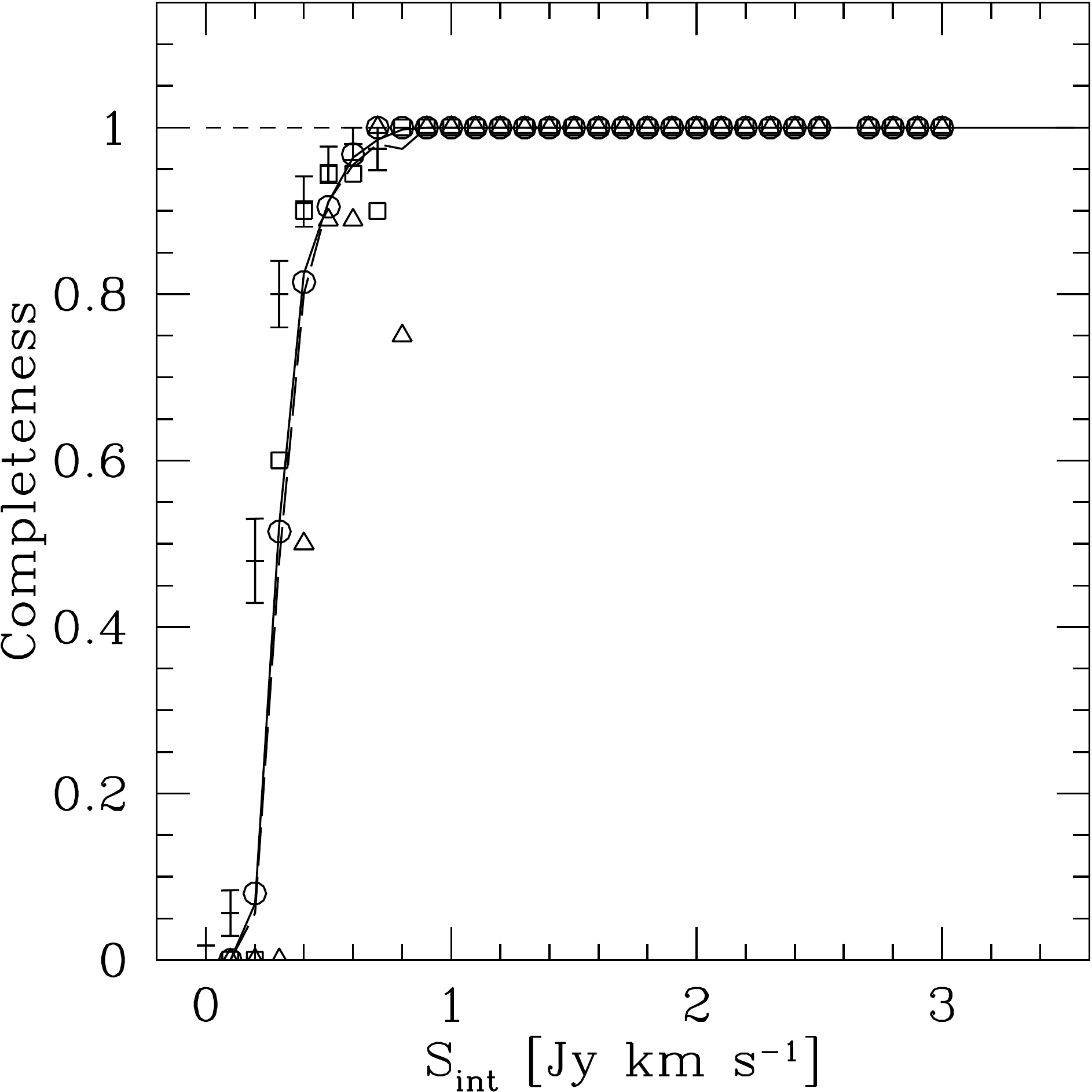}
  \includegraphics[width=0.44\textwidth]{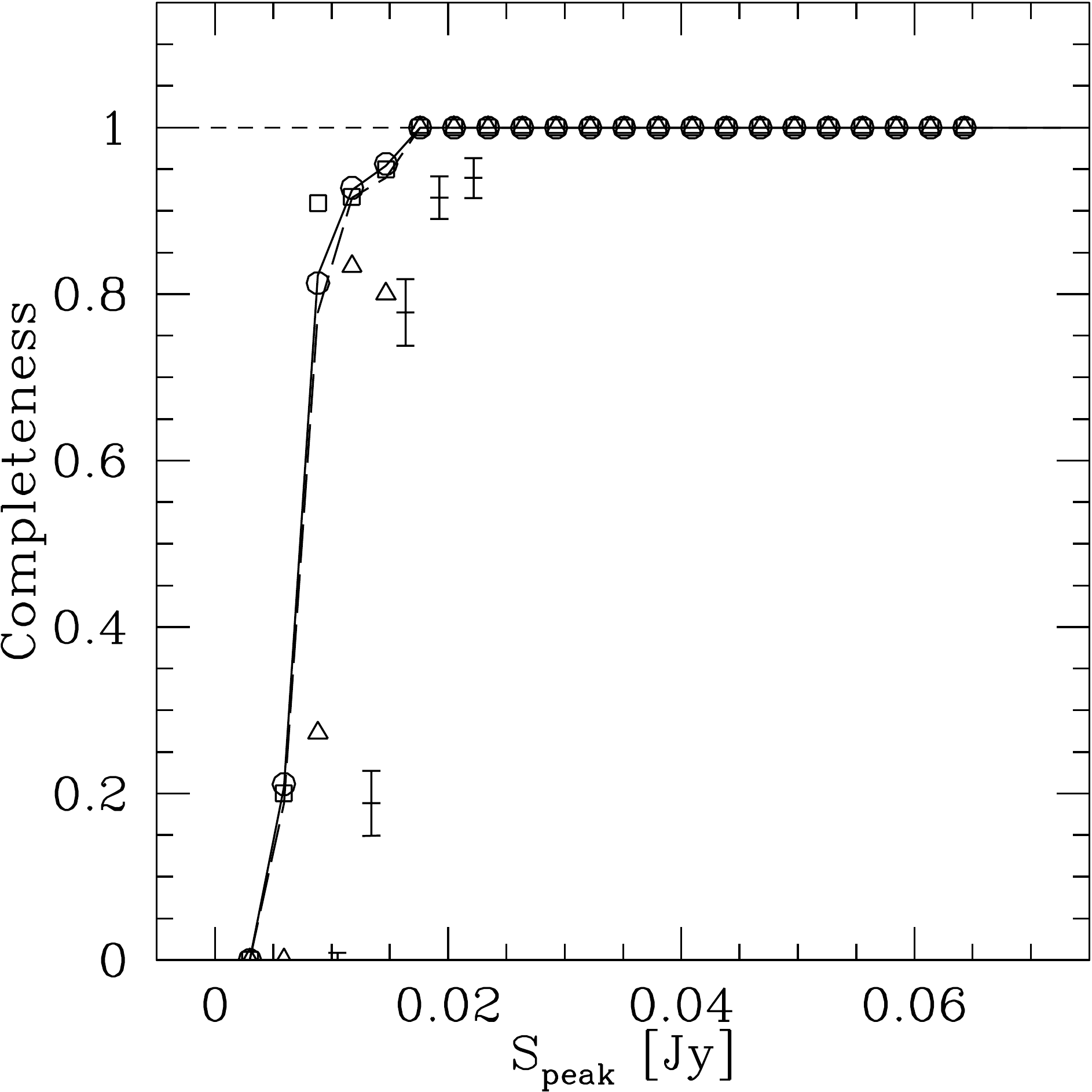}

\caption{\small Completeness  of the WSRT CVn survey.  Circles  represents  the
fraction  of objects  detected  in the  datacubes  within the  survey,
without an edge.  Squares correspond to the fraction of detected objects
residing  in the  datacubes  with  one edge  and  triangles represent  the
fraction  of objects  detected in  the datacubes  with two  edges. The
continuous  line  connects  the  bins with a fraction  of
detected objects weighted with the relative abundance of a type of the
datacube in  which the detection  resides. The long  dashed line  connects 
fractions of all objects detected in  the simulation regardless of the 
 type  of the  datacube in  which the detection  was inserted.  The short
dashed horizontal  line represents  the line for  which the  survey is
complete. The errors are estimated from the binomial statistics. For clarity, the errors are presented only for the completeness based on all datacubes without weighting for the datacube type, and they are offset along x-axis.}
\label{completeness}
\end{figure}

The completeness  of the survey is  estimated as a  function of \sint\
and \speak\ values.   To take into account all  possible sources which
would  be detected  in the  real survey,  datacubes with  the inserted
synthetic sources  were smoothed in  the spatial and  velocity domain.
The smoothing  was identical to  the smoothing of the  real datacubes.
The smoothed  datacubes were searched in all  resolutions applying the
criteria as  defined in Subsection~\ref{searching}.   This resulted in
the completeness  corrections shown in  Figure~\ref{completeness} that
were later applied in deriving the \HI\ mass function.

The  fraction  of  datacubes  with  one  or  two  edges  used  in  the
simulations was much larger than  the fraction of datacubes with edges
in  the real  survey. To  account for  this, the  completeness  of the
survey was estimated for each  type of the datacube independently. The reason for including the datacubes with edges and testing them separately was that the noise distribution in the edge cubes is much more inhomogeneous. The
completeness of the whole WSRT CVn survey was calculated weighting the
number of detected objects with  the relative abundance of the type of
datacubes (in the  survey) in which these objects  were detected.  The
weighted completeness  is considered  to be the  best estimate  of the
completeness of the whole survey.  It is presented with the continuous
line in Figure~\ref{completeness}.   From the simulations carried out,
it  follows that  the WSRT  CVn survey  is complete, at least in a statistical sense,  for  objects with
approximately \sint\ $> 0.9$ Jy  \kms\ and \speak\ $> 0.0175$ Jy. From
70 objects  detected in the  WSRT CVn survey,  12 of them  have \sint\
values in  the range for  which the survey  is incomplete. For  only 2
detections  the  incompleteness  is  larger than  50\%.   The  minimum
integrated flux  of an object  has to be  0.2 Jy \kms\ (centre  of the
first bin with a non-zero completeness) in order to be detected in the
WSRT CVn survey.

\section{Properties of the \HI\ detections}
\label{sec_prop}

\subsection{Comparison of the detections with previous observations}
\label{sub_crosscorr}

The cross--correlation  of the objects  detected in the WSRT  CVn blind
survey  with   known  objects   was  conducted  using   the  NASA/IPAC
extragalactic   Database  (NED).    Both,   positional  and   velocity
information, if available, was used to identify a counterpart of each of the detected
\HI\  sources.  In  addition, the  Lyon/Meudon  Extragalactic Database
(HYPERLEDA)  was used  and the  second generation  Digital  Sky Survey
(DSS) images,  centred on  the position of  the WSRT  CVn detections,
were visually  inspected. We inspected also the  SDSS images. We used the SDSS object identification and their redshift, if obtained, only in combination with the visual inspections of the images of cross-correlated galaxies, or parts of them, because  of the not yet fully resolved problem with deblending of the extended sources (e.g. \citealt{West.2005}, http://www.sdss.org/dr7/products/catalogs/index.html$\#$cav$\_$lowlat).

In total  67
objects detected in the WSRT  CVn survey were identified as galaxies
previously  detected  in  one   of  the  optical  wavebands. The cross-correlation is based on both the positional and velocity information for 58 objects. Objects
WSRT--CVn--67A  and  WSRT--CVn--67B  were  cross-identified  with  two
galaxies based on  position and redshift information in the  literature:
NGC4490 and NGC4485, respectively. Using  our data, we were not able to
separate  the  21-cm  emission  detected  around  WSRT--CVn--67A  and
WSRT--CVn--67B  into two  individual detections.  Object WSRT--CVn--34
(UGCA290) looks  like an interacting binary system,  addressed as such
in  some references.  We consider  it as  one object,  a  patchy dwarf
galaxy,  based  on the  resolved  stellar  photometry  carried out  by
\citet{Makarova.etal.1998}.

For the remaining 9 \HI\ detections (WSRT CVn 8, 13, 17, 19, 25, 30, 31, 47 and 55) the cross-correlation with the previous detections is based solely on the positional information of the assigned optical counterparts. Given  that these galaxies are best  visible in  blue
light  and $\sim$  10-20 arcsec  in size,  it is  most  probable, even
without   knowing  their   redshifts,  that   they  are   the  optical
counterparts  of  the \HI\  detections. Close to the position of two additional \HI\ detections, galaxies  are visible  in  both DSS and SDSS images. One of these galaxies (which overlaps WSRT CVn 42) is detected in the SDSS splitted into multiple detections (in the currently last available SDSS data release 7), therefore there does not exist a uniquely previusly identified galaxy for this \HI\ object. Two blue galaxies are detected in the area of the projected \HI\ of WSRT CVn 40 (see Appendix~\ref{app_atlas}). One of these galaxies has the SDSS measured redshift $z=0.029$, which completely disqualifies it as a possible contributor to the detected \HI. We cross-correlate therefore WSRT CVn 40 with the other galaxy. This galaxy is not detected by the SDSS pipeline, probably due to its proximity to a star.

An important question related to the cross-correlation of objects when using only positions and some properties of galaxies is what is the probability that this \HI-optical pair is only a chance projection. The geometric probability $P$ that a galaxy of magnitude $m$ and at  angular distance $\theta$ from the studied galaxies is only a chance projection (neglecting the correlation properties of galaxies) is given by \citep[e.g.][]{Wu&Keel.1998}

\be   
P(\theta, m) = 1 - \exp(- \pi \rho(m) \theta^2)
\ee

\noindent
where $\rho(m)$ is the surface number density of galaxies brighter than $m$. We obtained the surface number density of galaxies from the SDSS database by counting the number of galaxies within 11 randomly placed pointings of a radius 30 arcmin in the WSRT CVn survey region. Properties of galaxies were selected to resemble the properties of the secure host galaxies of the faint \HI\ detections. Roughly, we have taken $16<r<20$ mag and $0.2<g-r<0.6$ \citep[see Chapter 5 in][]{Kovac.2007}. We measured the density of galaxies with these properties to be about 52 per a pointing, leading to a probability that the cross-correlated opticaly identified galaxy is only a chance projection of 0.056 or 0.00014, if the optical detection is at a distance of 1 or 0.5 arcmin, respectively from the \HI\ detection. From the inspection of the images of the \HI\ overlayed on the top of the optical counterpart, it is clear that the projected distances between the optical detection and the maximum in the \HI\ surface density are less than 1 arcmin. We conclude that also our cross-correlations without the known distances of the optically detected galaxies are pretty secure. Moreover, our measurements provide first measurements of distances to these galaxies.

Finally, based  on the inspection of the DSS and SDSS images, there  was  one object  
detected in  \HI\
without an optical counterpart  (WSRT--CVn--61).  This object is found
a  few  arcmin  away  and  within  $\sim  110$  \kms\  from  NGC  4288
(WSRT--CVn--62).   This object has  already been  detected in  \HI\ by
\citet{Wilcots.etal.1996}  in \HI\  observations of  a sample  of five
barred    Magellanic    spiral    type    galaxies.     Interestingly,
\citet{Wilcots.etal.1996}  detected similar  \HI\  clouds, without  an
obvious optical counterpart on the Digital Sky Survey images, and with
the \HI\  mass $\sim$ 10$^7$  \Msol\ in four  out of five  galaxies in
their  sample. In  our  data, WSRT--CVn--61  is  barely resolved  (see
Appendix~\ref{app_atlas}), but  clearly distinguished from NGC 4288 in
the velocity.  It has  a single-peaked global \HI\ profile, consistent
with a  very weak rotation (W$_{50} \sim$  20 \kms). 

Potentially,  detecting  a dark  object  is  very  interesting in  the
context  of this  paper.   We  have carried  out  a follow-up  optical
observations in the field of NGC  4288, using the Wide Field Camera on
the   Isaac   Newton   2.5   meter   telescope,   La   Palma,   Canary
Islands. The observations did not reveal any sign of the stellar
light  nearby  the  position  of  WSRT--CVn--61 down  to  the  surface
brightness  limit  of 26.3  mag  arcsec$^{-2}$  in  $R$ and  27.4  mag
arcsec$^{-2}$   in   $B$  \citep[][Chapter 5]{Kovac.2007}.   The
observations will be presented in more details in a future paper. To
conclude, the nature of WSRT--CVn--61 is not entirely clear, but given
its proximity to NGC 4288, it  is likely a very LSB companion to this
galaxy. We treat  it as a separate object.

The homogenised \HI\  data (HOMHI) catalogue \citep{Paturel.etal.2003}
represents  a compilation  of \HI\  detections from  611  papers. This
catalogue was used to inspect  the whole volume covered by this survey
for previous \HI\ detections.   According to the HOMHI catalogue there
are 47 objects  which have been observed in the  21-cm line inside the
volume of the  WSRT CVn survey and  4 objects are at the  edges of the
observed region (objects  with WSRT--CVn indexes 1, 2,  60 and 62). Of
the  47  HOMHI  detections  inside   the  survey  volume,  44  can  be
cross-correlated  uniquely with  the WSRT  CVn detections  using their
position  on the  sky and  their heliocentric  velocities. One  of the
detections of the WSRT  CVn survey (WSRT--CVn--23) is cross-correlated
with two objects  in the HOMHI catalogue.  These  two HOMHI detections
have  the same heliocentric  velocity, and  their positions  differ by
0.015  deg   and  0.14  deg   in  right  ascension   and  declination,
respectively.   Their profile  widths are  identical and  their \sint\
values are almost the same.  From tracing these detections back in the
literature it  follows that their  names have been confused;  there is
only one object detected with the given \HI\ properties and velocity.

One of  the  HOMHI
detections does not  have a counterpart in the  WSRT CVn survey.  That
is  MAPS$\_$NGP O$\_$218$\_$0783987,  with  heliocentric velocity  636
\kms.   \citet{Huchtmeier.etal.2000}  measured  \sint\ =  0.66 Jy  \kms,
\speak\ = 0.026 $\pm$ 0.0046 Jy, $W_{50}$ = 27 \kms\ and $W_{20}$ = 34
\kms\ for this object, using  the single dish 100-m radio telescope at
Effelsberg. The  WSRT CVn survey  is slightly incomplete ($C  > 90\%$)
for  the \sint\  value  and complete  for  the \speak\  value of  this
object. We  carefully examined the  datacube from the WSRT  CVn survey
produced at the position  of MAPS$\_$NGP O$\_$218$\_$0783987. There is
no sign of  the 21-cm emission at that position in  our data. Based on
the cross-correlation  with the previous  observations, all detections
from the  WSRT CVn survey are  real.  We are inclined  to believe that
the detection in the  \citet{Huchtmeier.etal.2000} sample is not real,
 but is, perhaps interference.

There is no entry in  the HOMHI catalogue for objects WSRT--CVn--9 and
WSRT--CVn-61. HYPERLEDA, however, provides \HI\ data for WSRT--CVn--9.
For a comparison with the literature, we  used  the measurement  from  \citet{Wilcots.etal.1996} for  object
WSRT--CVn-61.   Object  WSRT--CVn--7  has  been listed  in  the  HOMHI
catalogue, but without a  measurement of integrated flux.  We obtained
the        \sint\       value       for        WSRT--CVn--7       from
\citet{Matthews&vanDriel.2000}. This  \sint\ value has  been corrected
for the finite size of  the Nan\c{c}ay telescope beam.  In total, from
70 detected \HI\ sources in the WSRT CVn survey, 19 have been detected
for the first time in the 21-cm emission line in this survey.

\begin{figure}
  \centering
  \includegraphics[width=0.44\textwidth]{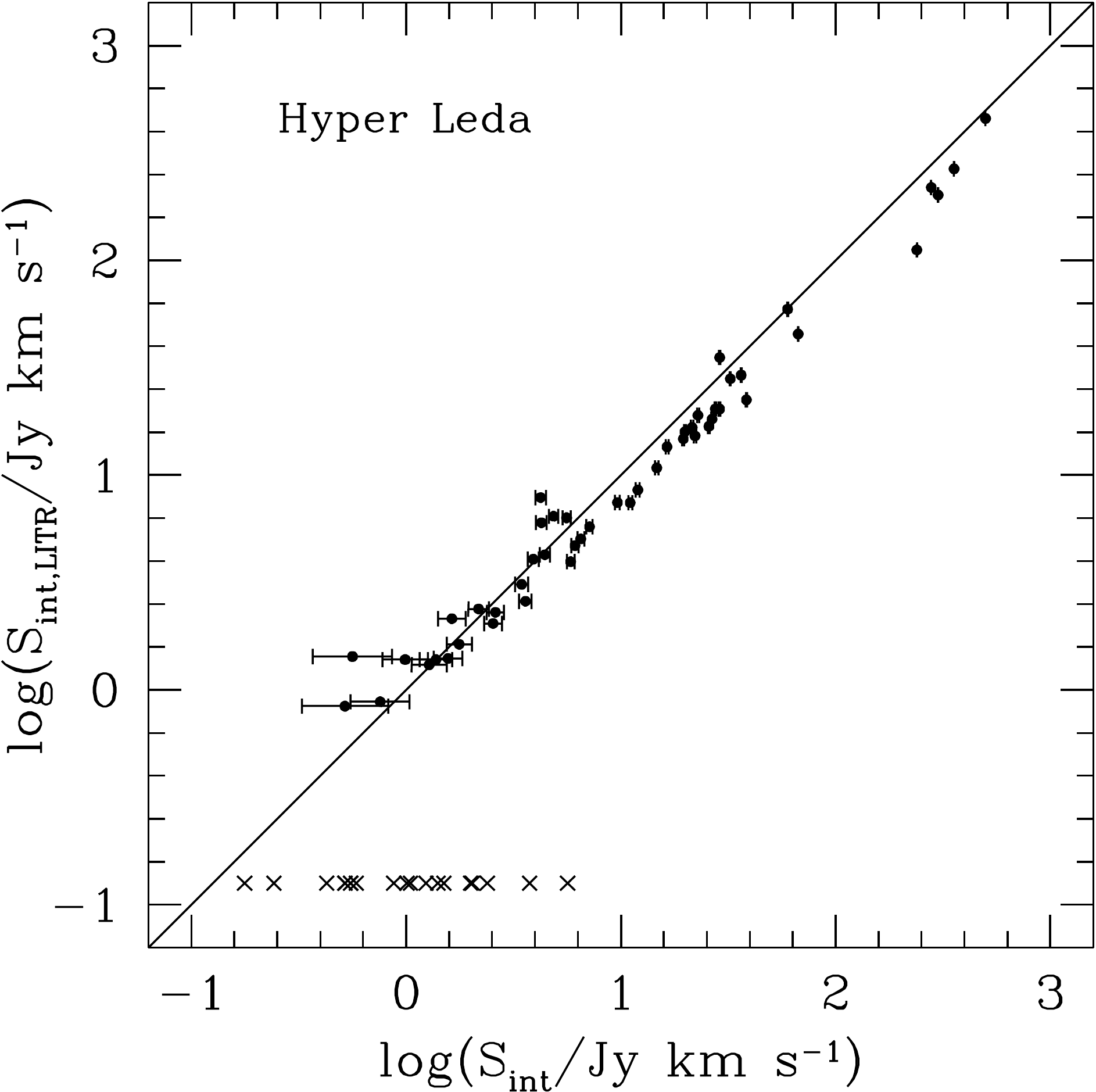}
  \includegraphics[width=0.44\textwidth]{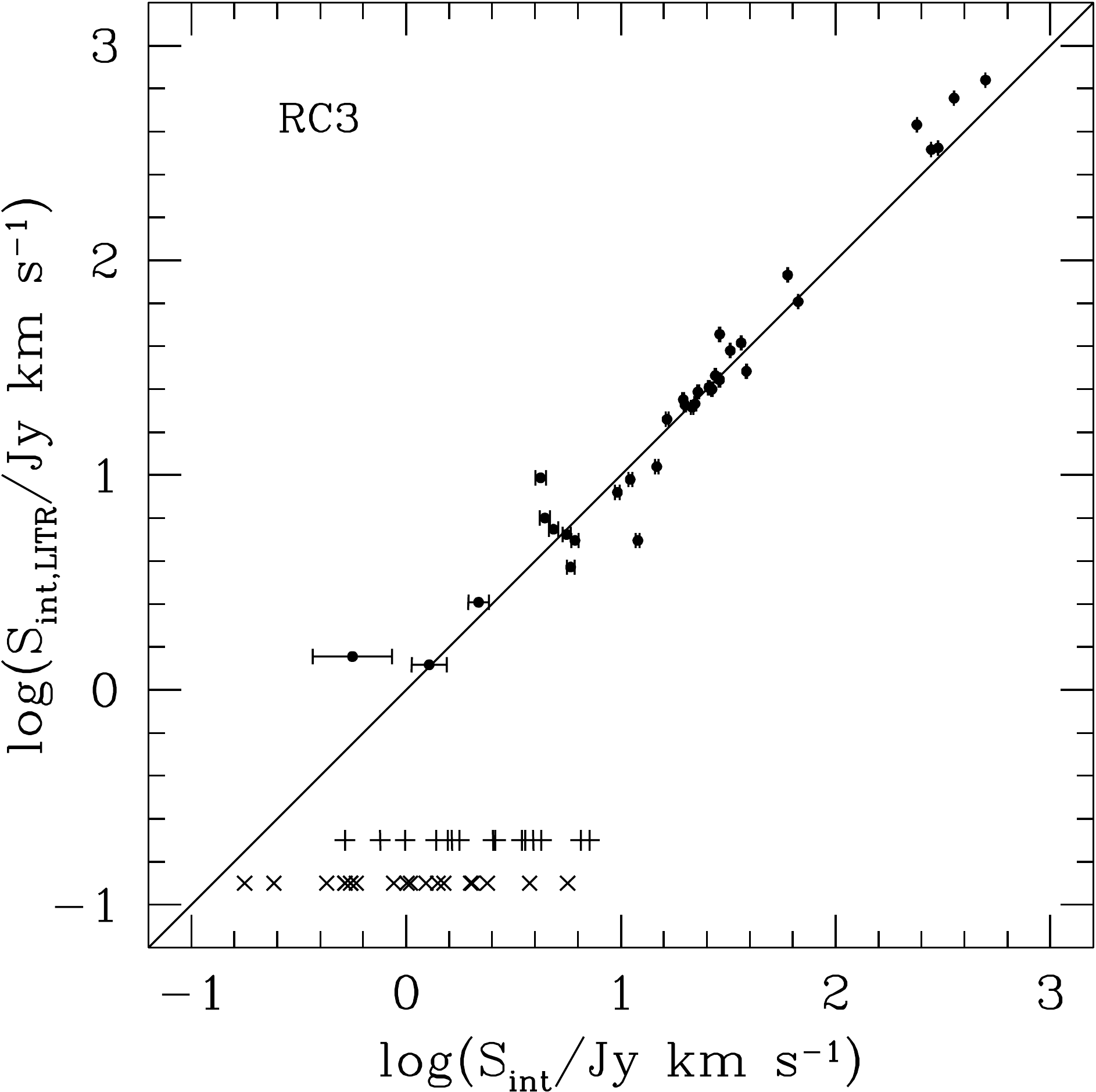}
  \caption{\small Comparison  of   the  WSRT  CVn  \sint\   values  with
literature  measurements.  For  WSRT--CVn--7  and WSRT--CVn-61  \sint,
literature measurements come from \citet{Matthews&vanDriel.2000} and 
\citet{Wilcots.etal.1996} respectively. In the top panel, literature
values  of  integrated  fluxes  obtained  from  HYPERLEDA  have  been
used. In the bottom panel  integrated fluxes published in RC3 have been
used. The integrated  flux values of objects for  which there is no
previous \HI\ measurement in the  literature has been indicated with an
 ``x''  symbol.   \sint\  values  of  detections  without  RC3
measurements, but for  which the \sint\ value in  the HOMHI catalogue is
available,  have been  indicated  in  the bottom  panel  with an  ``+''
symbol. We plot the errors only for our measurements.}
\label{comp_sint}
\end{figure}

In Figure~\ref{comp_sint}, the comparison of \sint\ values measured in
the WSRT CVn survey and \sint\ values available from the literature is
presented.  For objects with the WSRT--CVn indexes 7 and 61 we use the
values from  the references given above. For  the comparison presented
in the top panel in Figure~\ref{comp_sint}, we used the \sint\ values
retrieved from HYPERLEDA  for the rest of the objects.  It is obvious that
the HYPERLEDA  integrated fluxes  are systematically smaller  than the
WSRT  CVn  integrated  fluxes  for  approximately  \sint\  $>  10$  Jy
\kms. The  \sint\ values  in HYPERLEDA come  from the  HOMHI catalogue
(with the exception  of object WSRT--CVn--9) and the  majority of them
have been  measured with a single  dish telescope. We  have not traced
back in  the literature the  references for the  individual detections
from the HOMHI catalogue.  Instead, we collected \sint\ values for the
WSRT    CVn    detections    from    the   earlier    RC3    catalogue
\citep{deVaucouleurs.etal.1991}. These two  catalogues (HOMHI and RC3)
are not independent. The comparison between the integrated flux values
measured  in the  WSRT CVn  survey  and the  literature \sint\  values
collected from the RC3 catalogue, and  for the objects with the id's 7
and 61 from the individual papers,  is presented in the bottom panel in
Figure~\ref{comp_sint}.  The RC3  catalogue contains \sint\ values for
fewer  objects  detected  in  the  WSRT  CVn  survey  than  the  HOMHI
catalogue. Still, most  of the objects with \sint\  values above 10 Jy
\kms\  are  present  in  both  catalogues  considered.   There  is  no
systematic difference  between the WSRT CVn integrated  fluxes and the
RC3 integrated fluxes. It is possible, that the systematic offset seen
between our  values of  the integrated fluxes  and those in  the HOMHI
catalogue  is due  to the  corrections applied  in  the homogenisation
process of the  \HI\ data used to create  the \HI\ parameters provided
in  the  HOMHI catalogue. This difference can arise also from the distribution of \HI\ in a galaxy. We use the interferometric data and we are restricted to the flux measurements in the galaxies. If a galaxy has a lot of outlying \HI, we are not sensitive to include it, while the single dish observation picks it up.  Given that our measurements agree with the RC3,  we  use  the  RC3 measurements  of  \sint,
heliocentric   velocity,    $W_{50}$   and   $W_{20}$    for   objects
WSRT--CVn--67A  and  WSRT--CVn-67B,  for  which we  can  not  properly
measure the \HI\ properties from the WSRT CVn survey data.

\subsection{Distributions of \HI\ properties of the detections}

\begin{figure}
  \centering
  \includegraphics[width=0.23\textwidth]{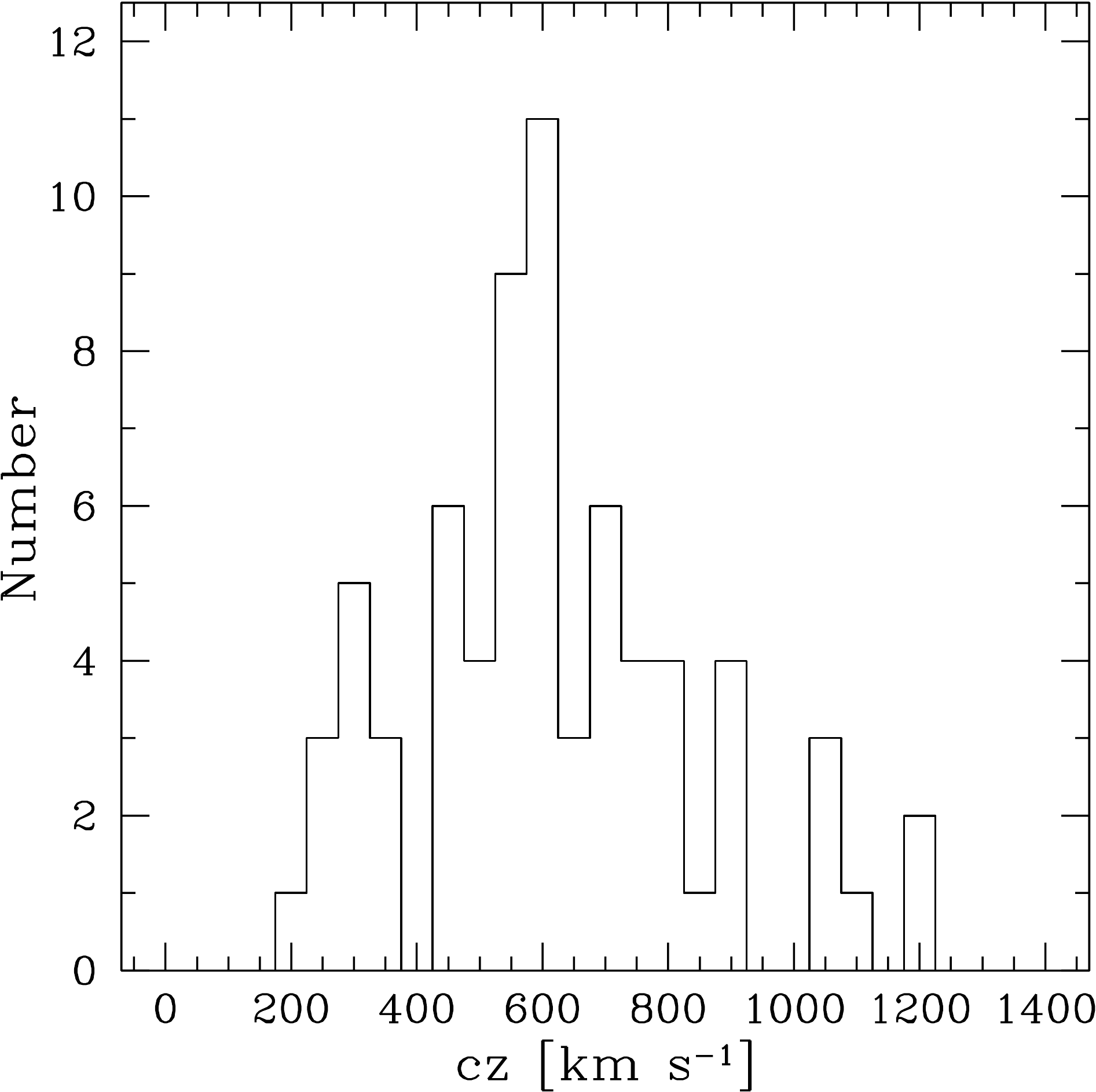}
  \includegraphics[width=0.23\textwidth]{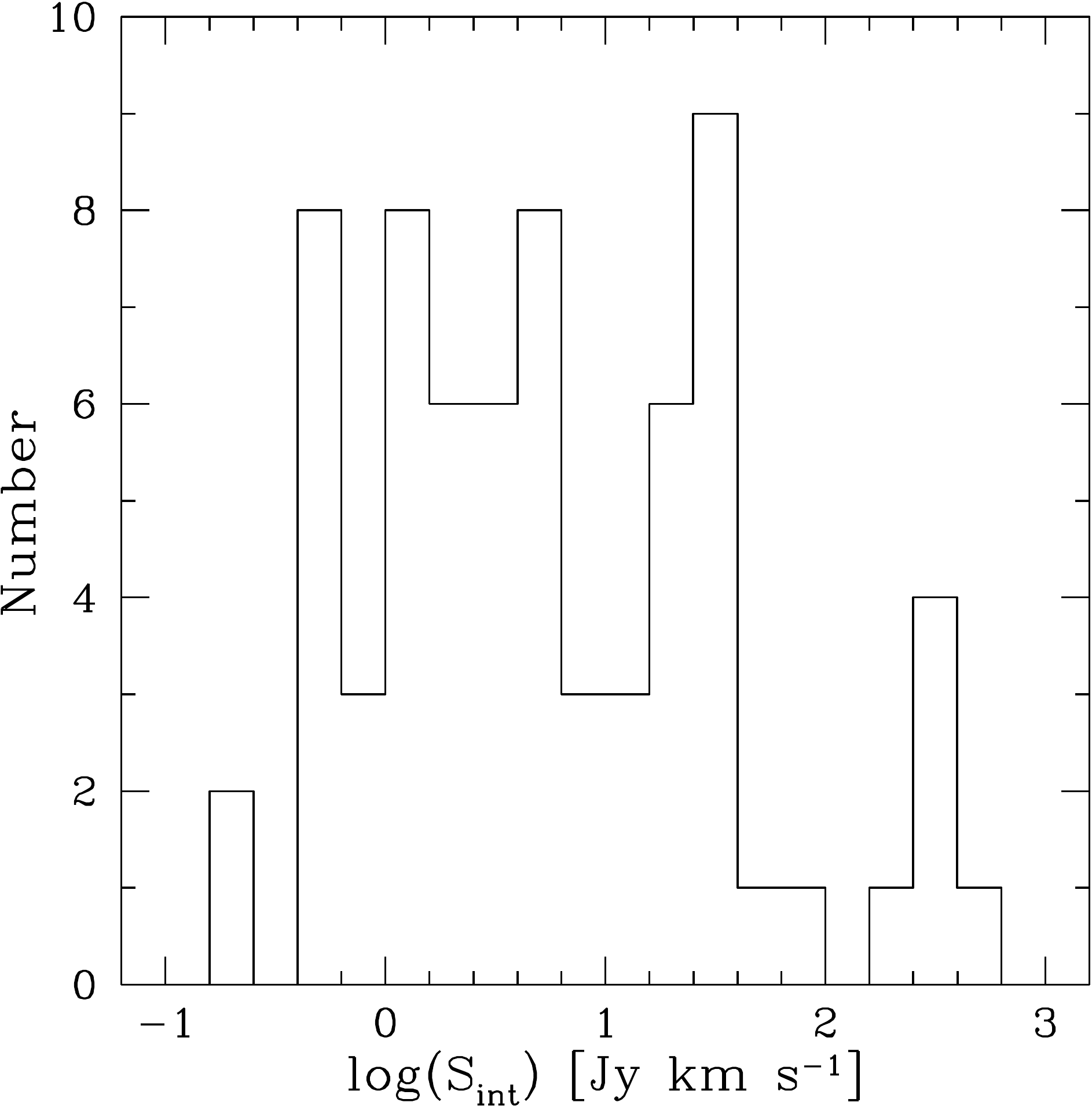}
  \includegraphics[width=0.23\textwidth]{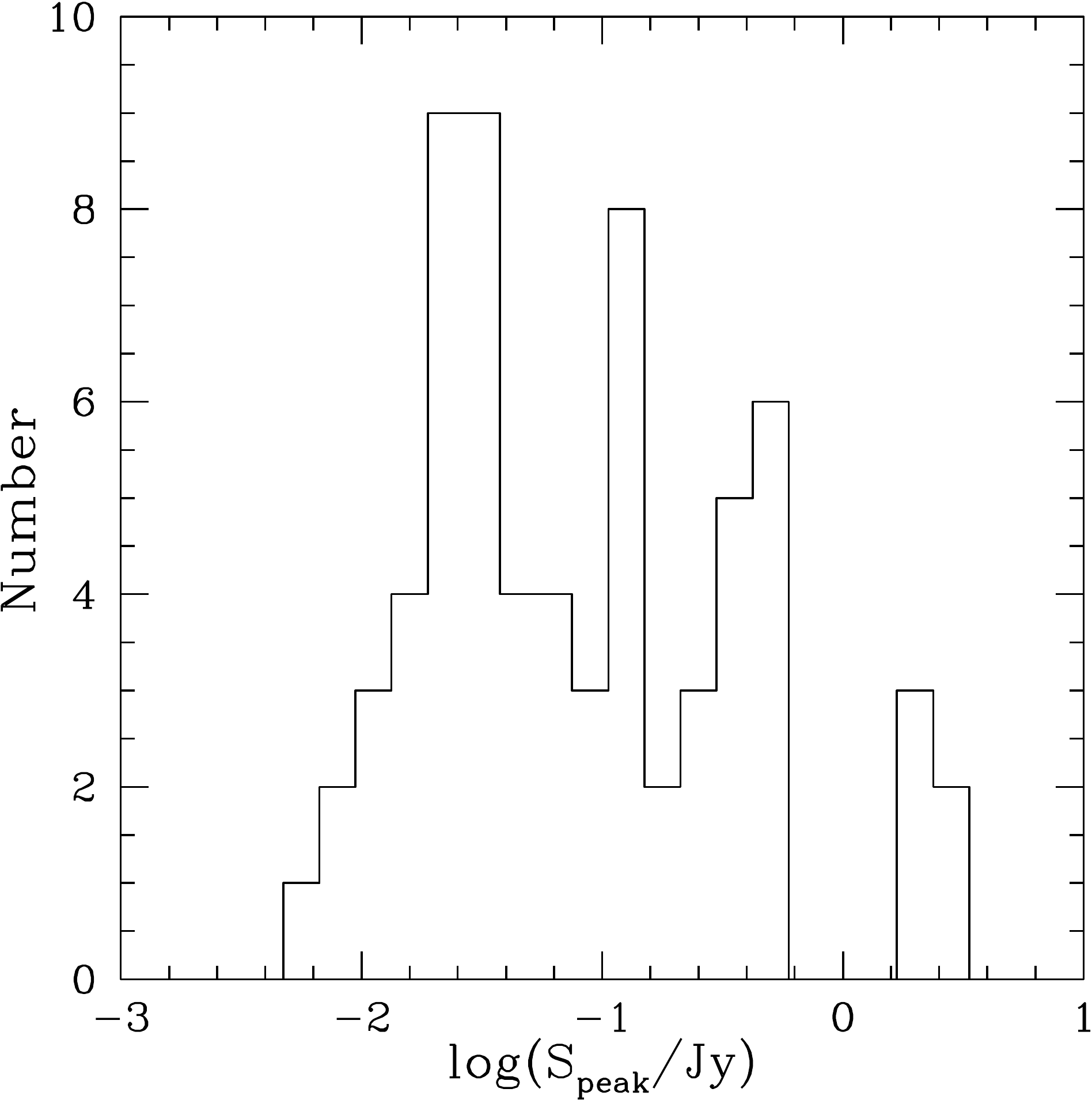}
  \includegraphics[width=0.23\textwidth]{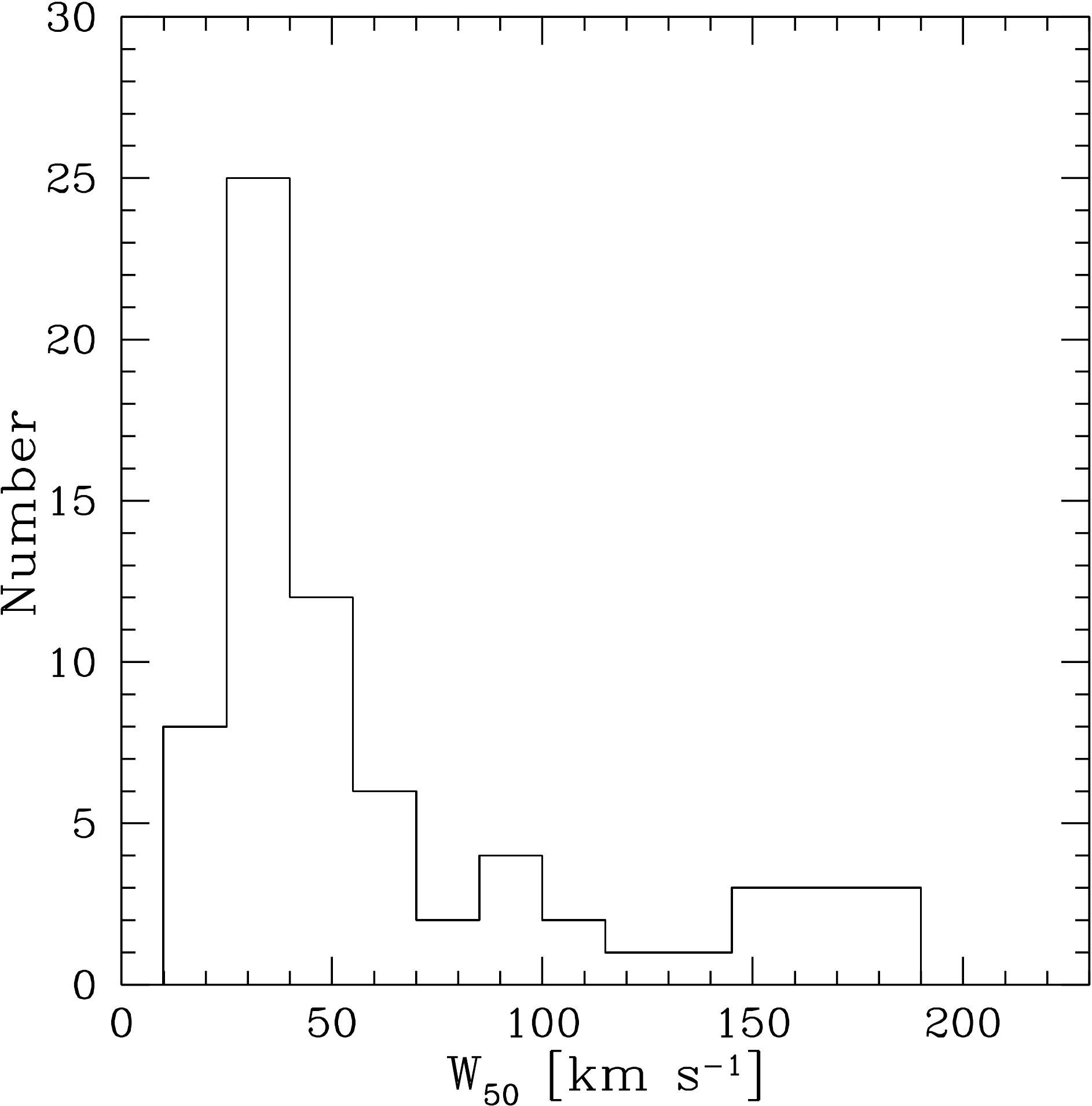}
\caption{\small Distributions of estimated \HI\ parameters. Histograms
show the  number distribution  of the WSRT  CVn detections in  bins of
$cz$  (top--left),  \sint\  (top--right),  \speak\  (bottom-left)  and
$W_{50}$ (bottom-right).}
\label{hist_prop}
\end{figure}

The  various  distributions  of  the measured  parameters  of  objects
detected  in \HI\  in this  survey can  be used  to examine  the basic
properties  of the  detected sample.  Histograms of  the distributions
with radial velocity, integrated flux, peak flux and profile width at
the  50$\%$ level  of the  maximum flux  in the  spectra are  shown in
Figure~\ref{hist_prop}.

The redshift distribution of the  detected objects is presented in the
top left histogram.  A fraction of 29$\%$ of all  detections fall in a
100 \kms\  wide interval  with velocities $525  \le cz \le  625$ \kms.
This   peak   coincides   with   the   peak   of   the   CVnII   cloud
\citep{Tully&Fisher.1987}.  The  peak of the  CVnI cloud is  at $\sim$
300 km s$^{-1}$ \citep{Tully&Fisher.1987}, clearly identifiable in the
histogram of observed redshifts.

The distribution of  measured \sint\ values is shown  in the top right
panel  of Figure~\ref{hist_prop}. The  detections with  values \sint\
$\ge  80$  Jy  \kms, 6  in  total,  are  excluded  from the  plot.  The
distribution of \speak\  values is presented in the  bottom left panel
of Figure~\ref{hist_prop}.  Most of the  objects detected in  the WSRT
CVn  survey have  the small  \sint\ and  \speak\ values  measured. For
example, a fraction of $63\%$ of the detected objects have \sint\ $\le
10$ Jy  \kms, while  72$\%$ of the  detections with  available \speak\
measurements, have $S_{\rm{peak}} \le 0.2$ Jy.

The last panel, bottom right in Figure~\ref{hist_prop}, corresponds to
the distribution  of the profile widths at  the 50$\%$ level  of the
maximum flux in the spectra of the detected objects, $W_{50}$ (corrected
for the  instrumental effects). This distribution
has  a prominent  peak around $W_{50}  \sim 35$  \kms. 86$\%$ of  the detected \HI\ objects have  $W_{50} \le
130$  \kms\ and can be considered as a candidate population of dwarf galaxies (\citealt{Duc.etal.1999} found that 75$\%$ of galaxies selected by the same profile width criteria are genuine dwarf galaxies). However, the observed profile widths are affected by the inclination of a galaxy, and we discuss this issue in more detail in Subsection~\ref{sub_hiincl}.

\begin{figure*}
  \centering
  \includegraphics[width=0.32\textwidth]{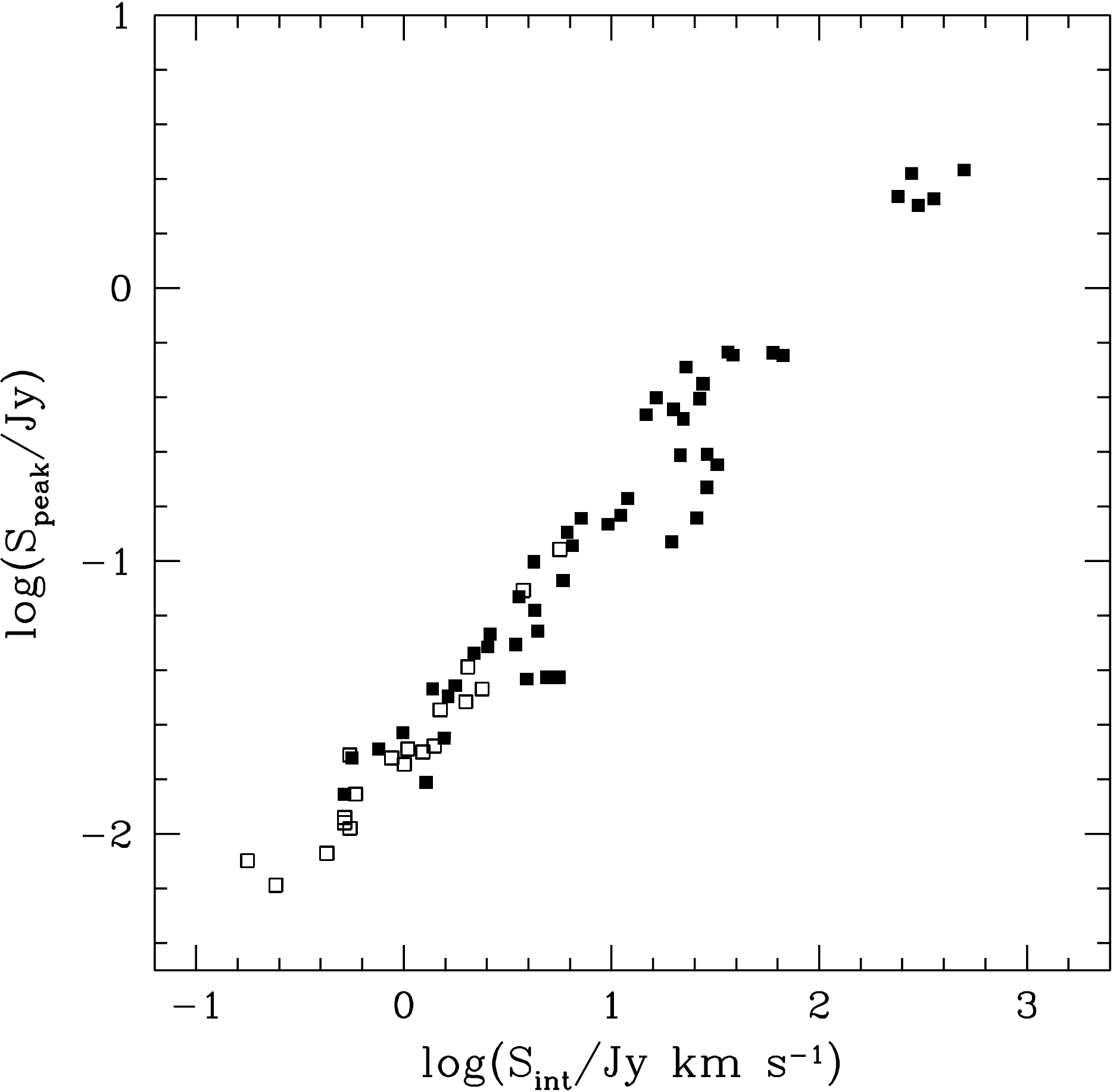}
  \includegraphics[width=0.32\textwidth]{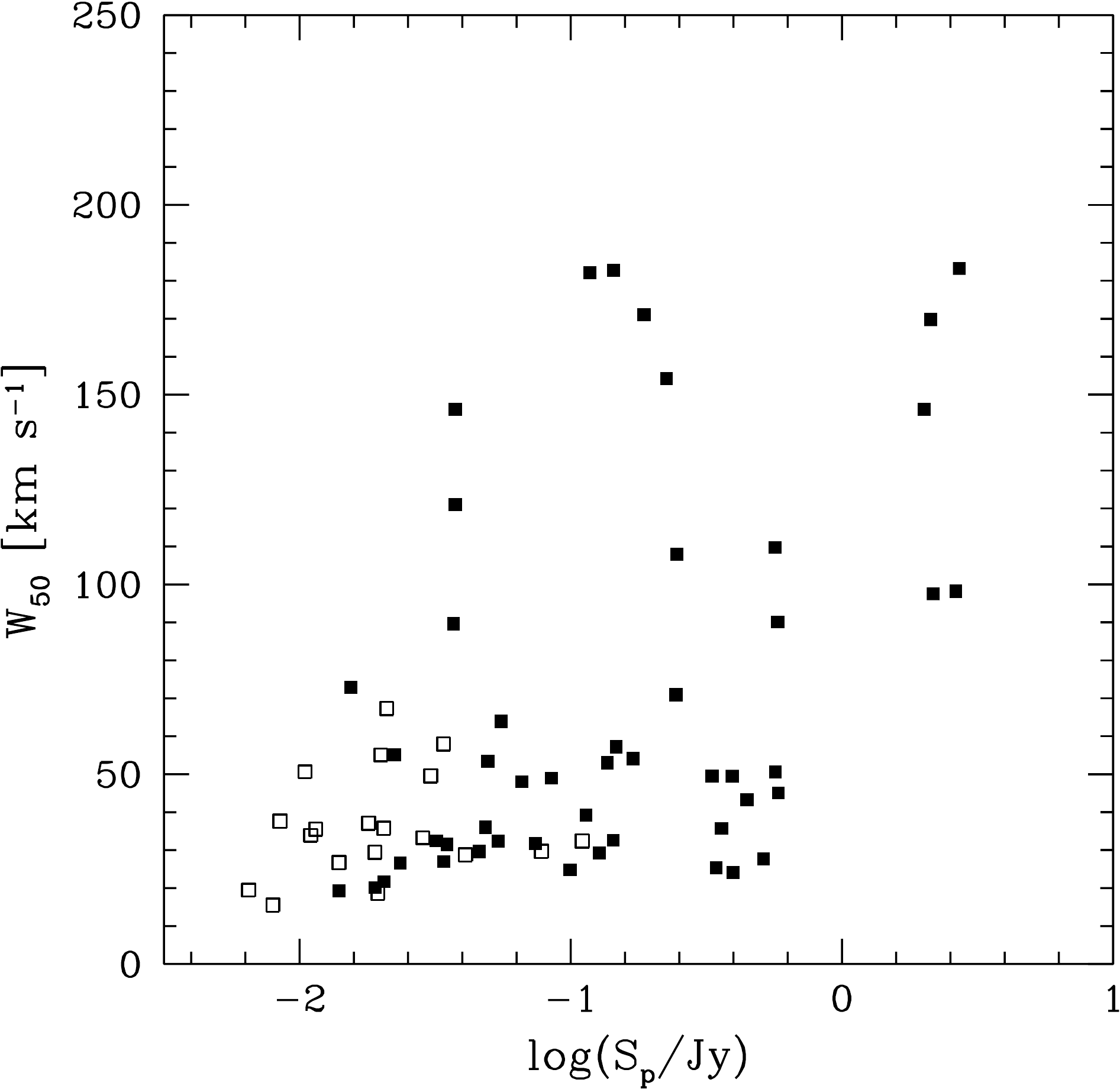}
  \includegraphics[width=0.32\textwidth]{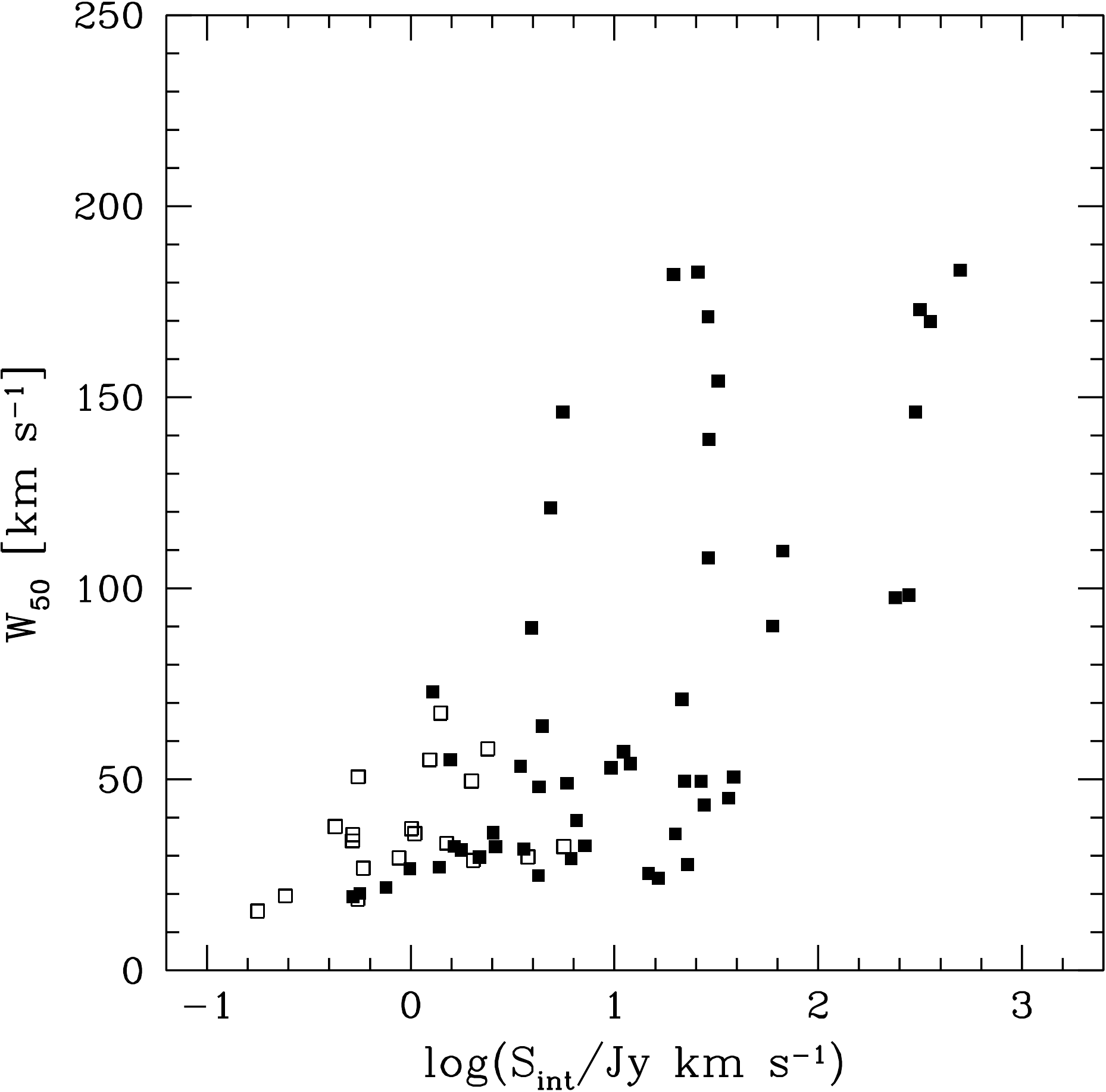}
   \includegraphics[width=0.32\textwidth]{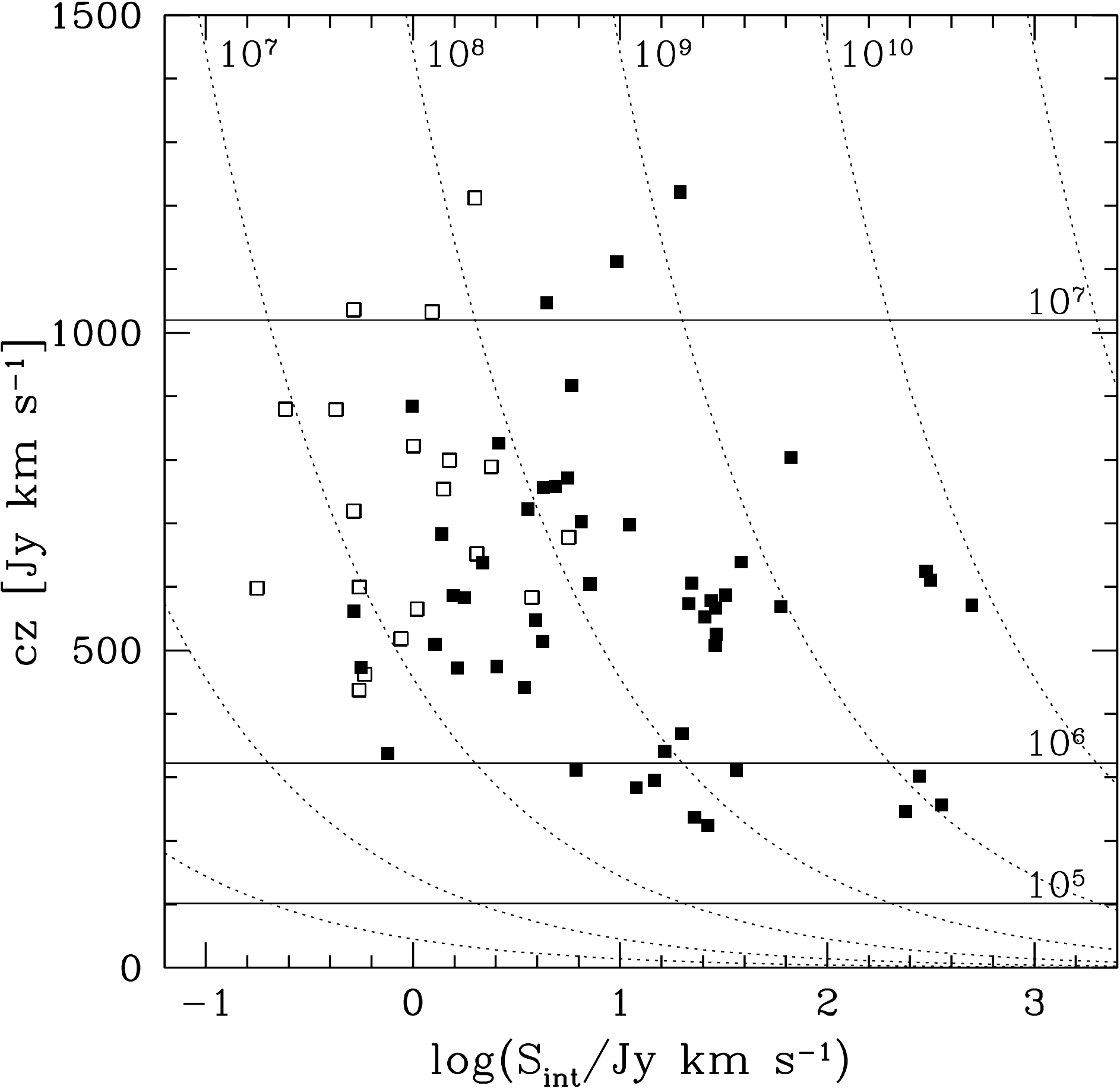}
  \includegraphics[width=0.32\textwidth]{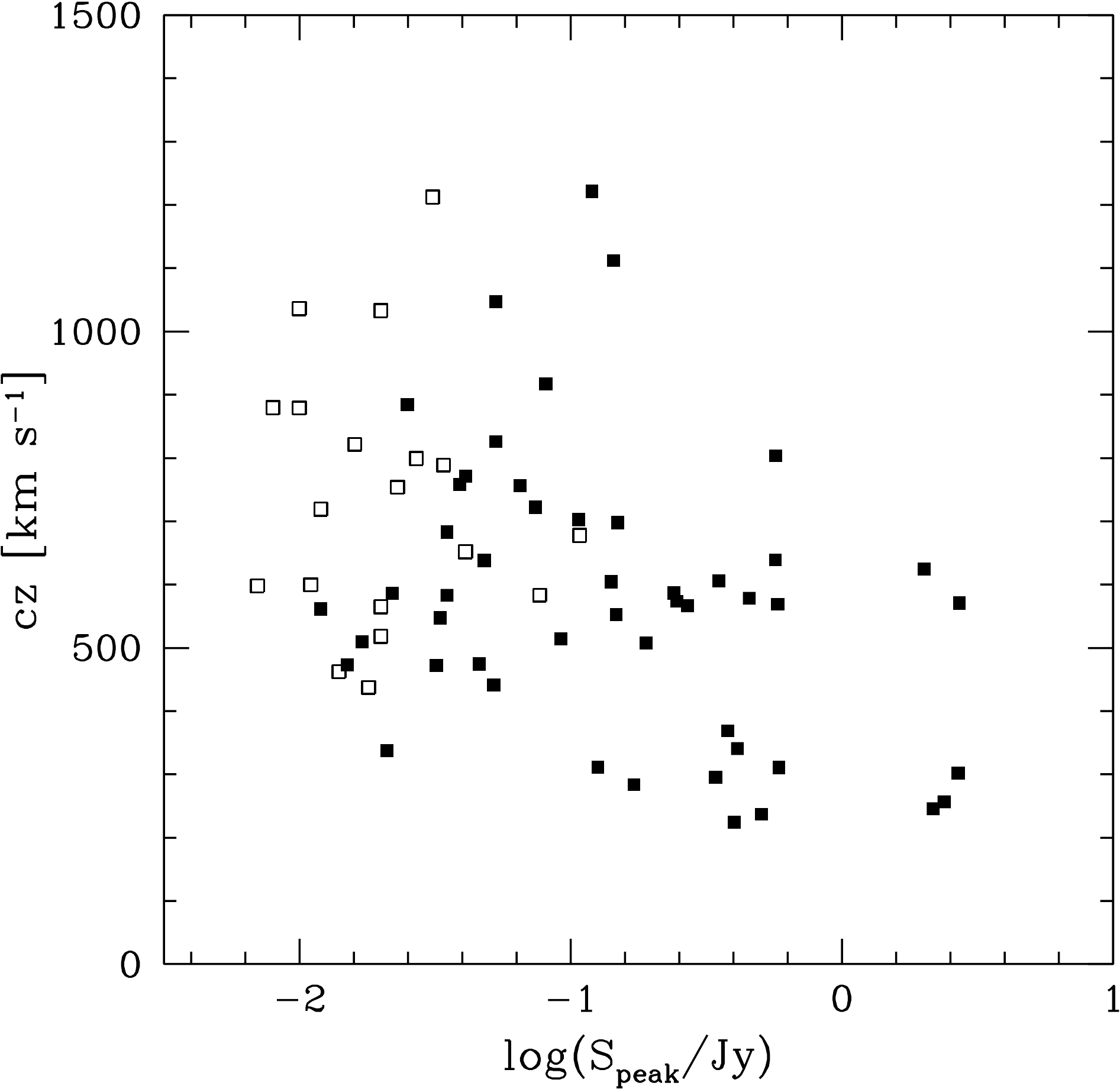}
  \includegraphics[width=0.32\textwidth]{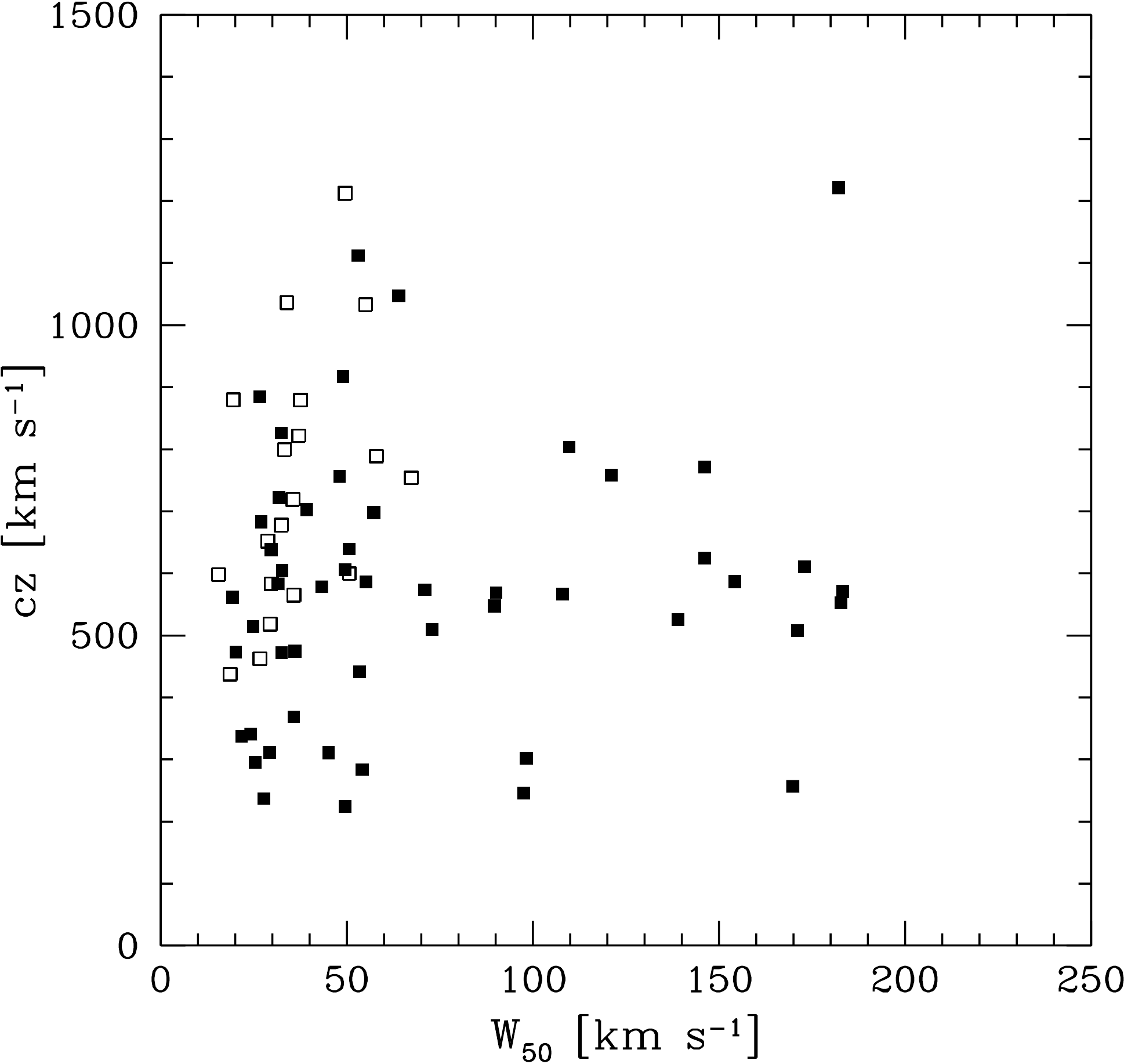}
  \includegraphics[width=0.32\textwidth]{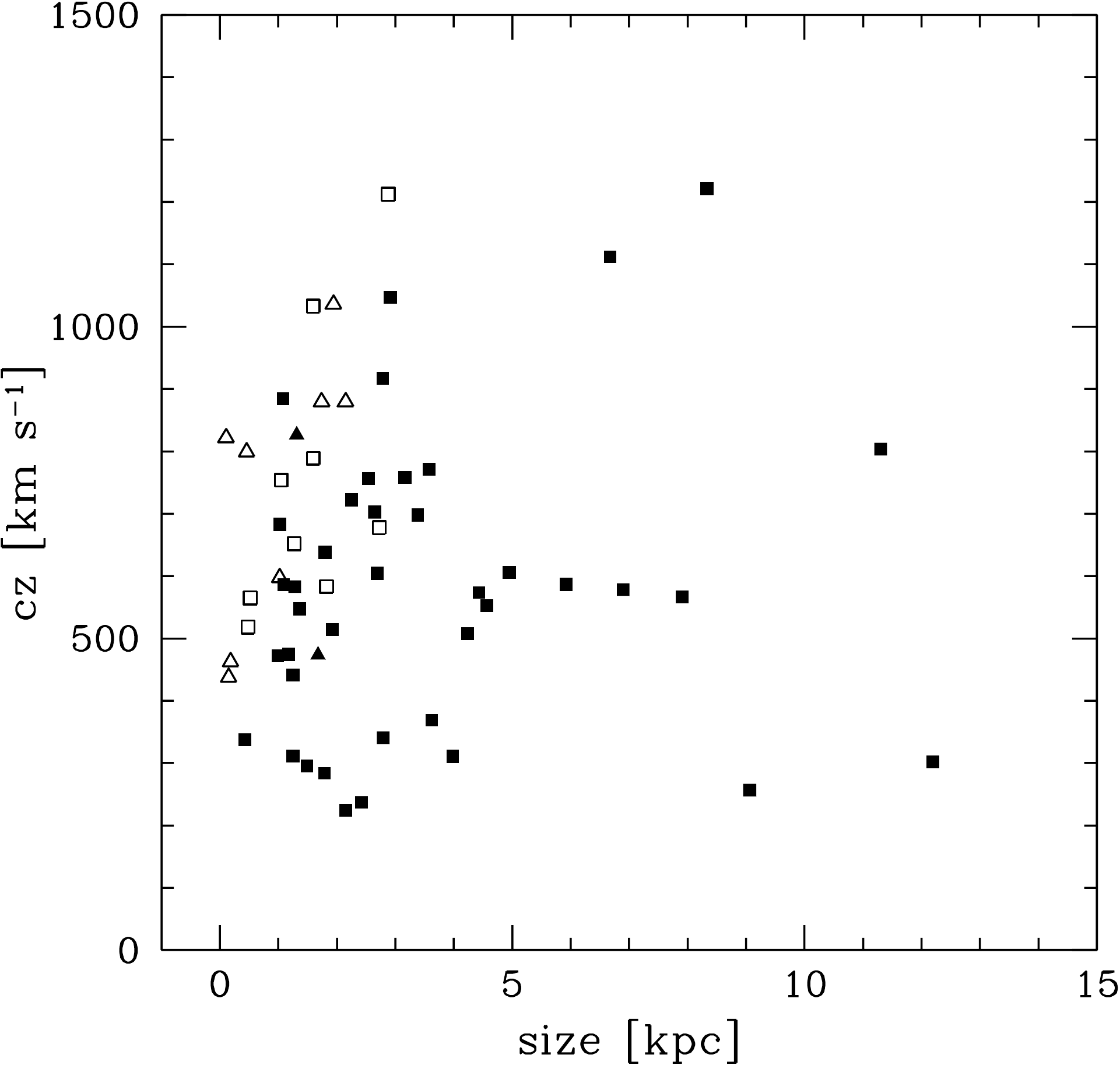}
\includegraphics[width=0.32\textwidth]{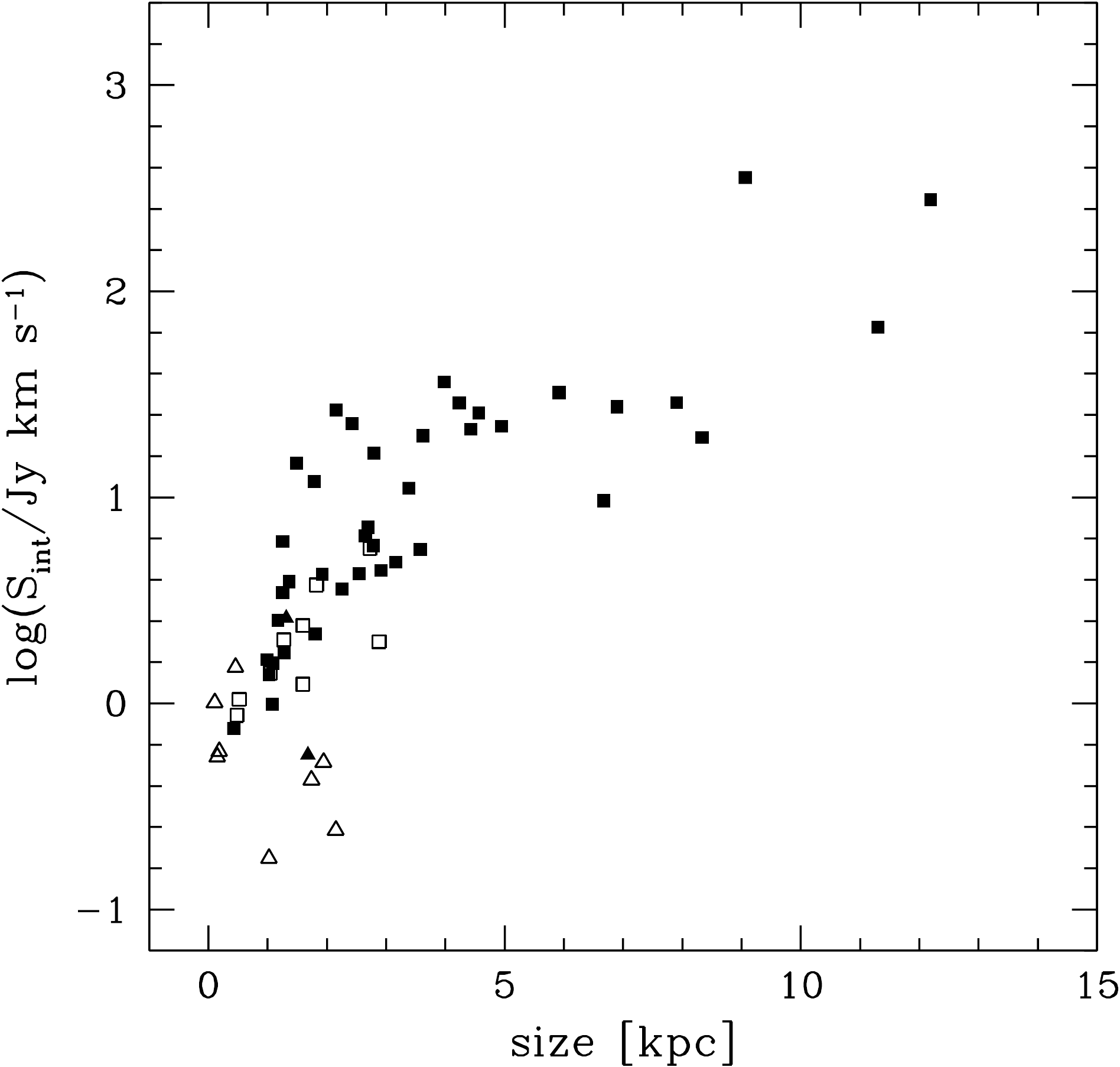}
\includegraphics[width=0.32\textwidth]{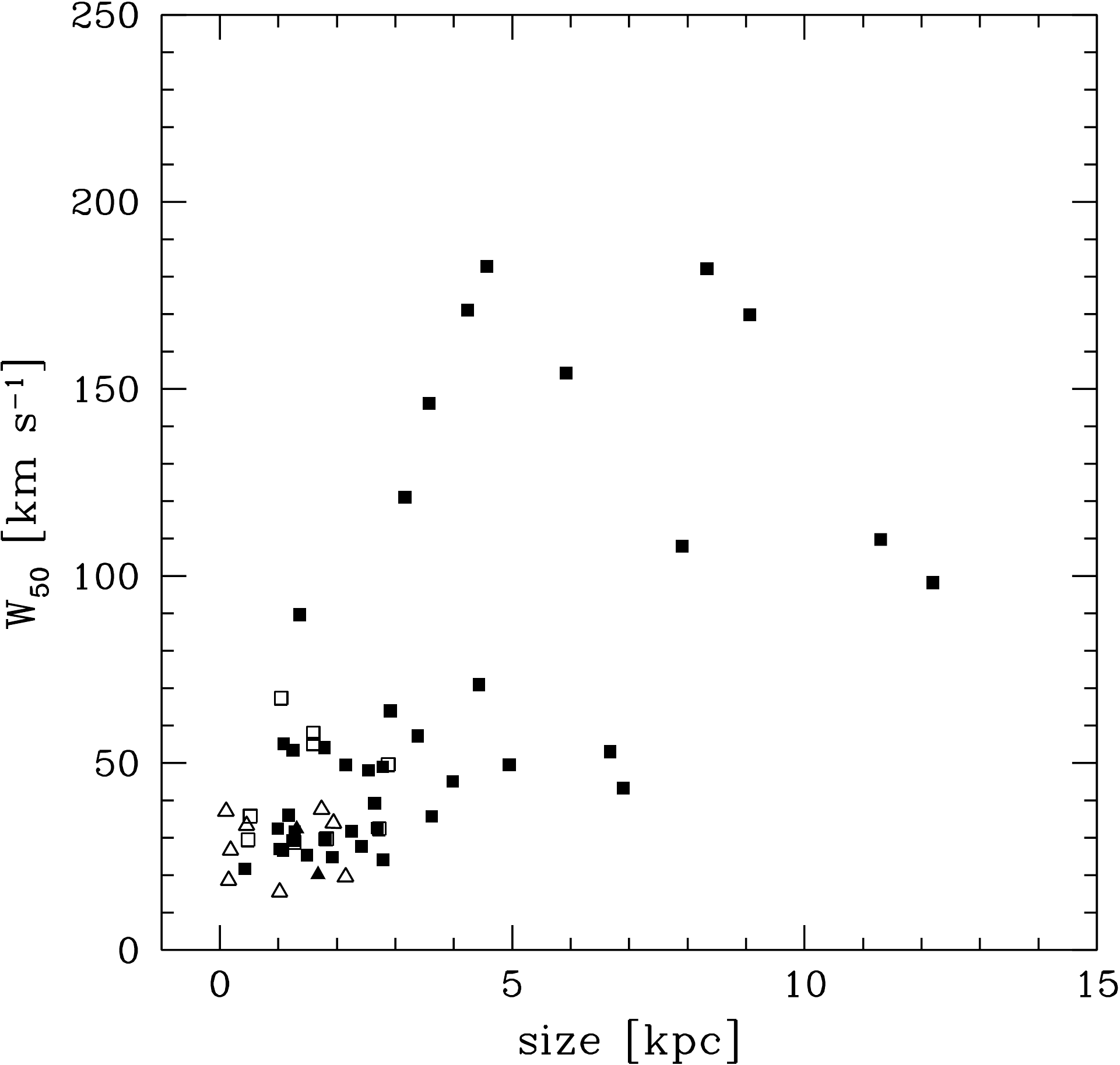}
\caption{\label{bin_prop} \small  Bi--variate  distributions  of  observed  parameters:
\sint, \speak, $W_{50}$, $cz$ and  \HI\ size. The objects detected for
the first time in the 21--cm emission line in the WSRT CVn survey have
been  presented with open  symbols; the  remaining objects  are marked
with filled symbols. In the  plots in the bottom-row, objects for which
size is estimated from the isophote  at a column density level of 1.25
$\times$ 10$^{20}$  atoms cm$^{-2}$ are marked with  squares; if their
size is estimated from a Gaussian fitted to the \HI\ distribution they
are marked with triangles. In the plot in the middle--left panel lines
of a constant  \HI\ mass are shown (dotted lines). The  \HI\ mass of a
detection (in \Msol)  has   been   calculated   assuming   an   optically   thin
approximation, M$_{HI}$ = 2.356  $\times$ 10$^5$ d$^2$ \sint, for \sint\
expressed  in Jy  \kms\ and  $d$ being  the distance  to an  object in
Mpc.  We assumed  $d =  cz/H_0$  and we  used $H_0$  = 70  $\mbox{\kms\
Mpc$^{-2}$}.$  The  lines of  constant \HI\ mass in \Msol\ are presented  with a
logarithmic step of 1.   The horizontal continuous lines represent the
maximum recession  velocity up  to which an  object of a  certain \HI\
mass  can be  detected  in the  WSRT  CVn survey.   We  used the  same
assumption about  the distance  to an object  as described  above. The
minimum detectable  integrated flux in the  WSRT CVn survey  is 0.2 Jy
\kms,  obtained  from the  calculations  of  the  completeness of  the
survey. However, we are complete only for \sint\ measurements above 0.9 Jy \kms.}
\end{figure*}

The bivariate  distributions of  velocity, profile width,  peak flux,
integrated flux  and size are shown  in Figure~\ref{bin_prop}. Objects
detected for the first time in \HI\ in this survey are marked with 
open  symbols. It  is clearly  visible  that the  newly detected  \HI\
objects  have small  integrated fluxes  and integrated peak
fluxes, small profile widths and small physical sizes.

The log-log  plot of \speak\  vs. \sint\ demonstrates the simple property
that  \HI\  detections  with  larger  integrated  fluxes  have  higher
integrated  peak  fluxes  and  vice  versa. It is interesting  that  this 
relation holds for the  detections over the whole range of 
observed integrated  fluxes and integrated peak  fluxes. The bivariate
distributions of W$_{50}$ vs. \sint\ or \speak\ show that there
are  no galaxies with  large W$_{50}$
with small integrated fluxes and low integrated peak fluxes in the volume probed. Objects with small
integrated  fluxes spread  over  large  profile  widths (if  they
exist) would be very difficult  to detect. {Similarly, low \speak\ galaxies with large line widths cannot be detected, because the flux is in the noise.} Smoothing in the velocity
domain increases the sensitivity to this  type of objects. However, 
smoothing  the datacubes  in the  WSRT CVn
survey in the velocity  domain, did not reveal any new detections.

The distributions  of measured  \HI\ parameters with  redshift ($cz$),
show  a  segregation  of  detections  in two  groups,  reflecting  the
positions  of the CVnI  and CVnII  clouds in  redshift space.  In the
redshift  distributions of \sint\  and \speak\ there appears to be an absence of
detections  with  small \sint\  (and  small  \speak)  at low  $cz$, also in that  part of parameter
space for which the survey  is complete. The survey by  itself does not have
any selection  effects which  would bias it  against the  detection of
objects  in the  nearby  Universe with  small  integrated fluxes.   As
already  discussed  in  Subsection~\ref{sub_crosscorr}, comparing  our
detections to previous \HI\ observations reveals that there are no
\HI\ objects  with $cz \le$  400 \kms\ which  have been missed  in the
WSRT CVn  survey.  Therefore, the  absence of an \HI\  population with
\sint\ $\le$ 6 Jy \kms\ (or  \speak\ $\le$ 0.1 Jy) is real. However, taking a flat \HI\ mass function and the number of detected objects in the higher mass bin 10$^7$-10$^8$ \Msol (5), one would expect $5 \pm \sqrt{5}$ objects in the 10$^6$-10$^7$ \Msol bin, which is not too inconsistent. Moreover, the volume  of the survey region
limited   with   $85  <   cz   <  322.5$   \kms\   is   less  than   1
h$_{70}^{-3}$Mpc$^3$. The observed redshift  distributions of \sint\  and \speak\ probably reflect just a peculiarity of the CVnI group.

In addition,  we present the  bivariate distributions of  $cz$, \sint\
and $W_{50}$ as a function of size of those objects for which the size
was estimated.  We used the average  value of the major and minor axis
estimated at  a column density level of  1.25 $\times$ 10$^{20}$ atoms  
cm$^{-2}$, or
only the FWHMs along the major  and minor axis of a Gaussian fitted to
the small detections.   To express the size in  kpc instead of arcsec
we have adopted distances $d$ to the objects calculated from $d = cz /
H_0$ using $H_0 = 70$ \kms\ Mpc$^{-2}$.

\begin{figure*}
  \centering
  \includegraphics[width=5.5cm]{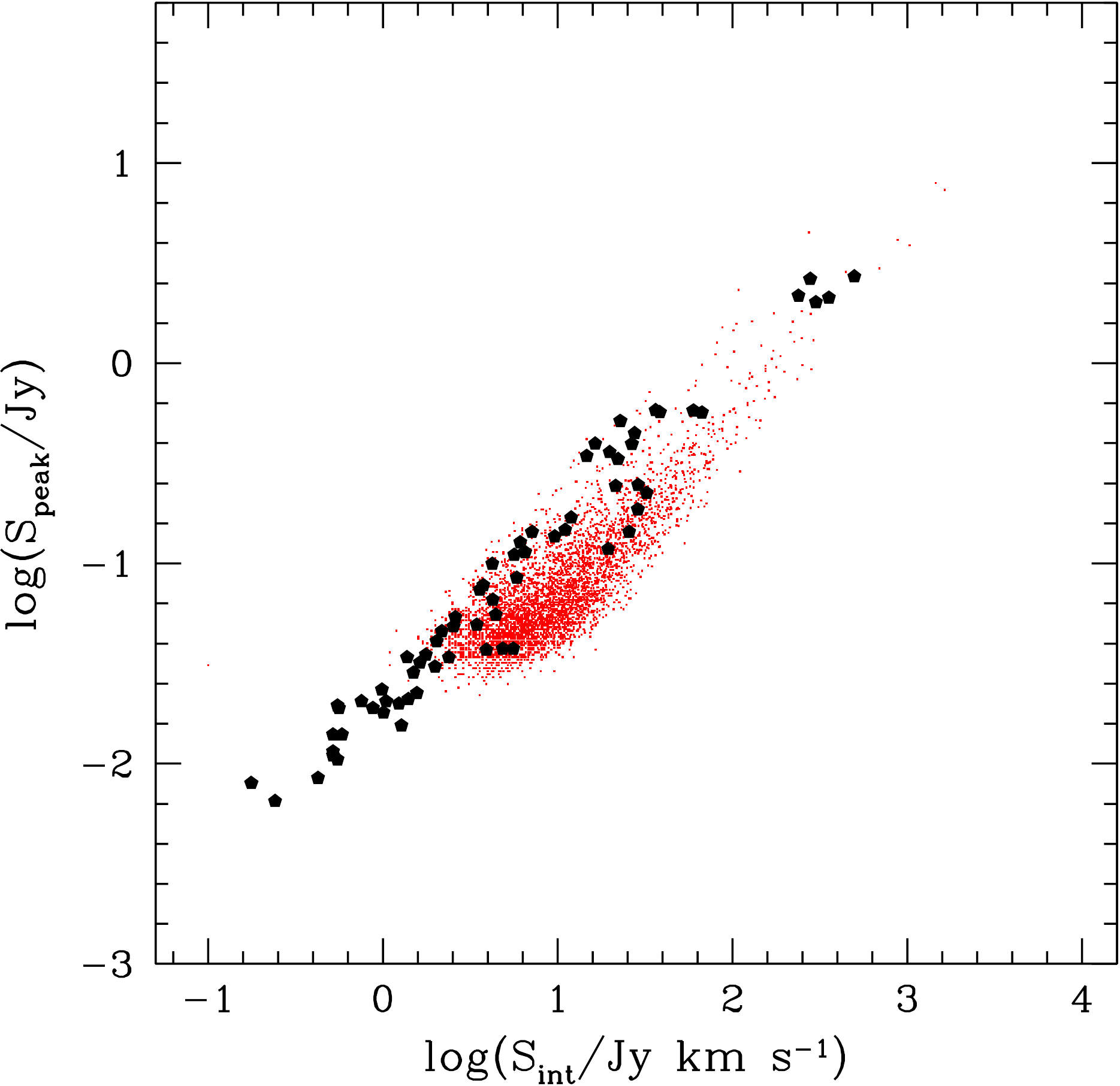}
  \includegraphics[width=5.5cm]{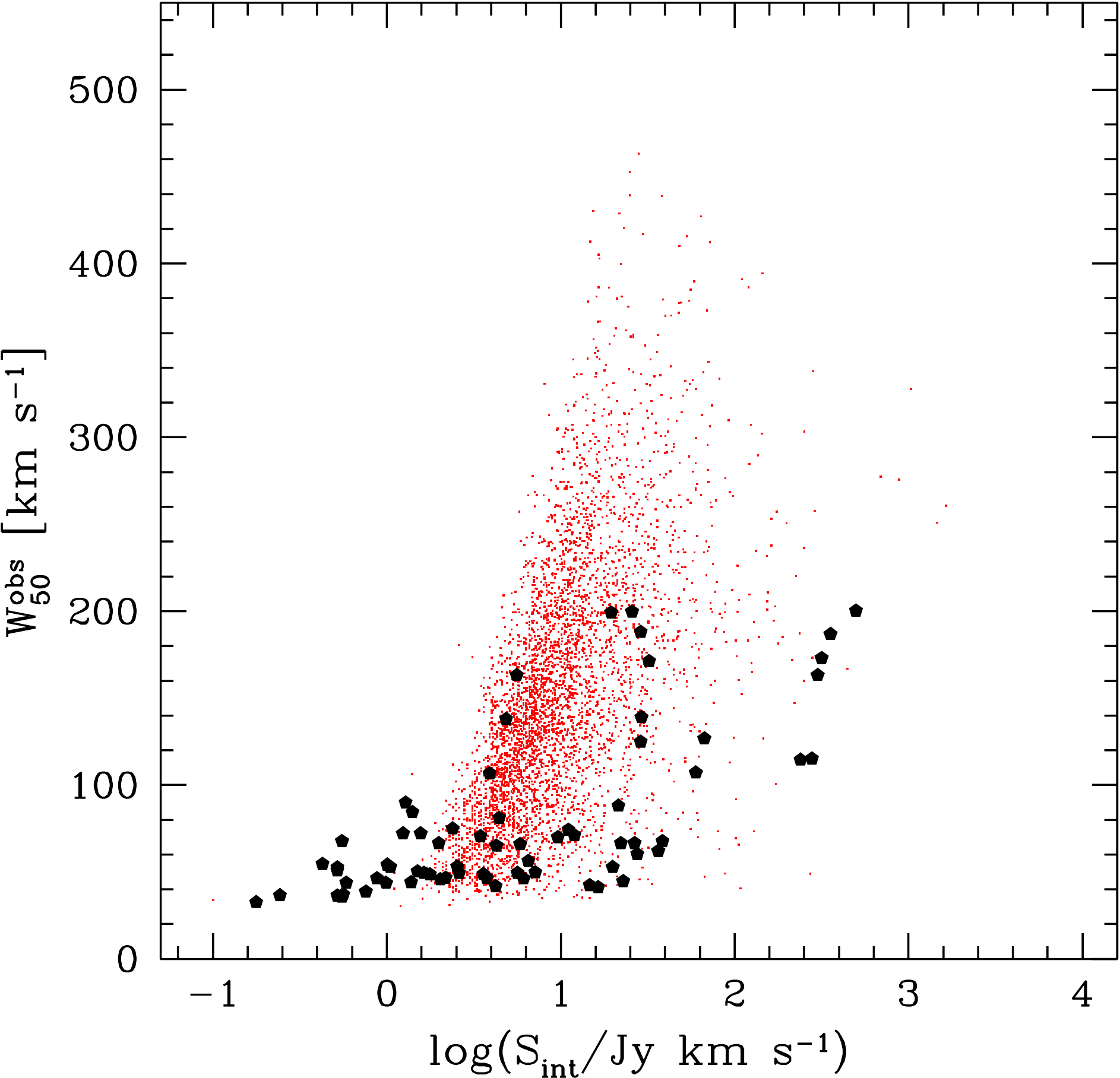}
  \includegraphics[width=5.5cm]{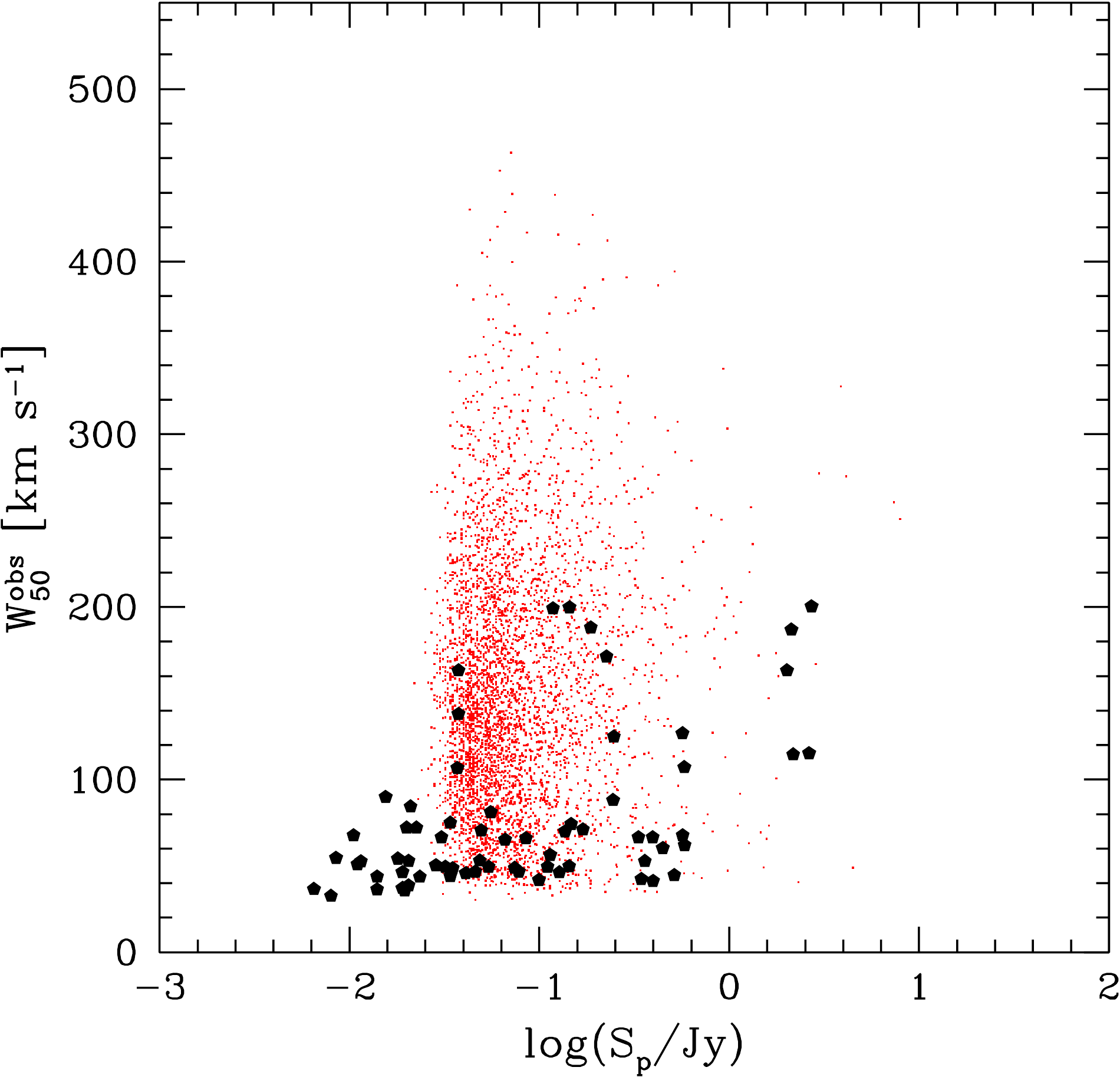}
  \includegraphics[width=5.5cm]{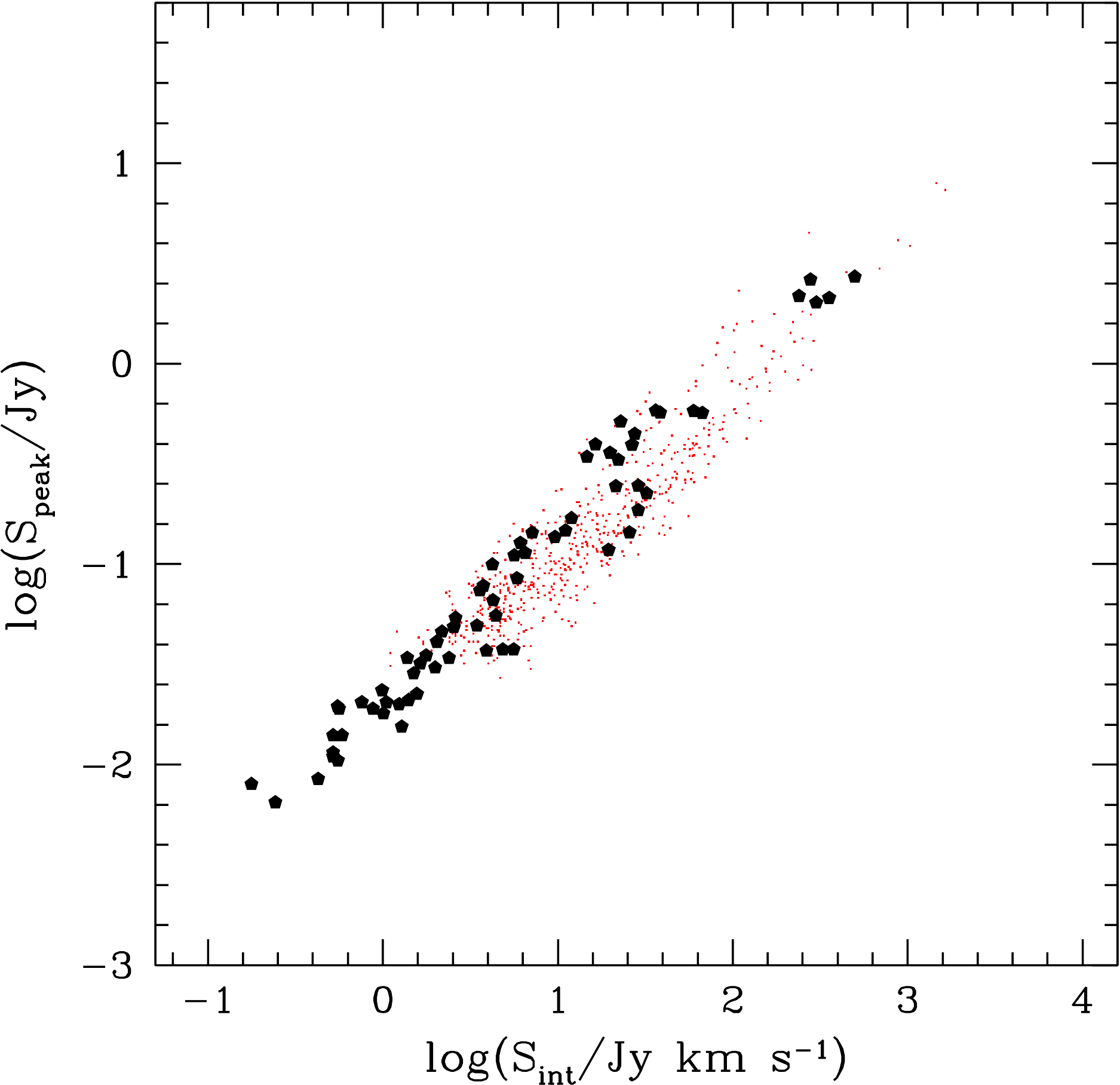}
  \includegraphics[width=5.5cm]{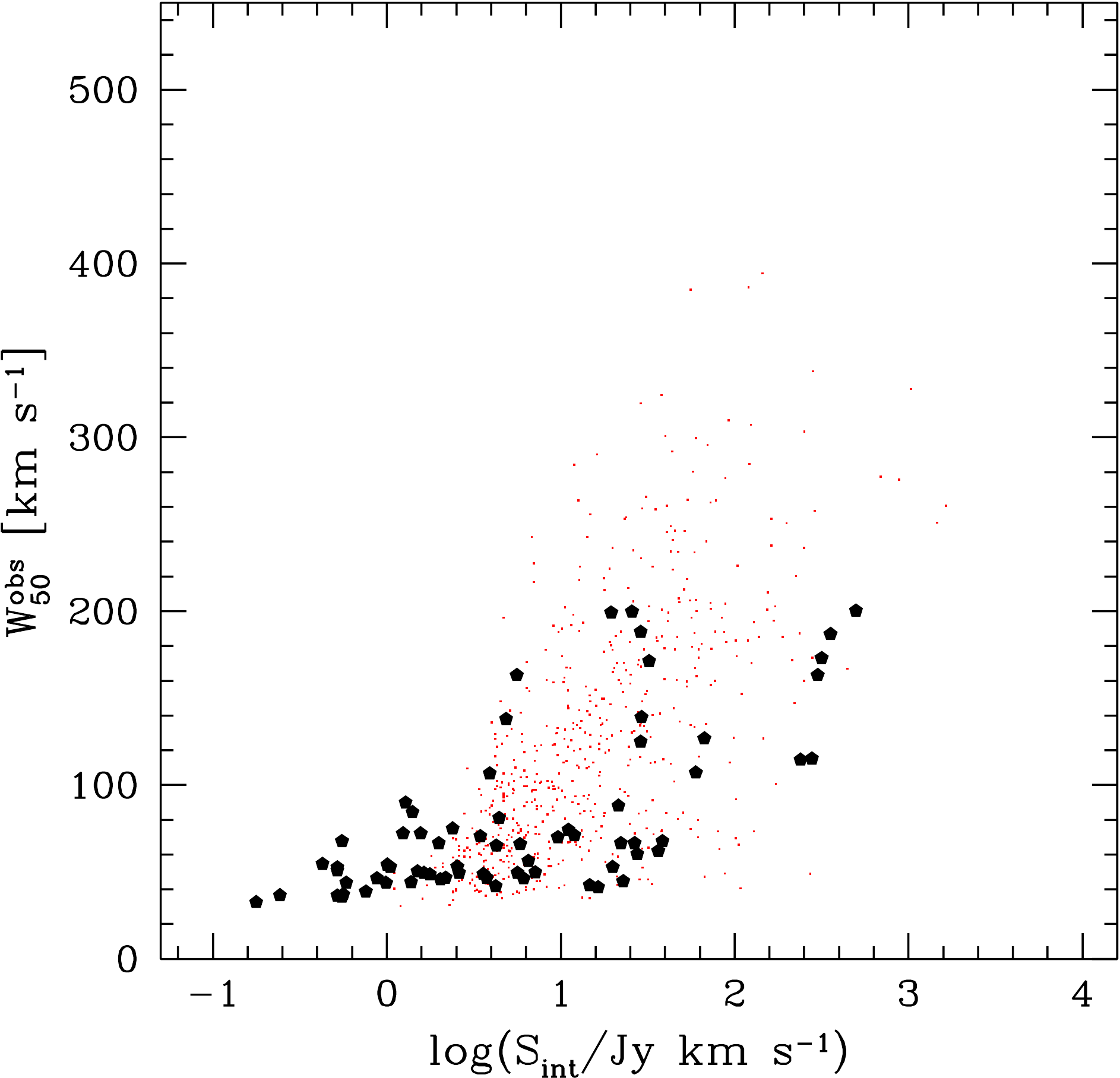}
  \includegraphics[width=5.5cm]{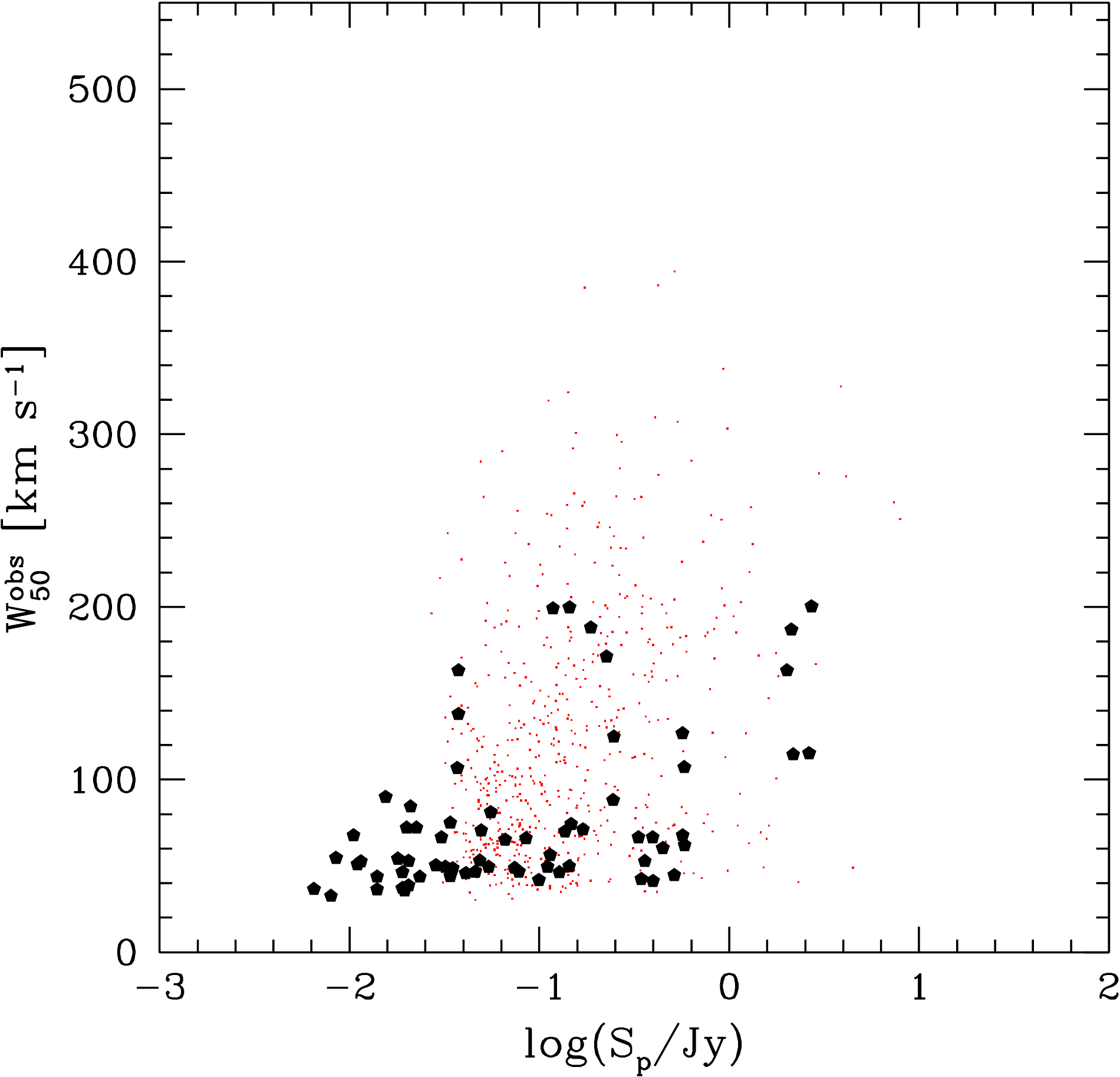}
  \caption{\small Comparison  with the HIPASS  detections.  Small
  dots  represent  the  HIPASS  measurements.   Big  filled  pentagons
  correspond to  the WSRT CVn  measurements.  Profile widths  have not
  been corrected for instrumental resolution,
  except for  the objects  WSRT--CVn--67A and WSRT--CVn-67B  for which
  the  literature  values of  profile  widths,  already corrected  for
  instrumental broadening,  have been used  in the middle  panels. The
  properties  of  the  WSRT  CVn  detections  are  compared  with  the
  properties of  the all HIPASS  detections in three upper  panels and
  with the  properties of  a subset of  the HIPASS  detections limited
  with $V_{LG} \le$ 1400 \kms\ in three lower panels, respectively. We detect very few big
  objects, i.e. galaxies with broad profiles and low \speak\ but large 
  \sint\ (when integrated over velocity). At larger distances these start 
  to dominate the sample. The comparison
  with the near galaxies in HIPASS in the bottom left panel clearly
  underlines this.} 
  \label{cvn_vs_hipass}
\end{figure*}

To get an idea which part  of the space of \HI\ parameters is explored
in the WSRT CVn survey, the measured properties of objects detected in
this survey are compared to the properties of objects detected in the southern $\delta < 2^0$ part of the \HI\
Parkes    All     Sky    Survey    \citep[HIPASS,][]{Barnes.etal.2001,
Meyer.etal.2004}.   Bivariate  distributions  of  \sint,  \speak\  and
W$_{50}^{obs}$ are  plotted in Figure~\ref{cvn_vs_hipass}.   Values of
$W_{50}^{obs}$  used in  the comparison  have not  been  corrected for
instrumental  resolution,  with  the  exception of  the  objects  with
WSRT-CVn  id's  67A  and  67B,  for which  the  $W_{50}^{res}$  values
obtained  from the  literature  have been  already  corrected for  the
finite spatial resolution of the  \HI\ observations.  The beam size of
the HIPASS survey  is 15.5 arcmin. It can be expected  that due to the
large size  of the  beam some of  the HIPASS detections  correspond to
multiple  objects.   From  the  comparison of  integrated  fluxes  and
integrated  peak fluxes  of the  detections  from the  two blind  \HI\
surveys it follows that the  WSRT CVn survey reveals \HI\ objects with
\sint\ and  \speak\ about  10 times smaller  than the  smallest HIPASS
detections.   This confirms our  ability to  detect objects  which are
faint in  \HI.  The relative  number of small  mass objects (objects
with  small   velocity  widths,  see   in  Subsection~\ref{sub_hiincl}
discussion on the effect of the inclination) in the WSRT CVn survey is
much larger when  compared to the relative amount  of low mass systems
detected in HIPASS.  On the  other hand, large (massive) galaxies have
not been detected  in the volume covered by the  WSRT CVn survey.  All
detections in  the WSRT CVn  survey have W$_{50}^{obs} \le  200$ \kms,
while  the detection  with  the  broadest profile  in  the HIPASS  has
W$_{50}^{obs} \sim 460  $ \kms. This is expected for the volume covered by the WSRT CVn survey because the \HI MF is a Schechter function with a flat faint end slope. The WSRT CVn  survey samples with high
sensitivity  a   specific  region  of   the  sky  up  to   a  distance
corresponding to $cz \sim 1330$  \kms, known to be populated by small,
gas-rich galaxies at small distances.  The HIPASS survey goes much deeper, up to $cz \sim
13000$ \kms, and  covers a variety of environments.  Therefore, for an
easier comparison, we include the same type of plots for the subset of
HIPASS detections limited with $V_{LG} \le$ 1400 \kms. Here, we mainly want to point out that the WSRT CVn survey has a much lower flux limit \sint\ than HIPASS.

\subsection{Effects of the inclination}
\label{sub_hiincl}

In the following section we  discuss the importance of the inclination
of  the objects  on  the results  presented. We use the  sizes of  the
objects measured from the \HI\ images, assuming intrinsically circular
simmetry, to obtain the inclination $i$ of an object according to
$\cos i = (b/a)$.  Because some objects are very  small or have
patchy   \HI\  distributions   the  derived   inclinations   are  only
indicative (i.e. they are very uncertain). However, our main goal here is to discuss the influence of the inclination of galaxies to the results presented so far: to the measured velocity widths and to understand if we have missed some \HI\ objects because of inclination effects. From the $b/a$ distribution (see discussion below) it appears that the errors in the measured inclinations are random (due to the noise in the data) and not systematic. We believe that for our purpose the inclinations measured from the \HI\ data should be sufficient.\footnote{We have also measured inclinations from the optical data. For example, the Spearman correlation coefficient between the inclinations measured from the WSRT CVn \HI\ data and the SDSS data is 0.41. We used finite disc thickness of 0.2.}

\begin{figure}

\includegraphics[width=0.43\textwidth]{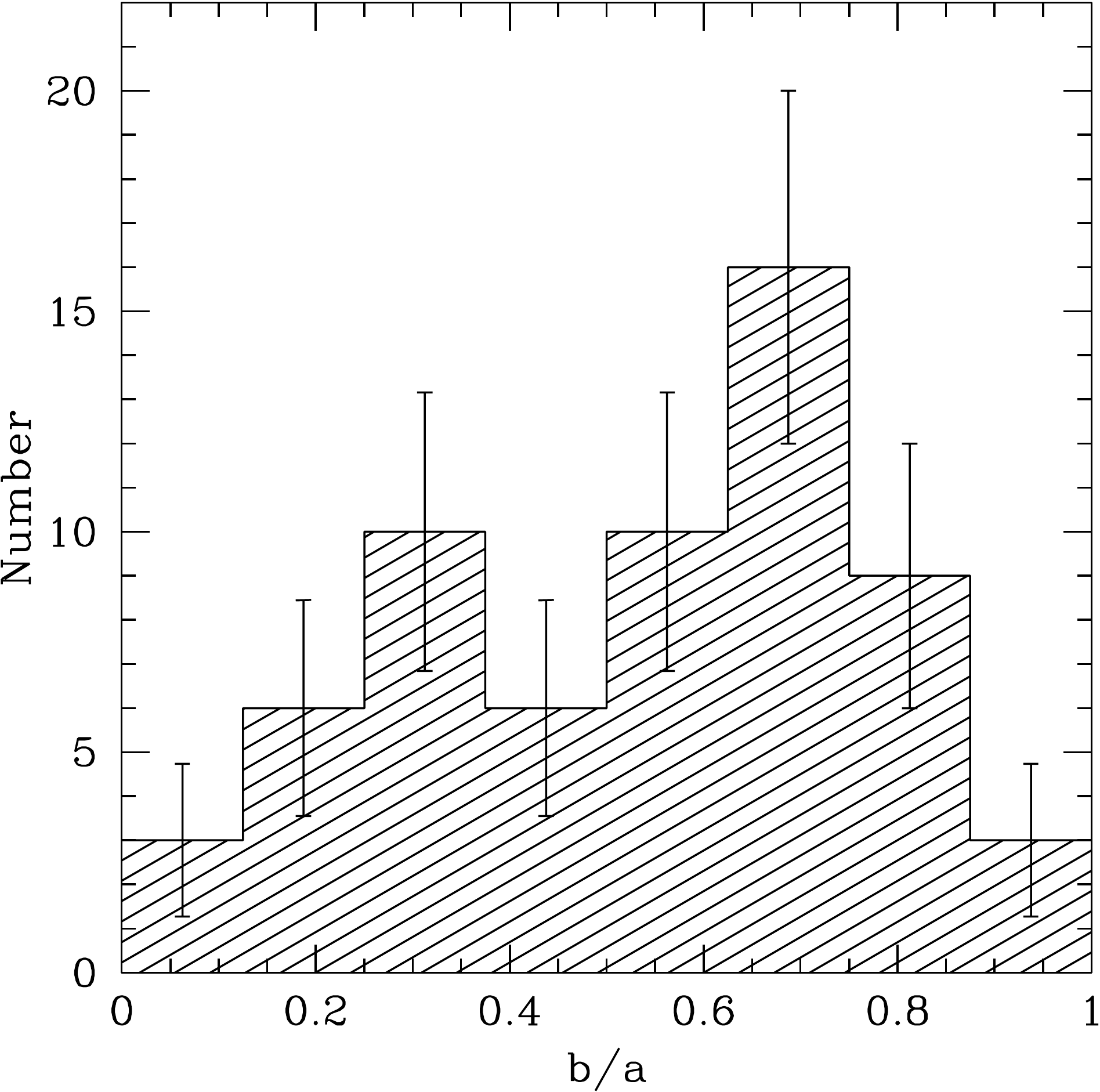}
  \caption{\small Histogram of $b/a$ ratios for 63 objects detected in
  the WSRT CVn  survey. $a$ and $b$ values are  obtained from the \HI\
  data.
\label{ba}}

\end{figure}

As  described in  Subsection~\ref{sub_hiparam}, we  estimated  the \HI\
sizes for 61 objects detected  in the survey. For 2 additional objects
we  used  the FWHM  of  a fitted  Gaussian,  not  deconvolved for  the
synthesised beam  as these objects  were too small.  The  histogram of
the  \HI\  axis  ratios  $b/a$  of  the  detections  is  presented  in
Figure~\ref{ba}.  The distribution  of  $b/a$ ratios  of  a sample  of
infitely thin  and round discs projected  randomly on the  sky will be
flat \citep{Hubble.1926}. It is  clear, that the distribution of $b/a$
values for  the galaxies in  this survey is  not flat, also when considering Poisson errors in the $b/a$ bins, and we  need to
understand why this is the case.

An important question is if there is a population of \HI\ galaxies which has
been missed in the survey due to their specific inclination. Could
the WSRT CVn survey miss galaxies with both small and large $b/a$
ratios? Apart from the errors in estimating the \HI\ diameters of the
detected objects, there are two additional effects which can cause the
observed $b/a$ distribution. First, $b/a$ can not equal 1 if disks are not circular. For example, the average face-on
system in the complete sample of Sb-Sc galaxies selected from ESO-LV
survey has $b/a \sim 0.7$ \citep{Valentijn.1994}. Second, if a galaxy has a finite thickness $z$, it means that there is a minimum $b/a$ value, and one should see an excess of galaxies at that value (equal to $z/a$) and a deficit below that value. The \HI\
disks are of finite thickness and the lack of detections with low $b/a$
ratios probably reflects this effect. So qualitatively, we can explain
both deficits in the histogram of $b/a$ ratios of the \HI\ distributions.
The conclusion then is that there is no population of \HI\ objects
with large and small $b/a$ ratios which has been systematically missed
in the survey.

\begin{figure}

\includegraphics[width=0.43\textwidth]{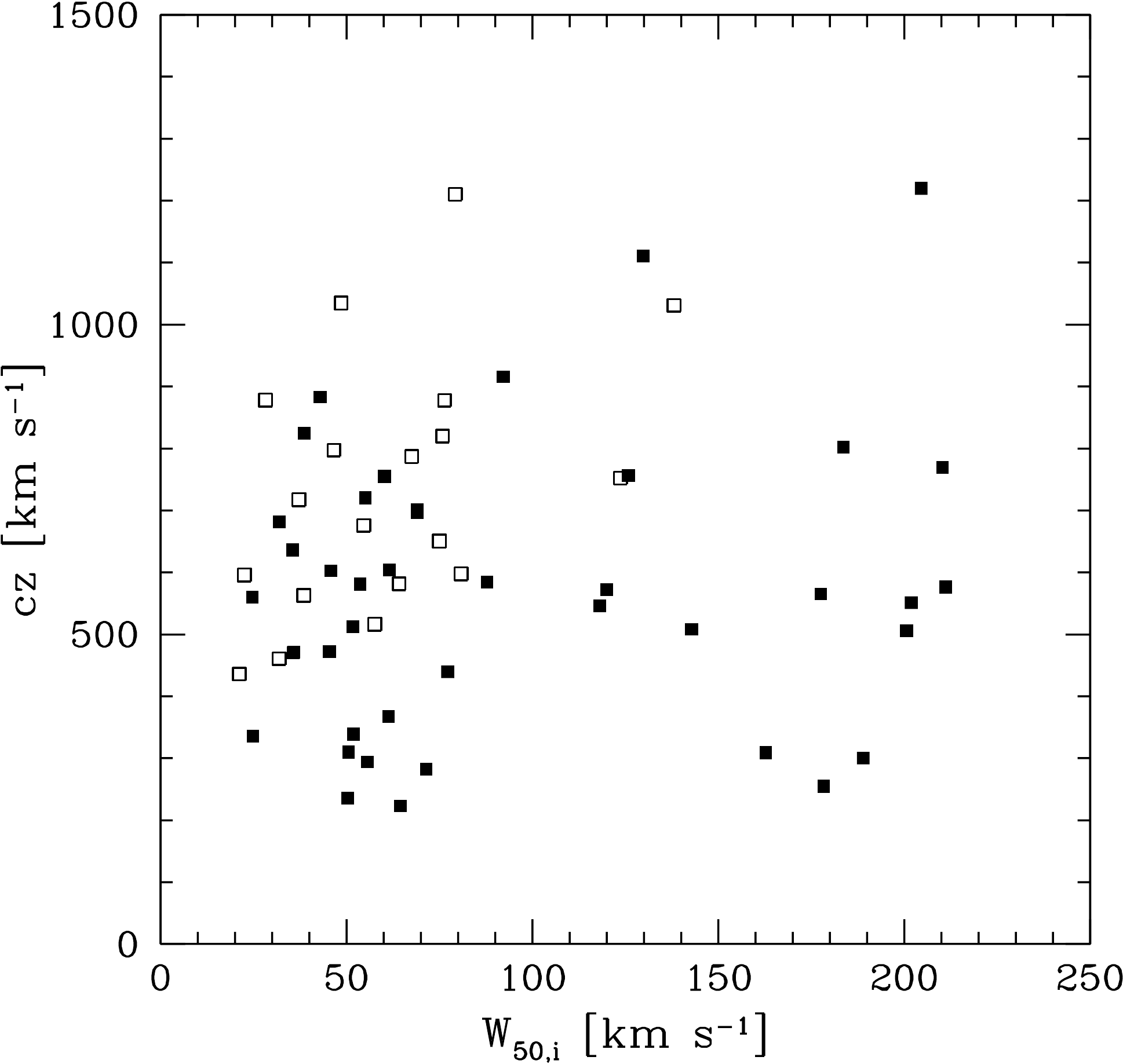}
\caption{\small Distribution of  the inclination corrected  $W_{50,i}$ values
with recession velocity.  The open symbols are
detections with \HI\ measurements obtained for the  first time in
this survey. The filled symbols  correspond to the detections with 
existing \HI\ data in the literature. 
\label{w50i}}

\end{figure}

Quantitatively, this conclusion  no systematically missed detections)
is confirmed by the  previous \HI\ observations carried out  in the
volume probed by  the WSRT CVn survey. The CVnI cloud  is nearby and a
large  number of  observations has  been carried  out in  this region,
sometimes with  much longer  integration times. However,  every single
\HI\ object  detected previously  in the nearby  Universe ($cz  < 600$
\kms)  has been  detected in  the WSRT  CVn survey.  According  to the
existing databases, only one of the known \HI\ sources has been (may be)
missed in the total volume of the survey.

From the \HI\ parameters measured for the detections from the WSRT CVn
survey, the  profile width is the  only quantity which  depends on the
inclination.  The  profile widths  presented in the  previous sections
are not corrected for the inclination - in reality, the profile widths
should be corrected with a factor $(\sin i)^{-1}$. Figure~\ref{w50i}
shows the distribution  of profile widths measured at  50\% of maximum
in the spectra and corrected for the inclination derived from the \HI\
maps, $W_{50,i}$.  In total, there  are 63 detections  with measurable
inclination  from the  \HI\ axis.  Due to  the small  inclination,
detection WSRT--CVn-35 is omitted from the plot. Generally, detections
spread  more  along  the  profile  width-axis  when  compared  to  the
corresponding  distribution  of   profile  widths  not  corrected  for
inclination     (right    panel     in    the     middle     row    in
Figure~\ref{bin_prop}).  However,  still  78\%  of all  galaxies  with
measured  $b/a$ ratios  have $W_{50,i}  \le  130$ \kms.   There is  no
drastic  change in  the  overall results  obtained  by correcting  the
profile widths for the inclination.

\section{Summary}
\label{sec_summary}

We have carried out a blind \HI\ survey with the WSRT in the volume of
the nearby  CVn groups of galaxies.  The survey covers an  area on the
sky of $\sim$  86 deg$^2$ and a velocity  range of approximately $-450
\le cz  \le 1330$ \kms\ wide. In  the volume probed by  the survey, we
detect  70  \HI\  objects.   Using  available  databases,  69  of  the
detections  can be  cross-correlated  with a  galaxy  detected in  the
optical wavelength.  Two galaxies  can not be resolved (WSRT--CVn--67A
and WSRT--CVn-67B).   One of  the detections (WSRT--CVn--61)  does not
have an optical counterpart. This  \HI\ cloud resides in the proximity
of NGC4288 (WSRT--CVn--62).

We did not detect any isolated \HI\ object without optical counterpart
 above  the   detection  limit  of   the  survey  (see  column   6  in
 Table~\ref{stat_all}). This  result is in agreement  with previous \HI\
 observations, each of them limited  by their own detection limit (see
 Table~\ref{hisumtab})  and with  the  recent
 theoretical        work        of        \citet{Taylor&Webster.2005}.
 According to their finding,
 all galaxies, whether dark or dim, would be detectable in \HI. In all
 but  one  model  studied,  dark  galaxies  would  undergo  some  star
 formation  and   therefore  they  would  be   detectable  at  optical
 wavelengths.

We estimated the various \HI\  parameters of the detected objects. The
uncertainties     in    the     parameters     are    discussed     in
Subsection~\ref{sub_hiparunc}.  The  parameters measured are presented
in    Tables~\ref{prop1}   and    ~\ref{prop2}.    The   columns    in
Table~\ref{prop1} are defined as follows:

\noindent
{\bf Column  (1)} Object's id in  the WSRT CVn survey.  \\ {\bf Column
(2)} Name of the galaxy associated with the \HI\ detection, taken from
NED.  \\  {\bf  Column (3)}  Morphology of  the  galaxy,  taken  from
HYPERLEDA. There  is no morphological classification  for objects with
 WSRT--CVn id's 29, 30 and  34 in HYPERLEDA. For these 3 objects we
provide  the  morphology  from  NED.  Object  WSRT--CVn--31  has  been
classified in NED  as a galaxy of type dE4,  which is obviously wrong,
and   we   omit  this   result.   Objects   without  a   morphological
classification either  in HYPERLEDA  or in NED  have been  marked with
``-''.  \\  {\bf Columns  (4)\&(5)}  Right  ascension and  declination
respectively, of  the galaxy  taken from NED.  For objects  without an
associated object in the NED database (objects with WSRT--CVn id's
25, 40,  42, 55  and 61), the  position of  the centre of  an ellipse,
fitted to the  \HI\ map of the object, has been  given. \\ {\bf Column
(6) \& (7)}  FWHM of a  two--dimensional Gaussian fitted to  the dirty
beam  of  a  datacube  which  contained the  detection,  expressed  in
arcsec.  \\   {\bf  Column(8)}  Position  angle,  in   degrees,  of  a
two--dimensional Gaussian  fitted to  the dirty beam  of a  cube which
contained  detection. \\  {\bf Column(9)}  Indicator of  previous \HI\
observations: 1 stands for previous \HI\ observations available in the
literature, 0 if otherwise. \\

Column description of the \HI\ properties listed in Table~\ref{prop2}:

\noindent
{\bf  Column (1)}  WSRT--CVn  id.  \\  {\bf  Column(2)} Profile  width
measured  at 50\%  of  the maximum  in  the integrated  spectrum of  a
detection,  corrected for  the instrumental  broadening, in \kms.   \\ 
{\bf Column(3)} Profile width  measured at 20\%
of the  maximum in the  integrated spectrum of a  detection, corrected
for the  instrumental broadening, in  \kms.  \\
{\bf Column(4)}  Integrated flux  in Jy \kms\  measured by  defining a
mask around  a detection.   \\ {\bf Column(5)}  Integrated flux  in Jy
\kms\ measured by  defining a box around a  detection using the MIRIAD
task MBSPECT. \\  {\bf Column(6)} Integrated peak flux in mJy measured
by defining a mask around a detection.  \\ {\bf Column (7)} Integrated
peak flux  in mJy measured by  defining a box around  a detection using
the MIRIAD task  MBSPECT. \\ {\bf Columns (8), (9)  \& (10)} Major and
minor axis and  position angle, respectively, of an  ellipse fitted to
the \HI\ distribution of a detection at a column density level of 1.25
$\times$  10$^{20}$ atoms  cm$^{-2}$. For  the objects  with WSRT--CVn
id's 7, 10,  11, 12, 15, 19, 22,  25, 30, 31, 42, 43, 47  and 61, only
the equivalent parameters of a two--dimensional Gaussian fitted to the
\HI\  map are presented.  The values  presented have  been deconvolved
with a beam  for all objects for which  measurements have been carried
out,  with the  exception of  objects with  the WSRT--CVn  id's  7 and
47 (we mark them with an $``\sim"$ sign). Major and minor axis have been given in arcseconds and position angles
are given in degrees. For the objects with an estimated size comparable to or smaller than the beam size we use an $``<"$ sign to indicate that these sizes are probably just upper limits.


\onecolumn
\begin{longtable}{l l c l l l r r r c}
\caption{\small Properties of the \HI\ detected object, part I} 
\label{prop1} \\
\hline
\hline
WSRT & NED name & Morph  & Ra (J2000)   & Dec (J2000)  & V$_{LG}$  & $Beam_a$ & $Beam_b$ & $Beam_{pa}$ & \HI\   \\
CVn ID & - & - & h\ m\ s & d\ m\ s & \kms\ & arcsec & arcsec & deg & status \\ 
\hline
\endhead
\hline
1 & NGC4359 & SBc & 12 24 11.1 & 31 31 18 & 1221.6 & 59.4 & 26.7 & 0.4 & 1 \\
2 & UGC07698 & IB & 12 32 54.4 & 31 32 28 & 310.9 & 57.4 & 27.0 & 1.6 & 1 \\
3 & UGC07428 & IB & 12 22 2.5 & 32 5 43 & 1112.1 & 59.6 & 26.4 & -0.1 & 1 \\
4 & NGC4509 & Sab & 12 33 6.8 & 32 5 30 & 917.3 & 54.9 & 27.3 & 0.3 & 1 \\
5 & MCG +06-28-022 & Sc & 12 43 7.1 & 32 29 26 & 884.6 & 53.4 & 29.0 & 0.3 & 1 \\
6 & KDG178 & - & 12 40 10.0 & 32 39 32 & 756.4 & 53.4 & 29.0 & -0.8 & 1 \\
7 & FGC1497 & Sd & 12 47 0.6 & 32 39 5 & 509.9 & 57.4 & 27.8 & 0.7 & 1 \\
8 & UGCA292 & I & 12 38 40.0 & 32 46 1 & 295.8 & 57.1 & 27.9 & 0.3 & 1 \\
9 & CG1042 & - & 12 41 47.1 & 32 51 25 & 683.1 & 57.1 & 27.9 & 0.3 & 1 \\
10 & MAPS-NGP O$\_$267$\_$0609178 & - & 12 20 25.8 & 33 14 32 & 1036.2 & 50.5 & 27.3 & -10.1 & 0 \\
11 & KUG1230+334A & Sbc & 12 32 35.9 & 33 13 23 & 799.2 & 69.7 & 24.6 & -11.6 & 0 \\
12 & KUG1230+336 & Scd & 12 33 24.9 & 33 21 3 & 826.1 & 69.7 & 24.6 & -11.6 & 1 \\
13 & MAPS-NGP O$\_$268$\_$1525572 & - & 12 36 49.4 & 33 36 48 & 518.2 & 53.4 & 27.5 & -1.8 & 0 \\
14 & NGC4395 & Sm & 12 25 48.9 & 33 32 48 & 301.9 & 46.1 & 30.0 & 0.0 & 1 \\
15 & MAPS-NGP O$\_$267$\_$0529325 & I & 12 22 52.7 & 33 49 43 & 561.7 & 54.1 & 31.8 & -5.1 & 1 \\
16 & UGC07916 & I & 12 44 25.1 & 34 23 12 & 606.0 & 55.8 & 28.7 & 0.6 & 1 \\
17 & KUG1216+353 & - & 12 19 0.6 & 35 5 36 & 753.8 & 53.6 & 29.1 & -8.2 & 0 \\
18 & UGC07427 & I & 12 21 55.0 & 35 3 4 & 722.1 & 53.6 & 29.1 & -8.1 & 1 \\
19 & MAPS-NGP O$\_$268$\_$1082578 & - & 12 44 25.5 & 35 11 48 & 879.1 & 51.8 & 29.7 & -4.6 & 0 \\
20 & NGC4534 & Sd & 12 34 5.4 & 35 31 8 & 803.6 & 51.8 & 29.6 & -4.0 & 1 \\
21 & UGC07605 & IB & 12 28 38.9 & 35 43 3 & 311.7 & 49.4 & 31.9 & 0.0 & 1 \\
22 & KUG1230+360 & - & 12 33 15.1 & 35 44 2 & 821.7 & 52.7 & 30.5 & 0.2 & 0 \\
23 & UGC07949 & I & 12 46 59.8 & 36 28 35 & 340.6 & 58.9 & 29.4 & 6.2 & 1 \\
24 & UGC07559 & I & 12 27 5.1 & 37 8 33 & 224.8 & 63.0 & 28.2 & -12.2 & 1 \\
25 & SDSSJ123226.18+365455.5 & - & 12 32 26.5 & 36 54 40 & 879.6 & 57.0 & 30.7 & -4.5 & 0 \\
26 & UGC07599 & Sm & 12 28 28.5 & 37 14 1 & 284.1 & 54.6 & 31.2 & -5.3 & 1 \\
27 & UGC07699 & SBc & 12 32 48.0 & 37 37 18 & 507.5 & 53.4 & 30.3 & -4.4 & 1 \\
28 & KDG105 & I & 12 21 43.0 & 37 59 14 & 582.9 & 49.3 & 29.1 & -1.6 & 1 \\
29 & BTS133 & ImIII & 12 24 7.5 & 37 59 35 & 652.2 & 48.3 & 29.5 & -1.5 & 0 \\
30 & BTS142 & dE2 & 12 33 6.6 & 38 7 4 & 719.2 & 45.9 & 30.4 & -2.3 & 0 \\
31 & BTS146 & dE4 & 12 40 2.1 & 38 0 2 & 462.5 & 45.2 & 30.8 & 1.8 & 0 \\
32 & KUG1218+387 & - & 12 20 54.9 & 38 25 49 & 583.1 & 46.9 & 33.0 & 19.2 & 0 \\
33 & IC3687 & IAB & 12 42 15.1 & 38 30 12 & 369.0 & 45.8 & 31.2 & 8.5 & 1 \\
34 & UGCA290 & Impec & 12 37 21.8 & 38 44 38 & 472.4 & 52.2 & 30.9 & -1.0 & 1 \\
35 & UGC07719 & Sd & 12 34 0.6 & 39 1 10 & 698.2 & 52.0 & 30.9 & -1.1 & 1 \\
36 & NGC4369 & Sa & 12 24 36.2 & 39 22 59 & 1047.1 & 50.5 & 31.6 & -1.5 & 1 \\
37 & MCG+07-26-024 & Sc & 12 33 53.0 & 39 37 39 & 677.7 & 50.6 & 30.0 & 0.3 & 0 \\
38 & UGC07678 & SABc & 12 32 0.4 & 39 49 55 & 702.7 & 45.0 & 30.1 & 0.1 & 1 \\
39 & UGC07774 & Sc & 12 36 22.5 & 40 0 19 & 552.9 & 47.5 & 29.0 & -0.5 & 1 \\
40 & - & - & 12 33 24.3 & 40 44 51 & 1032.9 & 49.1 & 28.2 & 2.3 & 0 \\
41 & UGC07751 & I & 12 35 11.7 & 41 3 39 & 638.0 & 46.4 & 31.5 & 0.3 & 1 \\
42 & - & - & 12 43 56.7 & 41 27 34 & 437.7 & 45.6 & 31.8 & -0.5 & 0 \\
43 & MAPS-NGP O$\_$218$\_$0298413 & - & 12 31 9.0 & 42 5 39 & 597.8 & 49.3 & 31.0 & 0.3 & 0 \\
44 & MCG+07-26-011 & Sd & 12 28 52.2 & 42 10 41 & 441.3 & 45.2 & 31.8 & 0.3 & 1 \\
45 & MCG+07-26-012 & Sc & 12 30 23.6 & 42 54 6 & 474.3 & 45.5 & 29.5 & 0.2 & 1 \\
46 & UGC07690 & I & 12 32 26.9 & 42 42 15 & 573.8 & 45.5 & 29.5 & 0.3 & 1 \\
47 & [KK98]133 & I & 12 19 32.8 & 43 23 11 & 599.3 & 49.1 & 31.8 & -1.3 & 0 \\
48 & UGC07608 & I & 12 28 44.2 & 43 13 27 & 578.2 & 46.7 & 33.0 & 0.0 & 1 \\
49 & UGC07577 & I & 12 27 40.9 & 43 29 44 & 237.3 & 43.8 & 31.6 & 0.0 & 1 \\
50 & LEDA166142 & I & 12 43 57.3 & 43 39 43 & 337.6 & 43.0 & 31.7 & -0.6 & 1 \\
51 & MAPS-NGP O$\_$172$\_$0310506 & - & 12 49 31.0 & 44 21 33 & 564.7 & 41.7 & 29.6 & -7.4 & 0 \\
52 & UGC07320 & Sd & 12 17 28.5 & 44 48 41 & 586.2 & 44.1 & 31.8 & -1.5 & 1 \\
53 & NGC4460 & S0-a & 12 28 45.5 & 44 51 51 & 547.7 & 45.3 & 31.2 & -1.5 & 1 \\
54 & UGC07827 & I & 12 39 38.9 & 44 49 14 & 604.4 & 45.3 & 31.2 & -1.1 & 1 \\
55 & SDSSJ124759.96+445851.4 & - & 12 48 0.0 & 44 59 0 & 1212.0 & 40.5 & 29.2 & 1.2 & 0 \\
56 & NGC4242 & Sd & 12 17 30.2 & 45 37 10 & 567.0 & 41.0 & 31.9 & -0.3 & 1 \\
57 & UGC07391 & Sc & 12 20 16.2 & 45 54 30 & 788.6 & 42.6 & 32.7 & 0.0 & 0 \\
58 & UGC07408 & I & 12 21 15.0 & 45 48 41 & 514.1 & 44.7 & 31.5 & 0.0 & 1 \\
59 & NGC4389 & SBbc & 12 25 35.1 & 45 41 5 & 771.2 & 42.5 & 32.8 & 0.0 & 1 \\
60 & UGC07301 & Scd & 12 16 42.1 & 46 4 44 & 758.3 & 47.8 & 30.1 & -5.0 & 1 \\
61 & - & - & 12 20 43.4 & 46 12 33 & 473.2 & 44.8 & 29.8 & -3.1 & 1 \\
62 & NGC4288 & SBcd & 12 20 38.1 & 46 17 30 & 586.7 & 44.8 & 29.8 & -3.1 & 1 \\
63 & NGC4656 & SBm & 12 43 57.7 & 32 10 5 & 624.7 & 61.9 & 28.6 & -1.2 & 1 \\
64 & NGC4631 & SBcd & 12 42 8.0 & 32 32 29 & 571.1 & 53.4 & 29.0 & -0.7 & 1 \\
65 & NGC4618 & SBm & 12 41 32.8 & 41 9 3 & 568.9 & 48.5 & 30.5 & 0.6 & 1 \\
66 & NGC4625 & SABm & 12 41 52.7 & 41 16 26 & 639.2 & 46.3 & 31.6 & 0.3 & 1 \\
67A & NGC4490 & SBcd & 12 30 36.4 & 41 38 37 & 618.3 & 46.2 & 31.6 & 0.1 & 1 \\
67B & NGC4485 & IB & 12 30 31.2 & 41 42 0 & 512.5 & 46.2 & 31.6 & 0.1 & 1 \\
68 & NGC4449 & Sc & 12 28 11.9 & 44 5 40 & 245.8 & 45.4 & 30.9 & 2.5 & 1 \\
69 & NGC4244 & I & 12 17 29.6 & 37 48 26 & 256.5 & 55.2 & 29.8 & -5.8 & 1 \\
\end{longtable}

\begin{longtable}{l l l l l l l r r r}
\caption{\small \label{prop2} Properties of the \HI\ detected object, part II} \\
\hline
\hline
WSRT & W$_{50}$ &  W$_{20}$ & S$_{\rm int,c}$ & S$_{\rm int,MBSPECT}$ & S$_{\rm peak,c}$ & S$_{\rm peak,MBSPECT}$ & a  & b & pa   \\
CVn ID & \kms\ & \kms\ & Jy \kms\ & Jy \kms\ & mJy & mJy & arcsec & arcsec & deg \\
\hline
\endhead
\hline
1 & 182.1 & 198.3 & 19.75 & 19.24 & 120 & 116 & 163.2 & 33.8 & -74.1 \\
2 & 45.1 & 60.0 & 36.54 & 36.13 & 585 & 580 & 192.4 & 177.6 & -26.8 \\
3 & 53.1 & 68.9 & 10.03 & 9.22 & 144 & 129 & 94.6 & 78.8 & 50.6 \\
4 & 49.0 & 74.5 & 5.46 & 6.21 & 81 & 89 & $<$51.1 & 36.6 & 49.8 \\
5 & 26.7 & 39.7 & 1.11 & 0.86 & 25 & 22 & $<$21.9 & $<$13.5 & -7.6 \\
6 & 48.1 & 71.5 & 4.32 & 4.21 & 65 & 67 & 71.3 & $<$25.9 & -0.3 \\
7 & 73.0 & 85.2 & 1.43 & 1.13 & 17 & 14 & $\sim$ 65.1 & $\sim$ 48.1 & $\sim$ 44.5 \\
8 & 25.4 & 38.1 & 14.36 & 15.01 & 342 & 346 & 81.2 & 64.2 & 83.0 \\
9 & 27.1 & 41.3 & 1.27 & 1.49 & 35 & 33 & $<$33.8 & $<$9.5 & -5.2 \\
10 & 33.9 & 60.2 & 0.32 & 0.72 & 10 & 12 & $<$35.8 & $<$18.4 & 3.4 \\
11 & 33.3 & 58.5 & 1.32 & 1.68 & 27 & 30 & $<$67.3 & 33.0 & 20.1 \\
12 & 32.4 & 51.3 & 2.36 & 2.85 & 53 & 55 & 74.8 & $<$22.3 & 56.5 \\
13 & 29.5 & 42.7 & 0.92 & 0.83 & 20 & 18 & $<$15.5 & $<$11.5 & -15.6 \\
14 & 98.3 & 117.1 & 290.69 & 266.43 & 2679 & 2589 & 674.7 & 492.1 & -49.4 \\
15 & 19.4 & 31.2 & 0.45 & 0.59 & 12 & 16 & $<$38.5 & $<$14.9 & -70.7 \\
16 & 49.6 & 65.6 & 23.57 & 20.73 & 352 & 312 & 174.6 & 61.4 & -1.6 \\
17 & 67.4 & 101.7 & 1.23 & 1.58 & 23 & 19 & $<$23.6 & $<$16.6 & 1.3 \\
18 & 31.8 & 47.2 & 3.59 & 3.59 & 74 & 74 & 54.0 & 36.1 & 27.4 \\
19 & 37.7 & 73.2 & 0.44 & 0.41 & 10 & 7 & $<$32.6 & $<$24.7 & 81.4 \\
20 & 109.8 & 124.3 & 67.58 & 66.17 & 570 & 564 & 247.3 & 159.0 & -70.2 \\
21 & 29.3 & 46.4 & 5.75 & 6.48 & 126 & 129 & 69.7 & 46.3 & -15.1 \\
22 & 37.2 & 96.1 & 0.66 & 1.36 & 16 & 20 & $<$42.2 & 32.1 & -43.7 \\
23 & 24.2 & 33.9 & 17.52 & 15.31 & 412 & 381 & 132.9 & 103.8 & 37.3 \\
24 & 49.5 & 67.0 & 27.06 & 26.11 & 401 & 388 & 196.1 & 80.8 & -47.8 \\
25 & 19.6 & 27.1 & 0.31 & 0.17 & 8 & 5 & $<$46.4 & $<$24.2 & -13.4 \\
26 & 54.2 & 73.1 & 11.86 & 12.11 & 171 & 168 & 127.5 & 54.2 & -52.6 \\
27 & 171.1 & 193.9 & 30.07 & 27.49 & 190 & 183 & 189.5 & 51.7 & 32.0 \\
28 & 31.6 & 46.0 & 1.79 & 1.75 & 35 & 35 & $<$38.4 & $<$25.1 & -73.3 \\
29 & 28.8 & 52.8 & 1.69 & 2.38 & 41 & 41 & $<$30.4 & $<$25.9 & -17.0 \\
30 & 35.6 & 67.9 & 0.41 & 0.62 & 12 & 11 & $<$45.9 & $<$4.1 & 21.9 \\
31 & 26.7 & 41.9 & 0.55 & 0.62 & 14 & 14 & 53.5 & $<$16.0 & -57.3 \\
32 & 29.7 & 47.3 & 3.51 & 4.00 & 77 & 79 & 50.6 & 39.8 & 24.5 \\
33 & 35.8 & 55.8 & 21.15 & 18.62 & 380 & 339 & 171.1 & 112.8 & -15.1 \\
34 & 32.5 & 62.2 & 1.46 & 1.80 & 32 & 32 & $<$51.8 & $<$8.9 & 58.0 \\
35 & 57.3 & 78.6 & 11.29 & 10.83 & 149 & 145 & 106.9 & 33.2 & -19.4 \\
36 & 64.0 & 109.5 & 3.68 & 5.17 & 53 & 58 & $<$40.6 & 39.8 & 24.1 \\
37 & 32.5 & 51.3 & 5.35 & 5.94 & 108 & 113 & 70.6 & 45.7 & 8.3 \\
38 & 39.3 & 60.5 & 5.93 & 7.08 & 107 & 121 & 64.9 & 43.9 & 65.5 \\
39 & 182.7 & 204.6 & 25.46 & 25.88 & 147 & 141 & 201.9 & 36.5 & -80.8 \\
40 & 55.1 & 75.4 & 1.06 & 1.42 & 20 & 20 & $<$24.3 & $<$20.4 & 17.9 \\
41 & 29.7 & 43.3 & 2.21 & 2.14 & 48 & 44 & 62.7 & $<$18.8 & 4.8 \\
42 & 18.7 & 28.4 & 0.38 & 0.72 & 18 & 21 & 52.0 & $<$11.8 & 48.9 \\
43 & 15.6 & 23.6 & 0.21 & 0.14 & 7 & 9 & $<$32.4 & $<$17.0 & -19.4 \\
44 & 53.5 & 76.1 & 3.46 & 3.45 & 52 & 47 & 53.9 & $<$28.0 & -18.1 \\
45 & 36.1 & 52.9 & 2.30 & 2.78 & 46 & 51 & 52.3 & $<$19.3 & -62.9 \\
46 & 71.0 & 88.0 & 21.49 & 21.45 & 246 & 243 & 135.2 & 87.9 & 30.9 \\
47 & 50.7 & 59.4 & 0.47 & 0.63 & 11 & 10 & $\sim$ 60.7 & $\sim$ 36.8 & $\sim$ 28.4 \\
48 & 43.3 & 61.3 & 28.45 & 26.64 & 456 & 438 & 176.1 & 168.7 & -8.2 \\
49 & 27.7 & 39.1 & 22.04 & 23.68 & 505 & 523 & 173.9 & 121.0 & -53.6 \\
50 & 21.8 & 31.1 & 0.75 & 0.76 & 21 & 20 & $<$29.7 & $<$7.0 & 5.9 \\
51 & 35.8 & 71.1 & 0.84 & 1.25 & 20 & 21 & $<$23.5 & $<$3.2 & 36.1 \\
52 & 55.2 & 74.4 & 1.49 & 1.64 & 22 & 23 & $<$33.5 & $<$20.3 & 51.5 \\
53 & 89.7 & 128.3 & 3.51 & 4.30 & 33 & 41 & 50.5 & $<$21.4 & 38.1 \\
54 & 32.7 & 48.9 & 6.80 & 7.50 & 141 & 146 & 86.3 & 42.4 & -37.8 \\
55 & 49.6 & 65.8 & 2.01 & 1.98 & 31 & 30 & 42.6 & $<$26.0 & -16.3 \\
56 & 108.0 & 125.2 & 31.00 & 26.70 & 270 & 223 & 247.0 & 155.7 & 30.9 \\
57 & 58.0 & 74.5 & 2.27 & 2.50 & 34 & 34 & 46.3 & $<$12.2 & -6.5 \\
58 & 24.9 & 39.6 & 3.69 & 4.78 & 92 & 107 & 61.1 & 47.0 & 9.0 \\
59 & 146.2 & 159.8 & 6.11 & 5.06 & 41 & 34 & 88.4 & 45.7 & -71.3 \\
60 & 121.1 & 137.1 & 5.25 & 4.46 & 39 & 36 & 112.3 & $<$ 8.2 & 82.0 \\
61 & 20.2 & 30.0 & 0.28 & 0.85 & 15 & 23 & 61.0 & 41.6 & -17.3 \\
62 & 154.3 & 181.9 & 34.27 & 30.27 & 240 & 211 & 179.5 & 112.1 & -49.3 \\
63 & 146.2 & 174.7 & -- & 299.98 & -- & 2015 & -- & -- & -- \\
64 & 183.3 & 300.9 & -- & 498.66 & -- & 2711 & -- & -- & -- \\
65 & 90.2 & 114.9 & -- & 59.80 & -- & 580 & -- & -- & -- \\
66 & 50.7 & 70.3 & -- & 38.43 & -- & 569 & -- & -- & -- \\
67A & 173.0 & 236.0 &  222.84 & 222.84 & -- & -- & -- & -- & -- \\
67B & 139.0 & 168.0 & 29.11  & 29.11 & -- & -- & -- & -- & -- \\
68 & 97.5 & 138.3 & -- & 238.97 & -- & 2173 & -- & -- & -- \\
69 & 169.8 & 189.4 & 400.49 & 311.45 & 2386 & 1863 & 933.9 & 87.0 & 48.1 \\
\end{longtable}


\twocolumn

\section*{Acknowledgements}

K.K. acknowledges financial support by The Netherlands
Organisation for Scientific Research (NWO), under Grant No.
614.031.014. 

We are  grateful to  Jacqueline van Gorkom,  Edwin Valentijn  and Marc
Verheijen for  helpful discussions and  the latter also for critical
reading of an advanced version of this manuscript. We thank the referee for providing constructive comments and help in improving the contents of this paper.

The
Westerbork  Synthesis  Radio  Telescope  is  operated  by  the  ASTRON
(Netherlands Foundation  for Research in Astronomy)  with support from
NWO.   This  research has  made  use  of  the NASA/IPAC  Extragalactic
Database  (NED) which is  operated by  the Jet  Propulsion Laboratory,
California Institute  of Technology, under contract  with the National
Aeronautics and Space Administration.  We acknowledge the usage of the
HyperLeda database (http://leda.univ-lyon1.fr).

\bibliographystyle{mn2e}
\bibliography{ref_rev1,../../bib/journals}

\appendix
\section{Atlas of \HI\ observations}
\label{app_atlas}

Here, we present the \HI\ images of the WSRT CVn detections
overlayed  over  the  DSS  $B$-band images,  their  global  \HI\
profiles and position-velocity (PV) diagrams.

For  objects with  the  WSRT-CVn id's  from  1 to  62  and for  object
WSRT-CVn-69, the  \HI\ contours  correspond to the \HI\ distribution integrated over velocity within the mask (masks  are  described   in  Subsection~\ref{sub_hiparam}).   For  the
extended  objects (objects  with the  WSRT-CVn  id's from  63 to  68),
presented  \HI\ contours  are based  on the  zeroth moment  image. Masks were not defined around the extended objects. The
contours are given at \HI\ column  density levels of 0.1, 0.5, 1, 2.5,
5, 15, 20,  25, ... with a step +5,  $\times$ 10$^{20}$ atoms cm$^{-2}$
for objects with the WSRT-CVn id's from  1 to 32, from 39 to 47, 49 to
52,  54, 55, 58,  61 and  62. An  additional contour  at level  of 7.5
$\times$  10$^{20}$ atoms  cm$^{-2}$  is added  for  objects with  the
following WSRT-CVn id's: from 33 to 38, 48, 53, 56, 57, 59 and 60. For
objects WSRT-CVn-63 and WSRT-CVn-64  \HI\ contours are given at levels
2.5, 5, 10, 25 and 50 $\times$ 10$^{20}$ atoms cm$^{-2}$.  For objects
WSRT-CVn-65 and WSRT-CVn-66 \HI\ contours  are given at levels 2.5, 5,
10, 15,  20 and 25  $\times$ 10$^{20}$ atoms cm$^{-2}$.   Similar, for
objects WSRT-CVn-67A, WSRT-CVn-67B  and WSRT-CVn-68 contours are given
at levels 3, 5, 10, 15,  20 and 25 $\times$ 10$^{20}$ atoms cm$^{-2}$.
For object WSRT-CVn-69, the contours  are given at \HI\ column density
levels of 0.1, 1, 10, 25 and 50 $\times$ 10$^{20}$ atoms cm$^{-2}$.

Global  profiles  are  obtained  with  the  MIRIAD  task  MBSPECT,  as
described in Subsection~\ref{sub_hiparam}.   The dotted vertical lines
mark the region in velocity in which the analysis was carried out. The
solid circle corresponds  to the peak in the  global profile. The open
squares and crosses are given at  the position at 50\% and 20\% of the
peak  maximum obtained in  the MBSPECT  processes of  maximisation and
minimisation of the profile  widths, respectively.  For reference, the
dotted horizontal line marks the level of zero flux density.

The  PV diagram  is a  2-dimensional slice  through  the 3-dimensional
datacube calculated  along the kinematic  axis and along  the velocity
axis. The kinematic axis  are chosen as the axis  passing through the \HI\
centre  of the  object  for  which the  velocity  gradient is  maximal
(obtained visually  using the KARMA  tool KPVSLICE). The  contours are
given at  levels -3  and -1.5  times rms noise  in the  whole datacube
(dashed black  contours) and 1.5, 3, 5,  7, 9, ... in steps +2, times  
 rms noise in
the  whole datacube  (white  continuous lines)  for  objects with  the
WSRT-CVn id's from 1 to  62 and WSRT-CVn-69. The positive contours for
objects with the WSRT-CVn id's from 63  to 66, 67A and 67B and 68 (two
PV diagrams)  are given at  1.5, 3,  5, 10, 20,  30,
... in steps +10, times rms noise in the datacube. 

For detections with  the WSRT CVn indexes from 1 to  60 and for object
WSRT-CVn-69, we present 3-panel  figures. Each of the figures contains
the \HI\ image on top of  the optical image in the left panel, the
\HI\ global profile  in the middle panel and the  PV diagram in the
right panel.  For objects WSRT-CVn-61 and WSRT-CVn-62,  the \HI\ image
on top of  the optical image is  in the top row (the  larger object is
WSRT-CVn-62). In the  middle row we present the  global profile and PV
diagram  of WSRT-CVn-61.  In  the  bottom row  we  present the  global
profile  and PV  diagram  of WSRT-CVn-62.   The  same distribution  of
panels  holds   also  for  the   pairs  of  objects   WSRT-CVn-63  and
WSRT-CVn-64, and  WSRT-CVn-65 and WSRT-CVn-66.   In the panels  in the
top  row,  WSRT-CVn-63 is  presented  in  the  lower left  corner  and
WSRT-CVn-65 is  presented in the lower  part of the  image. The middle
rows  correspond to  object WSRT-CVn-63  and WSRT-CVn-65.   The bottom
rows  correspond  to objects  WSRT-CVn-64  and  WSRT-CVn-66. The  \HI\
distribution  around WSRT--CVn--63 (NGC  4565) and  WSRT--CVn--64 (NGC
4631) is too complex for making a  high quality image of the \HI\ with
the  present  $uv$   coverage.  For  more  detail  we   refer  to  the
observations            of            \citet{Rand&vanderHulst.1993,
Rand&vanderHulst.1994}.

For the objects WSRT-CVn-67A  and WSRT-CVn-67B, only one global profile
is obtained  (left in the  middle row). In  the panel in the  top row,
WSRT-CVn-67A is the larger object in  the optical. The panel in the middle
row (right)  shows the PV diagram of  WSRT-CVn-67A, the PV  diagram in the
bottom row corresponds to object WSRT-CVn-67B.

Object WSRT-CVn-68 is very extended in  \HI. The panel in the top row
shows  the \HI\  contours  of this  detection  on top  of its  optical
counterpart. In the middle row in the left the global profile panel is
presented. The kinematic axis used for the PV diagram presented in the
right panel  in the middle row  represents the kinematics  of the \HI\
emission concentrated on top of the optical counterpart. The kinematic
axis used  to produce the  PV diagram in  the bottom row  are obtained
taking into consideration all extended \HI\ emission.

\clearpage

\onecolumn
\begin{figure}

\centering
WSRT-CVn-1
\vskip 2mm 
\includegraphics[width=0.25\textwidth]{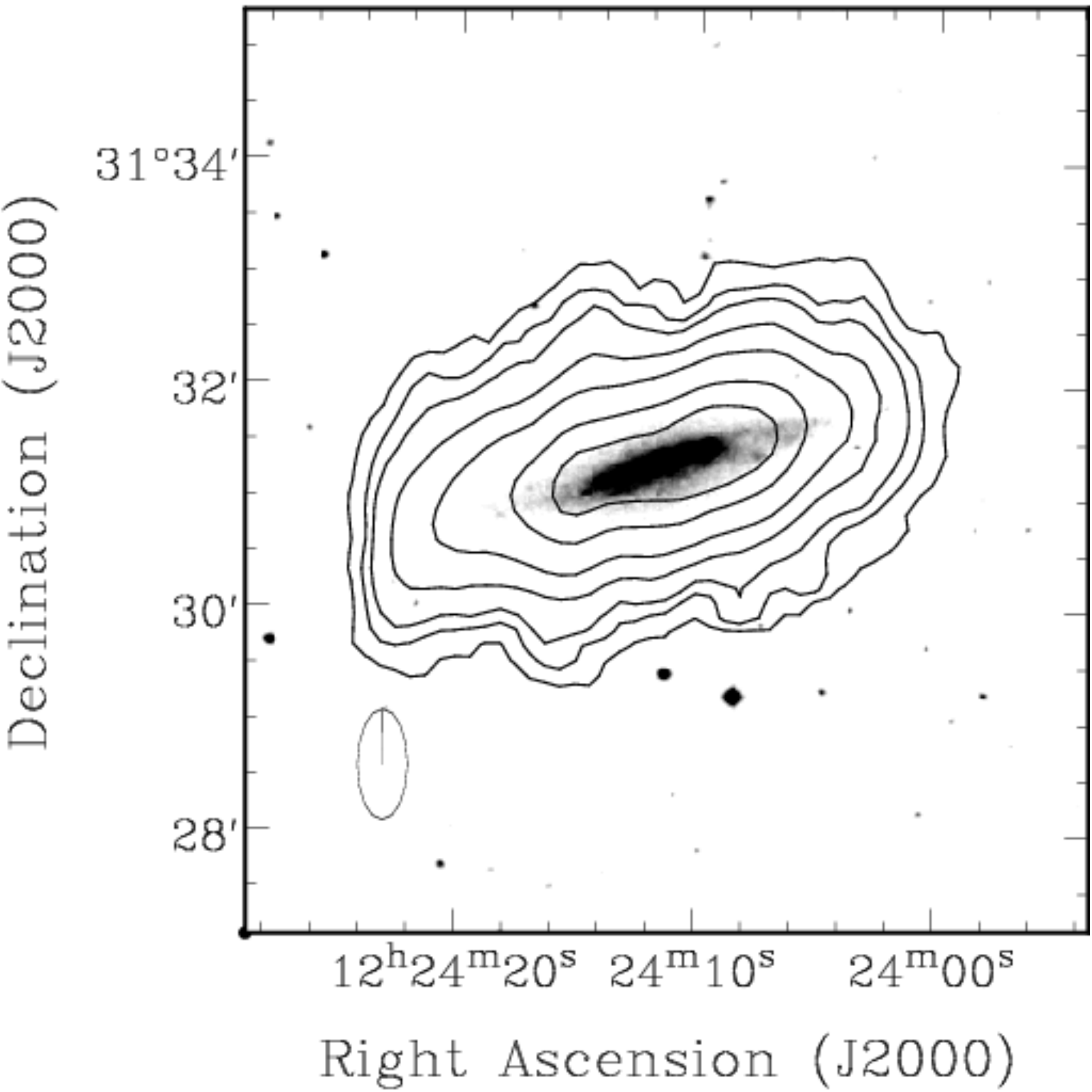}
\hskip 5mm
\includegraphics[height=0.17\textheight]{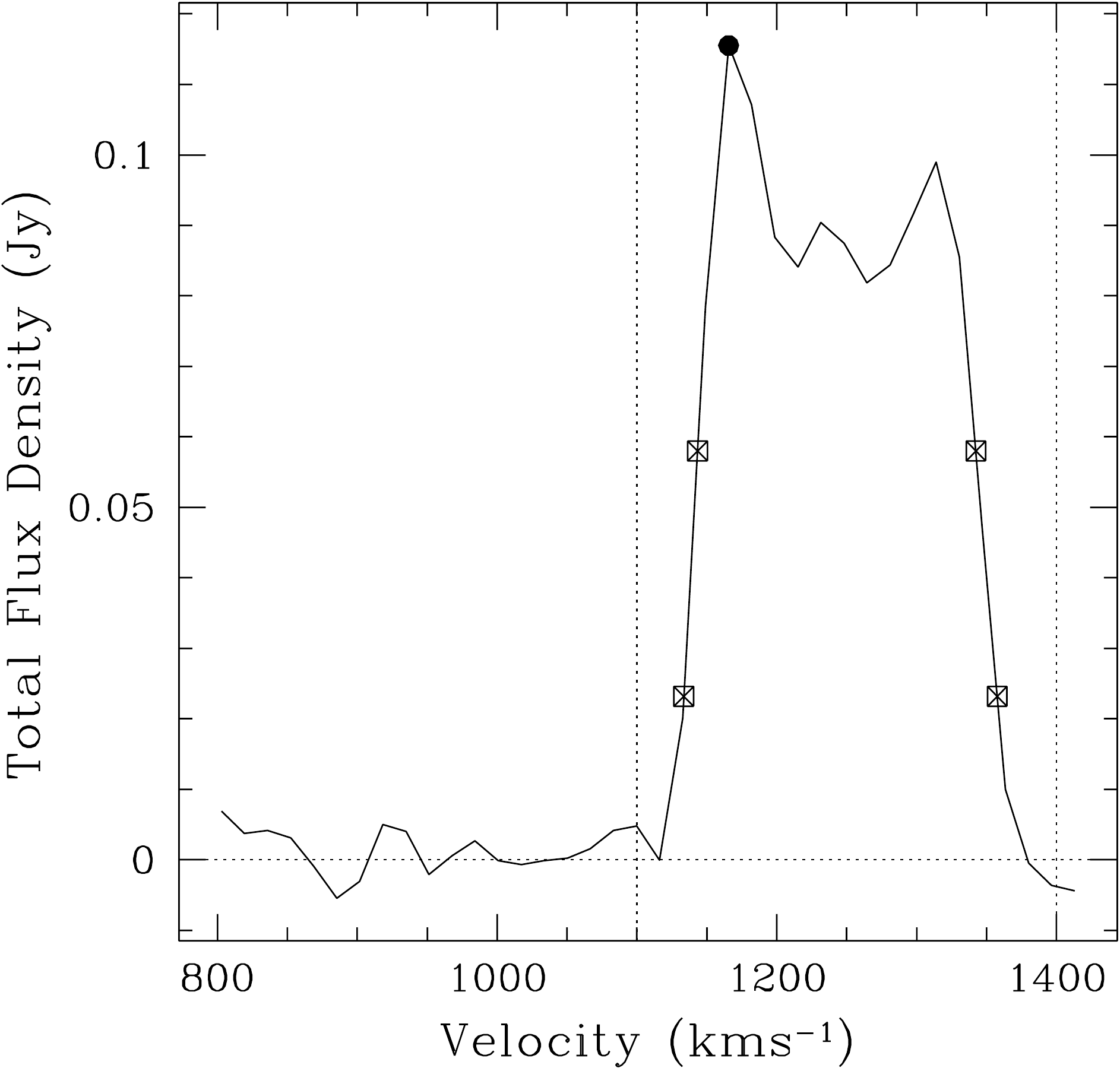}
\includegraphics[height=0.17\textheight]{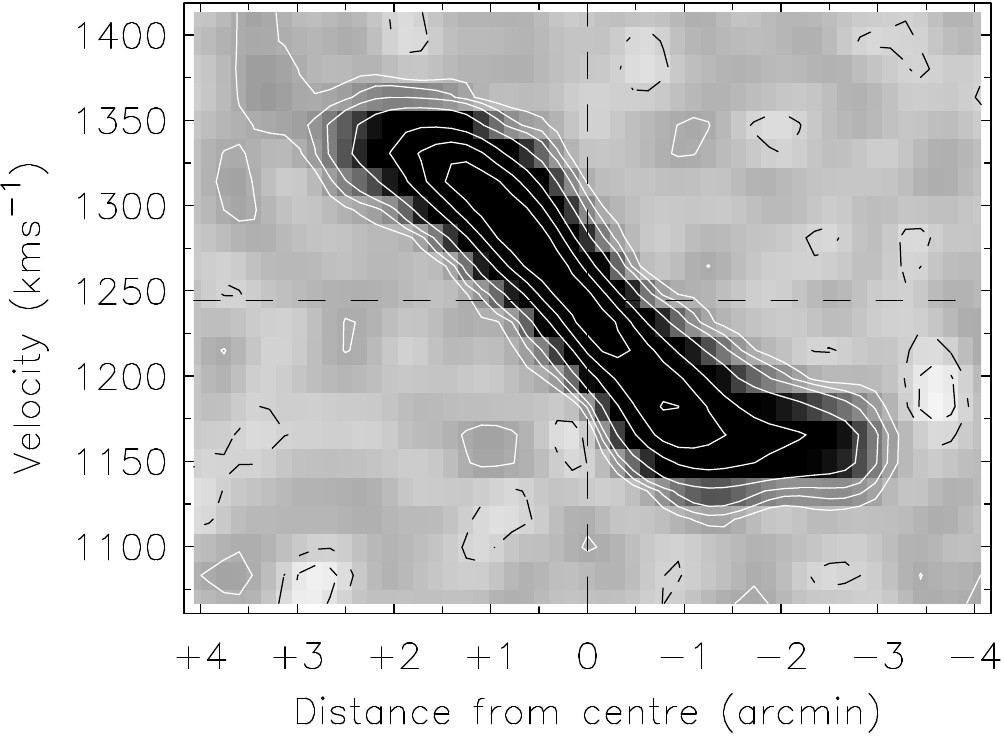}

\vskip 2mm
\centering
WSRT-CVn-2
\vskip 2mm
\includegraphics[width=0.25\textwidth]{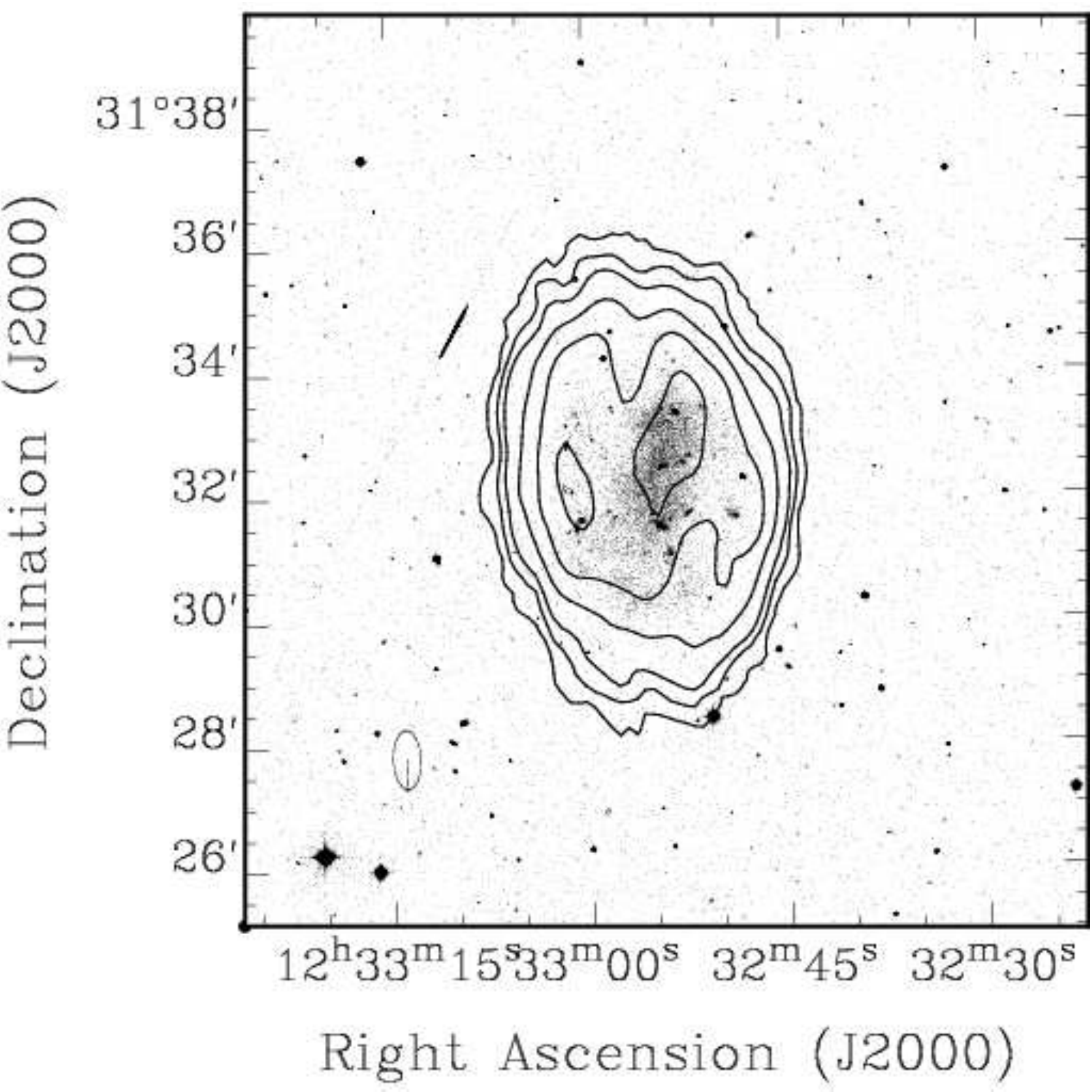}
\hskip 5mm
\includegraphics[height=0.17\textheight]{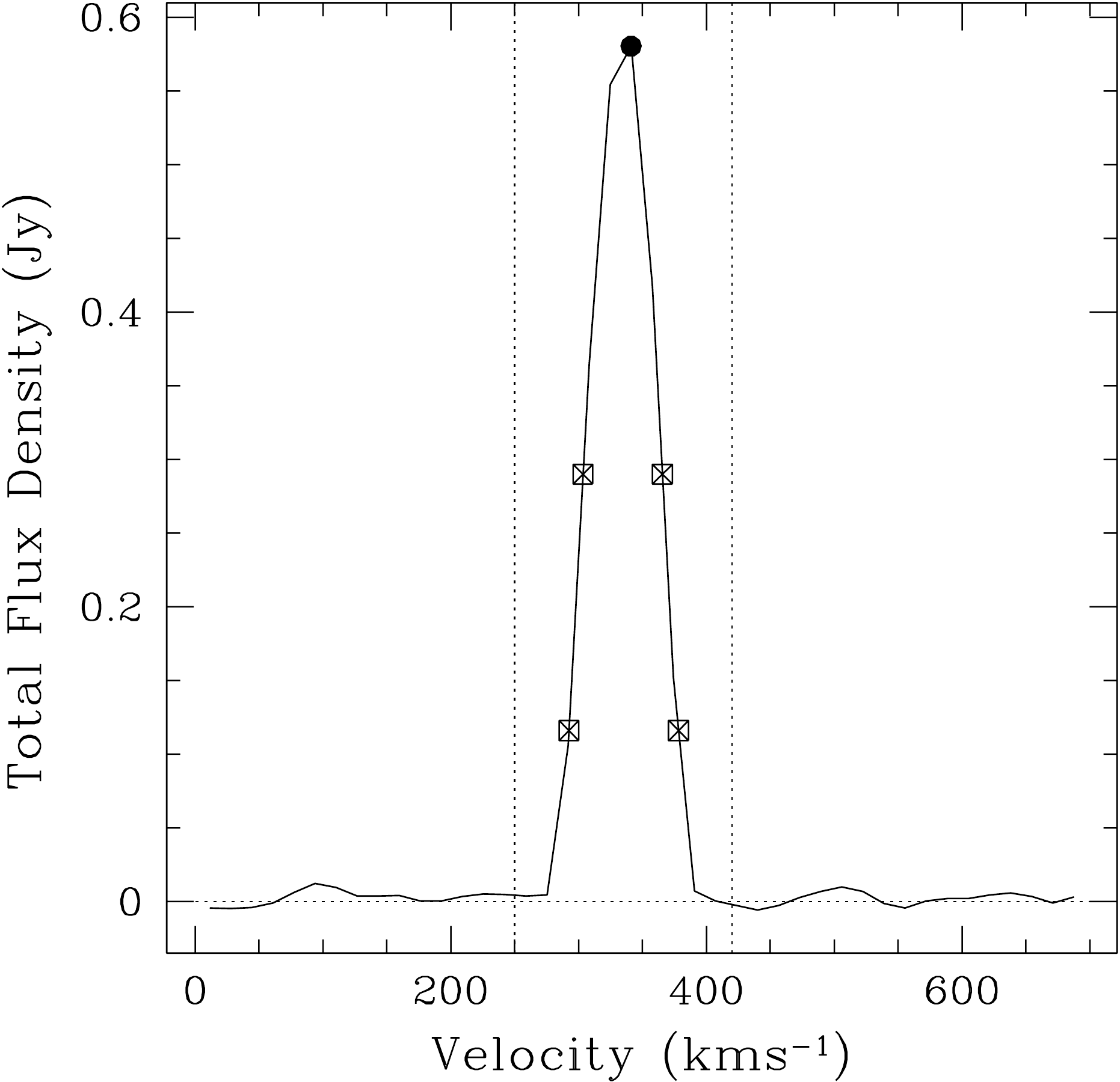}
\includegraphics[height=0.17\textheight]{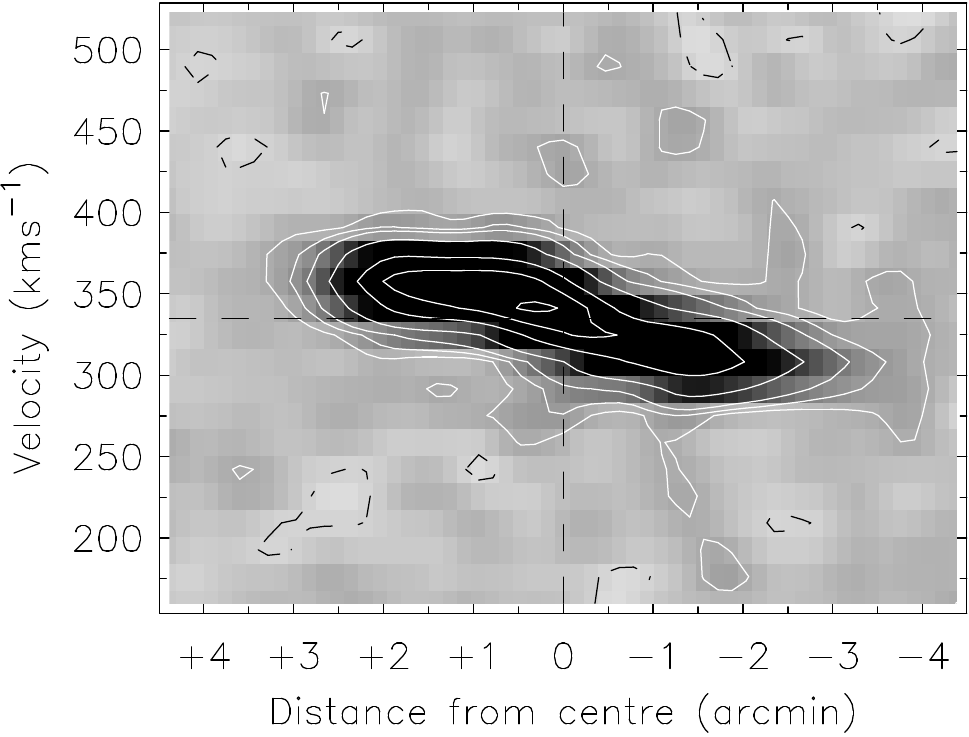}

\vskip 2mm
\centering
WSRT-CVn-3
\vskip 2mm
\includegraphics[width=0.25\textwidth]{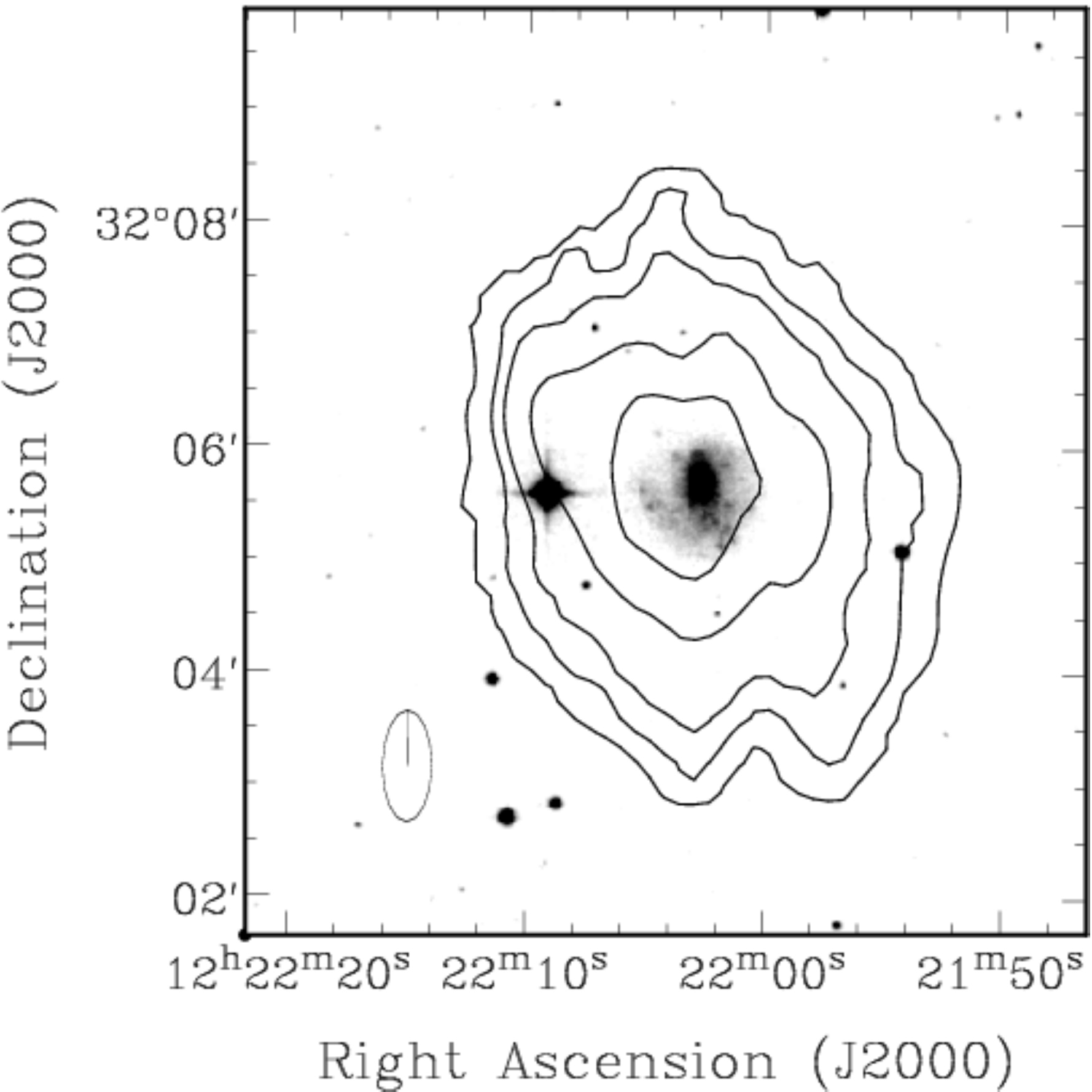}
\hskip 5mm
\includegraphics[height=0.17\textheight]{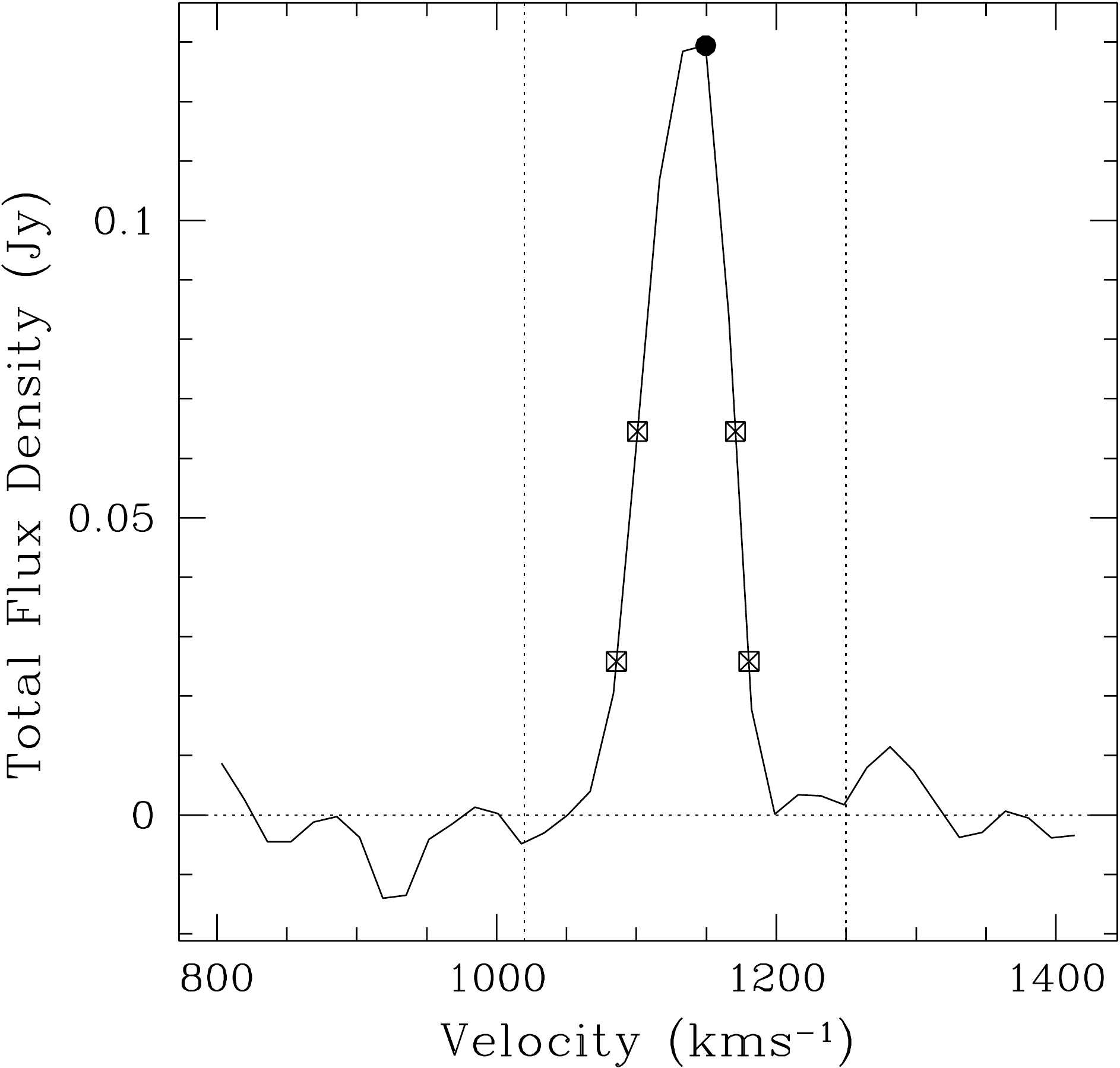}
\includegraphics[height=0.17\textheight]{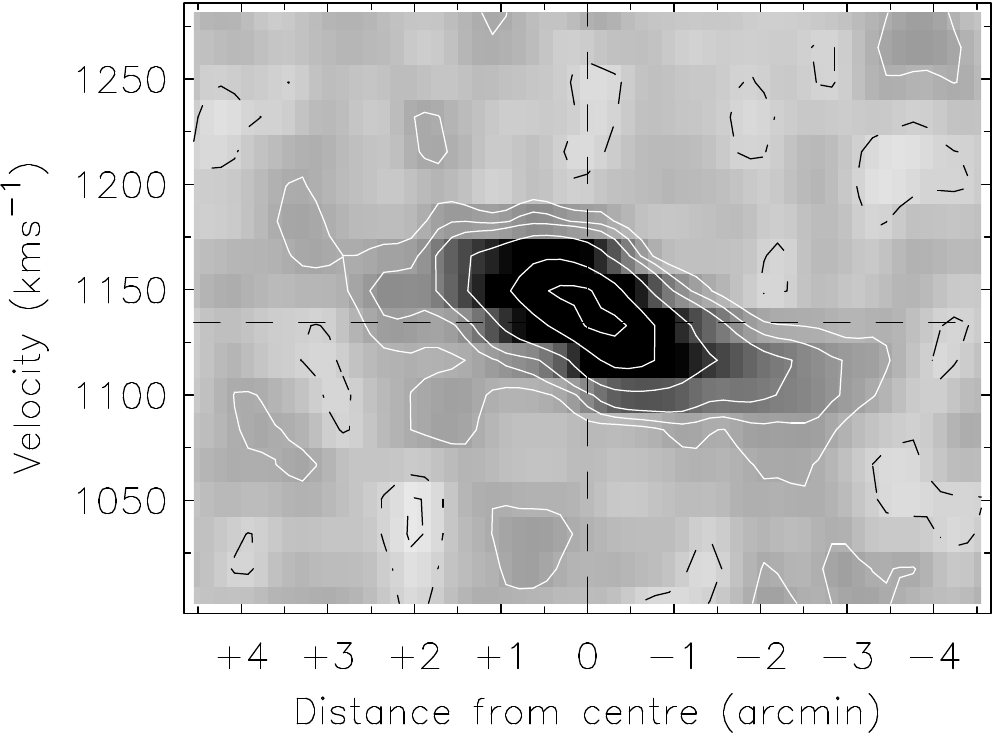}

\vskip 2mm
\centering
WSRT-CVn-4
\vskip 2mm
\includegraphics[width=0.25\textwidth]{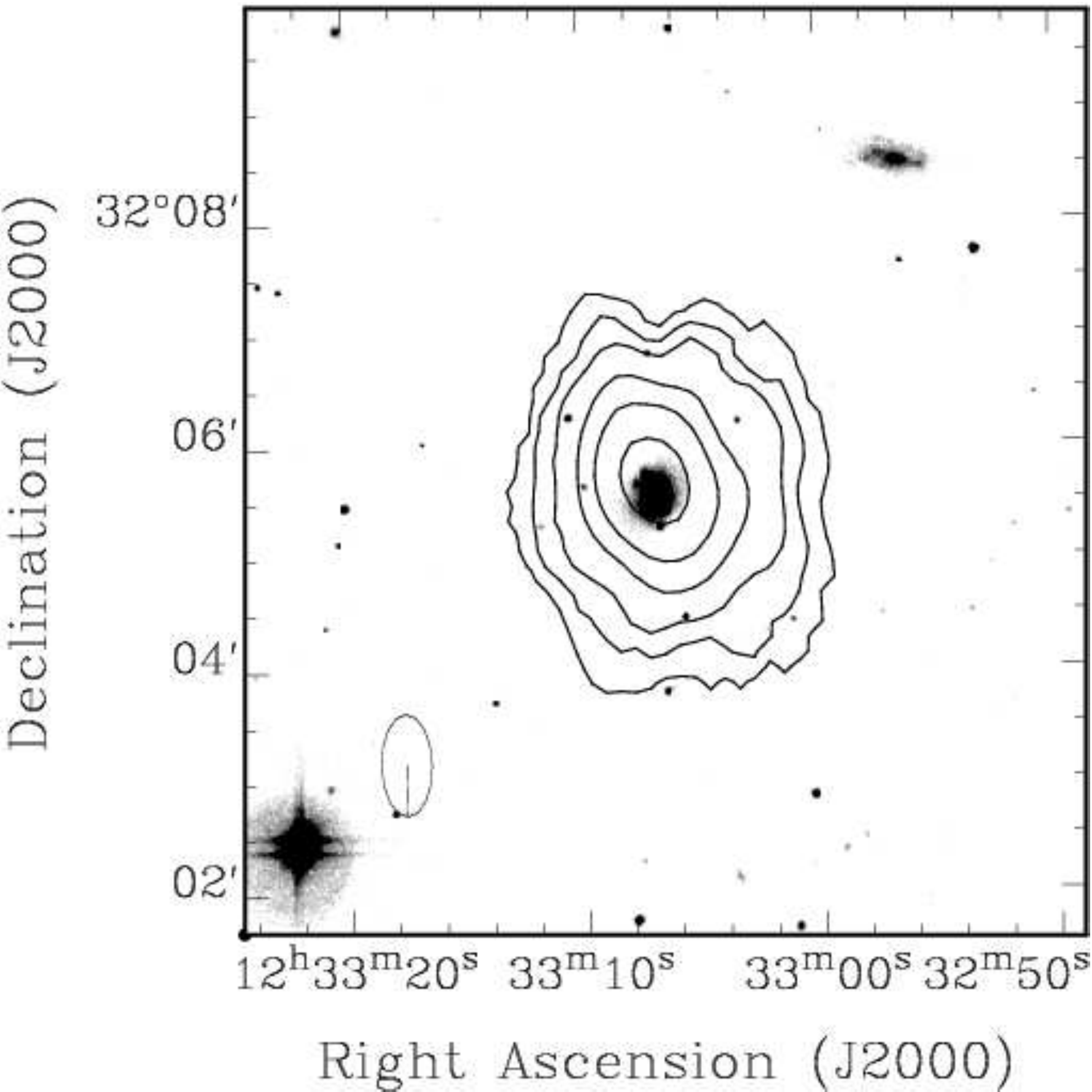}
\hskip 5mm
\includegraphics[height=0.17\textheight]{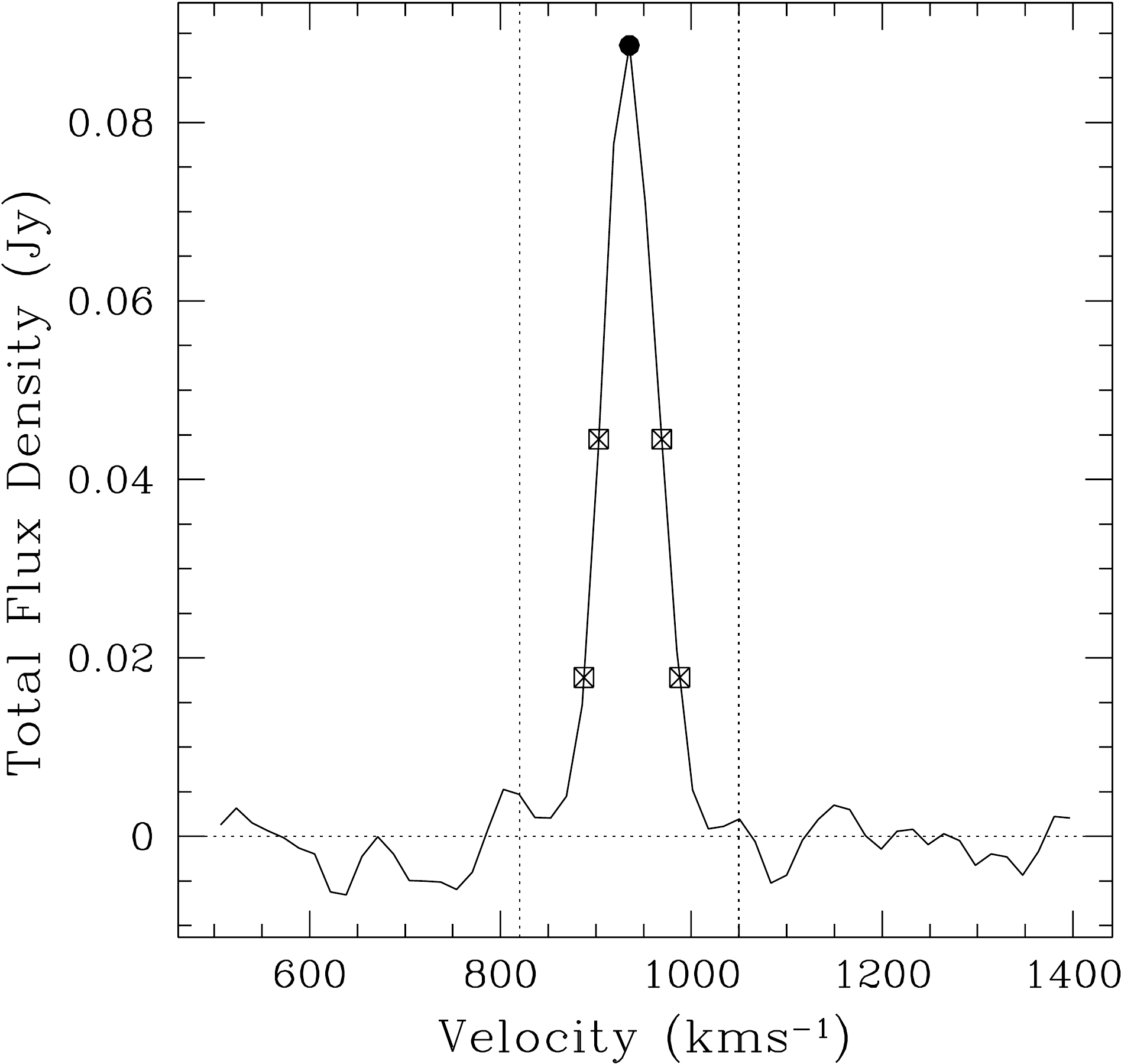}
\includegraphics[height=0.17\textheight]{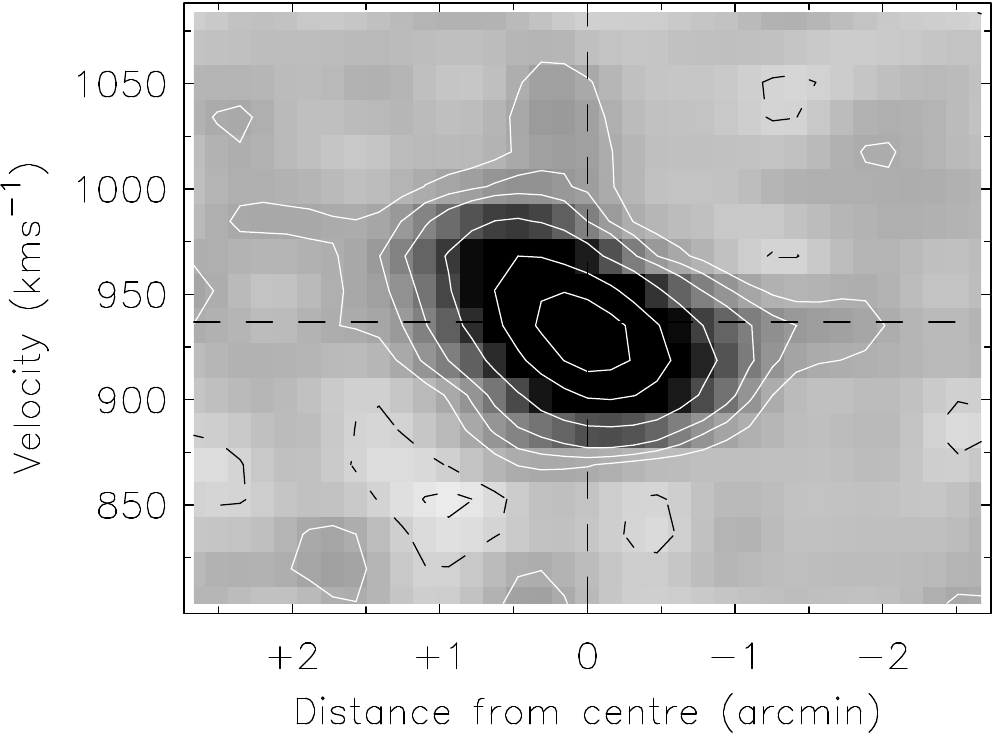}

\end{figure}

\addtocounter{figure}{-1}
\begin{figure}

\vskip 2mm
\centering
WSRT-CVn-5
\vskip 2mm
\includegraphics[width=0.25\textwidth]{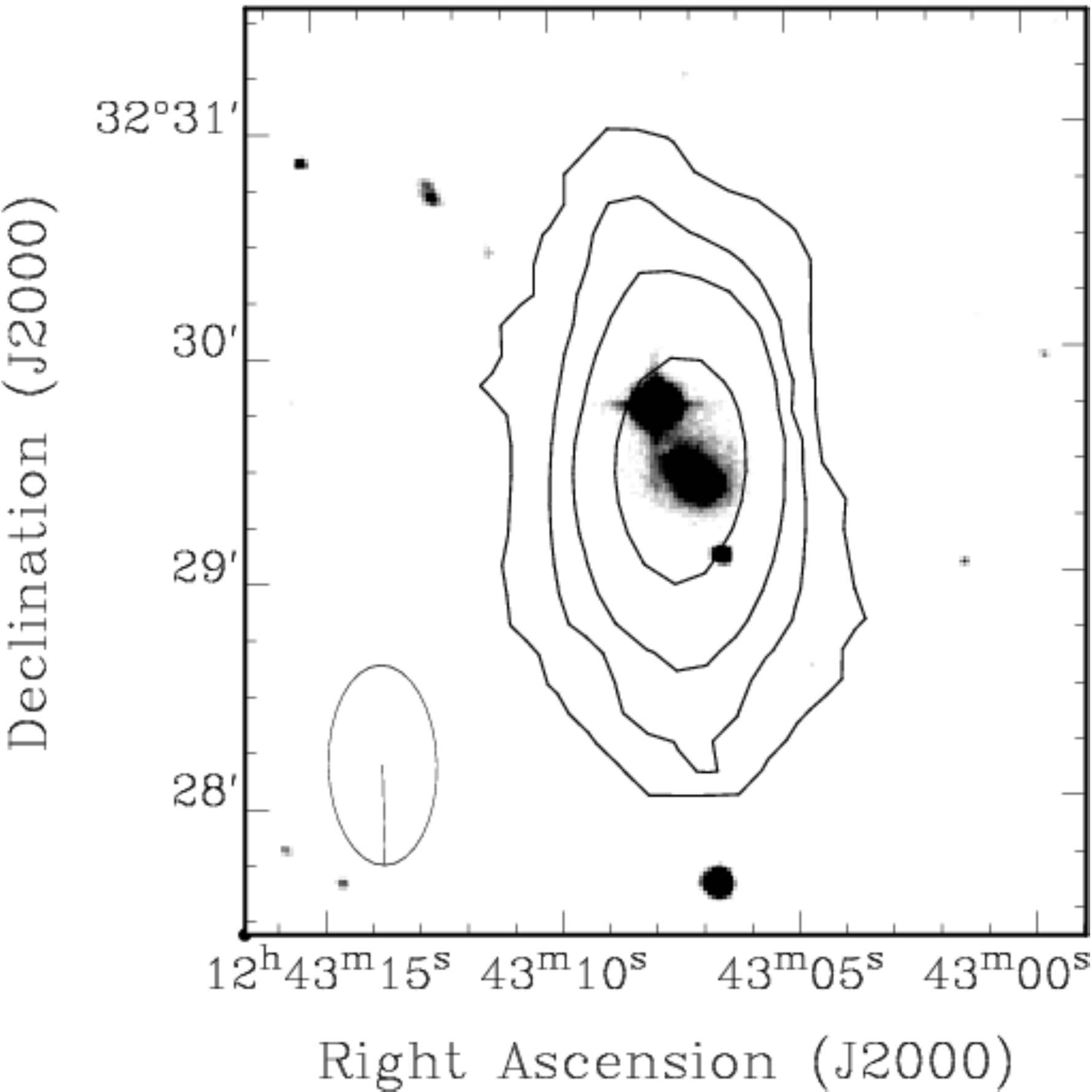}
\hskip 5mm
\includegraphics[height=0.17\textheight]{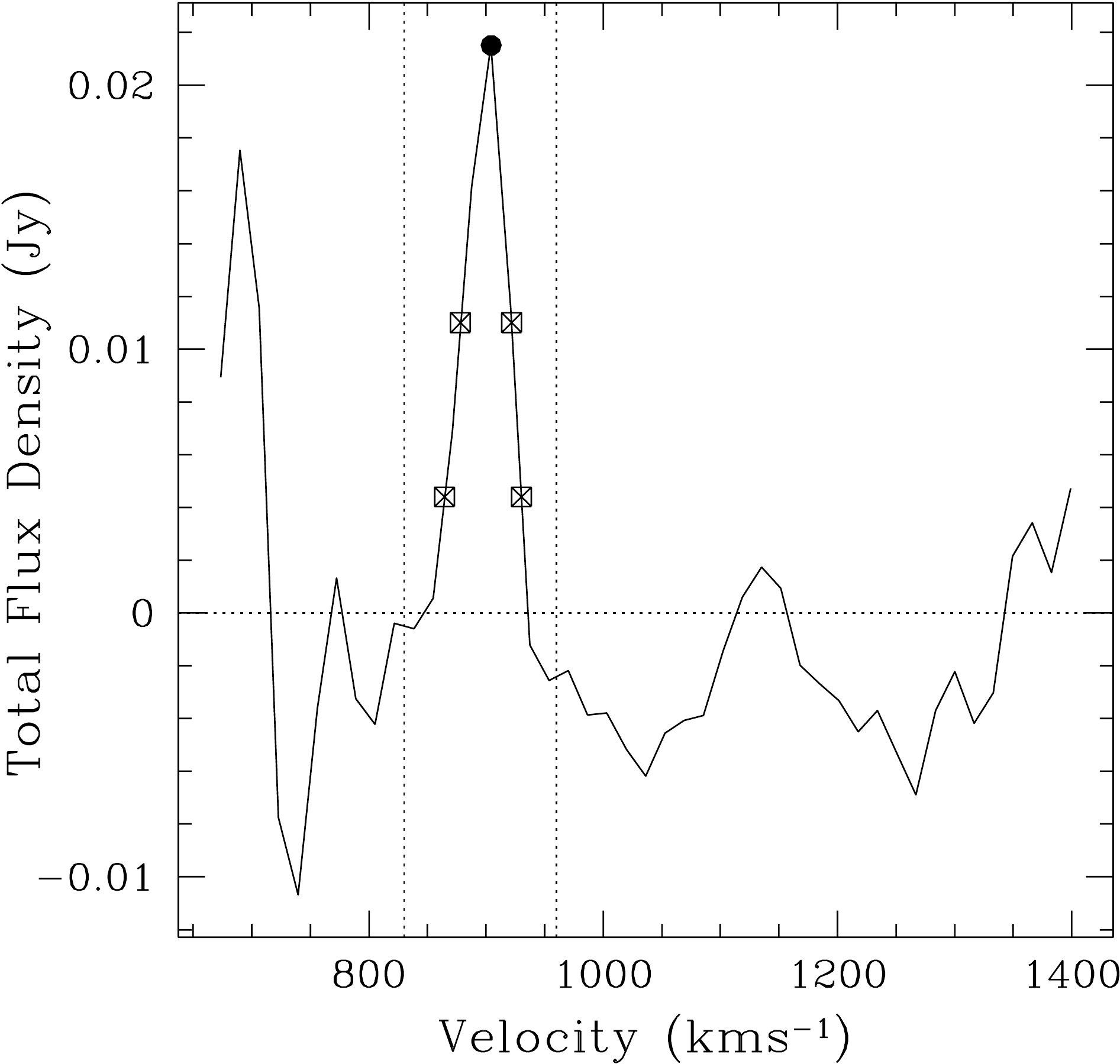}
\includegraphics[height=0.17\textheight]{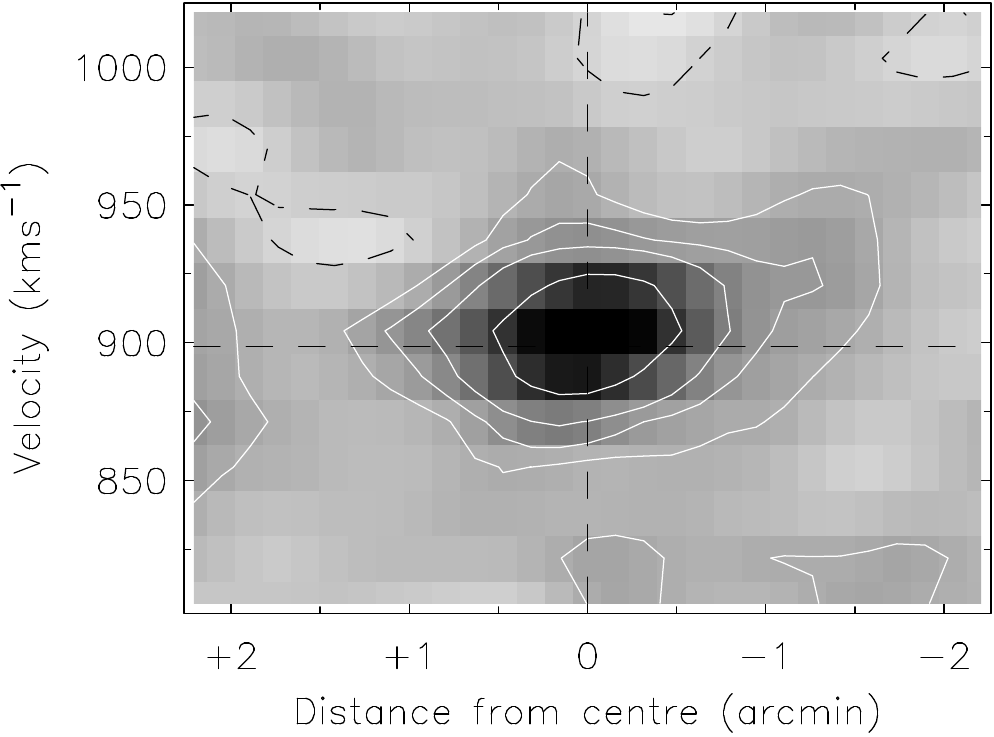}

\vskip 2mm
\centering
WSRT-CVn-6
\vskip 2mm
\includegraphics[width=0.25\textwidth]{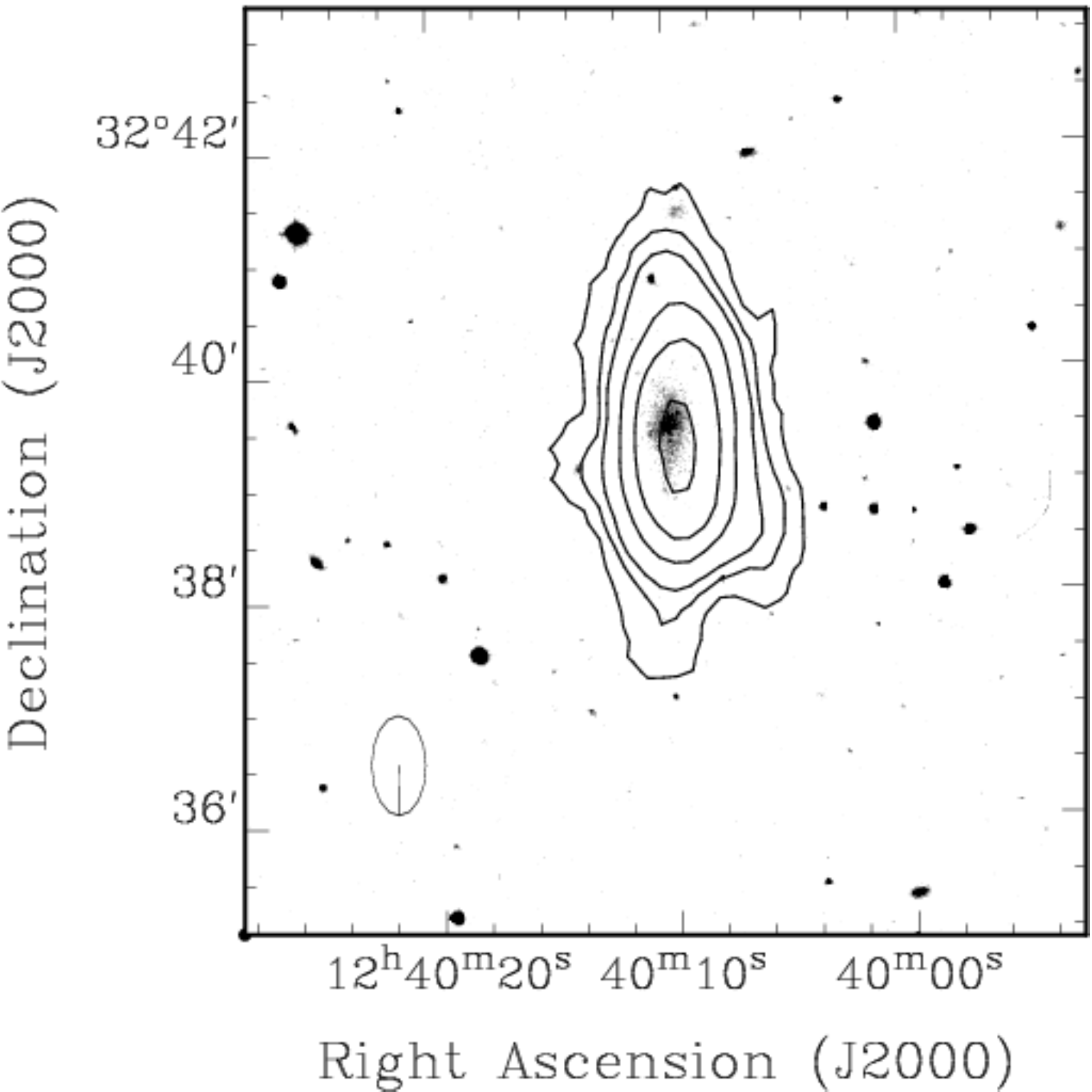}
\hskip 5mm
\includegraphics[height=0.17\textheight]{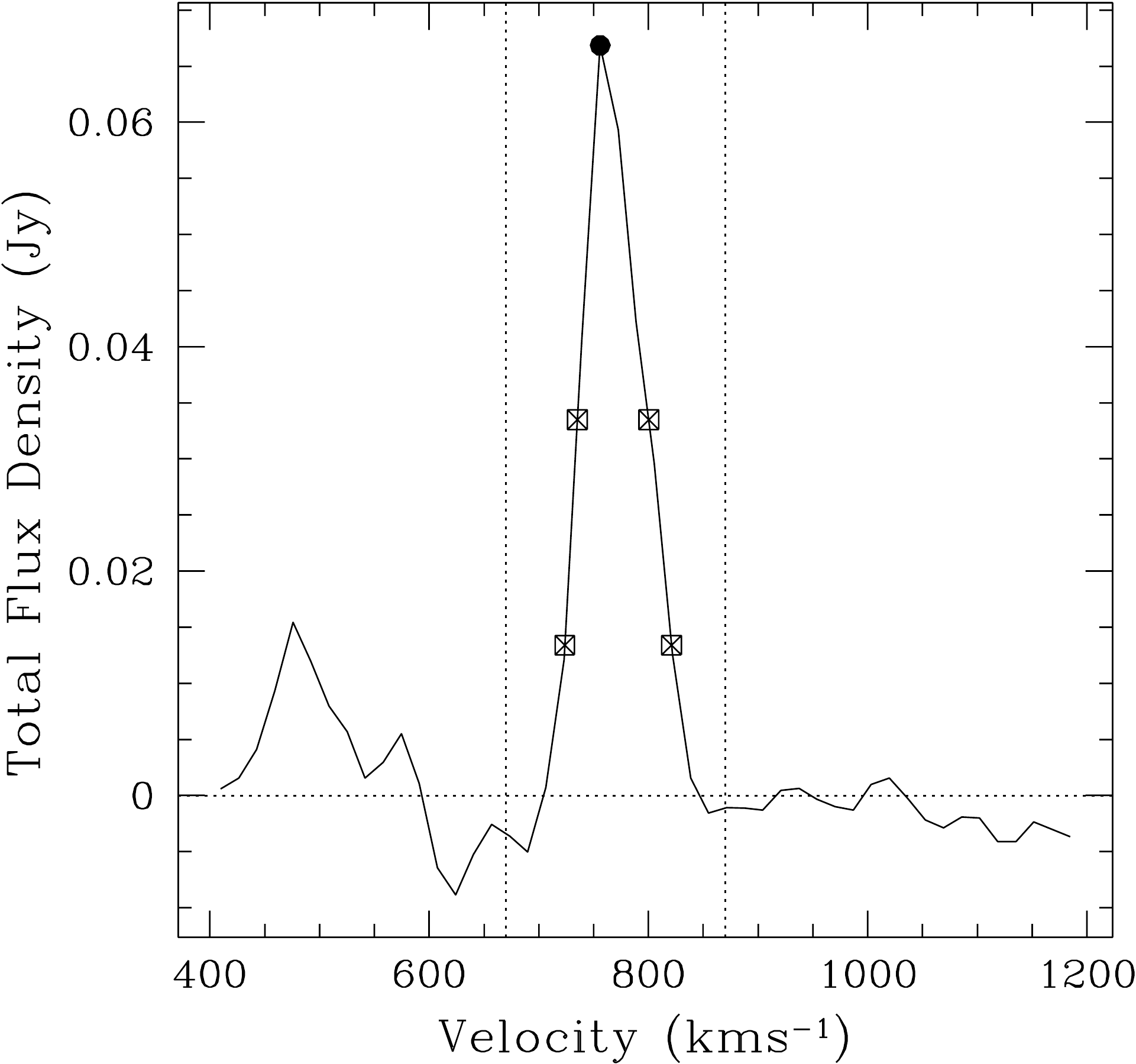}
\includegraphics[height=0.17\textheight]{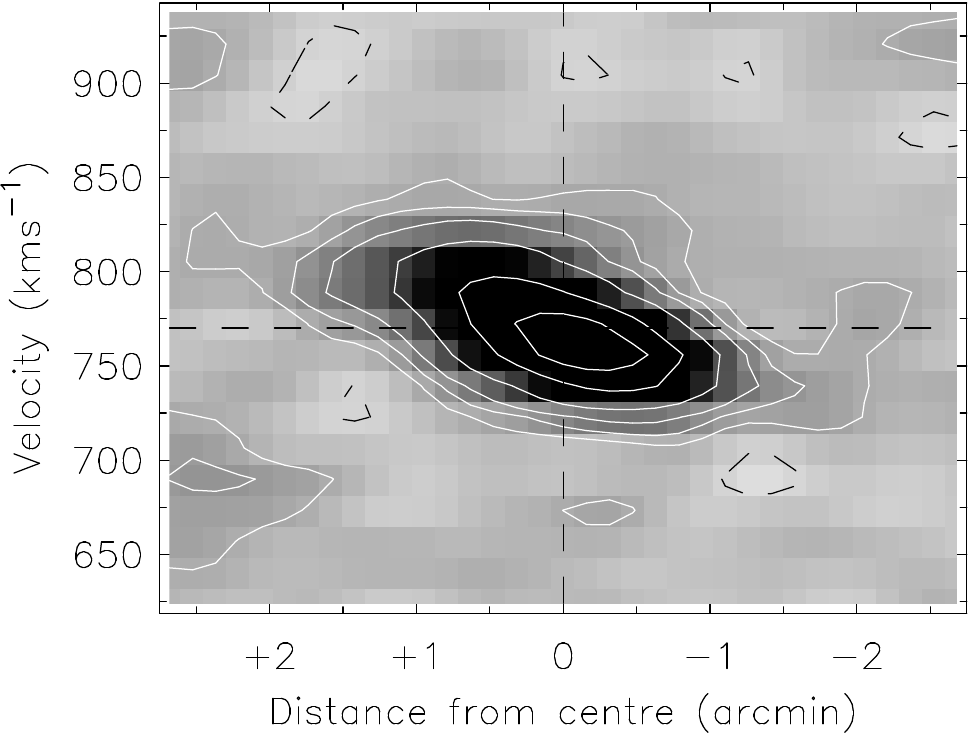}

\vskip 2mm
\centering
WSRT-CVn-7
\vskip 2mm
\includegraphics[width=0.25\textwidth]{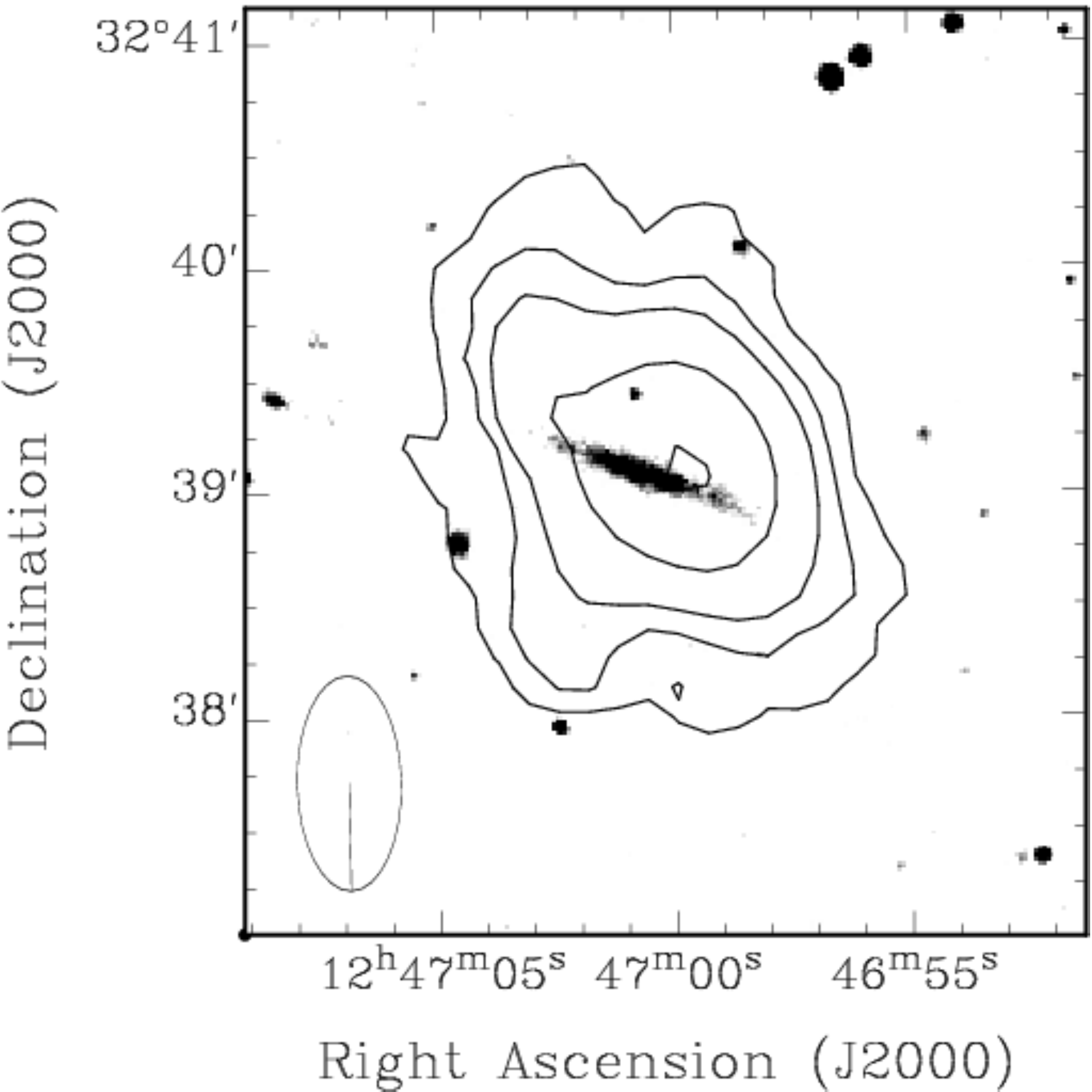}
\hskip 5mm
\includegraphics[height=0.17\textheight]{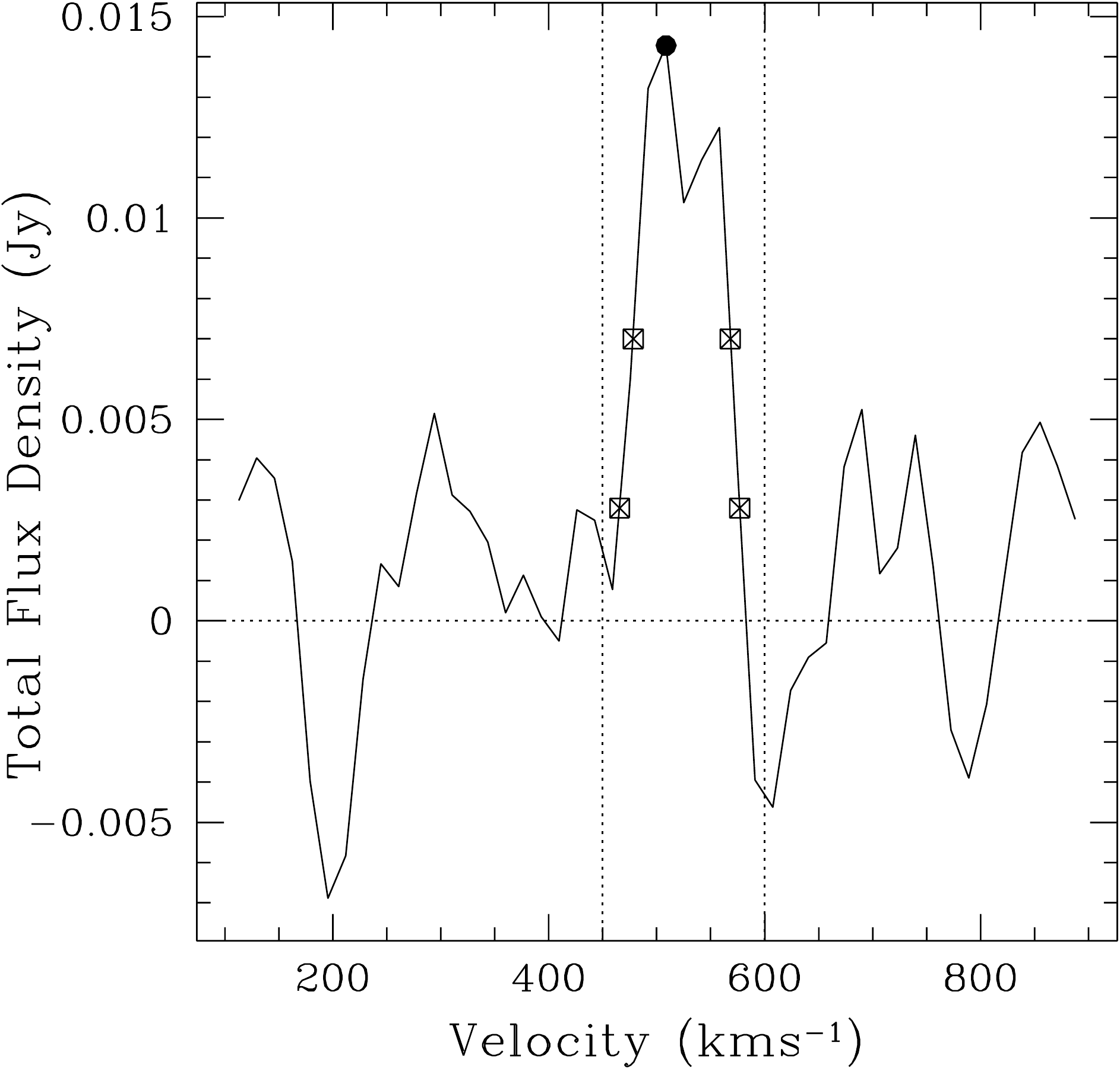}
\includegraphics[height=0.17\textheight]{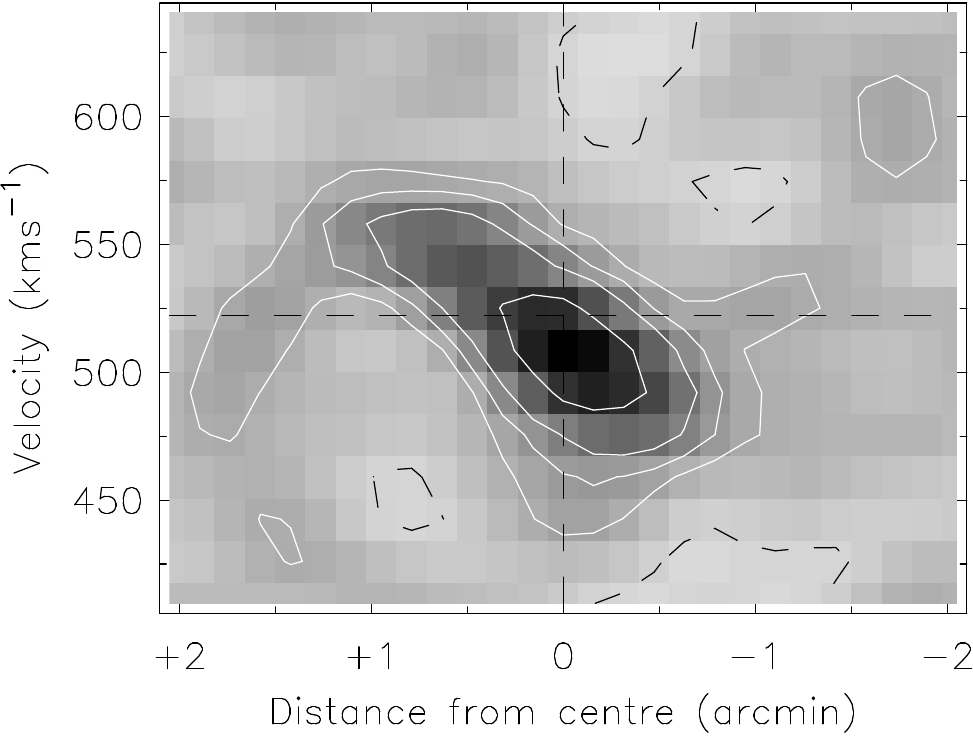}

\vskip 2mm
\centering
WSRT-CVn-8
\vskip 2mm
\includegraphics[width=0.25\textwidth]{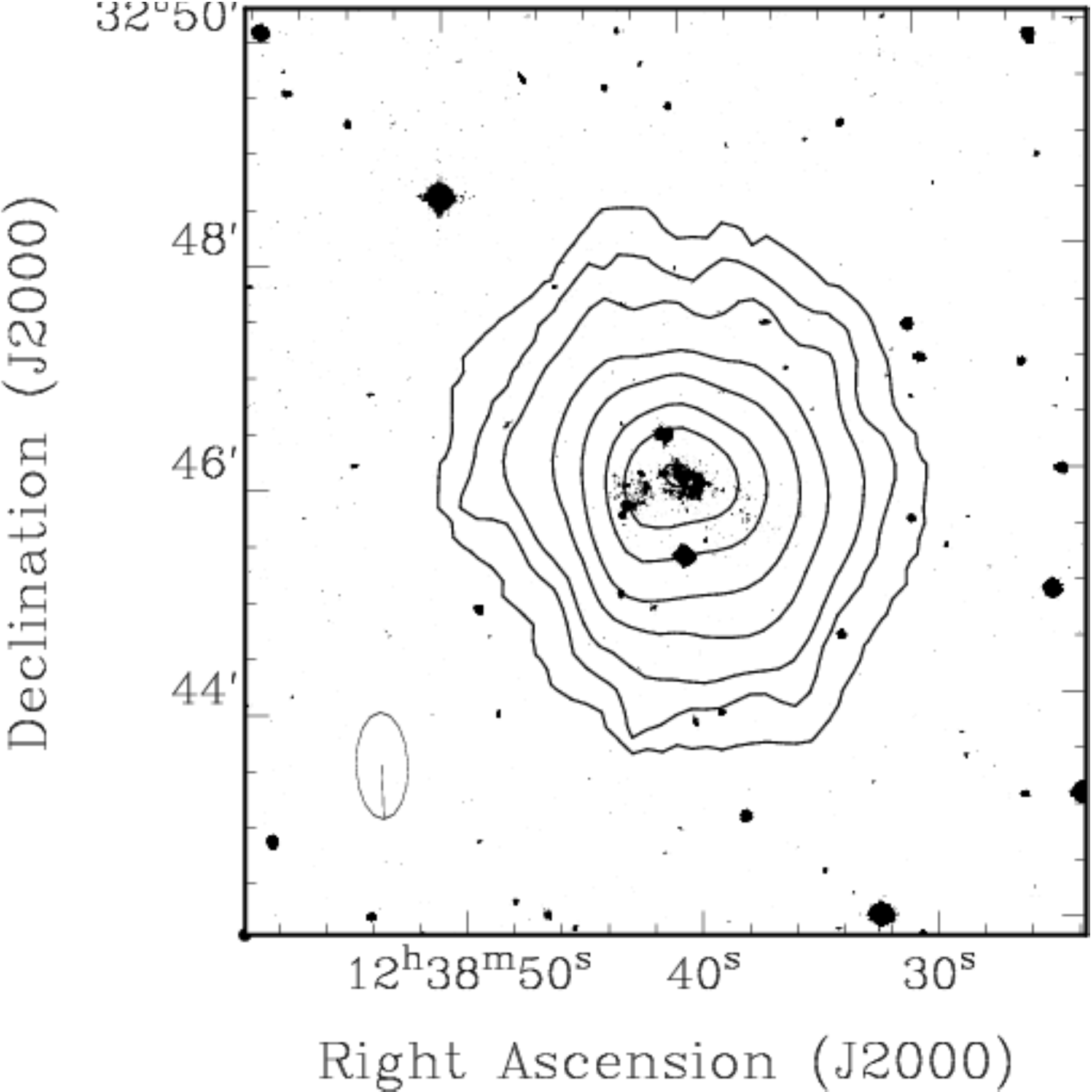}
\hskip 5mm
\includegraphics[height=0.17\textheight]{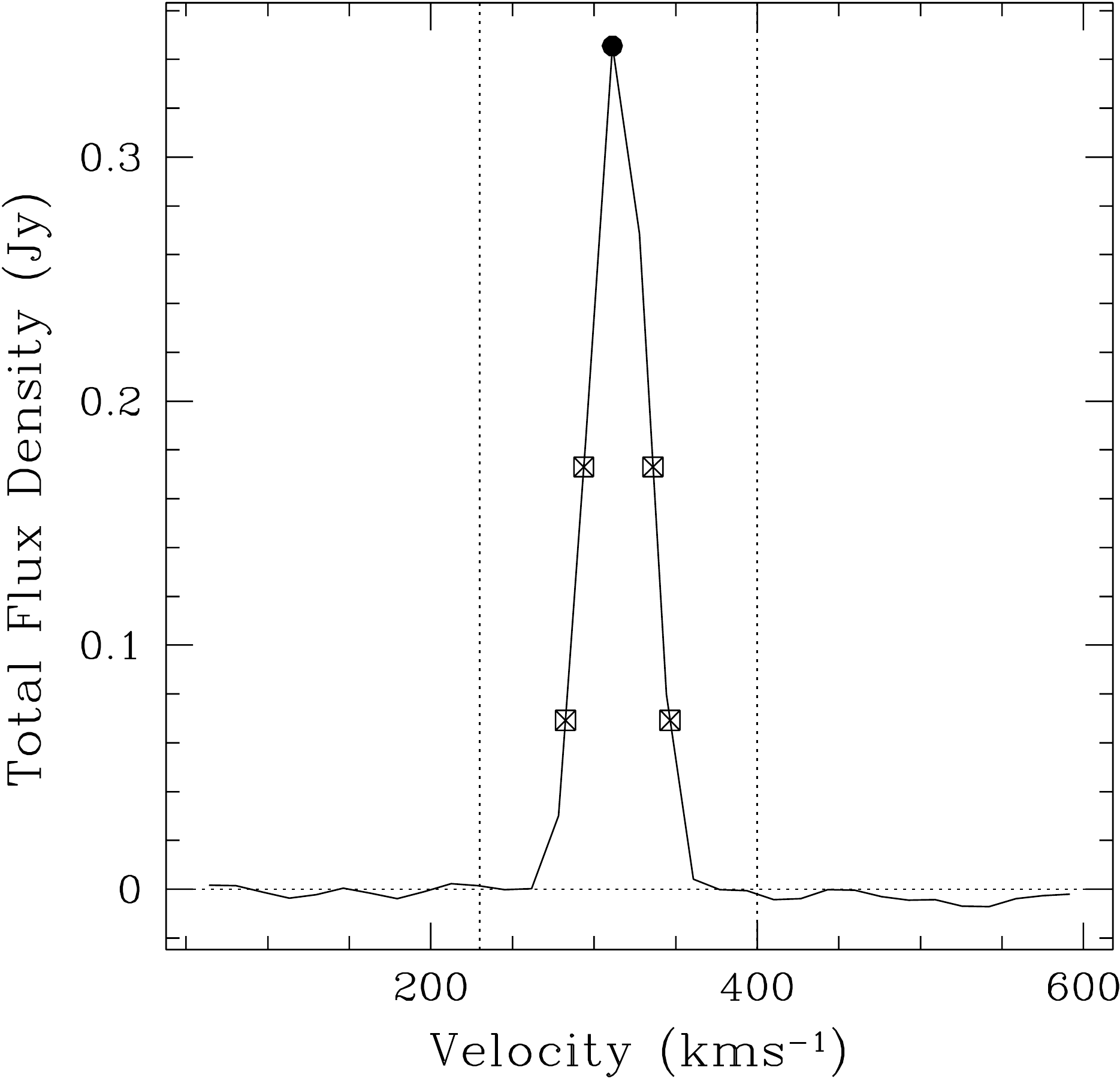}
\includegraphics[height=0.17\textheight]{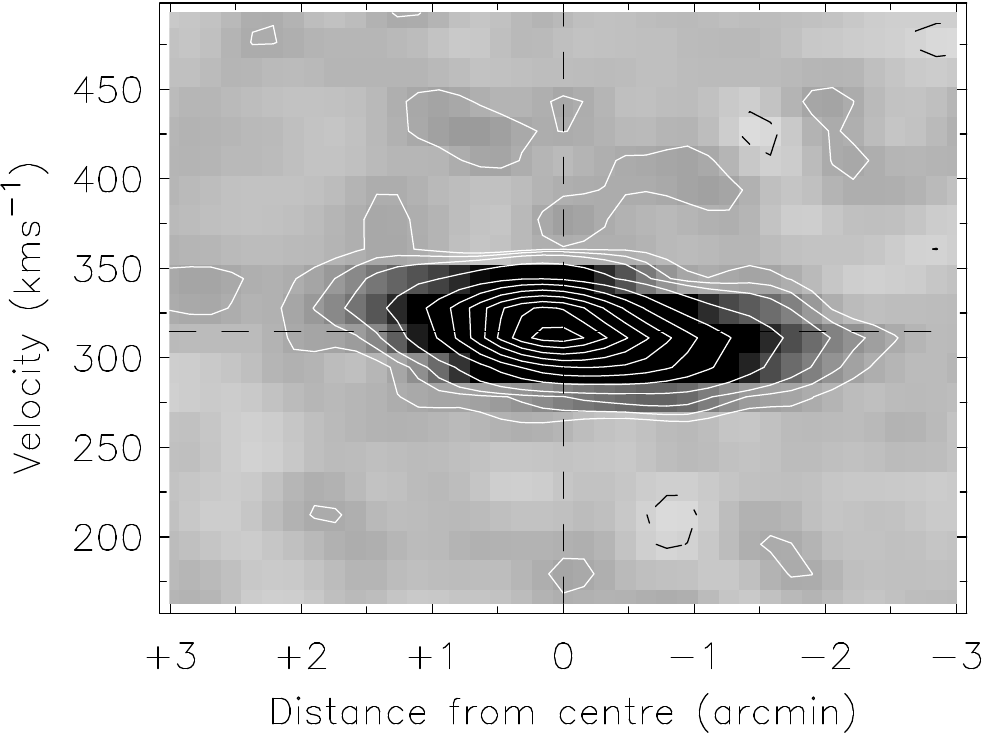}

\end{figure}

\clearpage

\addtocounter{figure}{-1}
\begin{figure}

\vskip 2mm
\centering
WSRT-CVn-9
\vskip 2mm
\includegraphics[width=0.25\textwidth]{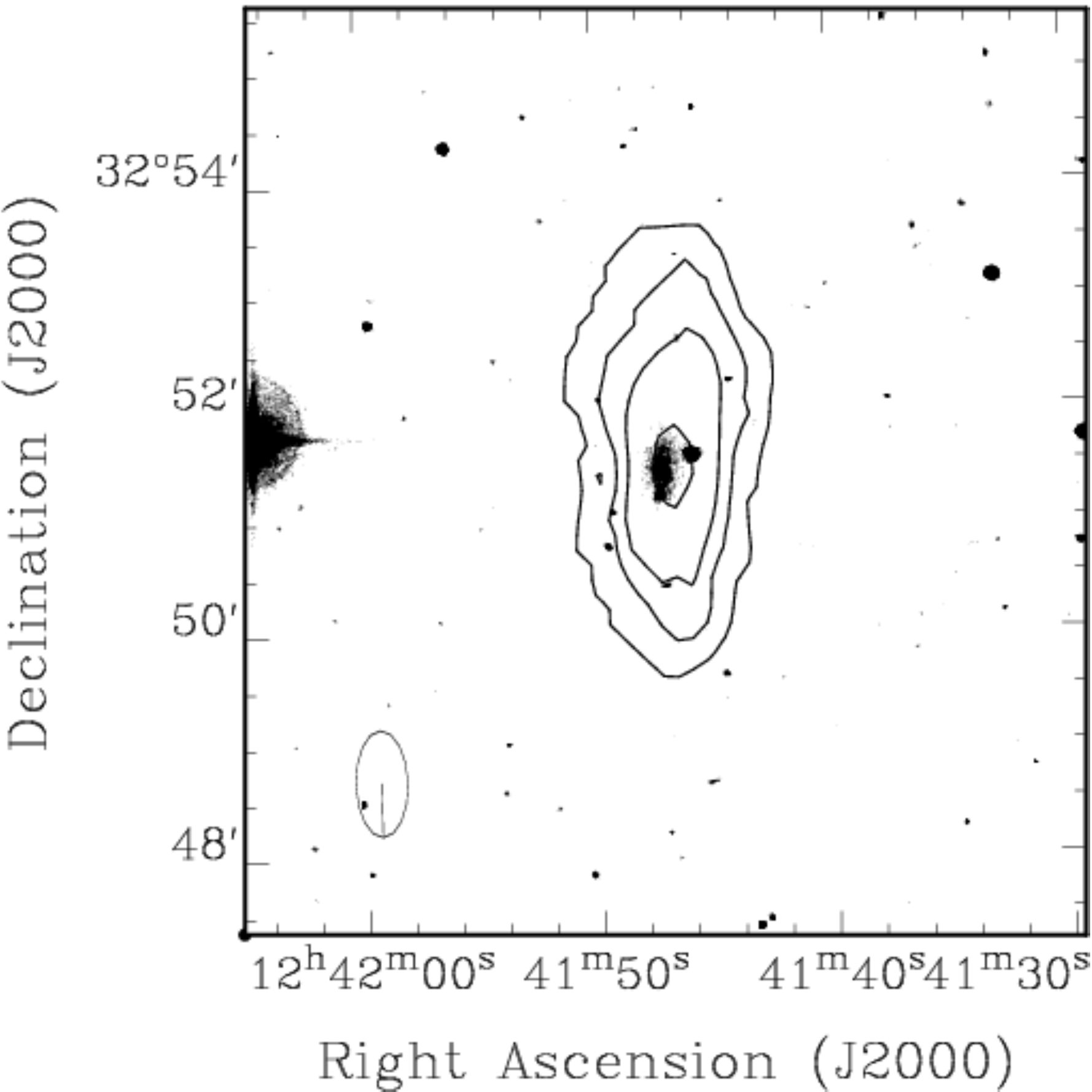}
\hskip 5mm
\includegraphics[height=0.17\textheight]{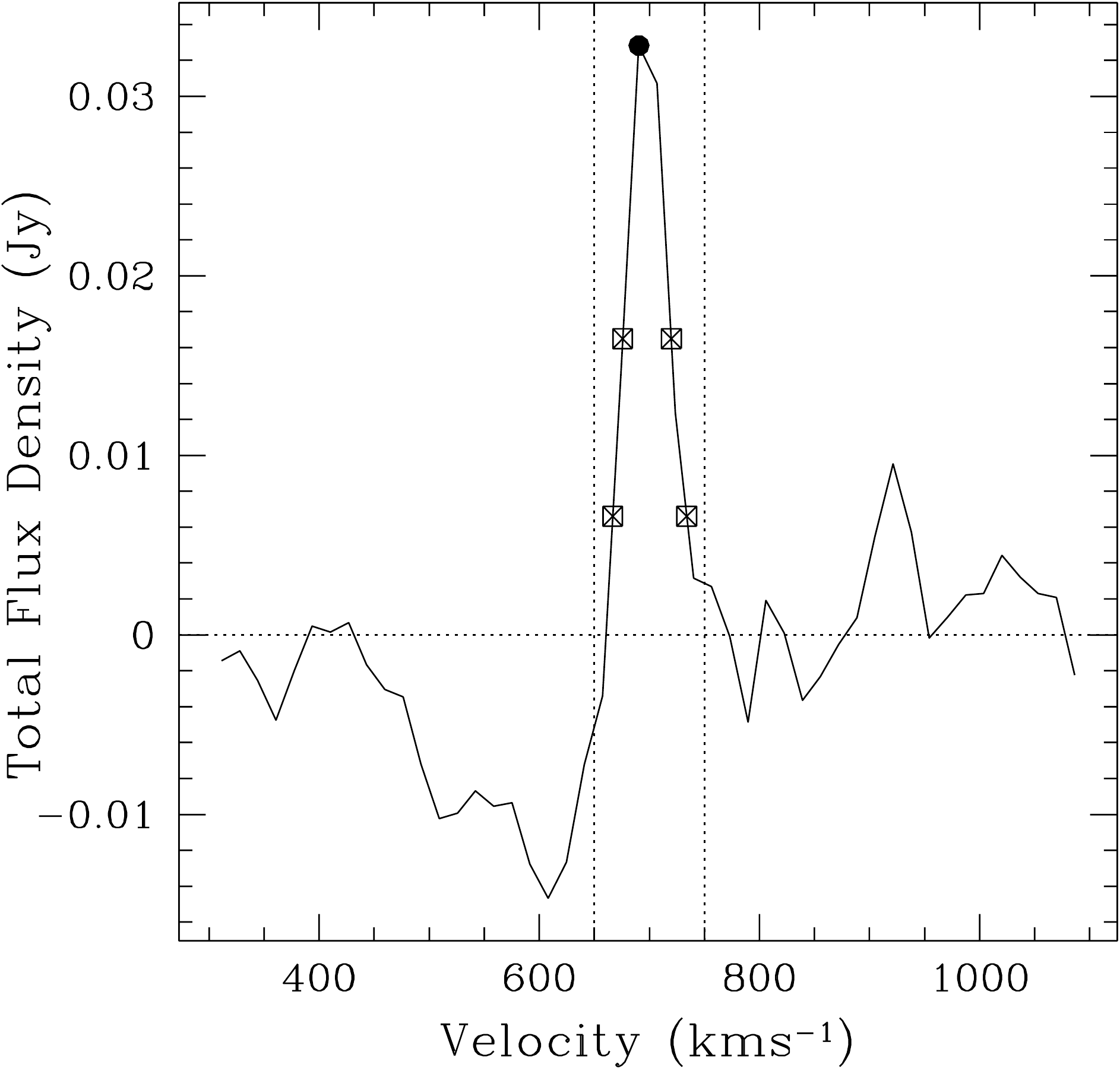}
\includegraphics[height=0.17\textheight]{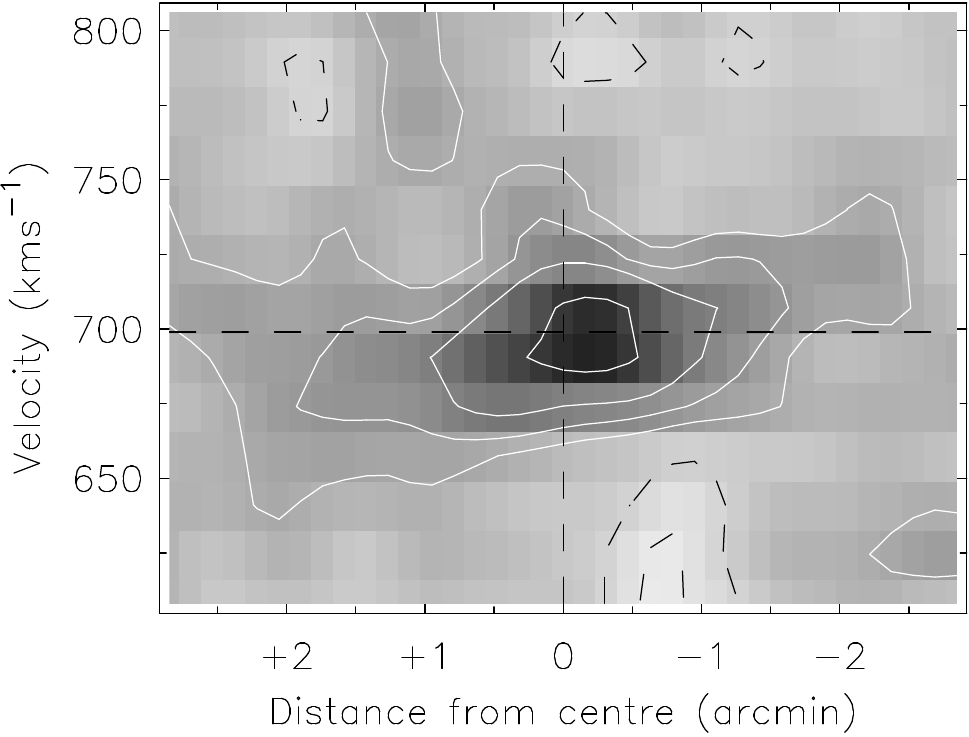}

\vskip 2mm
\centering
WSRT-CVn-10
\vskip 2mm
\includegraphics[width=0.25\textwidth]{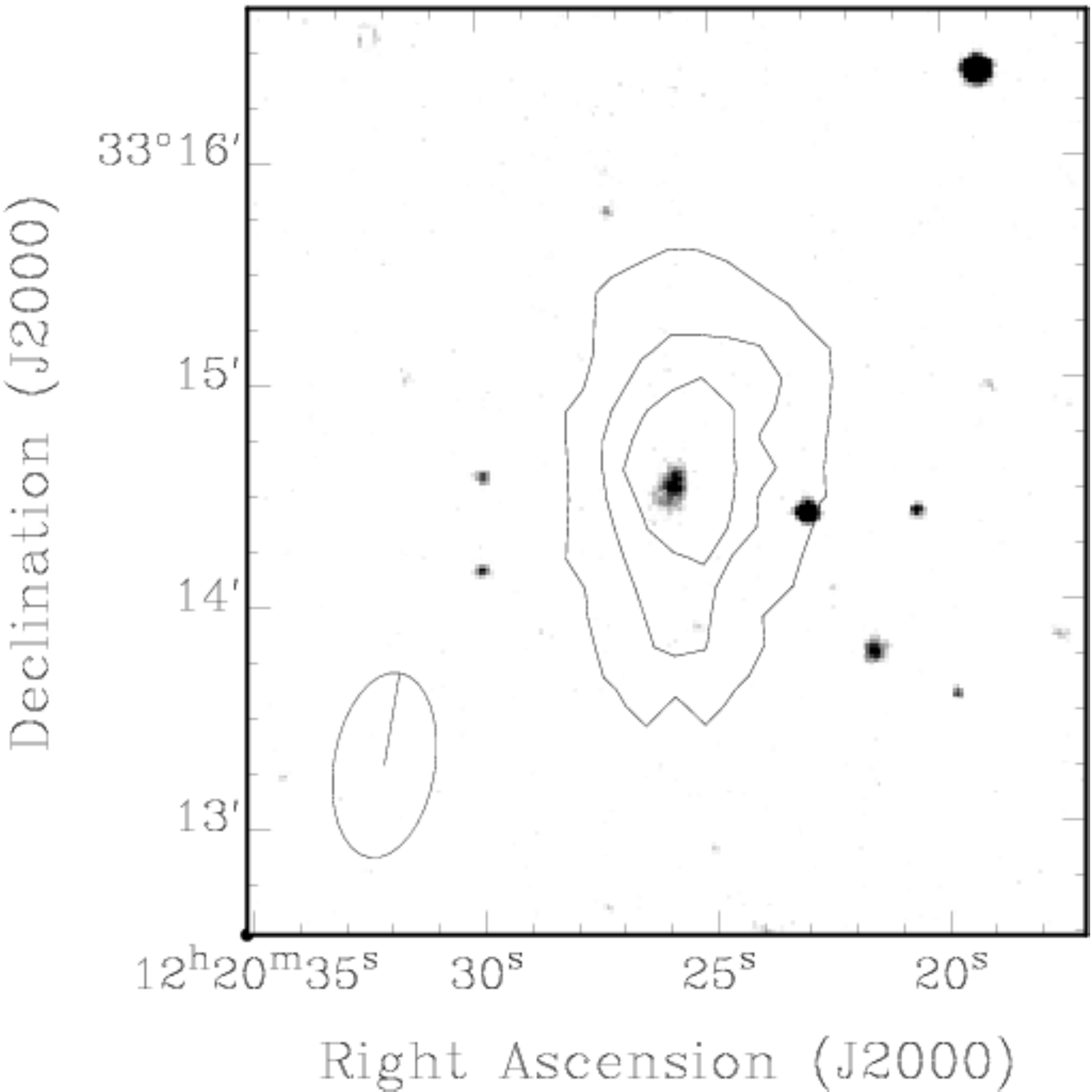}
\hskip 5mm
\includegraphics[height=0.17\textheight]{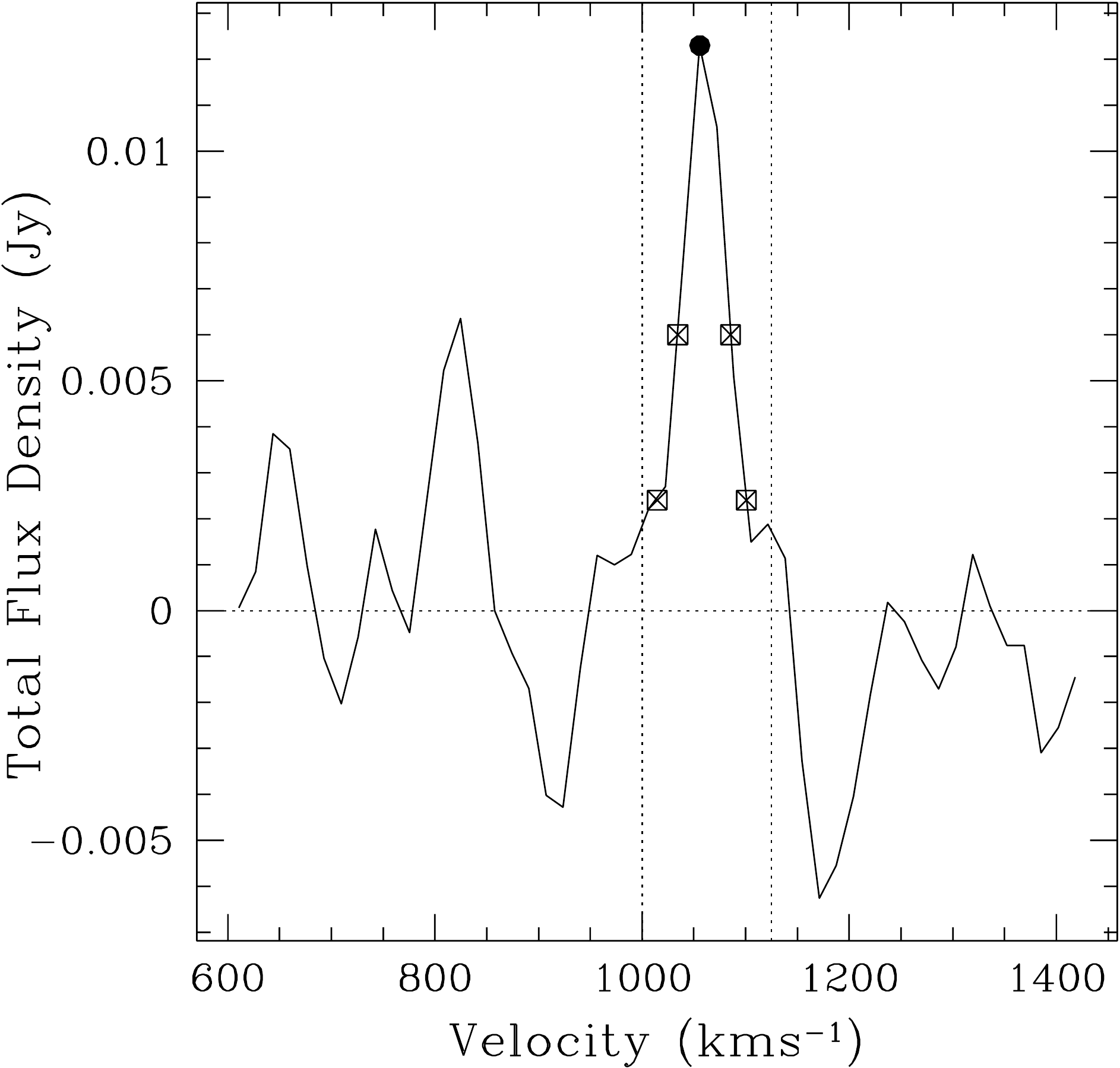}
\includegraphics[height=0.17\textheight]{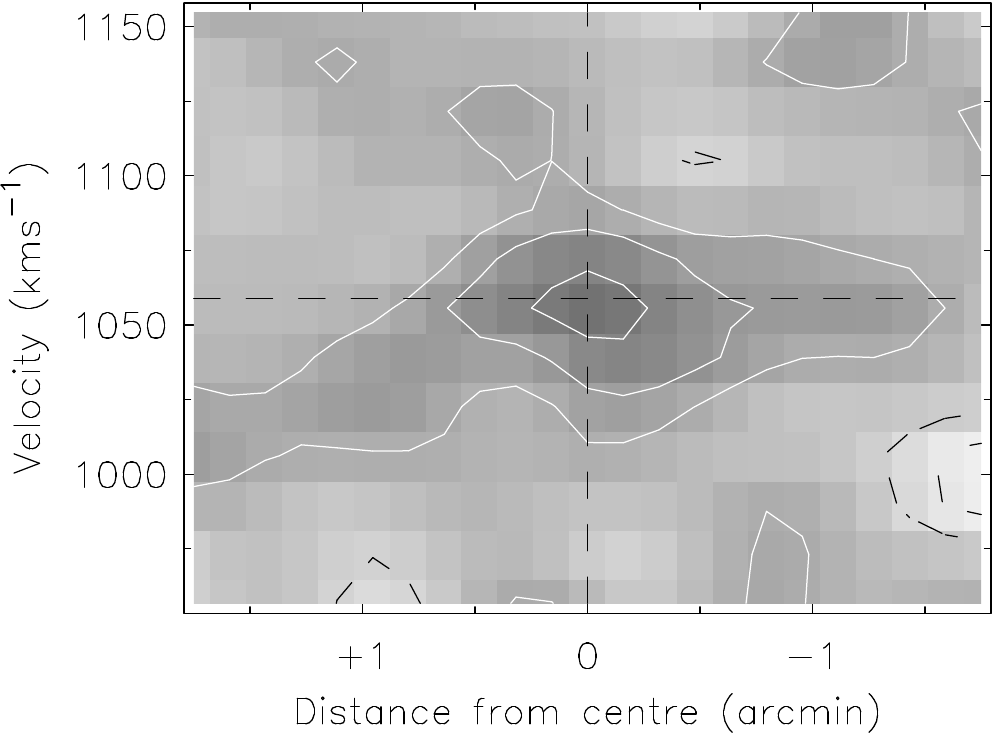}

\vskip 2mm
\centering
WSRT-CVn-11
\vskip 2mm
\includegraphics[width=0.25\textwidth]{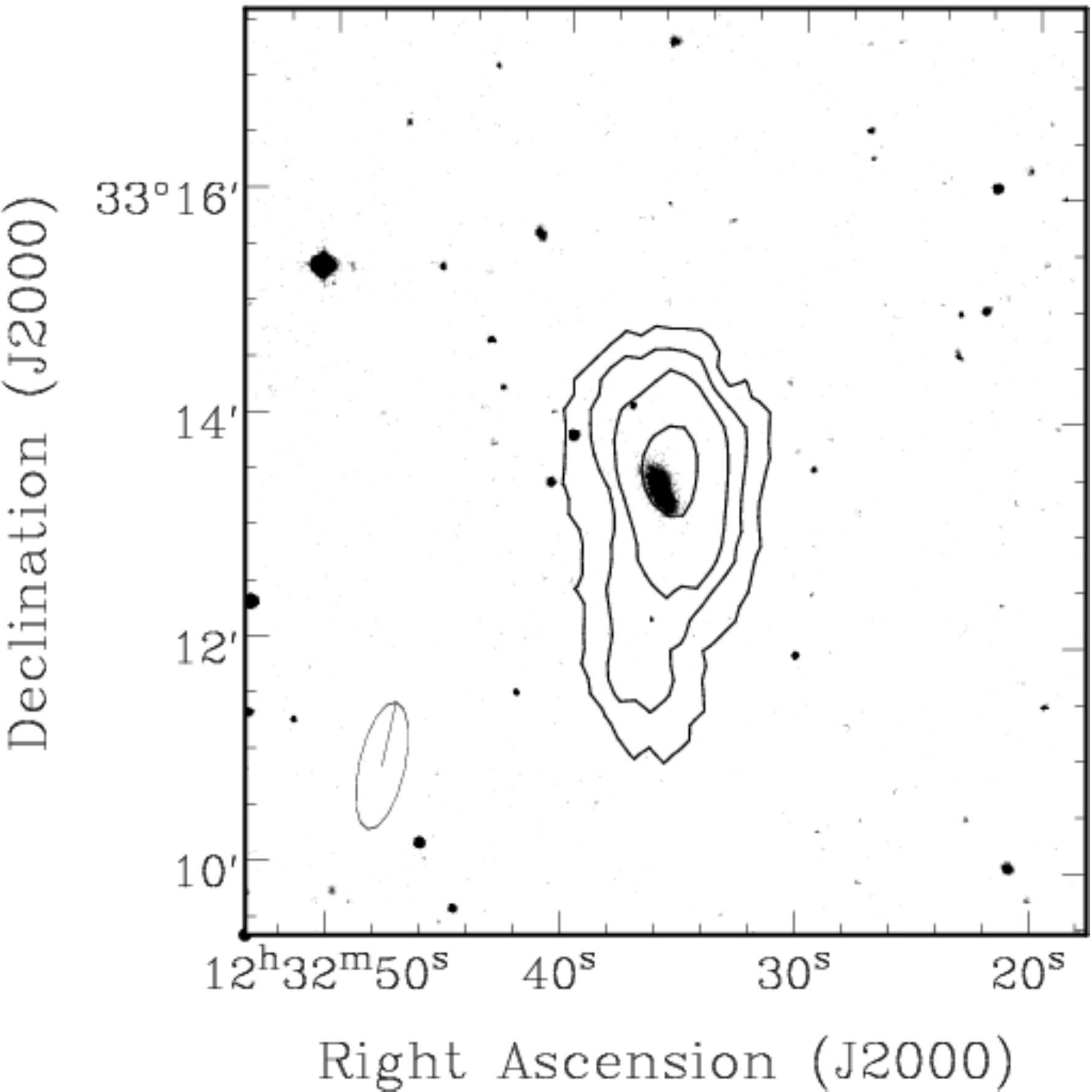}
\hskip 8mm
\includegraphics[height=0.17\textheight]{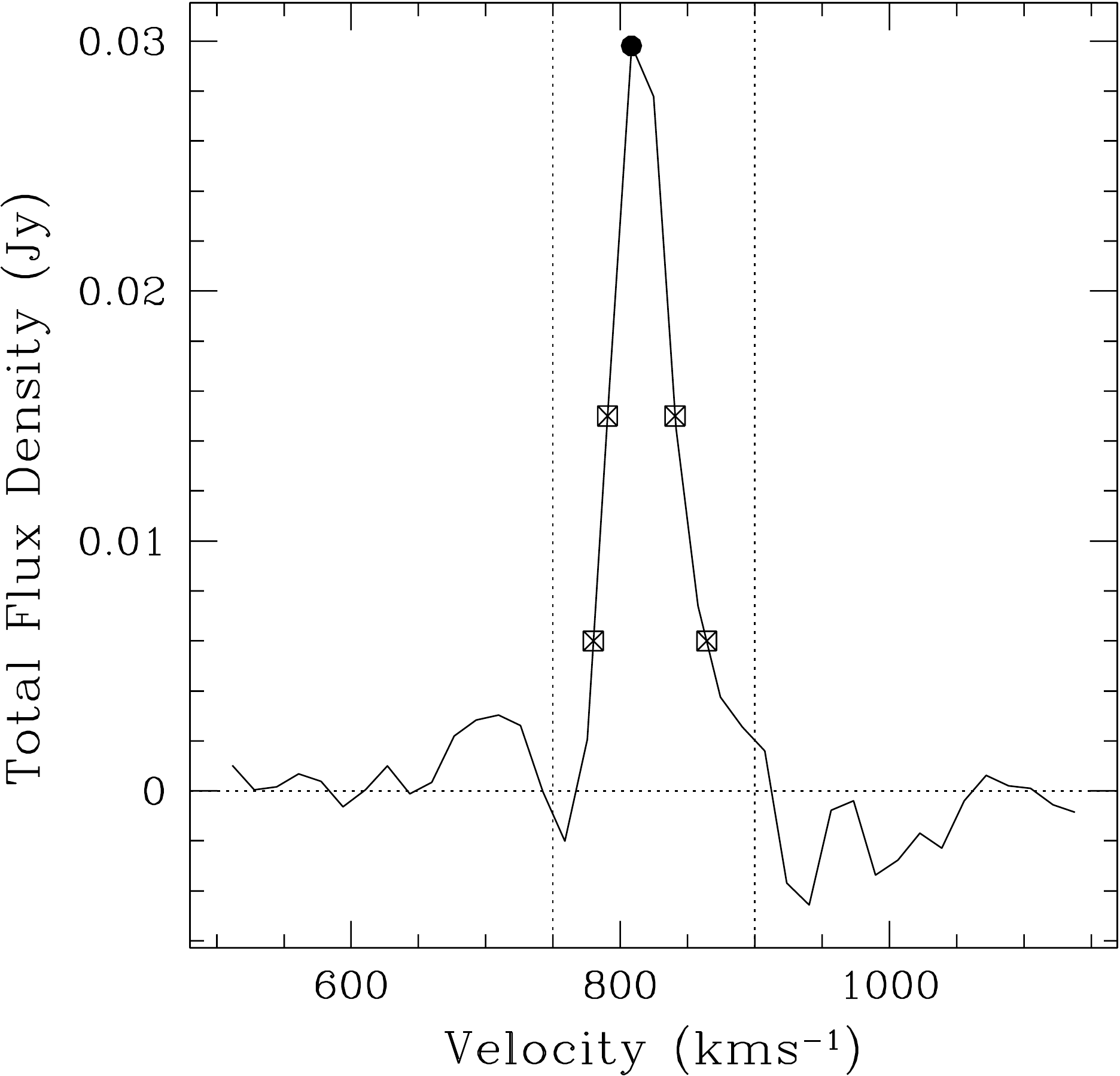}
\includegraphics[height=0.17\textheight]{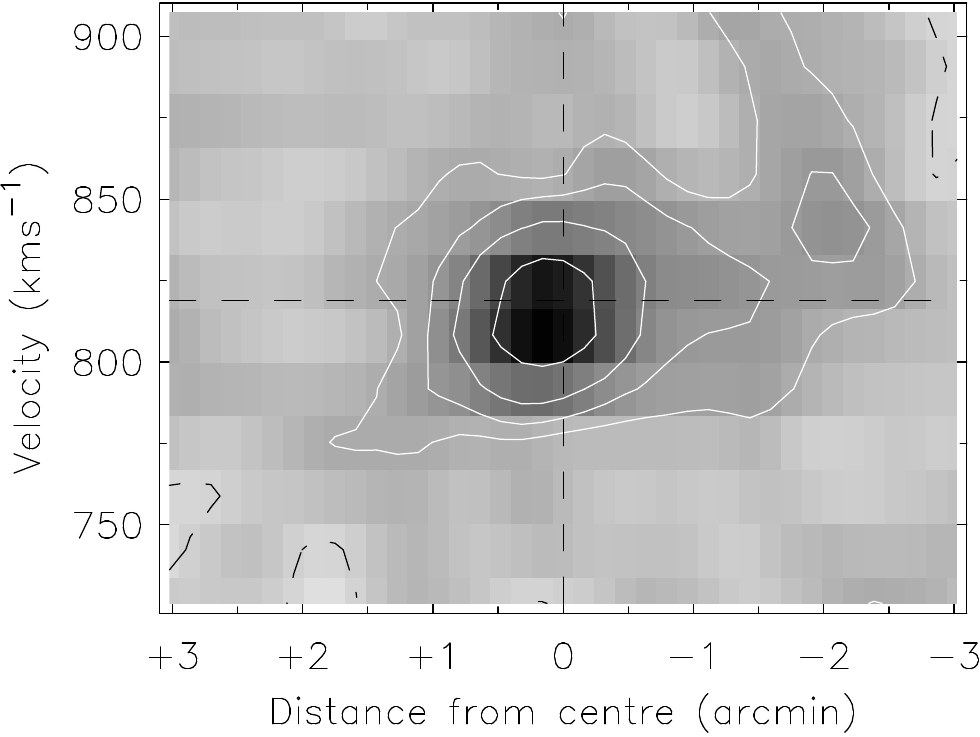}

\vskip 2mm
\centering
WSRT-CVn-12
\vskip 2mm
\includegraphics[width=0.25\textwidth]{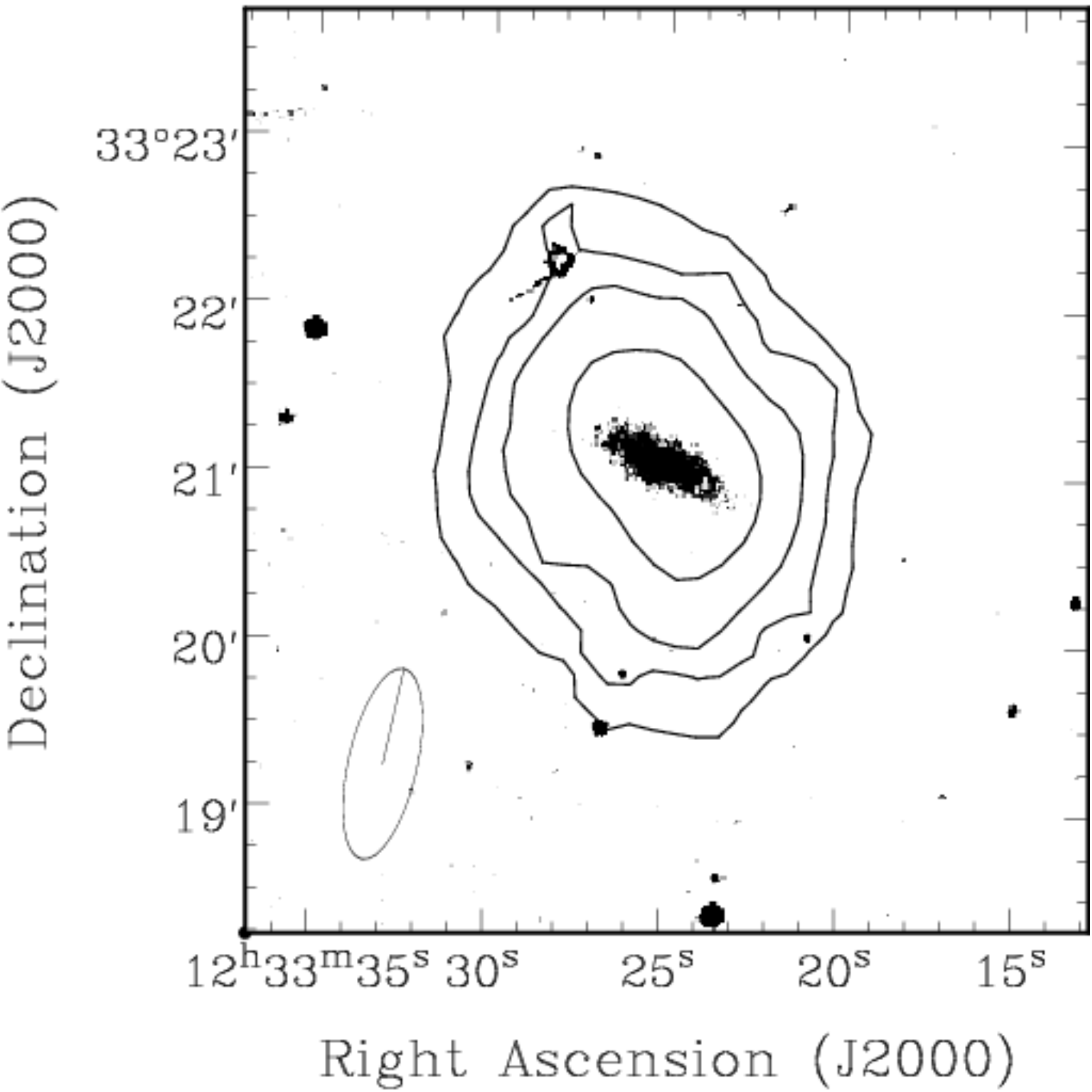}
\hskip 5mm
\includegraphics[height=0.17\textheight]{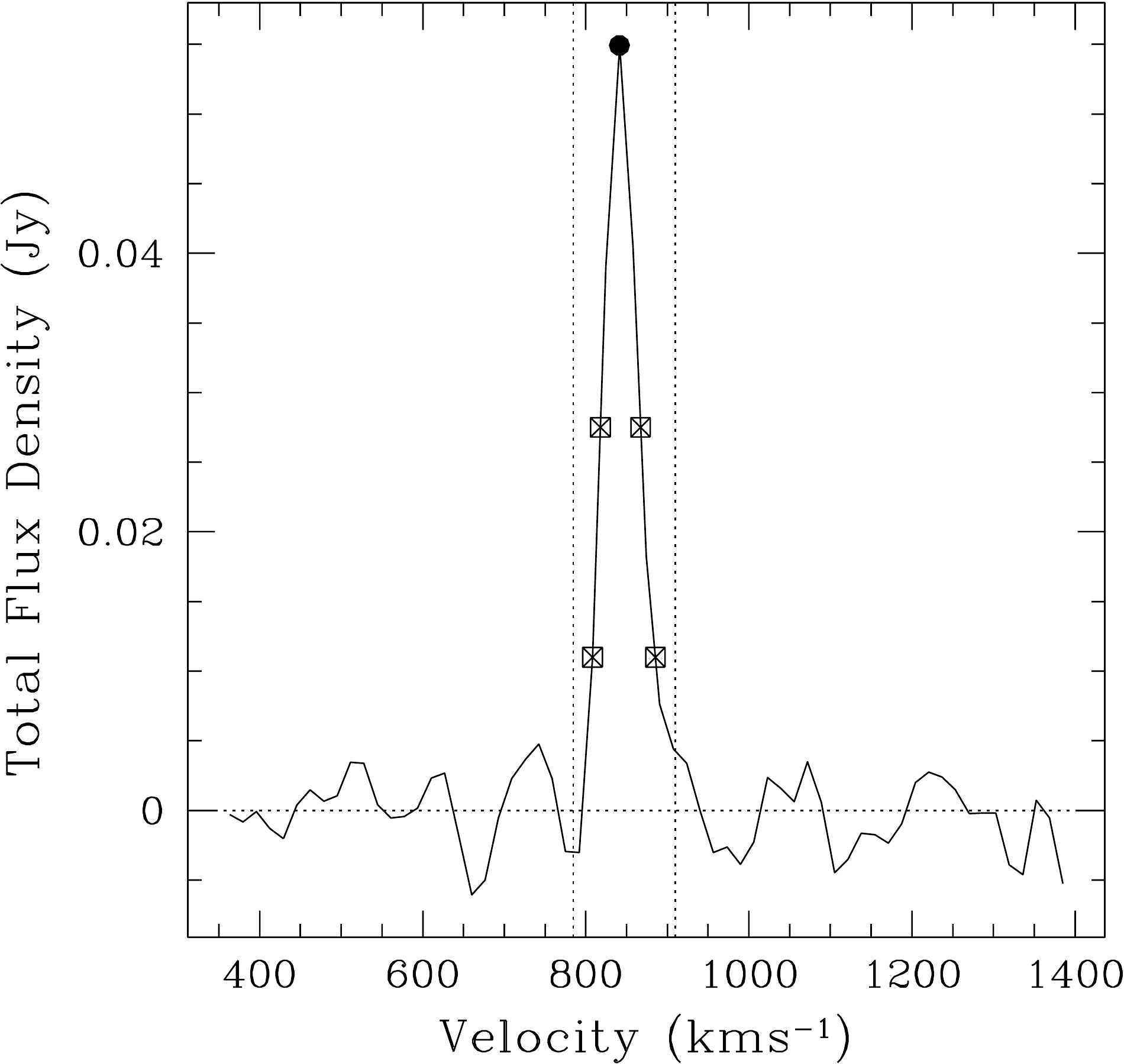}
\includegraphics[height=0.17\textheight]{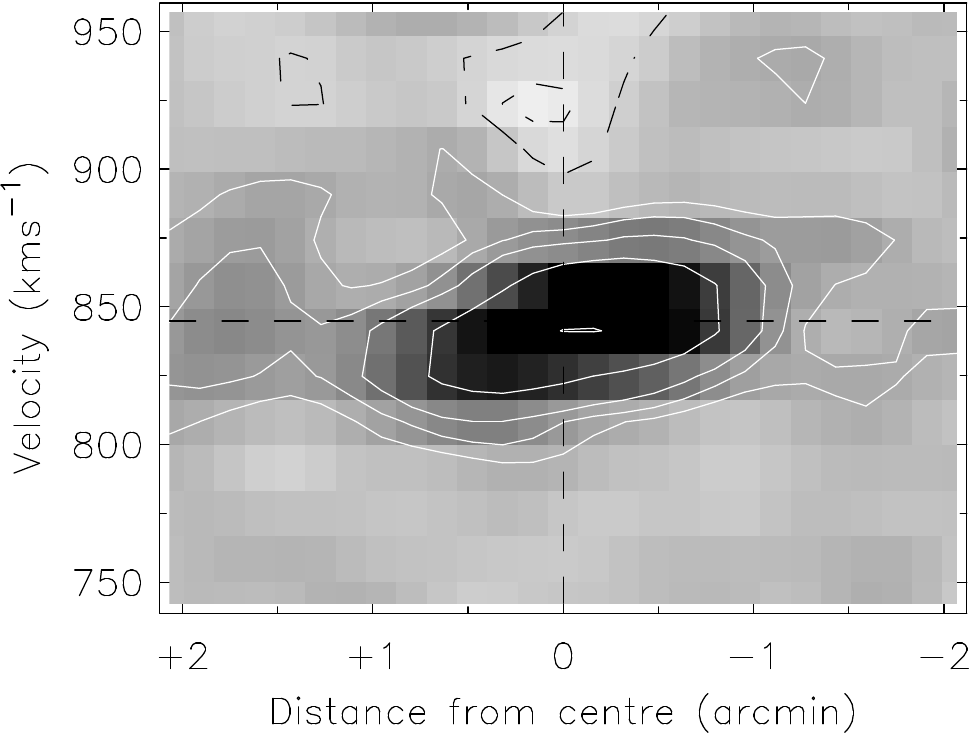}

\end{figure}

\clearpage

\addtocounter{figure}{-1}
\begin{figure}

\vskip 2mm
\centering
WSRT-CVn-13
\vskip 2mm
\includegraphics[width=0.25\textwidth]{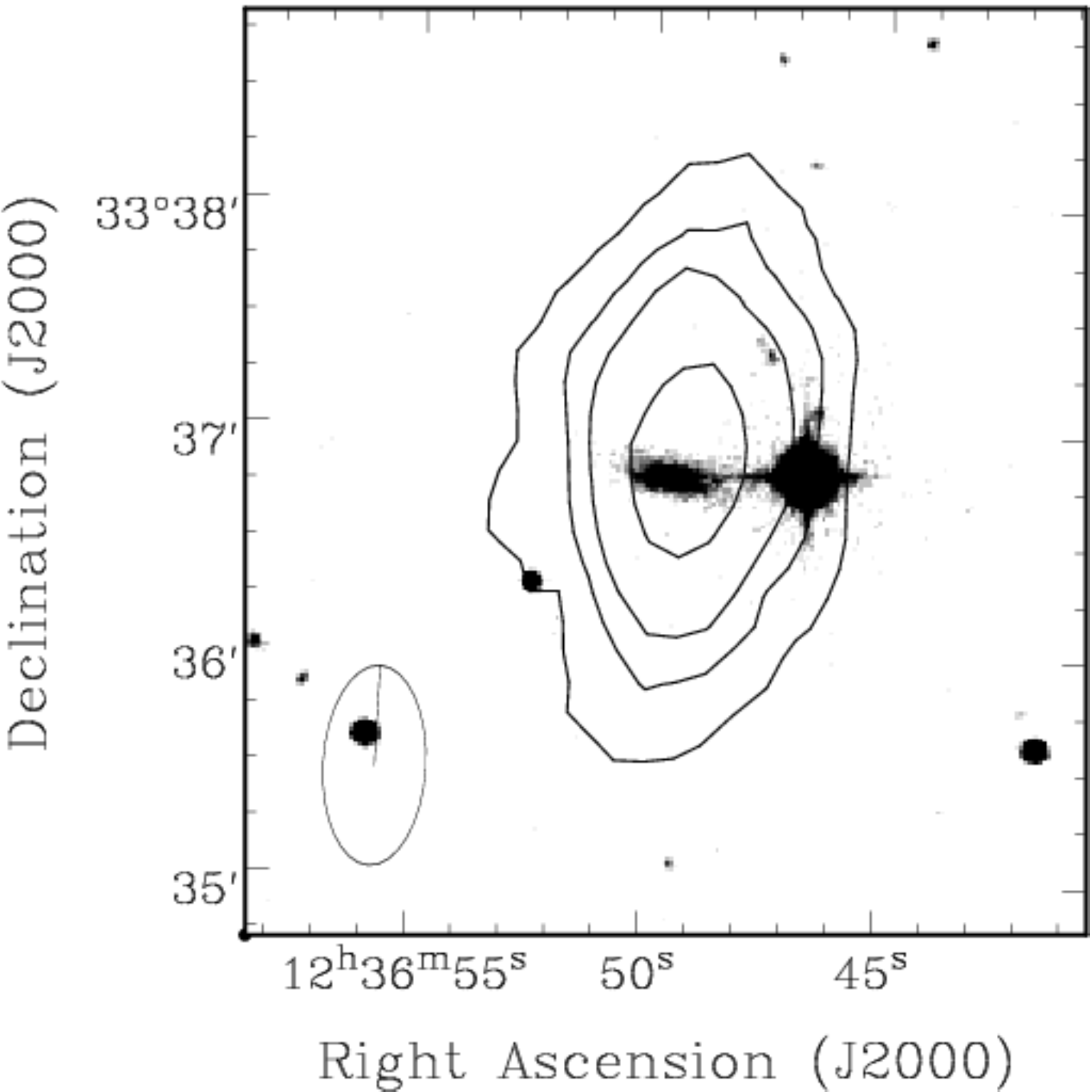}
\hskip 5mm
\includegraphics[height=0.17\textheight]{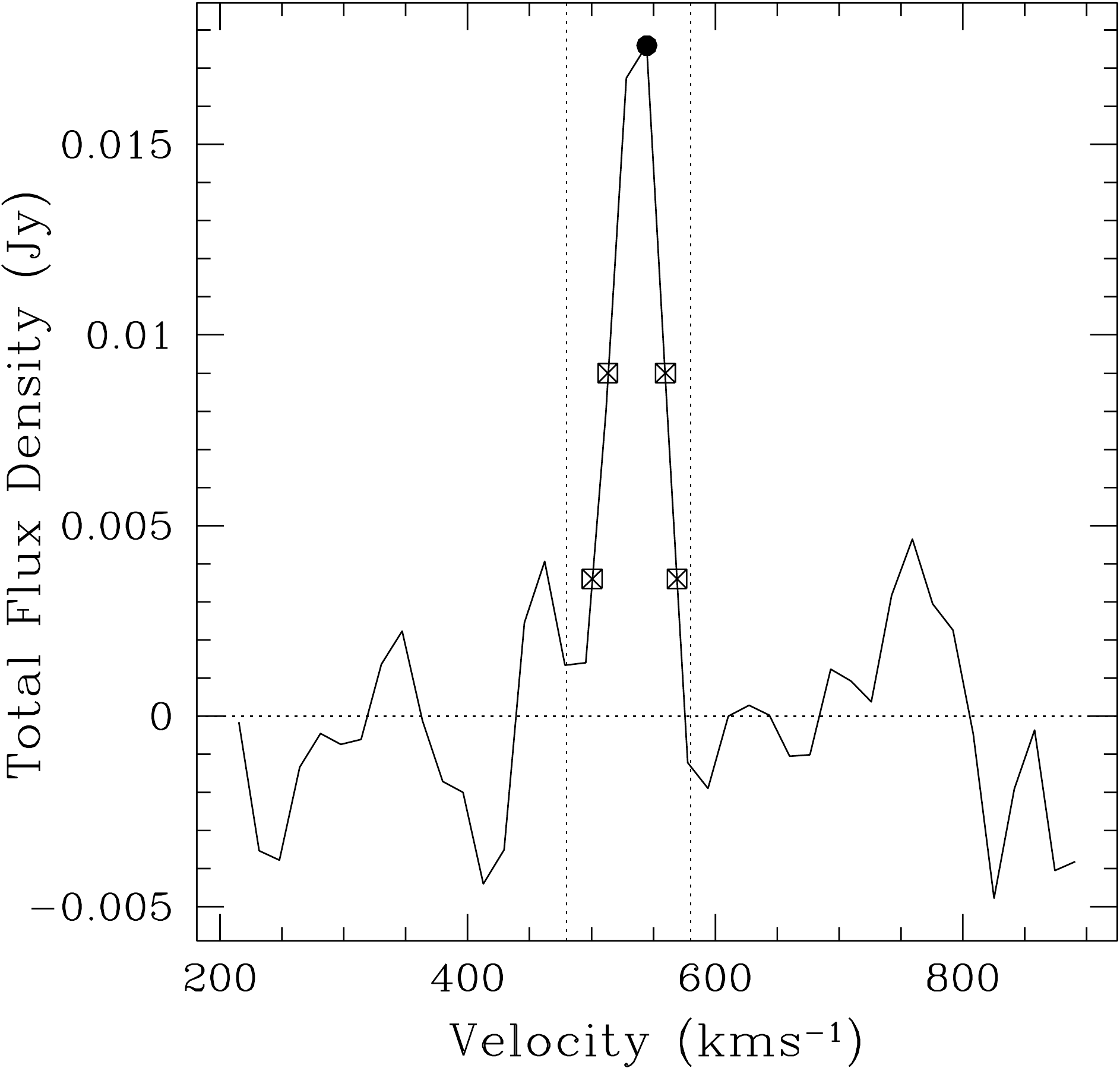}
\includegraphics[height=0.17\textheight]{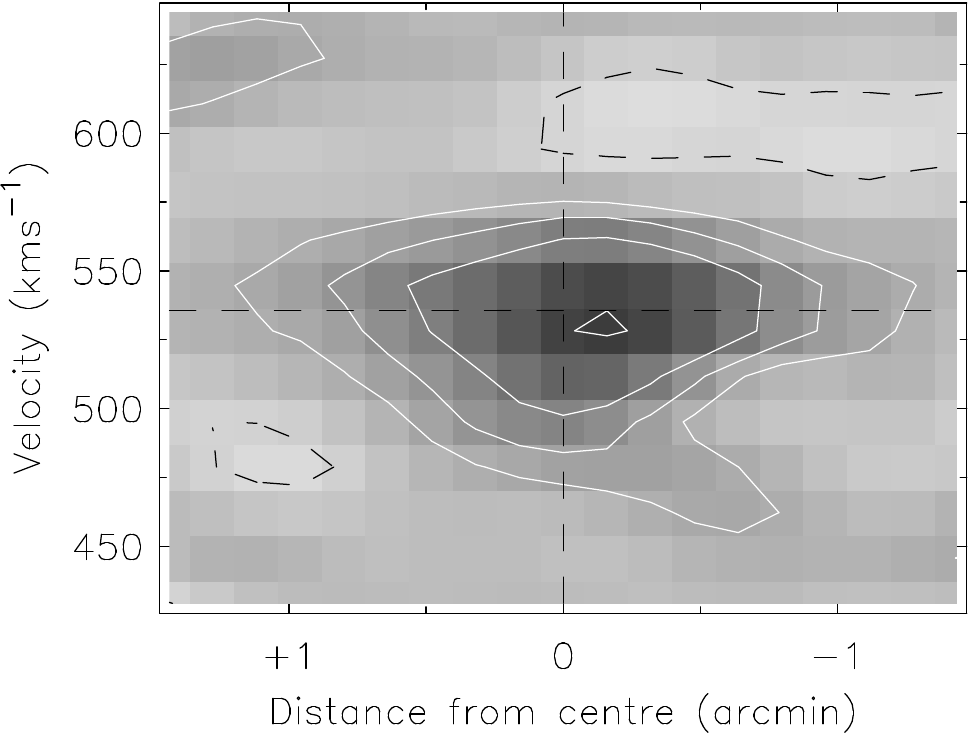}

\vskip 2mm
\centering
WSRT-CVn-14
\vskip 2mm
\includegraphics[width=0.25\textwidth]{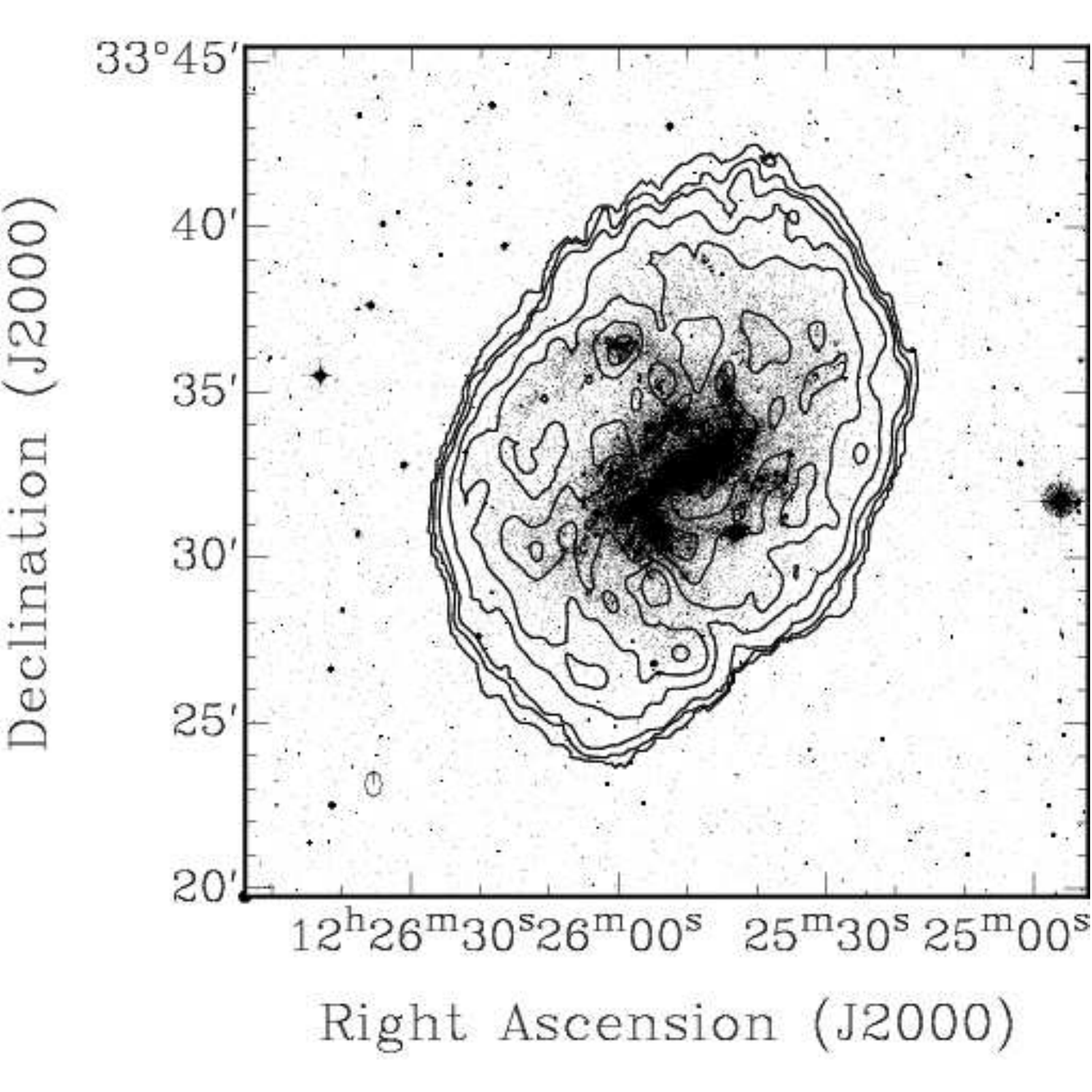}
\hskip 5mm
\includegraphics[height=0.17\textheight]{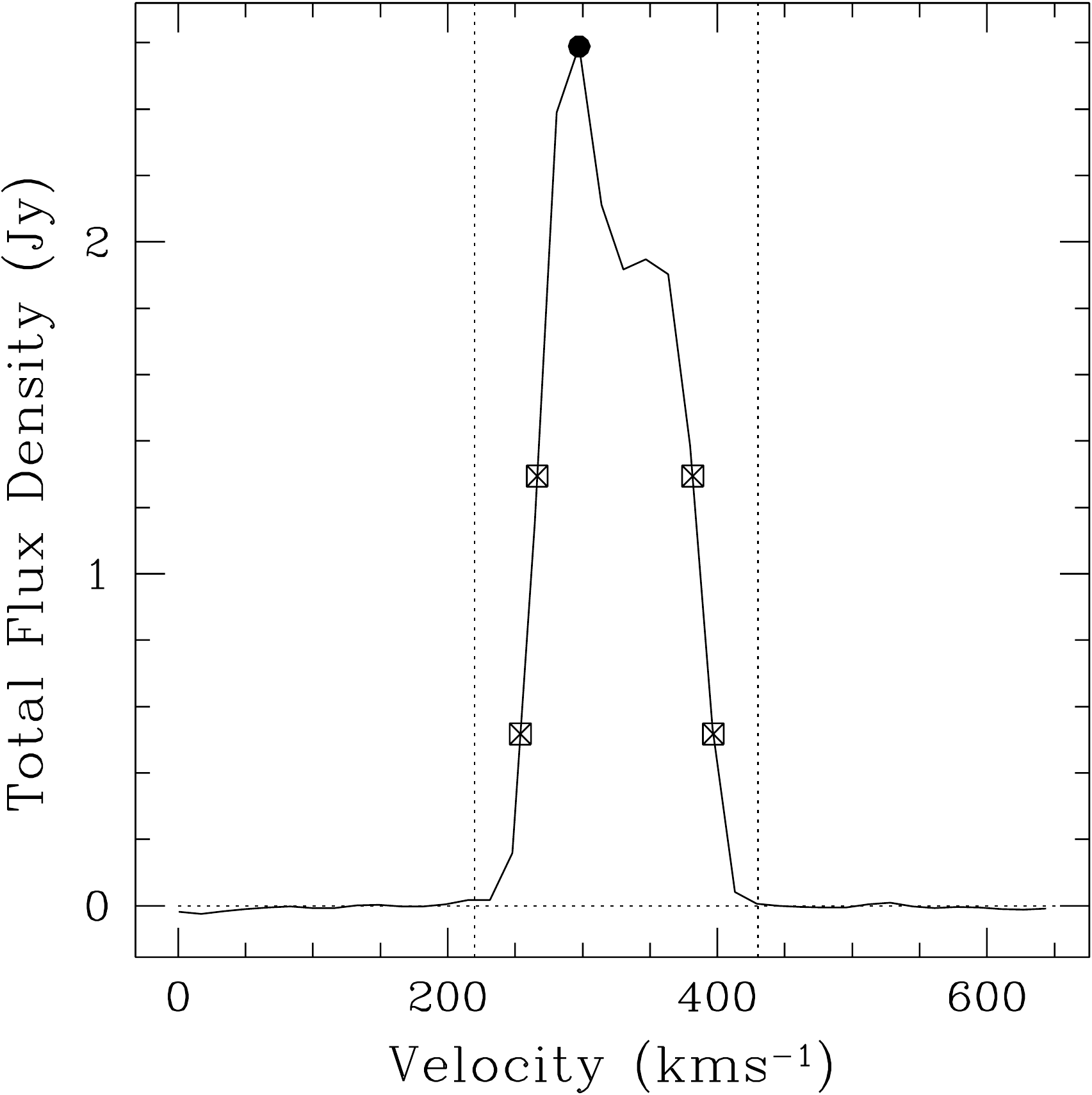}
\includegraphics[height=0.17\textheight]{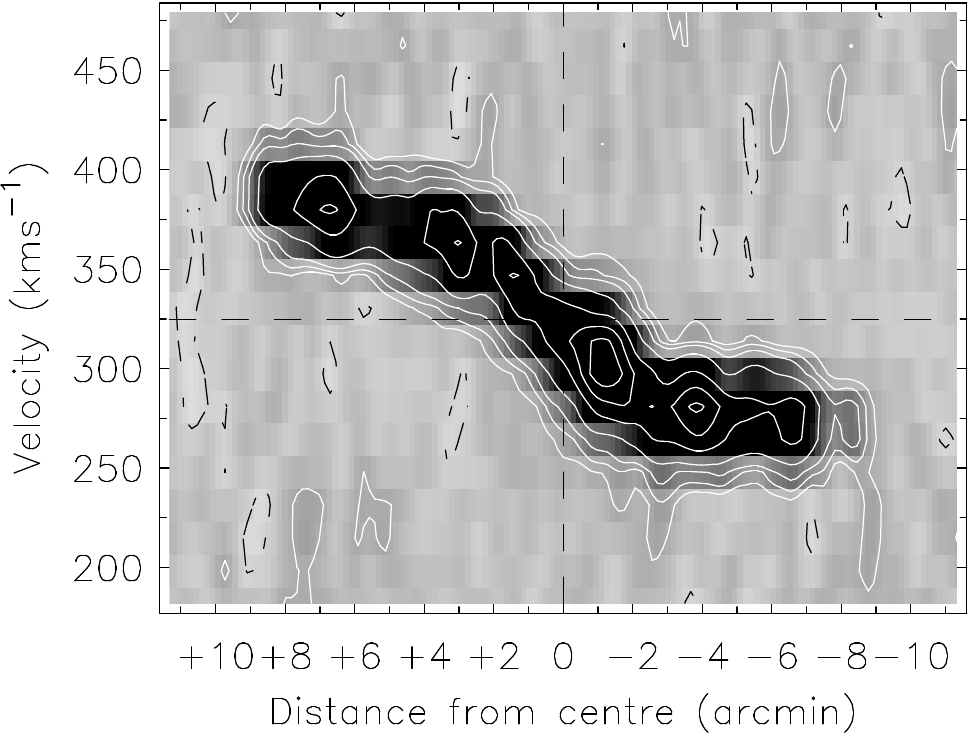}

\vskip 2mm
\centering
WSRT-CVn-15
\vskip 2mm
\includegraphics[width=0.25\textwidth]{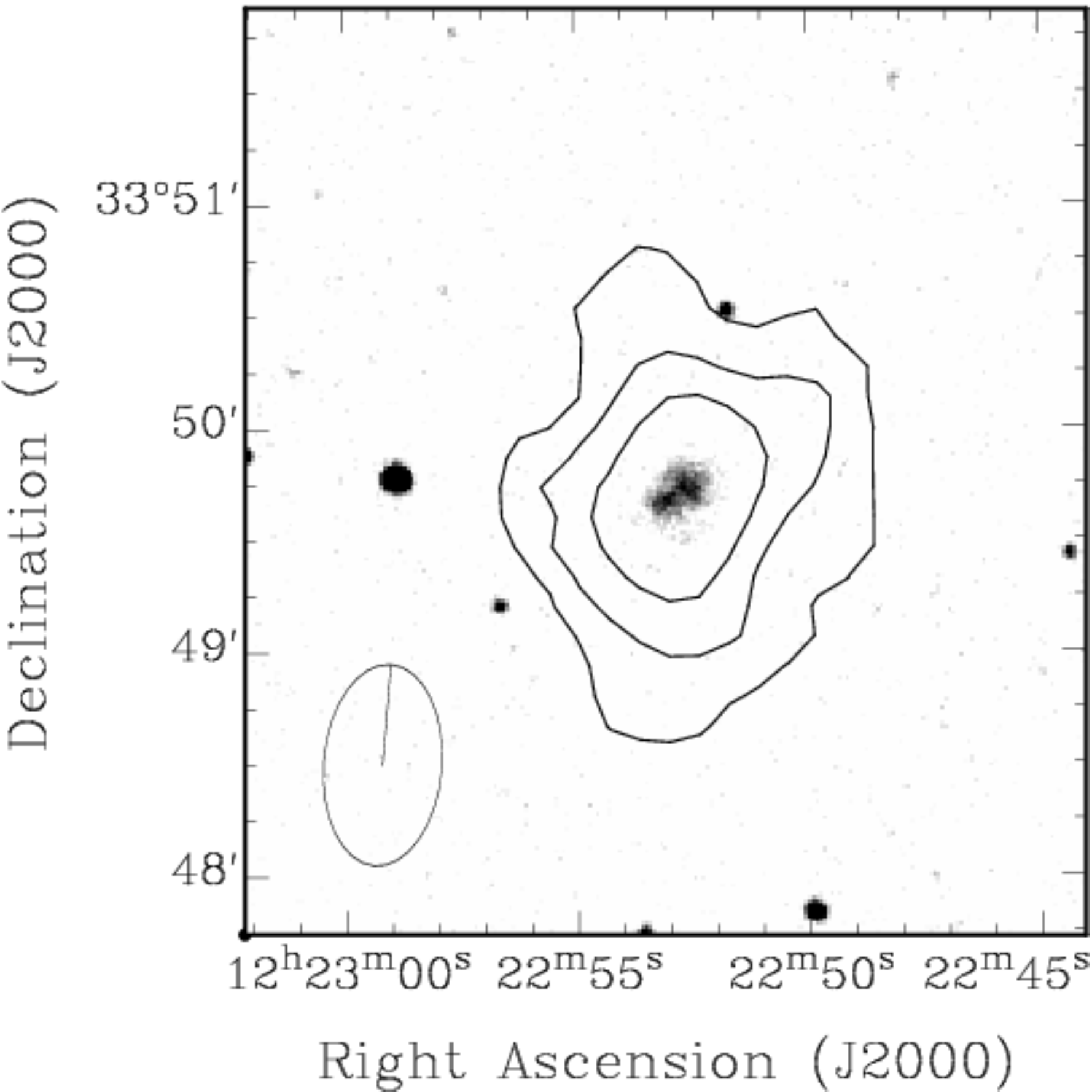}
\hskip 5mm
\includegraphics[height=0.17\textheight]{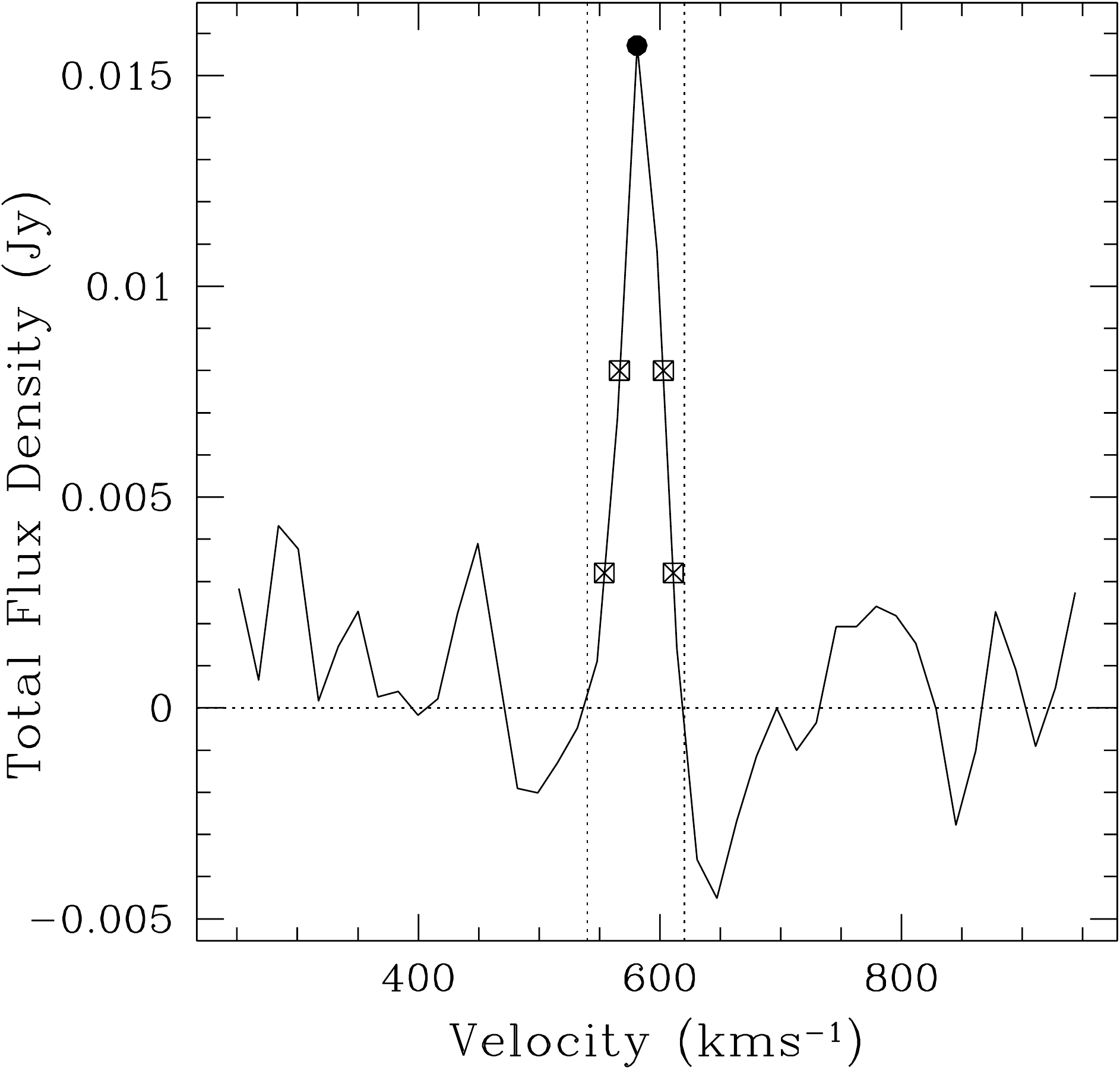}
\includegraphics[height=0.17\textheight]{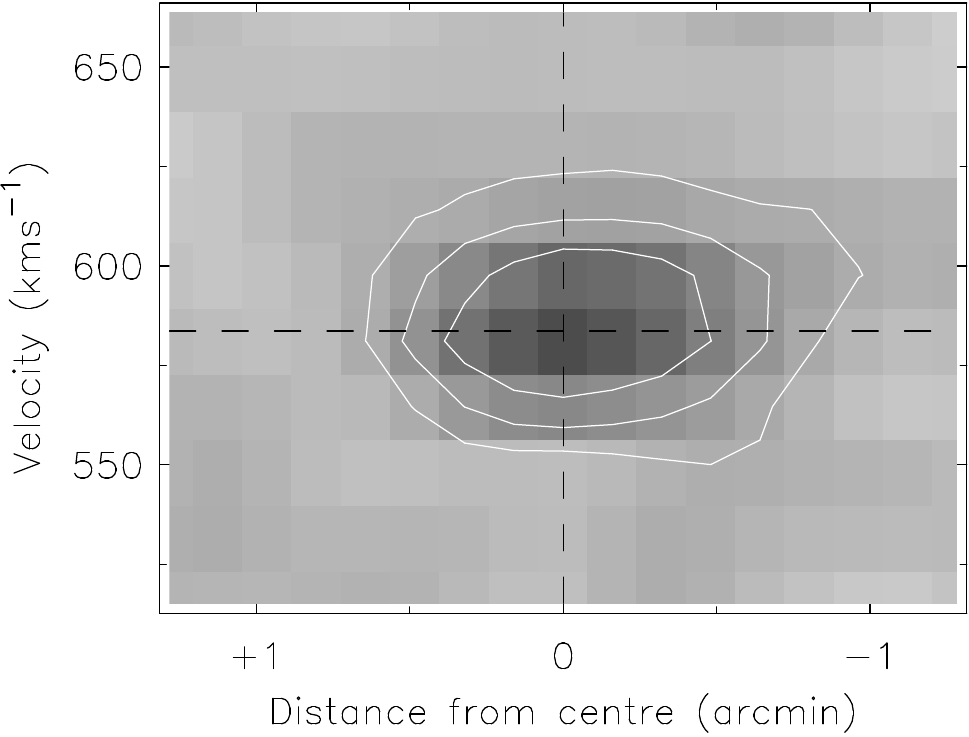}

\vskip 2mm
\centering
WSRT-CVn-16
\vskip 2mm
\includegraphics[width=0.25\textwidth]{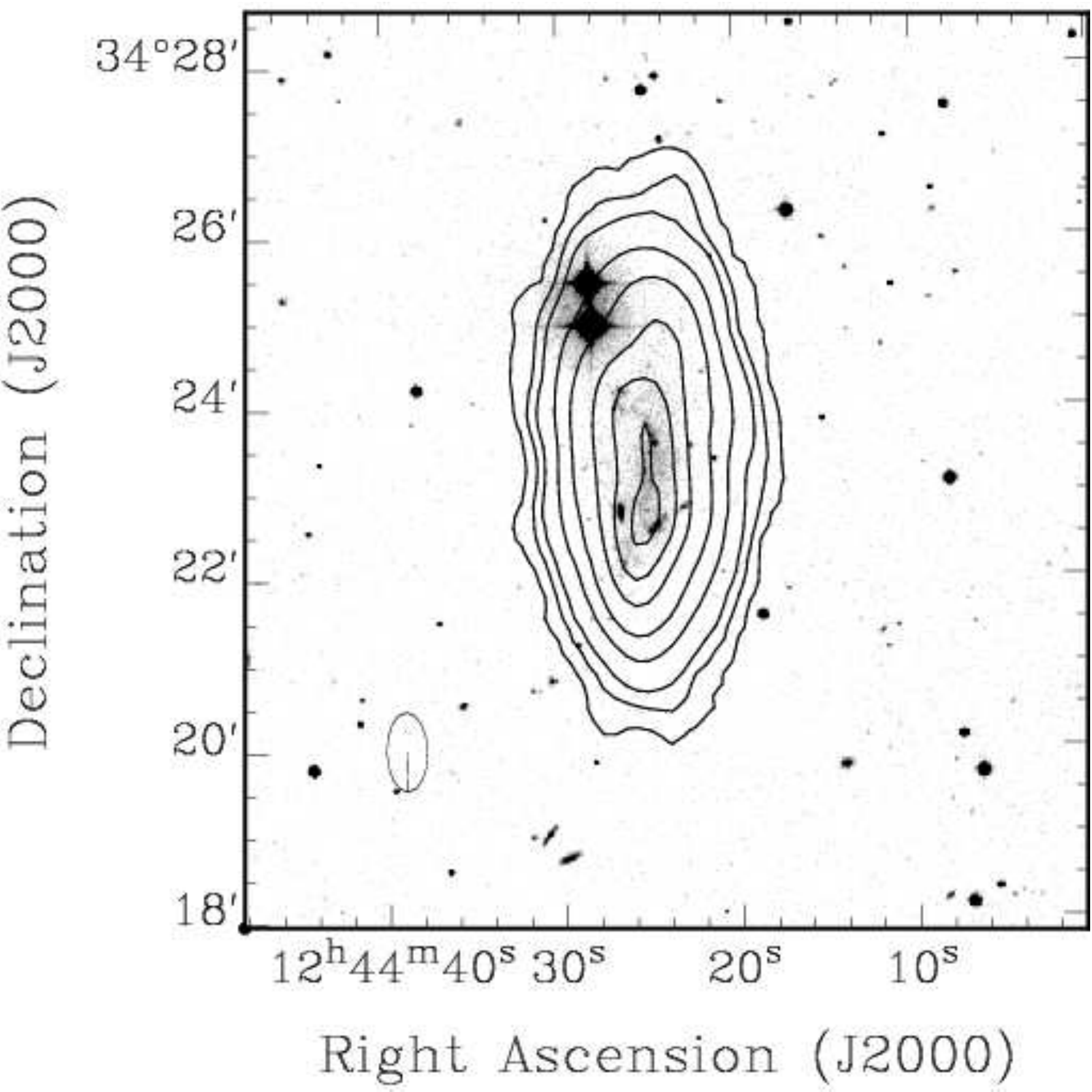}
\hskip 5mm
\includegraphics[height=0.17\textheight]{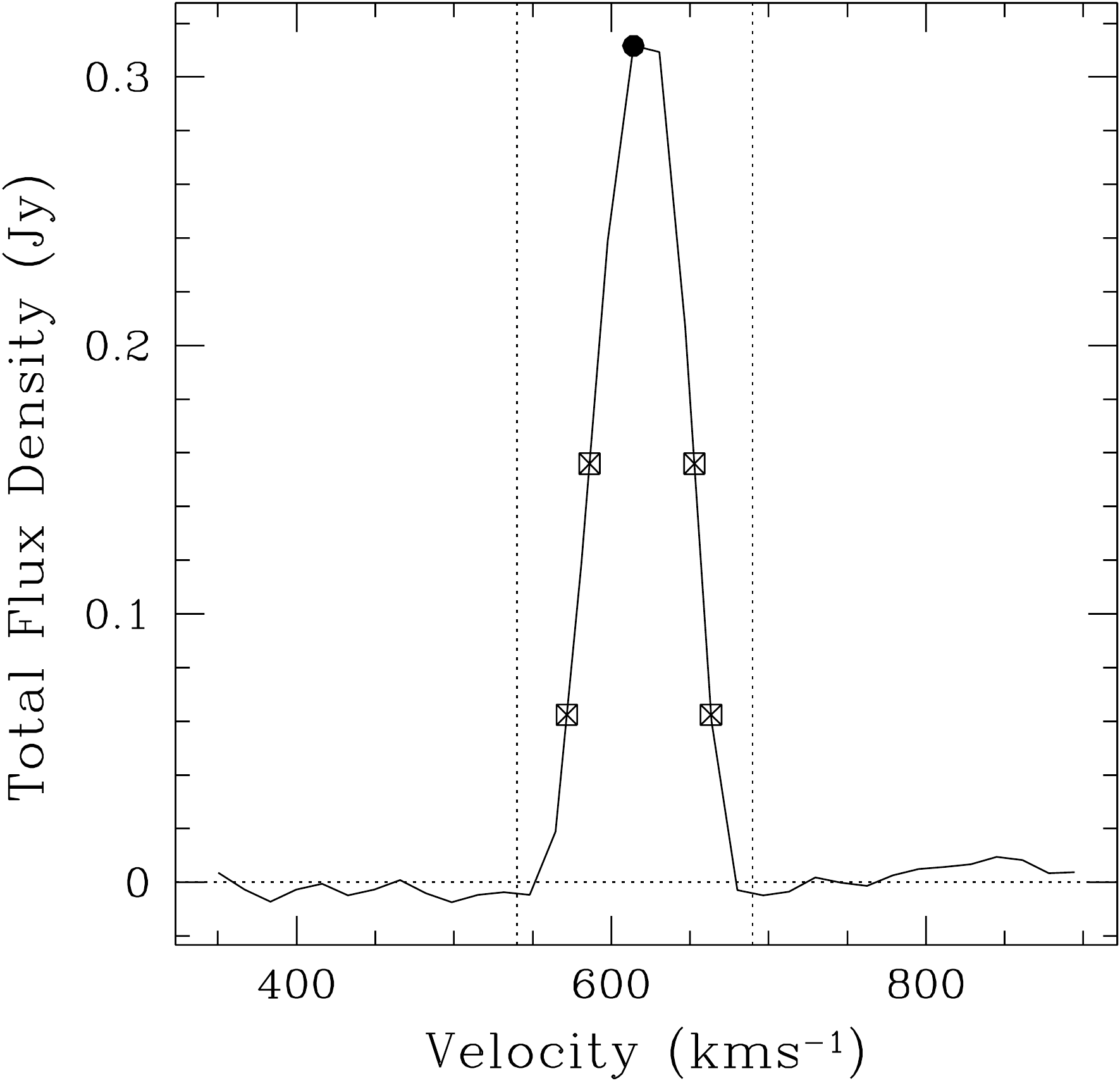}
\includegraphics[height=0.17\textheight]{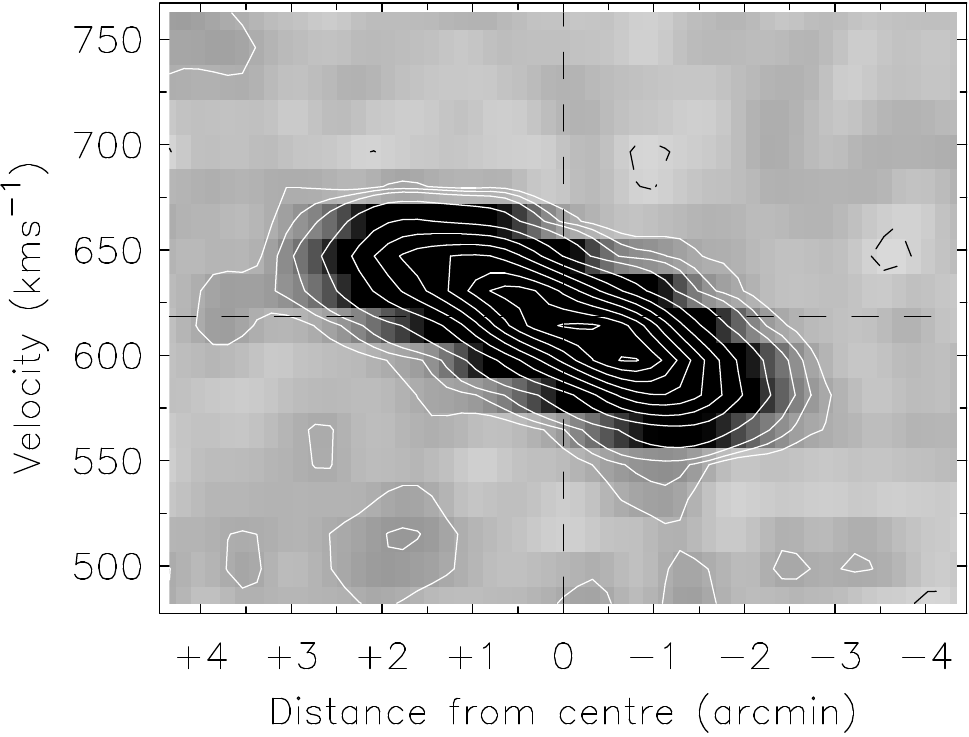}

\end{figure}

\clearpage

\addtocounter{figure}{-1}
\begin{figure}

\vskip 2mm
\centering
WSRT-CVn-17
\vskip 2mm
\includegraphics[width=0.25\textwidth]{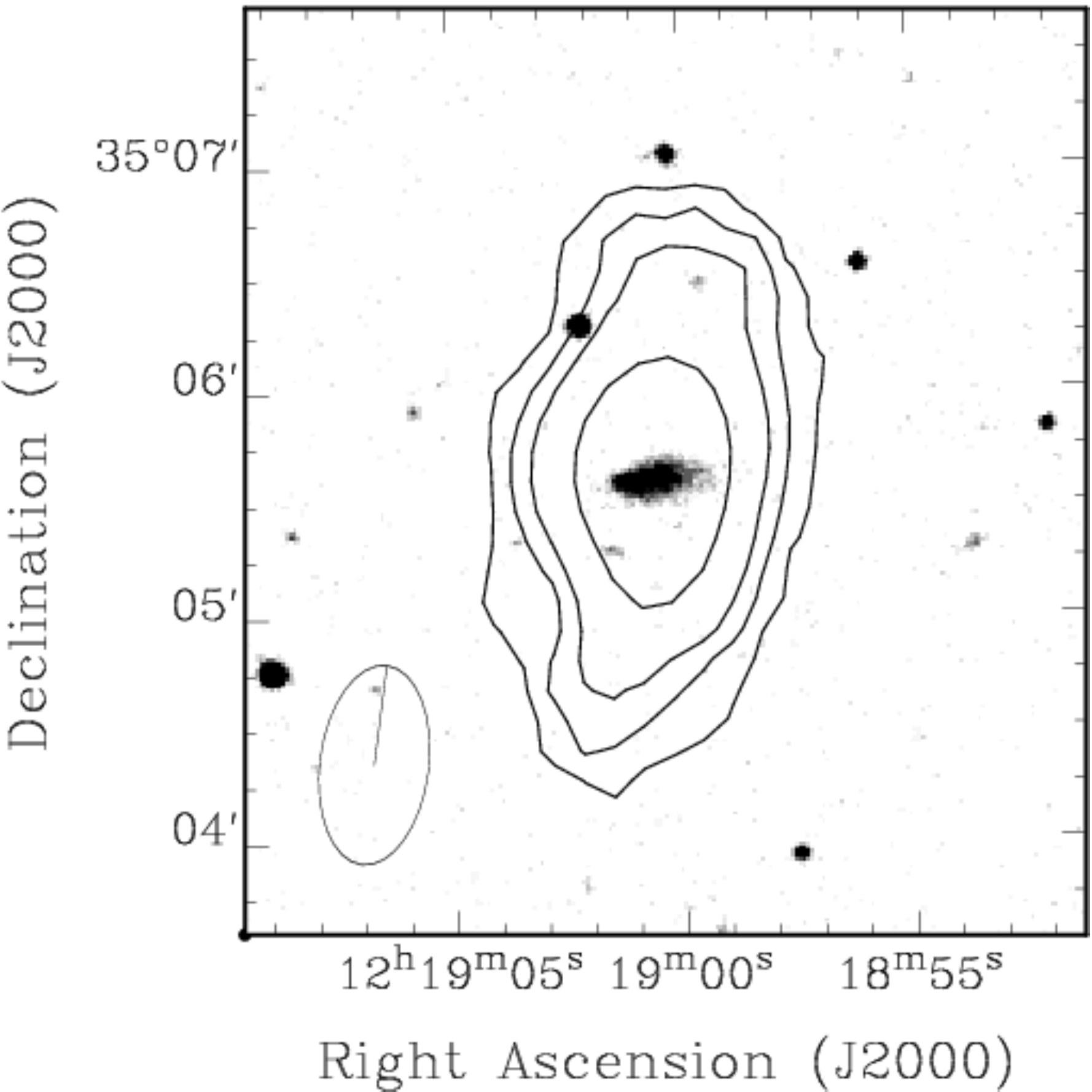}
\hskip 5mm
\includegraphics[height=0.17\textheight]{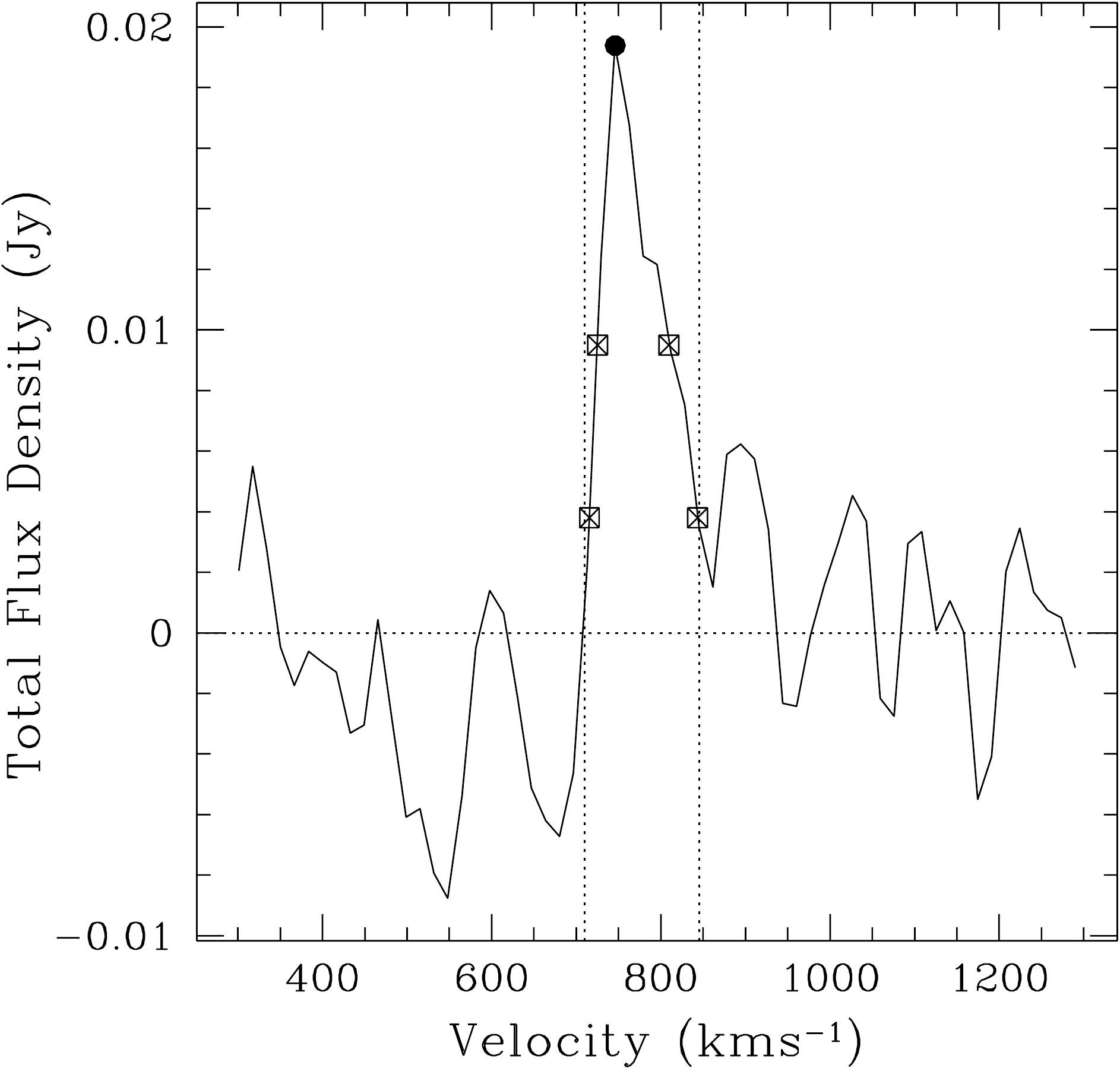}
\includegraphics[height=0.17\textheight]{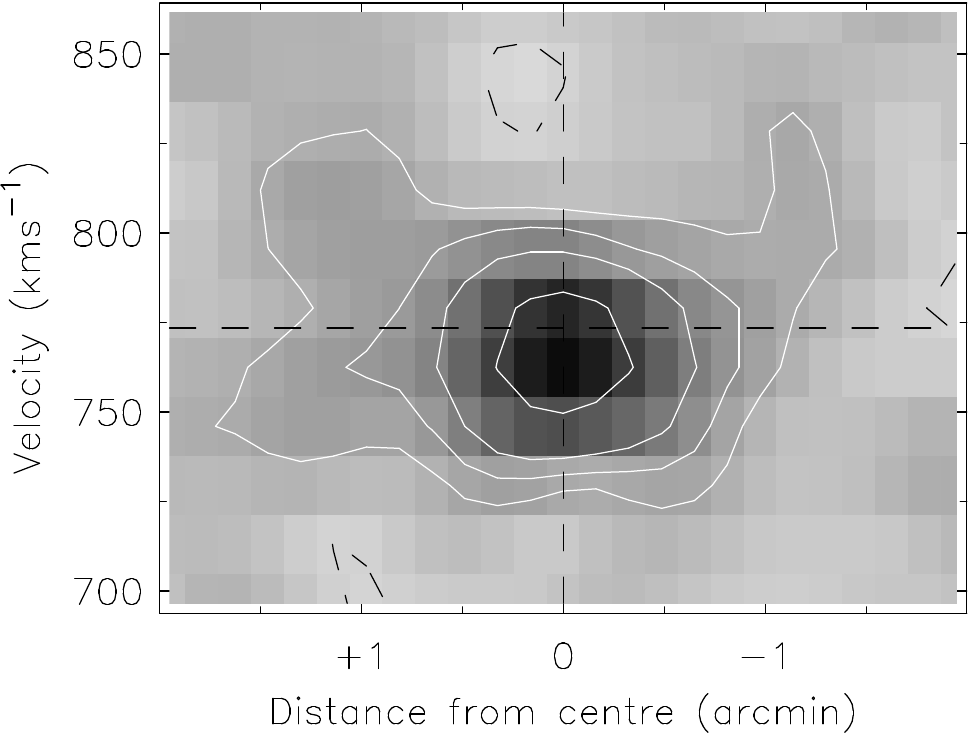}

\vskip 2mm
\centering
WSRT-CVn-18
\vskip 2mm
\includegraphics[width=0.25\textwidth]{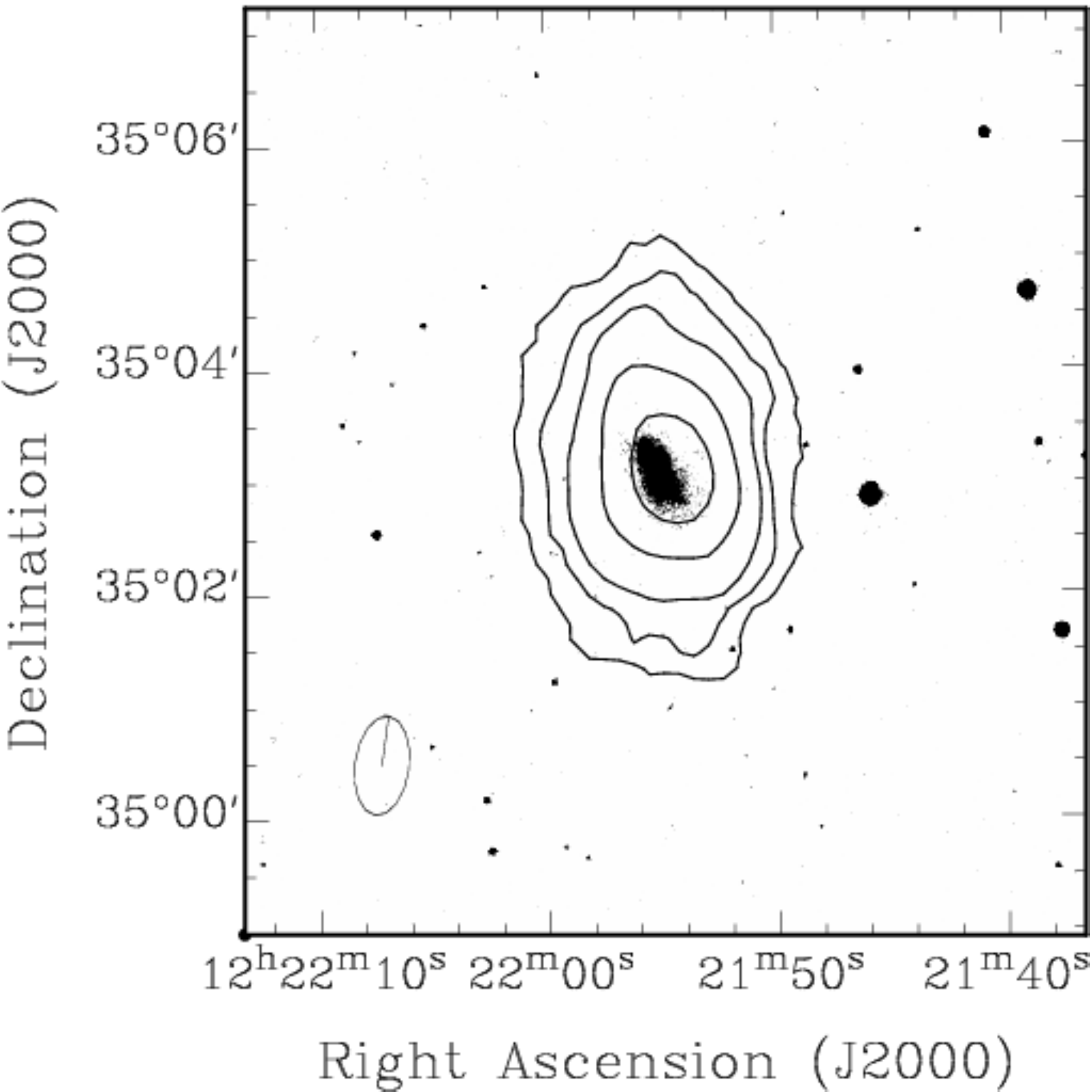}
\hskip 5mm
\includegraphics[height=0.17\textheight]{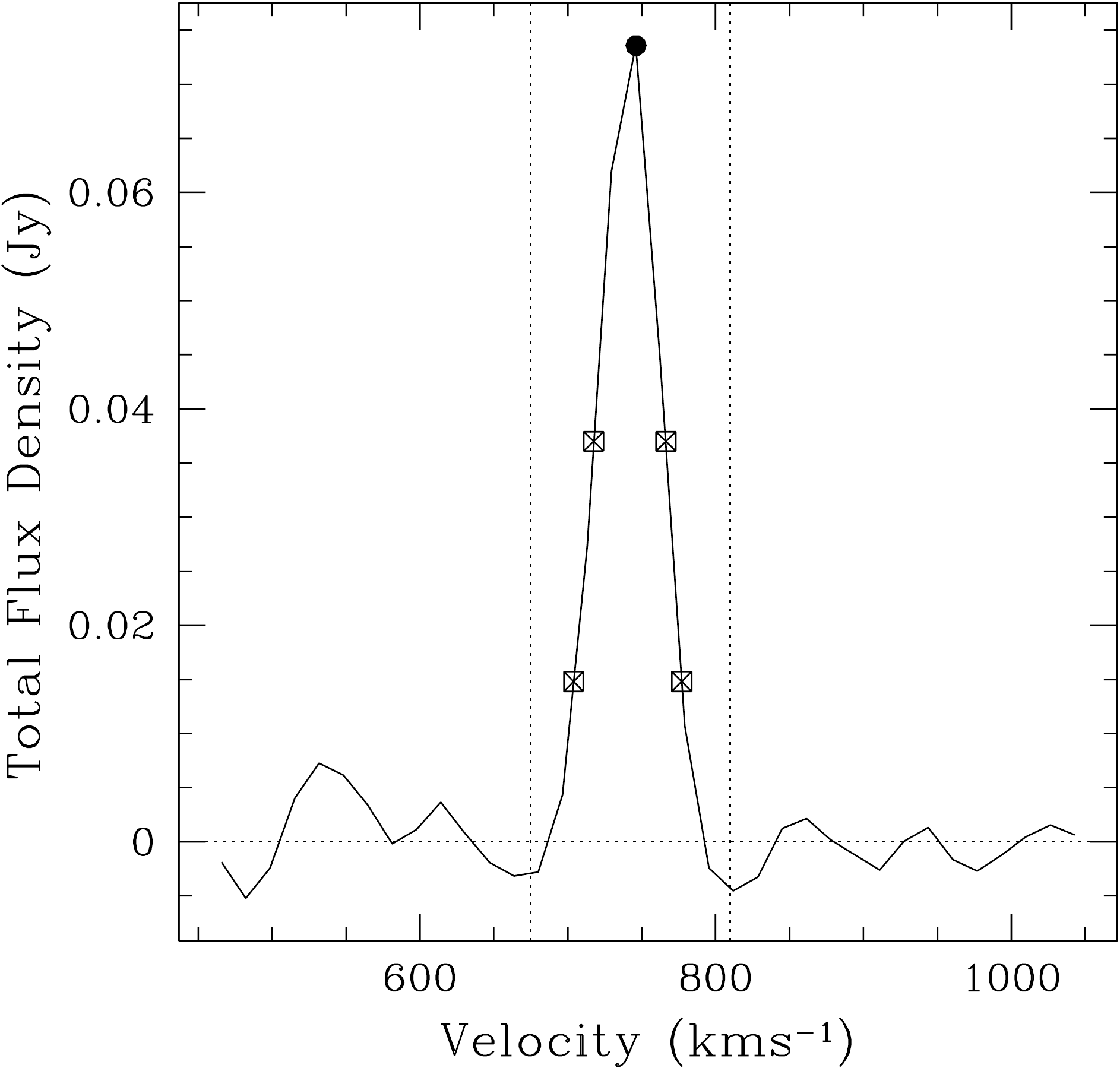}
\includegraphics[height=0.17\textheight]{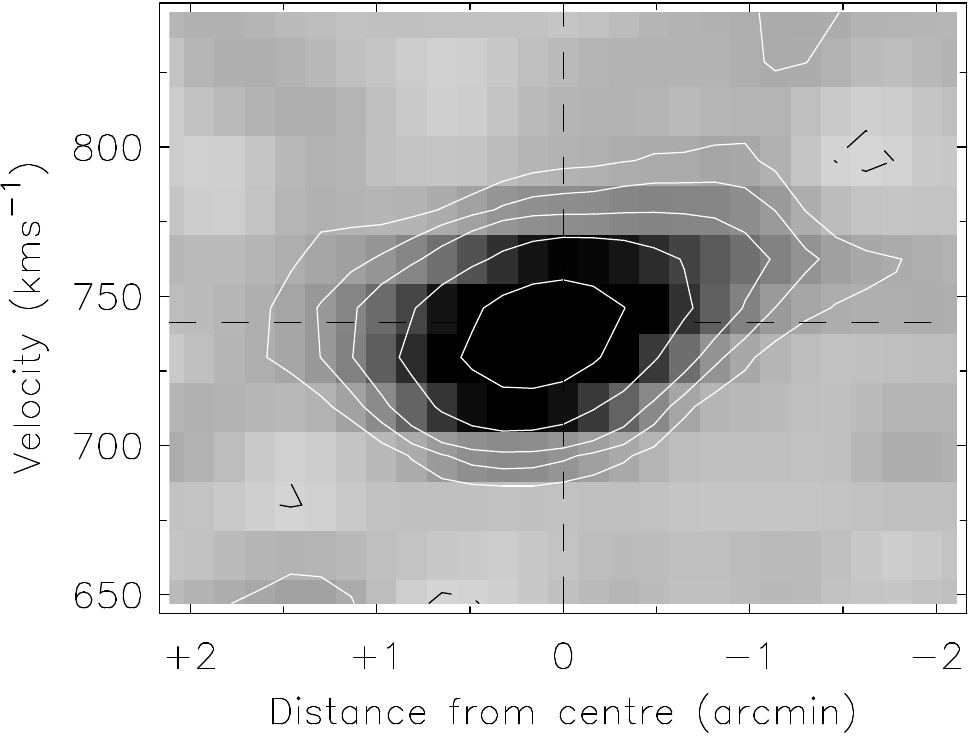}

\vskip 2mm
\centering
WSRT-CVn-19
\vskip 2mm
\includegraphics[width=0.25\textwidth]{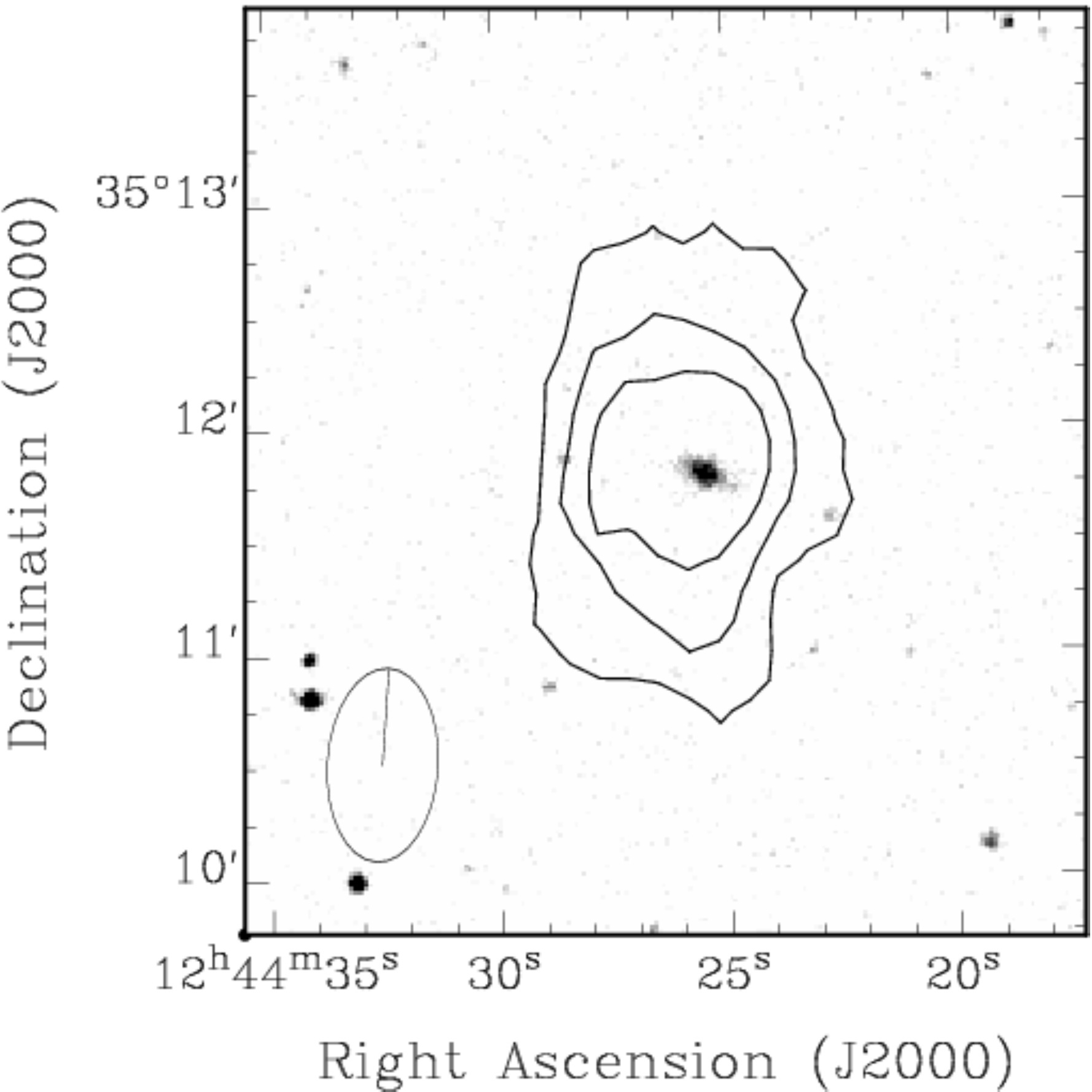}
\hskip 5mm
\includegraphics[height=0.17\textheight]{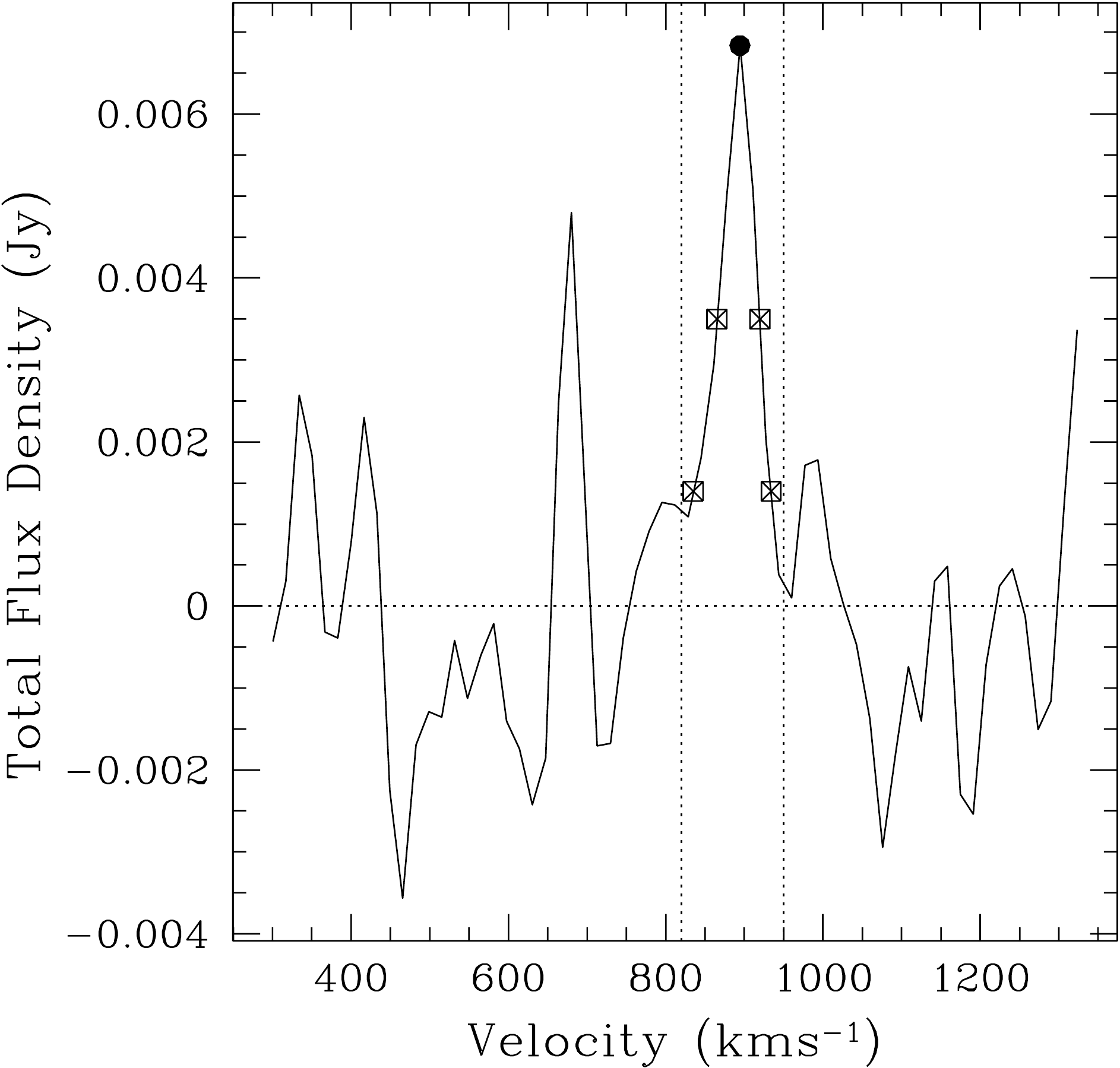}
\includegraphics[height=0.17\textheight]{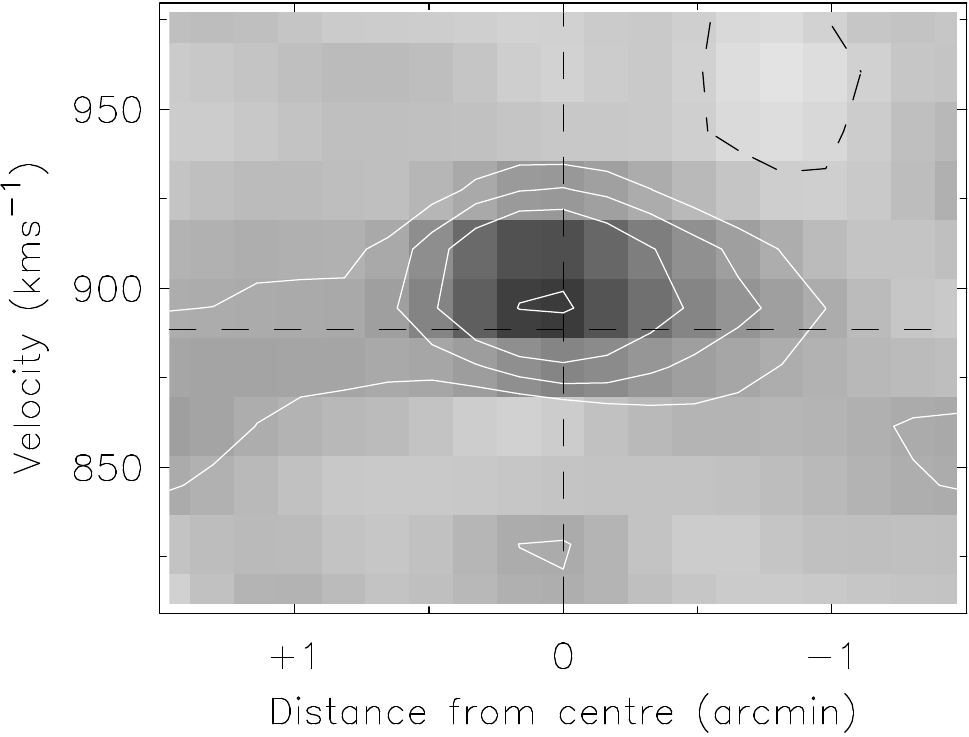}

\vskip 2mm
\centering
WSRT-CVn-20
\vskip 2mm
\includegraphics[width=0.25\textwidth]{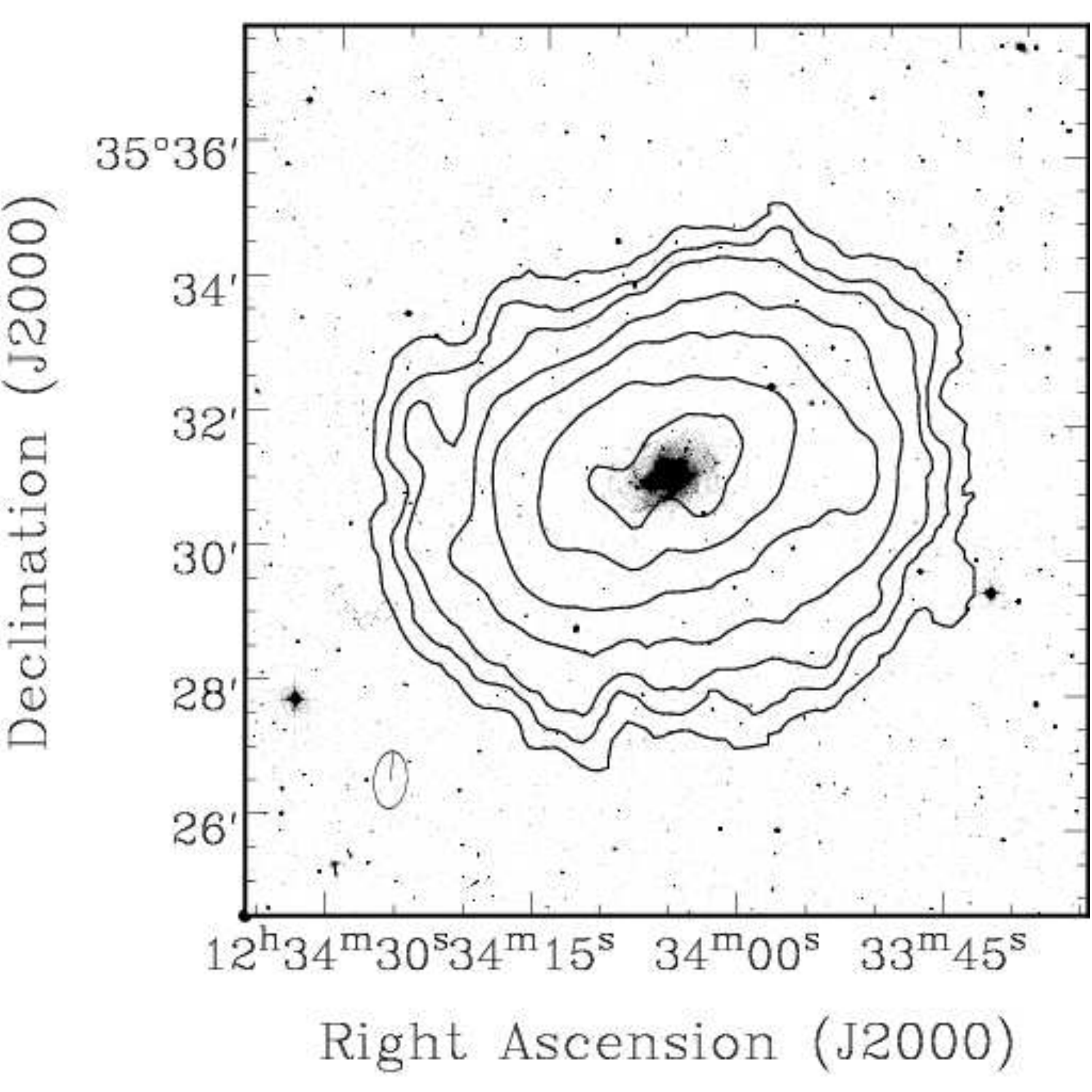}
\hskip 5mm
\includegraphics[height=0.17\textheight]{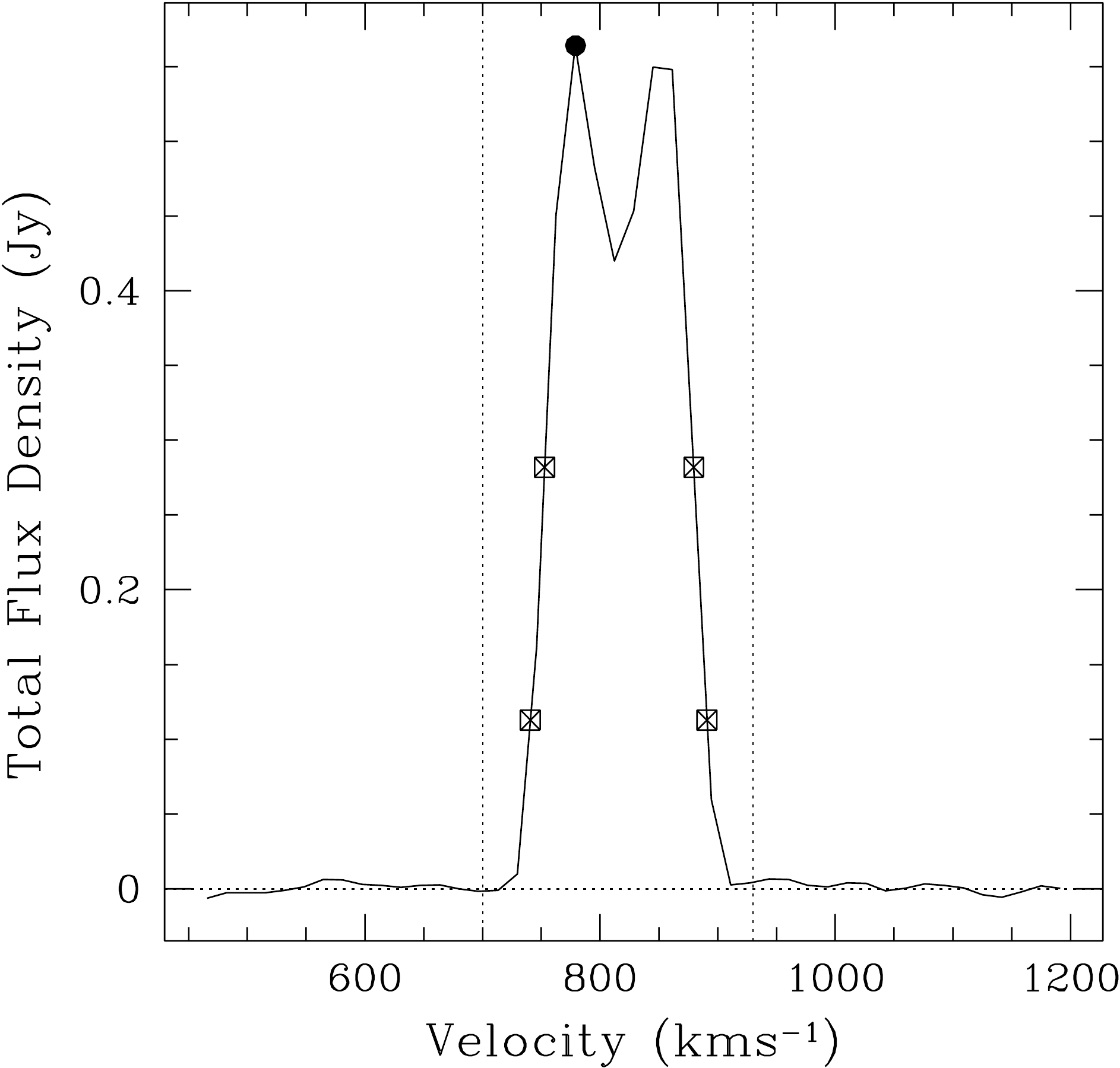}
\includegraphics[height=0.17\textheight]{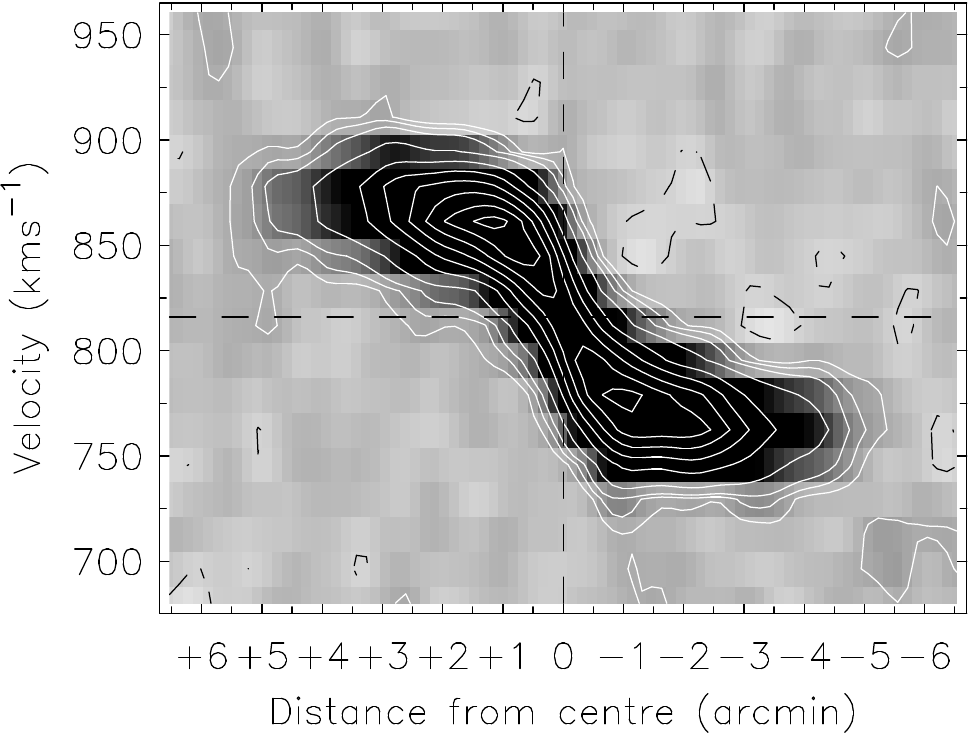}

\end{figure}

\clearpage

\addtocounter{figure}{-1}
\begin{figure}

\vskip 2mm
\centering
WSRT-CVn-21
\vskip 2mm
\includegraphics[width=0.25\textwidth]{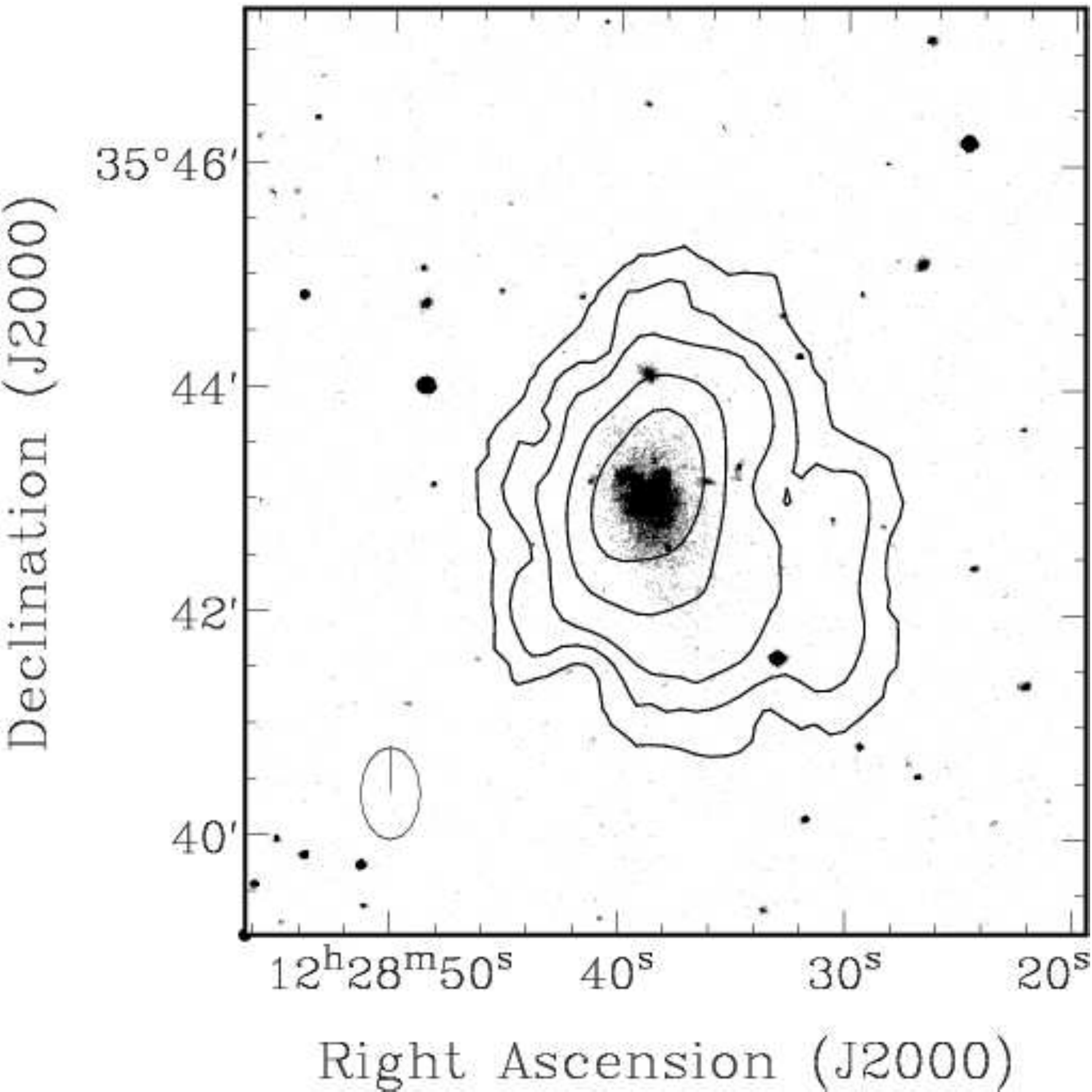}
\hskip 5mm
\includegraphics[height=0.17\textheight]{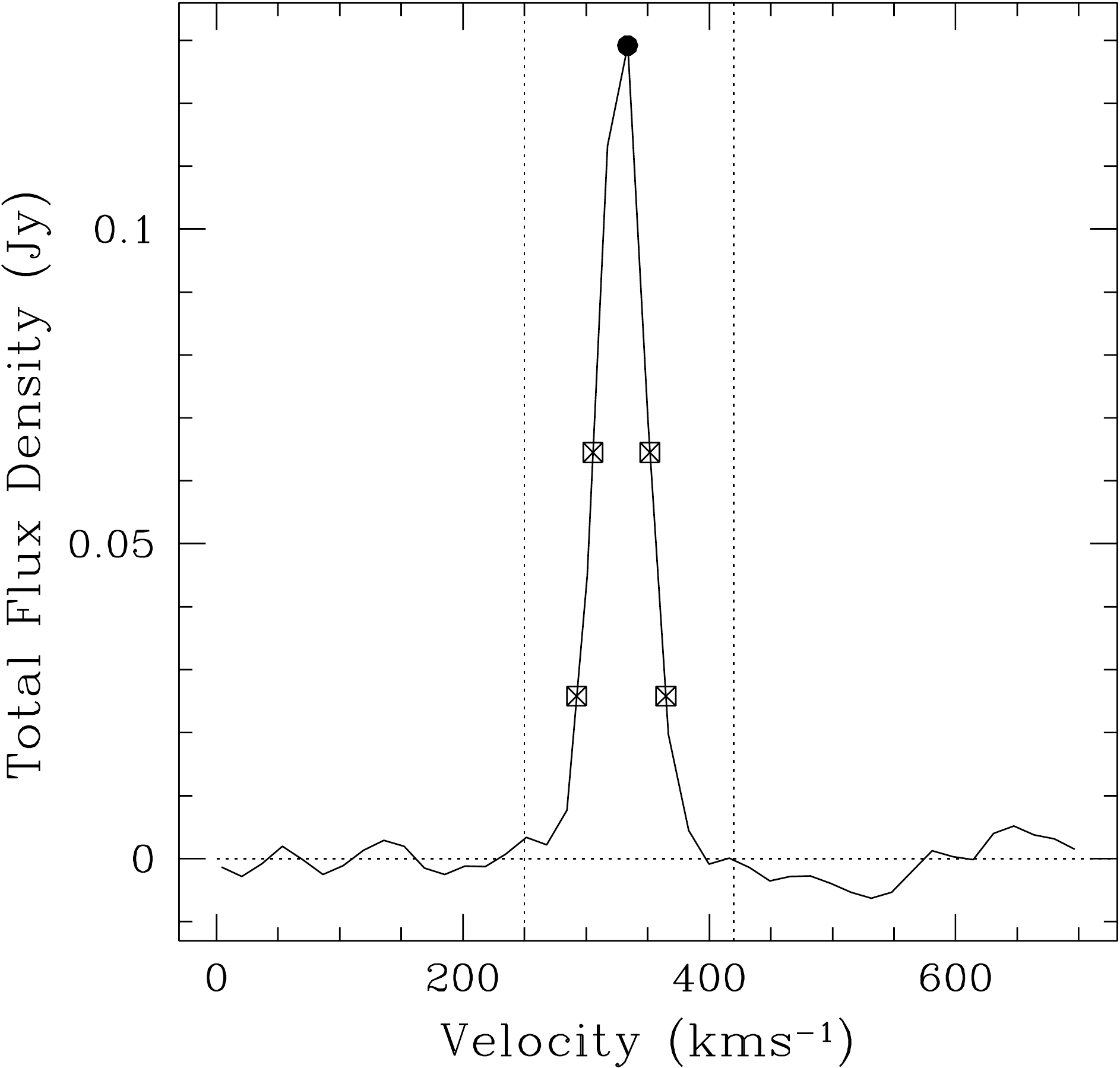}
\includegraphics[height=0.17\textheight]{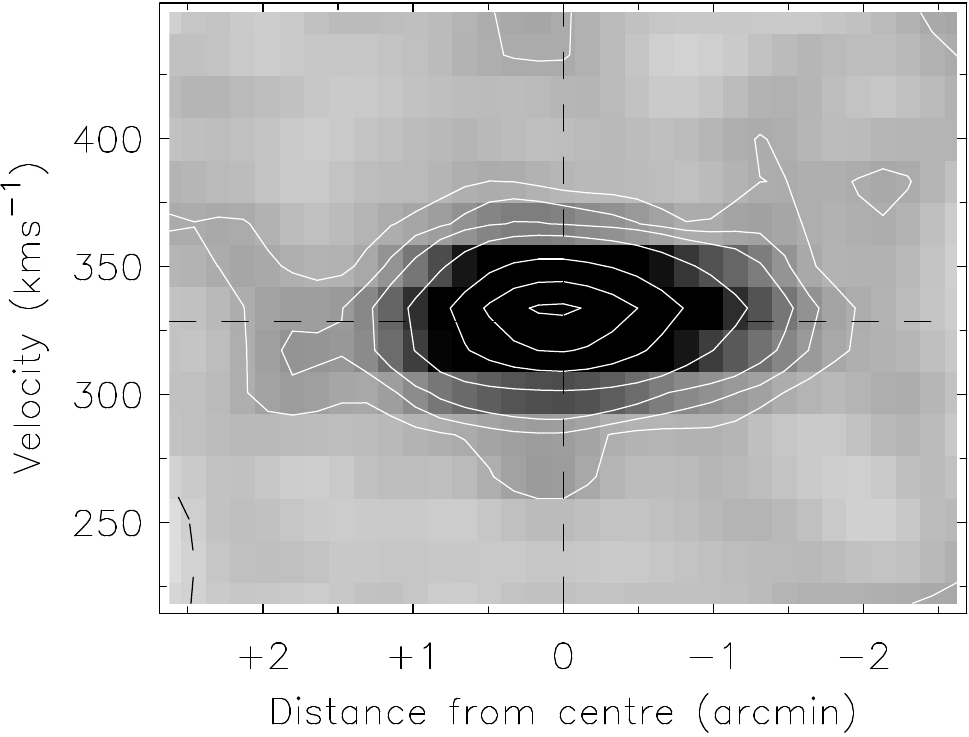}

\vskip 2mm
\centering
WSRT-CVn-22
\vskip 2mm
\includegraphics[width=0.25\textwidth]{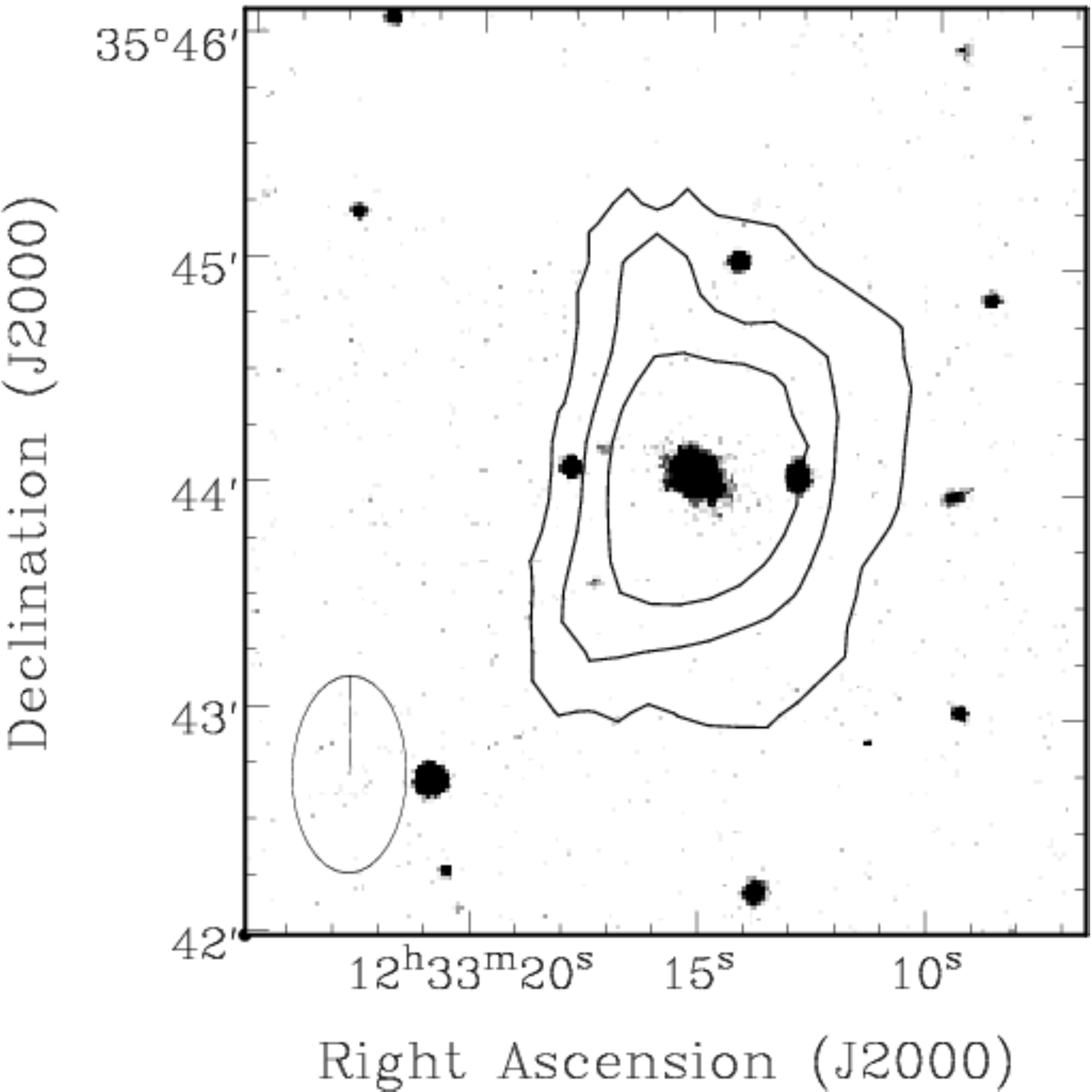}
\hskip 5mm
\includegraphics[height=0.17\textheight]{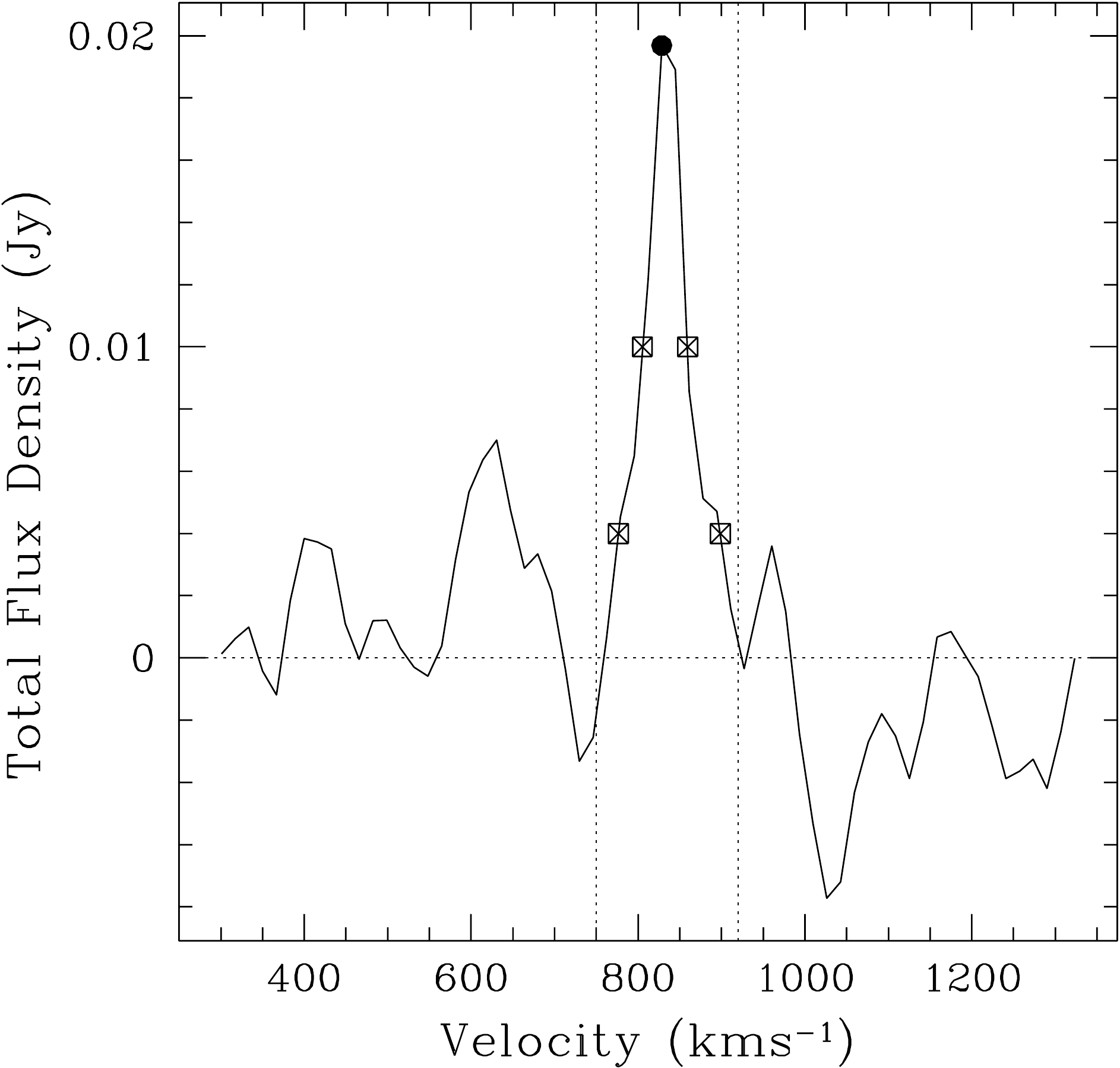}
\includegraphics[height=0.17\textheight]{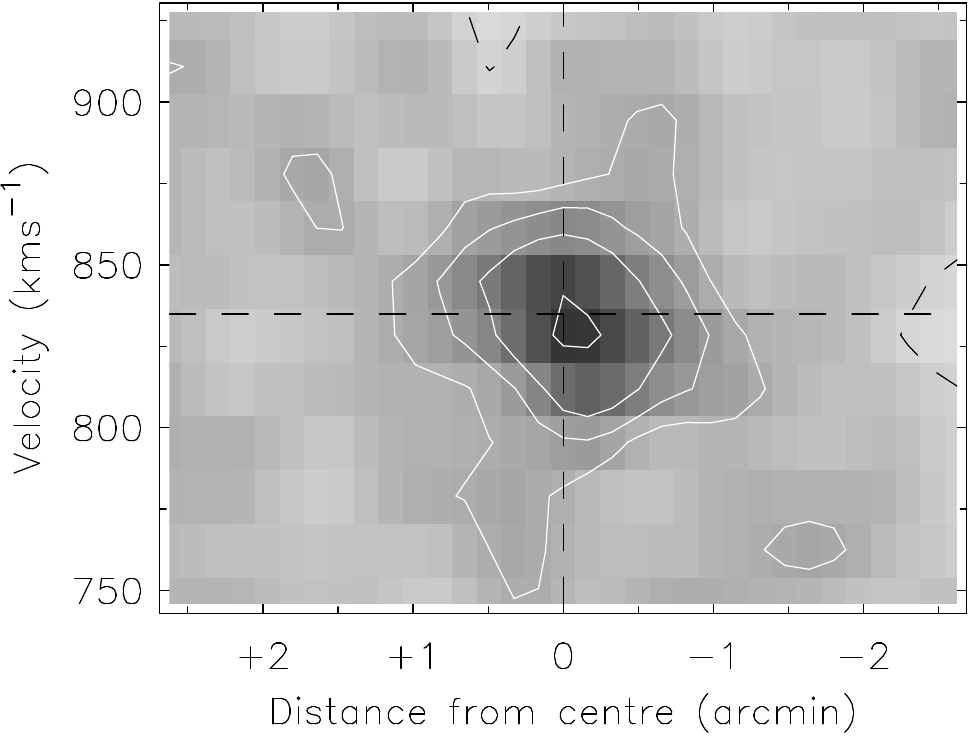}

\vskip 2mm
\centering
WSRT-CVn-23
\vskip 2mm
\includegraphics[width=0.25\textwidth]{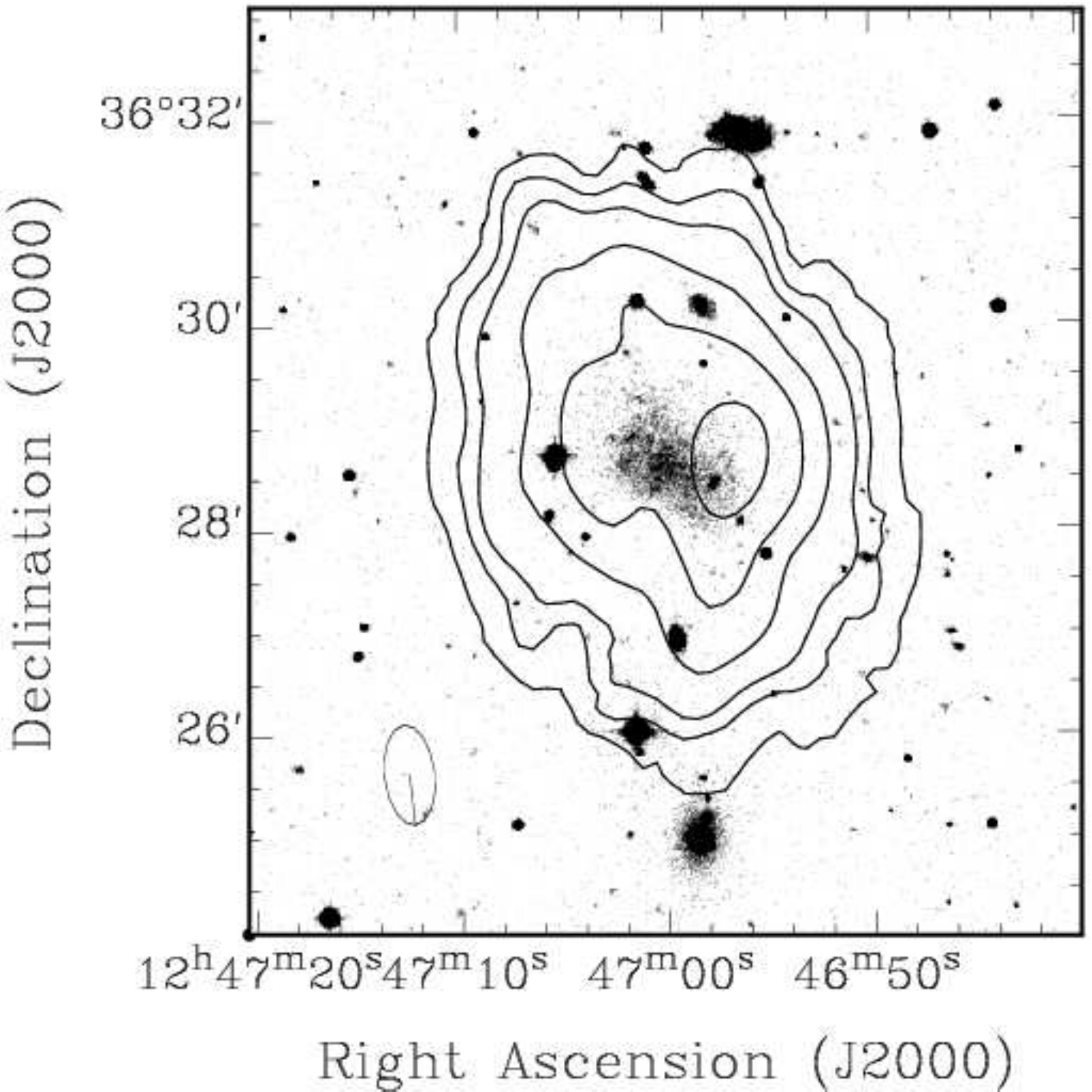}
\hskip 5mm
\includegraphics[height=0.17\textheight]{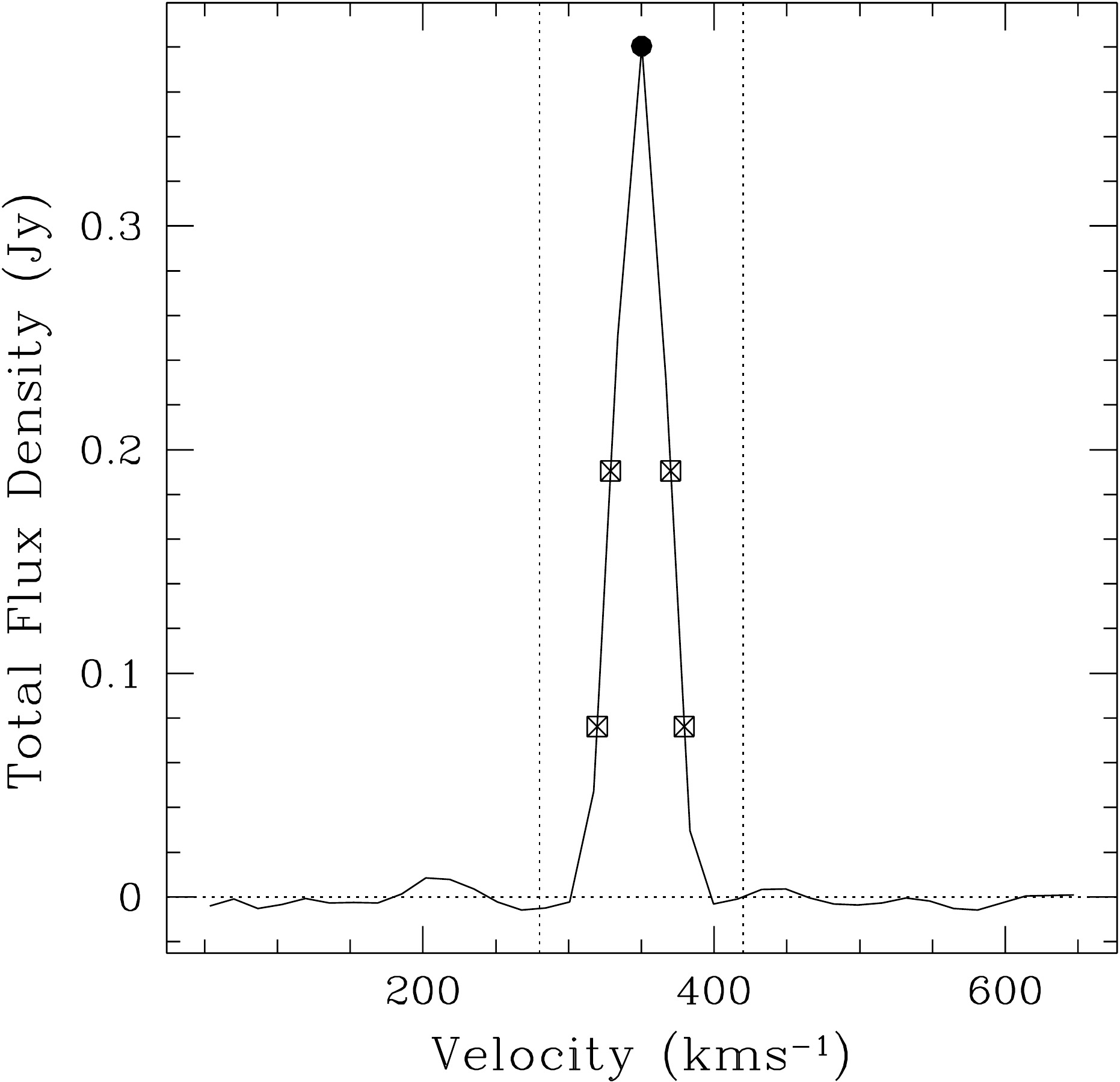}
\includegraphics[height=0.17\textheight]{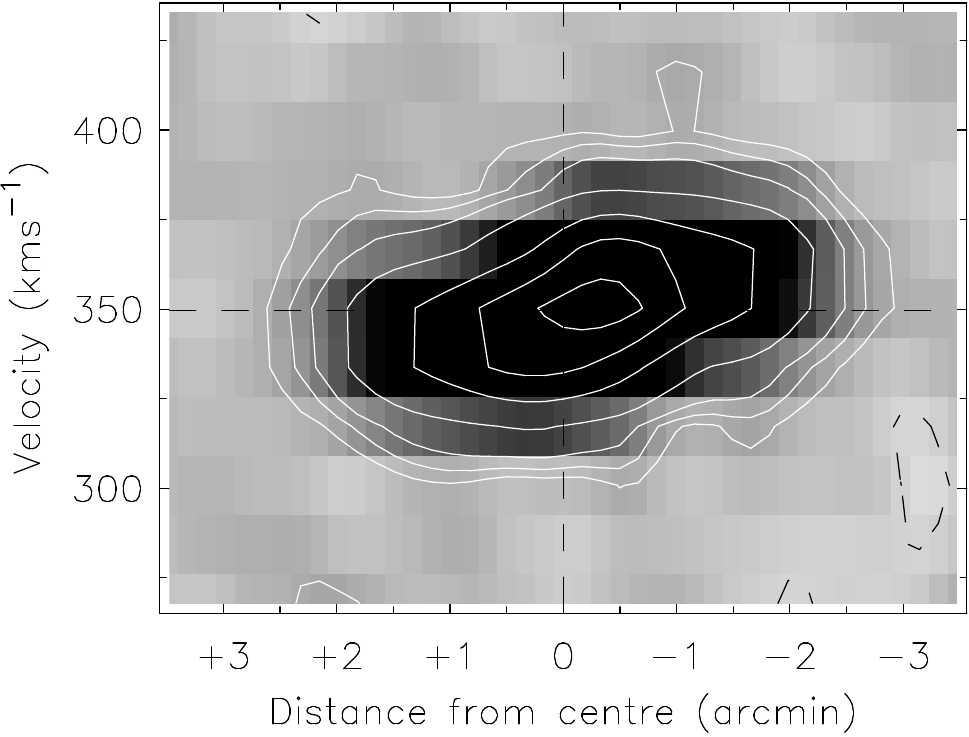}

\vskip 2mm
\centering
WSRT-CVn-24
\vskip 2mm
\includegraphics[width=0.25\textwidth]{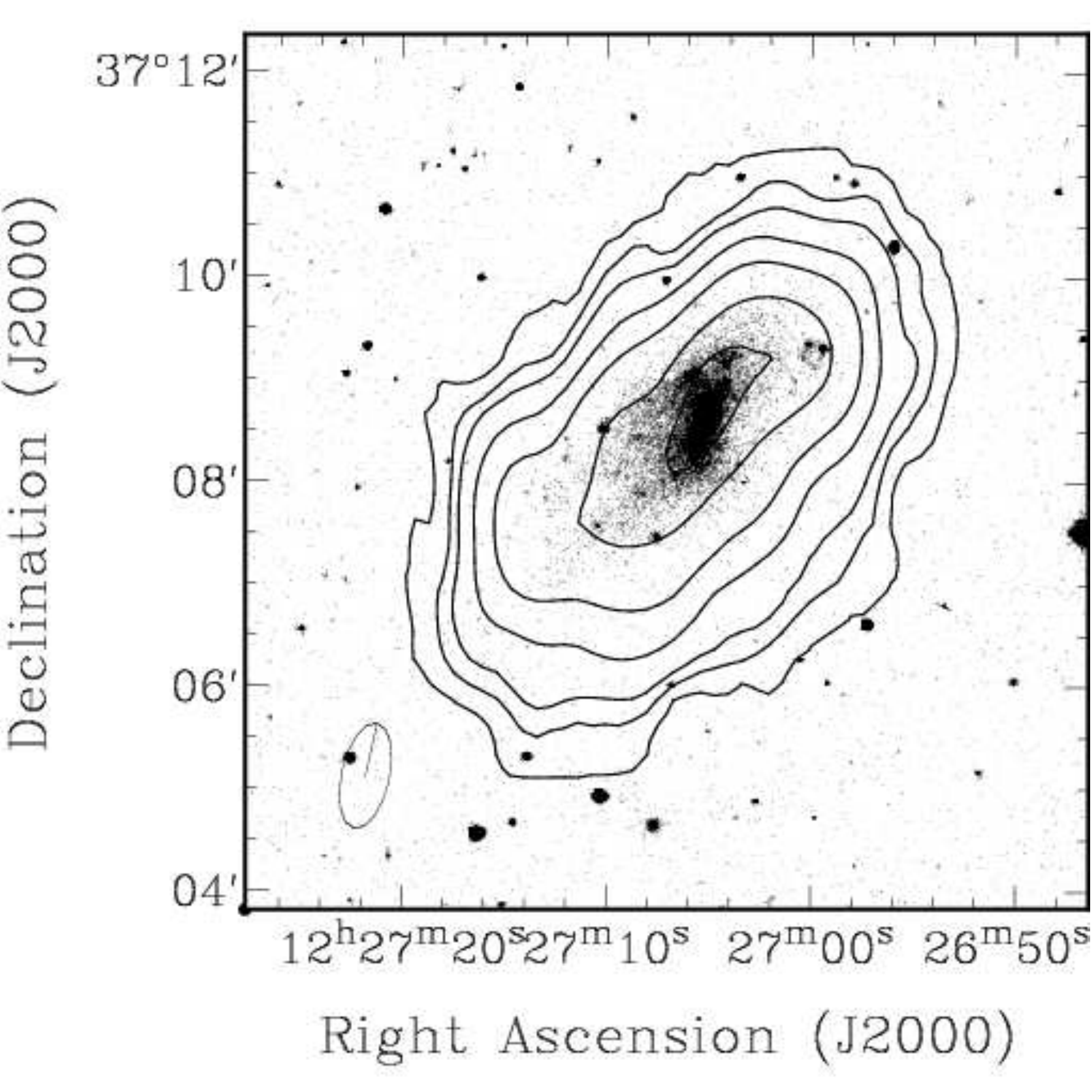}
\hskip 5mm
\includegraphics[height=0.17\textheight]{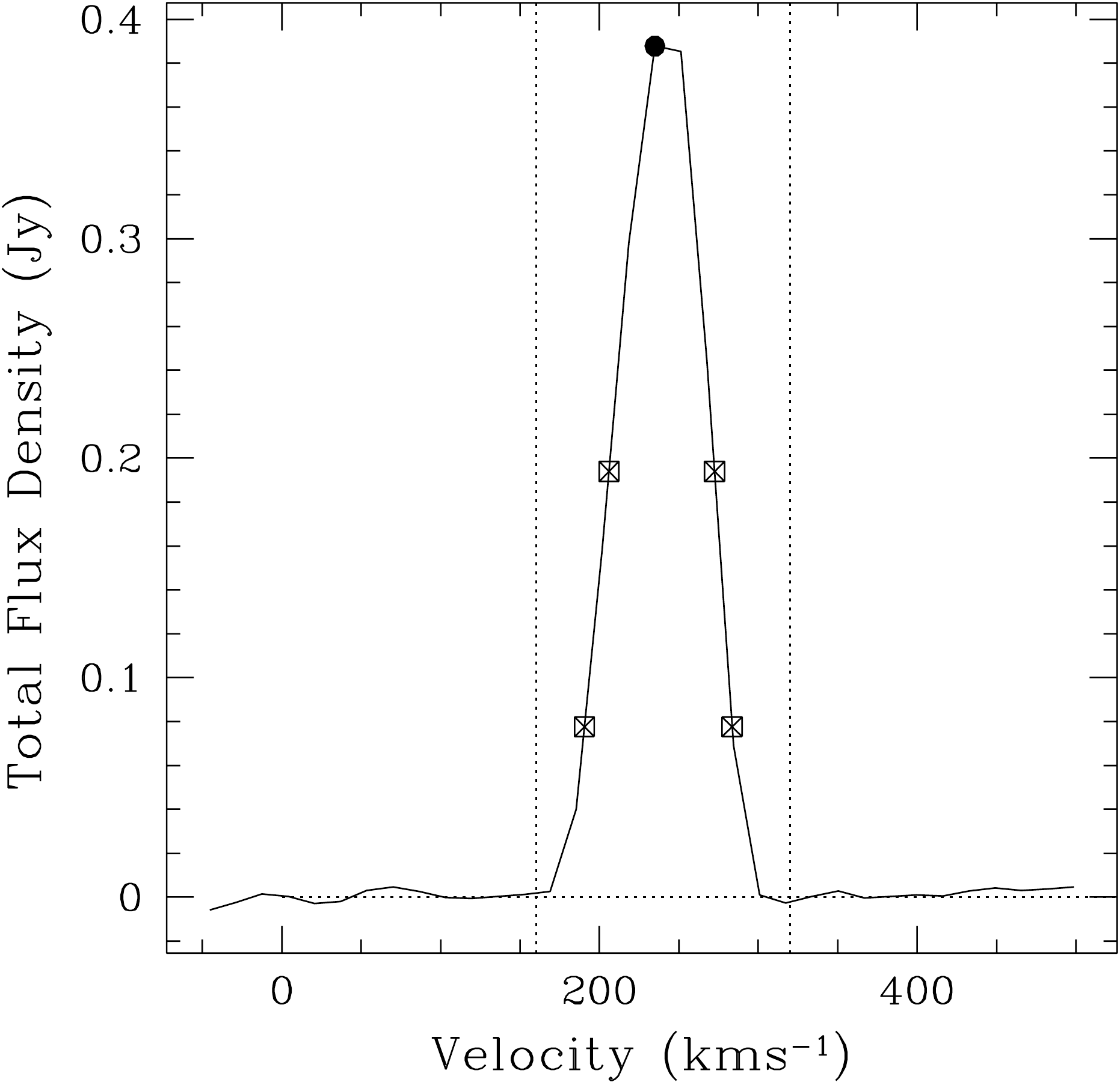}
\includegraphics[height=0.17\textheight]{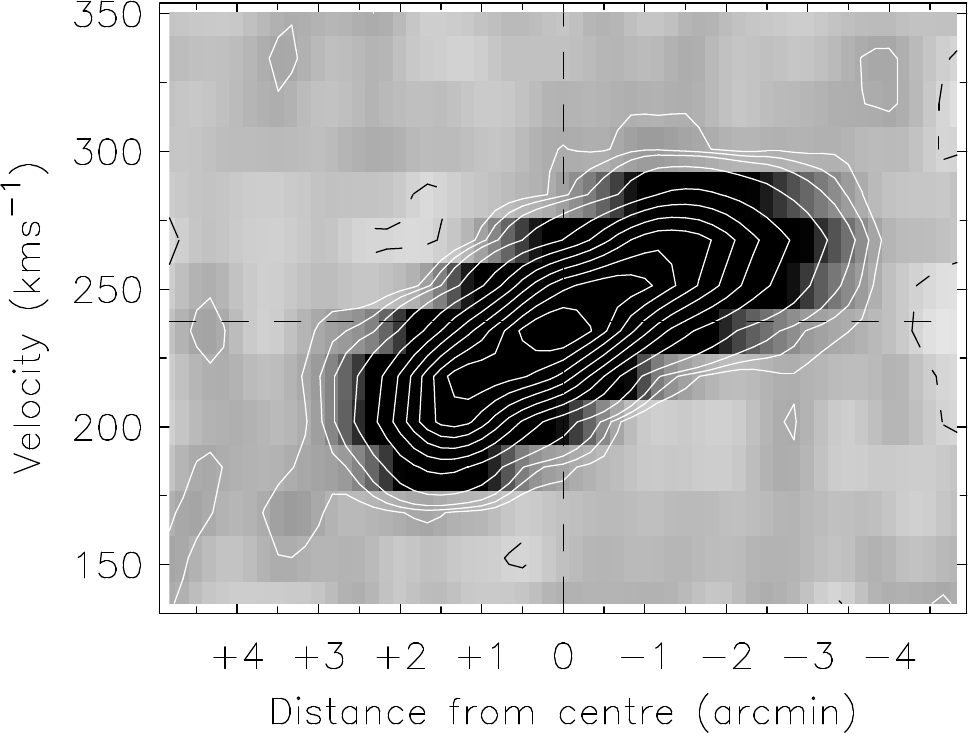}

\end{figure}

\clearpage

\addtocounter{figure}{-1}
\begin{figure}

\vskip 2mm
\centering
WSRT-CVn-25
\vskip 2mm
\includegraphics[width=0.25\textwidth]{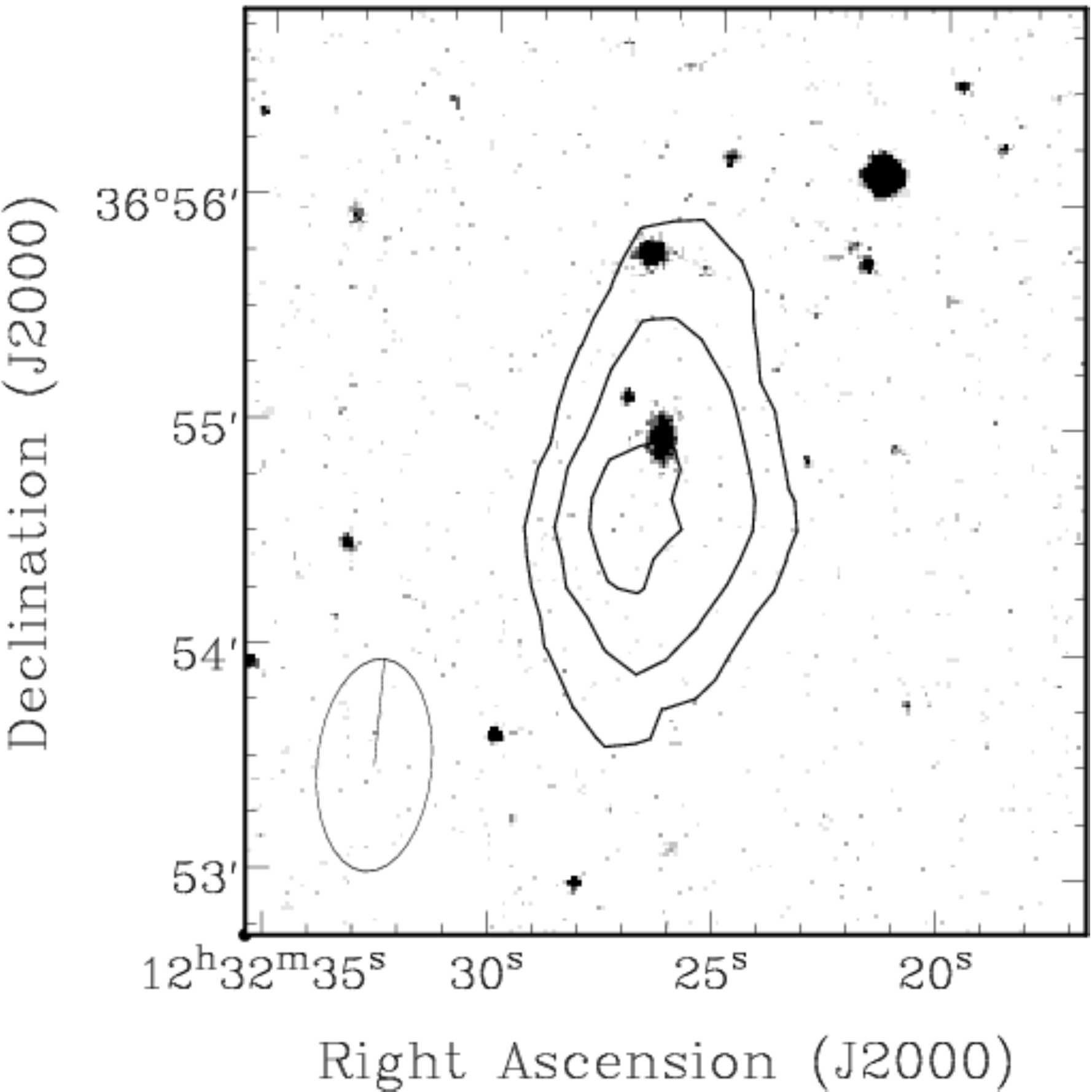}
\hskip 5mm
\includegraphics[height=0.17\textheight]{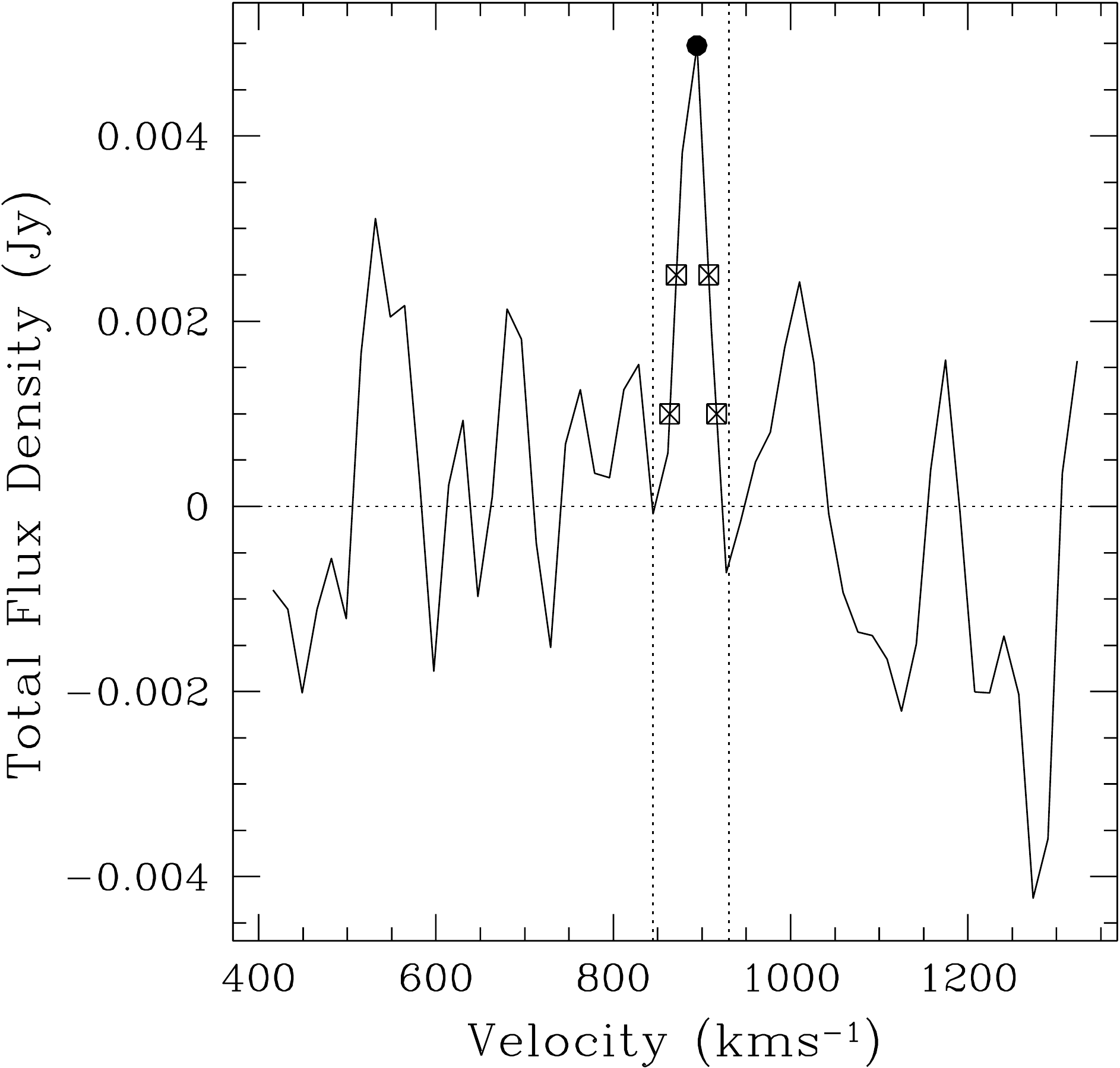}
\includegraphics[height=0.17\textheight]{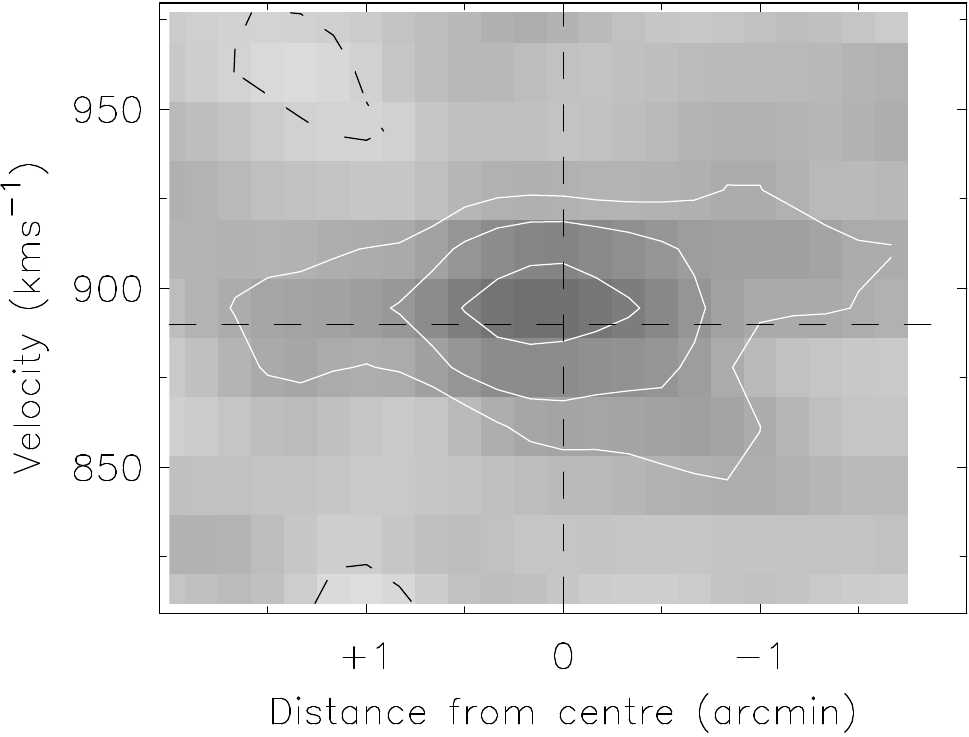}

\vskip 2mm
\centering
WSRT-CVn-26
\vskip 2mm
\includegraphics[width=0.25\textwidth]{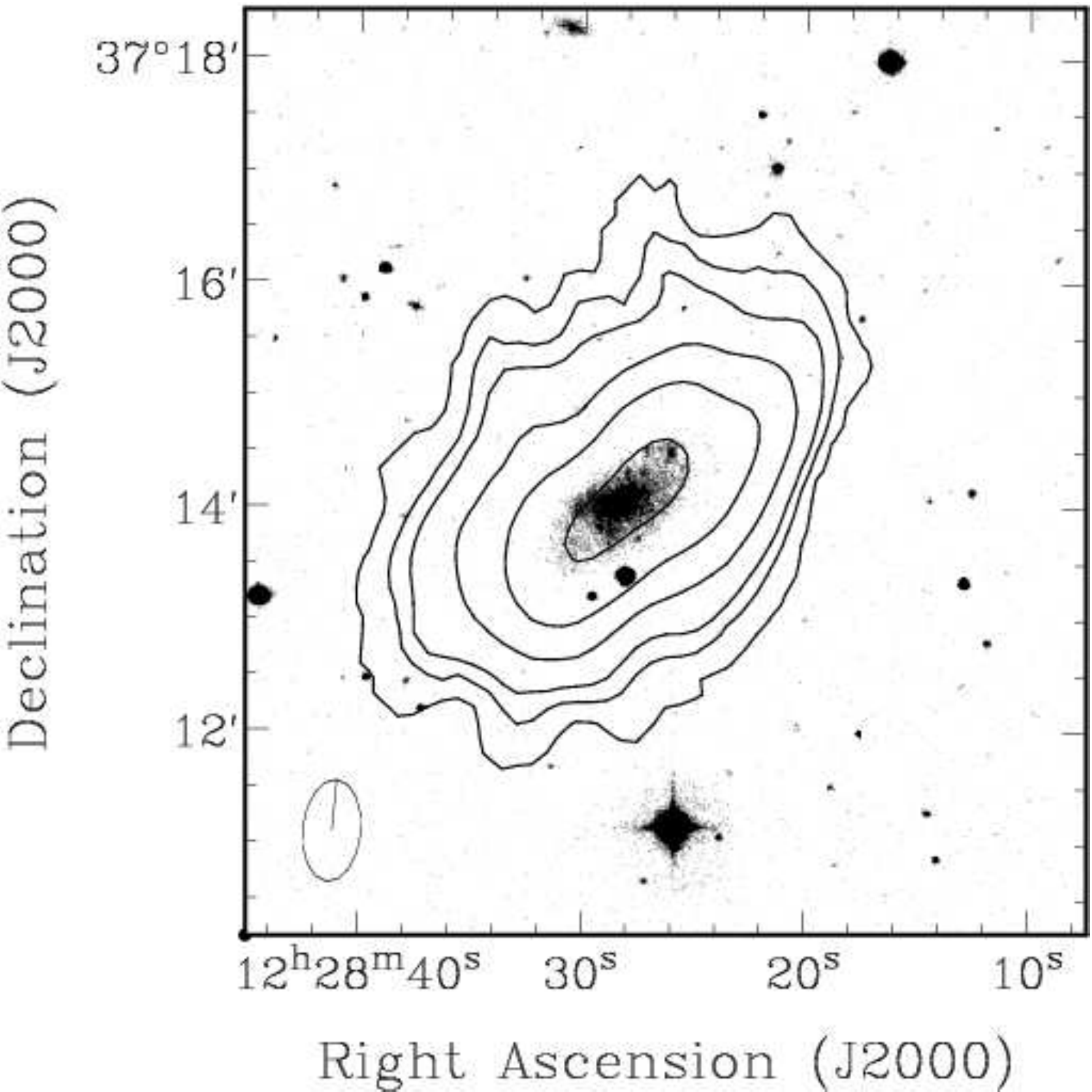}
\hskip 5mm
\includegraphics[height=0.17\textheight]{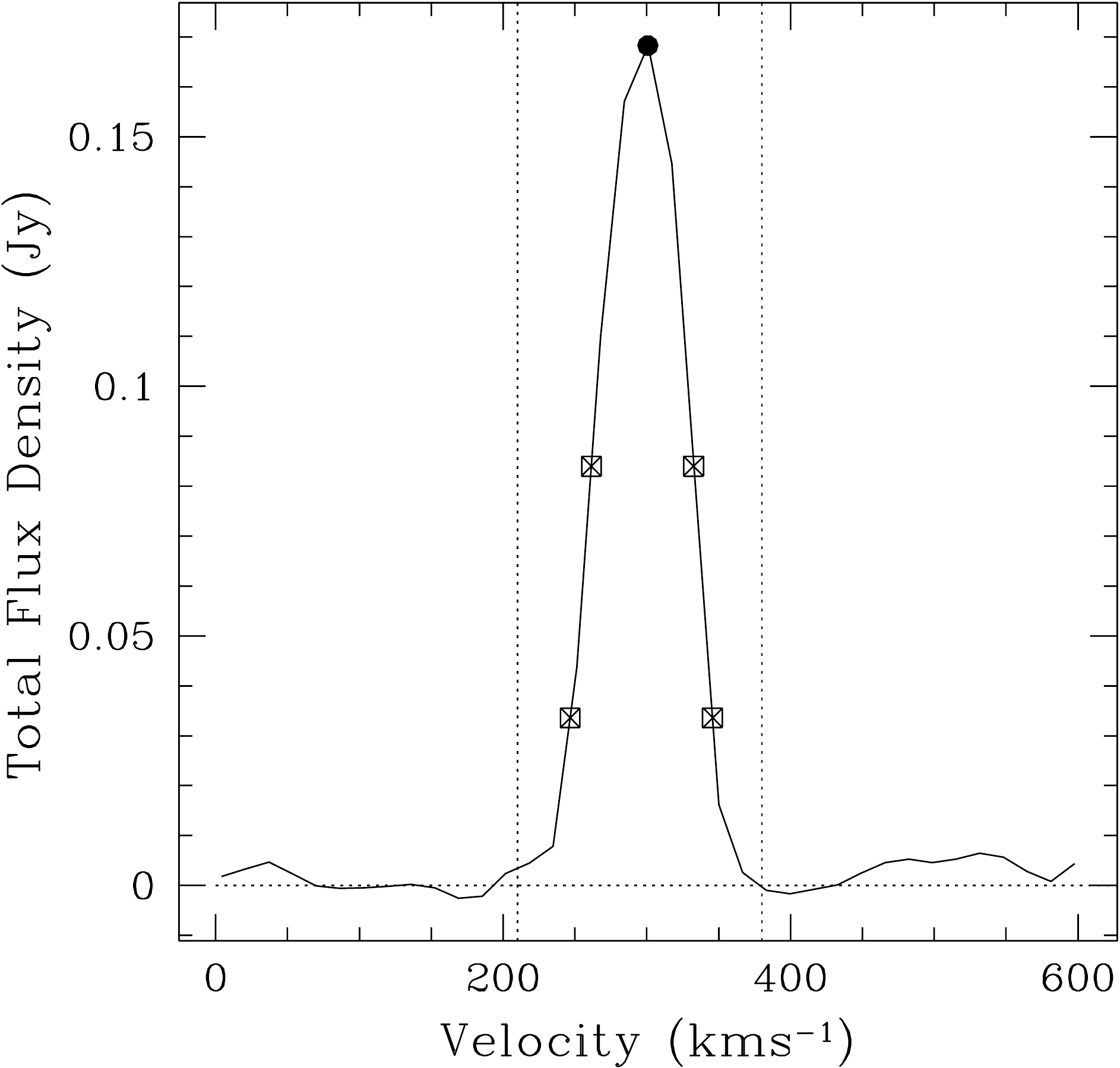}
\includegraphics[height=0.17\textheight]{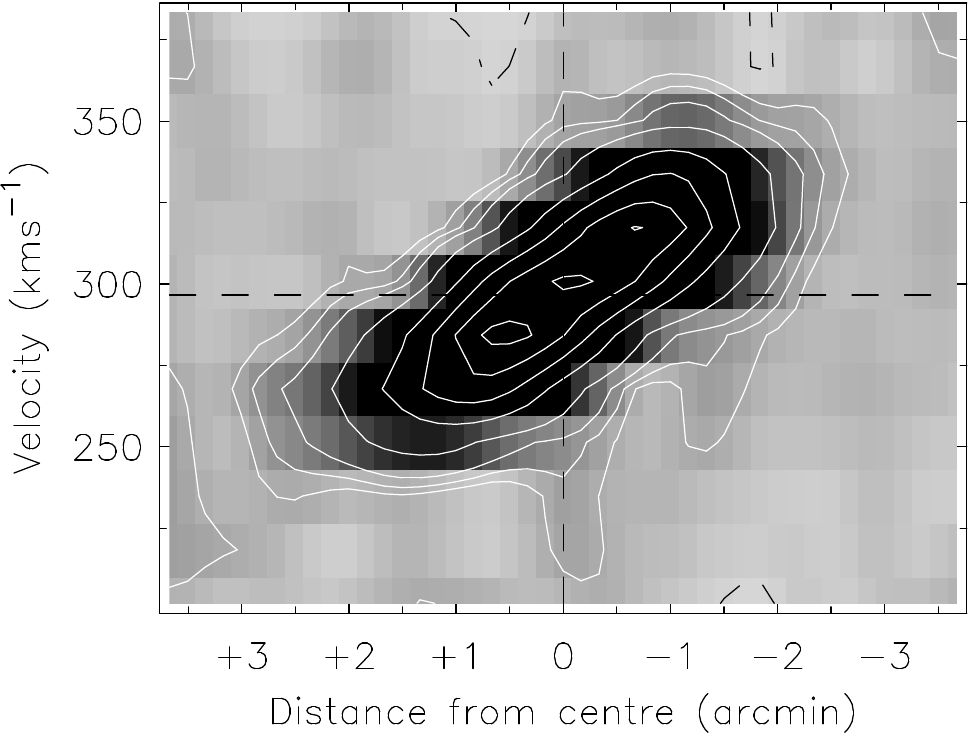}

\vskip 2mm
\centering
WSRT-CVn-27
\vskip 2mm
\includegraphics[width=0.25\textwidth]{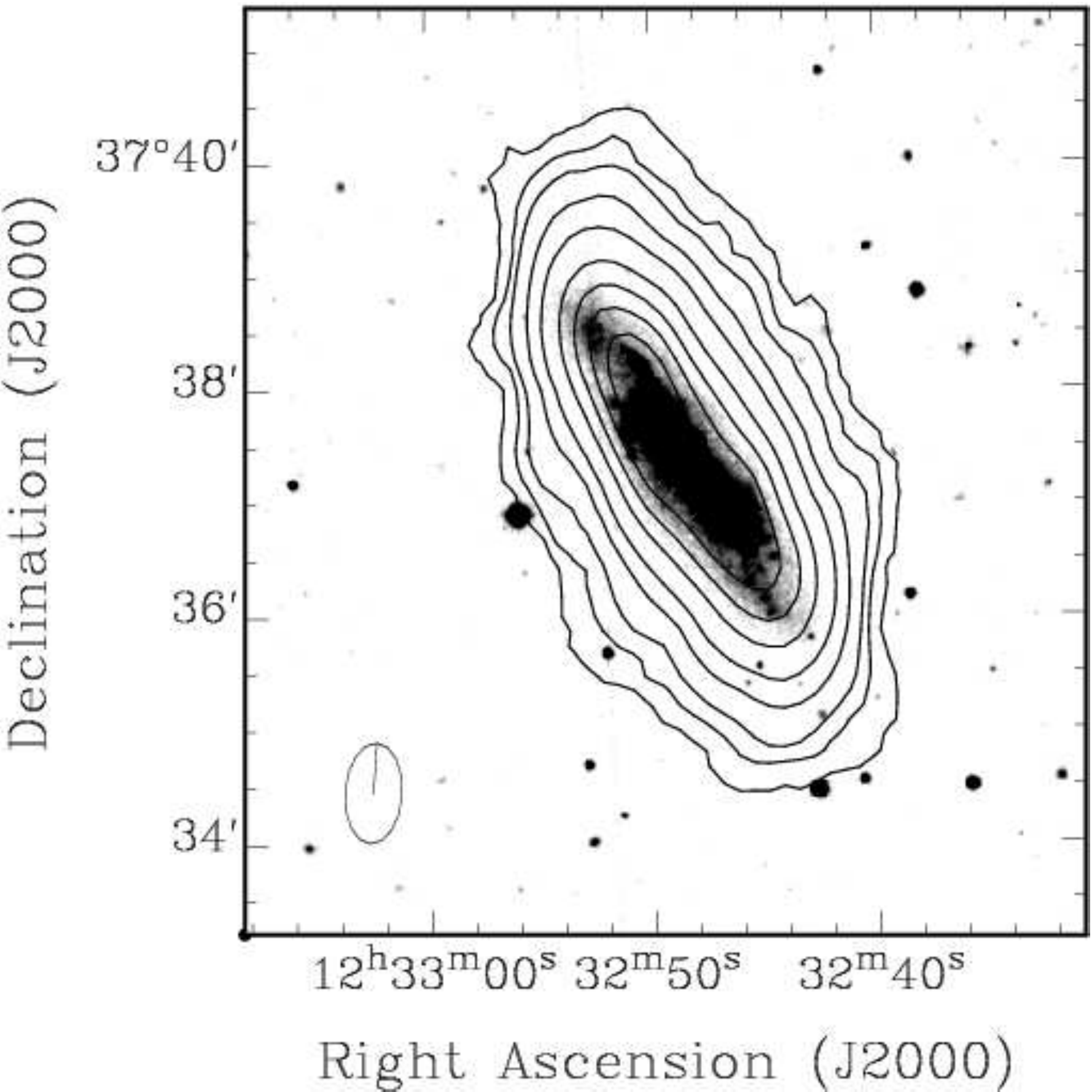}
\hskip 5mm
\includegraphics[height=0.17\textheight]{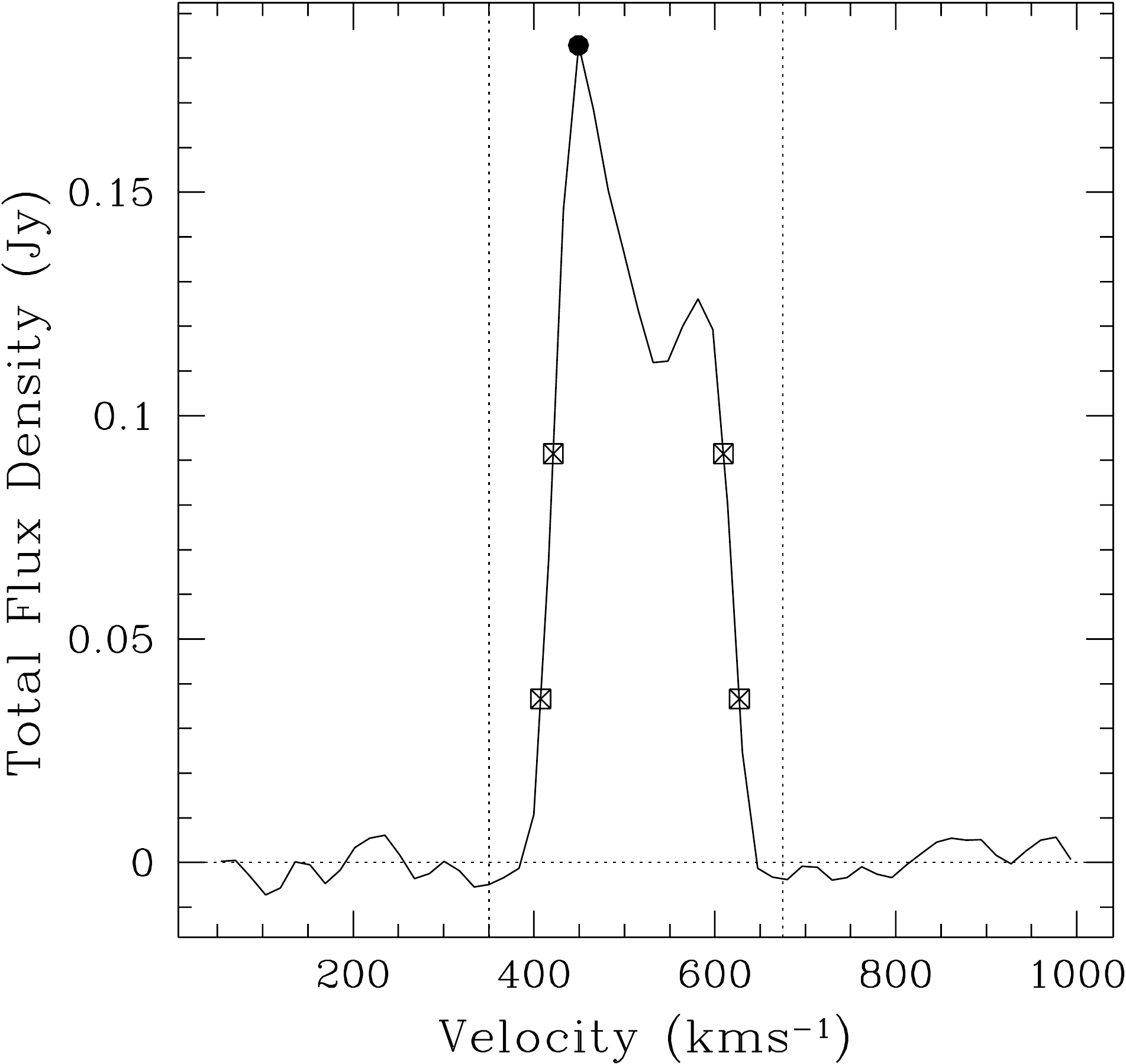}
\includegraphics[height=0.17\textheight]{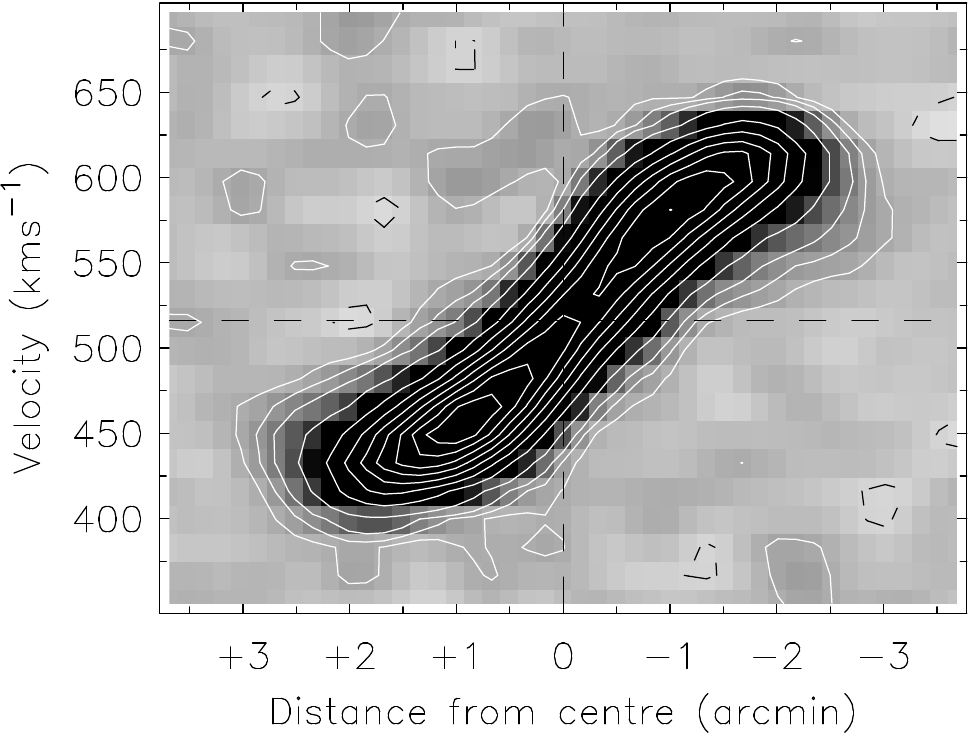}

\vskip 2mm
\centering
WSRT-CVn-28
\vskip 2mm
\includegraphics[width=0.25\textwidth]{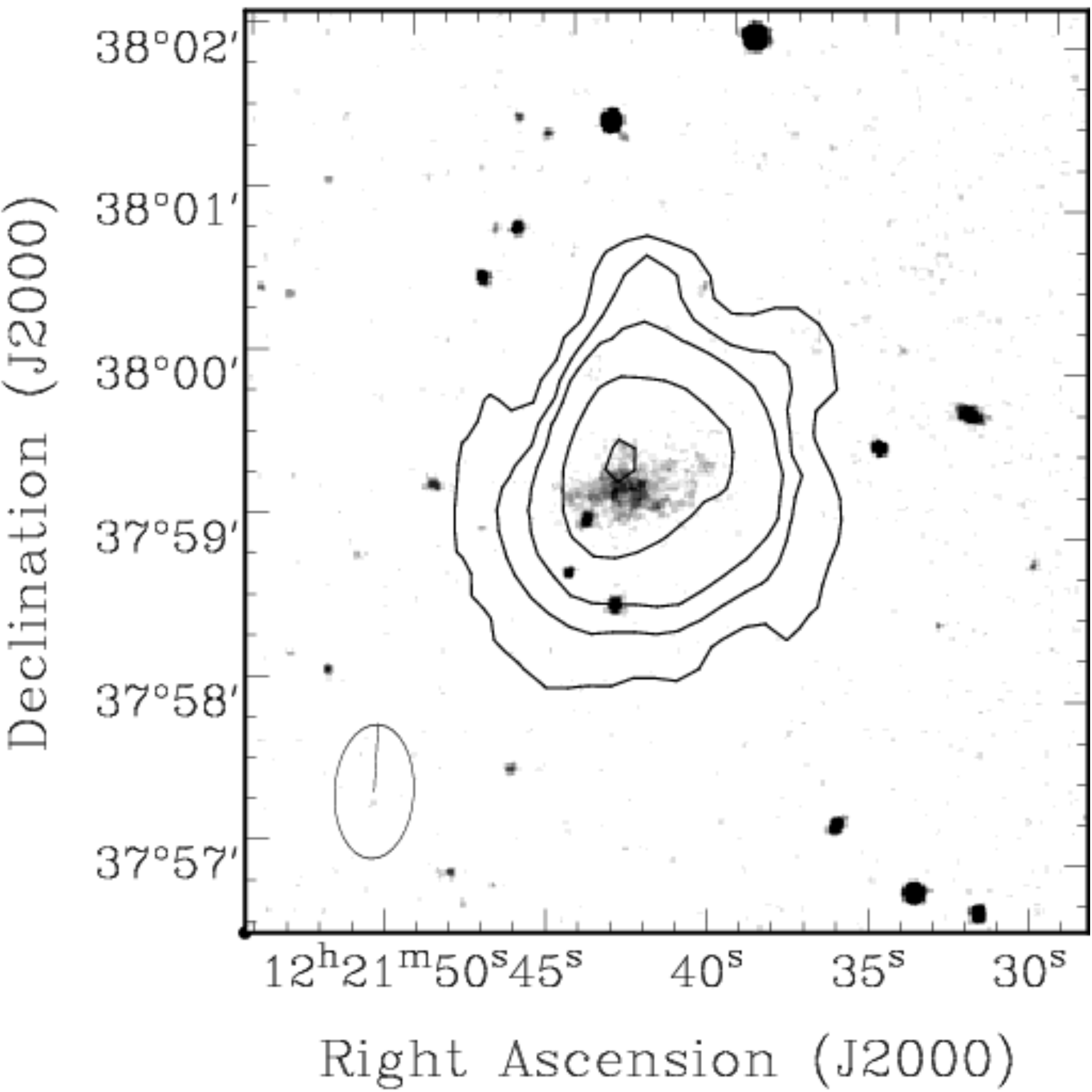}
\hskip 5mm
\includegraphics[height=0.17\textheight]{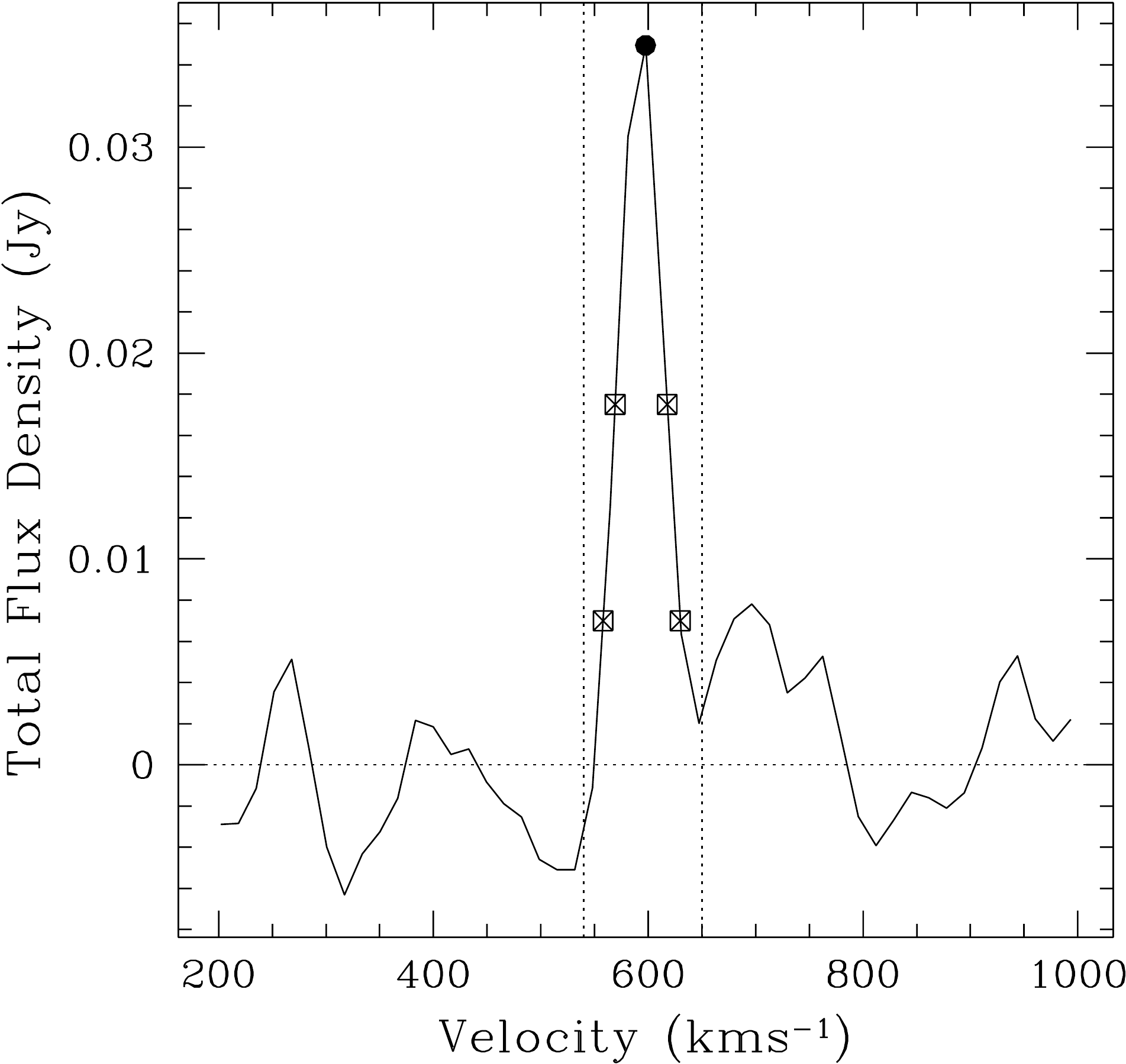}
\includegraphics[height=0.17\textheight]{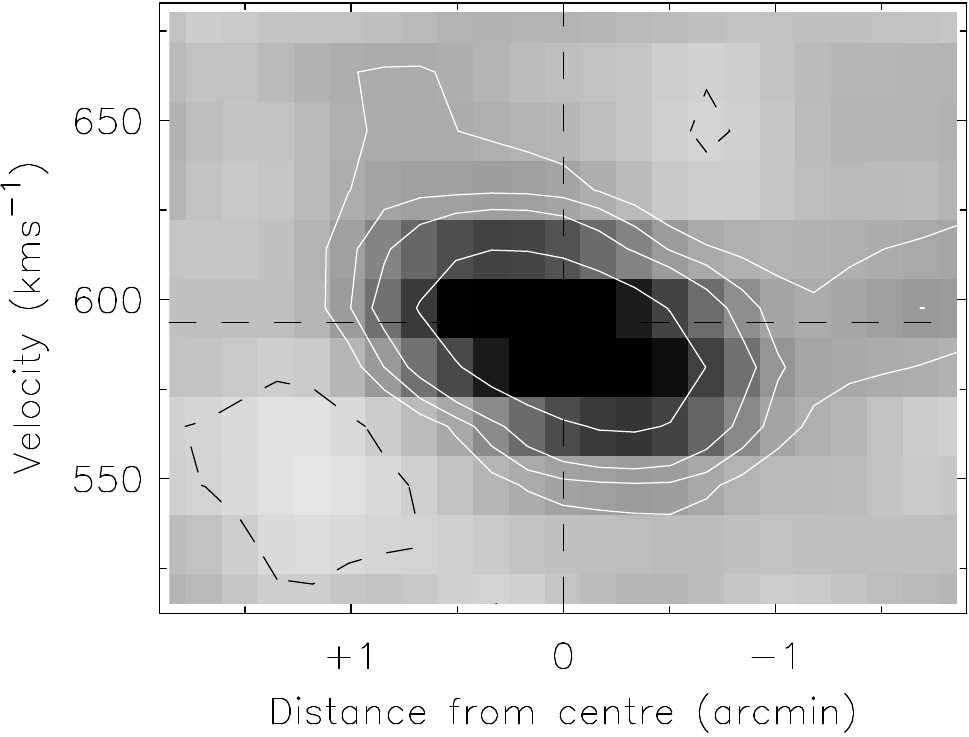}

\end{figure}

\clearpage

\addtocounter{figure}{-1}
\begin{figure}

\vskip 2mm
\centering
WSRT-CVn-29
\vskip 2mm
\includegraphics[width=0.25\textwidth]{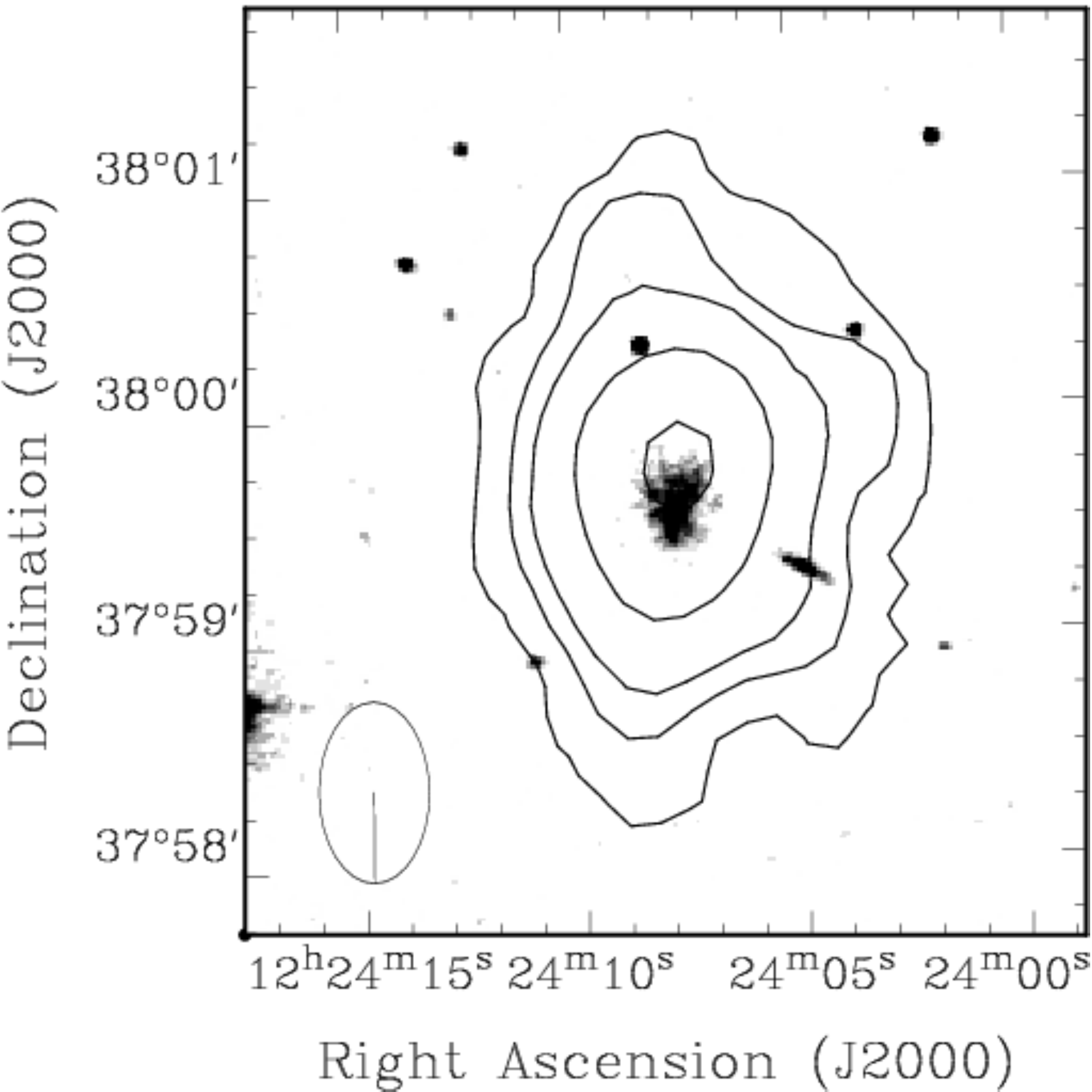}
\hskip 5mm
\includegraphics[height=0.17\textheight]{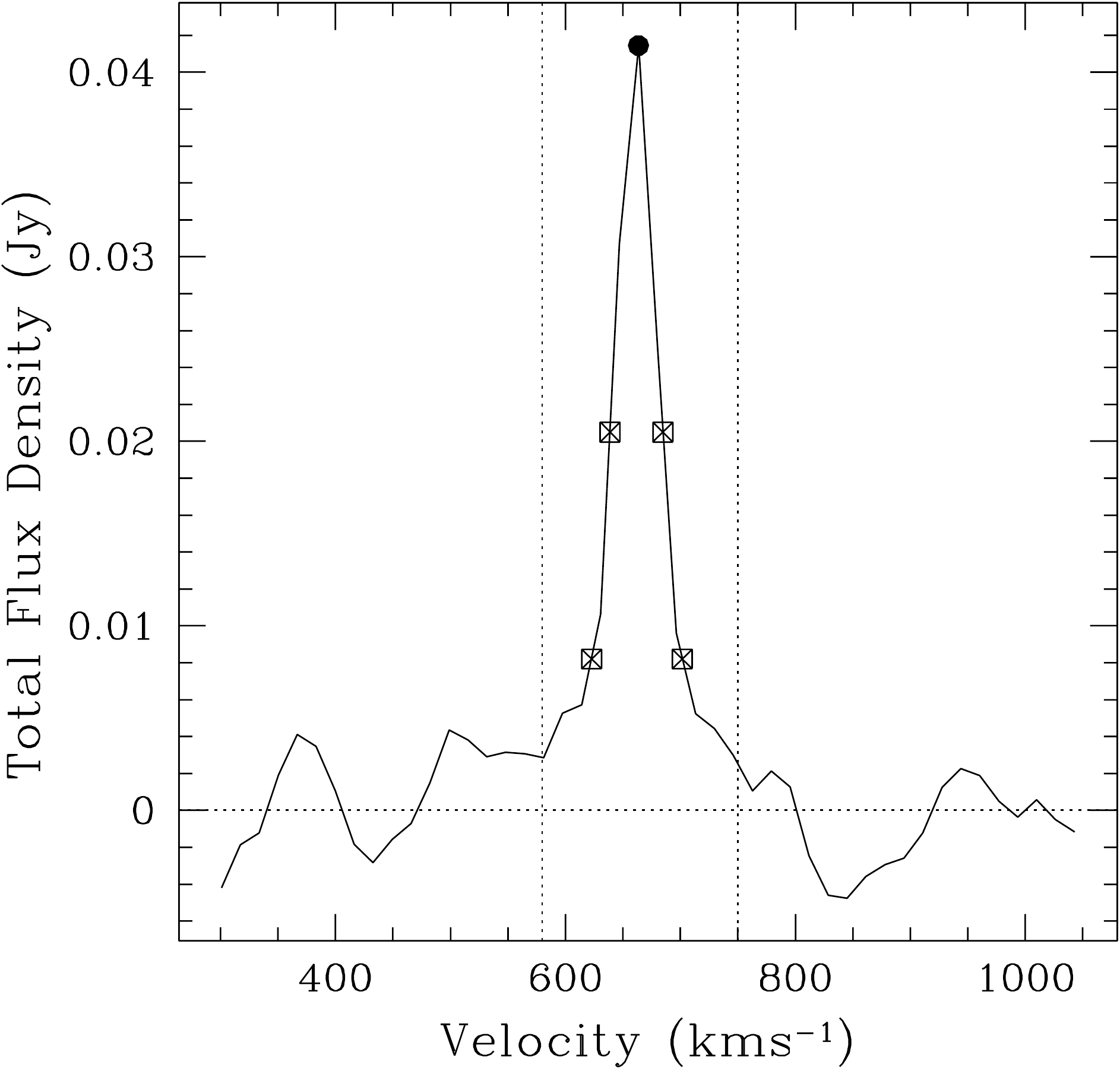}
\includegraphics[height=0.17\textheight]{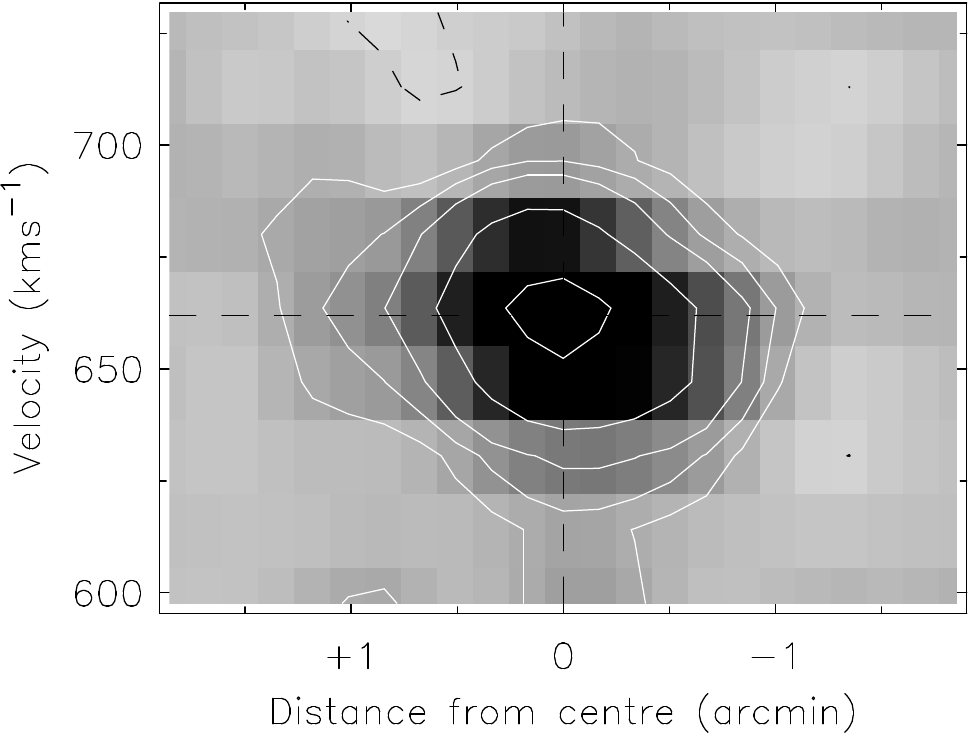}

\vskip 2mm
\centering
WSRT-CVn-30
\vskip 2mm
\includegraphics[width=0.25\textwidth]{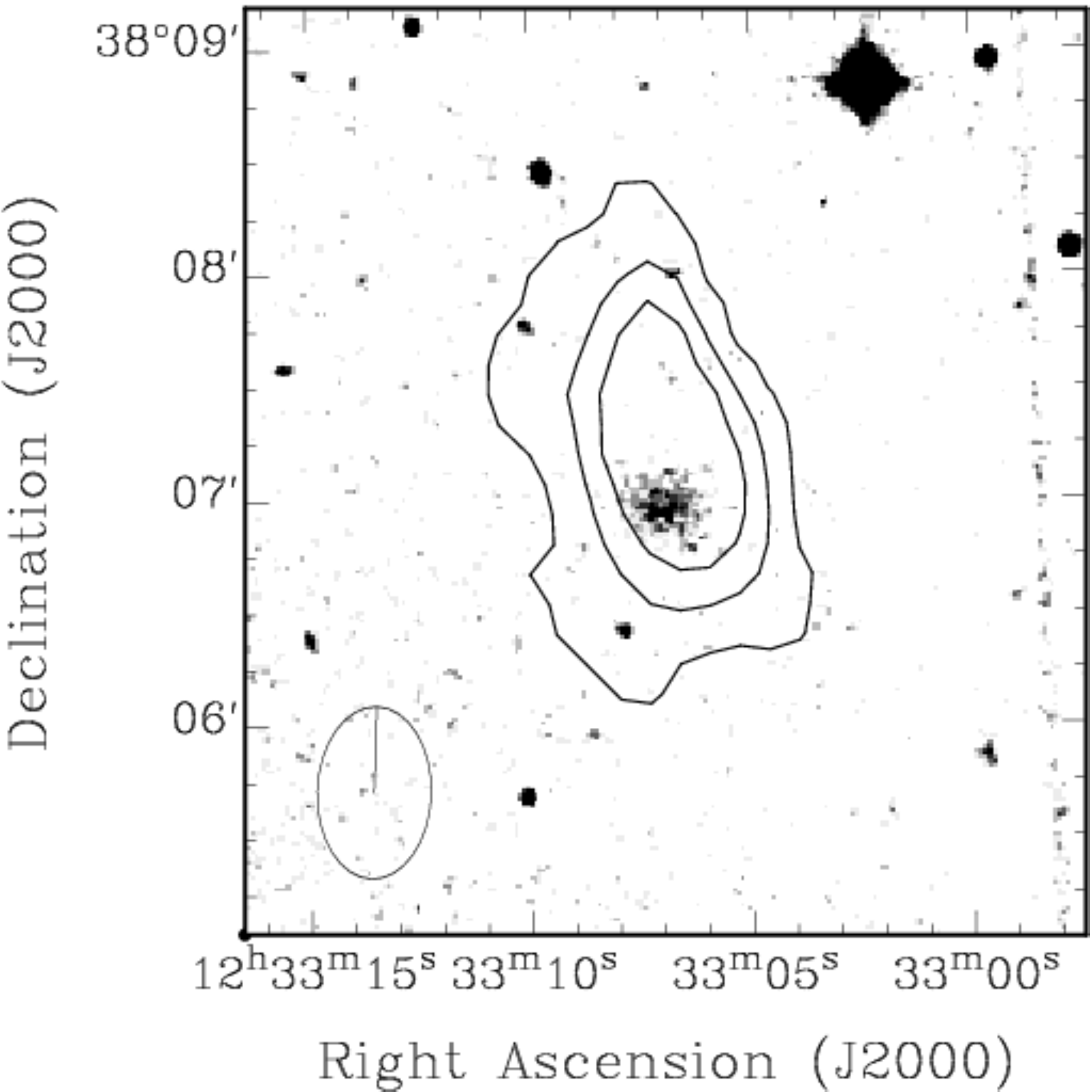}
\hskip 5mm
\includegraphics[height=0.17\textheight]{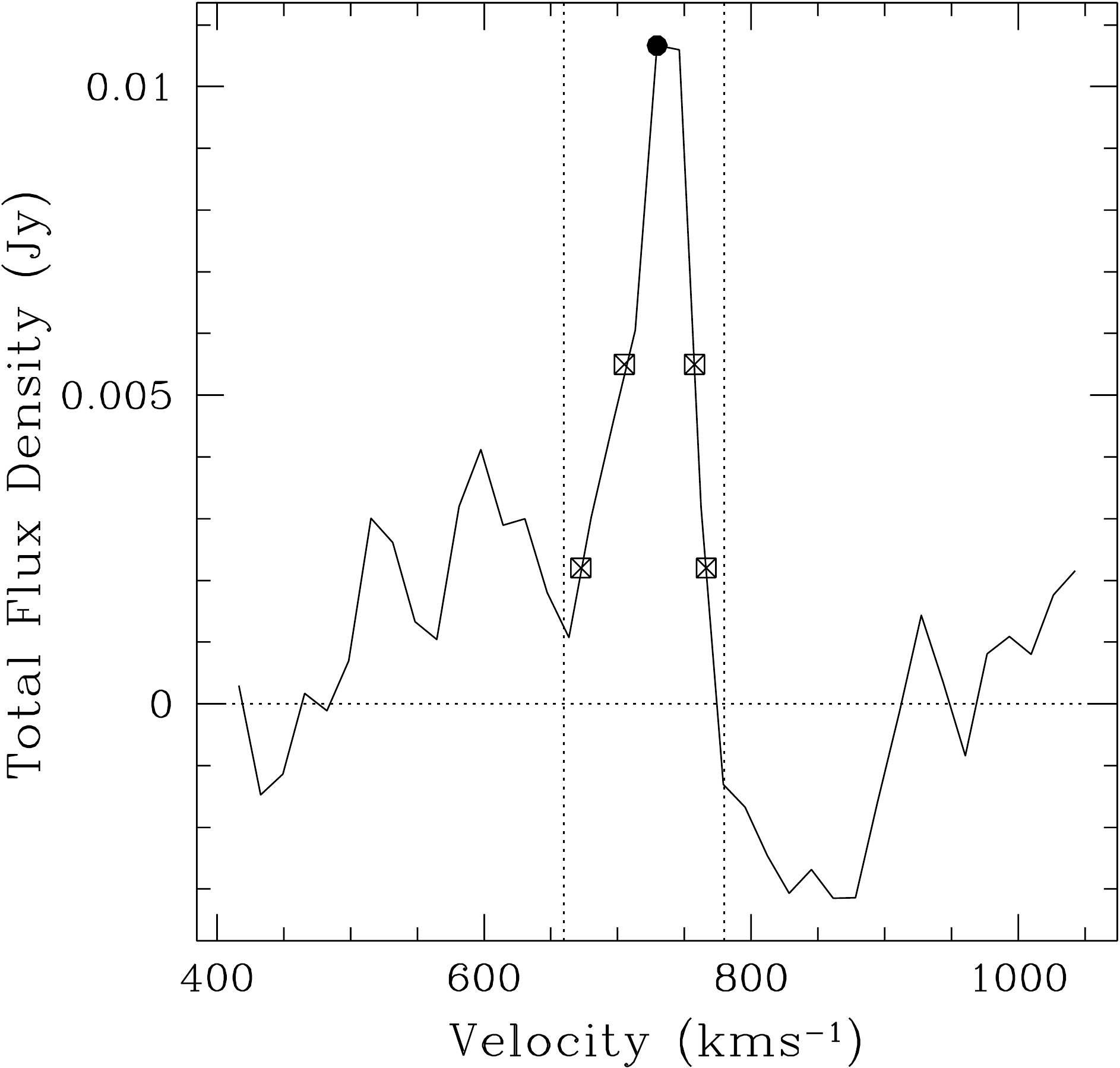}
\includegraphics[height=0.17\textheight]{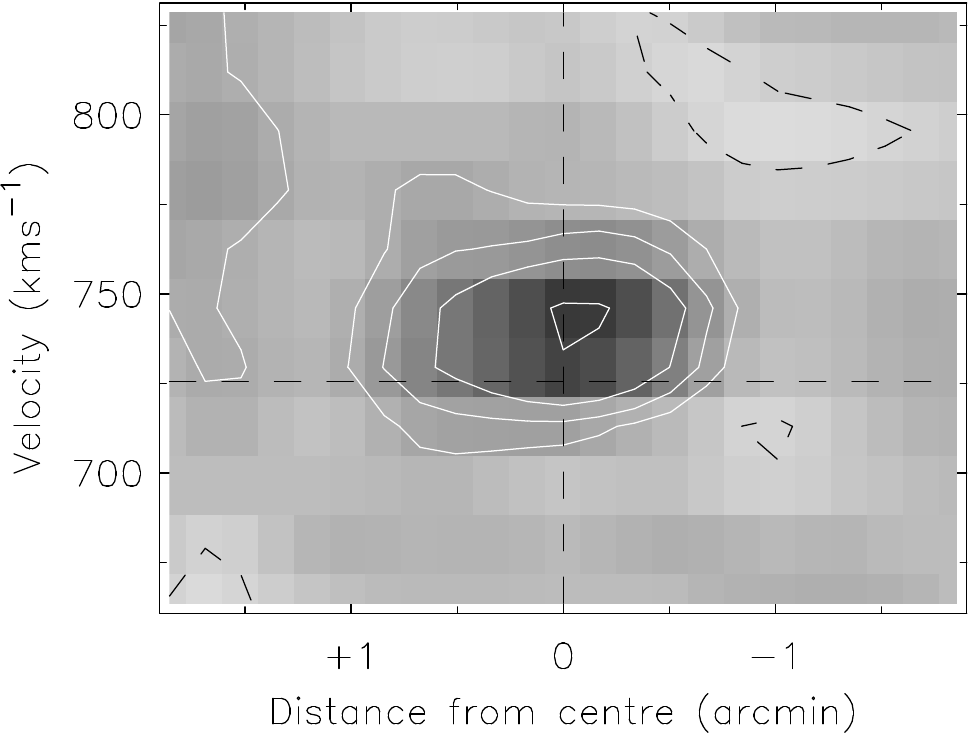}

\vskip 2mm
\centering
WSRT-CVn-31
\vskip 2mm
\includegraphics[width=0.25\textwidth]{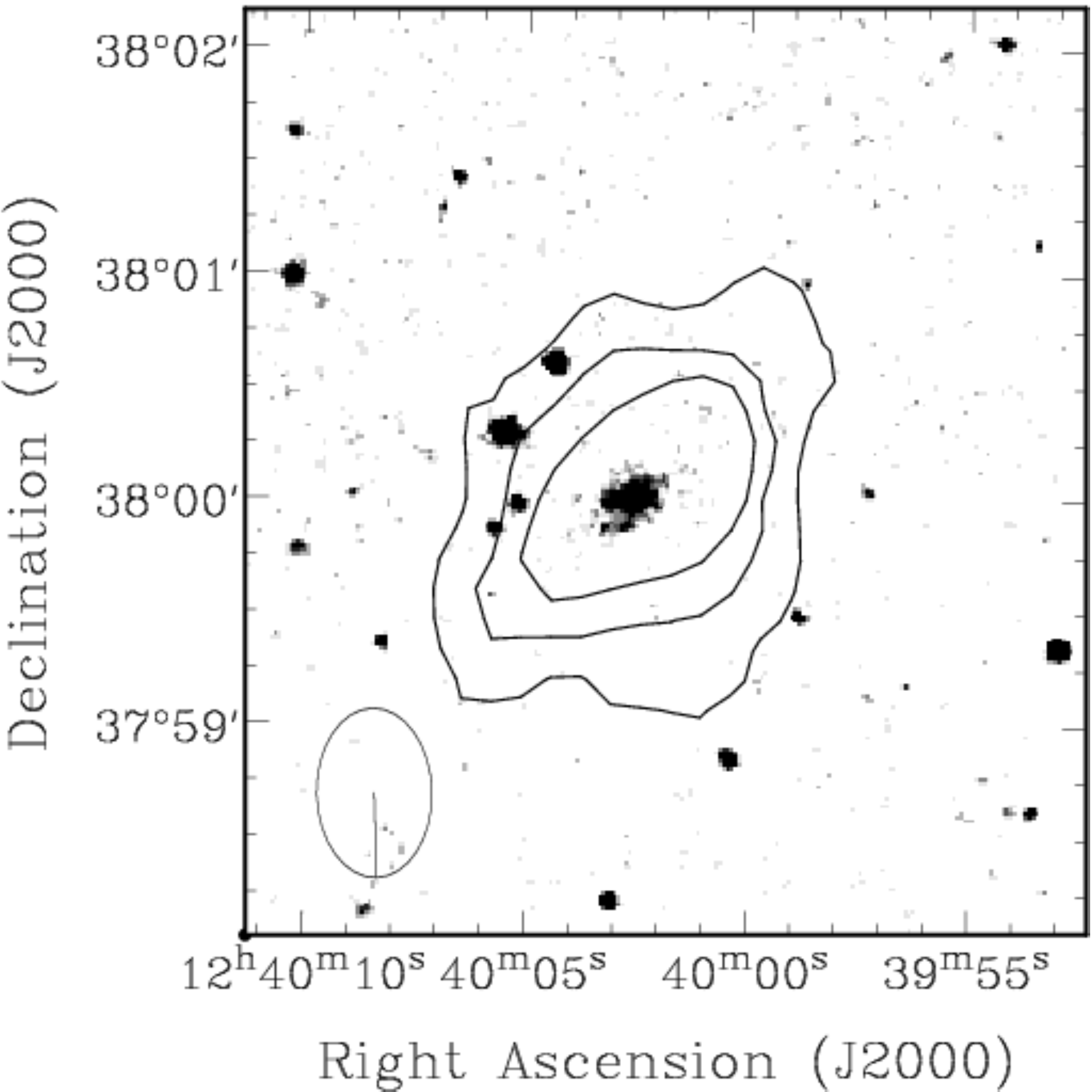}
\hskip 5mm
\includegraphics[height=0.17\textheight]{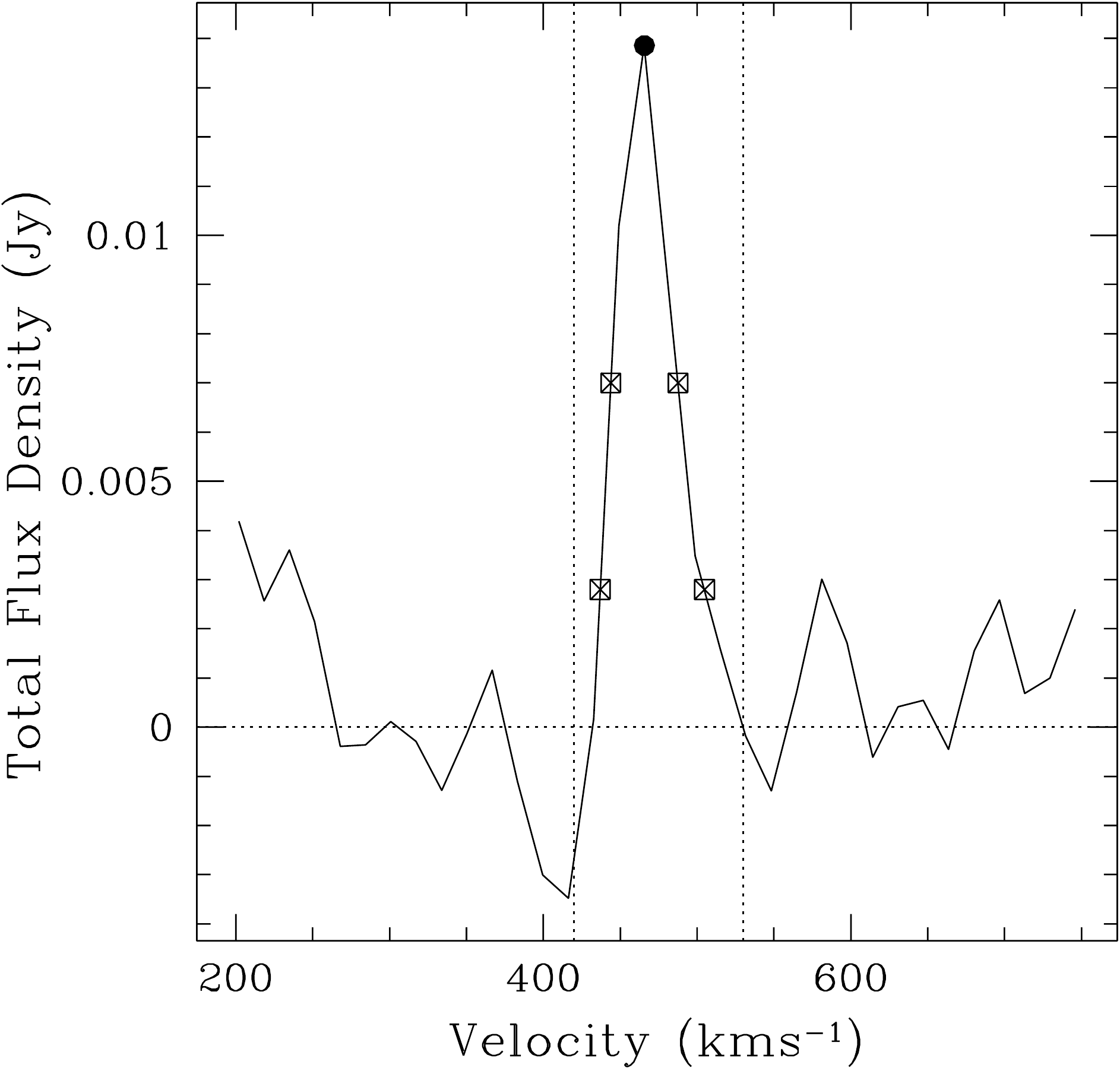}
\includegraphics[height=0.17\textheight]{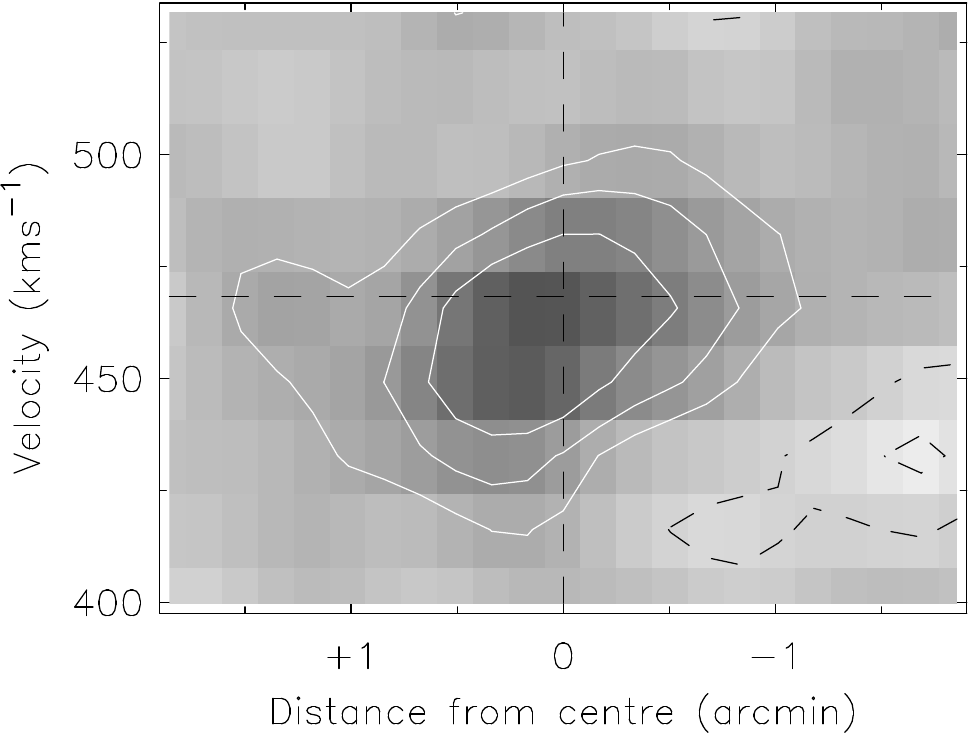}

\vskip 2mm
\centering
WSRT-CVn-32
\vskip 2mm
\includegraphics[width=0.25\textwidth]{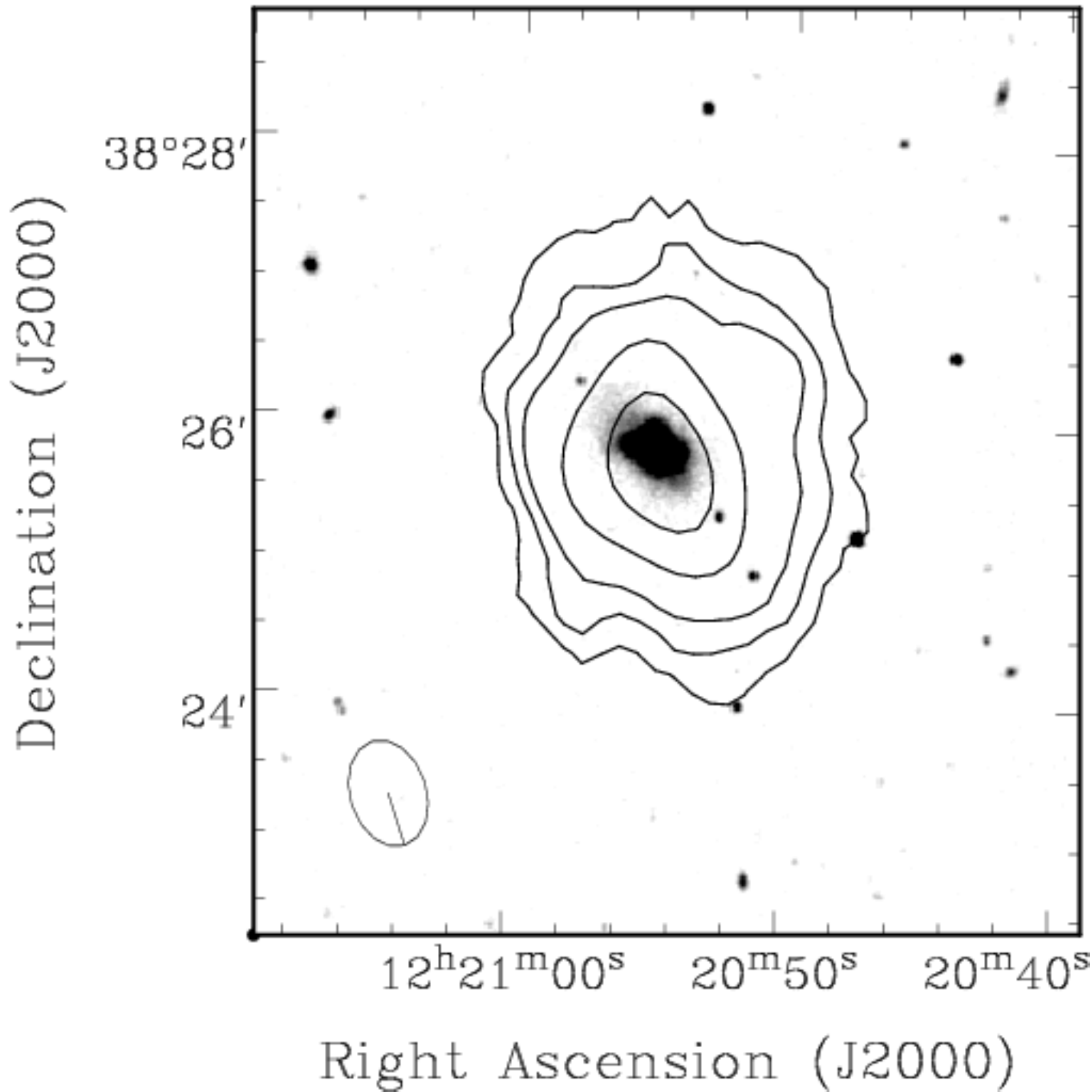}
\hskip 5mm
\includegraphics[height=0.17\textheight]{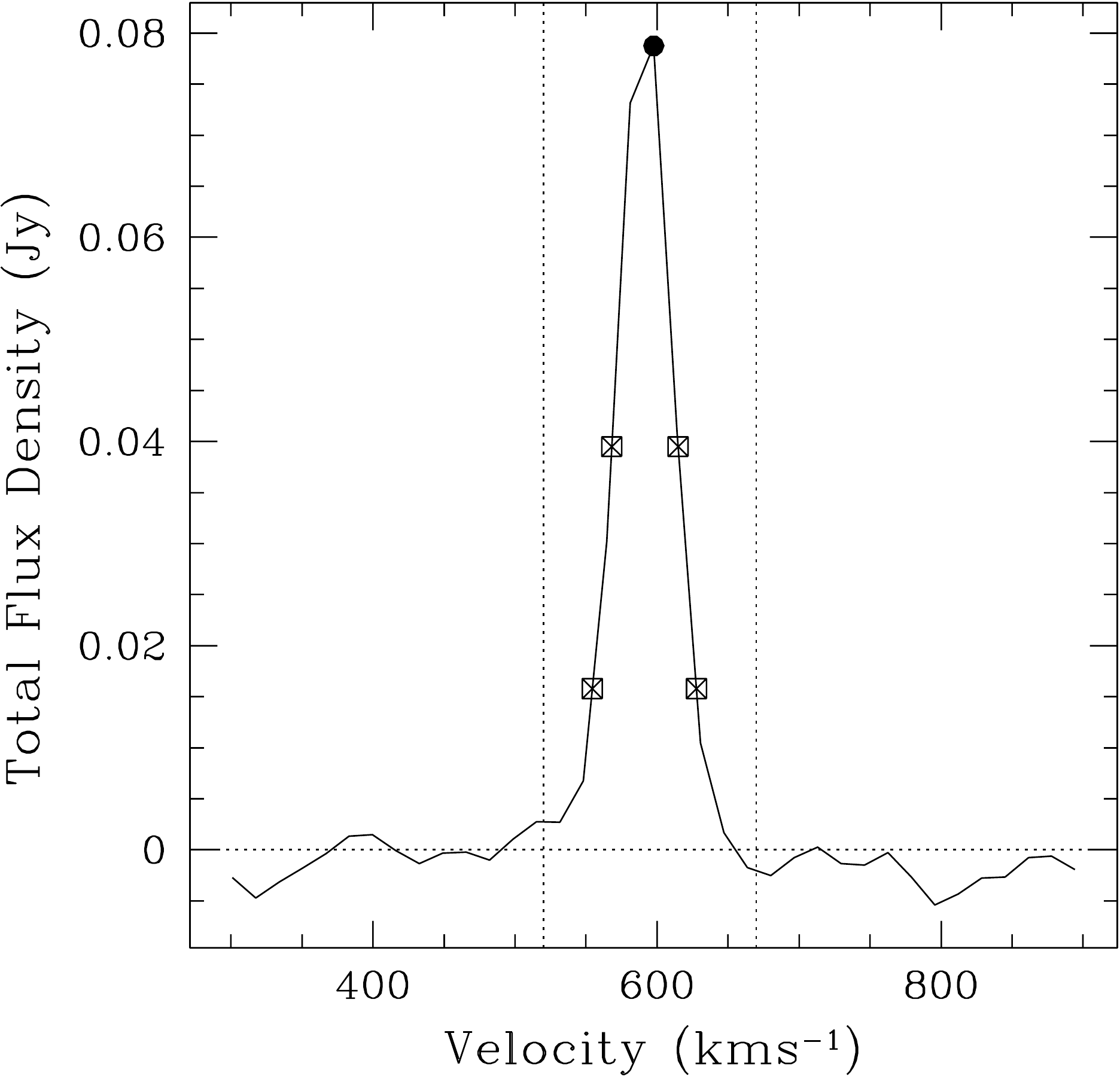}
\includegraphics[height=0.17\textheight]{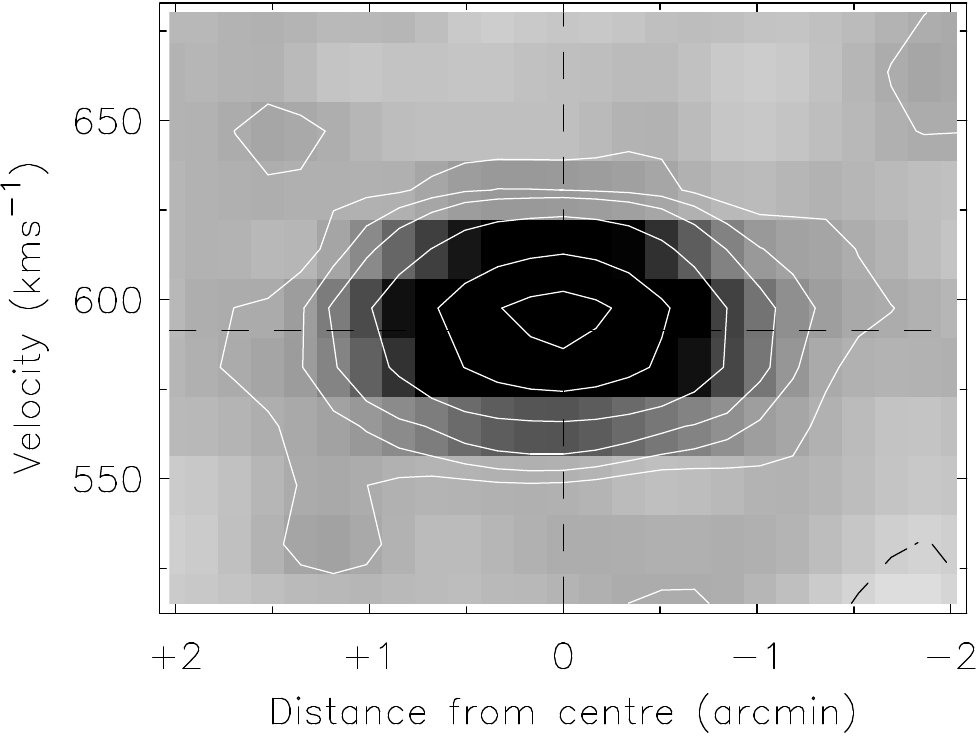}

\end{figure}

\clearpage

\addtocounter{figure}{-1}
\begin{figure}

\vskip 2mm
\centering
WSRT-CVn-33
\vskip 2mm
\includegraphics[width=0.25\textwidth]{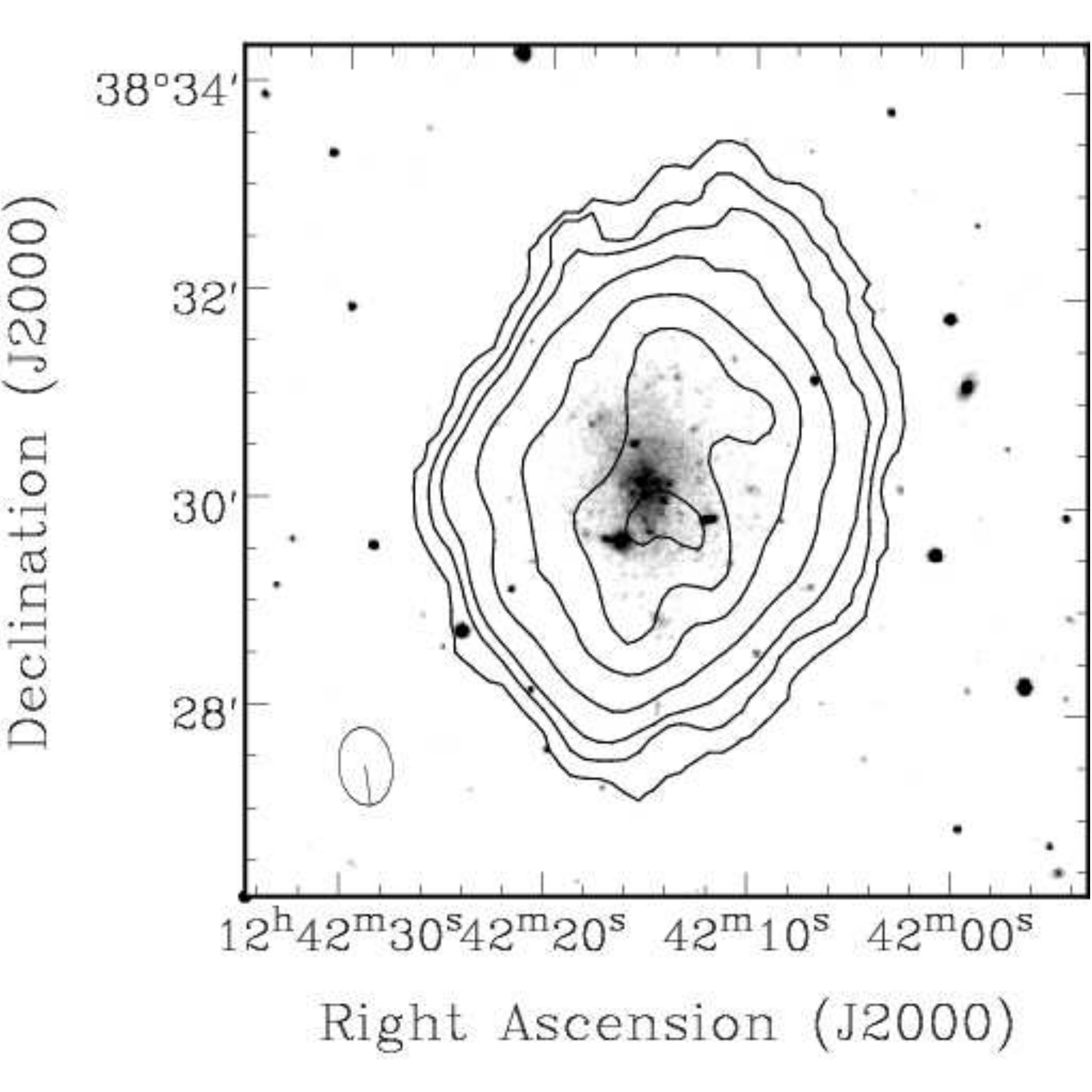}
\hskip 5mm
\includegraphics[height=0.17\textheight]{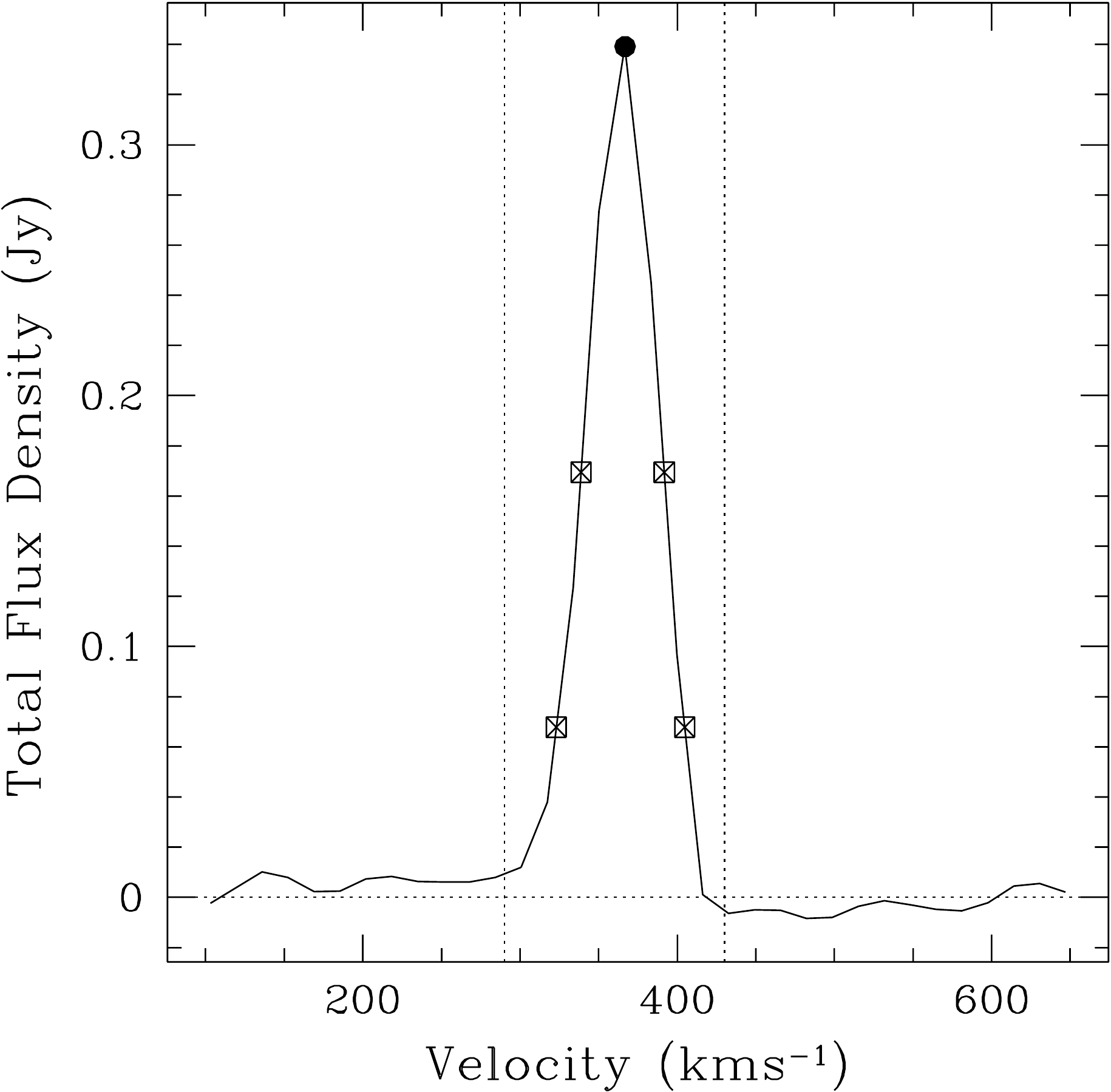}
\includegraphics[height=0.17\textheight]{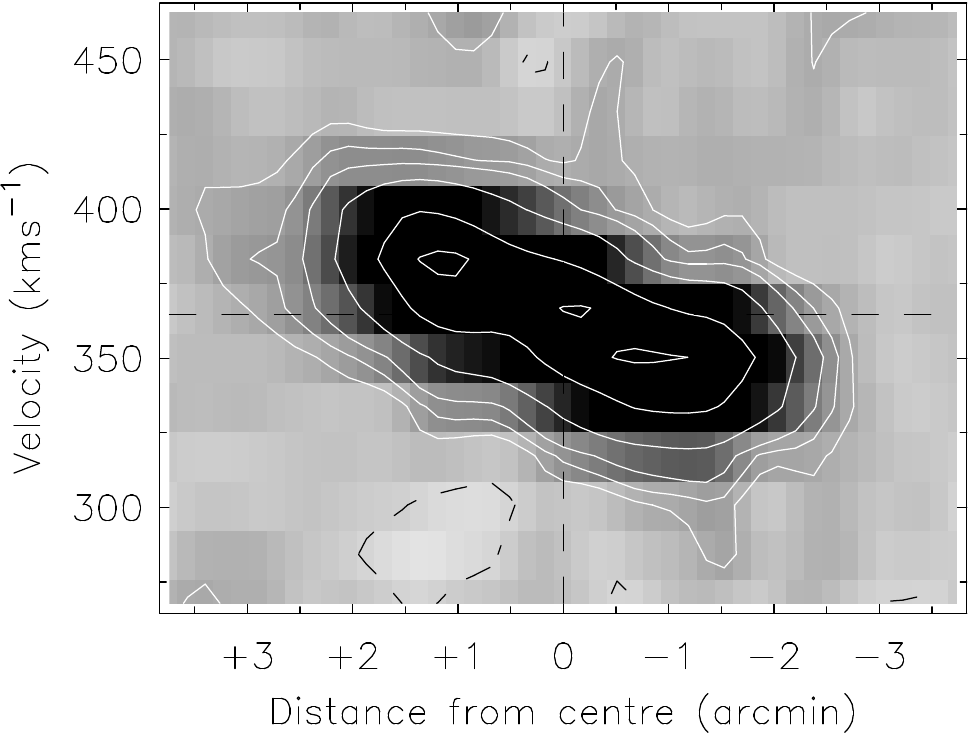}

\vskip 2mm
\centering
WSRT-CVn-34
\vskip 2mm
\includegraphics[width=0.25\textwidth]{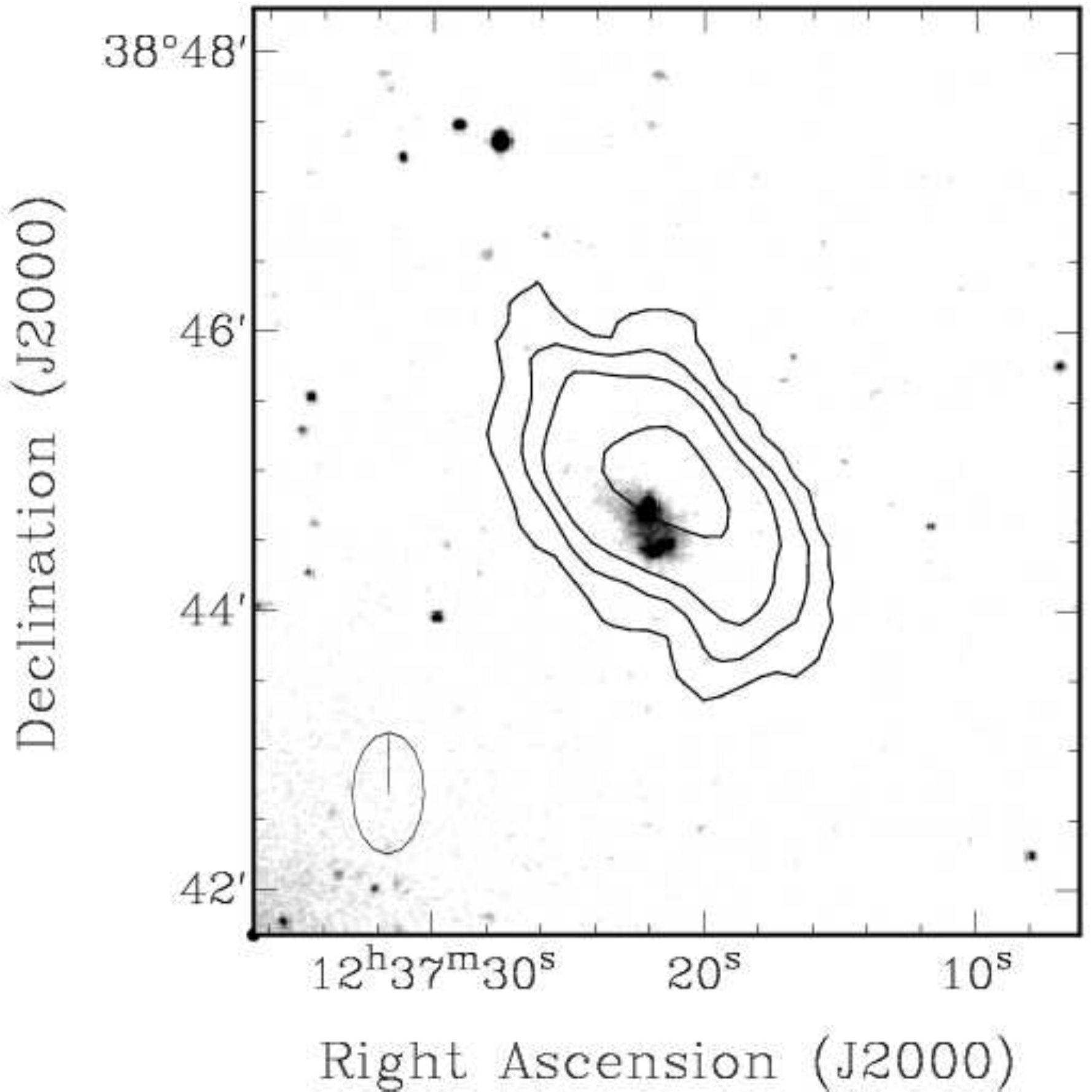}
\hskip 5mm
\includegraphics[height=0.17\textheight]{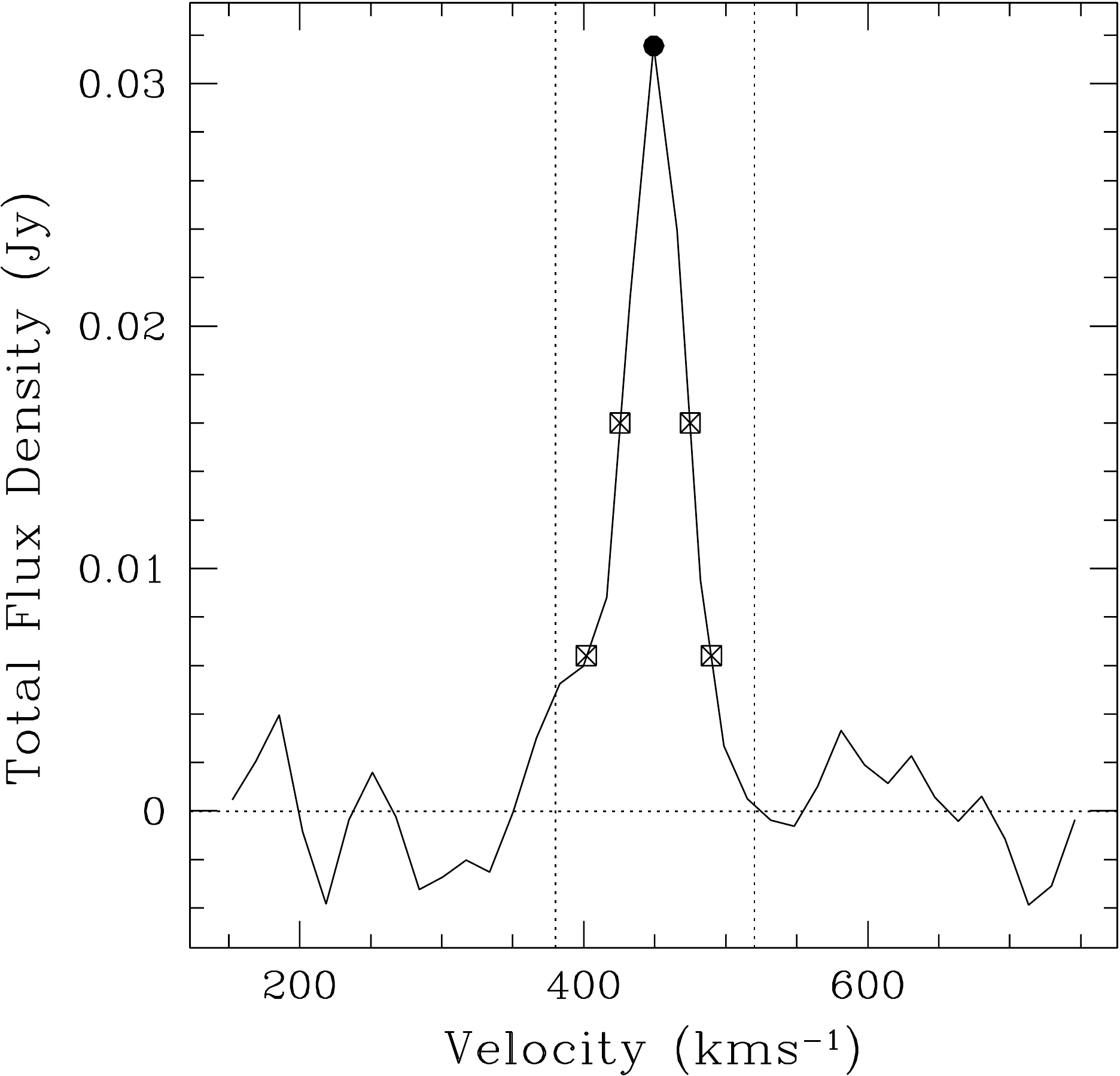}
\includegraphics[height=0.17\textheight]{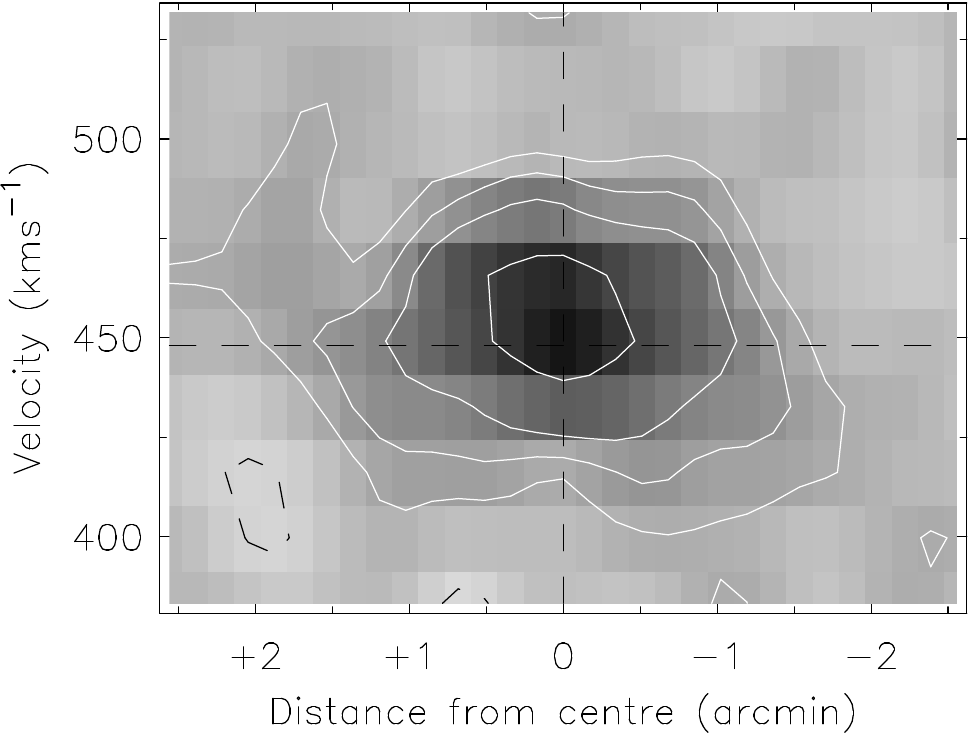}

\vskip 2mm
\centering
WSRT-CVn-35
\vskip 2mm
\includegraphics[width=0.25\textwidth]{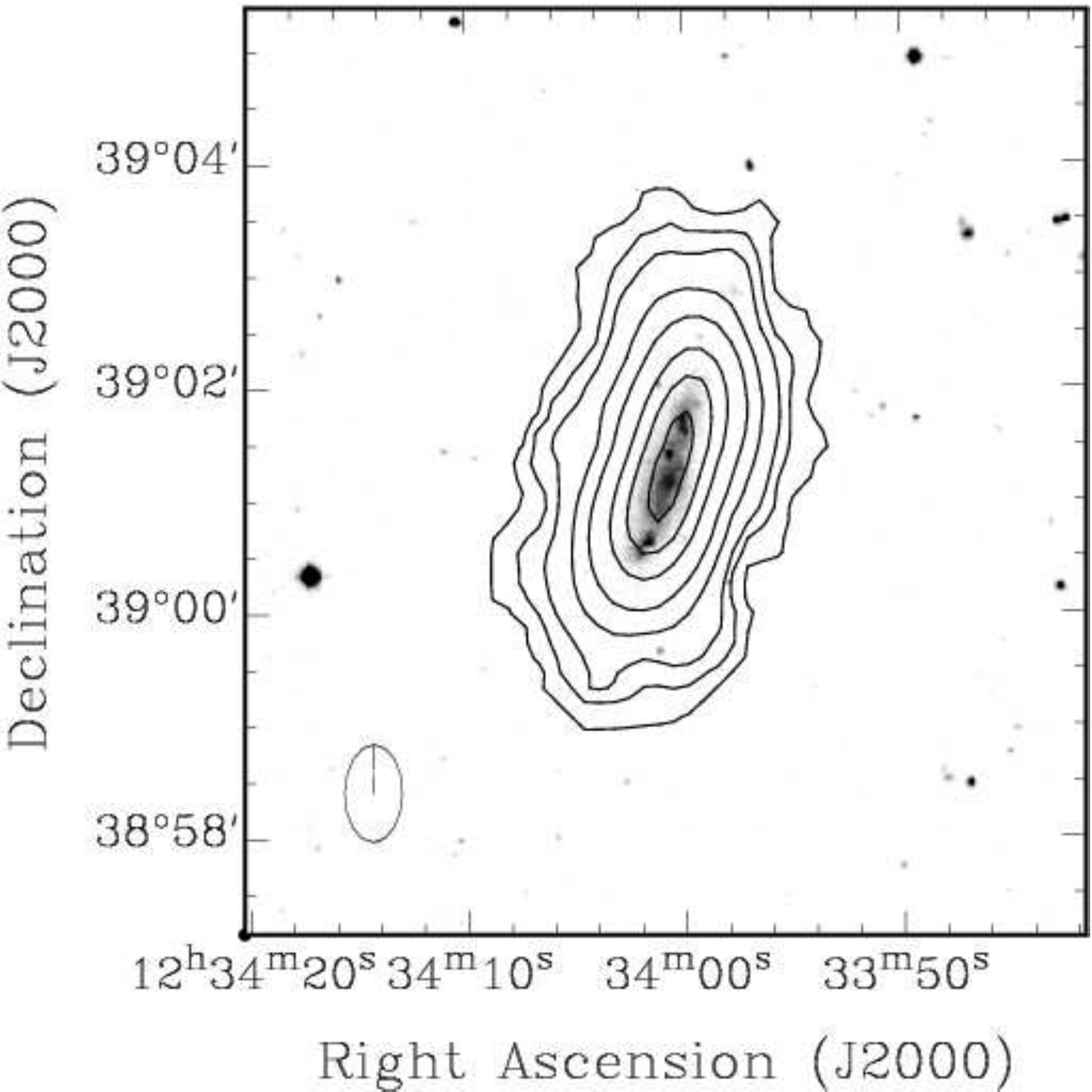}
\hskip 5mm
\includegraphics[height=0.17\textheight]{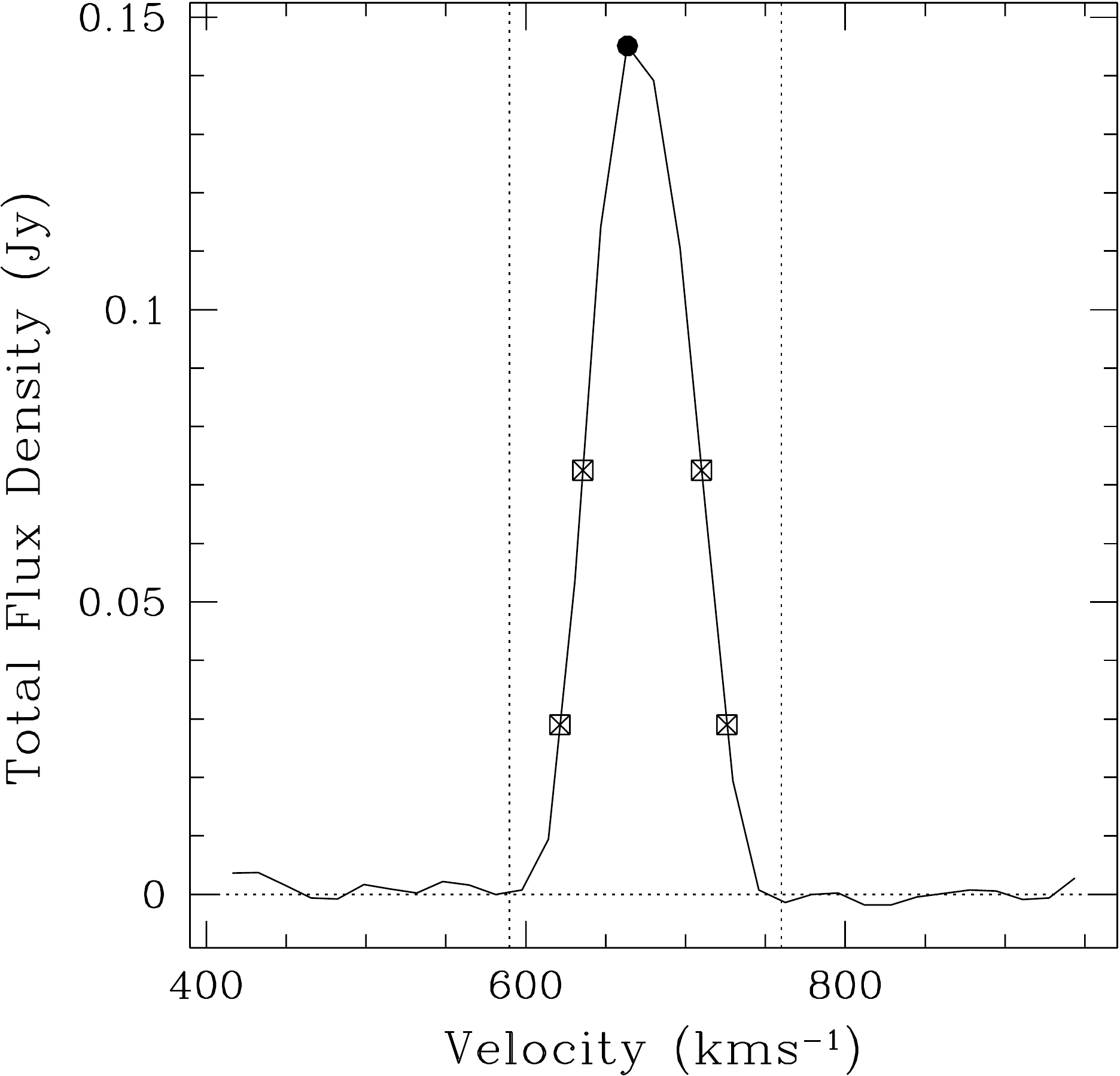}
\includegraphics[height=0.17\textheight]{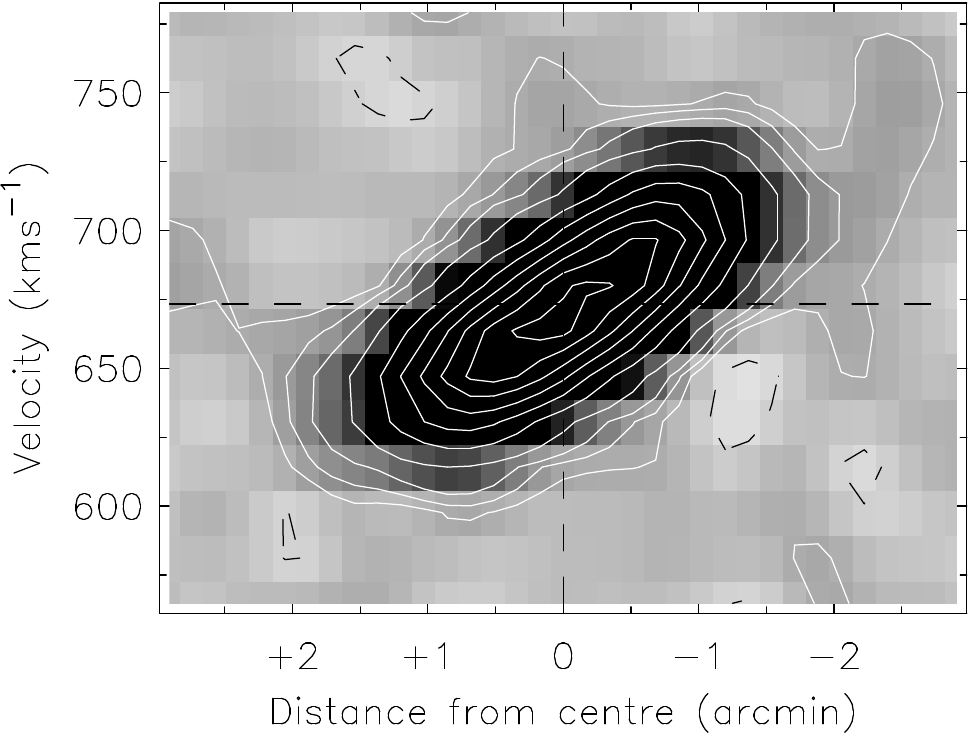}

\vskip 2mm
\centering
WSRT-CVn-36
\vskip 2mm
\includegraphics[width=0.25\textwidth]{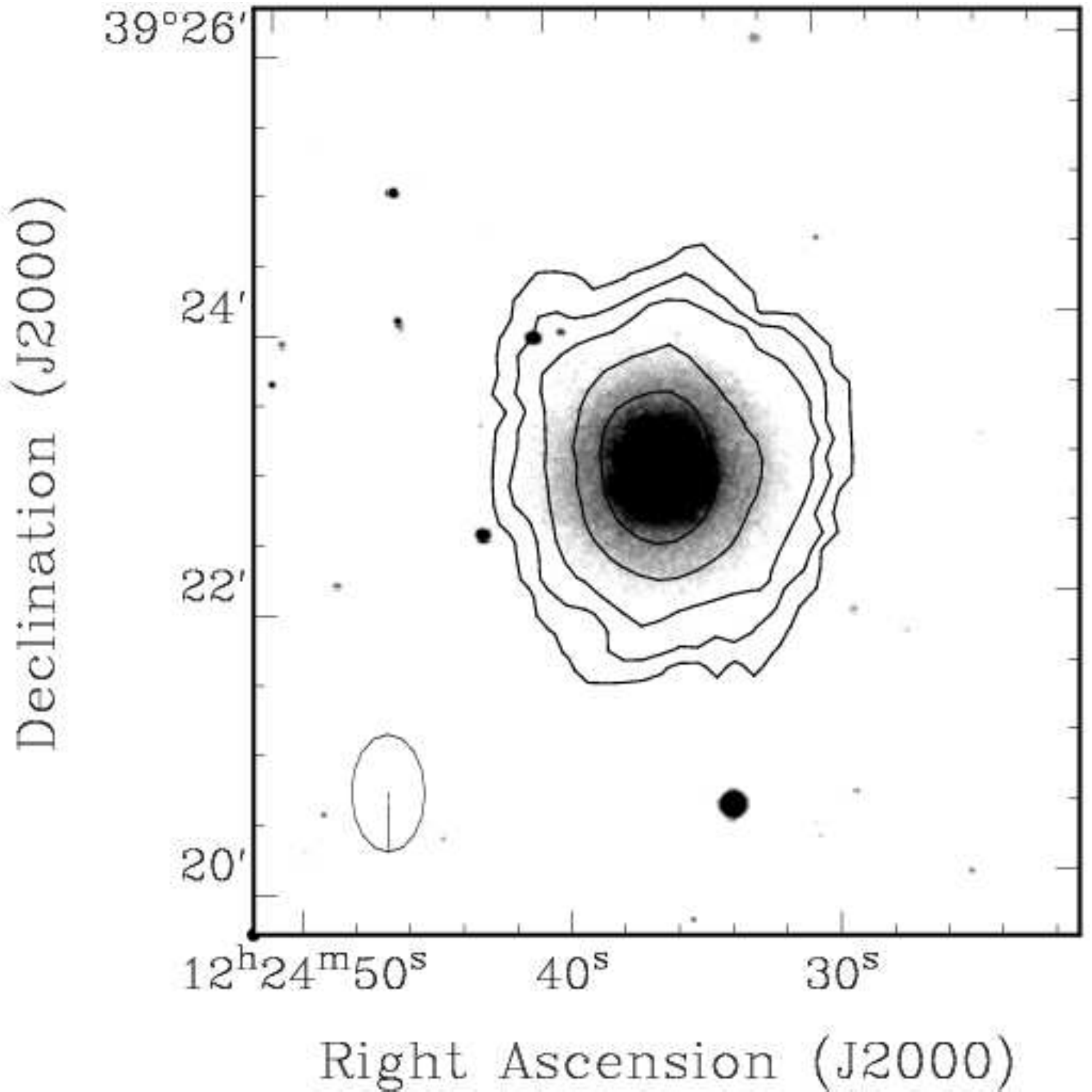}
\hskip 5mm
\includegraphics[height=0.17\textheight]{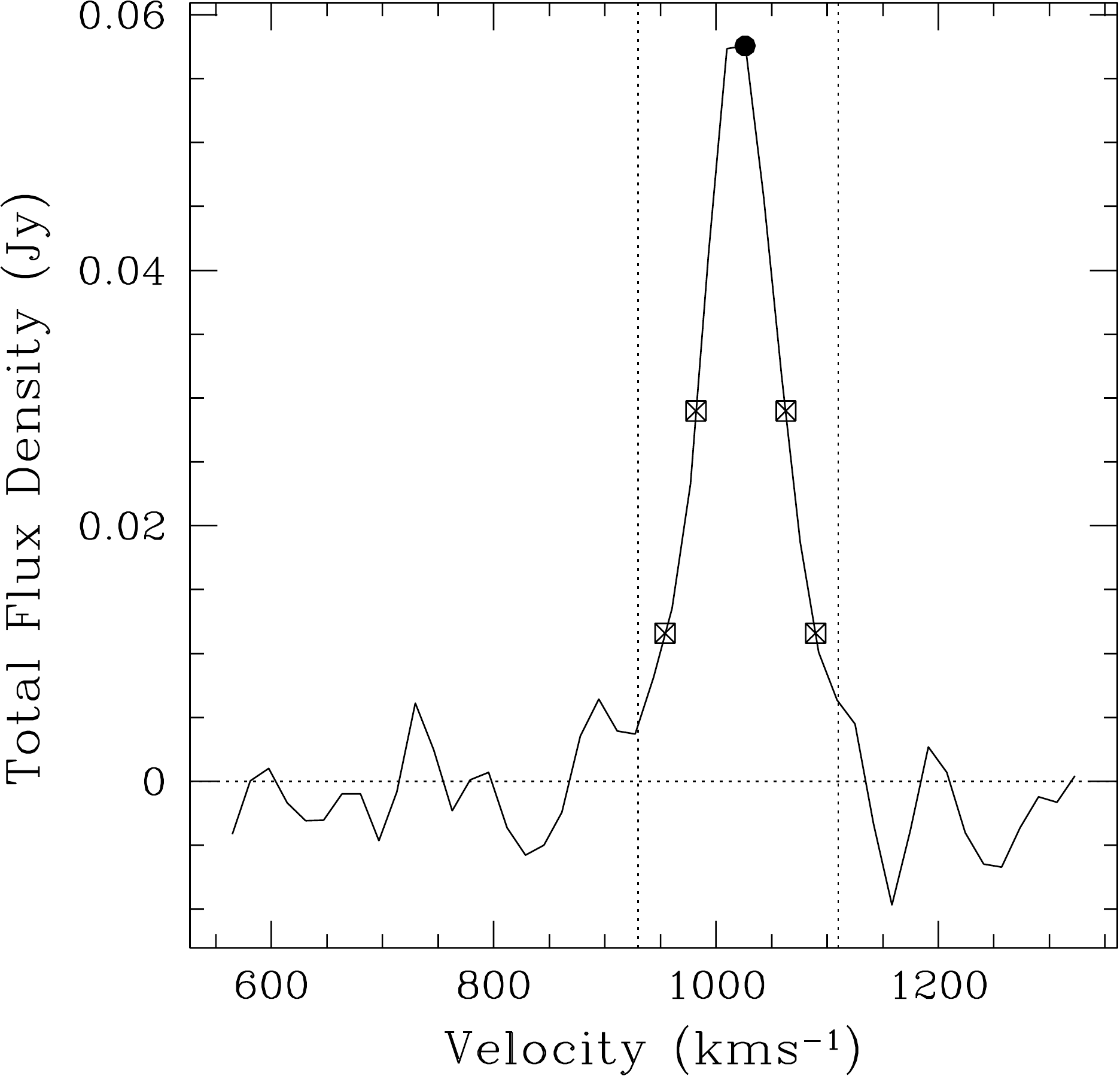}
\includegraphics[height=0.17\textheight]{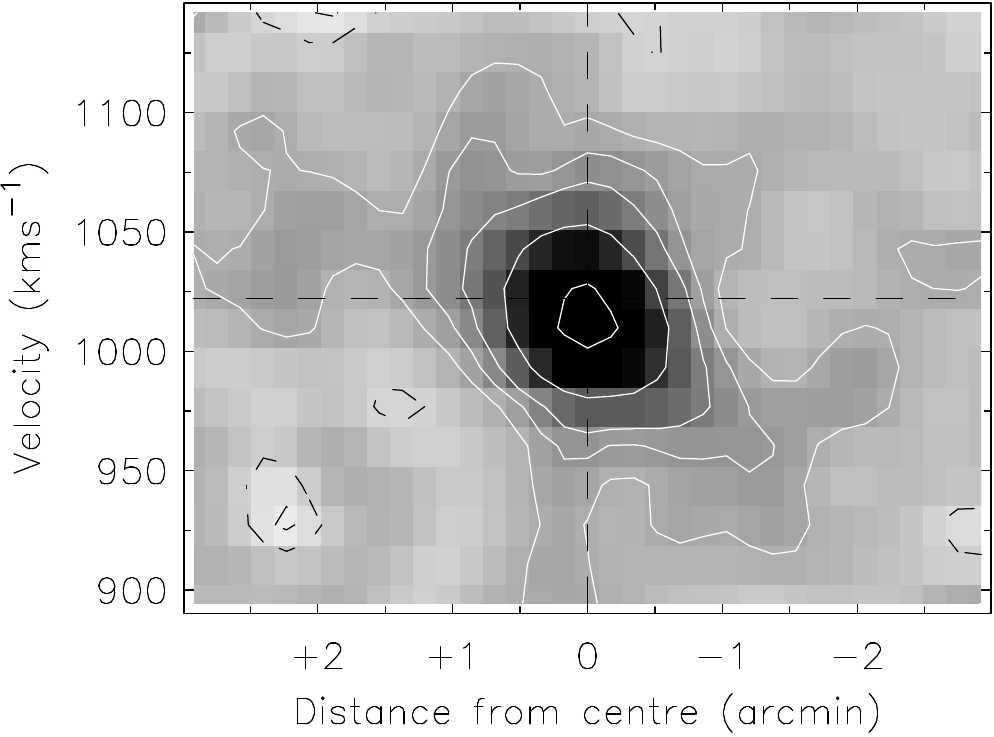}

\end{figure}

\clearpage

\addtocounter{figure}{-1}
\begin{figure}

\vskip 2mm
\centering
WSRT-CVn-37
\vskip 2mm
\includegraphics[width=0.25\textwidth]{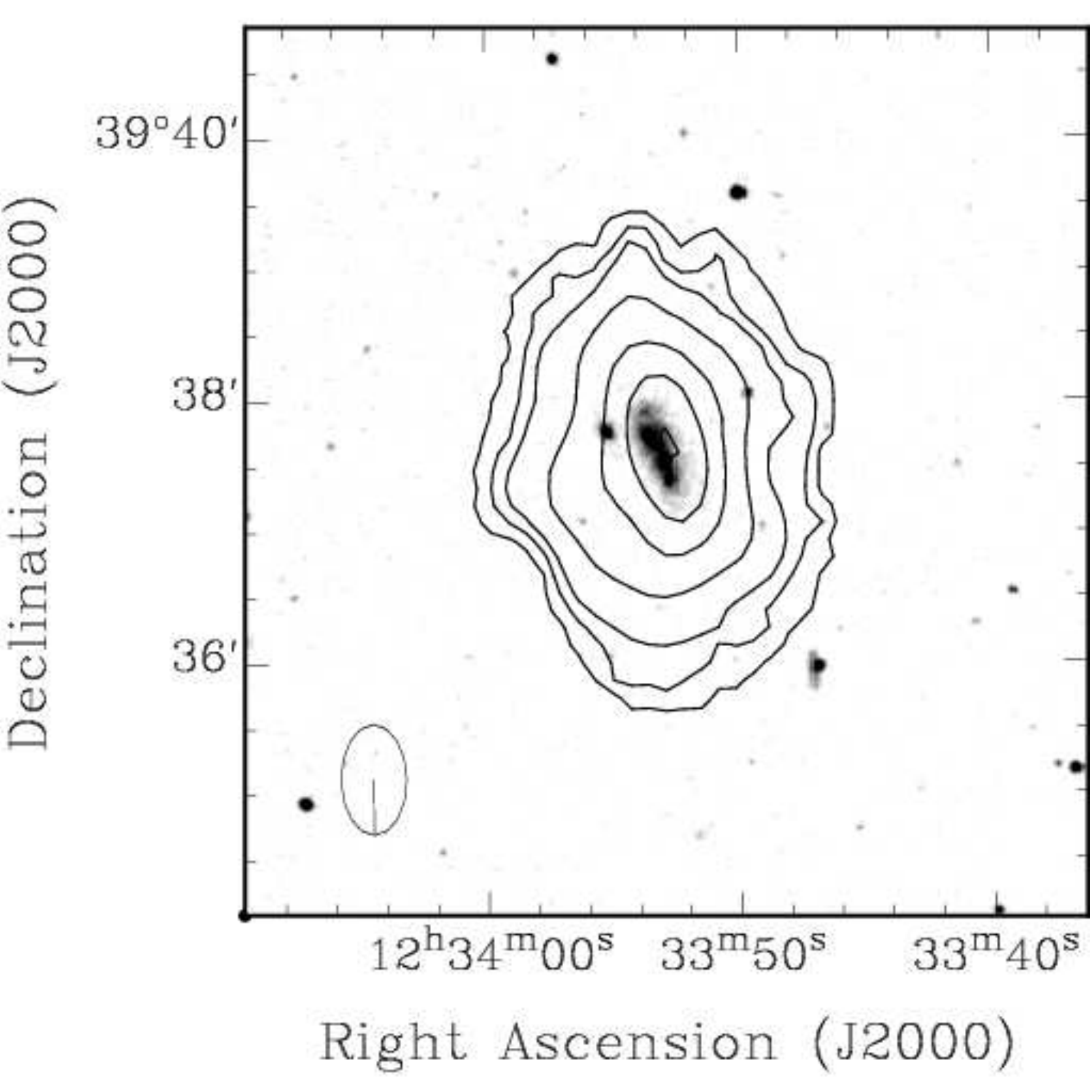}
\hskip 5mm
\includegraphics[height=0.17\textheight]{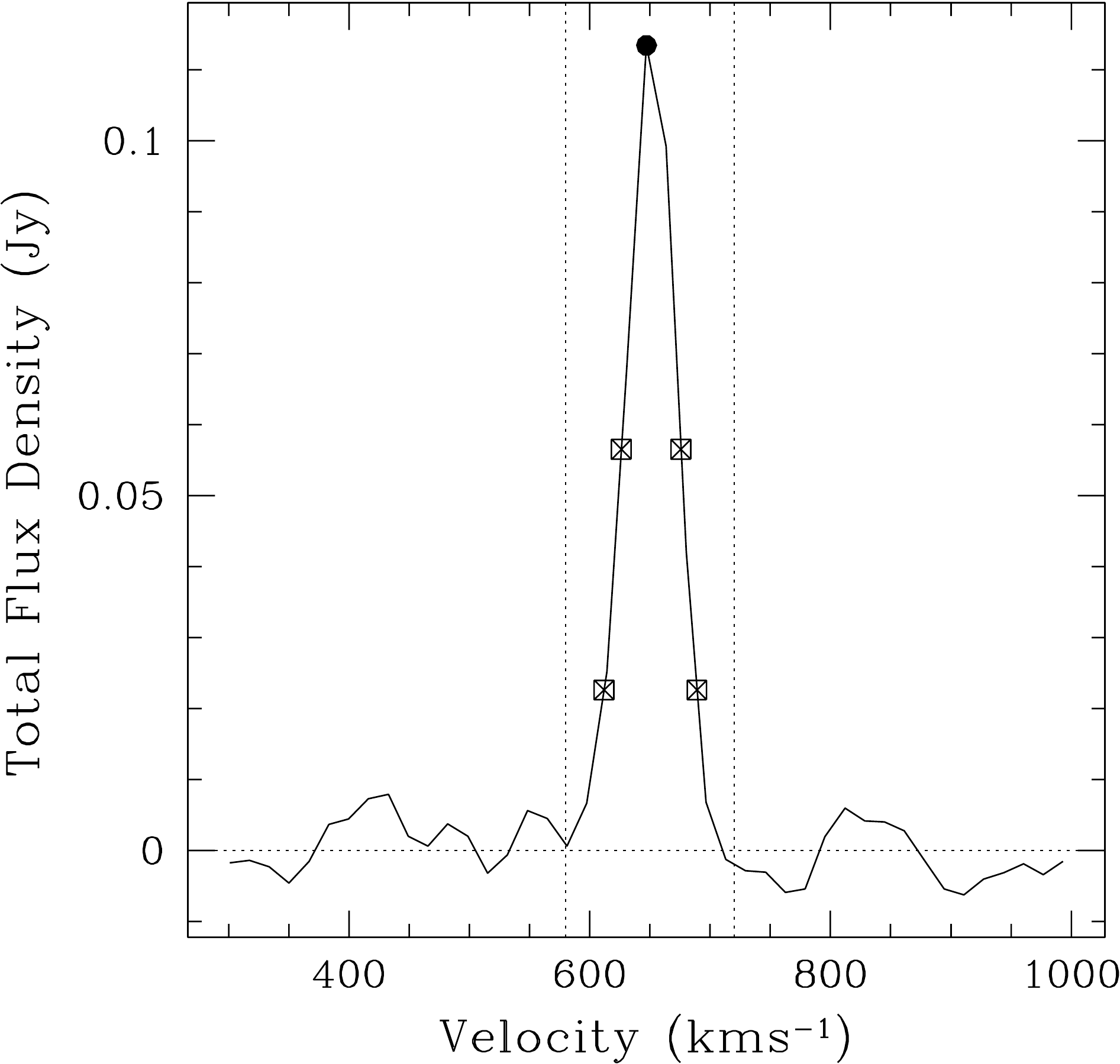}
\includegraphics[height=0.17\textheight]{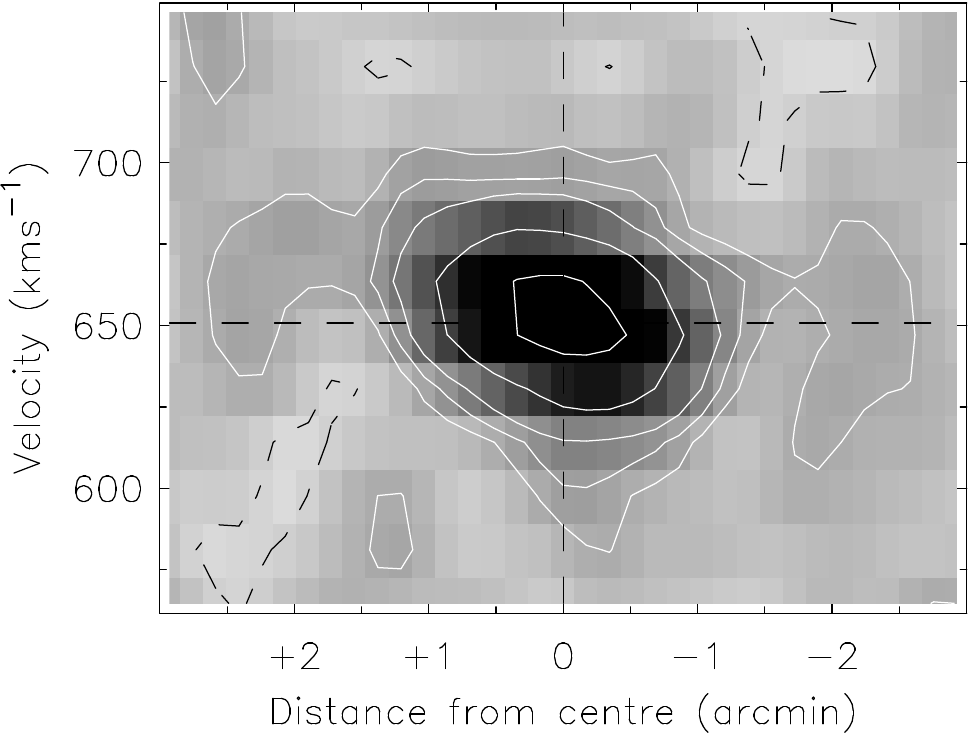}

\vskip 2mm
\centering
WSRT-CVn-38
\vskip 2mm
\includegraphics[width=0.25\textwidth]{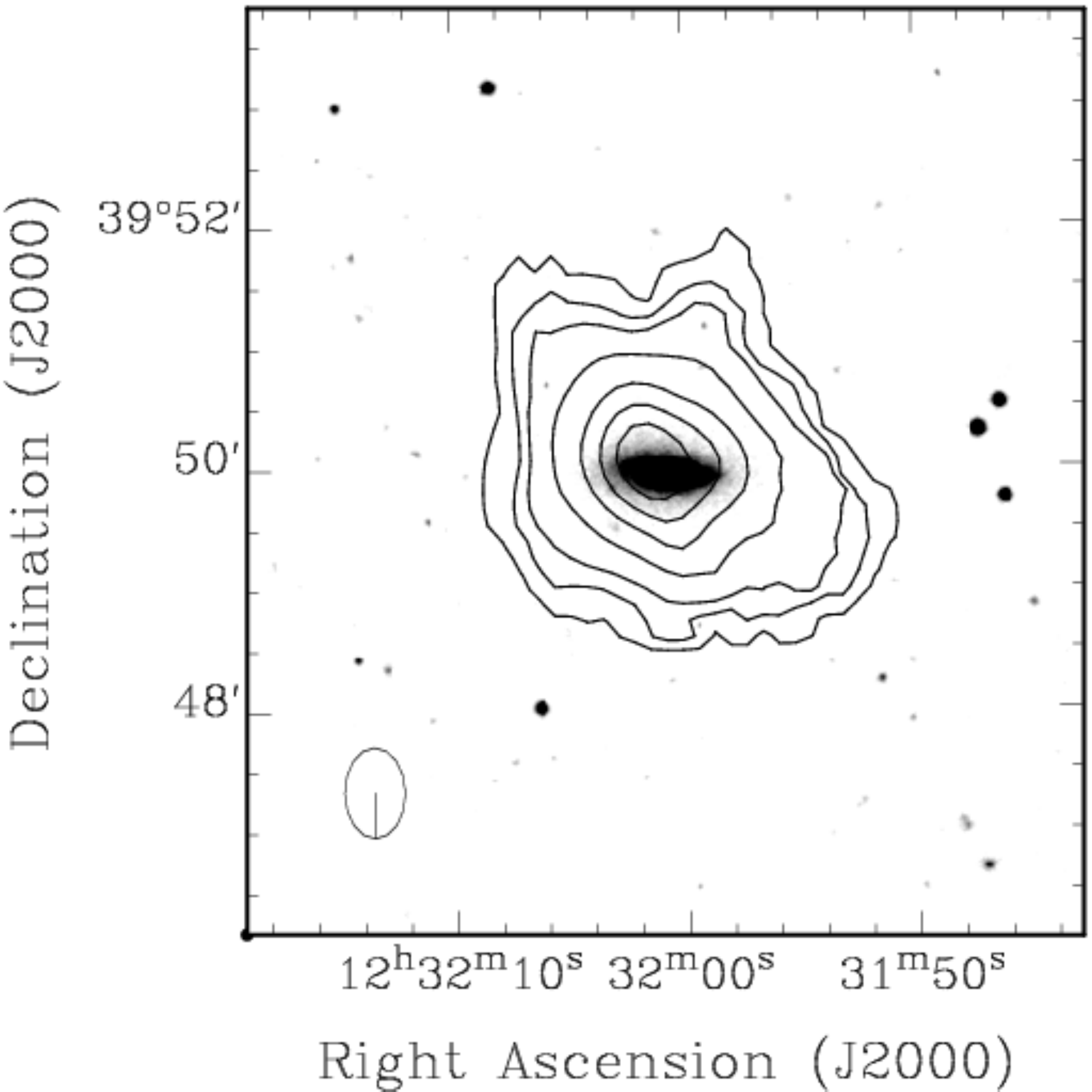}
\hskip 5mm
\includegraphics[height=0.17\textheight]{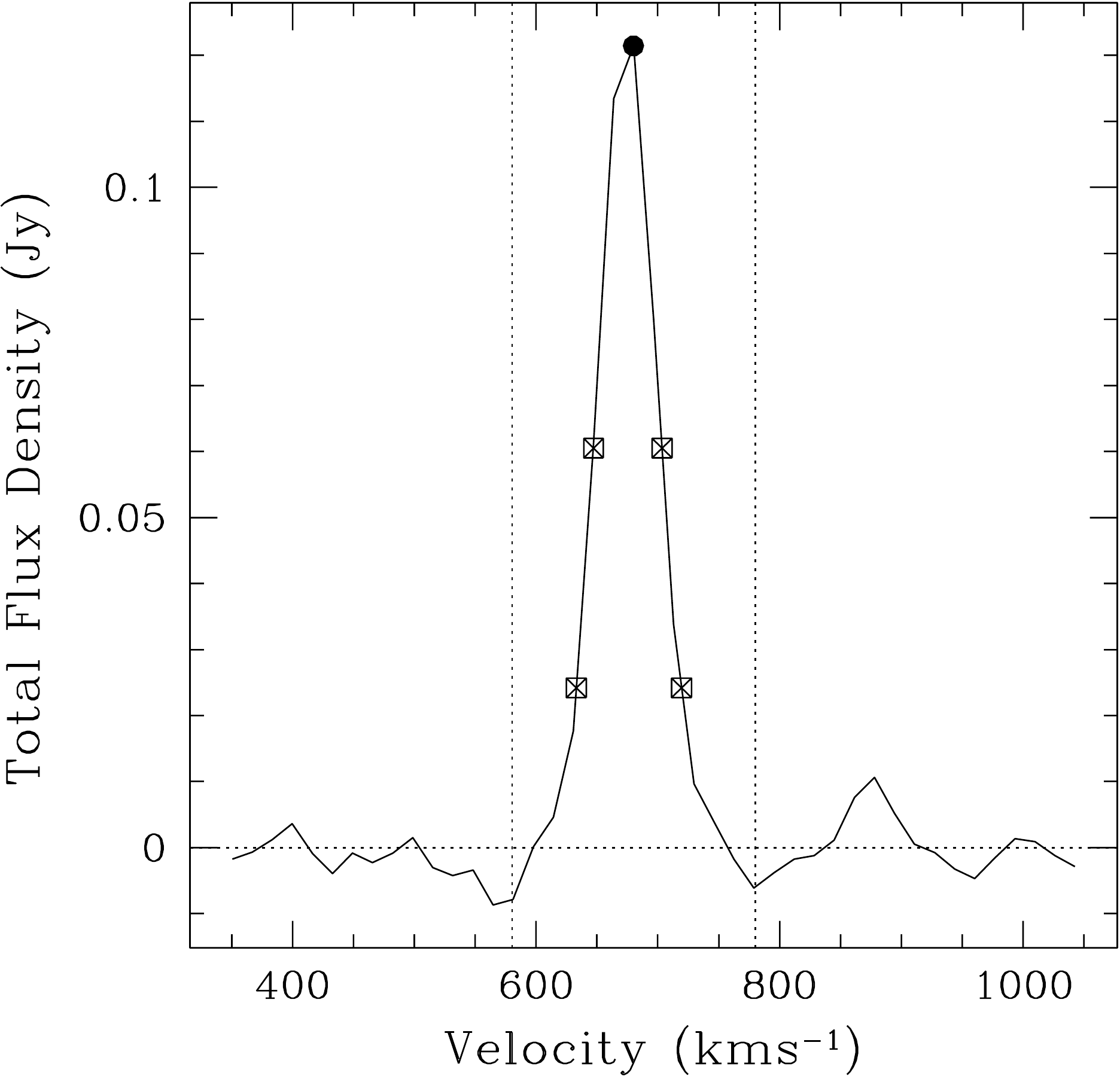}
\includegraphics[height=0.17\textheight]{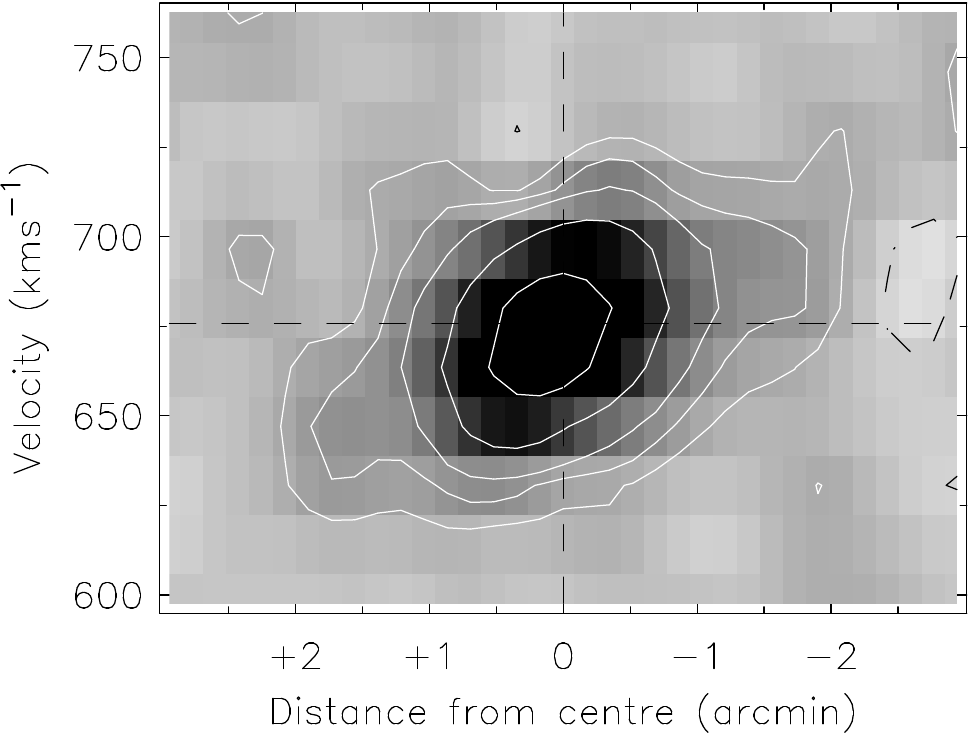}

\vskip 2mm
\centering
WSRT-CVn-39
\vskip 2mm
\includegraphics[width=0.25\textwidth]{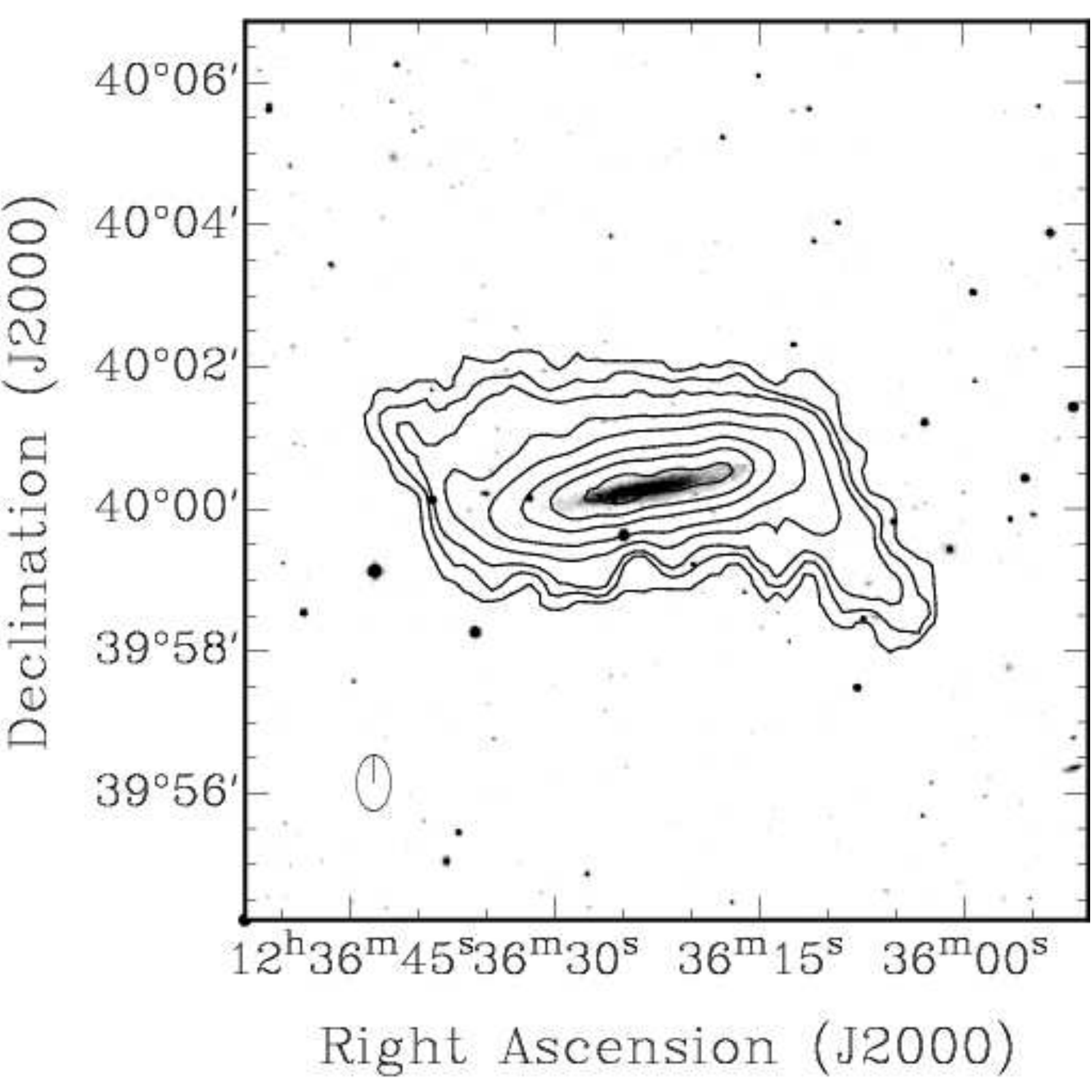}
\hskip 5mm
\includegraphics[height=0.17\textheight]{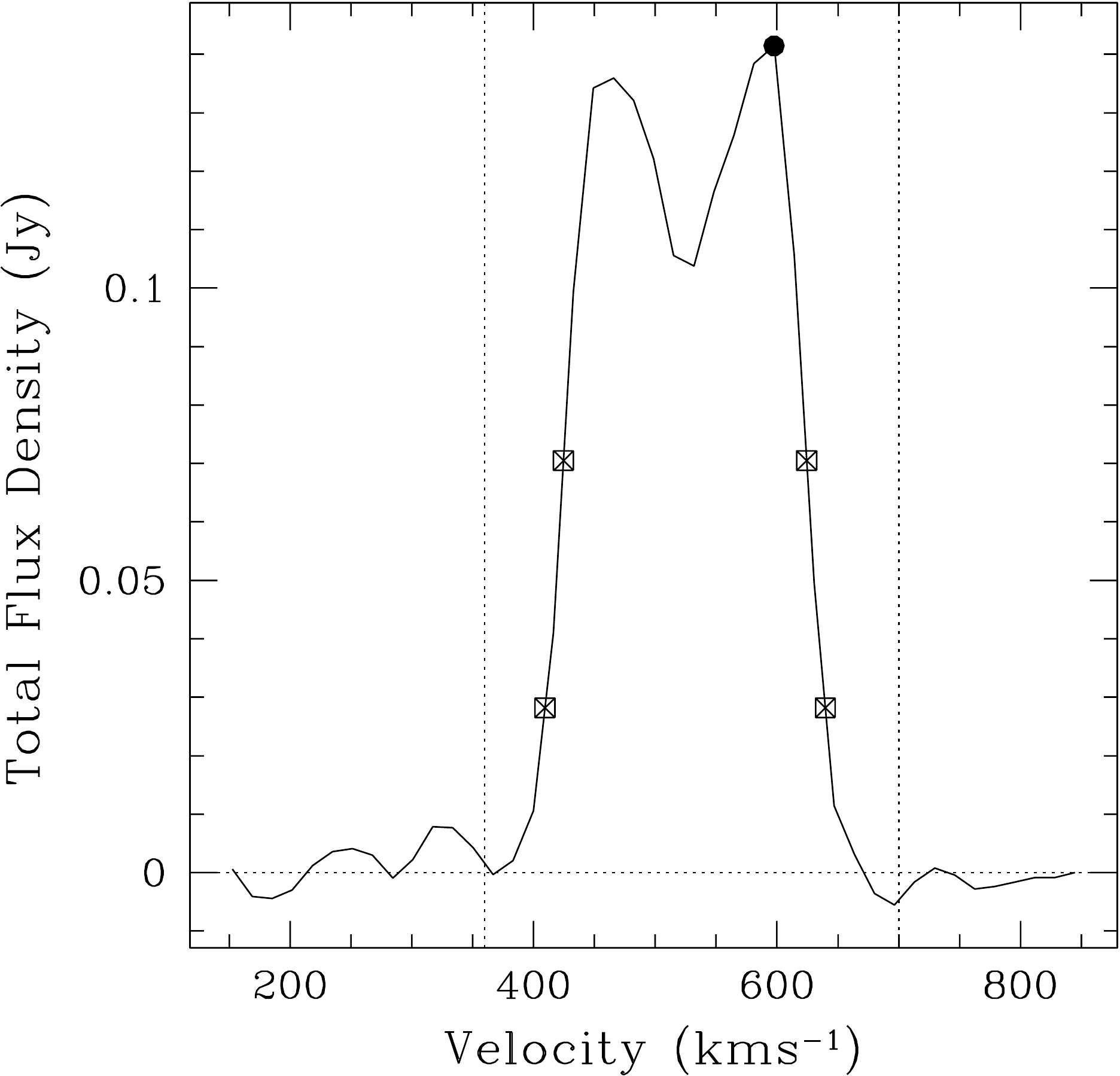}
\includegraphics[height=0.17\textheight]{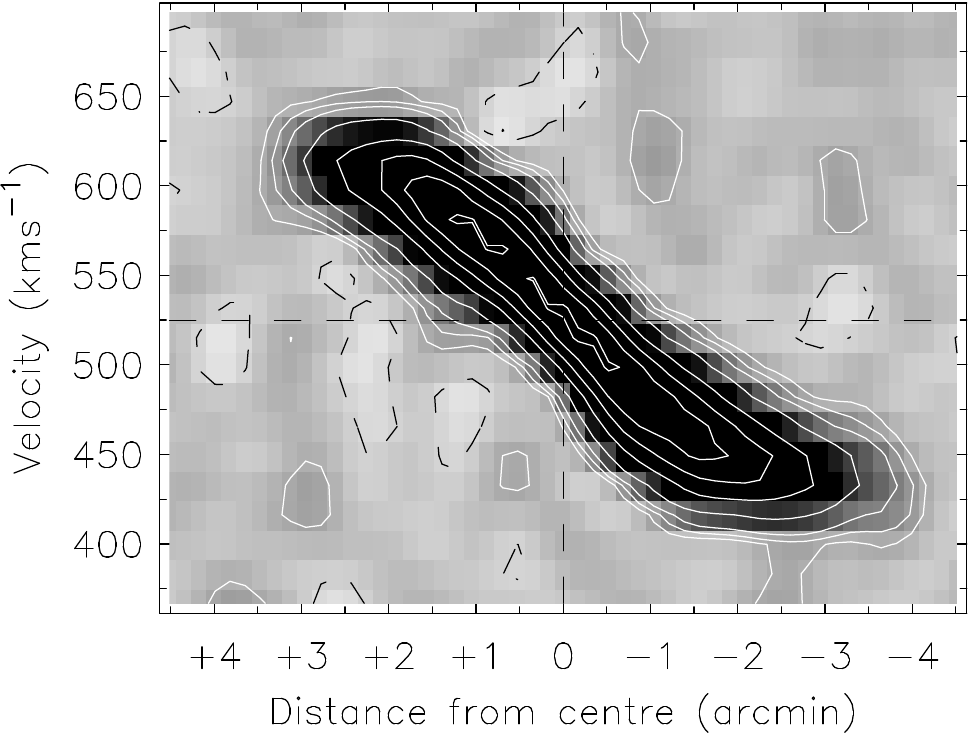}

\vskip 2mm
\centering
WSRT-CVn-40
\vskip 2mm
\includegraphics[width=0.25\textwidth]{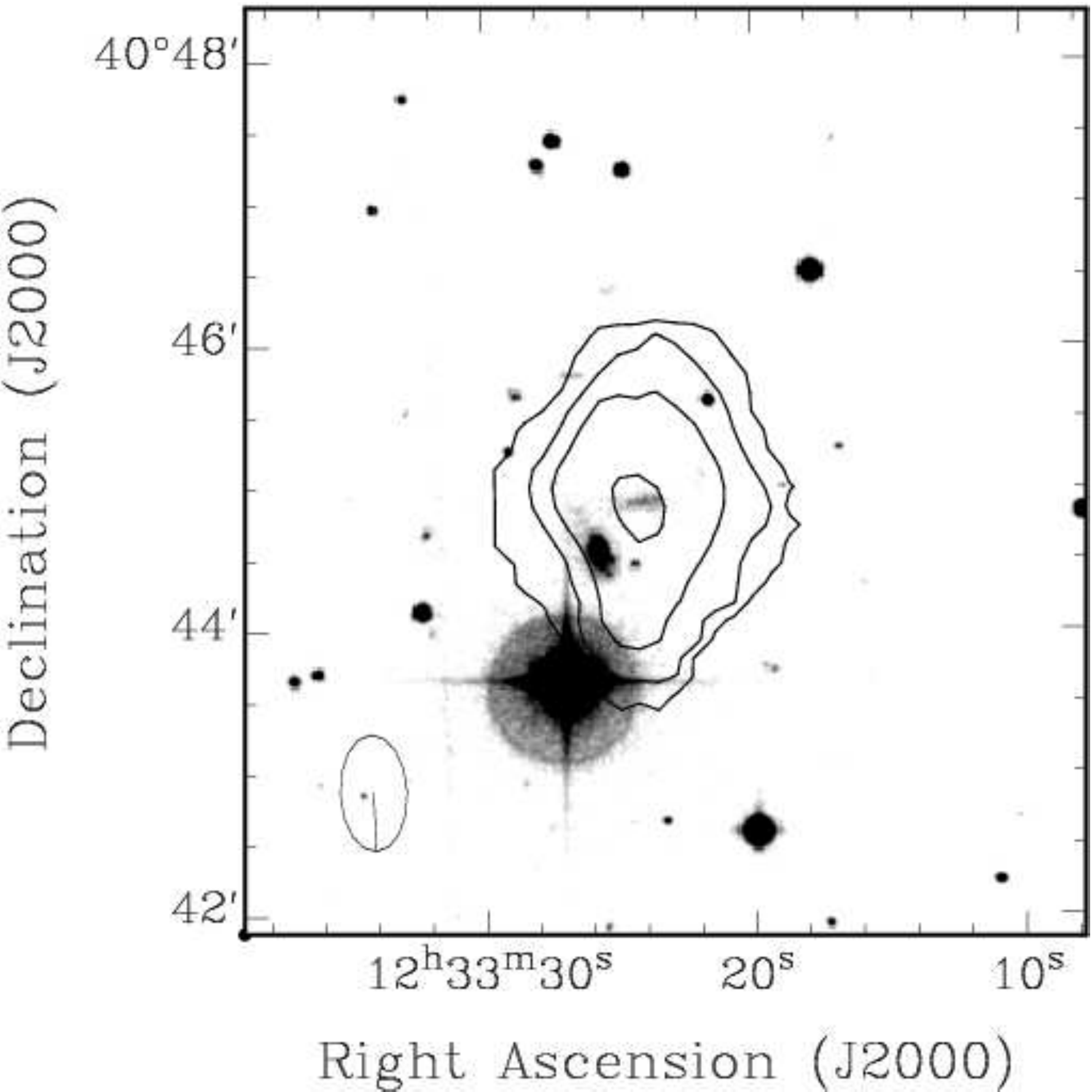}
\hskip 5mm
\includegraphics[height=0.17\textheight]{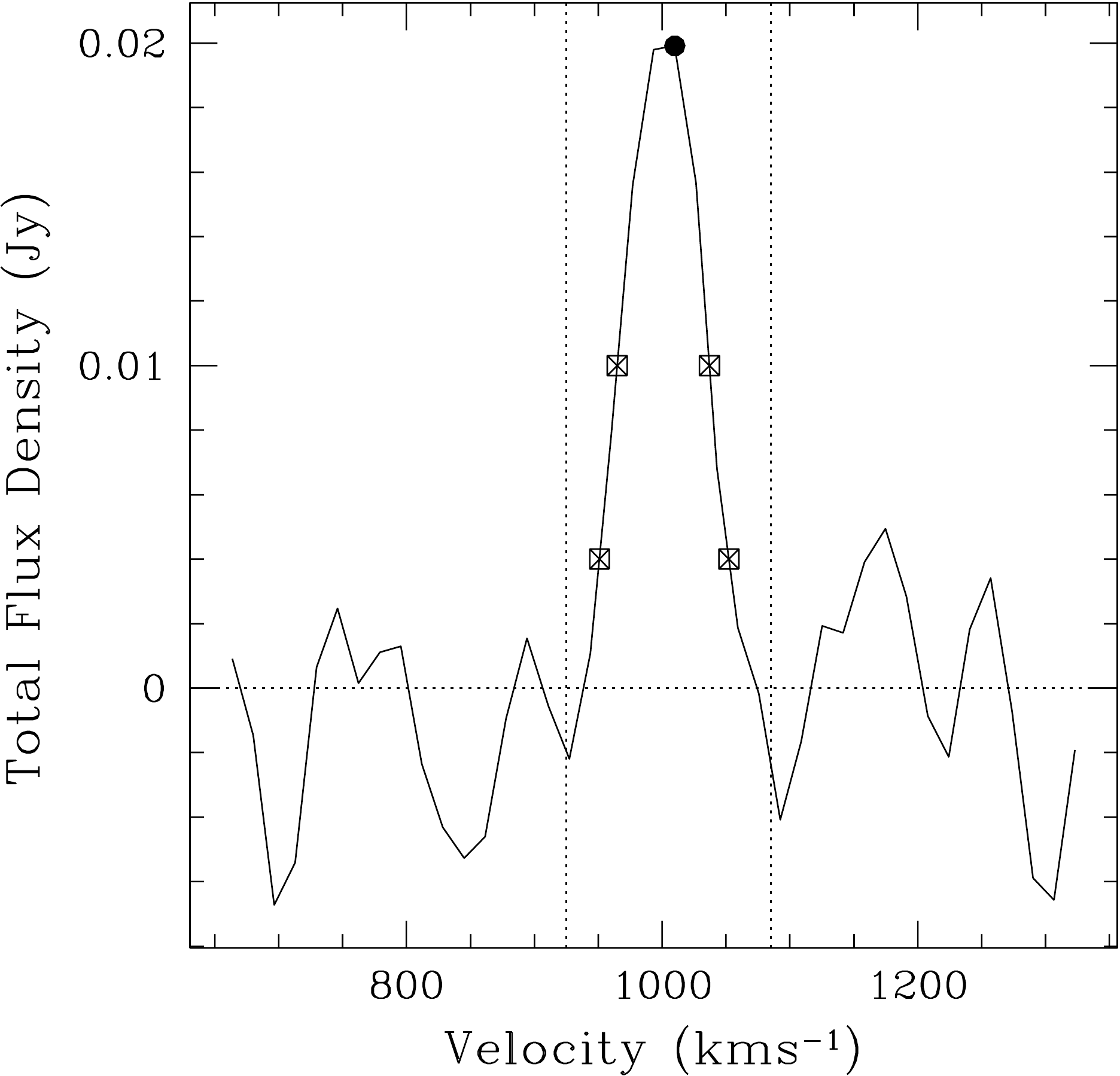}
\includegraphics[height=0.17\textheight]{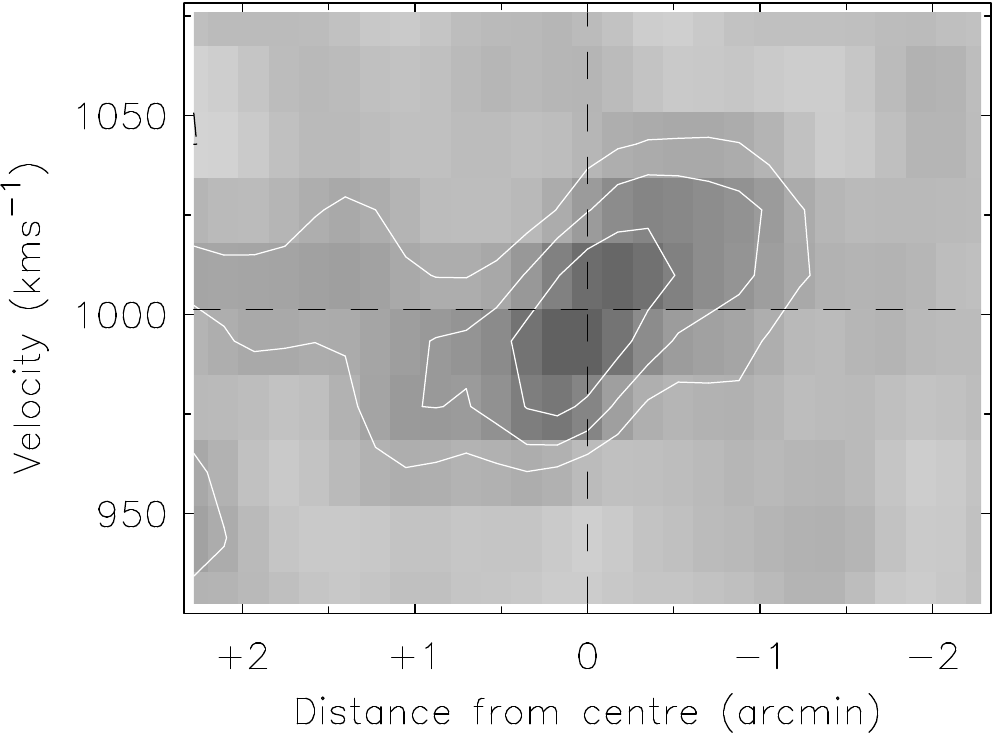}

\end{figure}

\clearpage

\addtocounter{figure}{-1}
\begin{figure}

\vskip 2mm
\centering
WSRT-CVn-41
\vskip 2mm
\includegraphics[width=0.25\textwidth]{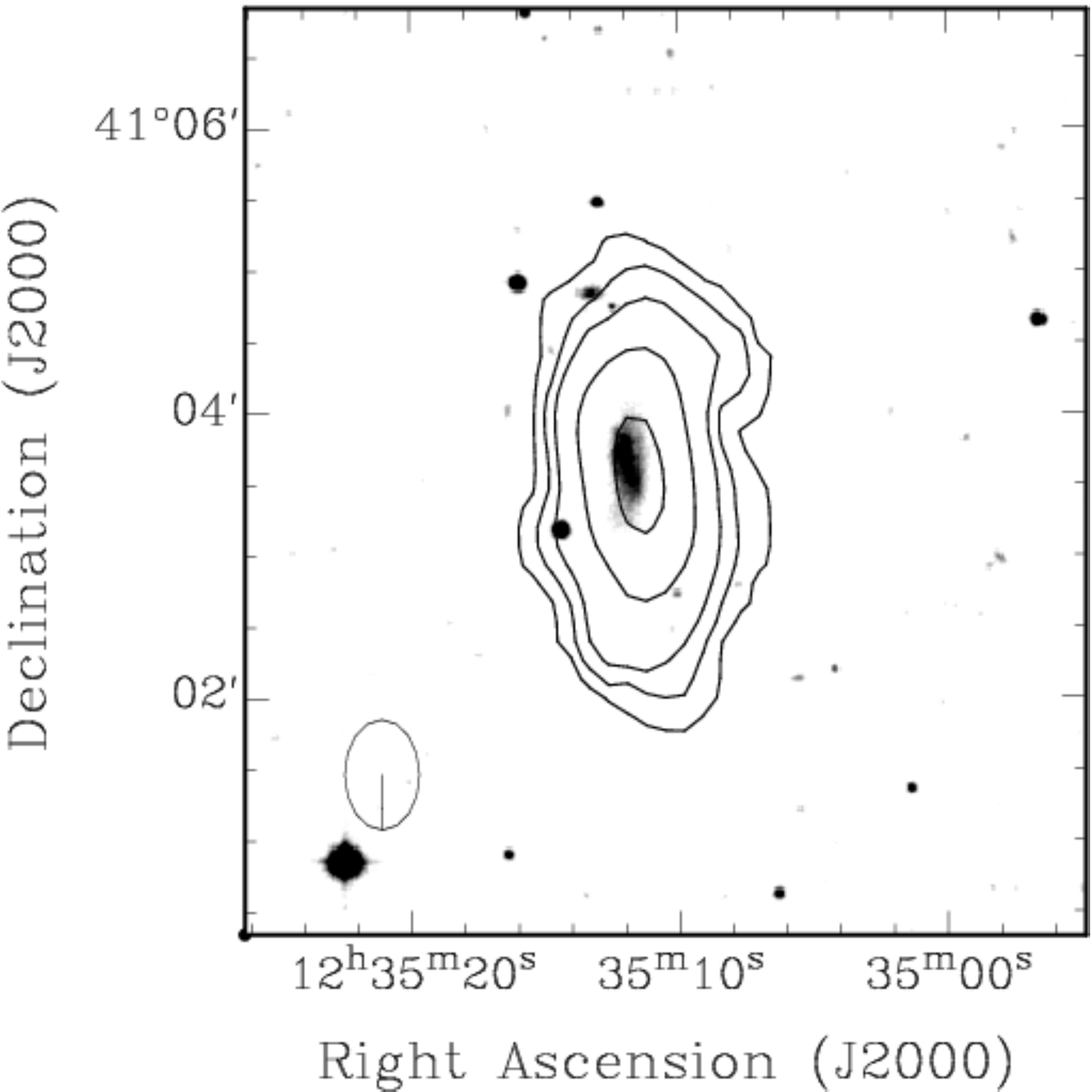}
\hskip 5mm
\includegraphics[height=0.17\textheight]{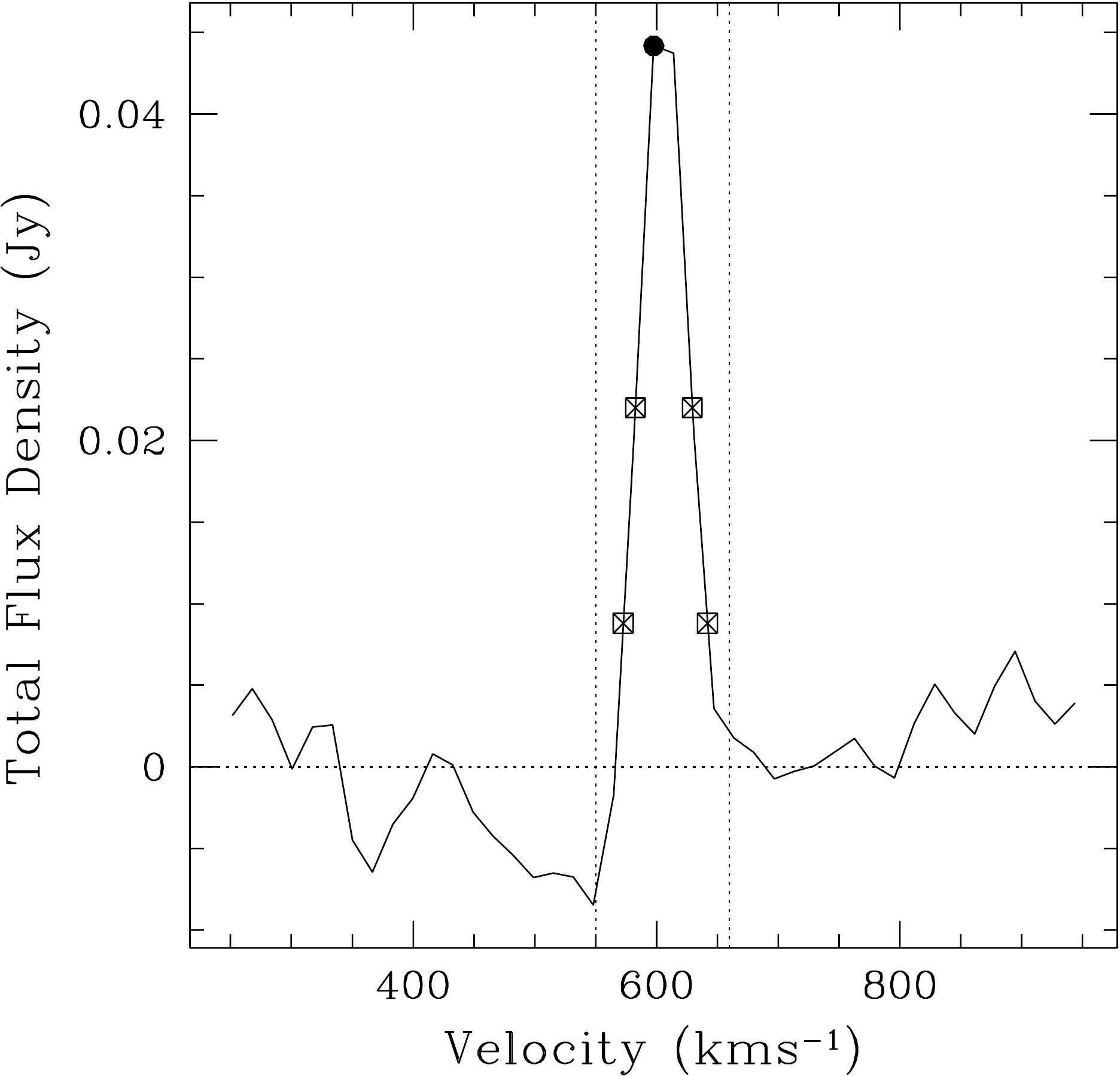}
\includegraphics[height=0.17\textheight]{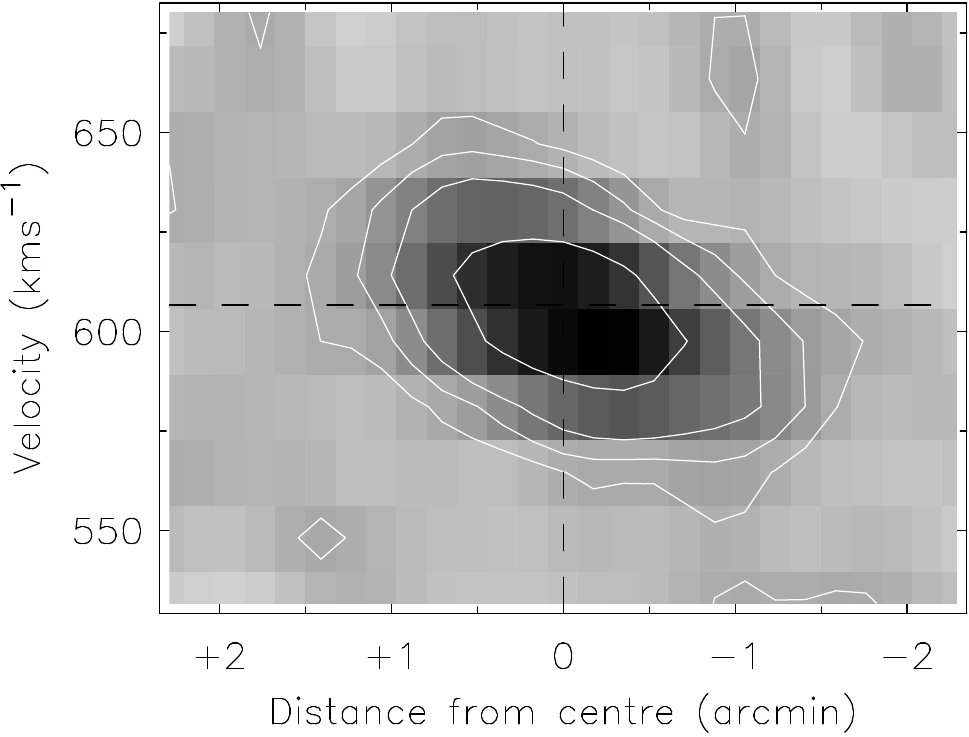}

\vskip 2mm
\centering
WSRT-CVn-42
\vskip 2mm
\includegraphics[width=0.25\textwidth]{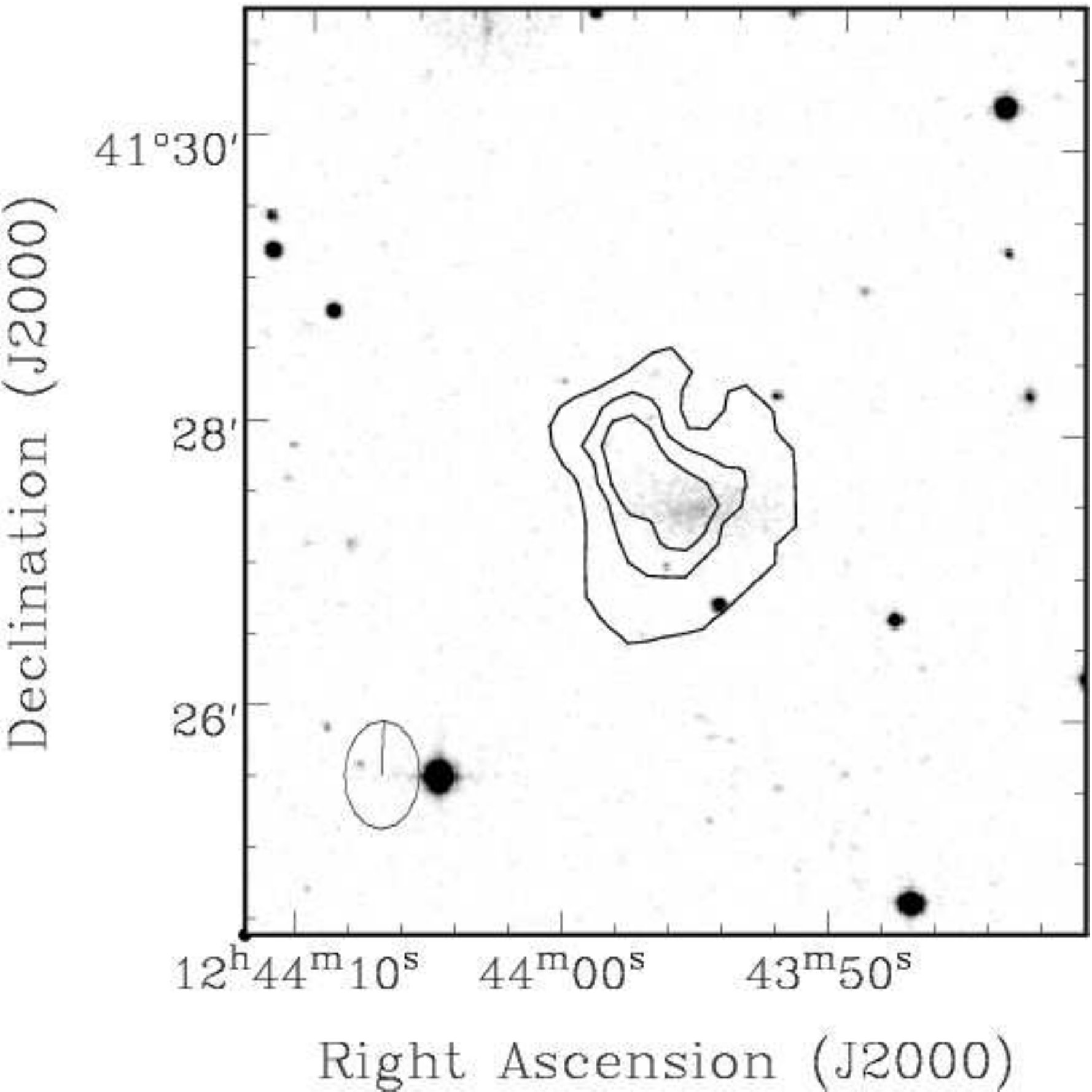}
\hskip 5mm
\includegraphics[height=0.17\textheight]{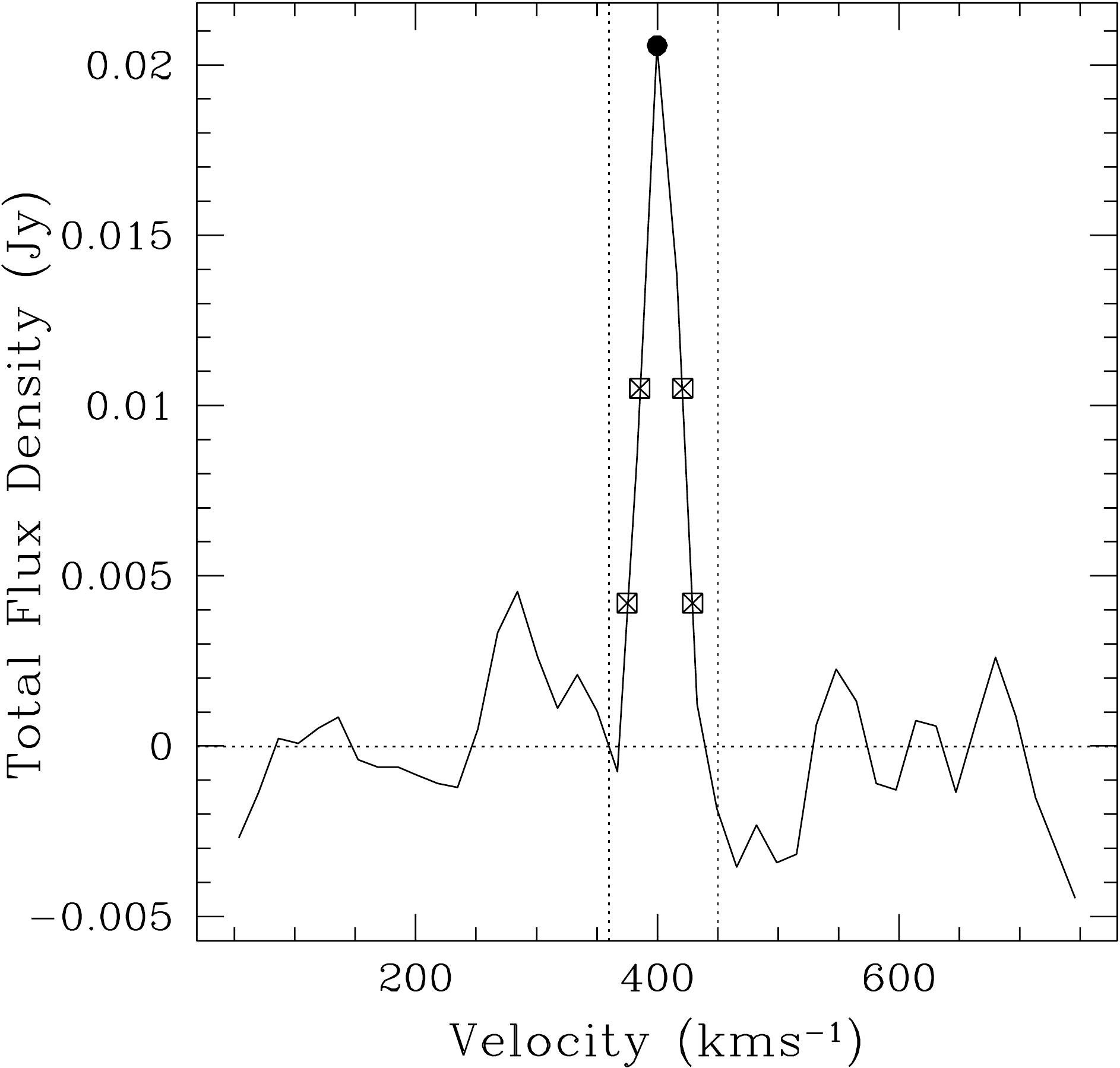}
\includegraphics[height=0.17\textheight]{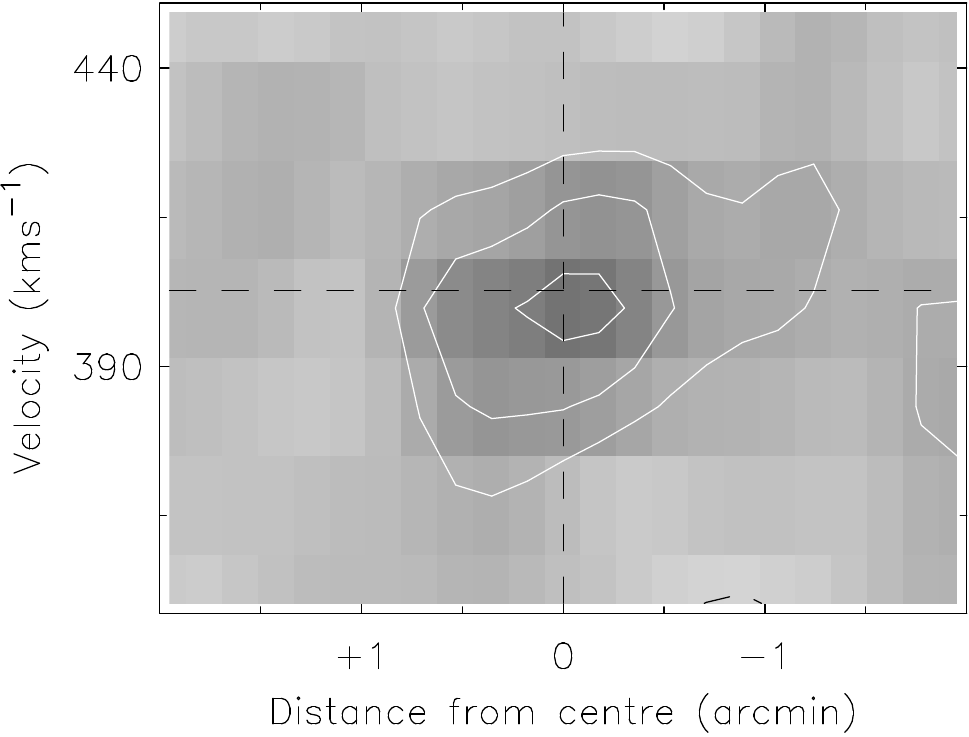}

\vskip 2mm
\centering
WSRT-CVn-43
\vskip 2mm
\includegraphics[width=0.25\textwidth]{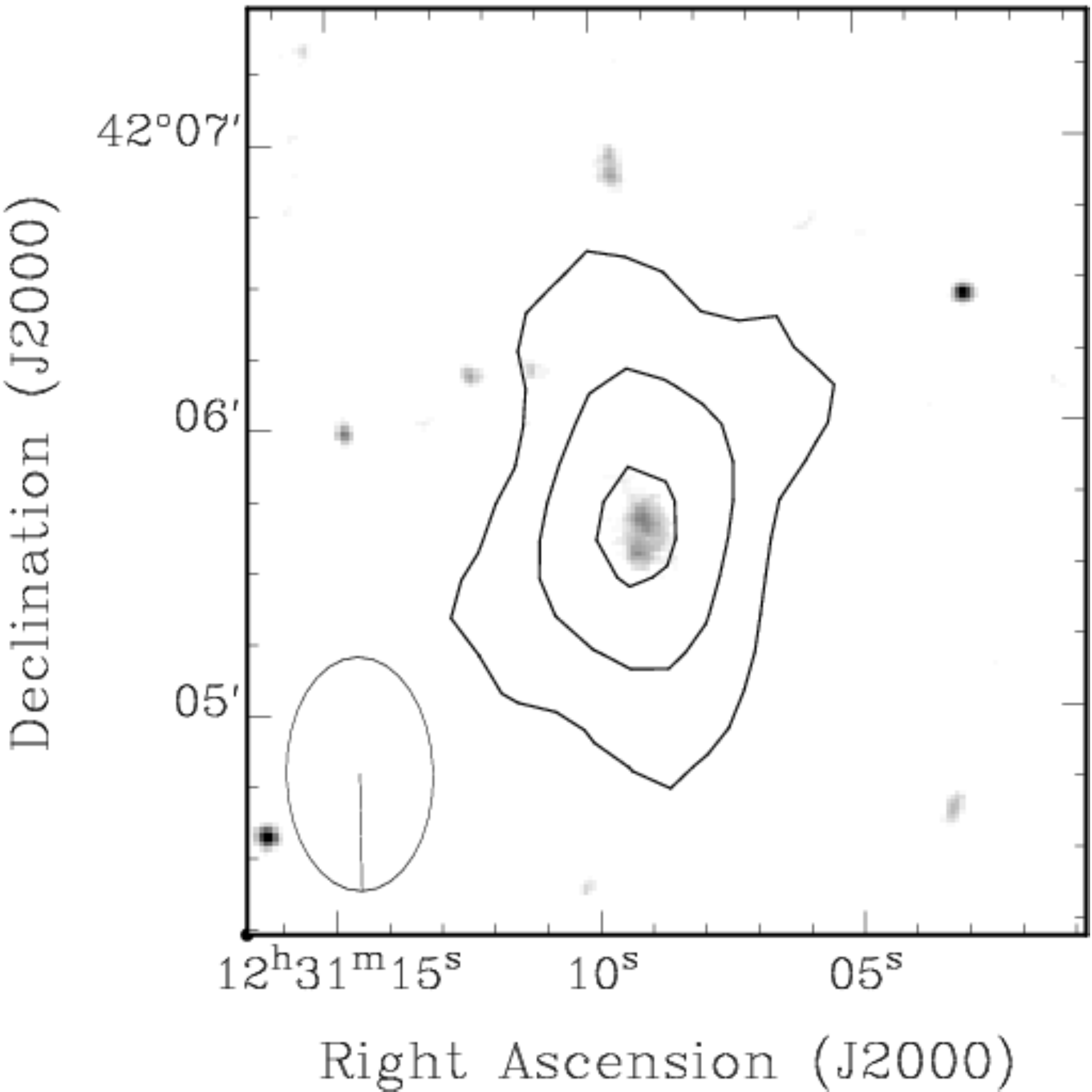}
\hskip 5mm
\includegraphics[height=0.17\textheight]{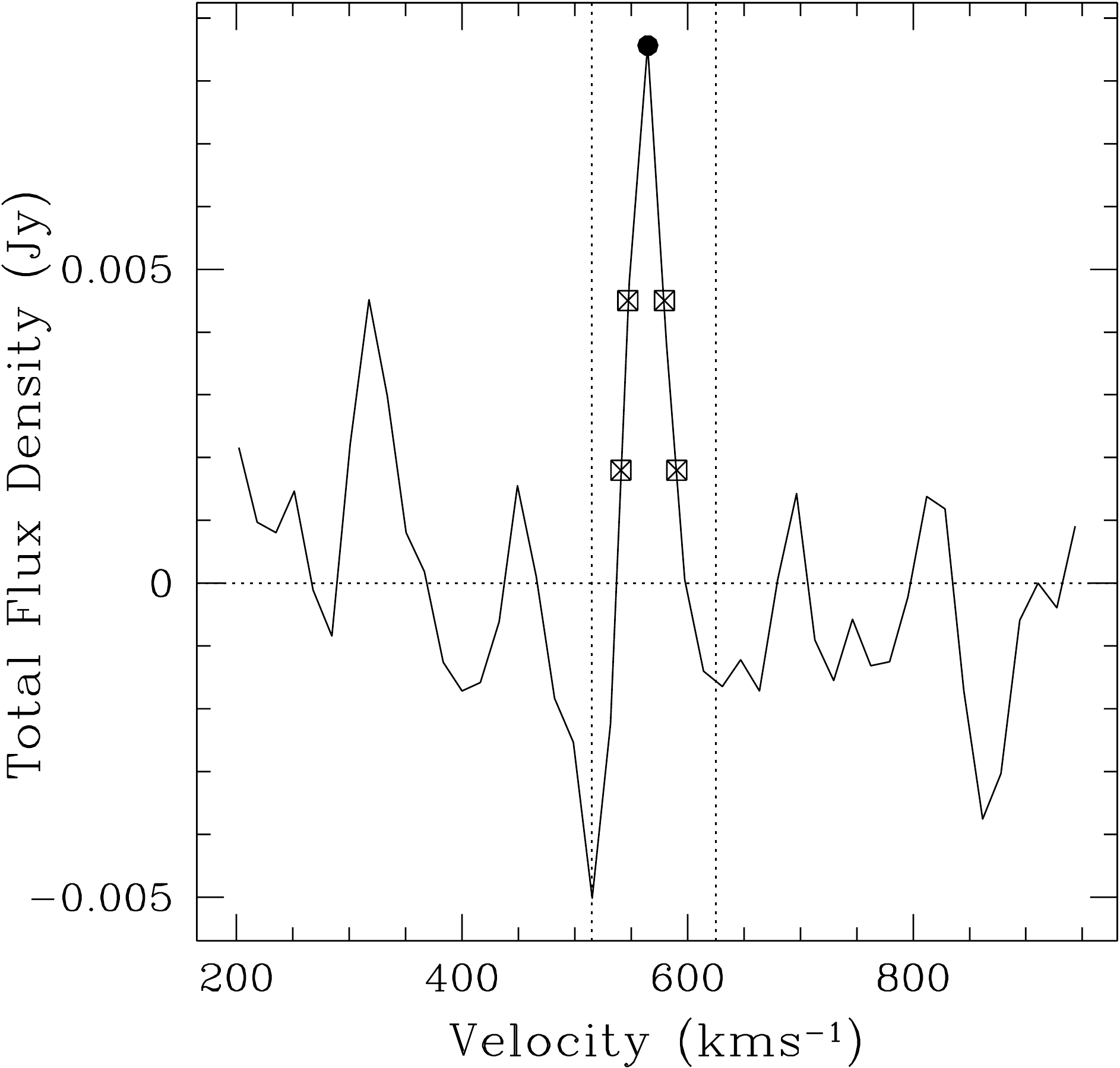}
\includegraphics[height=0.17\textheight]{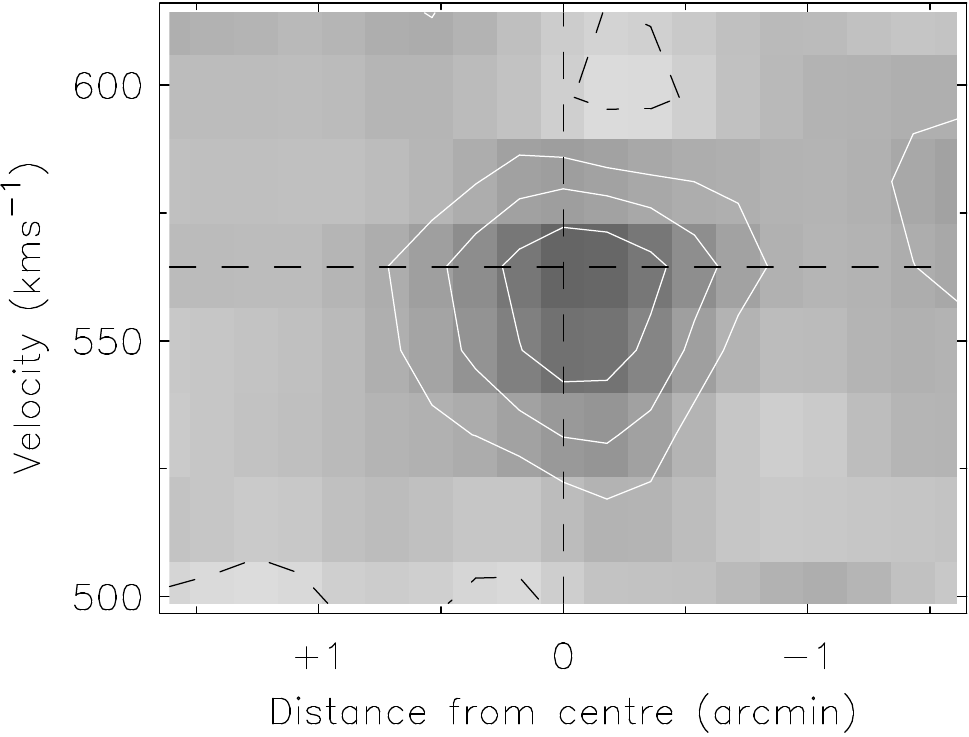}

\vskip 2mm
\centering
WSRT-CVn-44
\vskip 2mm
\includegraphics[width=0.25\textwidth]{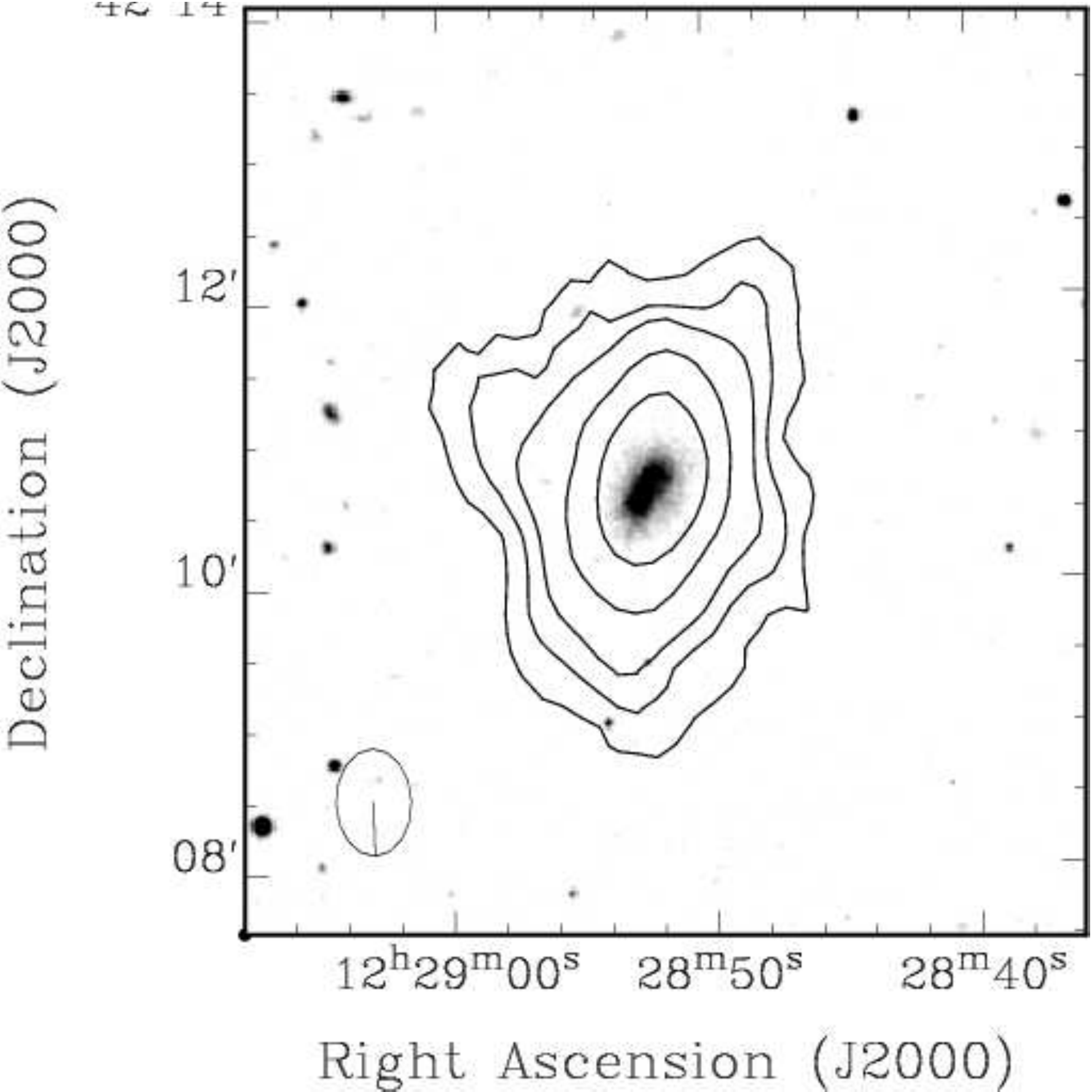}
\hskip 5mm
\includegraphics[height=0.17\textheight]{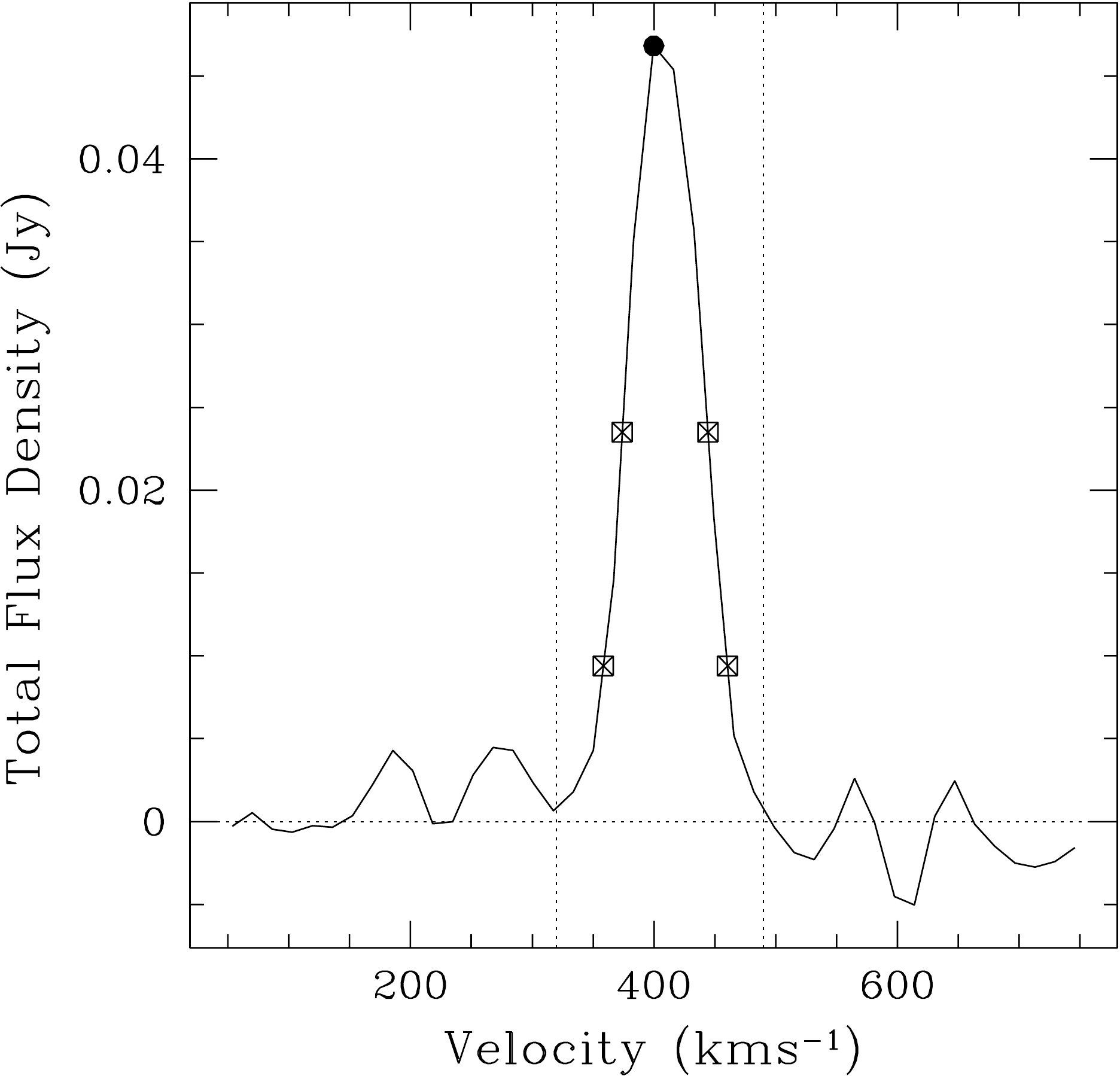}
\includegraphics[height=0.17\textheight]{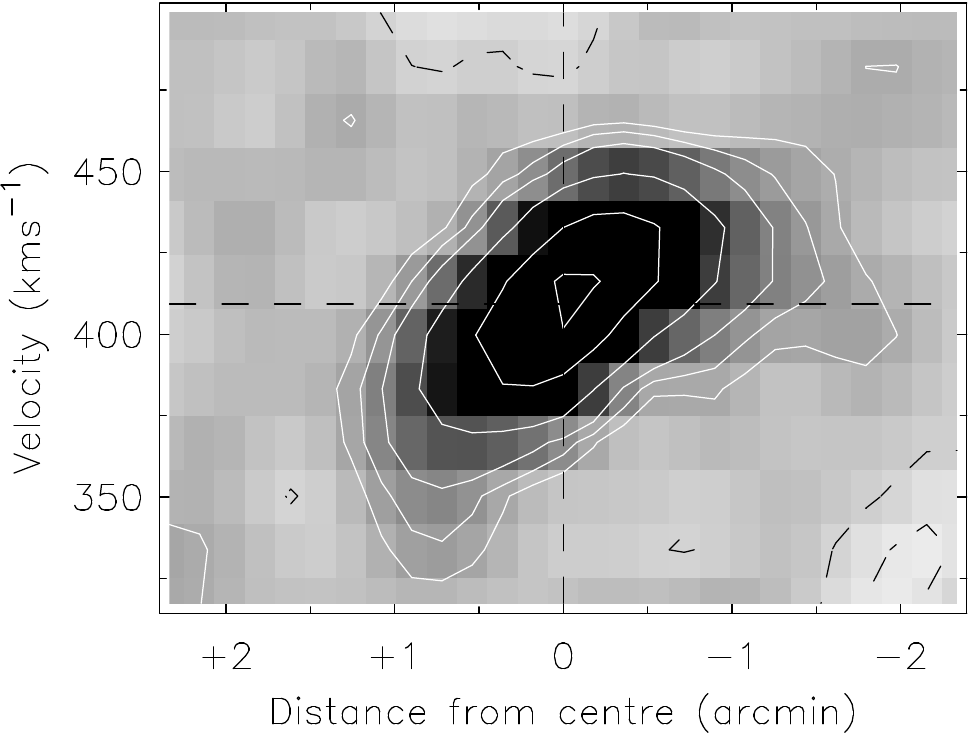}

\end{figure}

\clearpage

\addtocounter{figure}{-1}
\begin{figure}

\vskip 2mm
\centering
WSRT-CVn-45
\vskip 2mm
\includegraphics[width=0.25\textwidth]{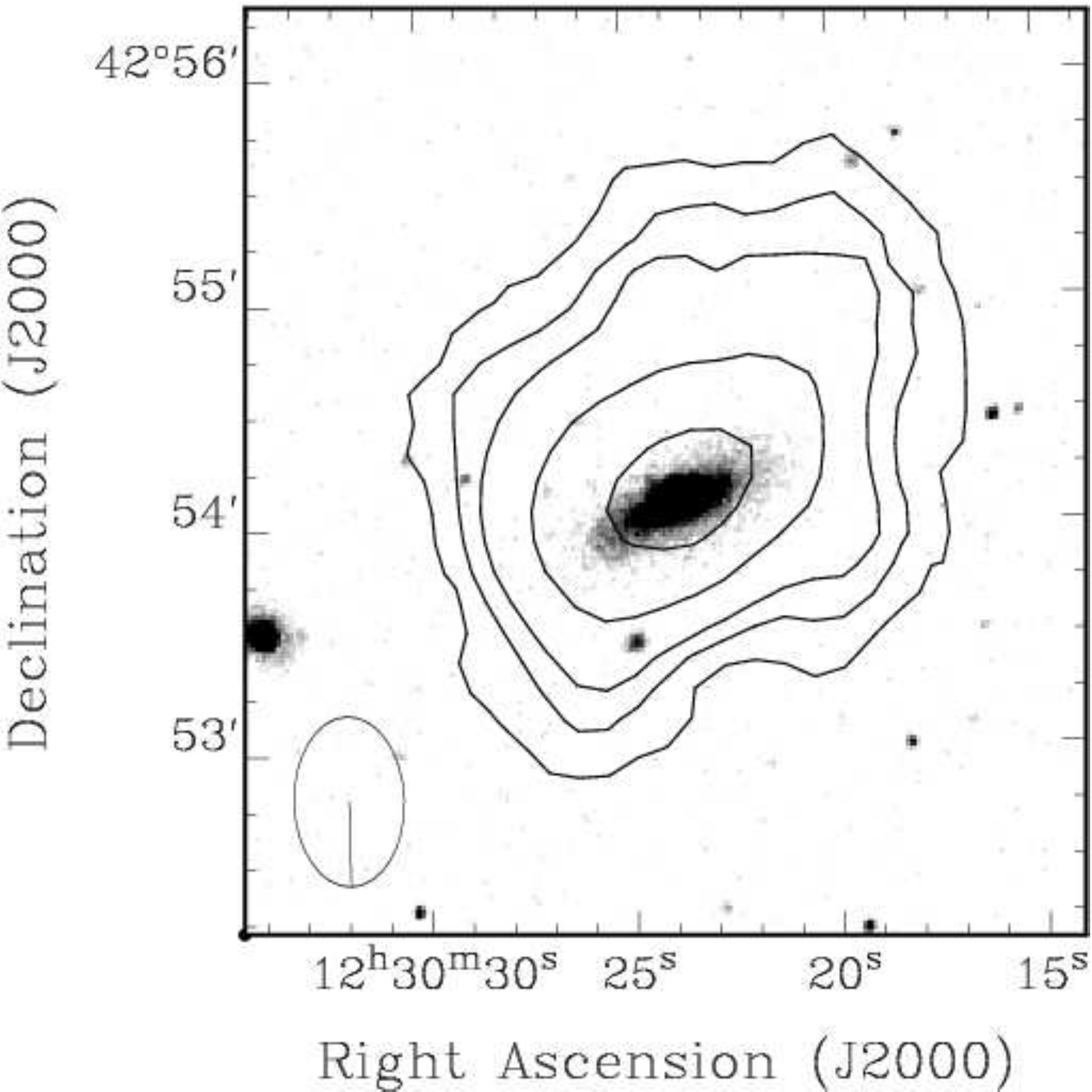}
\hskip 5mm
\includegraphics[height=0.17\textheight]{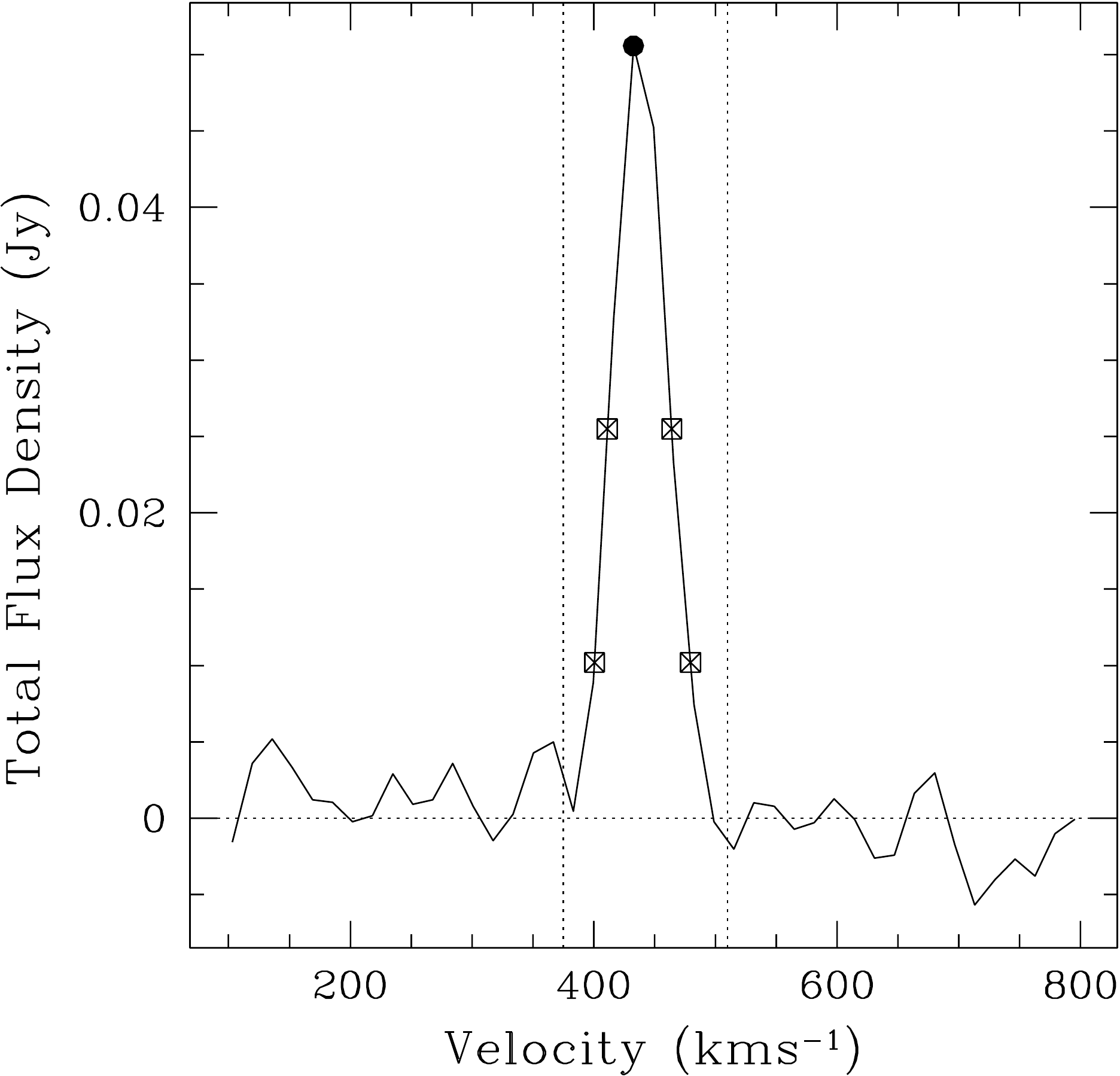}
\includegraphics[height=0.17\textheight]{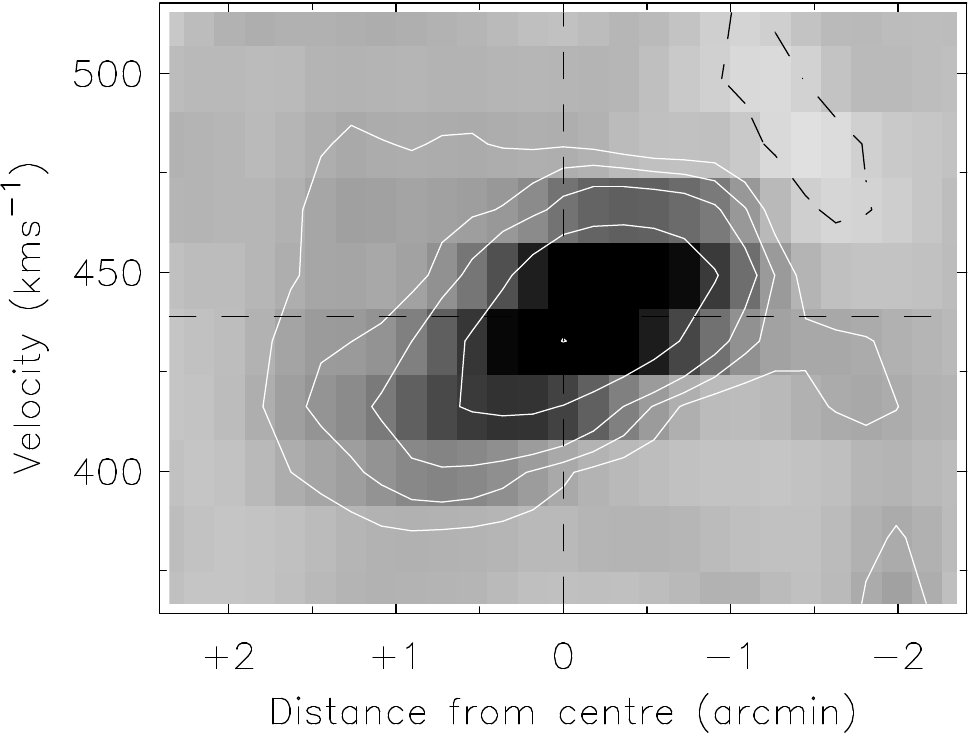}

\vskip 2mm
\centering
WSRT-CVn-46
\vskip 2mm
\includegraphics[width=0.25\textwidth]{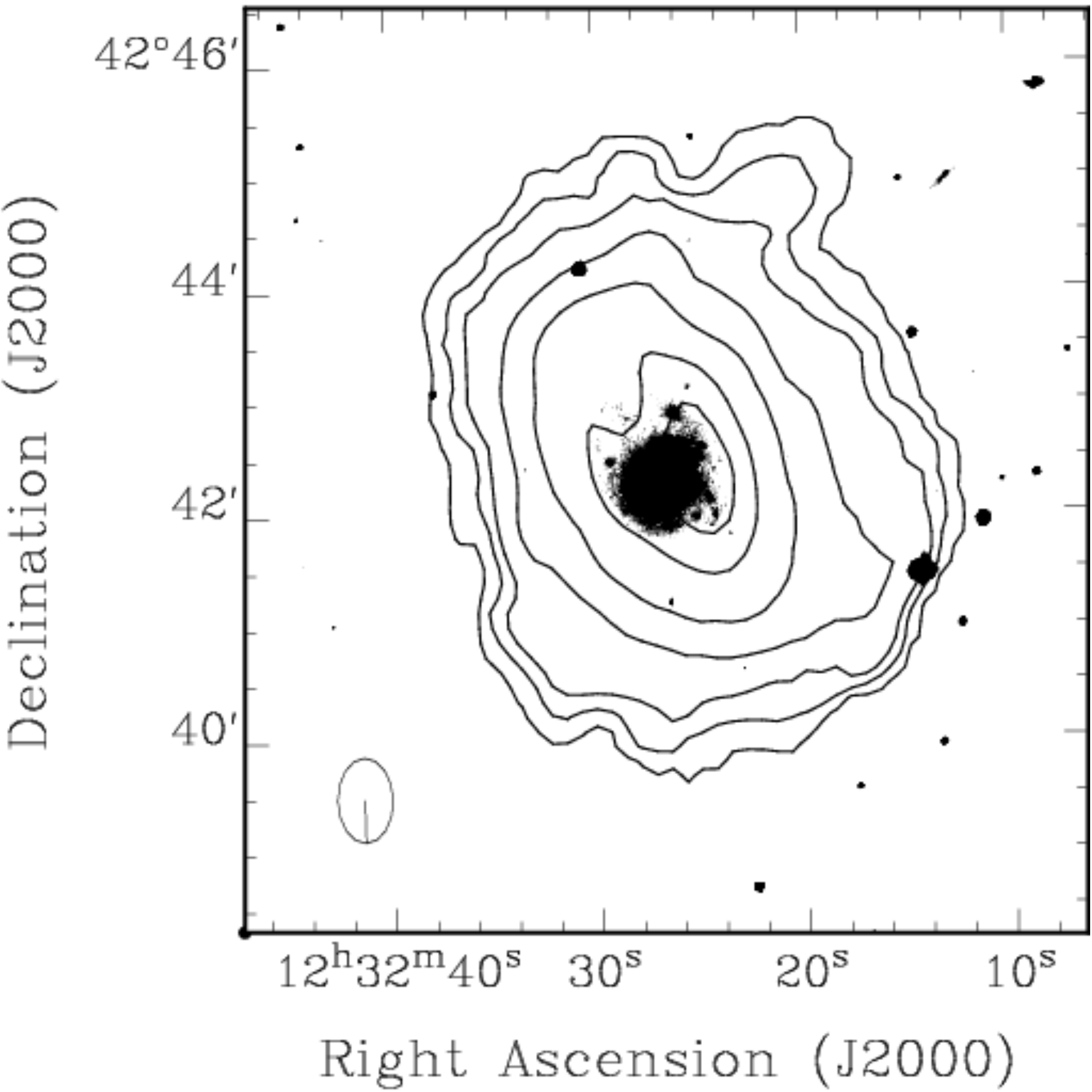}
\hskip 5mm
\includegraphics[height=0.17\textheight]{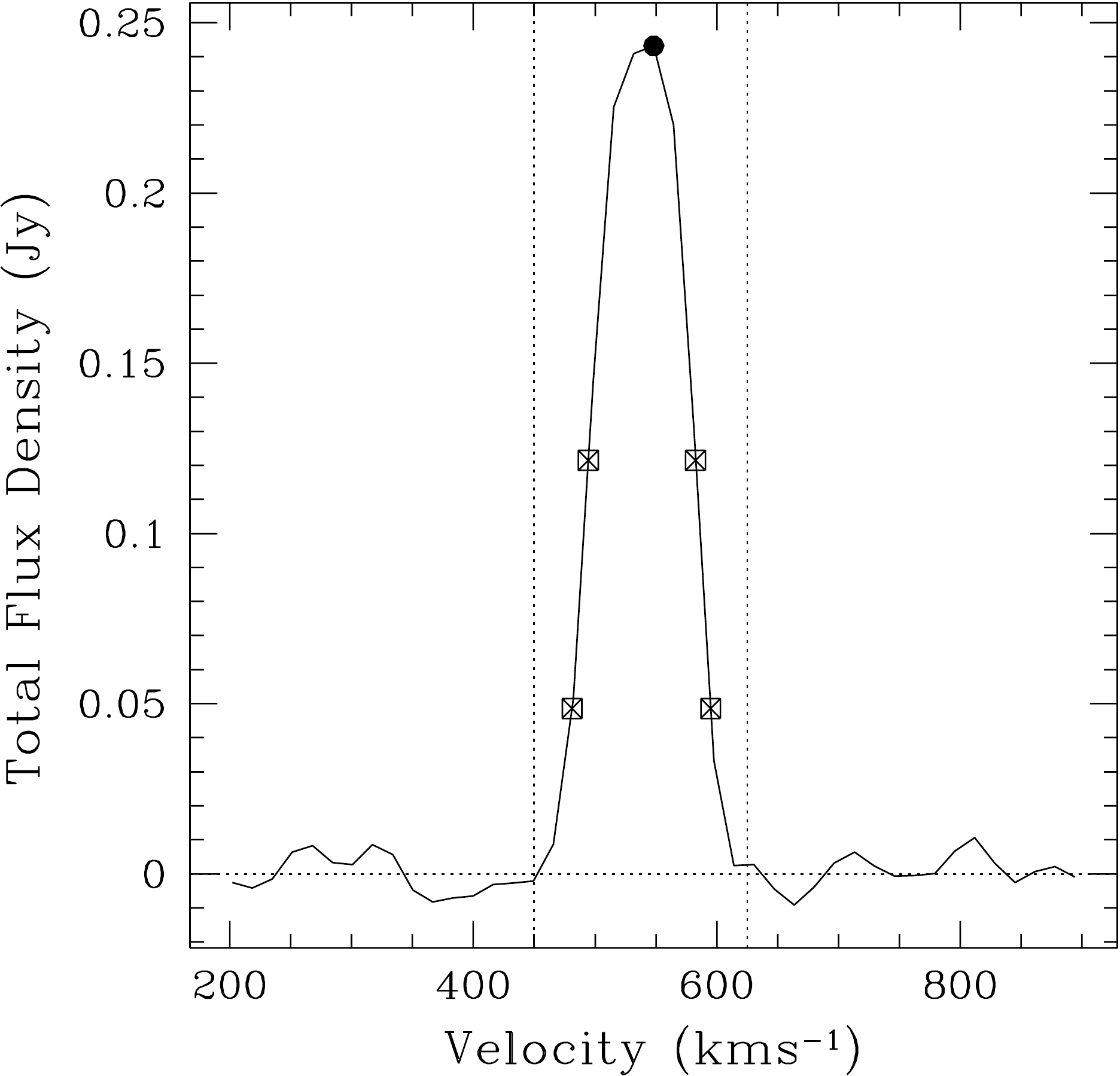}
\includegraphics[height=0.17\textheight]{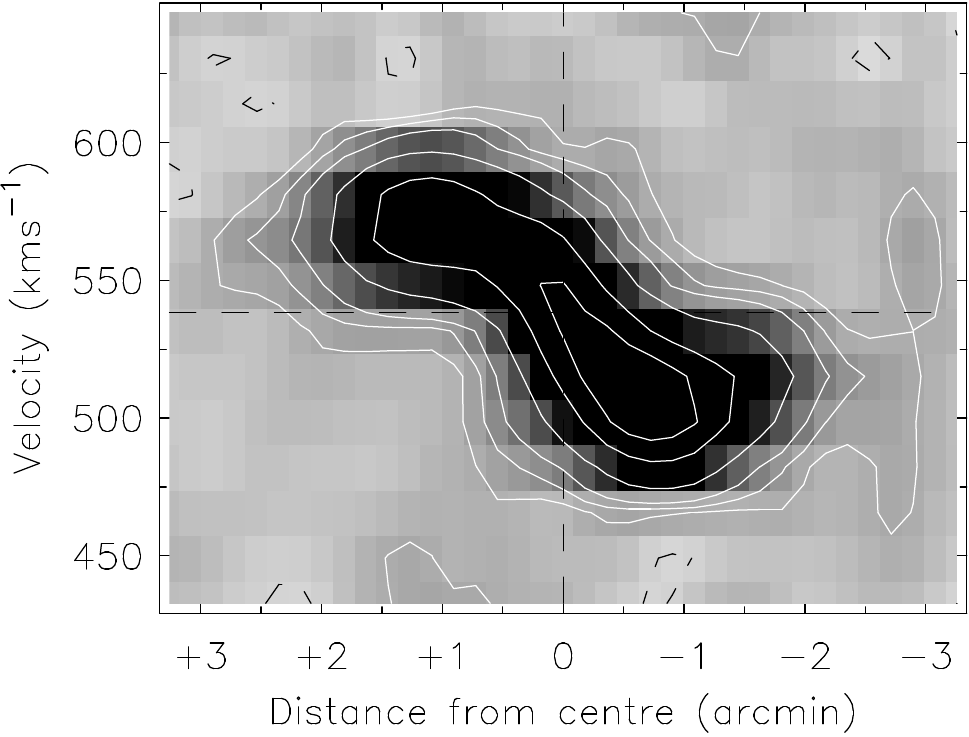}

\vskip 2mm
\centering
WSRT-CVn-47
\vskip 2mm
\includegraphics[width=0.25\textwidth]{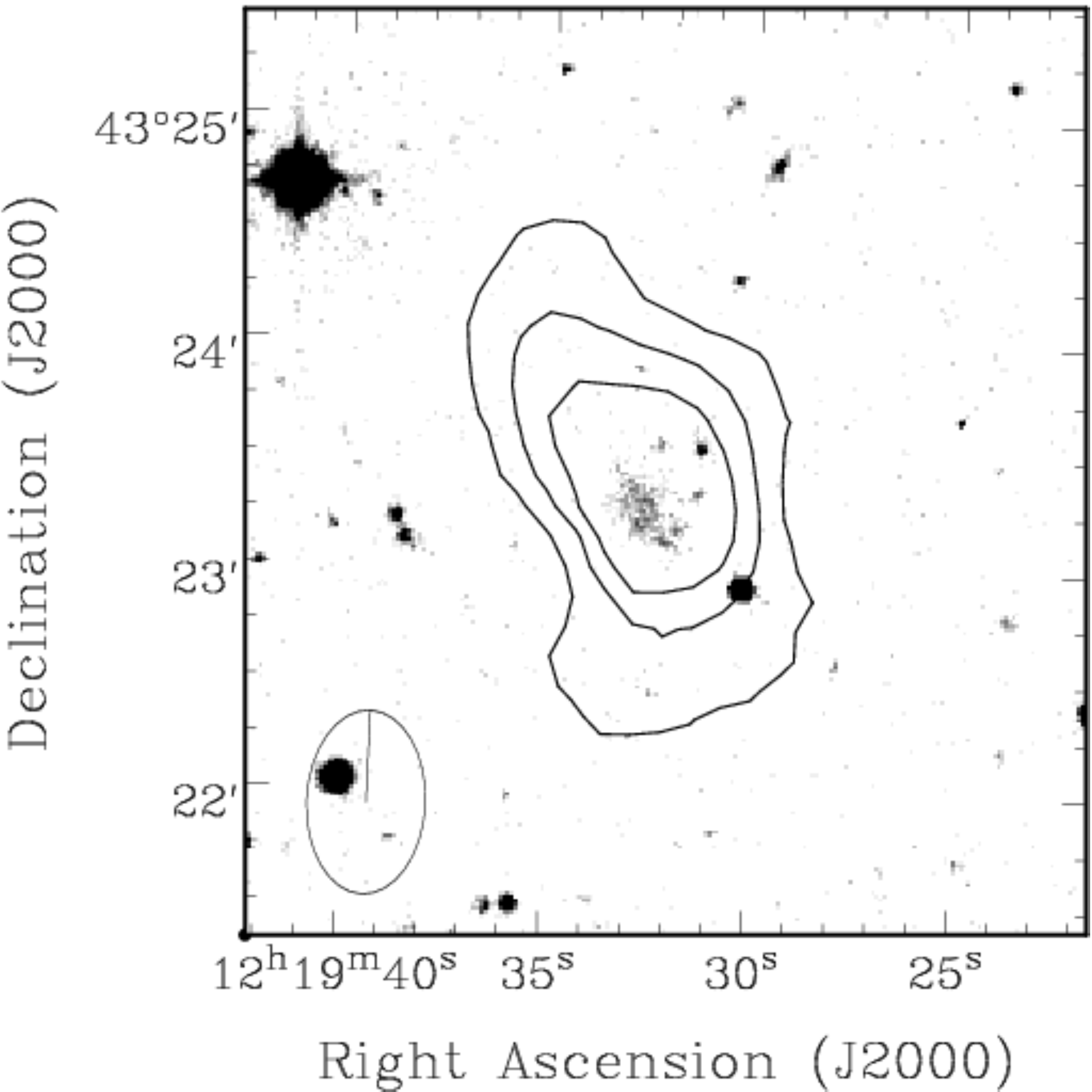}
\hskip 5mm
\includegraphics[height=0.17\textheight]{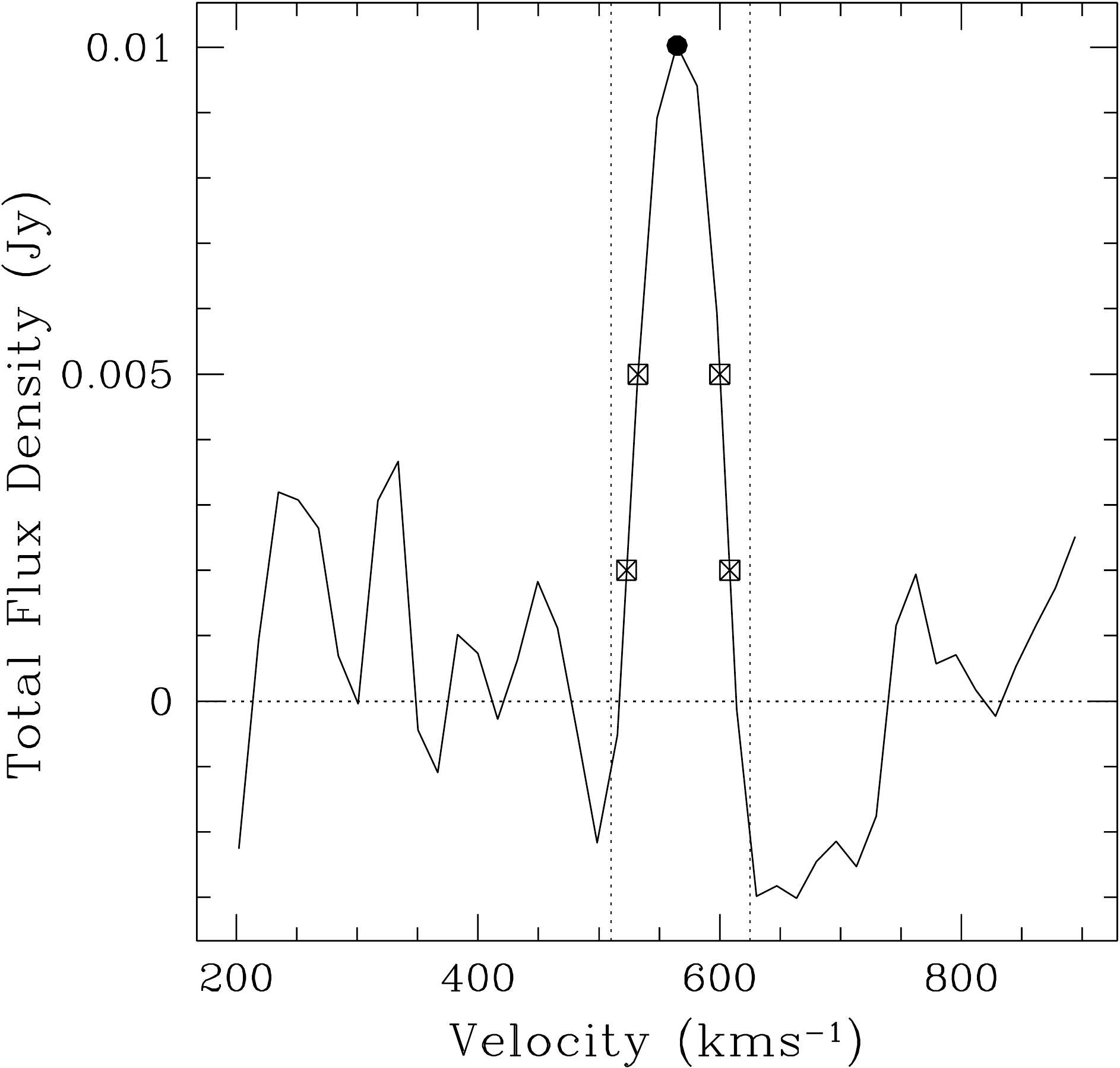}
\includegraphics[height=0.17\textheight]{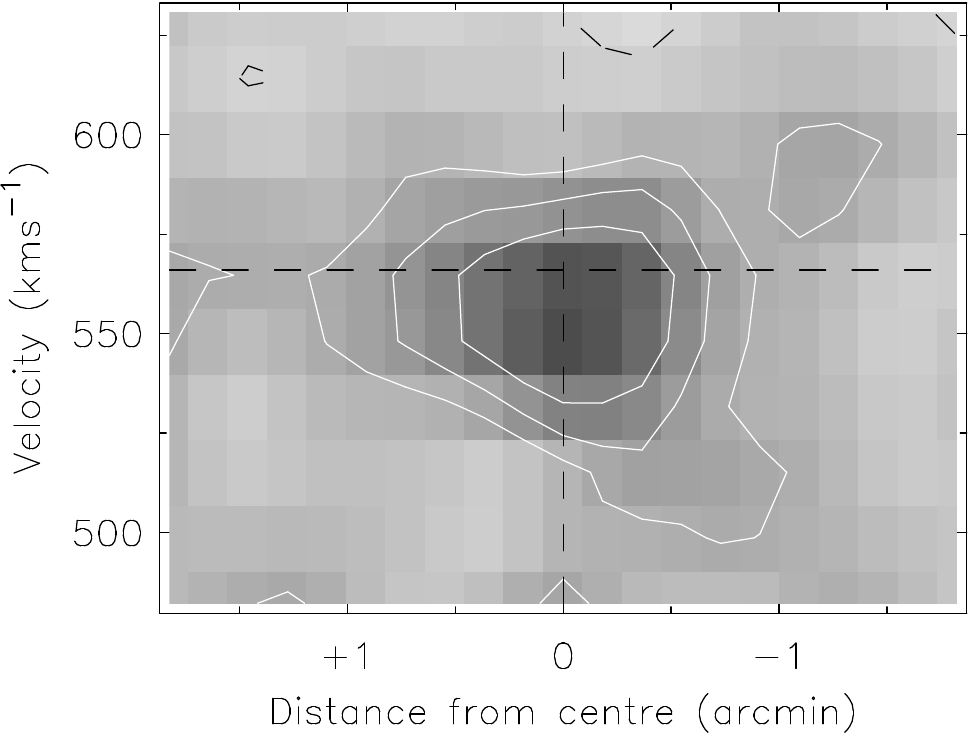}

\vskip 2mm
\centering
WSRT-CVn-48
\vskip 2mm
\includegraphics[width=0.25\textwidth]{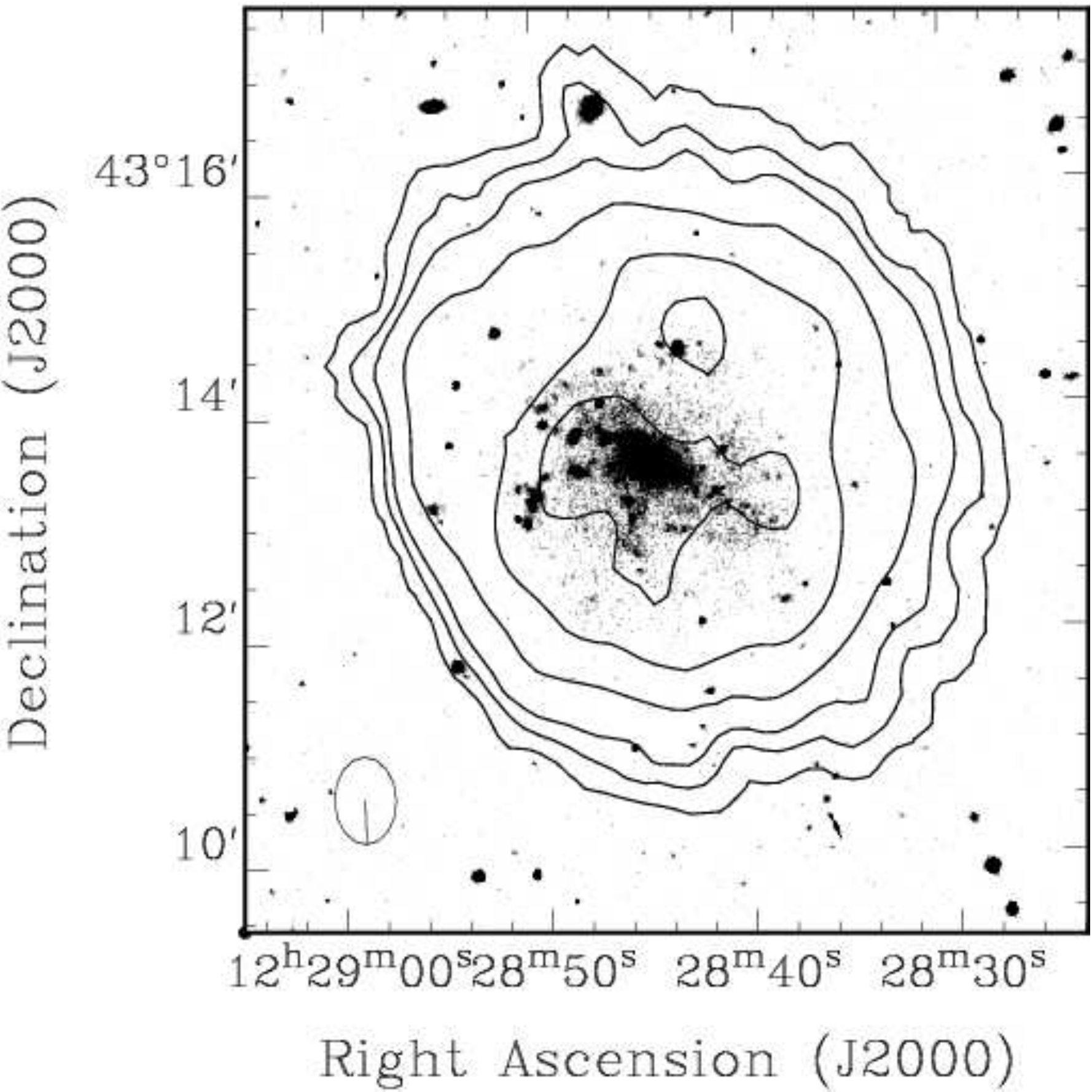}
\hskip 5mm
\includegraphics[height=0.17\textheight]{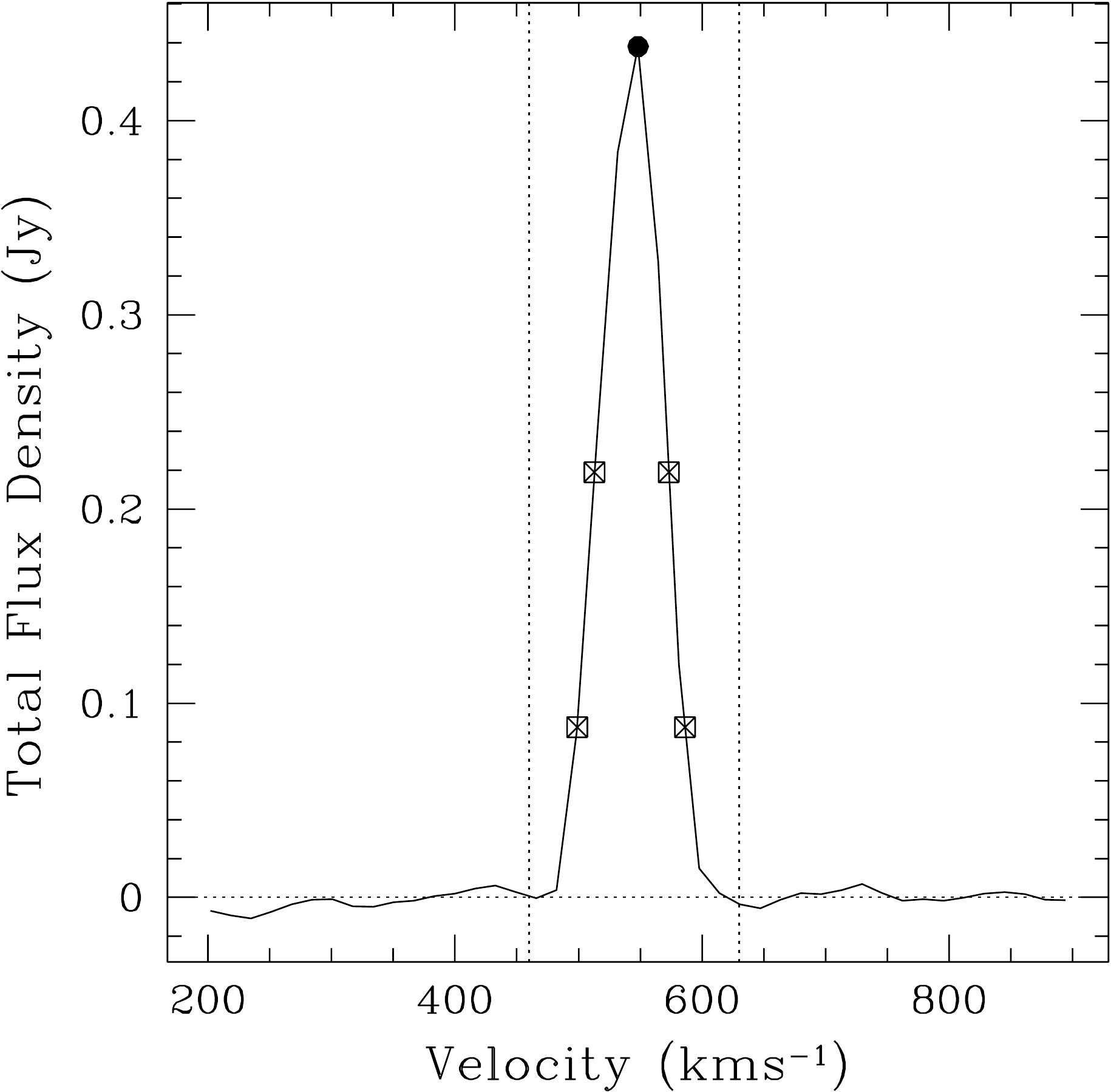}
\includegraphics[height=0.17\textheight]{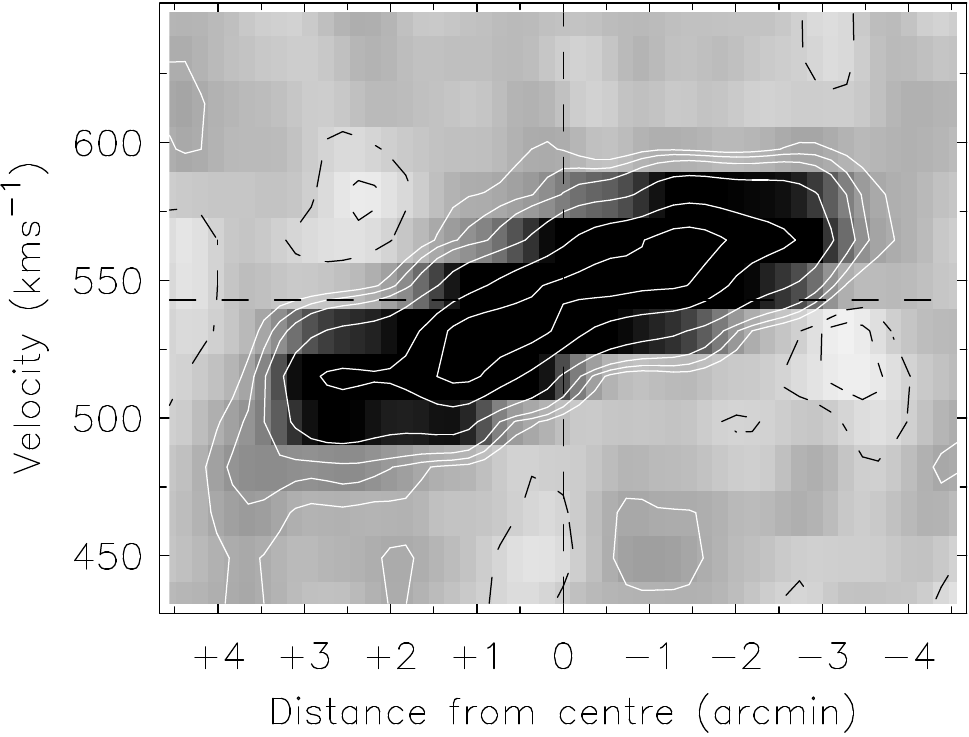}

\end{figure}

\clearpage

\addtocounter{figure}{-1}
\begin{figure}

\vskip 2mm
\centering
WSRT-CVn-49
\vskip 2mm
\includegraphics[width=0.25\textwidth]{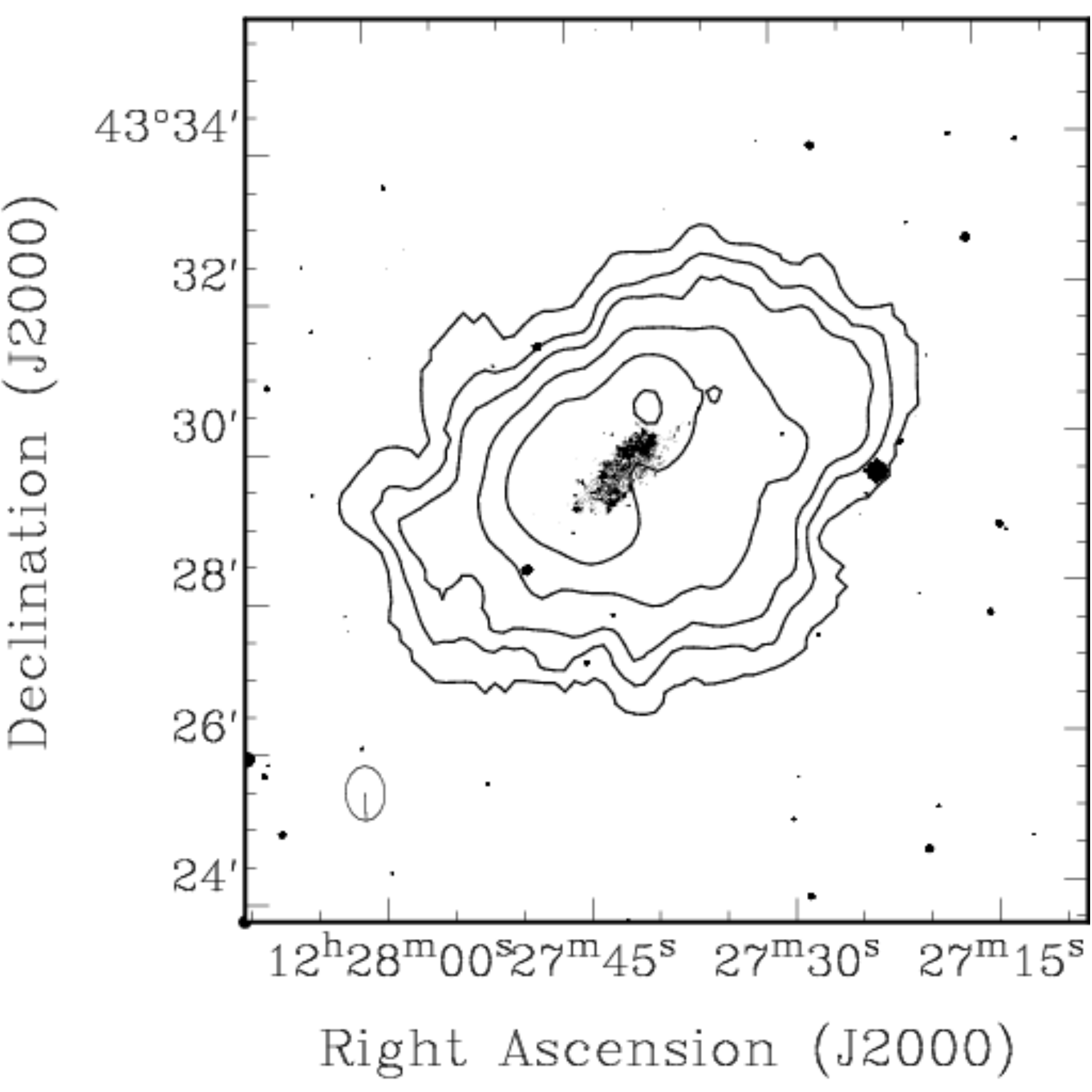}
\hskip 5mm
\includegraphics[height=0.17\textheight]{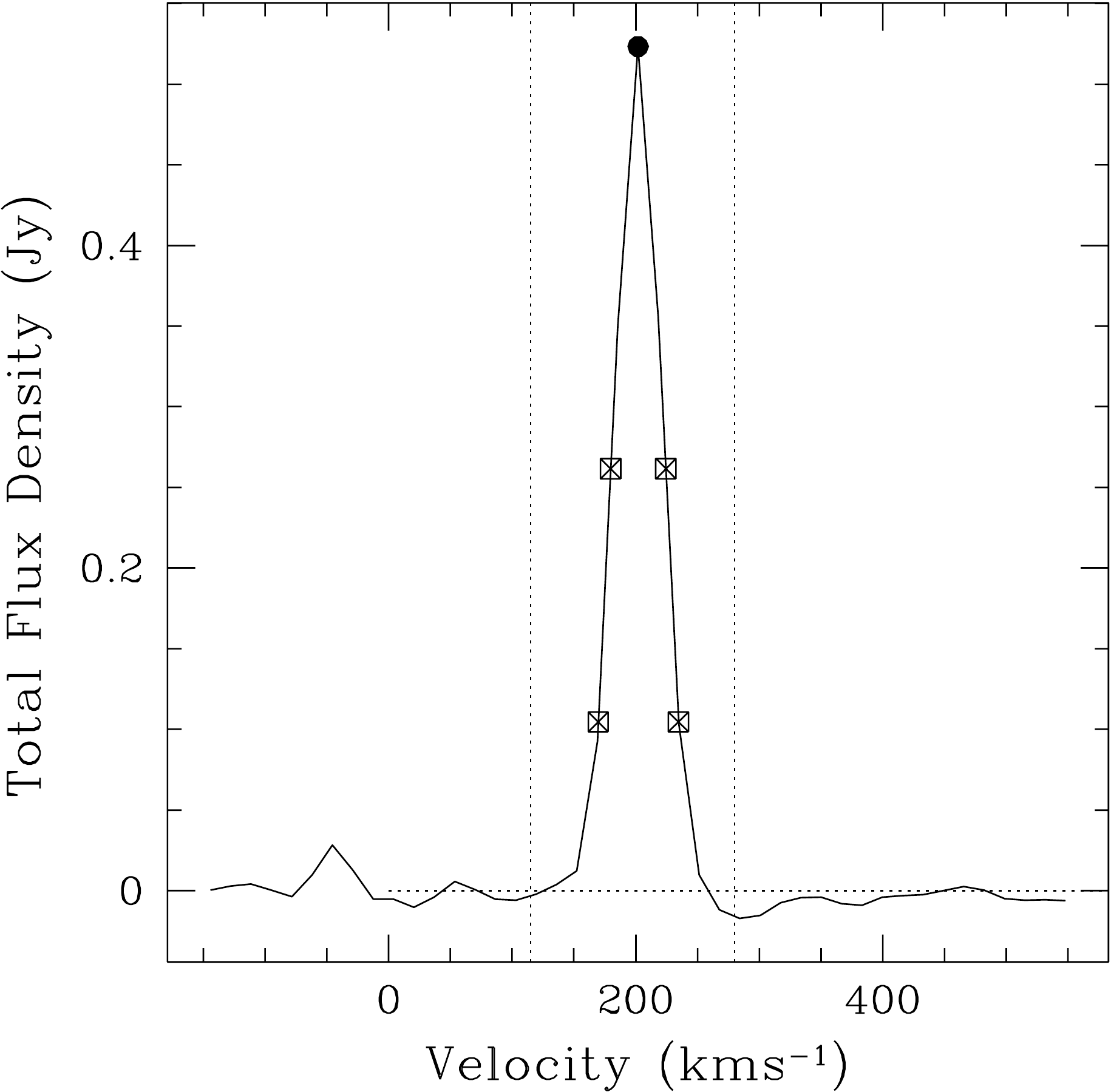}
\includegraphics[height=0.17\textheight]{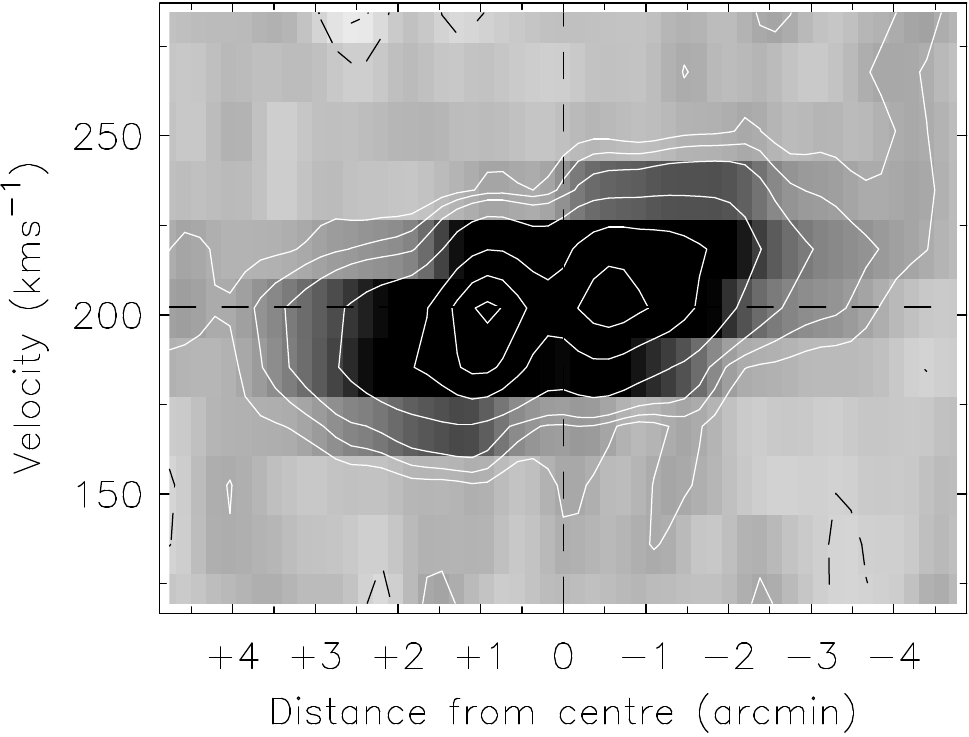}

\vskip 2mm
\centering
WSRT-CVn-50
\vskip 2mm
\includegraphics[width=0.25\textwidth]{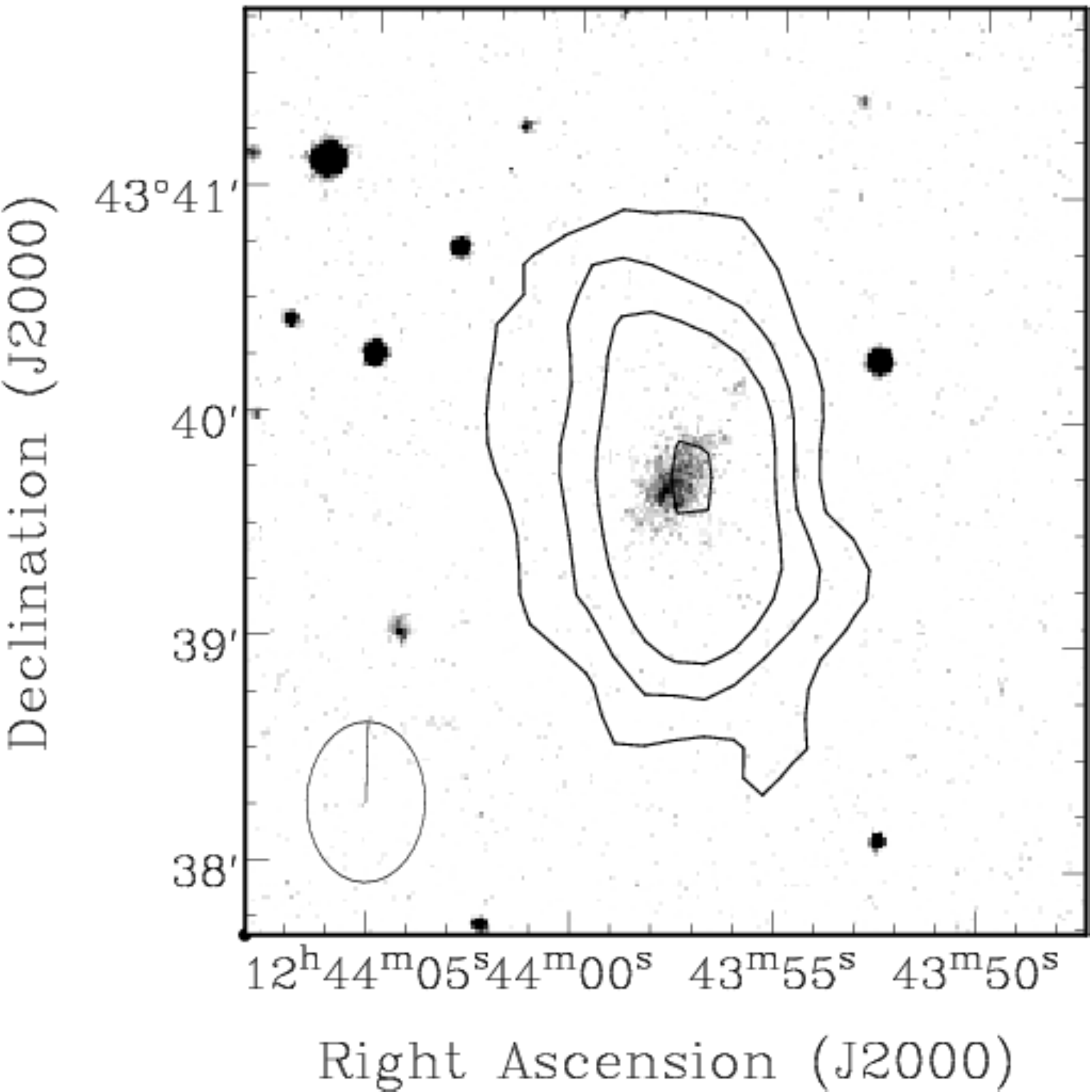}
\hskip 5mm
\includegraphics[height=0.17\textheight]{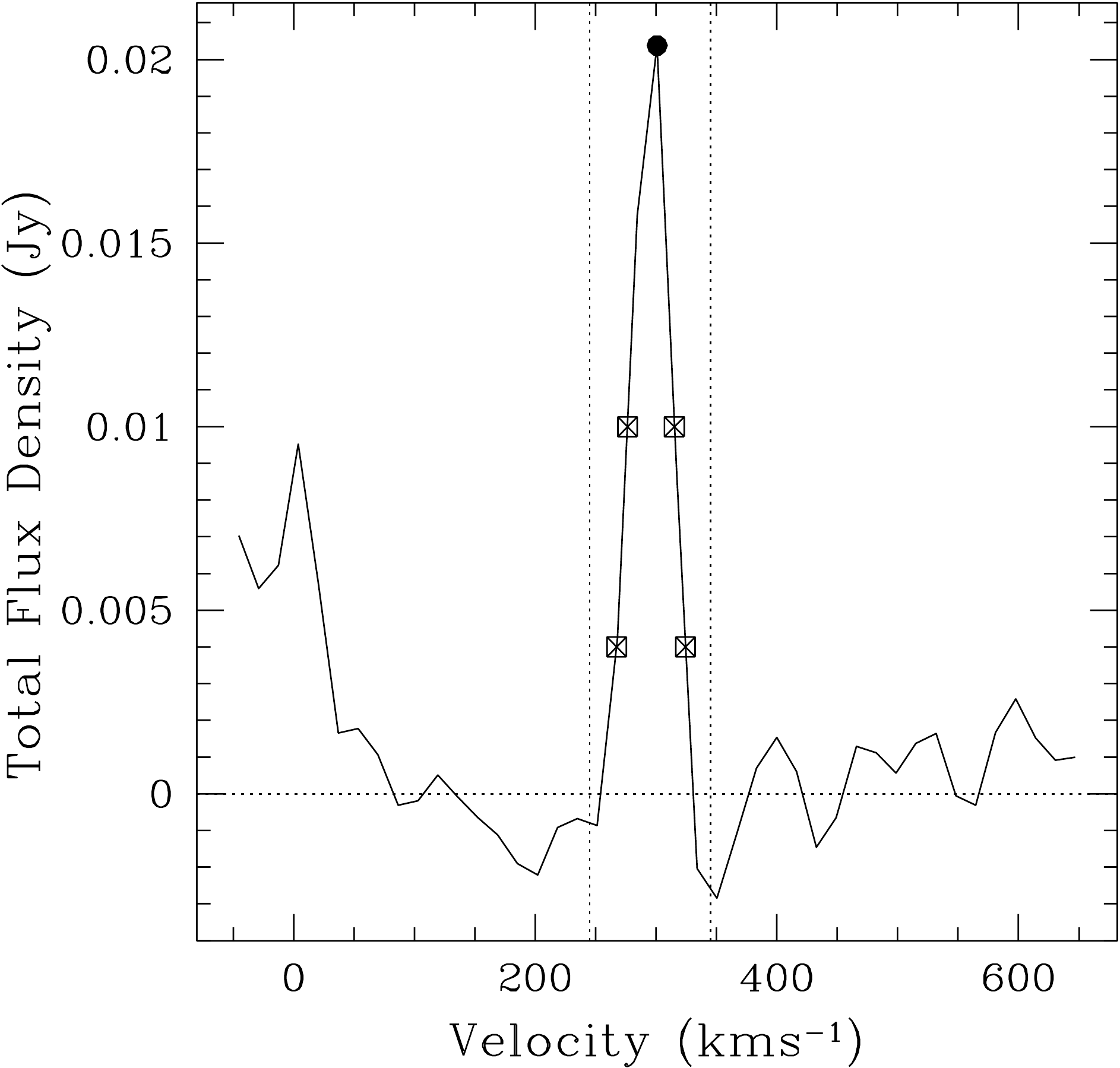}
\includegraphics[height=0.17\textheight]{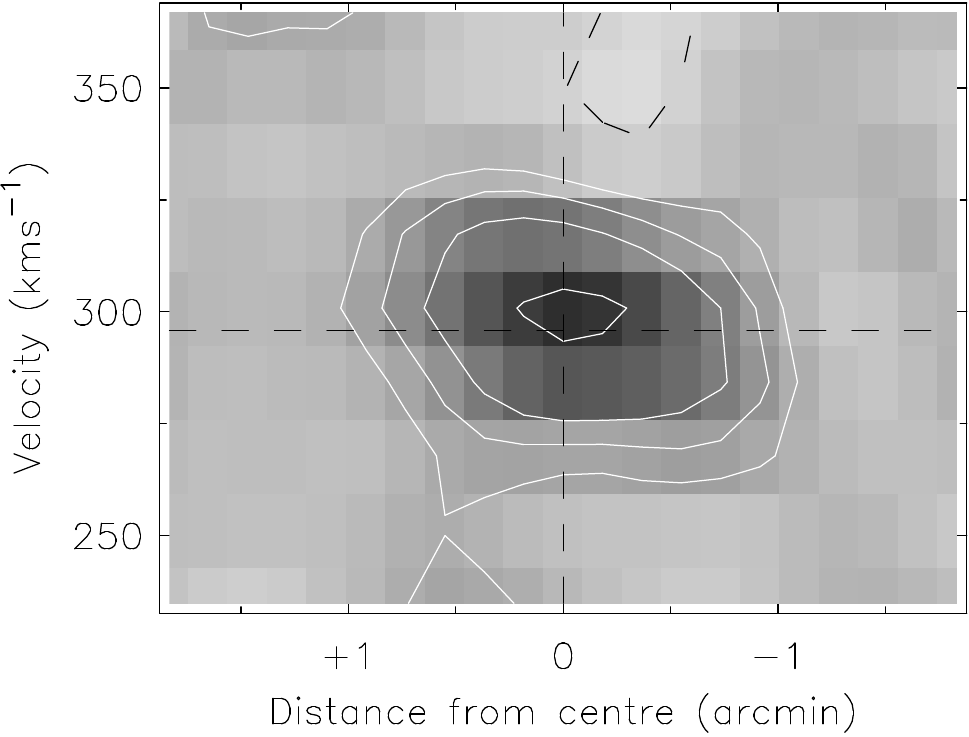}

\vskip 2mm
\centering
WSRT-CVn-51
\vskip 2mm
\includegraphics[width=0.25\textwidth]{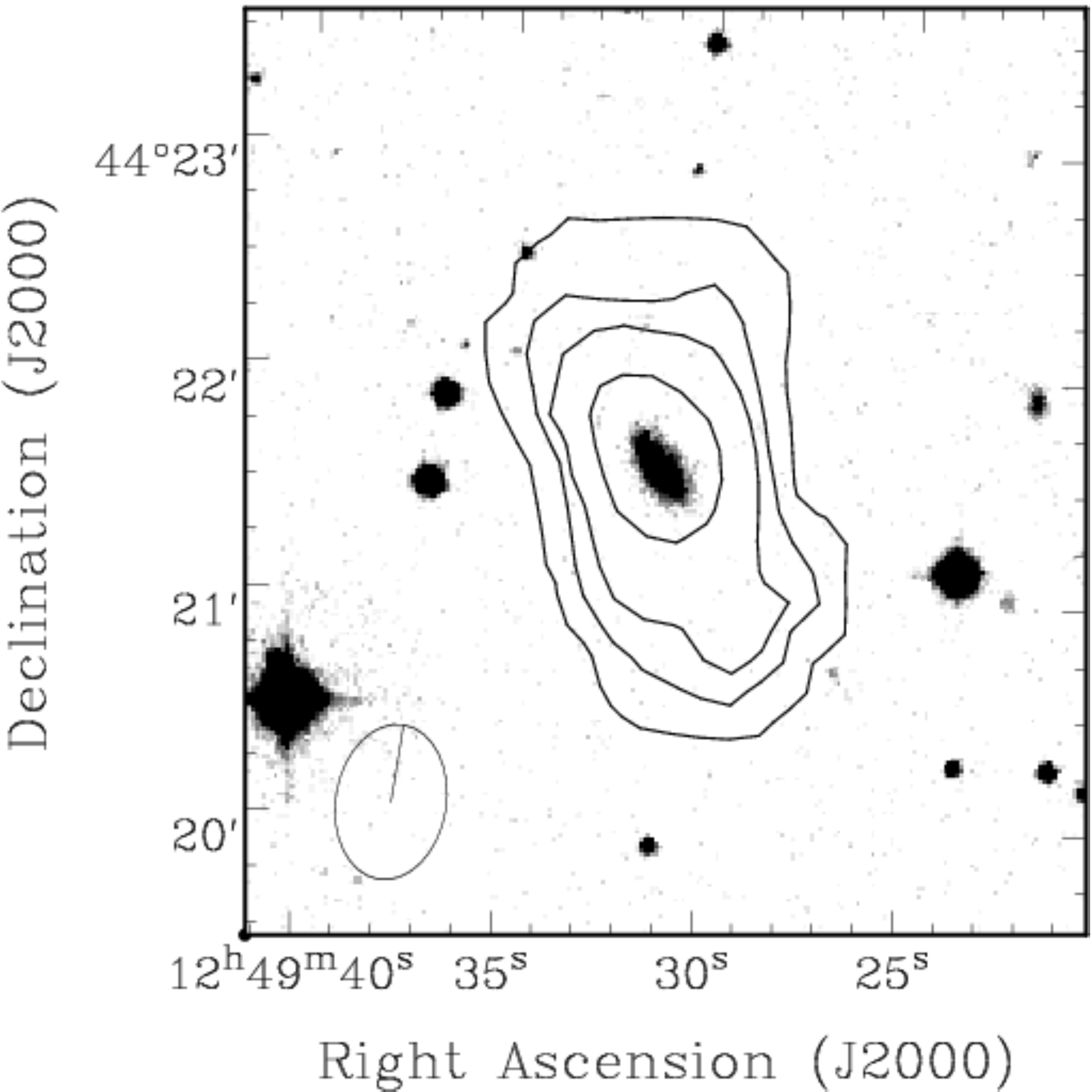}
\hskip 5mm
\includegraphics[height=0.17\textheight]{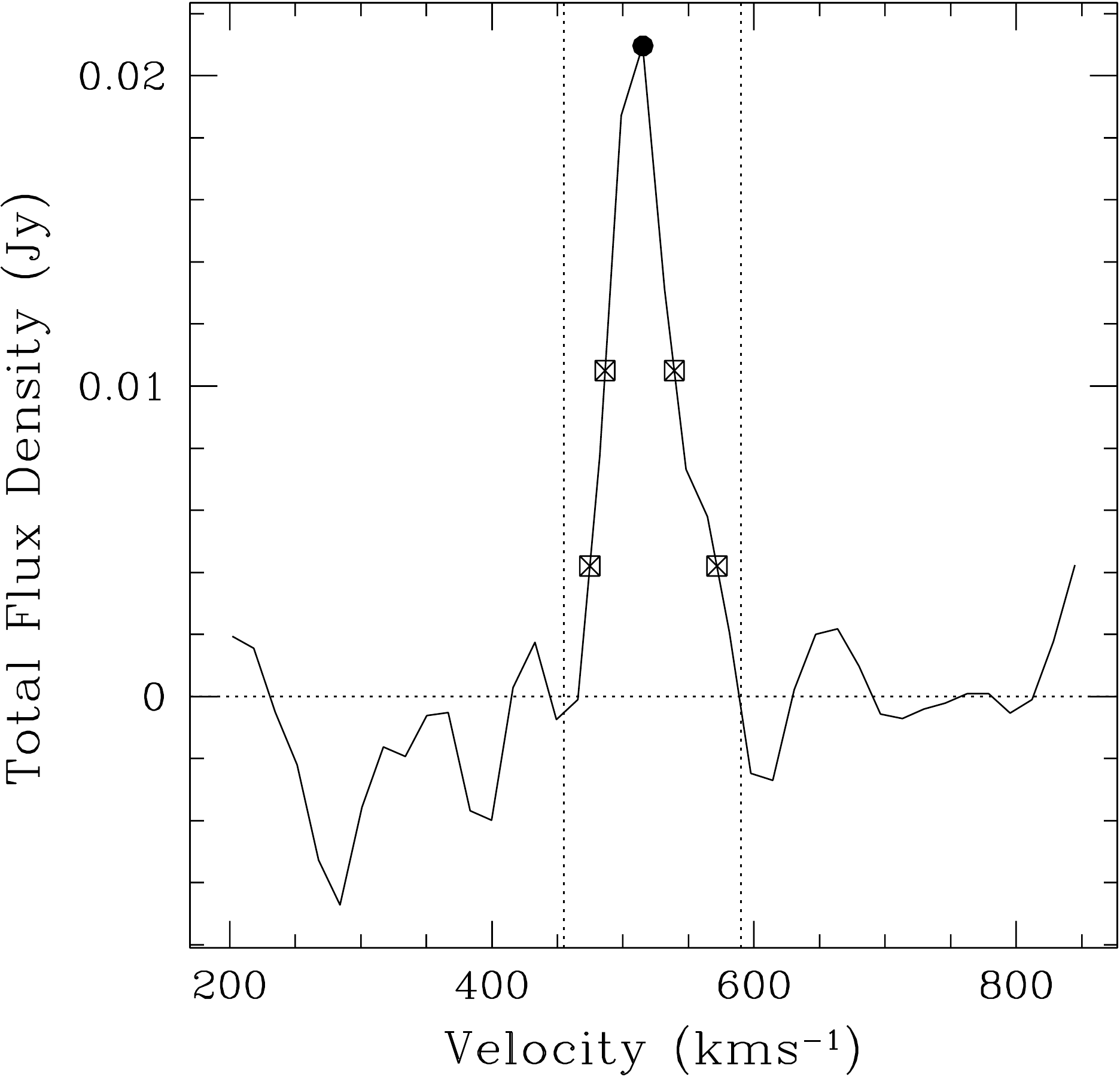}
\includegraphics[height=0.17\textheight]{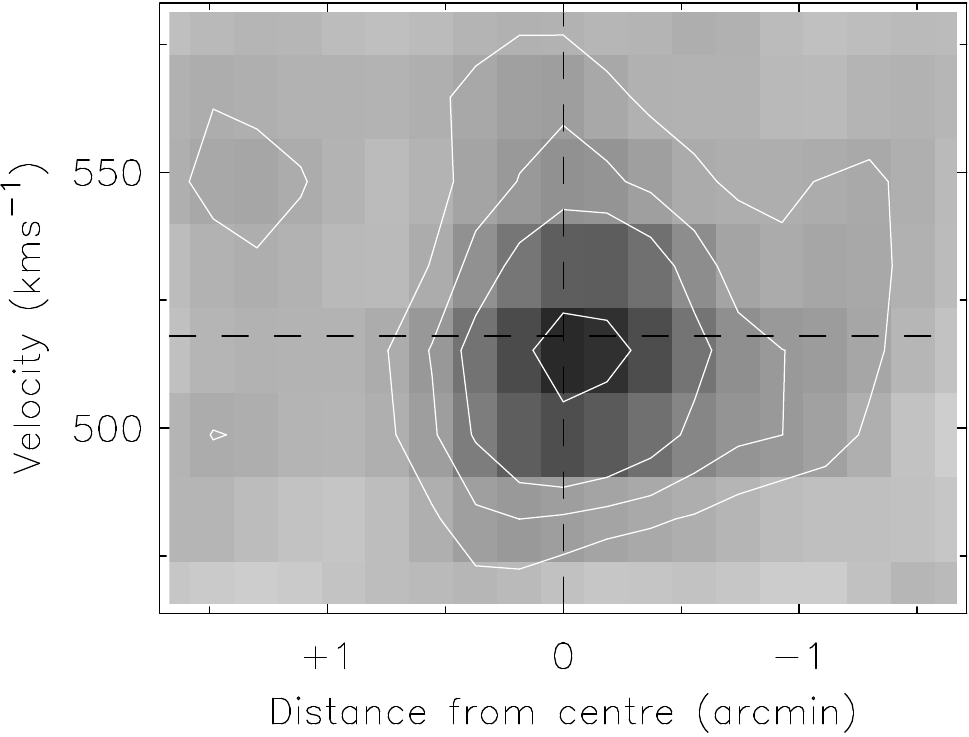}

\vskip 2mm
\centering
WSRT-CVn-52
\vskip 2mm
\includegraphics[width=0.25\textwidth]{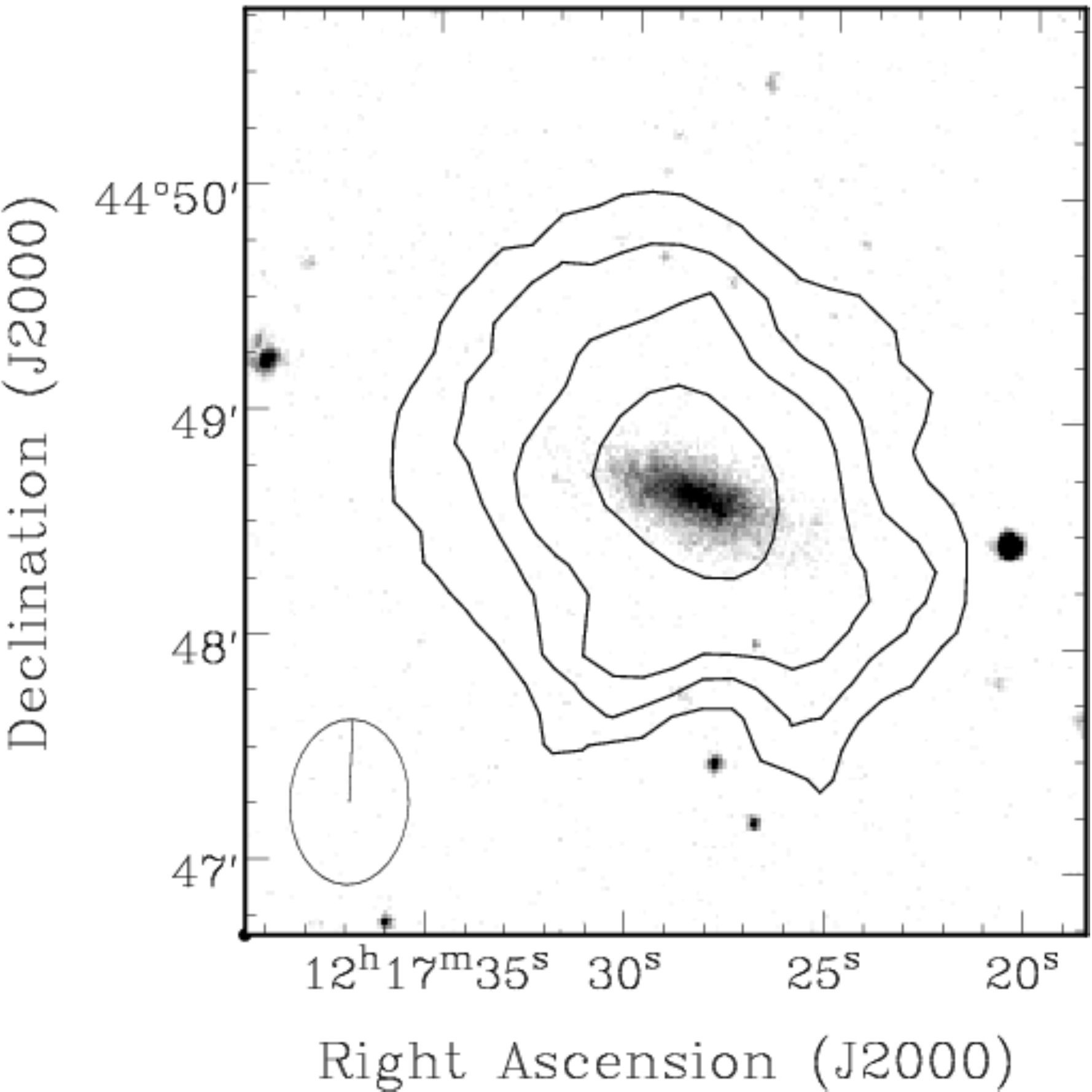}
\hskip 5mm
\includegraphics[height=0.17\textheight]{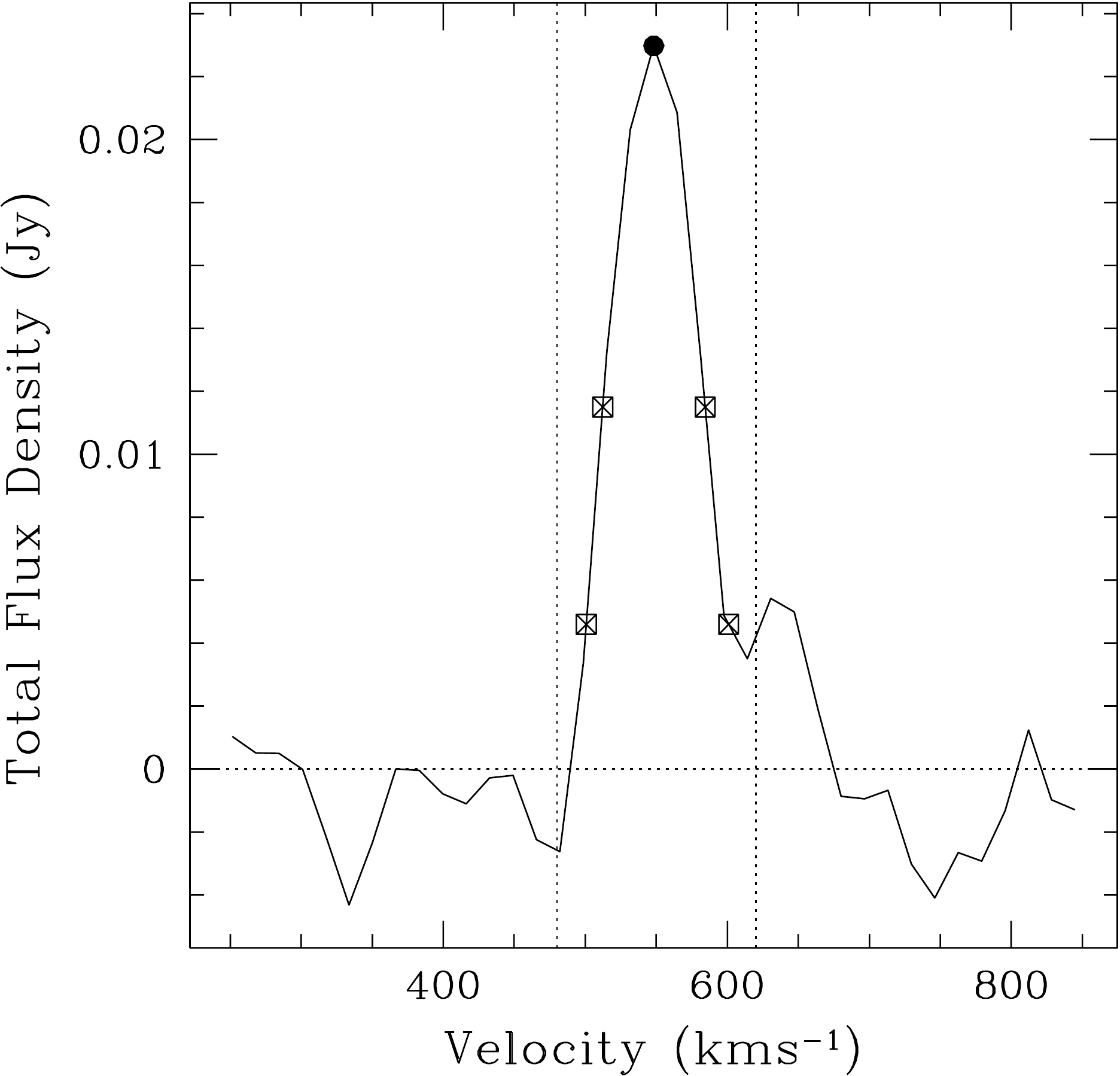}
\includegraphics[height=0.17\textheight]{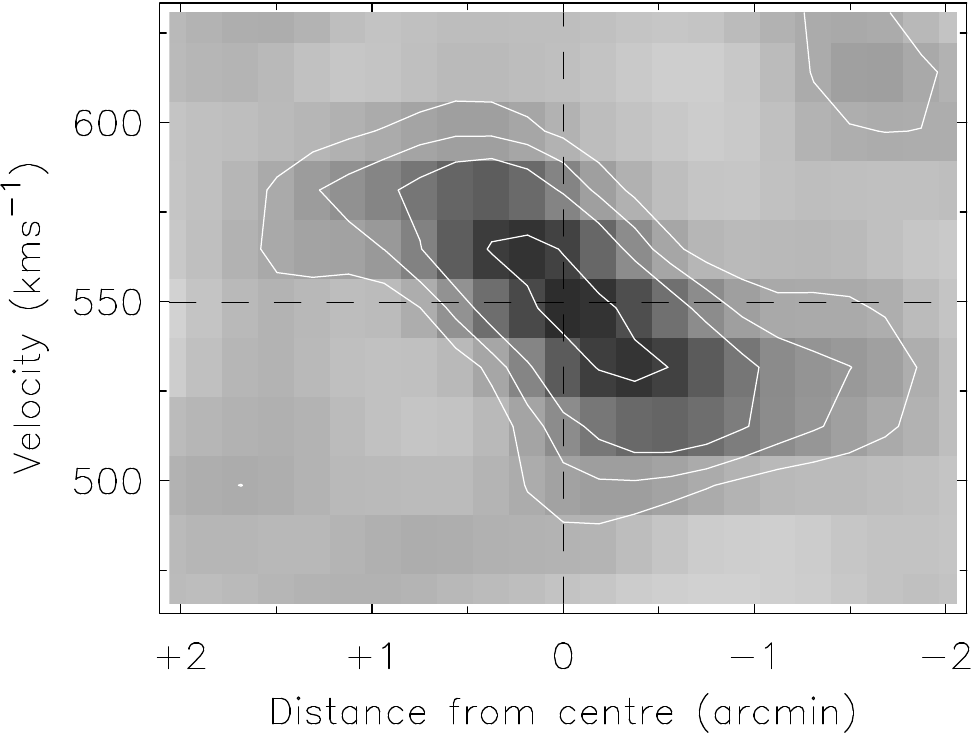}

\end{figure}

\clearpage

\addtocounter{figure}{-1}
\begin{figure}

\vskip 2mm
\centering
WSRT-CVn-53
\vskip 2mm
\includegraphics[width=0.25\textwidth]{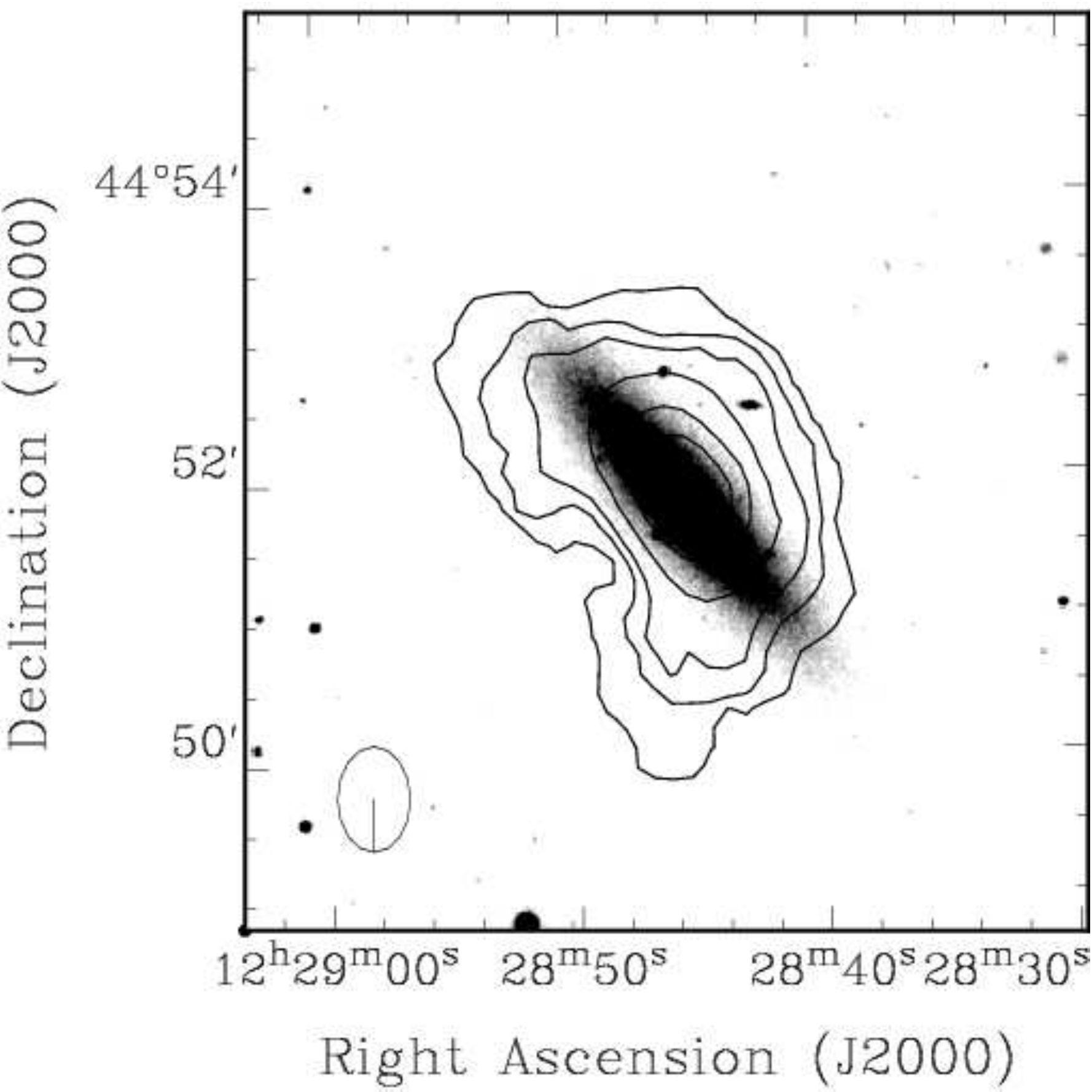}
\hskip 5mm
\includegraphics[height=0.17\textheight]{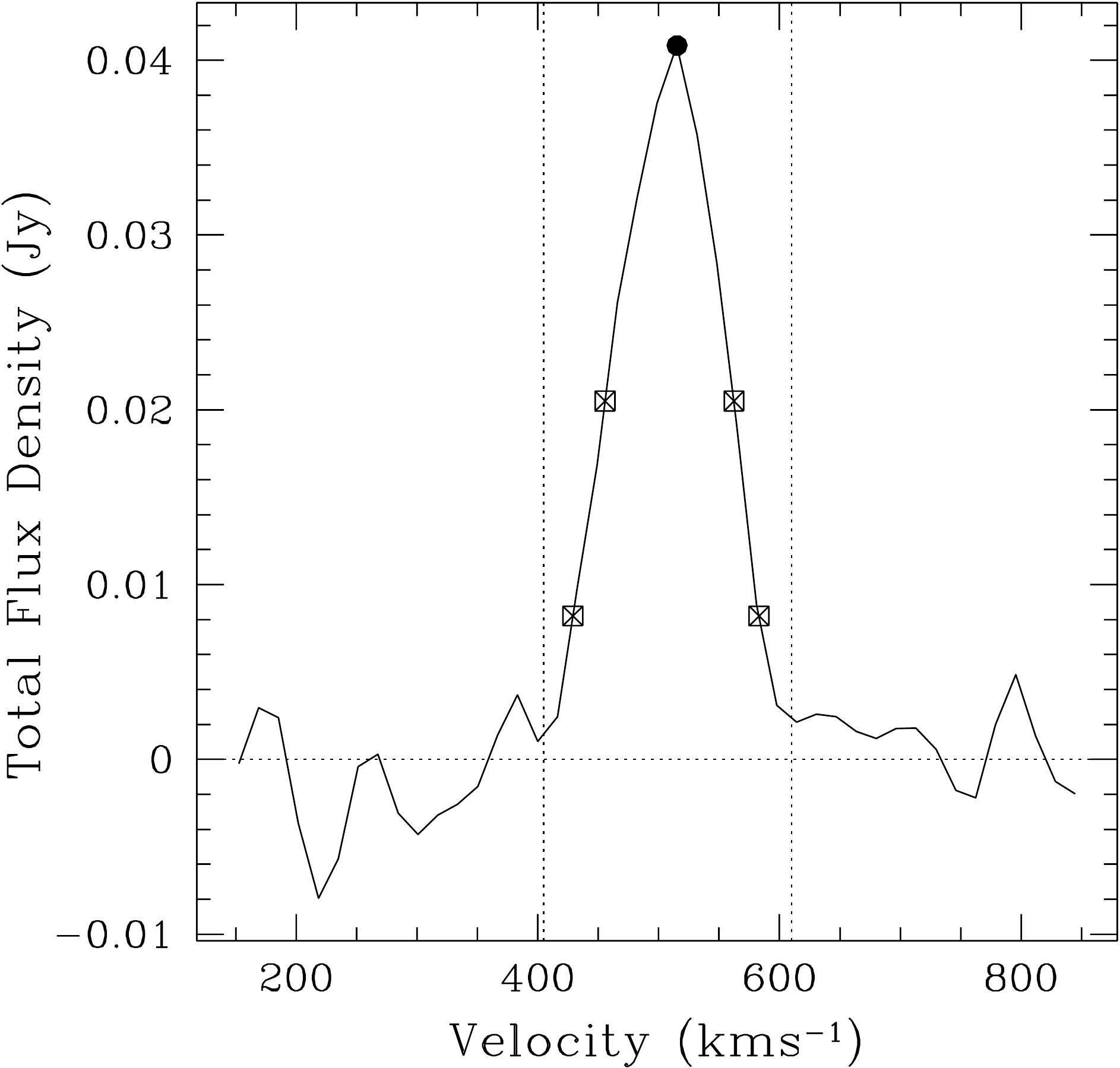}
\includegraphics[height=0.17\textheight]{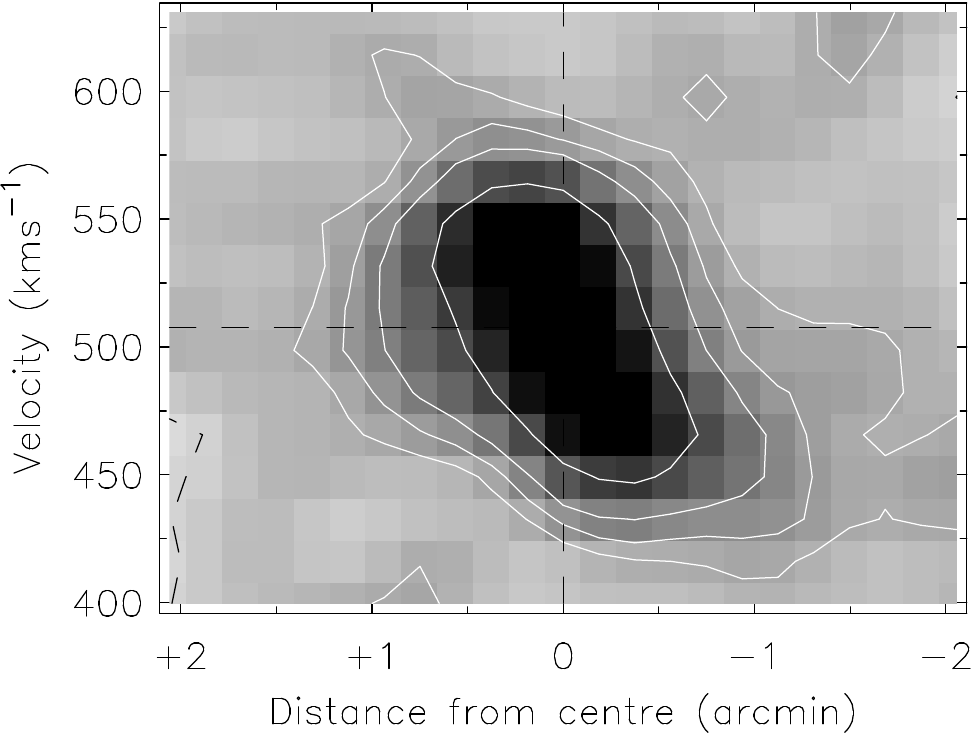}

\vskip 2mm
\centering
WSRT-CVn-54
\vskip 2mm
\includegraphics[width=0.25\textwidth]{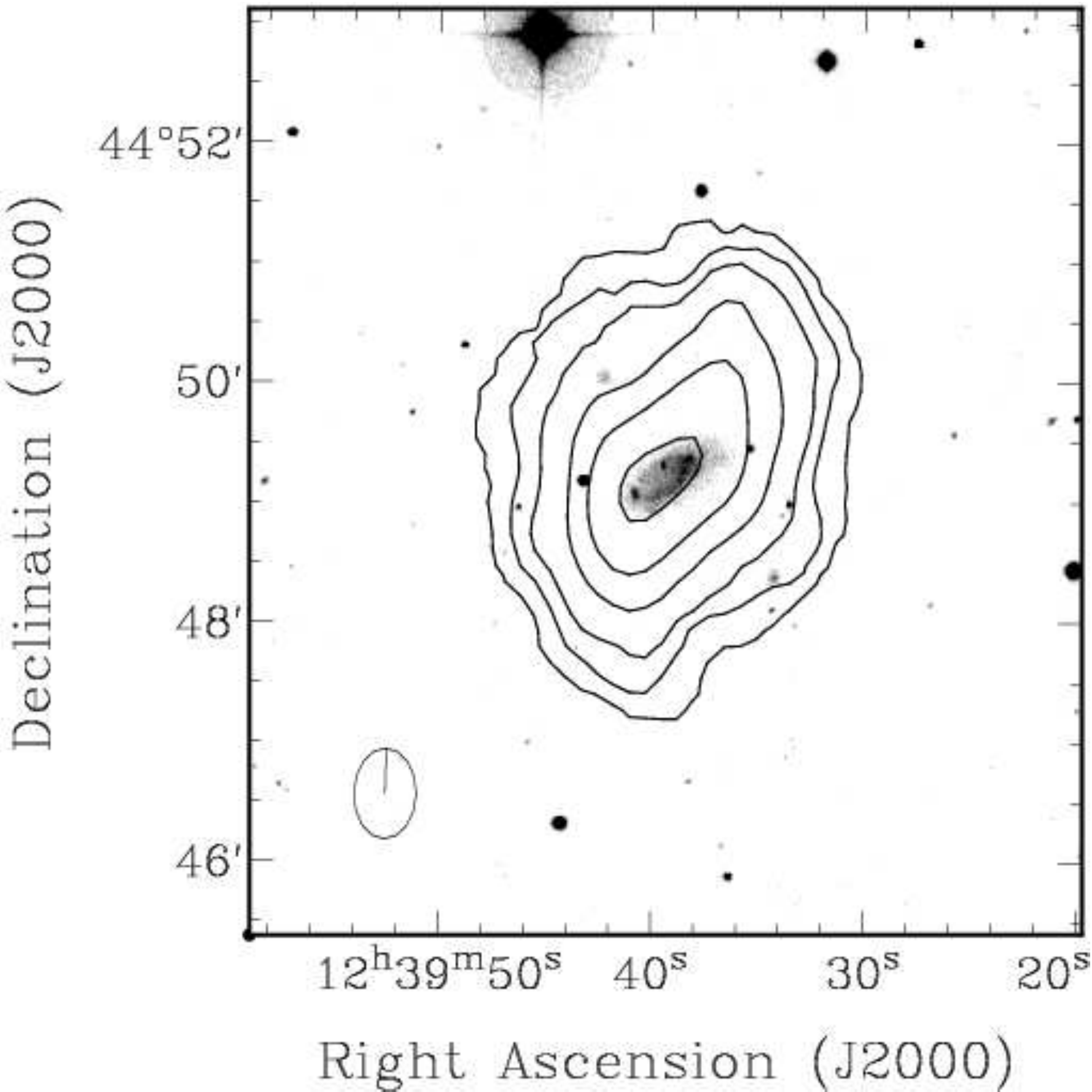}
\hskip 5mm
\includegraphics[height=0.17\textheight]{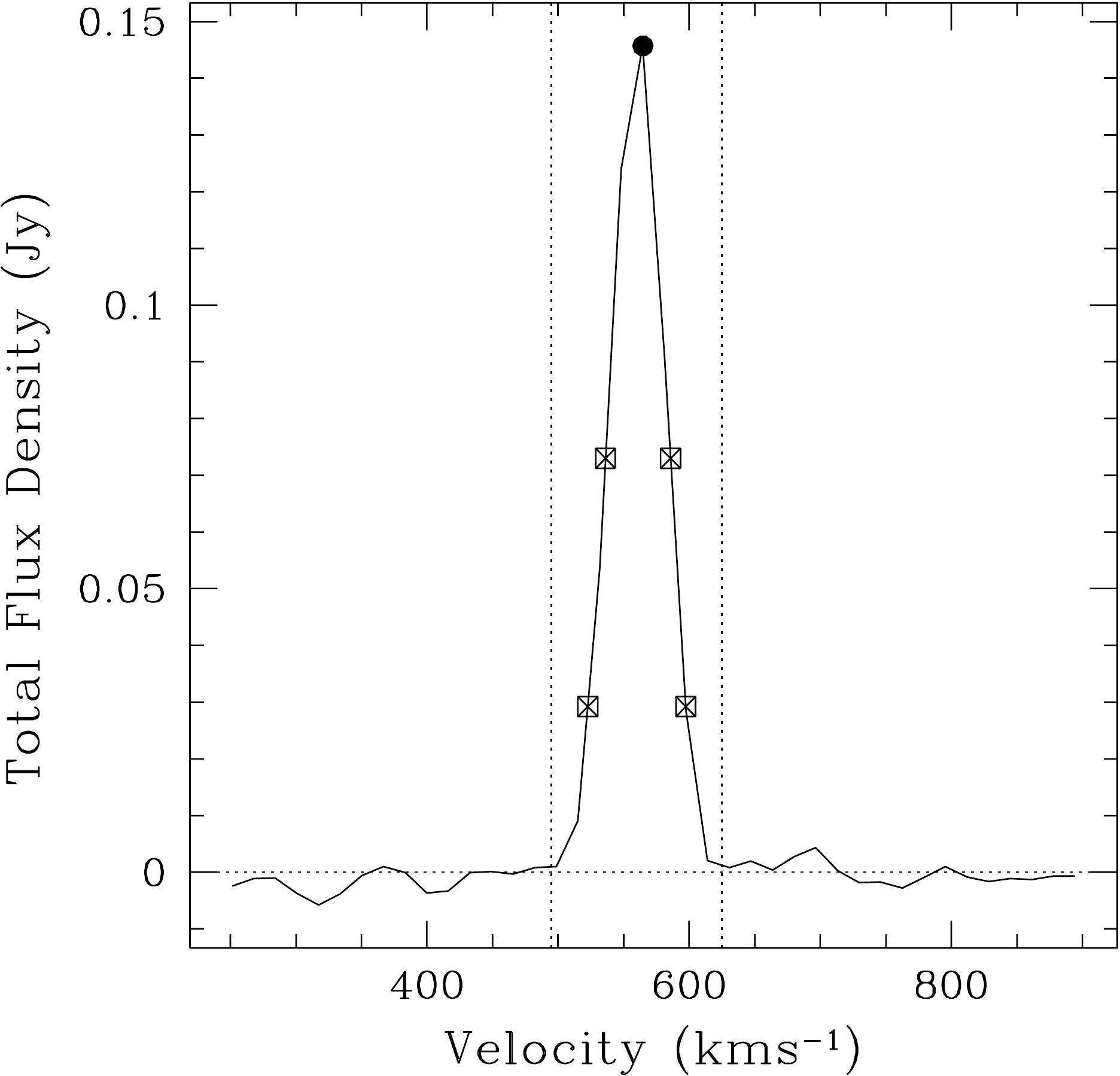}
\includegraphics[height=0.17\textheight]{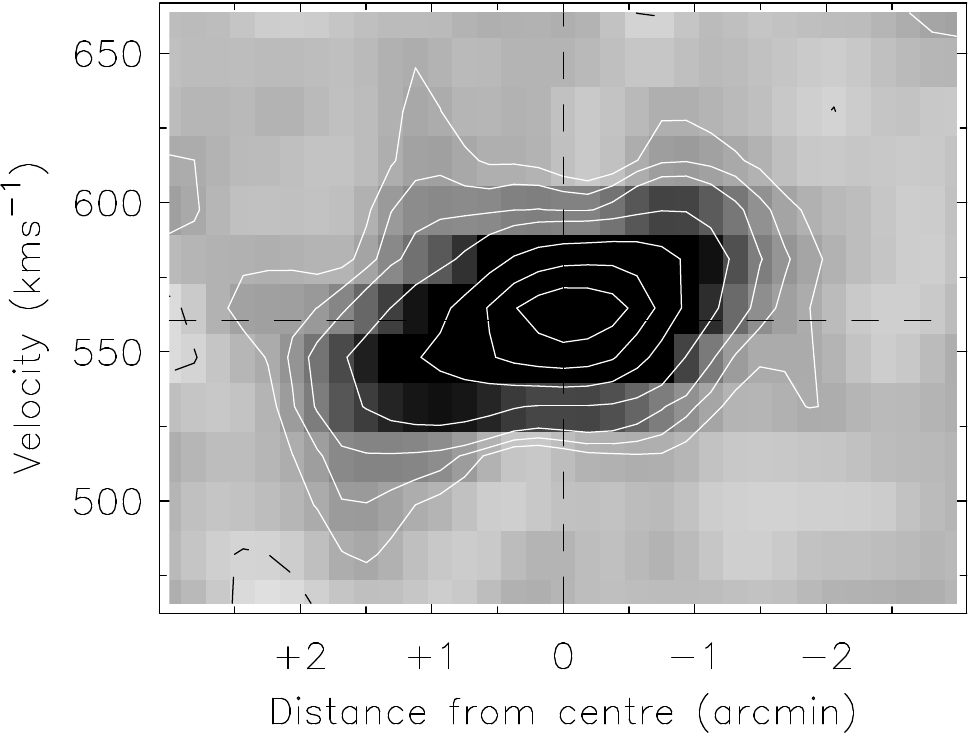}

\vskip 2mm
\centering
WSRT-CVn-55
\vskip 2mm
\includegraphics[width=0.25\textwidth]{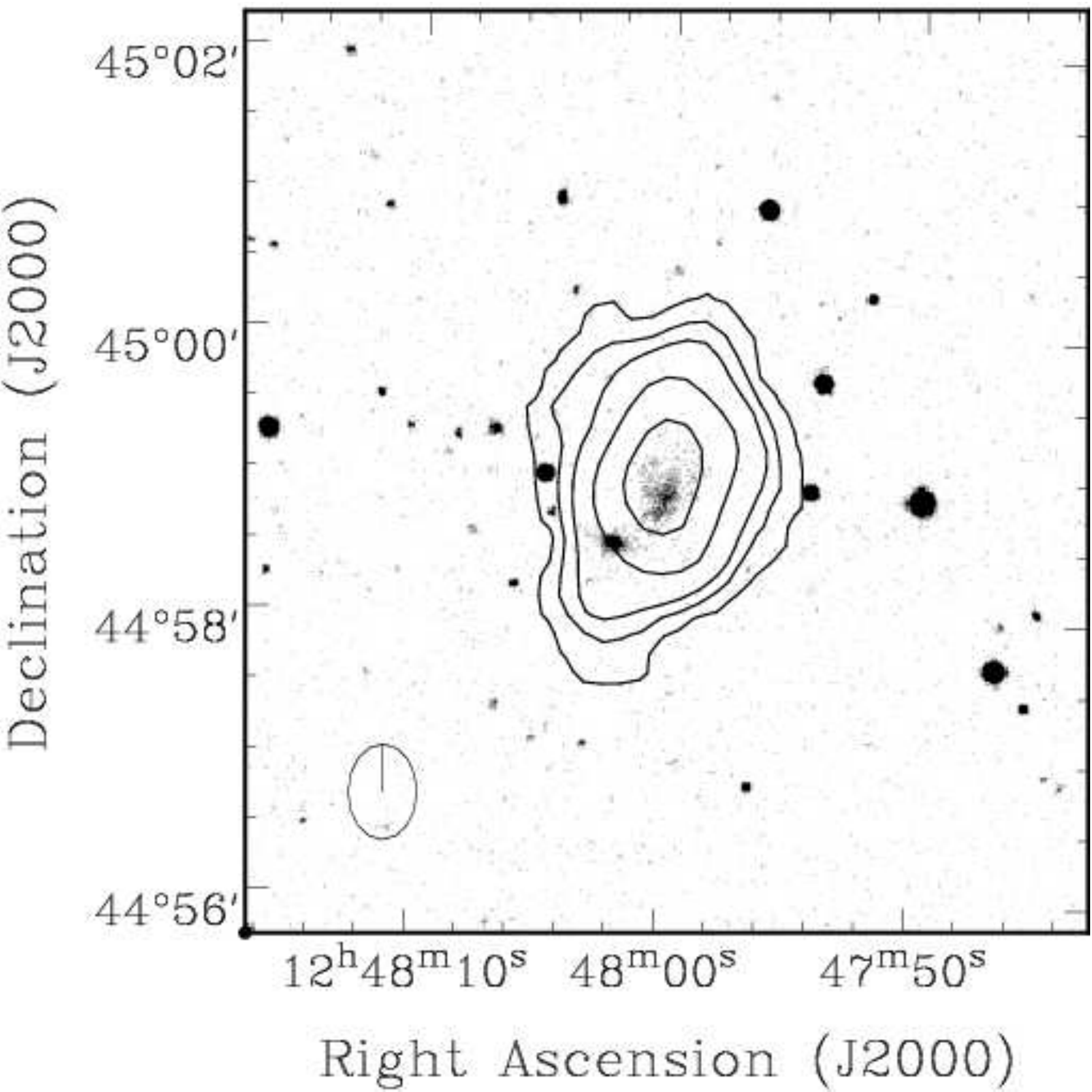}
\hskip 5mm
\includegraphics[height=0.17\textheight]{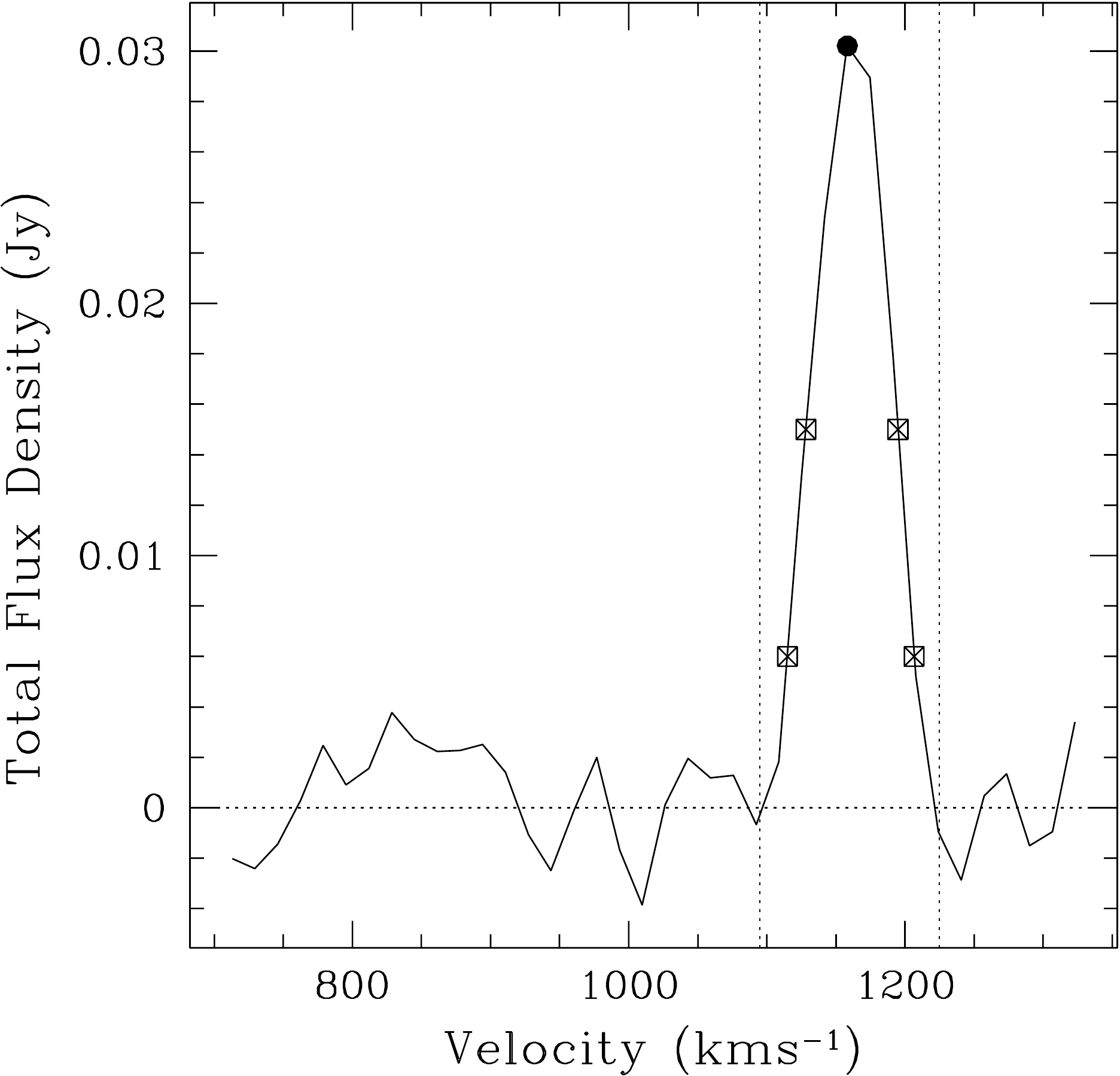}
\includegraphics[height=0.17\textheight]{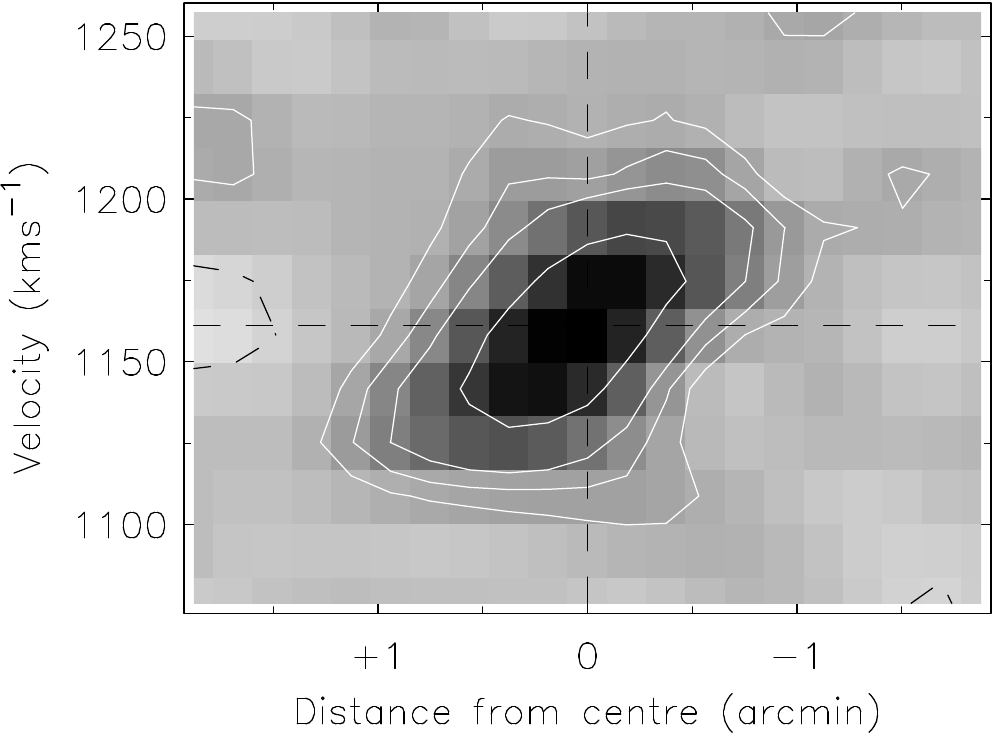}

\vskip 2mm
\centering
WSRT-CVn-56
\vskip 2mm
\includegraphics[width=0.25\textwidth]{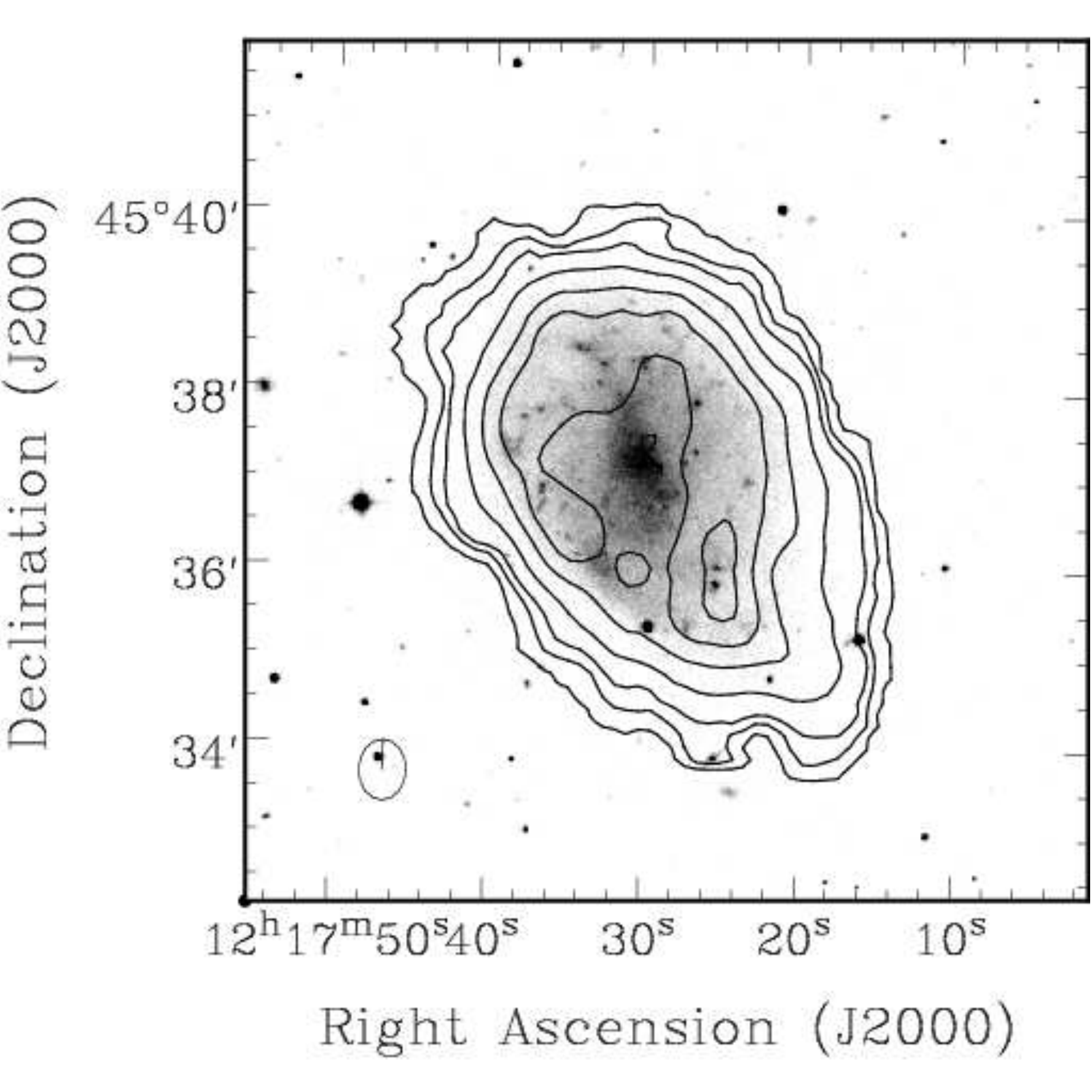}
\hskip 5mm
\includegraphics[height=0.17\textheight]{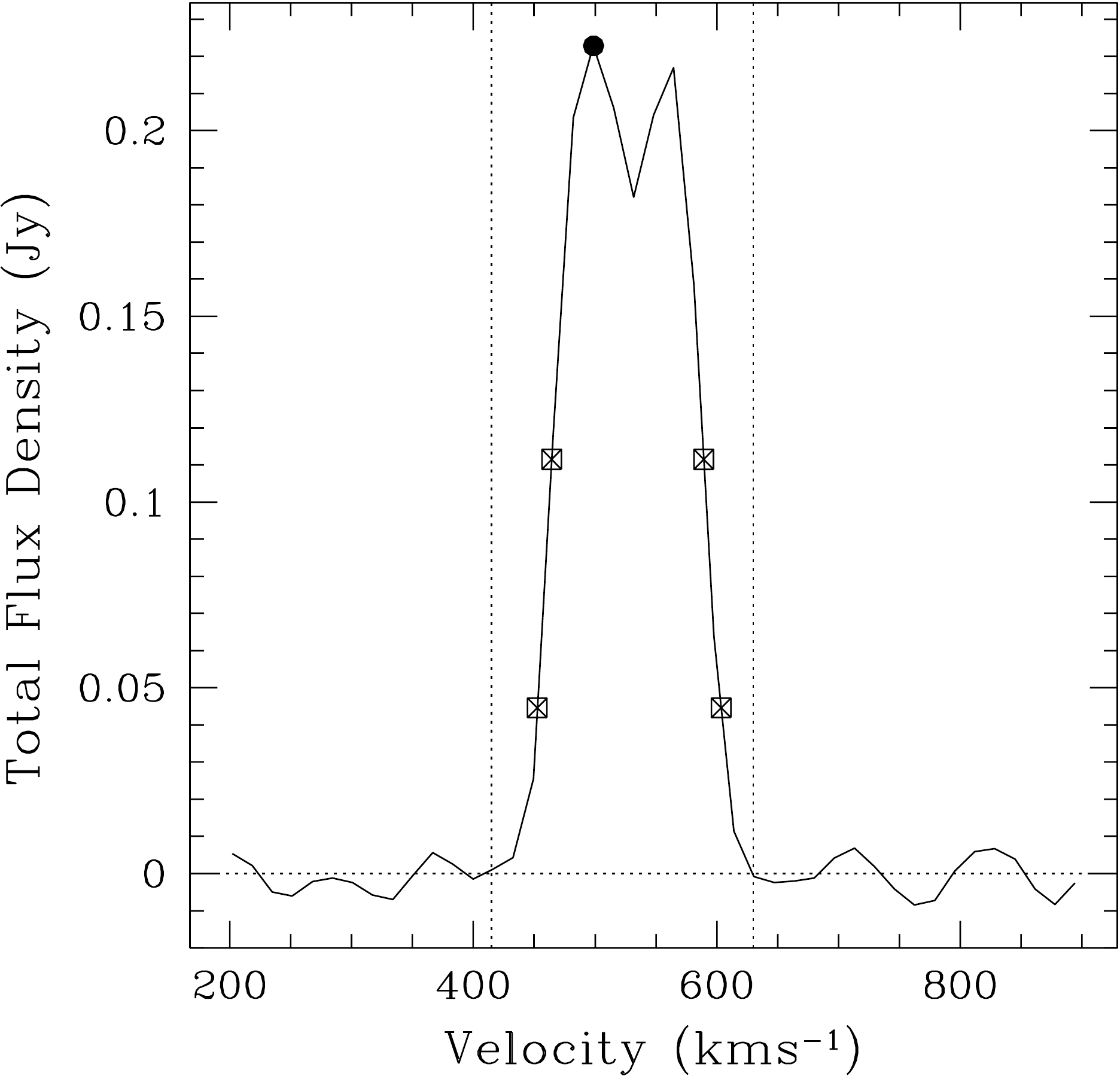}
\includegraphics[height=0.17\textheight]{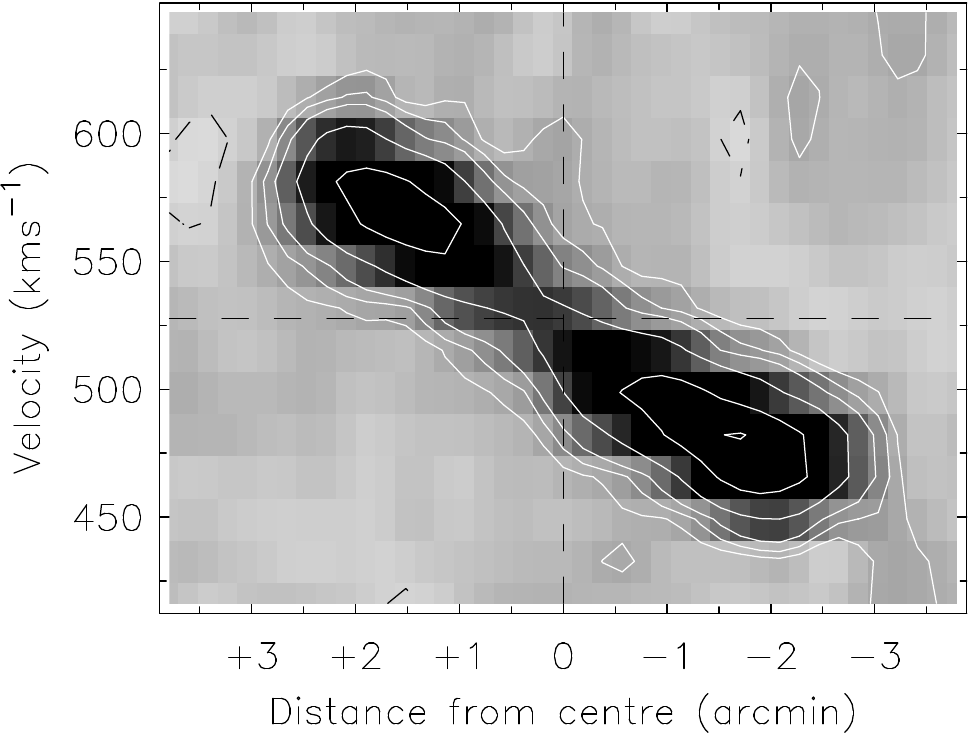}

\end{figure}

\clearpage

\addtocounter{figure}{-1}
\begin{figure}

\vskip 2mm
\centering
WSRT-CVn-57
\vskip 2mm
\includegraphics[width=0.25\textwidth]{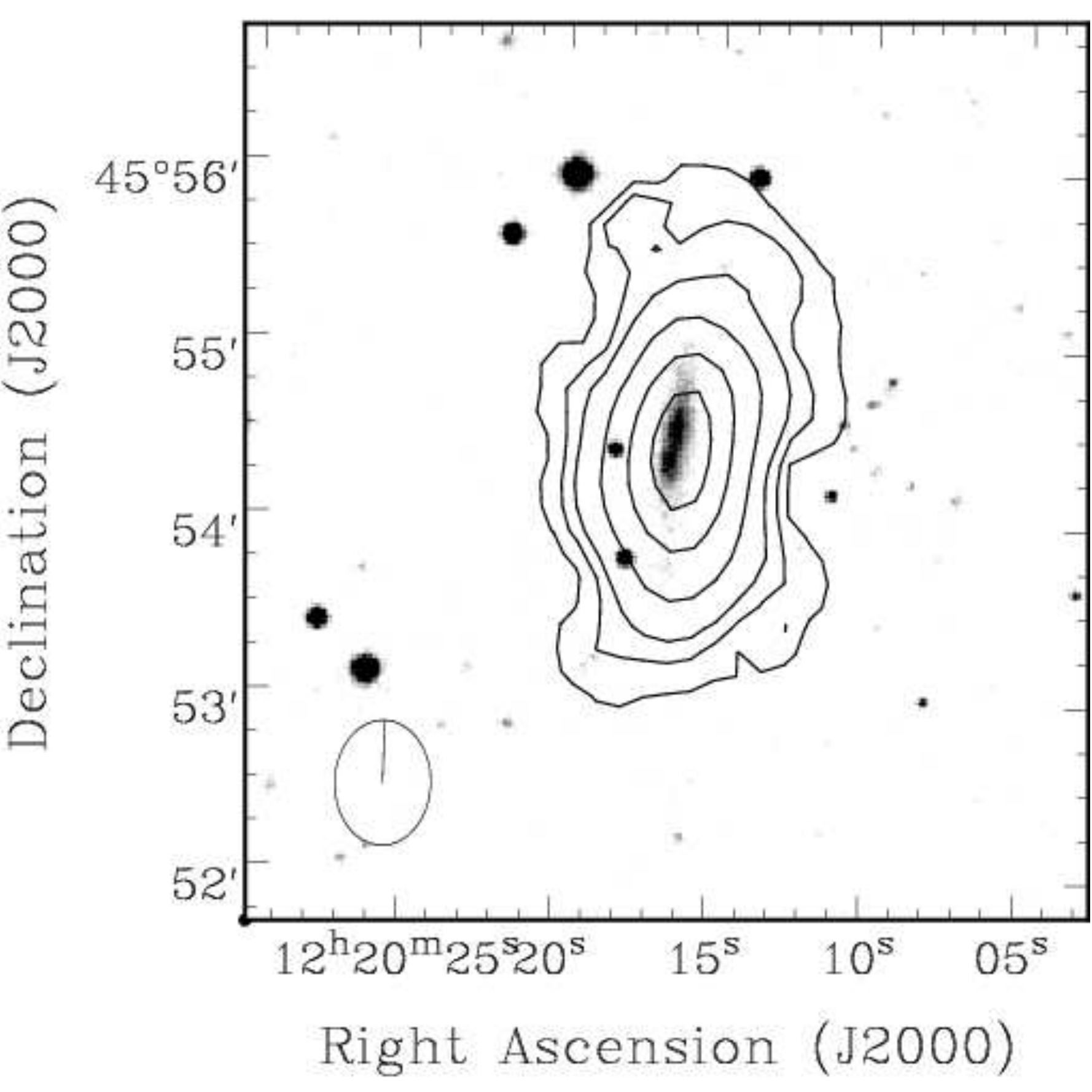}
\hskip 5mm
\includegraphics[height=0.17\textheight]{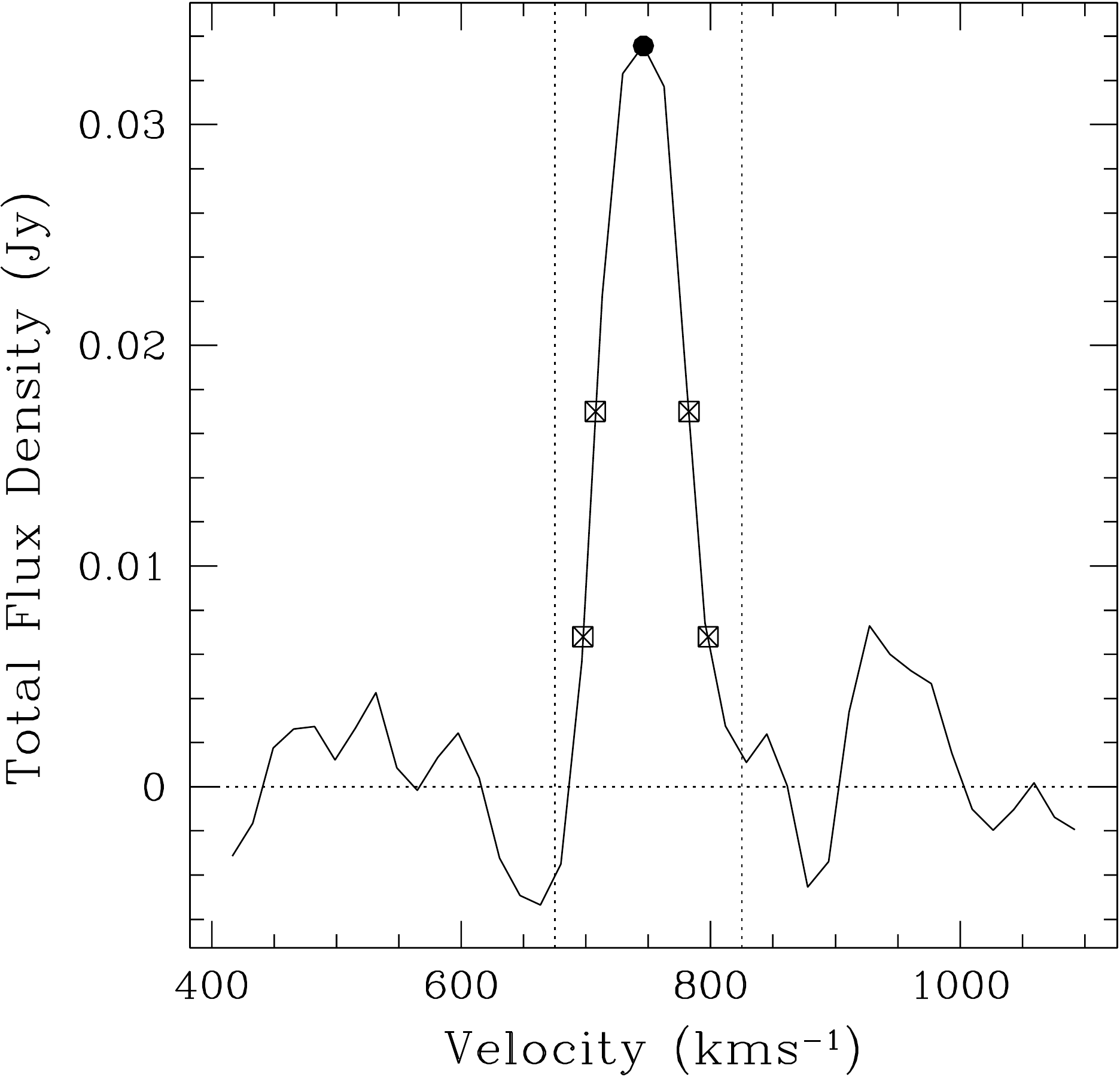}
\includegraphics[height=0.17\textheight]{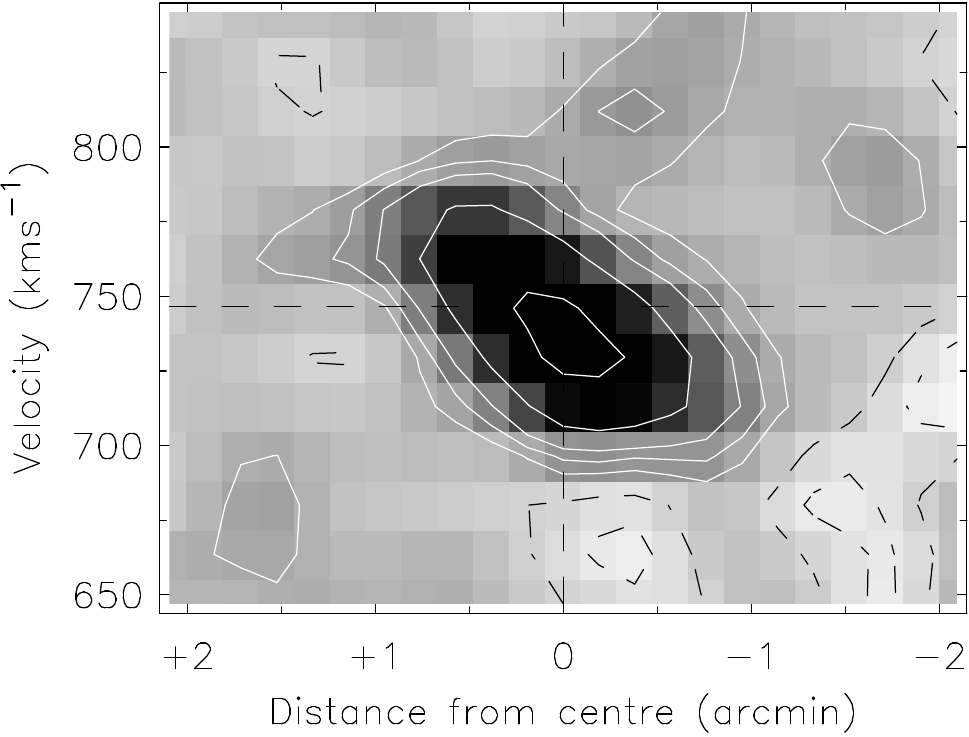}

\vskip 2mm
\centering
WSRT-CVn-58
\vskip 2mm
\includegraphics[width=0.25\textwidth]{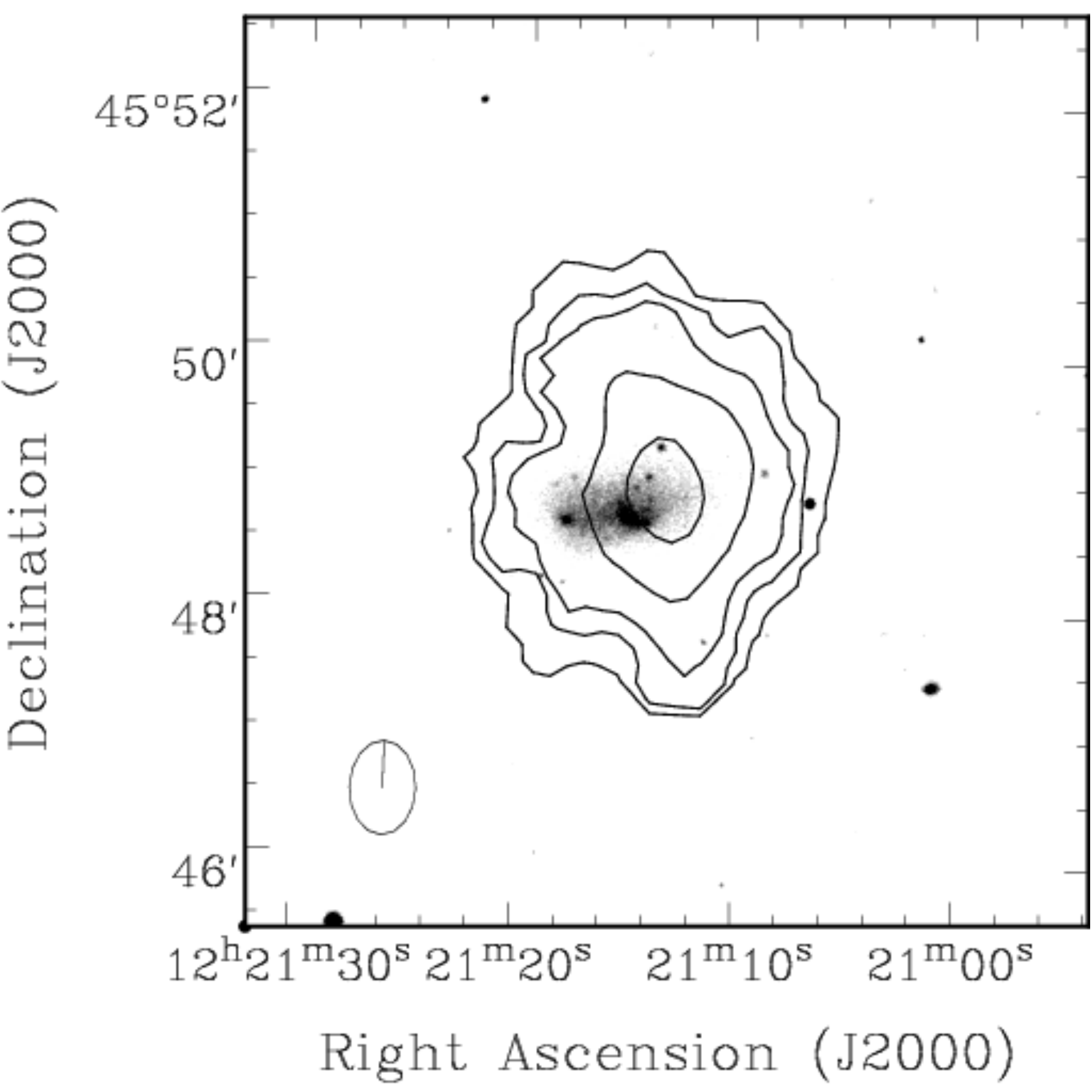}
\hskip 5mm
\includegraphics[height=0.17\textheight]{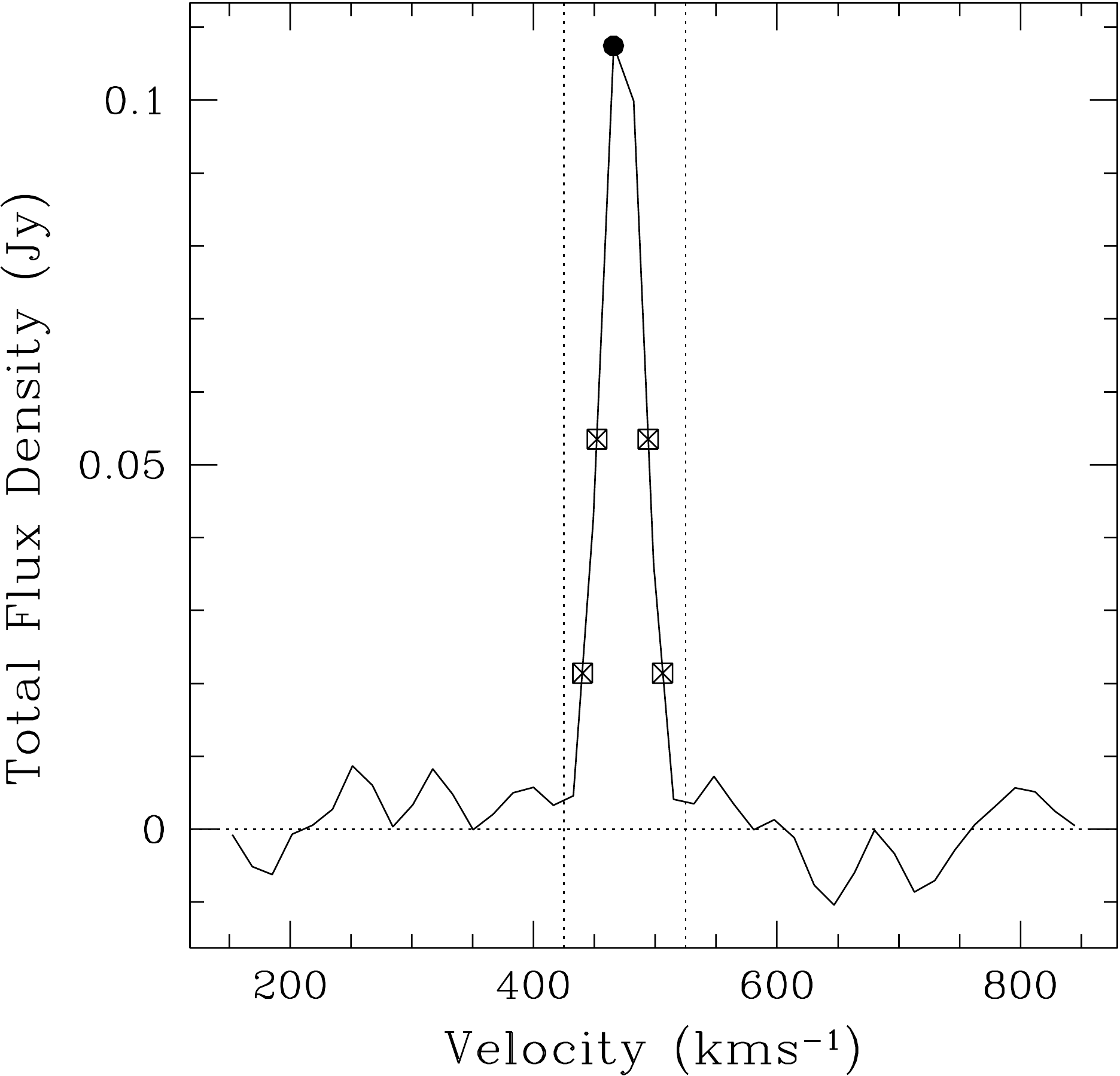}
\includegraphics[height=0.17\textheight]{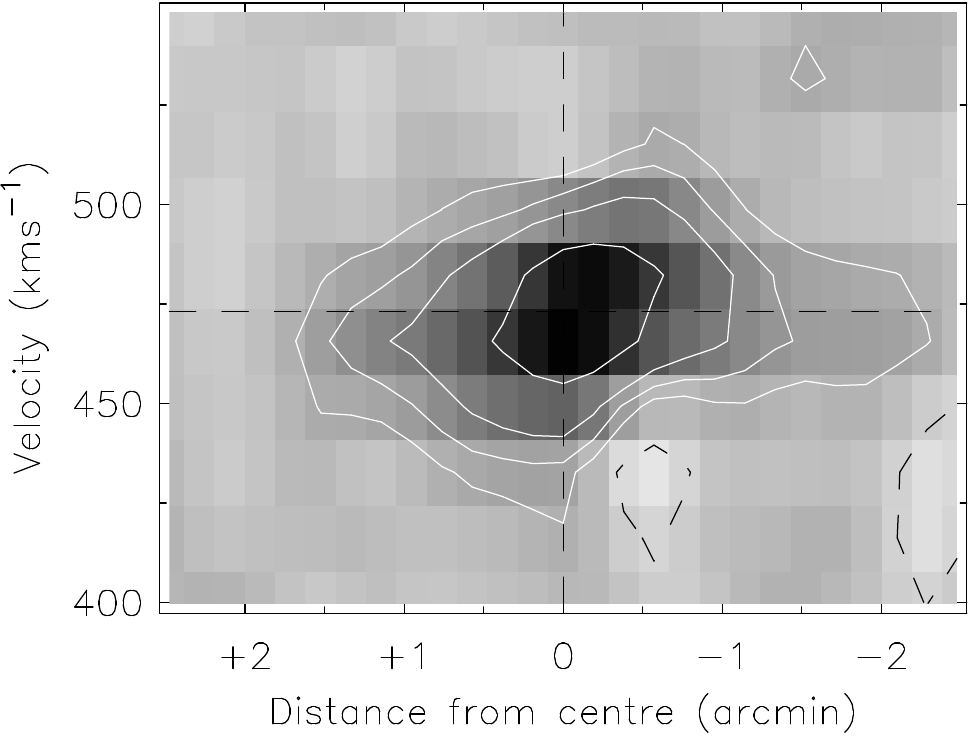}

\vskip 2mm
\centering
WSRT-CVn-59
\vskip 2mm
\includegraphics[width=0.25\textwidth]{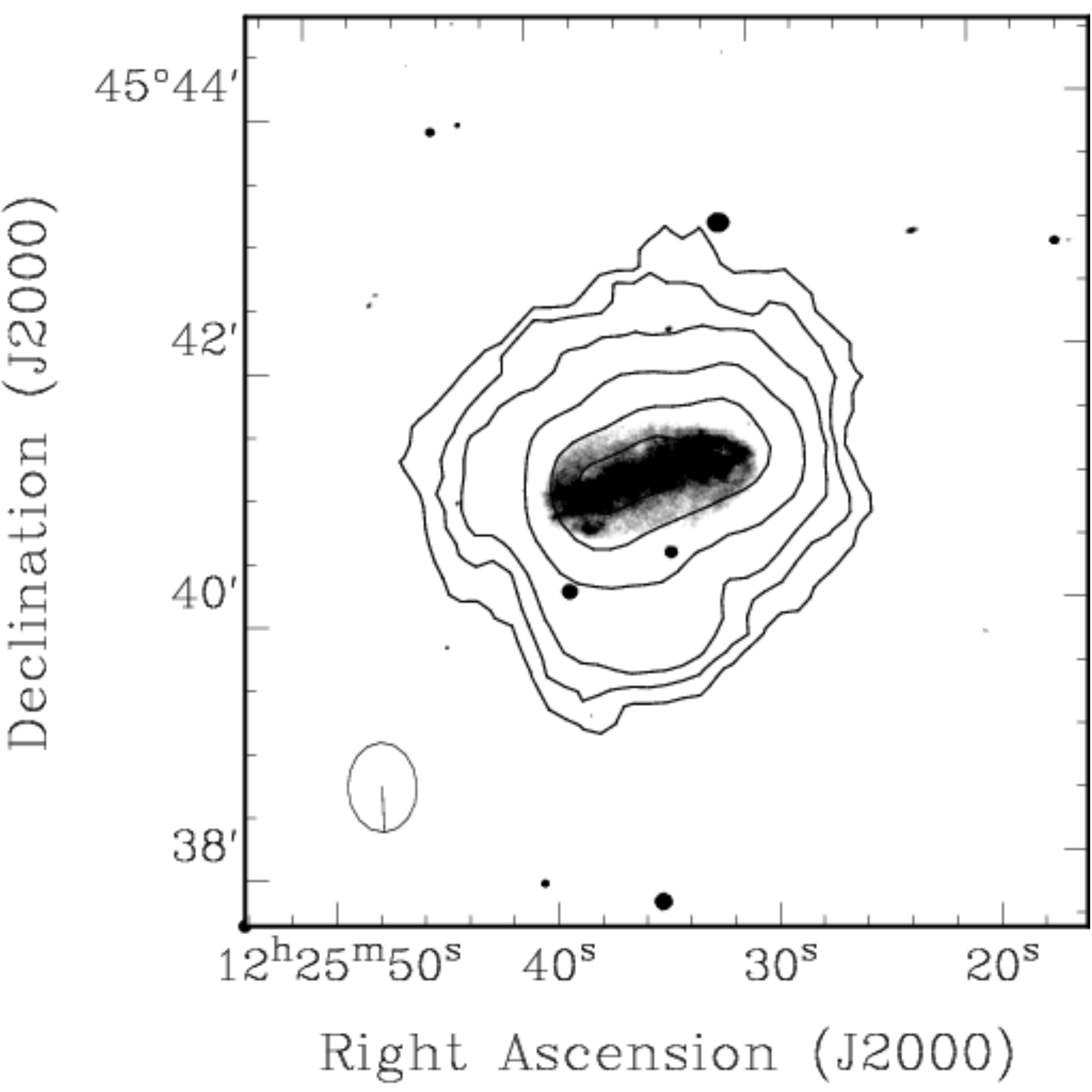}
\hskip 5mm
\includegraphics[height=0.17\textheight]{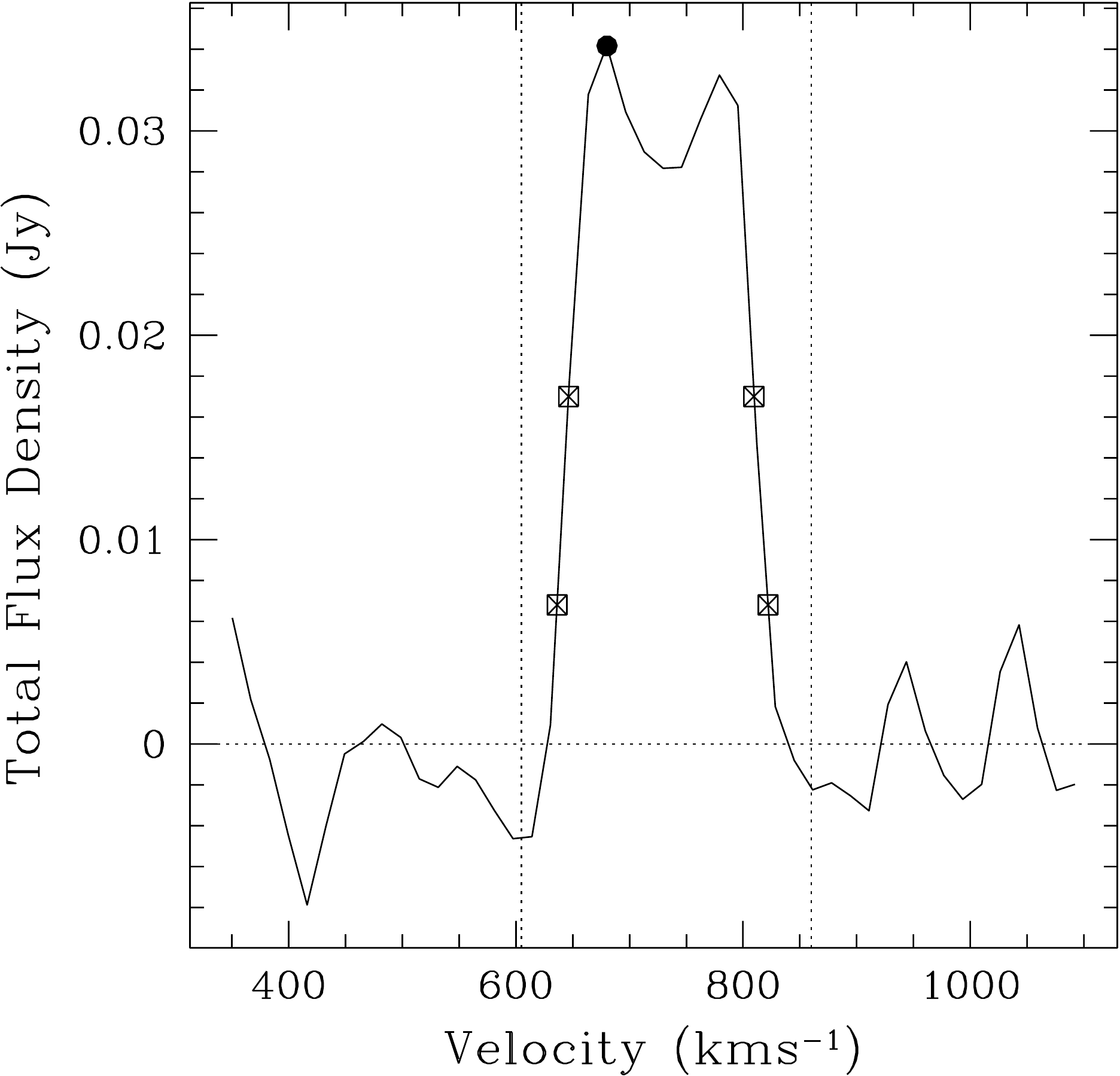}
\includegraphics[height=0.17\textheight]{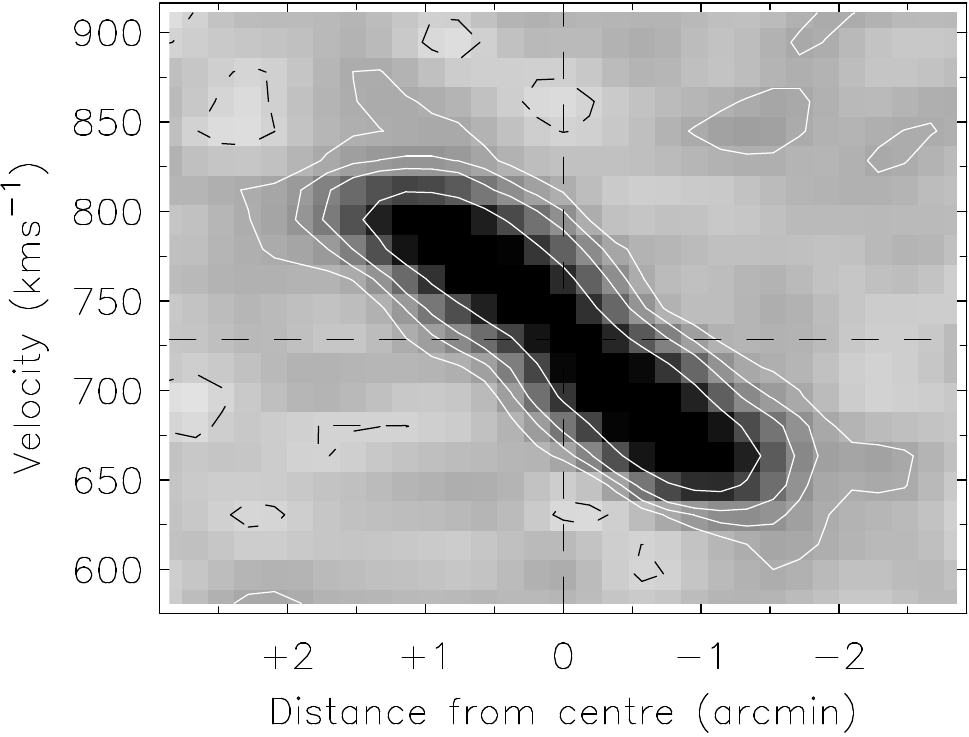}

\vskip 2mm
\centering
WSRT-CVn-60
\vskip 2mm
\includegraphics[width=0.25\textwidth]{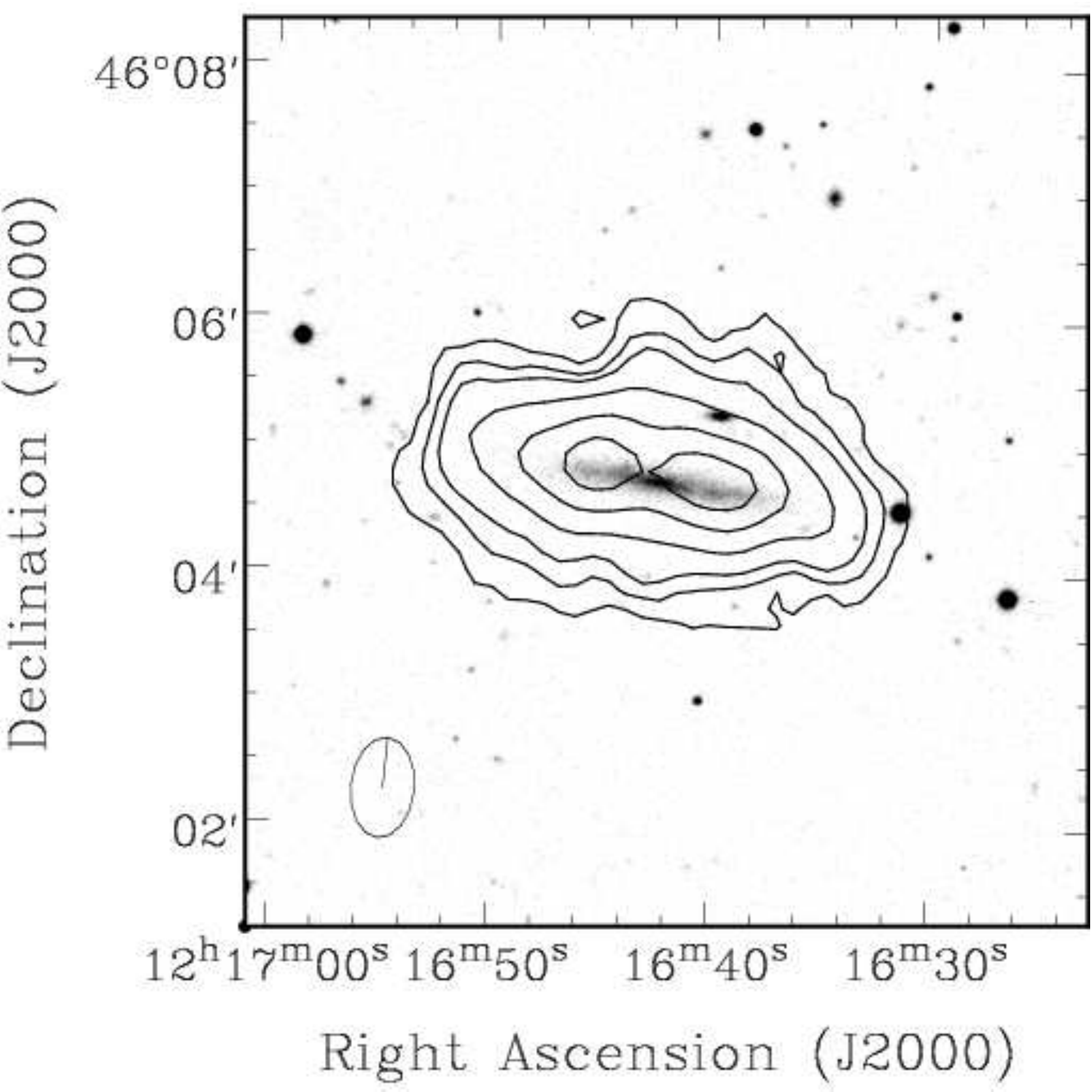}
\hskip 5mm
\includegraphics[height=0.17\textheight]{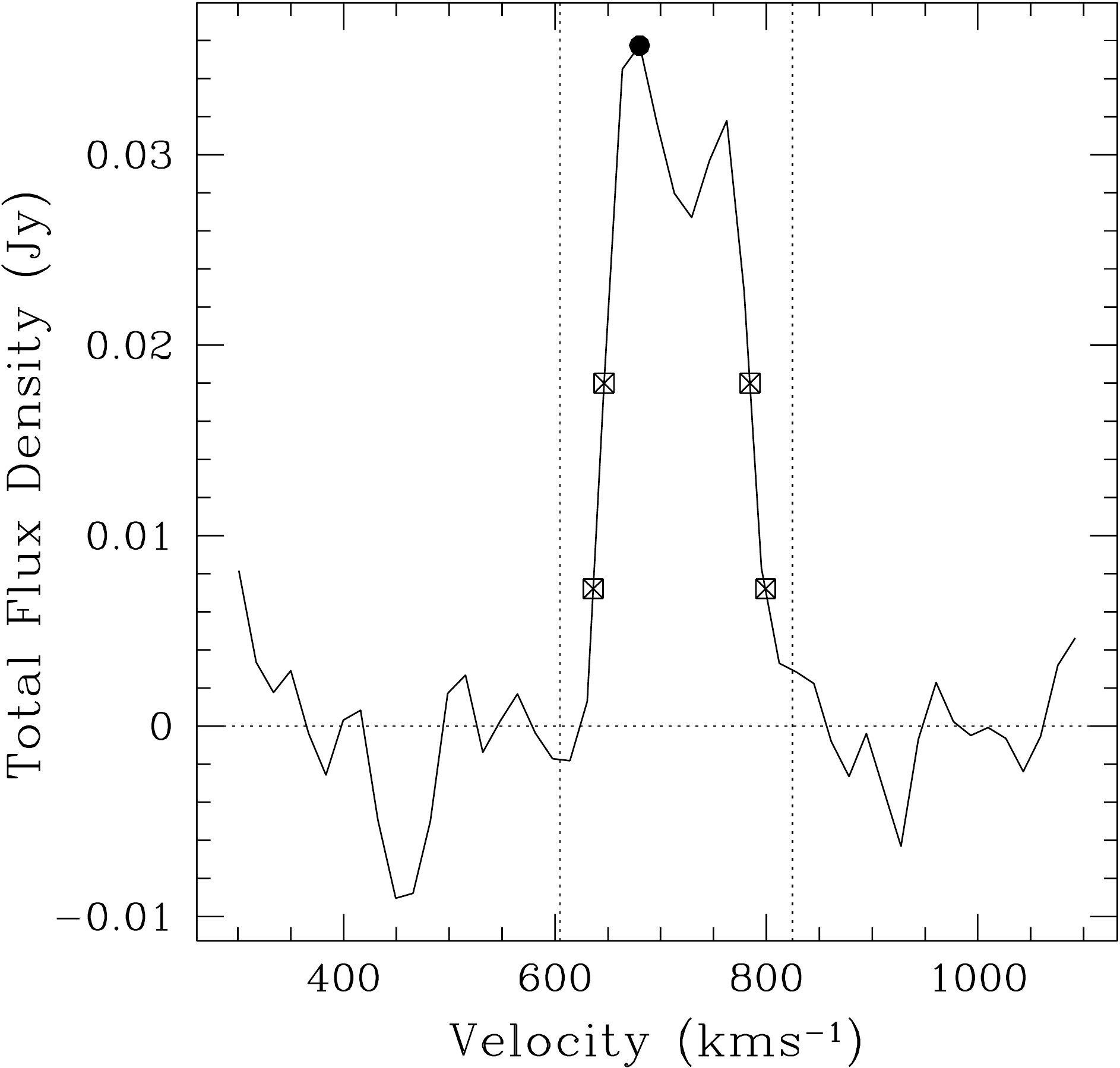}
\includegraphics[height=0.17\textheight]{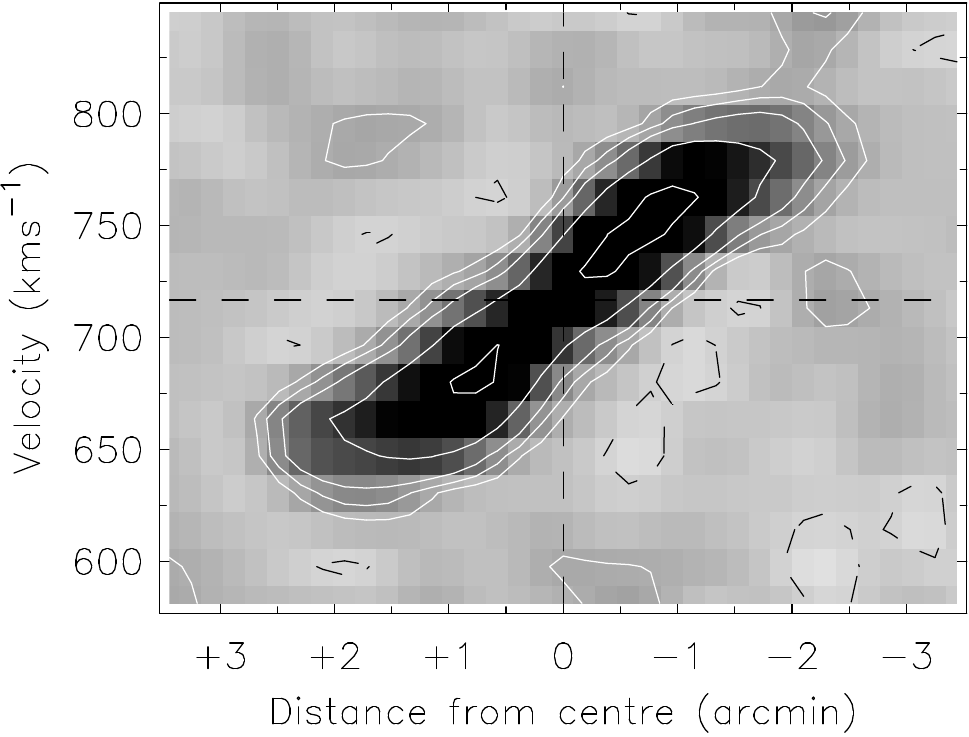}

\end{figure}

\clearpage

\addtocounter{figure}{-1}
\begin{figure}

\vskip 2mm
\centering
WSRT--CVn--61 and WSRT--CVn--62
\vskip 2mm
\begin{center}
\includegraphics[height=0.4\textheight]{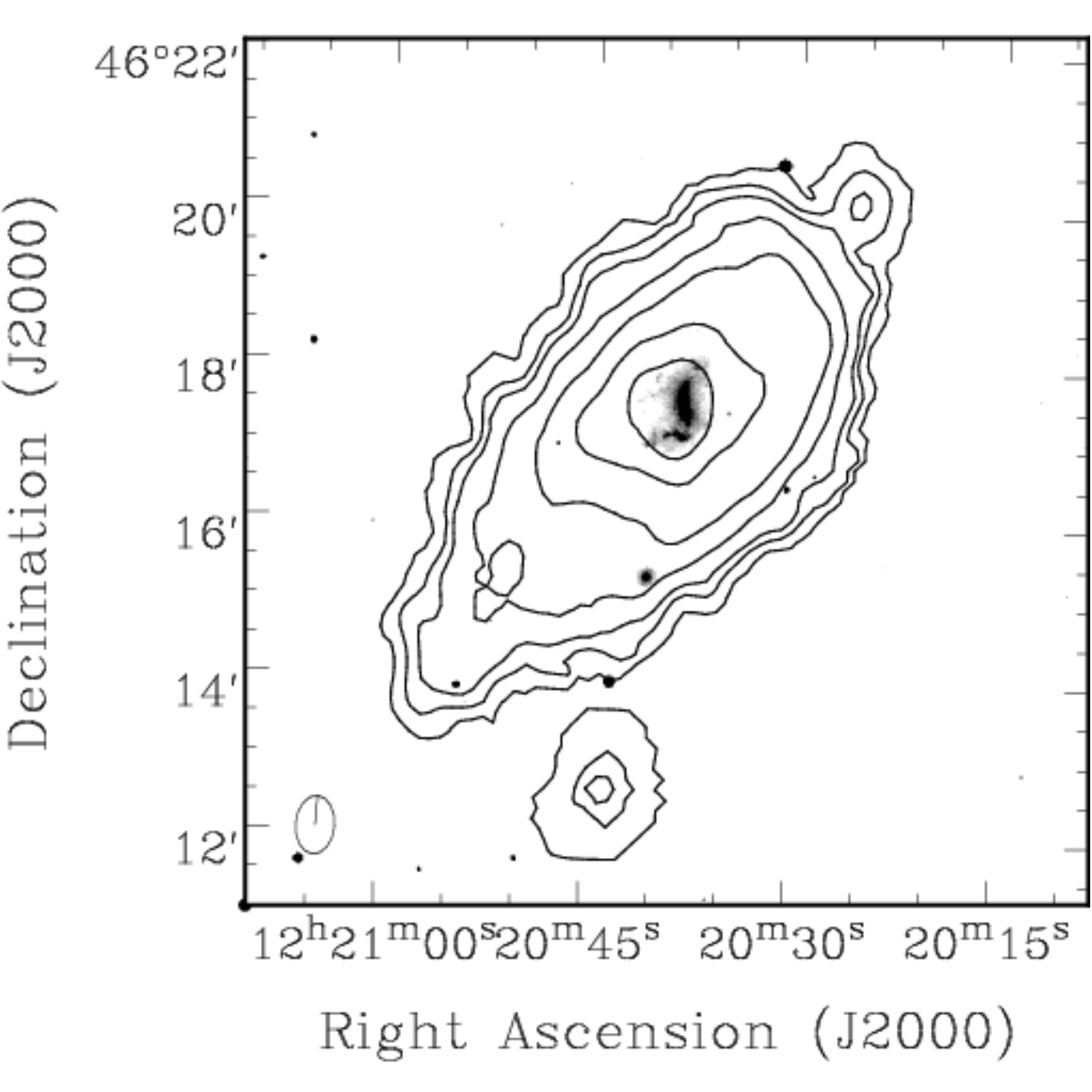}
\end{center}
\begin{center}
\includegraphics[height=0.21\textheight]{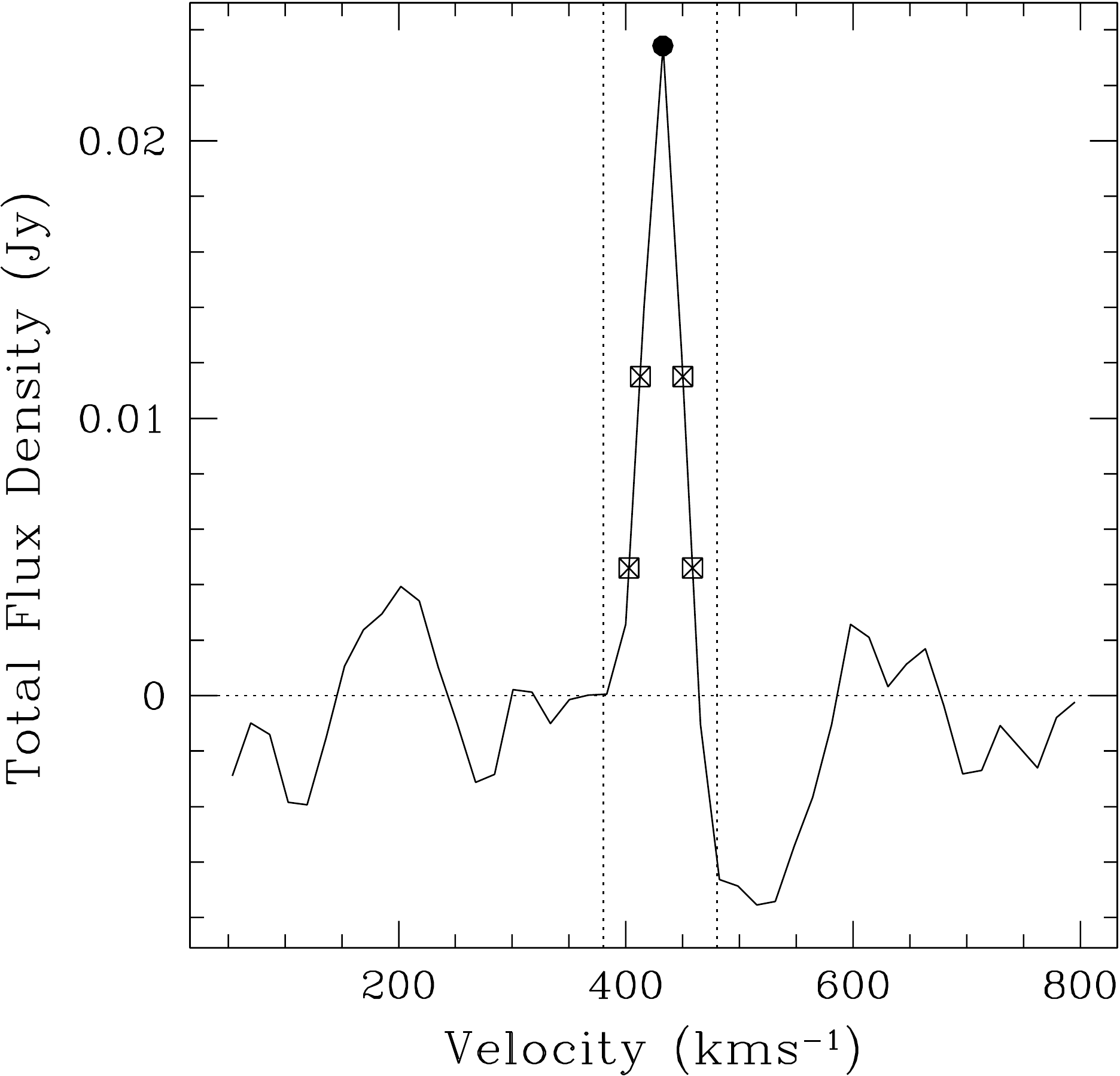}
\includegraphics[height=0.21\textheight]{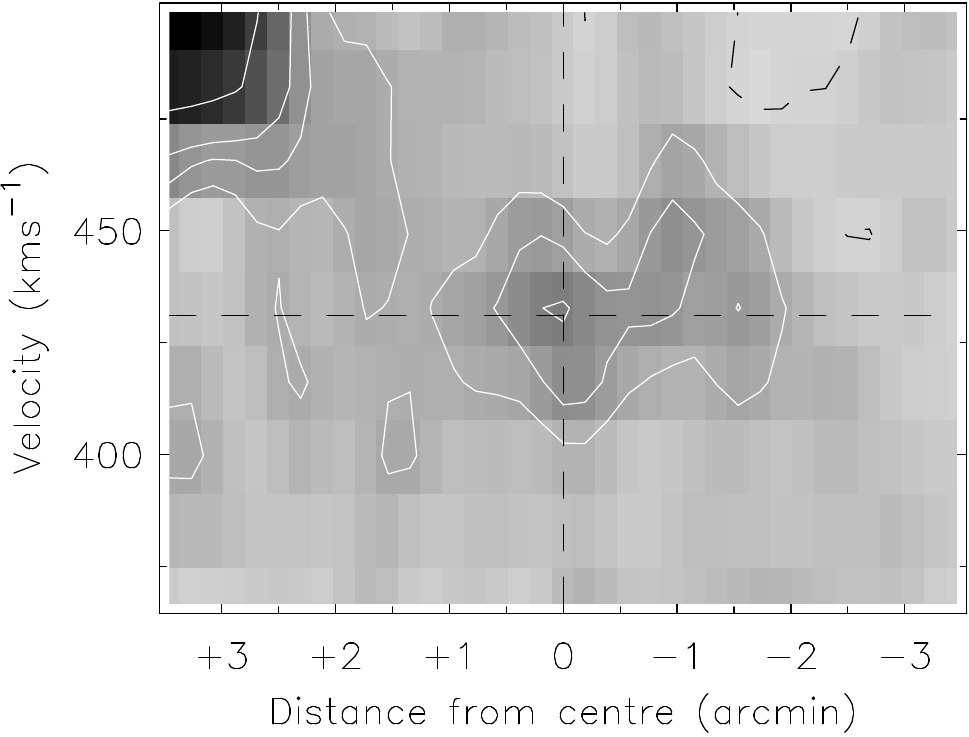}
\end{center}
\vskip 2mm
\begin{center}
\includegraphics[height=0.21\textheight]{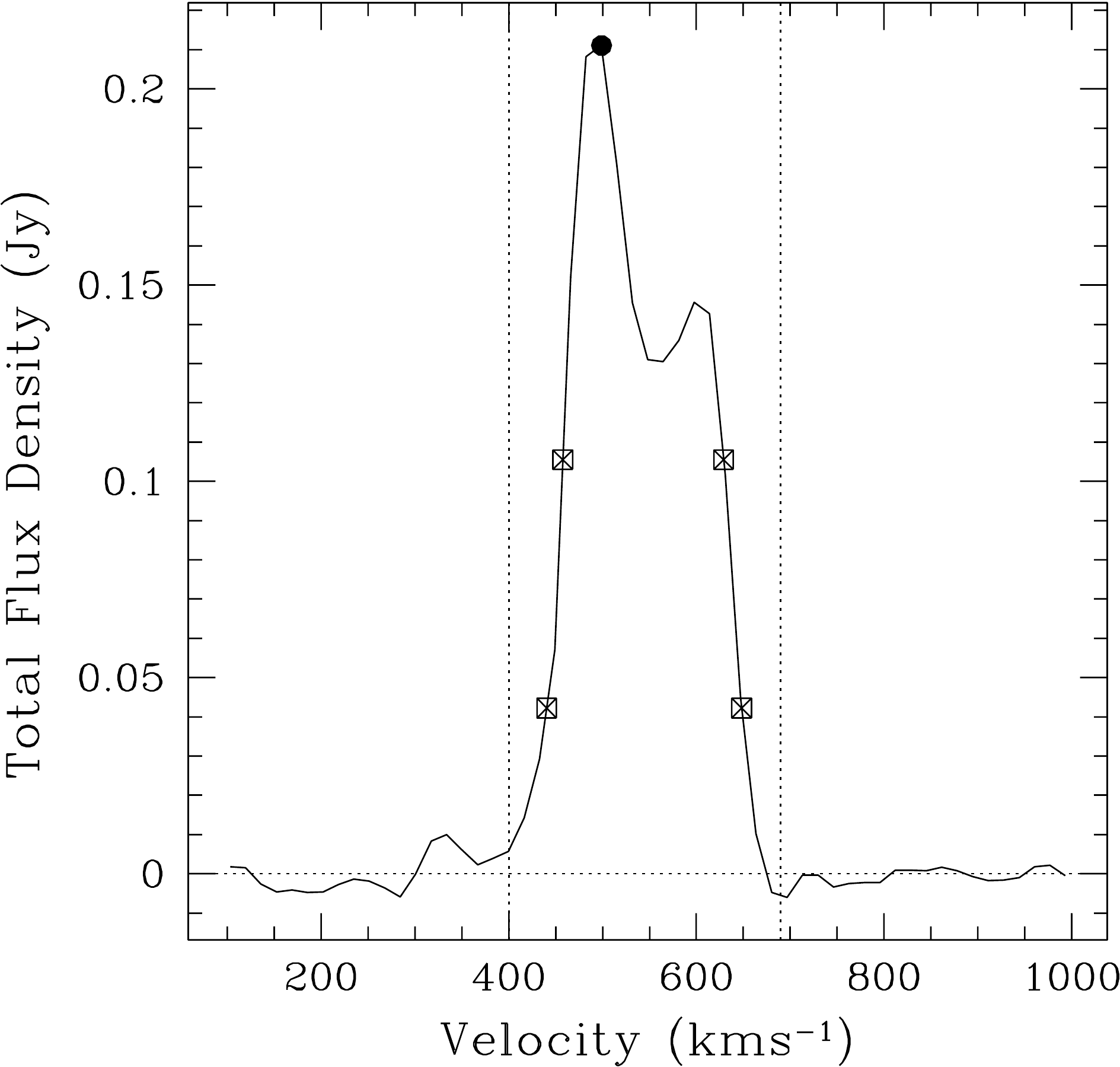}
\includegraphics[height=0.21\textheight]{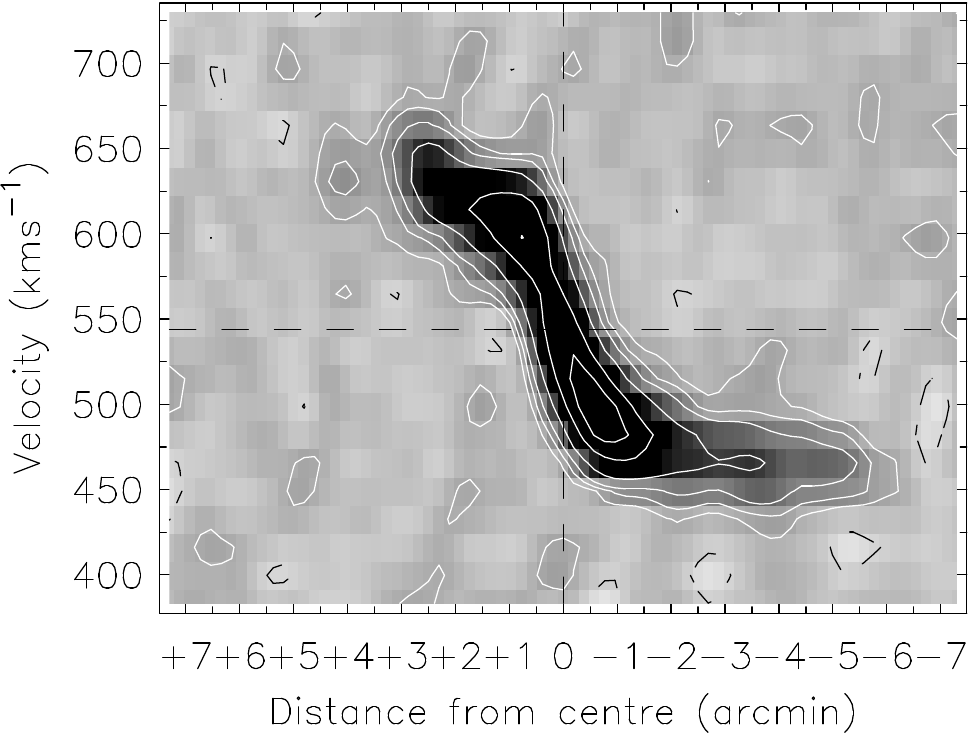}

\end{center}

\end{figure}

\clearpage

\addtocounter{figure}{-1}
\begin{figure}

\vskip 2mm
\centering
WSRT--CVn--63 and WSRT--CVn--64
\vskip 2mm
\begin{center}
\includegraphics[height=0.4\textheight]{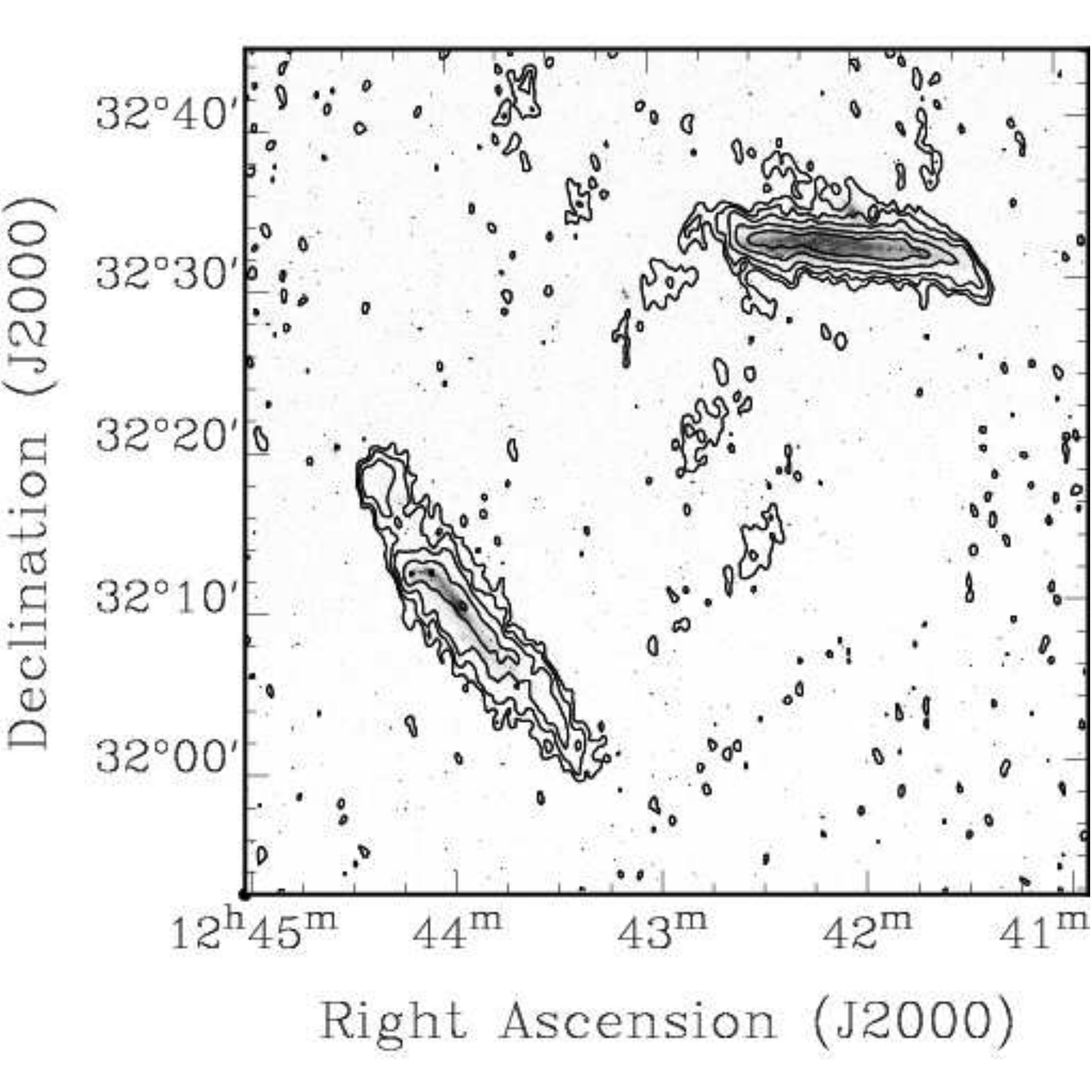}
\end{center}
\begin{center}
\includegraphics[height=0.21\textheight]{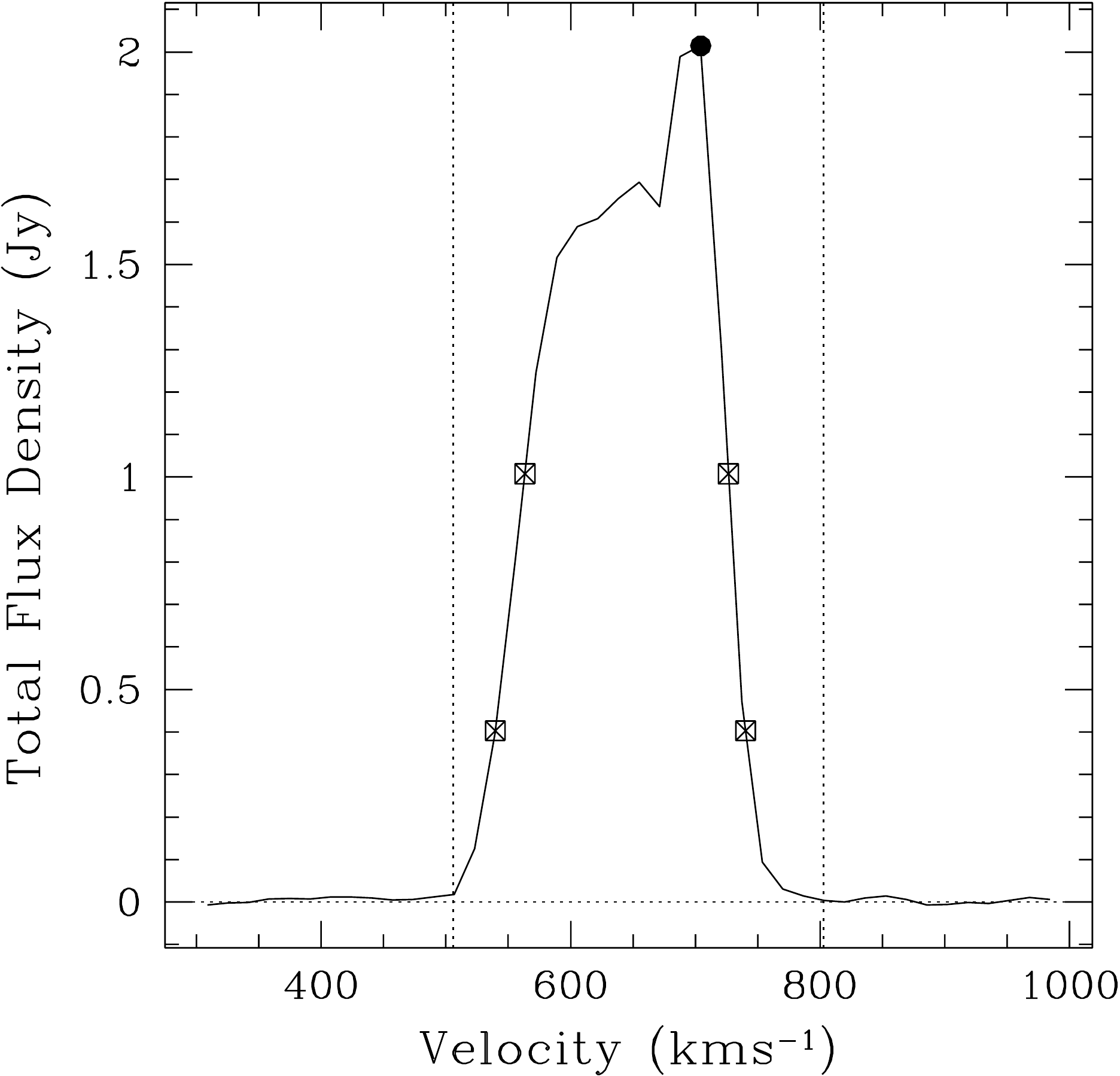}
\includegraphics[height=0.21\textheight]{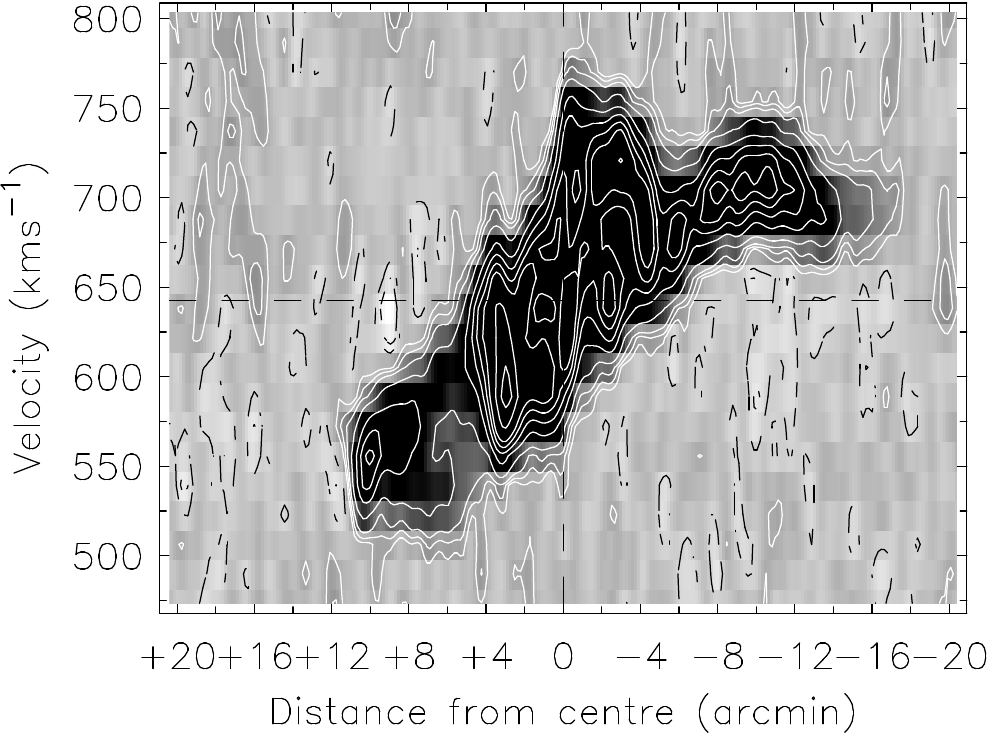}
\end{center}
\vskip 2mm
\begin{center}
\includegraphics[height=0.21\textheight]{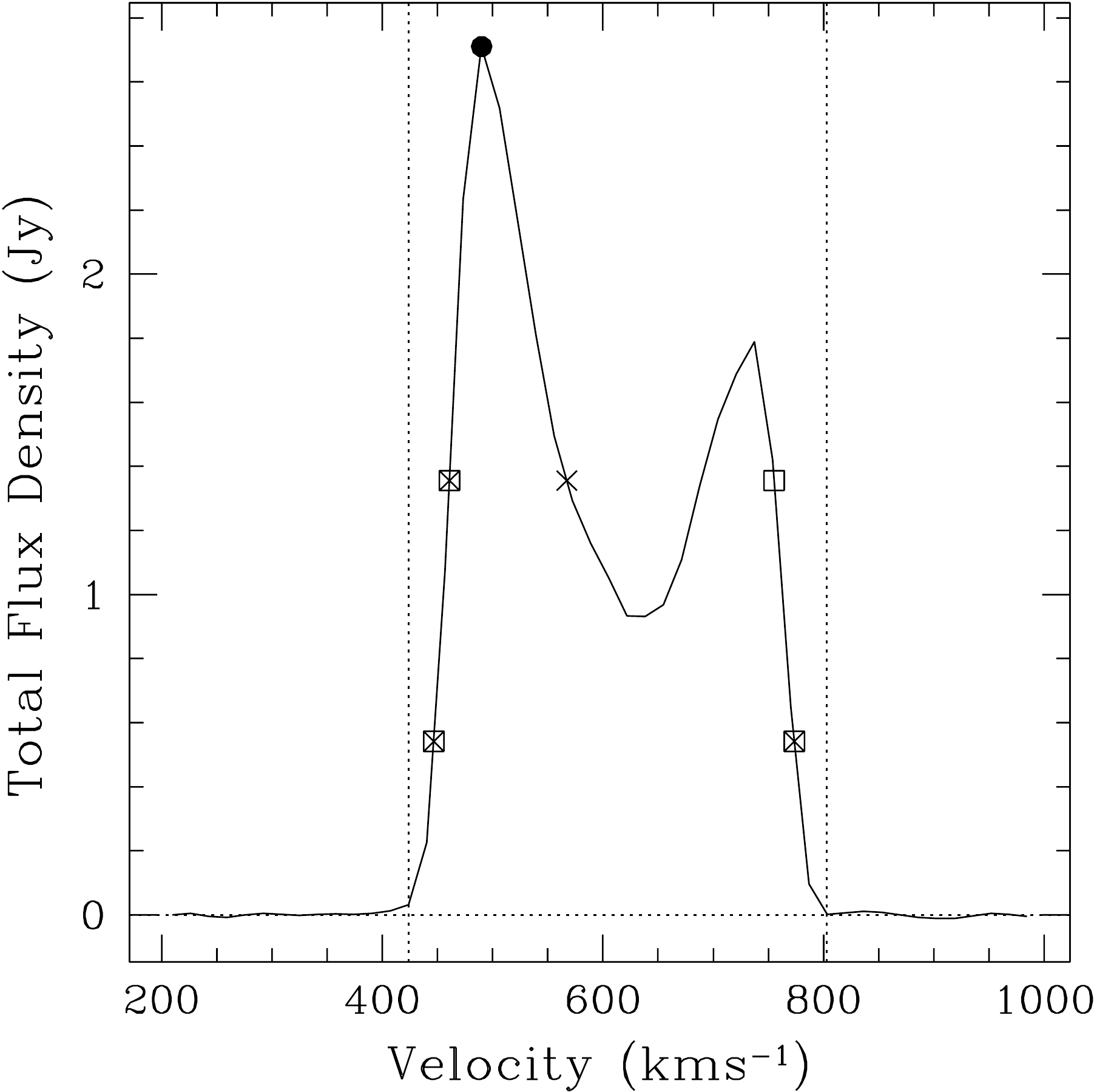}
\includegraphics[height=0.21\textheight]{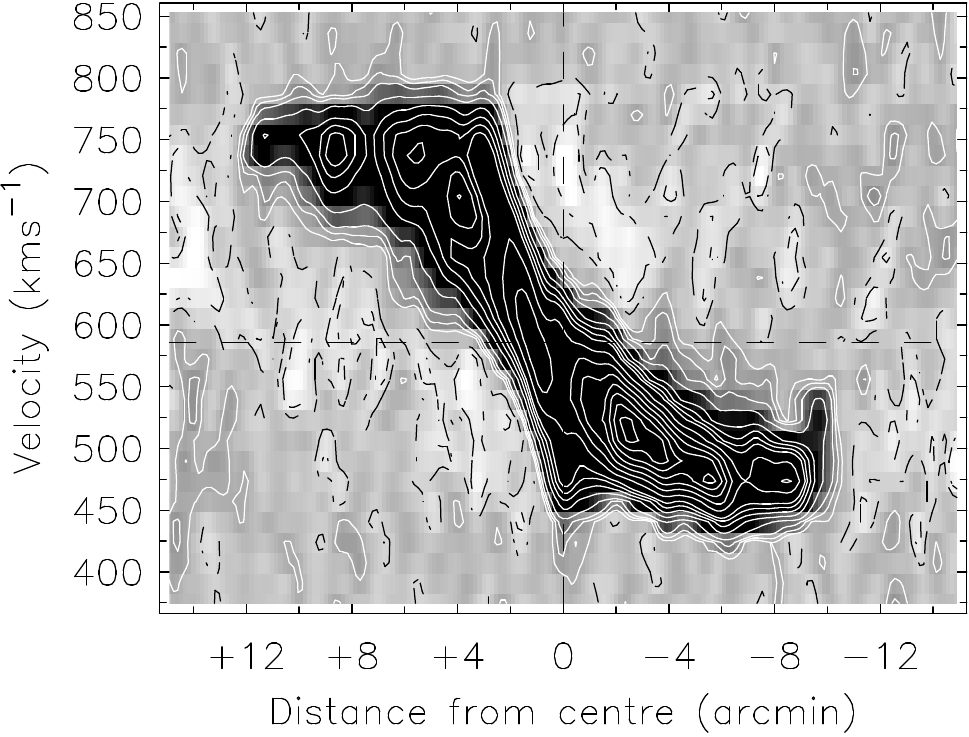}
\end{center}

\end{figure}

\clearpage

\addtocounter{figure}{-1}
\begin{figure}

\vskip 2mm
\centering
WSRT--CVn--65 and WSRT--CVn--66
\vskip 2mm
\begin{center}
\includegraphics[height=0.4\textheight]{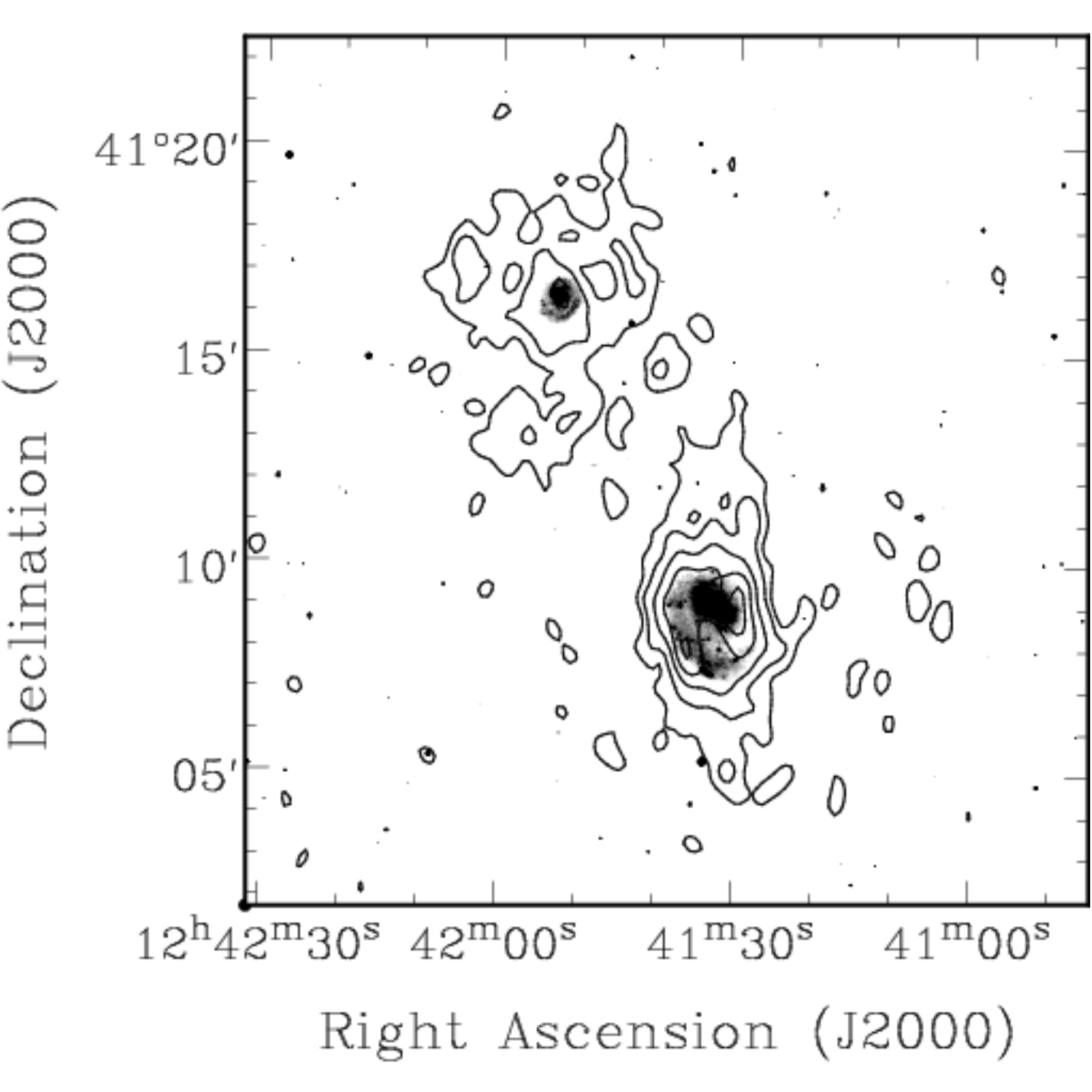}
\end{center}
\begin{center}
\includegraphics[height=0.21\textheight]{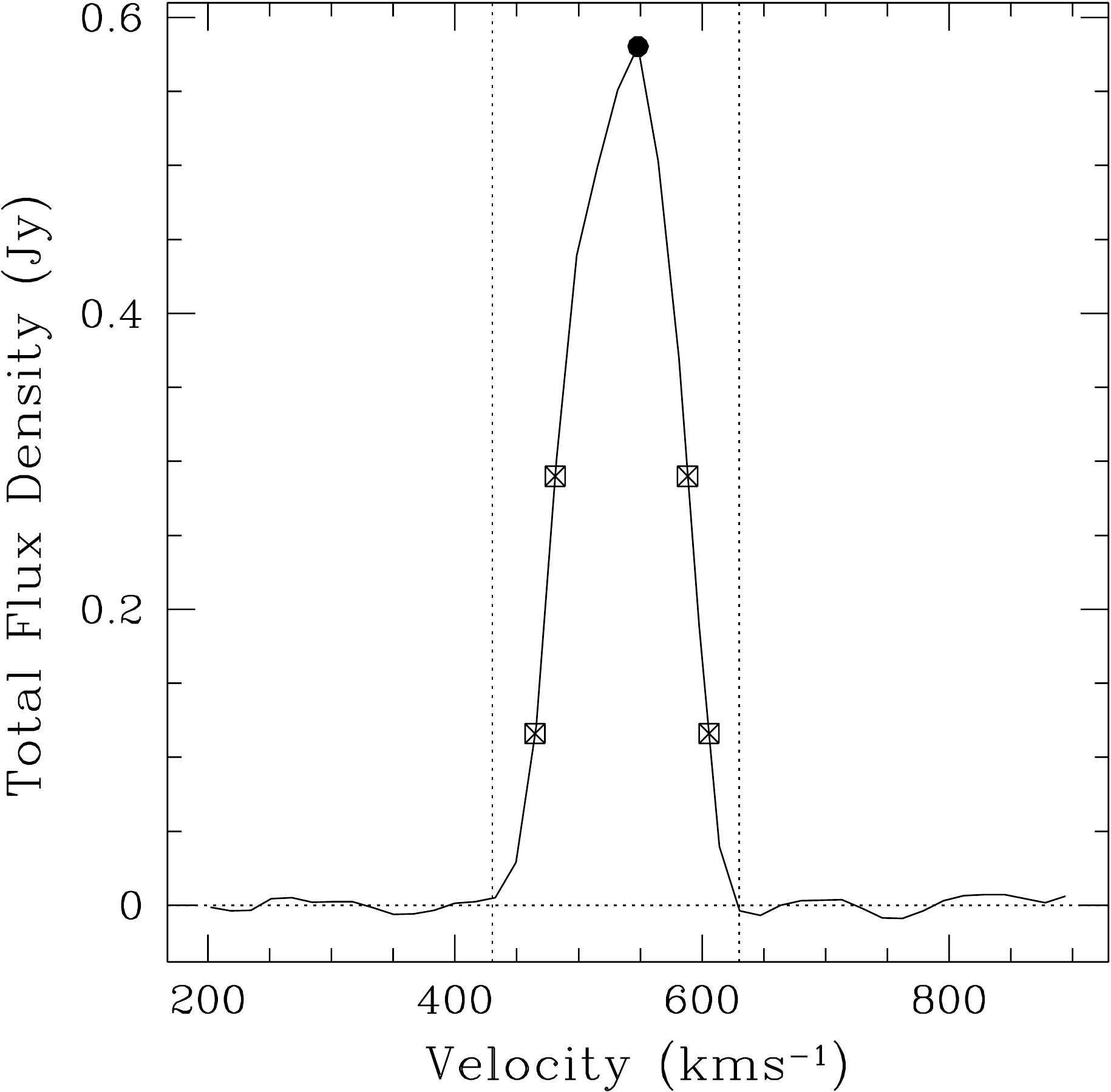}
\includegraphics[height=0.21\textheight]{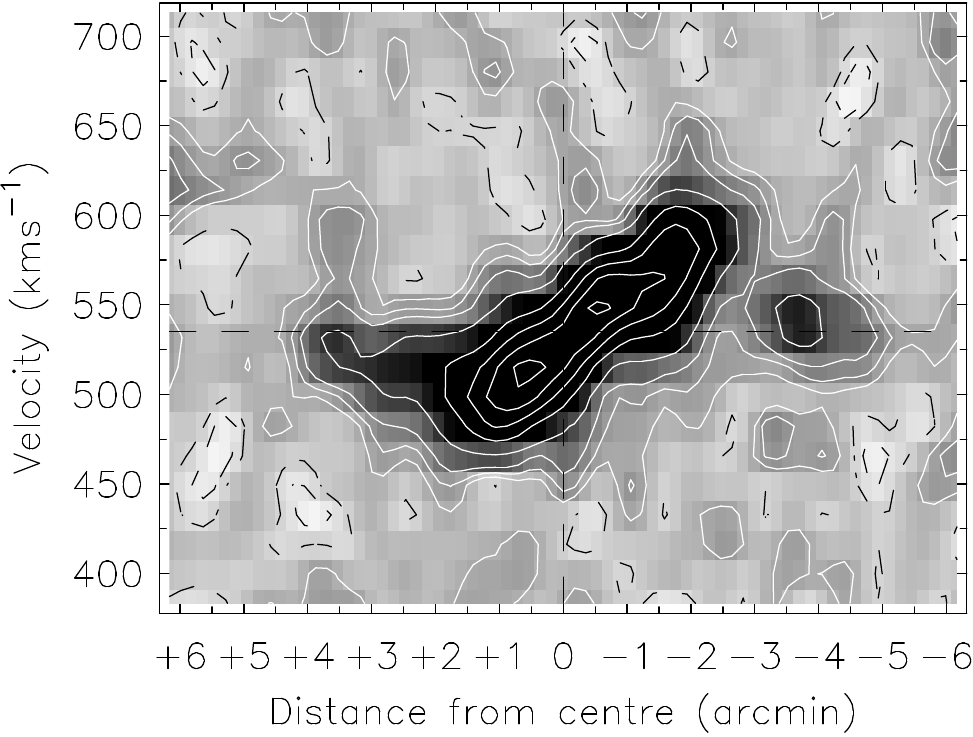}
\end{center}
\vskip 2mm
\begin{center}
\includegraphics[height=0.21\textheight]{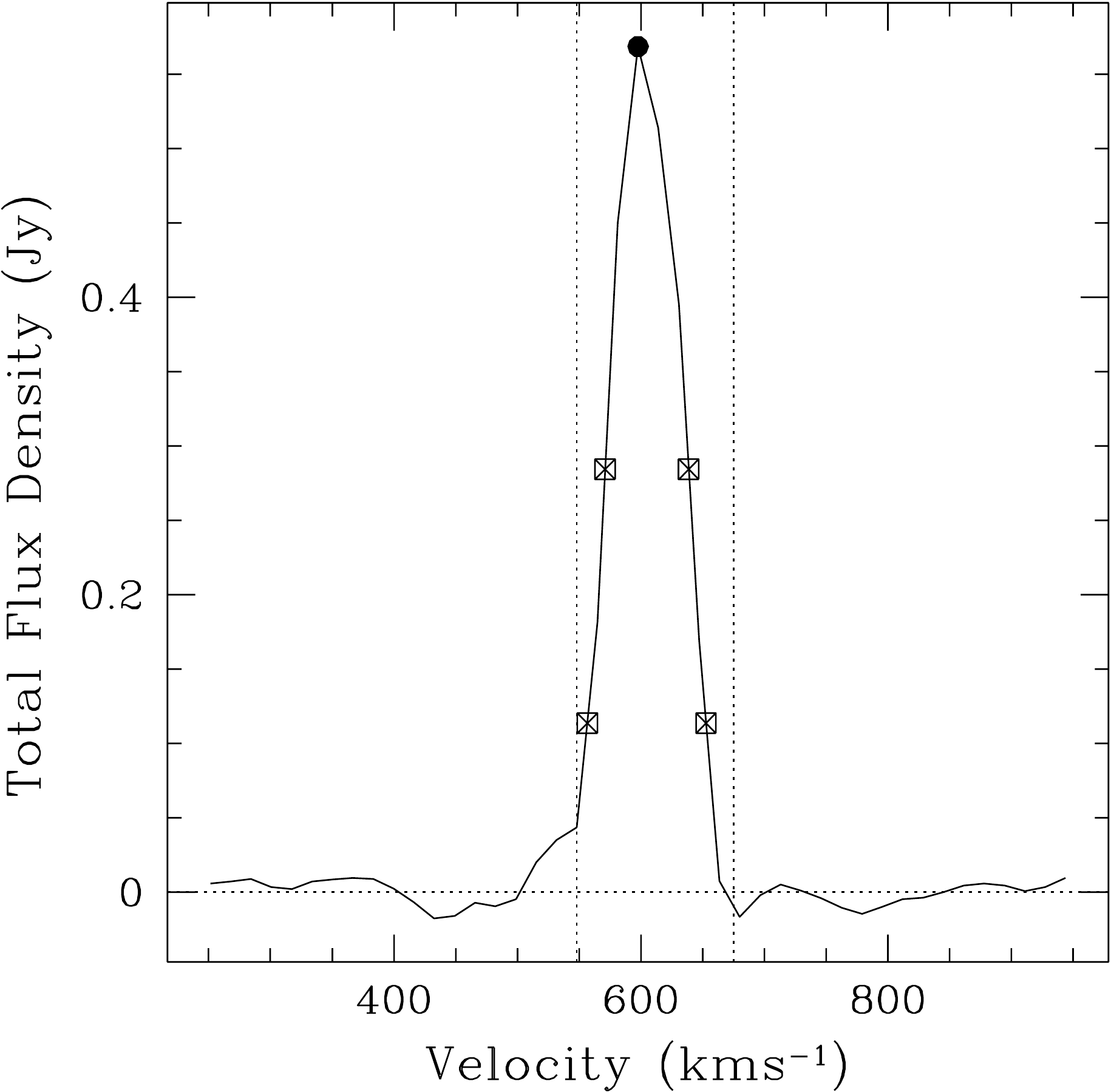}
\includegraphics[height=0.21\textheight]{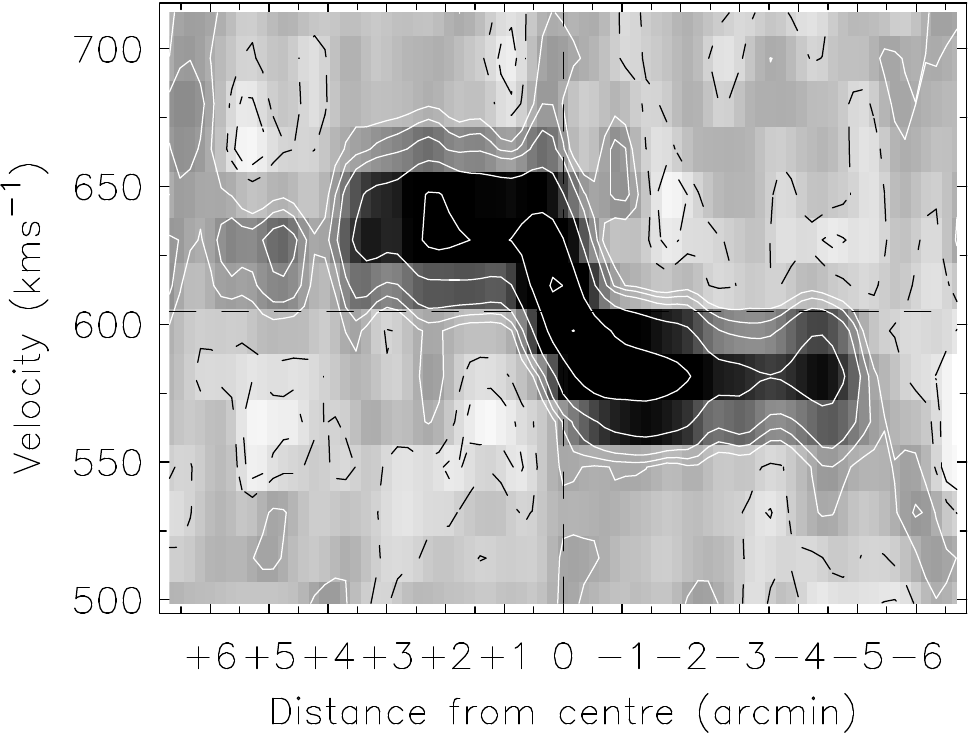}
\end{center}

\end{figure}

\clearpage

\addtocounter{figure}{-1}
\begin{figure}

\vskip 2mm
\centering
WSRT-CVn-67A and WSRT-CVn-67B
\vskip 2mm
\begin{center}
\includegraphics[height=0.4\textheight]{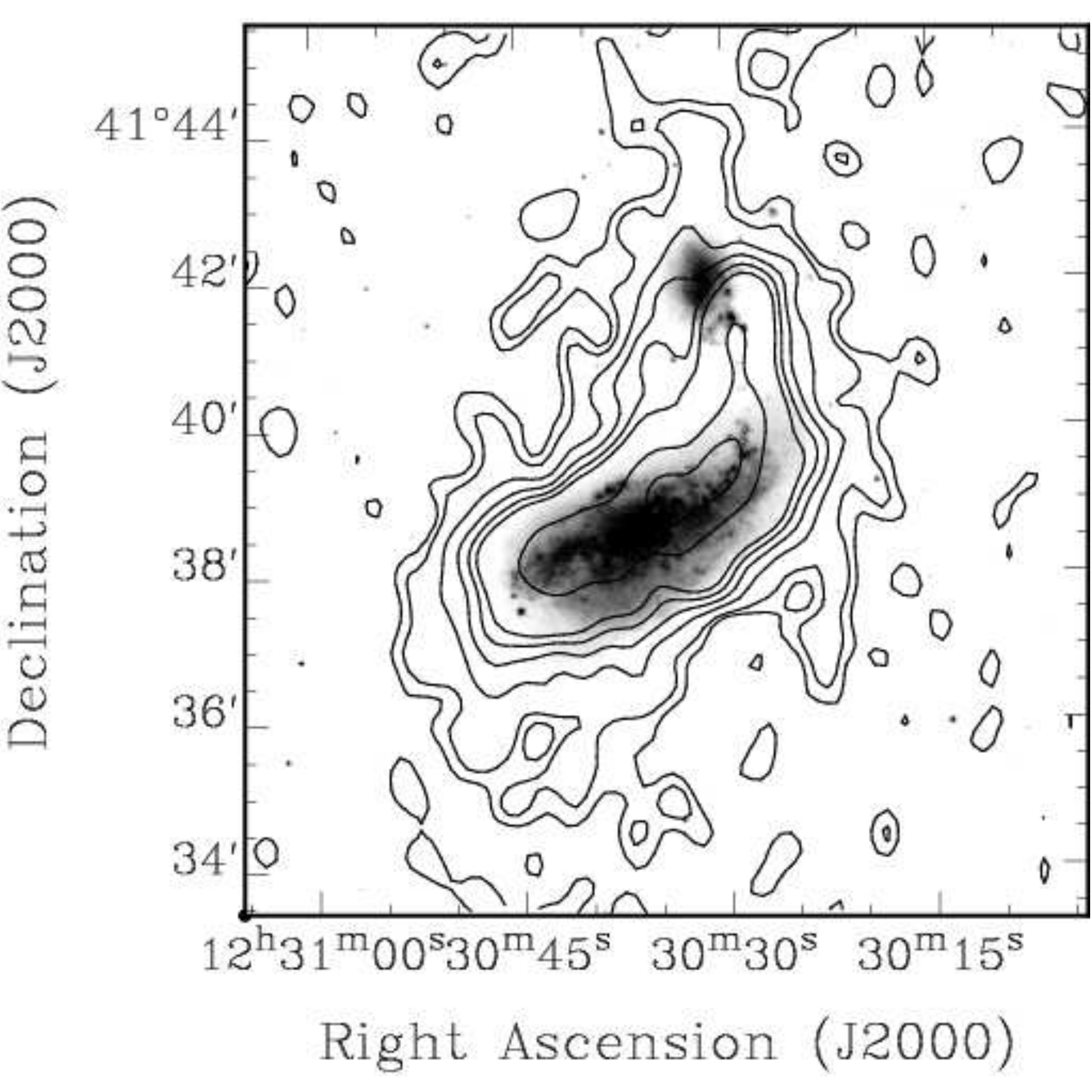}
\end{center}
\begin{center}
\includegraphics[height=0.21\textheight]{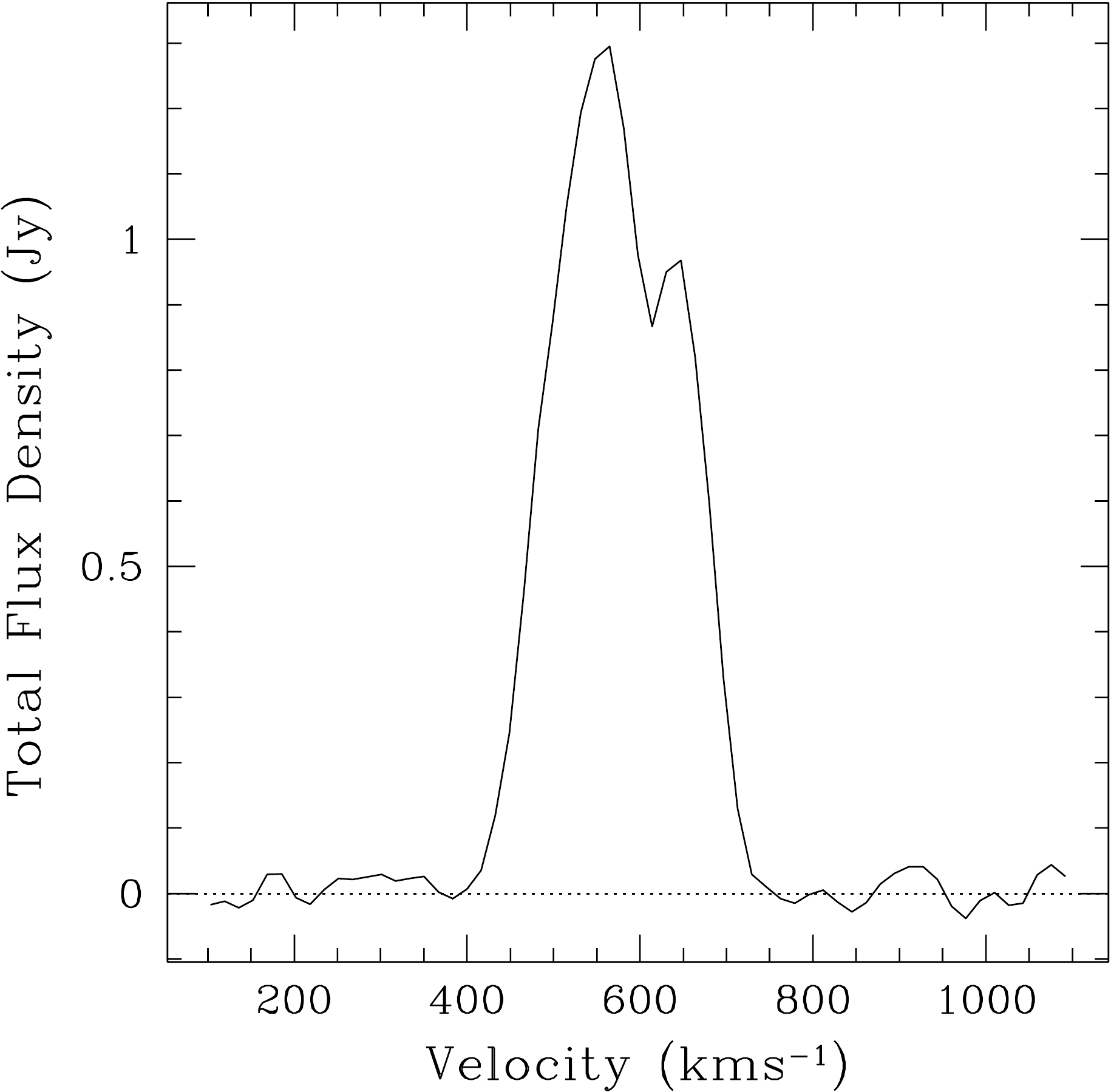}
\includegraphics[height=0.21\textheight]{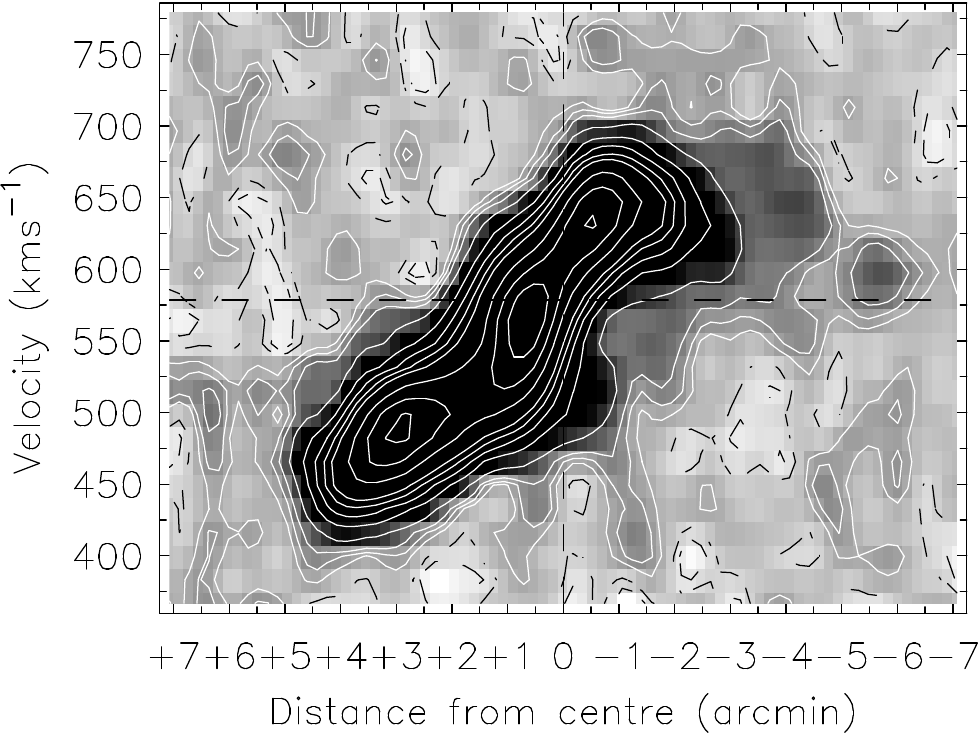}
\end{center}
\vskip 2mm
\begin{center}
\includegraphics[height=0.21\textheight]{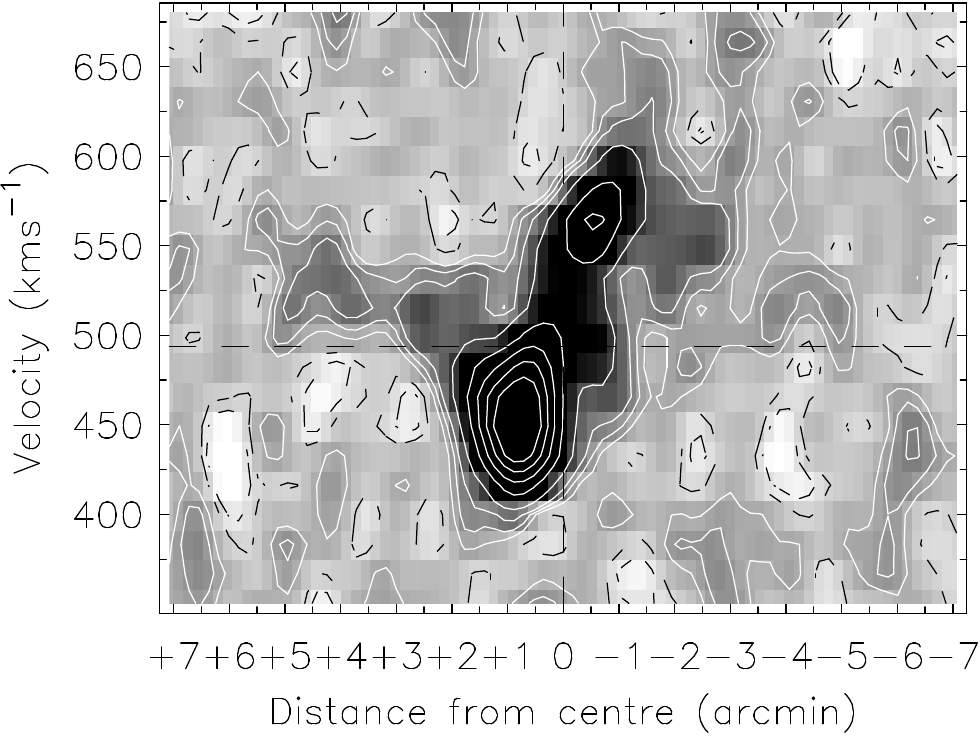}
\end{center}

\end{figure}

\clearpage

\addtocounter{figure}{-1}
\begin{figure}

\vskip 2mm
\centering
WSRT-CVn-68
\vskip 2mm
\begin{center}
\includegraphics[height=0.4\textheight]{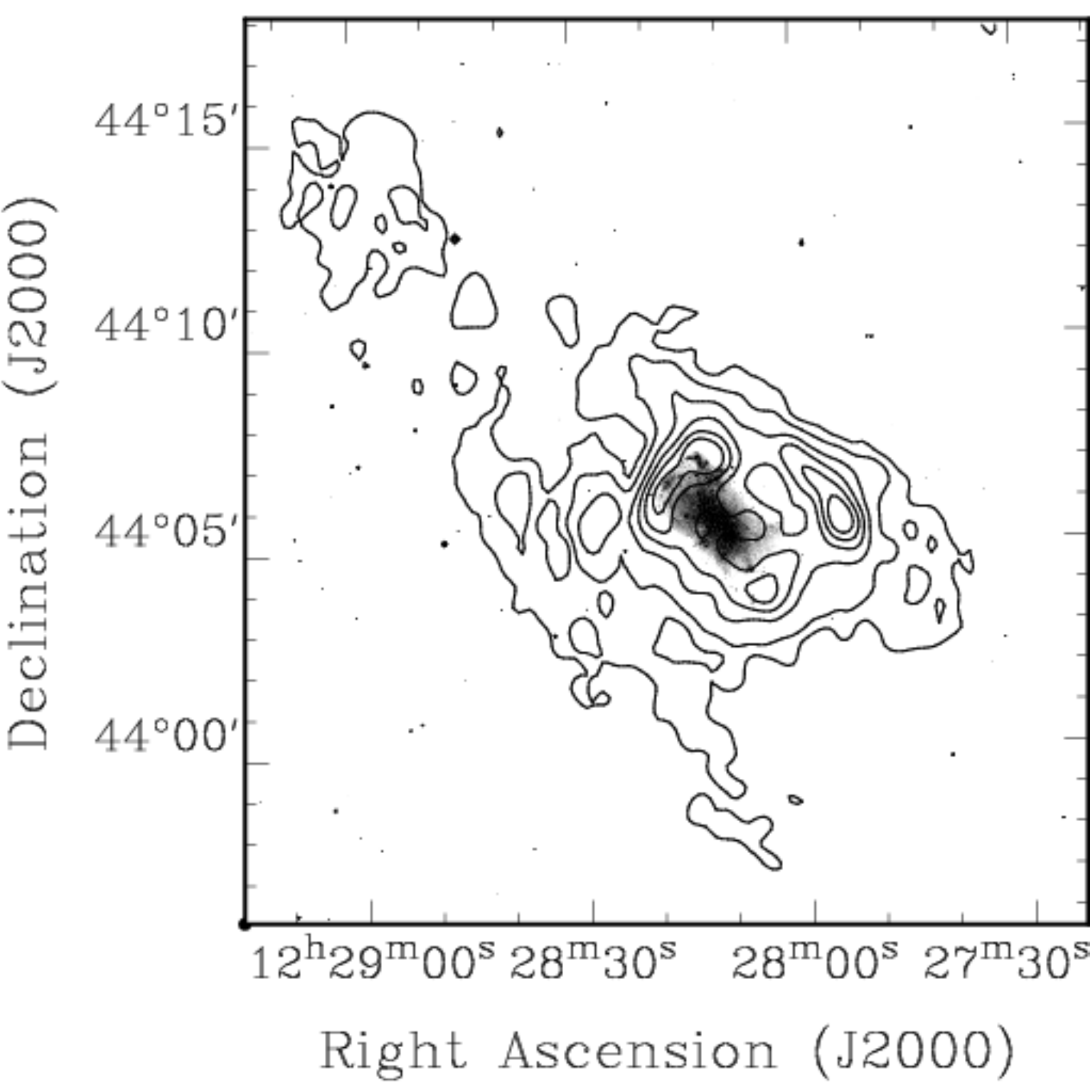}
\end{center}
\begin{center}
\includegraphics[height=0.21\textheight]{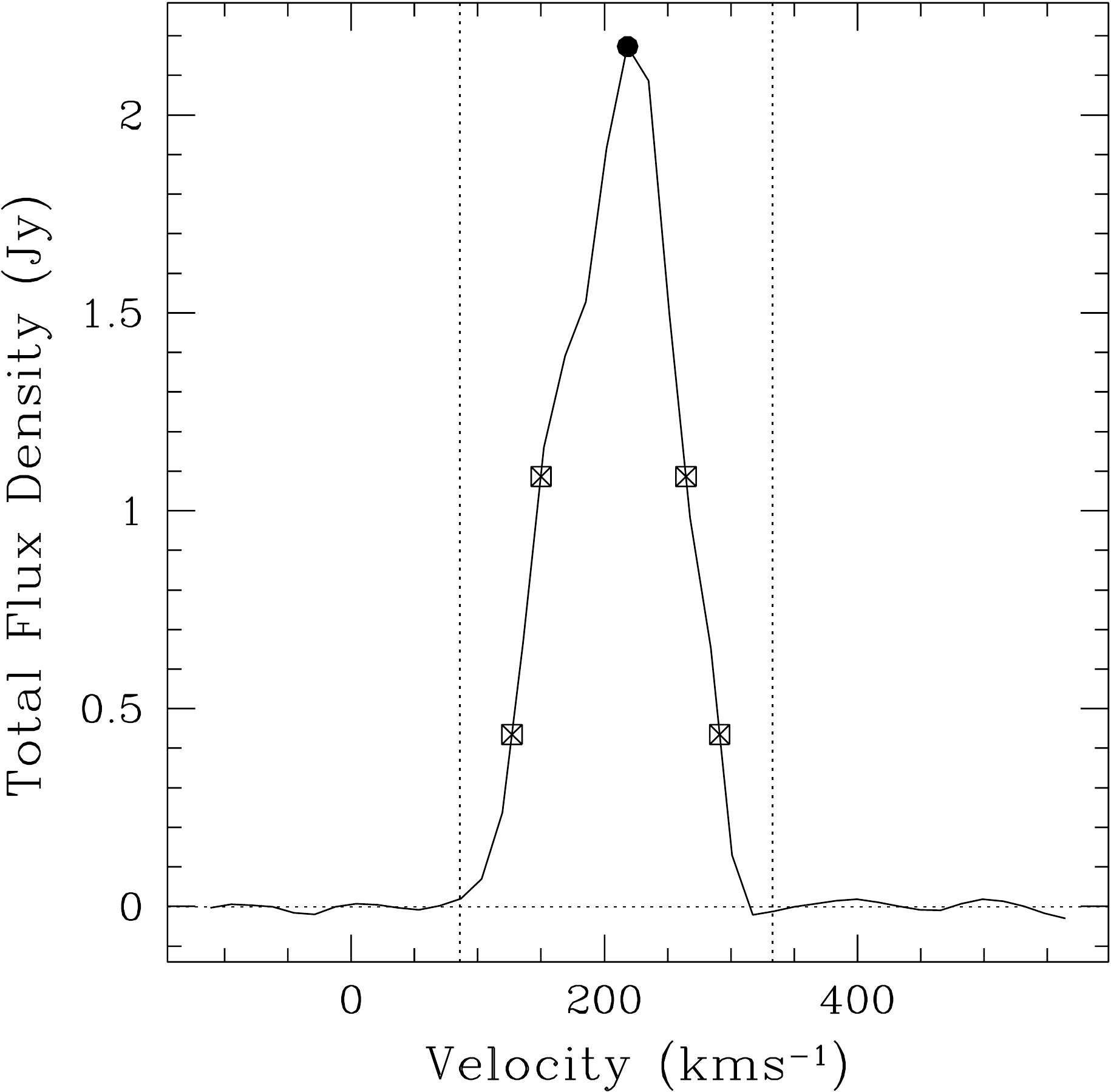}
\includegraphics[height=0.21\textheight]{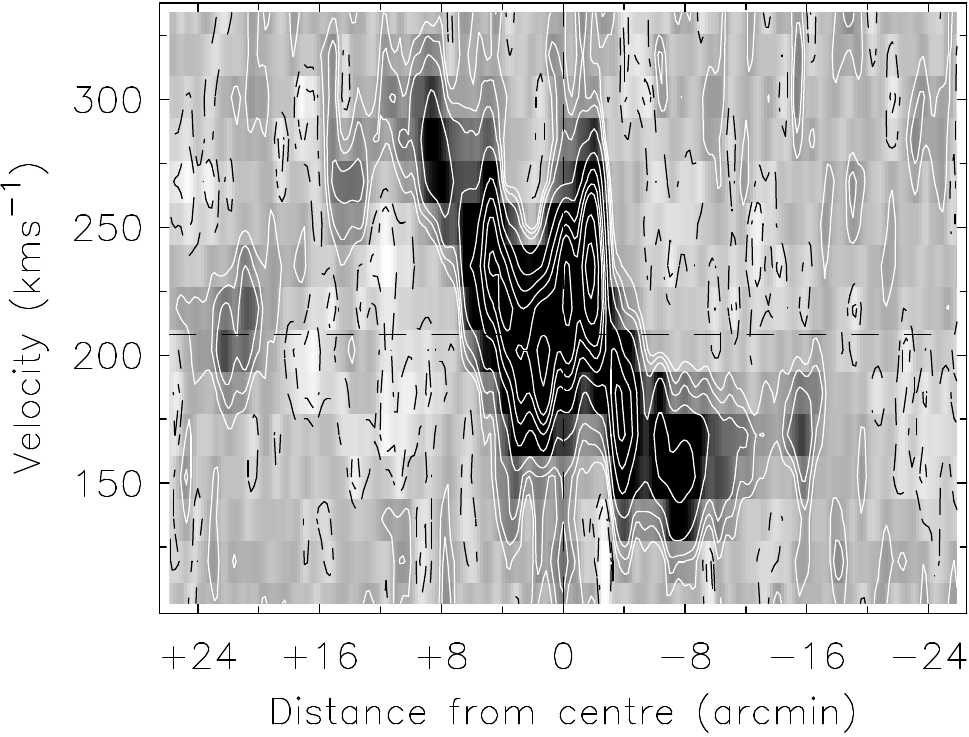}
\end{center}
\vskip 2mm
\begin{center}
\includegraphics[height=0.21\textheight]{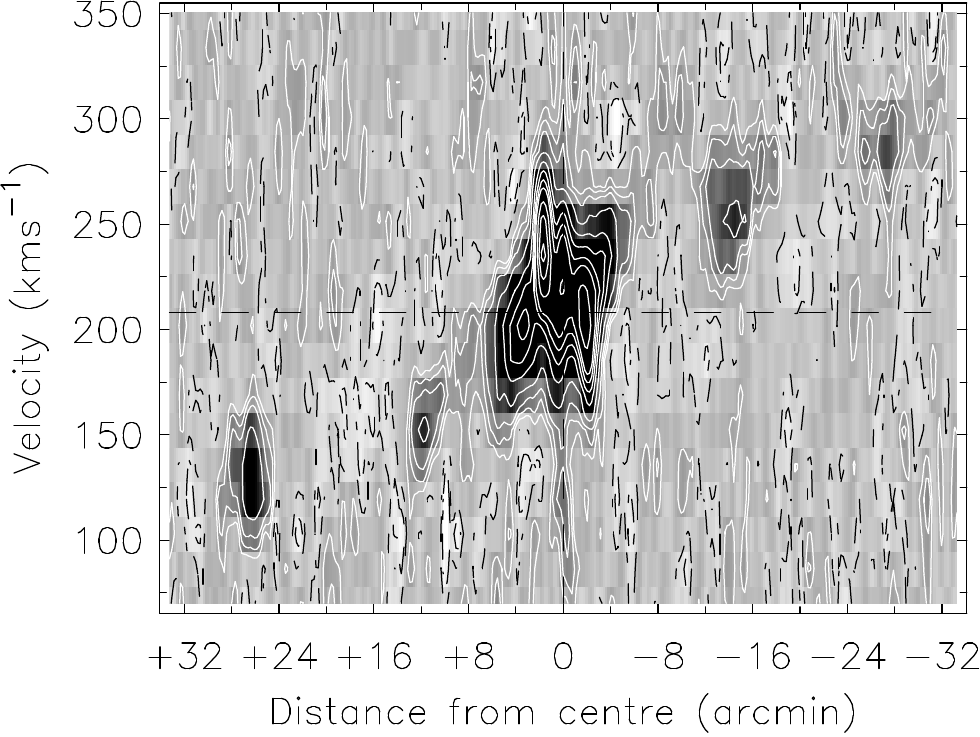}
\end{center}

\end{figure}

\clearpage

\addtocounter{figure}{-1}
\begin{figure}

\vskip 2mm
\centering
WSRT-CVn-69
\vskip 2mm
\includegraphics[width=0.25\textwidth]{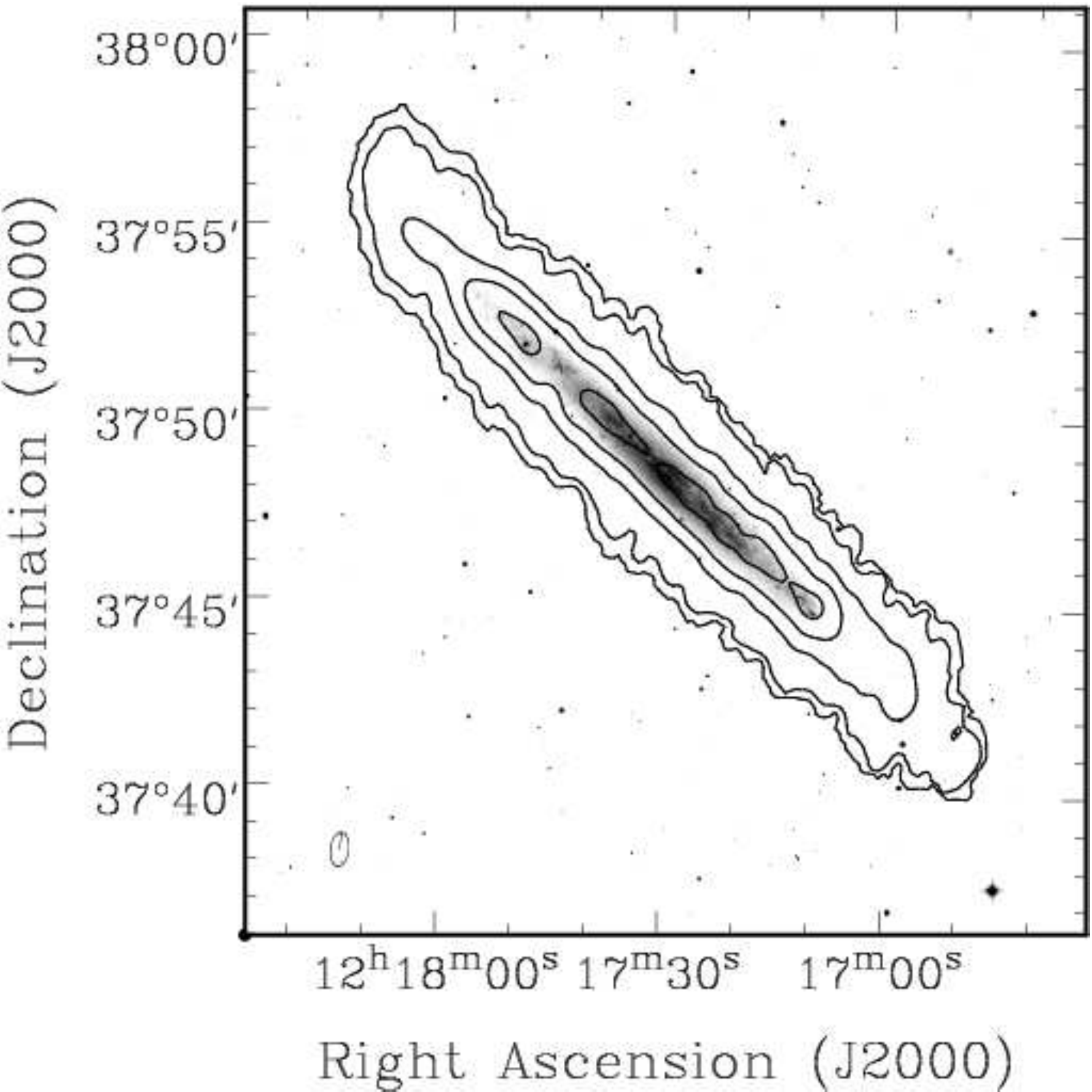}
\hskip 5mm
\includegraphics[height=0.17\textheight]{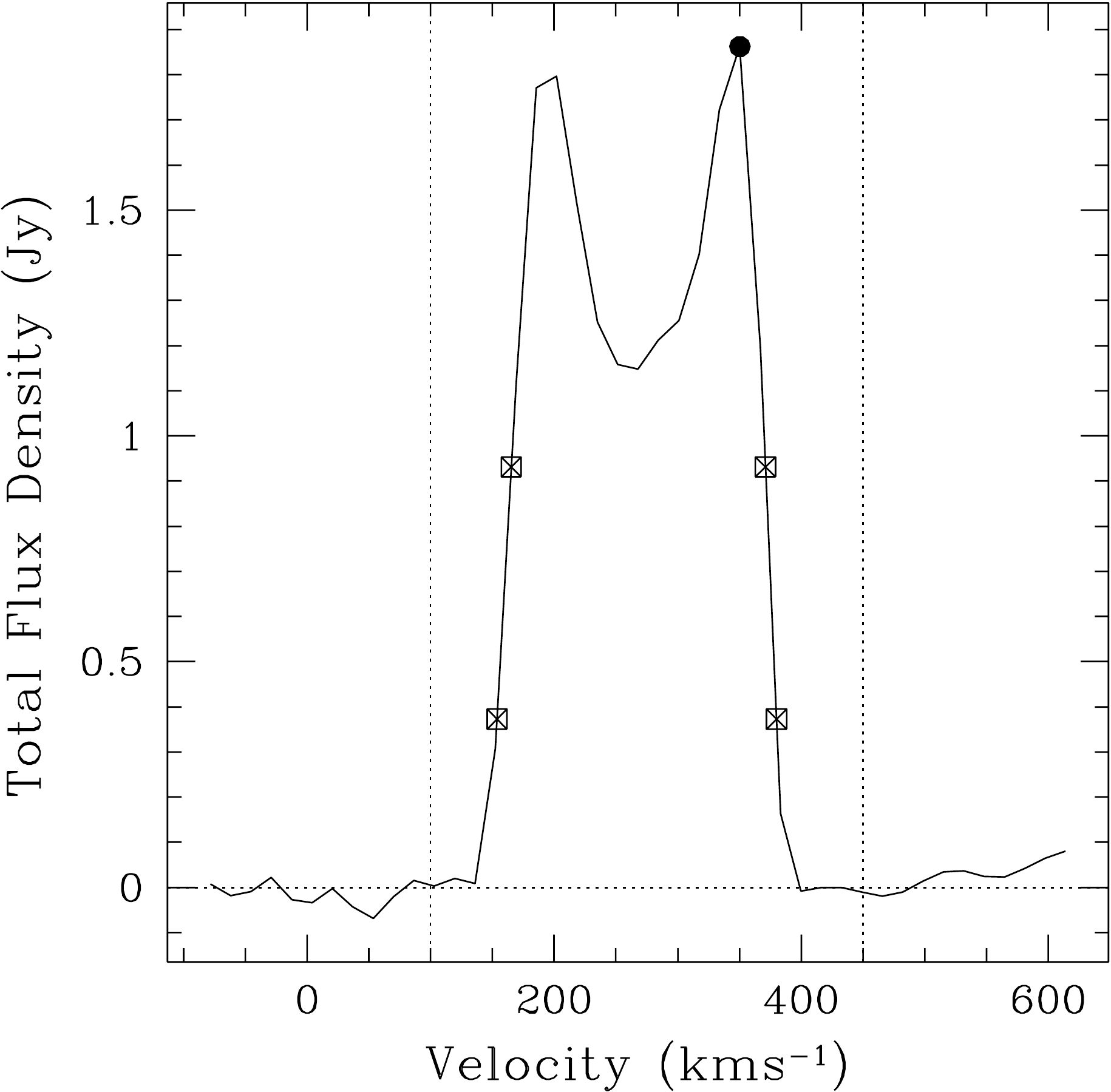}
\includegraphics[height=0.17\textheight]{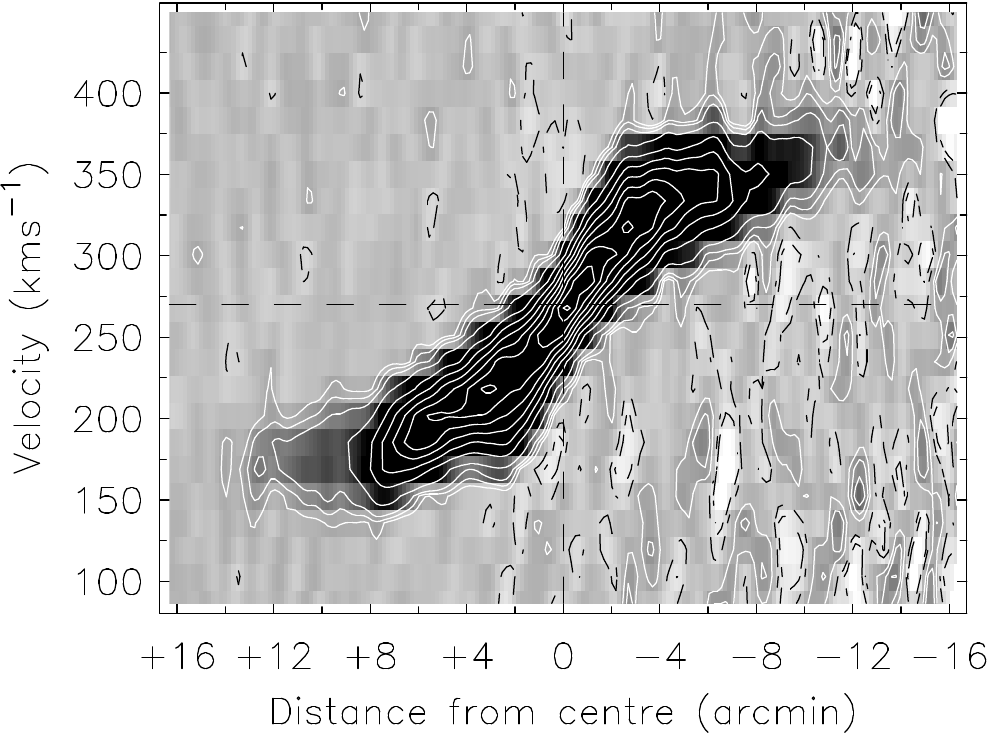}
\end{figure}

\bsp

\label{lastpage}

\end{document}